\documentclass[11pt,a4paper]{article}
\pdfoutput=1
\usepackage{subfiles}
\usepackage{jinstpub}
\usepackage[utf8]{inputenc}
\usepackage[english]{babel}
\usepackage{graphicx}
\usepackage[dvipsnames]{xcolor} 
\graphicspath{ {Figures/} }
\usepackage[permil]{overpic} % overlay pictures.
\usepackage{amsmath}
\usepackage{multirow}
\usepackage{tikz}
\usepackage{subfigure}
\usepackage{rotating}
\usepackage{tabularx}
\usepackage{multibib}
\newcites{Company}{Company references}

\usepackage{soul}% for underlines

\usepackage{array}
\usepackage{booktabs, multirow} % for borders and merged ranges
\usepackage{gensymb} % provide some additional symbols

\usepackage{lineno}
\nolinenumbers
\usepackage[toc,acronym]{glossaries}
\makeglossaries

\usepackage[inline]{enumitem} 
\setlist[itemize]{noitemsep}
\setlist[enumerate]{noitemsep}

\usepackage{siunitx}
\newcommand{\lvlonesection}[1]{\section{#1}}
\newcommand{\lvltwosection}[1]{\subsection{#1}}
\newcommand{\lvlthreesection}[1]{\subsubsection{#1}}
\newcommand{\editor}[1]{\relax}    % Remove all editor indications
\newcommand{\inlinelist}[1]{\emph{(#1)}~}
\input{svd-sym.tex}
\newacronym{label}{acronym}{Long form}

\newacronym{ADC}{ADC}{Analog to Digital Converter}
\newacronym{ASIC}{ASIC}{Application-specific integrated circuit}
\newacronym{basf}{\basf}{Belle II Analysis Software Framework}
\newacronym{BEAST}{BEAST}{Beam Exorcism for A STable Belle II experiment}
\newacronym{CCD}{CCD}{Charge-Coupled Device}
\newacronym{CDC}{CDC}{Central Drift Chamber}
\newacronym{CTE}{CTE}{Coefficient of Thermal Expansion}
\newacronym{CFRP}{CFRP}{Carbon Fiber Reinforced Plastic}
\newacronym{CMM}{CMM}{Coordinate Measurement Machine}
\newacronym{CMS}{CMS}{Compact Muon Solenoid experiment}
\newacronym{CMC}{CMC}{Common Mode Correction}
\newacronym{COPPER}{COPPER}{COmmon Pipelined Platform for Electronics Readout}
\newacronym{CVD}{CVD}{Chemical Vapour Deposition}
\newacronym{DAC}{DAC}{Digital to Analog Converter}
\newacronym{DAQ}{DAQ}{Data Acquisition}
\newacronym{DATCON}{DATCON}{DATa CONcentrator}
\newacronym{DCU}{DCU}{Diamond Control Unit}
\newacronym{DEPFET}{DEPFET}{Depleted P-Channel Field Effect Transistor}
\newacronym{DQM}{DQM}{Data Quality Monitoring}
\newacronym{DSSD}{DSSD}{Double-sided Silicon Strip Detector}
\newacronym{ELMB}{ELMB}{Embedded Local Monitor Board}
\newacronym{ENC}{ENC}{Equivalent Noise Charge}
\newacronym{EPICS}{EPICS}{Experimental Physics and Industrial Control System}
\newacronym{FADC}{FADC}{Flash Analog to Digital Converter}
\newacronym{FBG}{FBG}{Fiber Bragg Grating}
\newacronym{FIFO}{FIFO}{First In First Out}
\newacronym{FIR}{FIR}{Finite Impulse Response}
\newacronym{FOS}{FOS}{Fiber Optic Sensor}
\newacronym{FPGA}{FPGA}{Field Programmable Gate Array}
\newacronym{FTB}{FTB}{Finesse Transmitter Board}
\newacronym{FTSW}{FTSW}{Front-end Timing Switch}
\newacronym{FTDI}{FTDI}{Future Technology Devices International}
\newacronym{HEPA}{HEPA}{High Efficiency Particulate Air}
\newacronym{HER}{HER}{High Energy Ring}
\newacronym{HLT}{HLT}{High Level Trigger}
\newacronym{HV}{HV}{High Voltage}
\newacronym{I2C}{I2C}{Inter-Integrated Circuit serial communication bus}
\newacronym{IOC}{IOC}{Input/Output Controller}
\newacronym{IP}{IP}{Interaction Point}
\newacronym{LER}{LER}{Low Energy Ring}
\newacronym{LVPS}{LVPS}{Low Voltage Power Supply}
\newacronym{ONSEN}{ONSEN}{ONline SElection Nodes}
\newacronym{NIEL}{NIEL}{Non-Ionising Energy Loss}
\newacronym{NTC}{NTC}{Negative Temperature Coefficient}
\newacronym{MC}{MC}{Monte Carlo method}
\newacronym{MIP}{MIP}{Minimum Ionizing Particle}
\newacronym{MOSFET}{MOSFET}{Metal-Oxide-Semiconductor Field-Effect Transistor}
\newacronym{MPV}{MPV}{Most Probable Value}
\newacronym{OPI}{OPI}{OPerator Interfaces}
\newacronym{PCB}{PCB}{Printed Circuit Board}
\newacronym{PLC}{PLC}{Programmable Logic Controller}
\newacronym{PT}{PT}{Positioning Tower}
\newacronym{PXD}{PXD}{Pixel Detector}
\newacronym{QCG}{QCG}{Quality Control and Assurance Group}
\newacronym{QCS}{QCS}{final-focus superconducting quadrupole}
\newacronym{RC}{RC}{Run Control}
\newacronym{ROI}{ROI}{Region Of Interest}
\newacronym{RF}{RF}{accelerator radiofrequency, 508.9 MHz}
\newacronym{RSA}{RSA}{Rib Sub-Assembly}
\newacronym{SBW}{SBW}{SVD Backward module}
\newacronym{SFW}{SFW}{SVD Forward module}
\newacronym{SNR}{\snrCL}{Signal-to-Noise Ratio}
\newacronym{SPA}{SPA}{Sensor-Pitch Adapter sub-assembly}
\newacronym{SVD}{SVD}{Silicon Vertex Detector}
\newacronym{TAS}{TAS}{Two-Arms Support}
\newacronym{TCT}{TCT}{Transient Current Technique}
\newacronym{VLHI}{VLHI}{\acrshort{VXD} Local Hardwired Interlock}
\newacronym{VSEP}{\vsep}{Separation Voltage}
\newacronym{VXD}{VXD}{Vertex Detector, including \acrshort{PXD} and \acrshort{SVD}}
\newacronym{VXDTF2}{VXDTF2}{Vertex Detector Track Finding}

\newglossaryentry{annealing}{
    name = annealing, 
    description ={heat treatment that alters the physical and/or chemical properties of a material. In silicon sensors, it results in doping atoms to diffuse into substitutional positions in the crystal lattice, where they can become donors or acceptors}
}

\newglossaryentry{cluster}{
    name = cluster, 
    description ={set of adjacent strips representing a 1-D \gls{hit} on the plane of sensor}
}

\newglossaryentry{common-mode noise}{
    name = common-mode noise, 
    description ={a type of noise that causes correlated fluctuations on all the channels}
}

\newglossaryentry{DOCK}{
    name = DOCK, 
    description ={space reserved for interconnections of cables and services, close to the \acrshort{CDC} end plates}
}

\newglossaryentry{e-hut}{
    name = e-hut, 
    description ={three-floor electronics hut, located at one side of the Belle II detector}
}

\newglossaryentry{end-cone}{ 
    name = end-cone,
    description = {part of the \acrshort{SVD} support structure, cone-shaped and realized in carbon fibre}
}

\newglossaryentry{end-flange}{
    name = end-flange,
    description = {part of the \acrshort{SVD} support structure, flange-shaped, realized in aluminum, and located at both ends of the SVD}
}
\newglossaryentry{end-mount}{
    name = end-mount,
    description = {aluminum part of the \acrshort{SVD} ladder, located at both ends of the ladder, and joined to the support \gls{end-ring}}
}

\newglossaryentry{end-ring}{ 
    name = end-ring,
    description = {part of the support structure for a given \acrshort{SVD} \gls{layer}, located on an \gls{end-cone}}
}

\newglossaryentry{hit}{
    name = hit, 
    description ={generic term indicating the signal or set of signals induced by a charged particle track crossing the sensor; 1-D hit: signal on a set of adjacent strips in the sensor plane, forming a \gls{cluster}; 2-D hit: combined signal from 1-D hits on the two opposite sides, in the sensor plane ; 3-D hit: complete information on the signal induced by a charged particle track, including space position, time and released ionization charge}
}

\newglossaryentry{keratherm}
{
    name=keratherm,
    description={thermally conductive material, available in sheets of different thickness, with an adhesive layer on both sides}
}

\newglossaryentry{ladder}
{
    name=ladder,
    description={mechanically and electrically independent subset of \acrshort{SVD} \gls{layer}s, resembling the stave of a barrel}
}

\newglossaryentry{layer}
{
    name=layer,
    description={part of the \acrshort{SVD} detector, located at approximately fixed radius}
}

\newglossaryentry{noise}{
    name = noise,
    description = {root mean square (\texttt{RMS}) of the each strip signal after pedestal subtraction and common-mode corrections}
}

\newglossaryentry{occupancy}{
    name = occupancy,
    description = {fraction of strips above threshold in a single triggered event}
}

\newglossaryentry{origamischeme}{
    name = origami scheme,
    description = {Chip-on-sensor concept, obtained by connecting DSSD strips to front-end APV25 ASICs via flex circuits, bent over the DSSD edge}
    } 

\newglossaryentry{origamimodule}{
	name = origami module,
	description = {One of the module types that use the origami scheme. Three types exist: O-Z, O-CE, O+Z}
} 

\newglossaryentry{origamipipe}{
	name = origami pipe,
	description = {The thin stainless steel pipes used to cool the \apv\ chips on the origami modules once assembled in ladders}
} 

\newglossaryentry{origamiboard}{
	name = origami board,
	description = {The flexible printed circuit board used in the orgami modules to distribute power and signals to the \apv\ chips}
} 
    
\newglossaryentry{pinhole}{
    name = pinhole,
    description = {broken AC decoupling capacitor between the strip implant and the metal readout electrode} 
}
\newglossaryentry{pedestal}{
    name = pedestal,
    description = {average of each strip signal}
}
\newglossaryentry{ROIgls}{
    name =  ROI,
    description = {Region Of Interest on the PXD sensors where a signal cluster is probably located}
}

\newglossaryentry{spacepoint}{
    name = spacepoint,
    description = {combination of a $u/P$ side cluster with a $v/N$ side cluster to make the 3-D \gls{hit}}
}

\newglossaryentry{badpinhole}{
    name = substrate pinhole,
    description = {broken AC decoupling capacitor between the strip implant and the detector substrate} 
}
\newglossaryentry{triggerbin}{
 name = trigger bin,
    description = {the arrival time of the L1 trigger registered by the FADC with a clock frequency that is four times larger than the APV25 clock frequency. 
    The trigger bin allows to move the hit time measured
    in the SVD reference frame to the trigger reference frame, common to all detectors, and vice-versa}
}

\newglossaryentry{kokeshi}{
 name = kokeshi-pin,
    description = {a precision pin with a groove allowing good planar coupling between two surfaces with a lateral screw}
}

\title{The Design, Construction, Operation and Performance of the Belle II Silicon Vertex Detector}
\collaboration{%
\includegraphics[height=17mm]{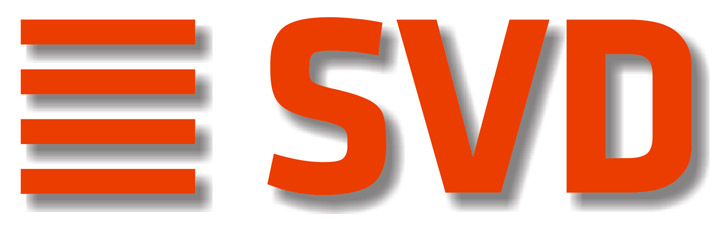}\\[6pt]
Belle II SVD collaboration}
\newcommand{\instCPPM}{Aix Marseille Universit\'{e}, CNRS/IN2P3, CPPM, 13288 Marseille, France}

\newcommand{\instBNL}{Brookhaven National Laboratory, Upton, New York 11973, U.S.A.}

\newcommand{\instPrague}{Faculty of Mathematics and Physics, Charles University, 121 16 Prague, Czech Republic}

\newcommand{\instDESY}{Deutsches Elektronen--Synchrotron, 22607 Hamburg, Germany}

\newcommand{\instFudan}{Key Laboratory of Nuclear Physics and Ion-beam Application (MOE) and Institute of Modern Physics, Fudan University, Shanghai 200443, China}

\newcommand{\instSOKENDAI}{The Graduate University for Advanced Studies (SOKENDAI), Hayama 240-0193, Japan}

\newcommand{\instKEK}{High Energy Accelerator Research Organization (KEK), Tsukuba 305-0801, Japan}

\newcommand{\instIITBhubaneswar}{Indian Institute of Technology Bhubaneswar, Satya Nagar 751007, India}

\newcommand{\instIITHyderabad}{Indian Institute of Technology Hyderabad, Telangana 502285, India}
\newcommand{\instIITMadras}{Indian Institute of Technology Madras, Chennai 600036, India}

\newcommand{\instHEPHYVienna}{Institute of High Energy Physics, Austrian Academy of Sciences, Vienna 1050, Austria}

\newcommand{\instPisaINFN}{INFN Sezione di Pisa, I-56127 Pisa, Italy}

\newcommand{\instRomaTreINFN}{INFN Sezione di Roma Tre, I-00146 Roma, Italy}

\newcommand{\instTriesteINFN}{INFN Sezione di Trieste, I-34127 Trieste, Italy}

\newcommand{\instISU}{Iowa State University, Ames, Iowa 50011, U.S.A.}
\newcommand{\instKyungpook}{Kyungpook National University, Daegu 41566, South Korea}

\newcommand{\instLMU}{Ludwig Maximilians University, 80539 Munich, Germany}

\newcommand{\instMNITJaipur}{Malaviya National Institute of Technology Jaipur, Jaipur 302017, India}

\newcommand{\instKrakow}{H. Niewodniczanski Institute of Nuclear Physics, Krakow 31-342, Poland}
\newcommand{\instNiigata}{Niigata University, Niigata 950-2181, Japan}

\newcommand{\instPanjab}{Panjab University, Chandigarh 160014, India}
\newcommand{\instPanjabPAU}{Punjab Agricultural University, Ludhiana 141004, India}

\newcommand{\instSeoul}{Seoul National University, Seoul 08826, South Korea}

\newcommand{\instTata}{Tata Institute of Fundamental Research, Mumbai 400005, India}

\newcommand{\instTohoku}{Department of Physics, Tohoku University, Sendai 980-8578, Japan}

\newcommand{\instTokyoMetropolitan}{Tokyo Metropolitan University, Tokyo 192-0397, Japan}

\newcommand{\instPisaUNIV}{Dipartimento di Fisica, Universit\`{a} di Pisa, I-56127 Pisa, Italy}

\newcommand{\instTriesteUNIV}{Dipartimento di Fisica, Universit\`{a} di Trieste, I-34127 Trieste, Italy}

\newcommand{\instIJCLab}{Universit\'{e} Paris-Saclay, CNRS/IN2P3, IJCLab, 91405 Orsay, France}
\newcommand{\instIPHC}{Universit\'{e} de Strasbourg, CNRS, IPHC, UMR 7178, 67037 Strasbourg, France}

\newcommand{\instMelbourne}{School of Physics, University of Melbourne, Victoria 3010, Australia}

\newcommand{\instUSTC}{University of Science and Technology of China, Hefei 230026, China}

\newcommand{\instUTokyo}{Department of Physics, University of Tokyo, Tokyo 113-0033, Japan}
\newcommand{\instIPMU}{Kavli Institute for the Physics and Mathematics of the Universe (WPI), University of Tokyo, Kashiwa 277-8583, Japan}

\newcommand{\instYonsei}{Yonsei University, Seoul 03722, South Korea}

\newcommand{\symCPPM}{a}	\affiliation[\symCPPM]{\instCPPM}
\newcommand{\symFudan}{b}	\affiliation[\symFudan]{\instFudan}
\newcommand{\symHEPHYVienna}{c}	\affiliation[\symHEPHYVienna]{\instHEPHYVienna}
\newcommand{\symIITBhubaneswar}{d}	\affiliation[\symIITBhubaneswar]{\instIITBhubaneswar}
\newcommand{\symIITMadras}{e}	\affiliation[\symIITMadras]{\instIITMadras}
\newcommand{\symIPHC}{f}	\affiliation[\symIPHC]{\instIPHC}
\newcommand{\symIPMU}{g}	\affiliation[\symIPMU]{\instIPMU}
\newcommand{\symKEK}{h}	\affiliation[\symKEK]{\instKEK}
\newcommand{\symKrakow}{i}	\affiliation[\symKrakow]{\instKrakow}
\newcommand{\symKyungpook}{j}	\affiliation[\symKyungpook]{\instKyungpook}
\newcommand{\symMelbourne}{k}	\affiliation[\symMelbourne]{\instMelbourne}
\newcommand{\symMNITJaipur}{l}	\affiliation[\symMNITJaipur]{\instMNITJaipur}
\newcommand{\symNiigata}{m}	\affiliation[\symNiigata]{\instNiigata}
\newcommand{\symPanjab}{n}	\affiliation[\symPanjab]{\instPanjab}
\newcommand{\symPanjabPAU}{o}	\affiliation[\symPanjabPAU]{\instPanjabPAU}
\newcommand{\symPisaINFN}{p}	\affiliation[\symPisaINFN]{\instPisaINFN}
\newcommand{\symPisaUNIV}{q}	\affiliation[\symPisaUNIV]{\instPisaUNIV}
\newcommand{\symPrague}{r}	\affiliation[\symPrague]{\instPrague}
\newcommand{\symSeoul}{s}	\affiliation[\symSeoul]{\instSeoul}
\newcommand{\symSOKENDAI}{t}	\affiliation[\symSOKENDAI]{\instSOKENDAI}
\newcommand{\symTata}{u}	\affiliation[\symTata]{\instTata}
\newcommand{\symTohoku}{v}	\affiliation[\symTohoku]{\instTohoku}
\newcommand{\symTriesteINFN}{w}	\affiliation[\symTriesteINFN]{\instTriesteINFN}
\newcommand{\symTriesteUNIV}{x}	\affiliation[\symTriesteUNIV]{\instTriesteUNIV}
\newcommand{\symUSTC}{y}	\affiliation[\symUSTC]{\instUSTC}
\newcommand{\symUTokyo}{z}	\affiliation[\symUTokyo]{\instUTokyo}
\newcommand{\symZZZBNL}{2}	\newcommand{\instZZZBNL}{now at \instBNL}	\affiliation[\symZZZBNL]{\instZZZBNL}
\newcommand{\symZZZCPPM}{3}	\newcommand{\instZZZCPPM}{now at \instCPPM}	\affiliation[\symZZZCPPM]{\instZZZCPPM}
\newcommand{\symZZZDESY}{4}	\newcommand{\instZZZDESY}{now at \instDESY}	\affiliation[\symZZZDESY]{\instZZZDESY}
\newcommand{\symZZZElettra}{5}	\newcommand{\instZZZElettra}{also at Elettra-Sincrotrone Trieste S.C.p.A., Basovizza 34149, Italy }	\affiliation[\symZZZElettra]{\instZZZElettra}
\newcommand{\symZZZICEPP}{6}	\newcommand{\instZZZICEPP}{now at ICEPP, University of Tokyo, Tokyo, Japan}	\affiliation[\symZZZICEPP]{\instZZZICEPP}
\newcommand{\symZZZIITHyderabad}{7}	\newcommand{\instZZZIITHyderabad}{now at \instIITHyderabad}	\affiliation[\symZZZIITHyderabad]{\instZZZIITHyderabad}
\newcommand{\symZZZIJCLab}{8}	\newcommand{\instZZZIJCLab}{now at \instIJCLab}	\affiliation[\symZZZIJCLab]{\instZZZIJCLab}
\newcommand{\symZZZIPHC}{9}	\newcommand{\instZZZIPHC}{now at \instIPHC}	\affiliation[\symZZZIPHC]{\instZZZIPHC}
\newcommand{\symZZZISU}{10}	\newcommand{\instZZZISU}{now at \instISU}	\affiliation[\symZZZISU]{\instZZZISU}
\newcommand{\symZZZIwate}{11}	\newcommand{\instZZZIwate}{now at Iwate University, Morioka, Japan}	\affiliation[\symZZZIwate]{\instZZZIwate}
\newcommand{\symZZZLMU}{12}	\newcommand{\instZZZLMU}{now at \instLMU}	\affiliation[\symZZZLMU]{\instZZZLMU}
\newcommand{\symZZZManchester}{13}	\newcommand{\instZZZManchester}{now at University of Manchester, Manchester, UK}	\affiliation[\symZZZManchester]{\instZZZManchester}
\newcommand{\symZZZNDU}{14}	\newcommand{\instZZZNDU}{now at Nippon Dental University, Niigata, Japan}	\affiliation[\symZZZNDU]{\instZZZNDU}
\newcommand{\symZZZNebraska}{15}	\newcommand{\instZZZNebraska}{now at University of Nebraska, Lincoln, USA}	\affiliation[\symZZZNebraska]{\instZZZNebraska}
\newcommand{\symZZZNISER}{16}	\newcommand{\instZZZNISER}{now at National Institute of Science Education and Research, Homi Bhabha National Institute, 752050 Jatni, India}	\affiliation[\symZZZNISER]{\instZZZNISER}
\newcommand{\symZZZRomaTreINFN}{17}	\newcommand{\instZZZRomaTreINFN}{now at \instRomaTreINFN}	\affiliation[\symZZZRomaTreINFN]{\instZZZRomaTreINFN}
\newcommand{\symZZZSeoul}{18}	\newcommand{\instZZZSeoul}{now at \instSeoul}	\affiliation[\symZZZSeoul]{\instZZZSeoul}
\newcommand{\symZZZSNS}{19}	\newcommand{\instZZZSNS}{now  at Scuola Normale Superiore, I-56127 Pisa, Italy }	\affiliation[\symZZZSNS]{\instZZZSNS}
\newcommand{\symZZZTIFPA}{20}	\newcommand{\instZZZTIFPA}{now at TIFPA-INFN, Dipartimento di Fisica, Universit\'{a} di Trento, I-38123 Trento, Italy}	\affiliation[\symZZZTIFPA]{\instZZZTIFPA}
\newcommand{\symZZZTokyoMetropolitan}{21}	\newcommand{\instZZZTokyoMetropolitan}{now at \instTokyoMetropolitan}	\affiliation[\symZZZTokyoMetropolitan]{\instZZZTokyoMetropolitan}
\newcommand{\symZZZYonsei}{22}	\newcommand{\instZZZYonsei}{now at \instYonsei}	\affiliation[\symZZZYonsei]{\instZZZYonsei}
\newcommand{\symZZZDeceased}{\dagger}	\newcommand{\instZZZDeceased}{Deceased}	\affiliation[\symZZZDeceased]{\instZZZDeceased}
\newcommand{\symZZZCorresponding}{1}	\newcommand{\instZZZCorresponding}{Corresponding author}
 
\author[\symKrakow]{K.~Adamczyk,}
\author[\symPanjab]{L.~Aggarwal,}
\author[\symUTokyo]{H.~Aihara,}
\author[\symTata]{T.~Aziz,}
\author[\symTata,\symZZZIPHC]{V.~Babu,}
\author[\symKrakow]{S.~Bacher,}
\author[\symIITBhubaneswar]{S.~Bahinipati,}
\author[\symTriesteINFN]{M.~Bari,}
\author[\symMelbourne]{Ti.~Baroncelli,}
\author[\symMelbourne]{To.~Baroncelli,}
% ...
\emailAdd{Francesco.Forti@pi.infn.it}

\abstract{
  The Silicon Vertex Detector of \belletwo\ is a state-of-the-art tracking and vertexing system based on double-sided silicon strip sensors,
  designed and fabricated by a large international collaboration in the period 2012--2018.
  Since 2019 it has been in operation providing high quality data with a small number of defective channels (<1\%), a large hit-finding efficiency (>99\%),
  a good signal-to-noise ratio (well in excess of 10 for all sensor configurations and tracks). Together with the good control over the alignment, these are all
  essential factors to achieve good tracking reconstruction and physics performance.
  In this extended paper we try to document all the aspects of the SVD challenges and achievements, in the spirit of providing information to the
  broader community and help the development of high quality detector systems, which are  essential tools to carry out physics research.
}

\keywords{Instrumentation for particle accelerators and storage rings - high energy;
  Particle tracking detectors;
  Solid state detectors;
  Si microstrip and pad detectors;
  Front-end electronics for detector readout;
  Radiation damage to detector materials (solid state);
  Detector alignment and calibration methods;
  Detector cooling and thermo-stabilization;
  Detector design and construction technologies and materials;
  Detector grounding;
  Overall mechanics design (support structures and materials, vibration analysis etc);
  Voltage distributions
  }
\begin{document}

\maketitle

%\onecolumn

%====================================================

\lvlonesection{Introduction} {\label{sec:intro}}
\editor{Christoph}
%\input Intro.tex
%\emph{Writing assignment: Christoph (ok)}

The Belle II Silicon Vertex Detector (\acrshort{SVD}) is an important part of the upgrade of the KEK $B$~factory consisting of the KEKB asymmetric-energy $e^+e^-$ collider~\cite{Kurokawa:2001nw} and the Belle experiment~\cite{Abashian:2000cg} to a Super $B$~factory with SuperKEKB and Belle II~\cite{Abe:2010gxa}. This upgrade significantly raises the physics requirements for the silicon tracking device. While KEKB collided 8~GeV electrons on 3.5~GeV positrons resulting in a Lorentz boost of $\beta\gamma=0.425$, the more constrained low-emittance optics and higher beam currents of SuperKEKB require a reduction of the beam asymmetry, and collisions of 7~GeV electrons on 4~GeV positrons imply a Lorentz boost factor of only 0.28, about two-thirds of that in KEKB. The resolution in the decay time difference $\Delta t$ of the two $B$~mesons produced in collisions at the center-of-mass energy of the $\Upsilon(4S)$ resonance is important for the study of time-dependent $CP$~violation and should not deteriorate due to this change. As $\Delta t$ is related to the Lorentz boost and the vertex separation $\Delta z$, $\Delta t\approx \Delta z/\beta\gamma c$, maintaining the same $\Delta t$ resolution as in Belle implies improving the vertexing performance by about two-thirds.

At a very early stage in the vertex detector design, it became apparent that this is only achievable by reducing the 30~mm beam pipe of Belle to 15~mm or even 10~mm. Early simulations indicated that beam-induced backgrounds would increase by a factor of 15 in Belle II compared to Belle~\cite{Adachi:2008da}. The \belletwo\ SVD must thus be considerably faster than the Belle SVD2 to limit hit occupancy in the innermost layers to a level acceptable for the reconstruction software. This determined the choice of the APV25 as the frontend readout \acrshort{ASIC}, which is capable of a 50~ns shaping time compared to the 800~ns of the VA1TA chip used in the Belle SVD2. Other guiding factors of the Belle II SVD design are the radiation tolerance and a drastic material reduction to avoid deterioration of the particle tracking performance due to multiple scattering in the device material. For radiation hardness, a total ionization dose (TID) of 10~Mrad  corresponding to ten years of operation is assumed in the SVD design.

\begin{figure}[hbt]
    \centering
    \includegraphics[width=\textwidth]{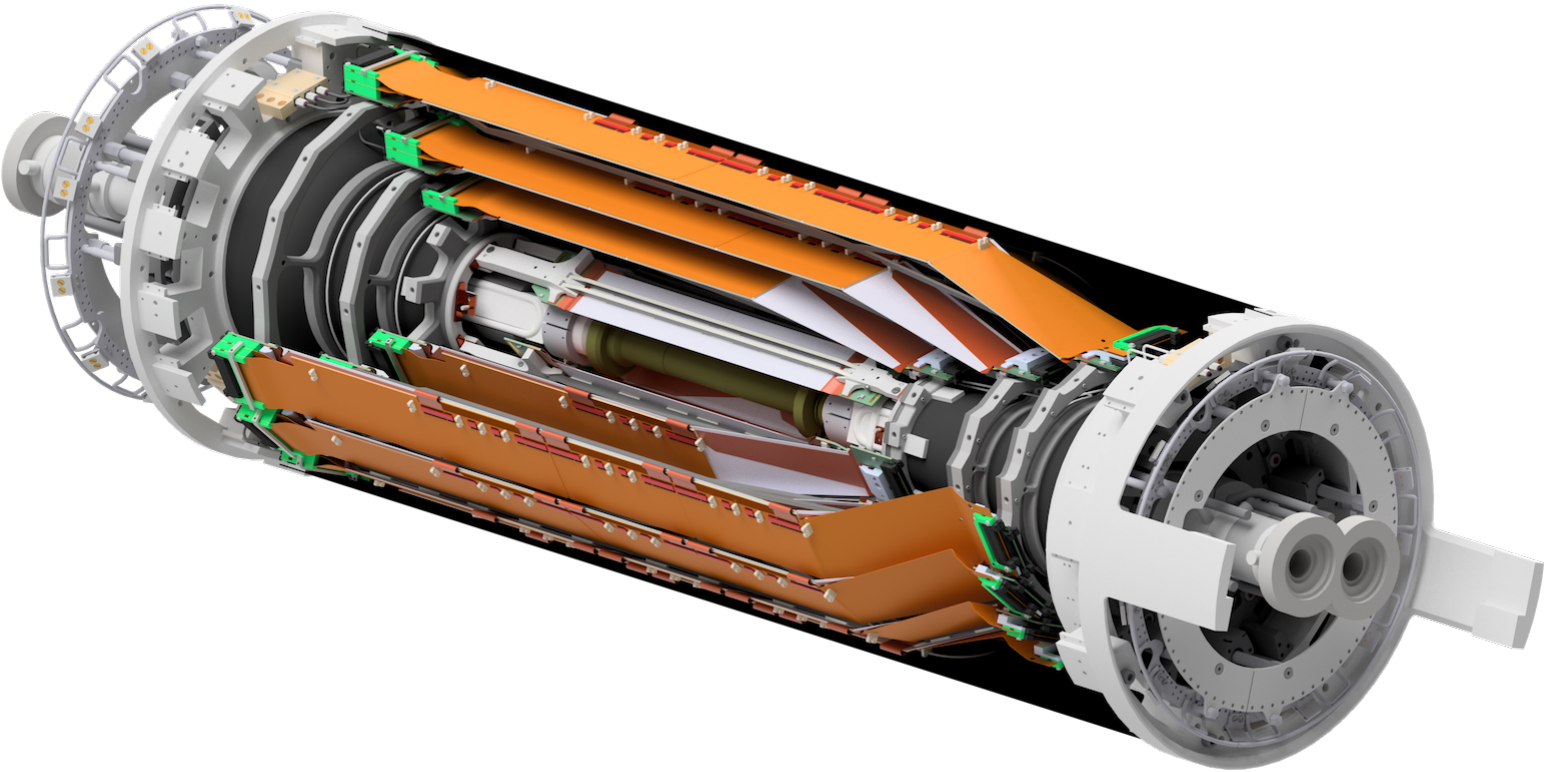}
    \caption{The Belle II vertex detector (VXD), composed of the pixel (PXD) and silicon strip (SVD) detectors.}
    \label{fig:C1_SVD_Rendering}
\end{figure}

\begin{figure}[hbt]
    \centering
    \includegraphics[width=\textwidth]{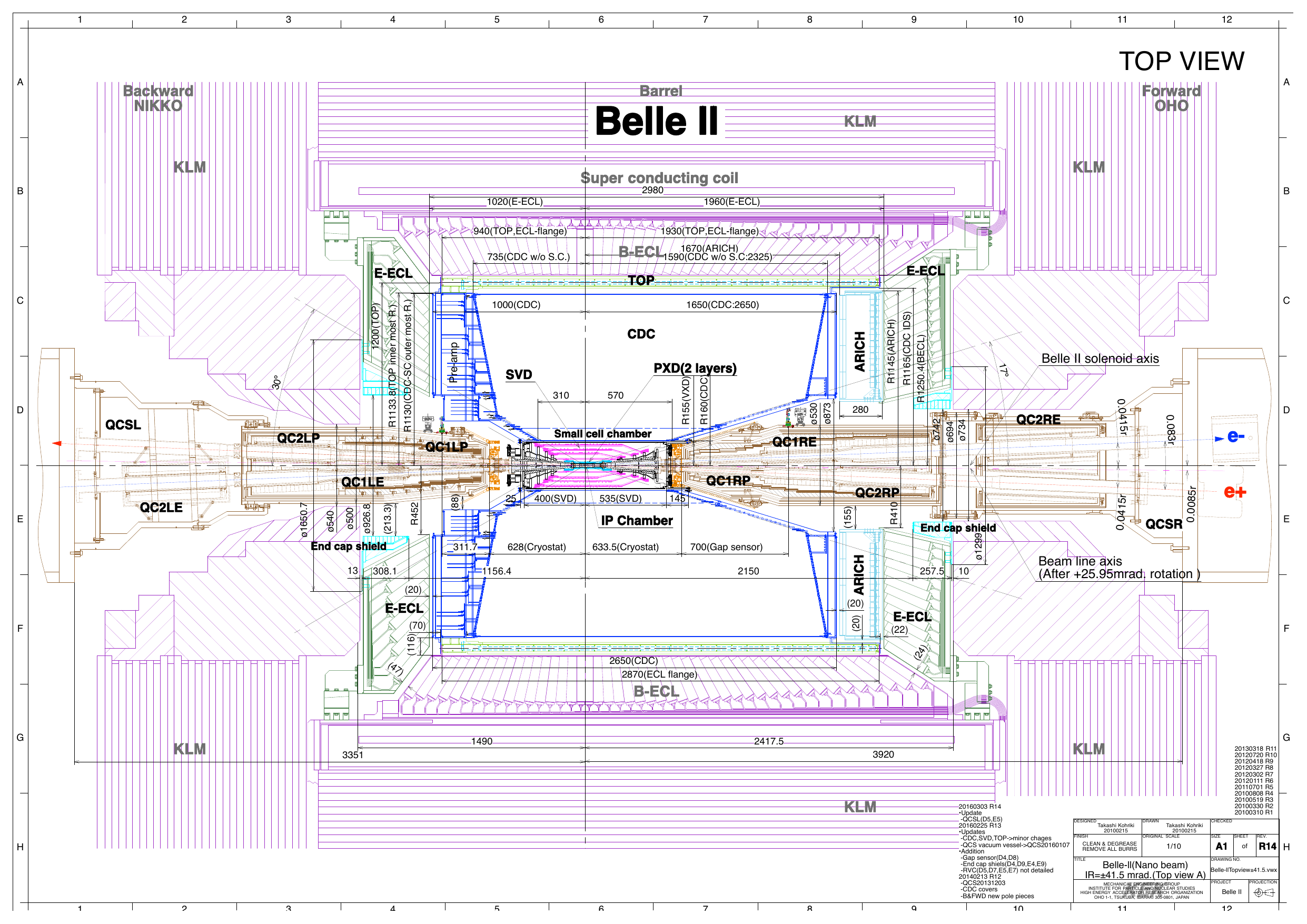}
    \caption{The Belle II detector, surrounding the electron-positron interaction region: charged particles tracking is performed by the PXD, SVD silicon sensors and the CDC drift chamber, occupying the inner volume.
    }
    \label{fig:C1_Belle2_Rendering}
\end{figure}

The result of this design and construction effort is the Belle II VerteX Detector (\acrshort{VXD}) shown in figure~\ref{fig:C1_SVD_Rendering}, which consists of two layers of DEpleted $P$-channel Field Effect Transistor (DEPFET) pixel sensors (\acrshort{PXD})~\cite{Kemmer:1986vh} and four layers of double-sided silicon strip sensors (SVD), arranged cylindrically around the beam pipe at radii between 14 and 135~mm. The cylindrical  beam pipe has an inner radius of \qty{10}{\mm} and incorporates a \qty{1}{\mm} paraffin cooling channel defined by  \qty{0.4}{\mm} and \qty{0.6}{\mm} inner and outer beryllium walls, resulting in an outer radius of \qty{12}{\mm}.
While the placement of the innermost layer is guided by the desired vertexing performance, the outer radius is set by mechanical considerations of the support structure. Four silicon strip layers are required to perform autonomous particle trajectory finding in the SVD, in order to improve the tracking performance at the low momentum end of the particle spectrum and to be able to provide regions of interest for a reduction of the PXD data size. The VXD is mounted directly on the Belle II beam pipe and installed in the central opening of the Central Drift Chamber (\acrshort{CDC}), the main tracking device of the Belle II detector, shown in figure~\ref{fig:C1_Belle2_Rendering}. It operates in the 1.5~T magnetic field of the Belle II solenoid and provides inner tracking, impact parameter measurement, decay vertex reconstruction, and low-momentum particle identification for the Belle II detector~\cite{Adachi:2018qme}. 

This paper describes the design, construction, operation, and performance of the SVD. %, {\it i.e.} the outer part of the VXD.
An initial appreciation of the overall project can be obtained by a %``first track'' 
reading of the overview paragraphs opening each of the following sections, where the main concepts are explained and the key features highlighted.

The paper is organised as follows. Section~\ref{sec:C2_design} contains the motivations and details of the SVD design. The construction of the detector is discussed in section~\ref{sec:C3_construction}.
Section~\ref{sec:environment} deals with environmental and radiation monitoring and related safety interlocks.
Section~\ref{sec:software_online} is dedicated to the SVD online and control software. The different steps of SVD installation and commissioning in the Belle II detector are reviewed in section~\ref{sec:C5_commissioning}, while section~\ref{sec:C6_operation} describes the operation of the detector during SuperKEKB physics data taking in the years 2019 and 2020. 
After an introduction to the SVD offline software (section~\ref{sec:software_offline}), the main performance parameters are discussed
in section~\ref{sec:C7_performance}. Finally, an outlook of future operations and developments is given in section~\ref{sec:conclusion}.

A glossary and acronym list are available at end of the main text to help interpret the many acronyms and technical terms. Active links are used when terms are used for the first time in a section to simplify browsing when reading the paper on a computer.
 % 1 page 

\lvlonesection{Detector Design}
\label{sec:C2_design}
%\editor{Christoph}

%\textcolor{red}{Preliminary ``track 1'' overview (Livio)}

The \acrshort{SVD} design is guided by the major physics goals described in section~\ref{sec:intro}, as part of the Belle II \acrshort{VXD}, operating in the challenging environment of the high-luminosity SuperKEKB electron-positron collider. The SVD contributes to inner tracking, impact parameter measurement, decay vertex reconstruction, and particle identification via energy loss measurements, with tight constraints on the material budget and strict requirements on radiation hardness and acceptable occupancy.

\begin{figure}[hbt]
    \centering
    \includegraphics[width=\textwidth]{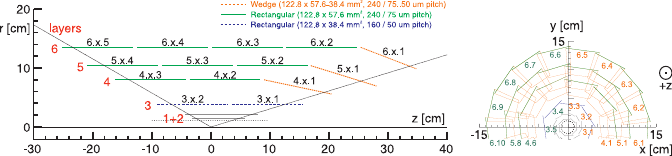}
    \caption{A schematic layout of the Belle II SVD: (left panel) longitudinal cut, identifying layers from 3 to 6, and individual sensors with their main properties; (right panel) transverse cut, showing a projection of layers and ladders, and the numbering scheme of the sensors. The reference frame is defined later in this section.}
    \label{fig:C2_schematic_layout}
\end{figure}

To fulfil these needs, four SVD \emph{\gls{layer}s} with cylindrical symmetry (figure~\ref{fig:C2_schematic_layout}) are inserted in the active volume between \acrshort{PXD} and \acrshort{CDC}. The windmill-like arrangement of the sensors around the beam direction causes a partial overlap of the sensor active regions (between 2\% and 8\% depending on the layer), which is needed for the internal alignment of the device. The Lorentz boost, originating from the asymmetric-energy collisions, imposes a forward-backward asymmetry to Belle II detectors. To cope with the inclination of the forward tracks as well as to minimize the material traversed in the forward part, a lamp-shade geometry was adopted for the forward end of the three outer \gls{layer}s of SVD, whose detectors are inclined with respect to the cylinder axis. SVD layers are conventionally numbered from 3 to 6, taking into account the first two inner PXD layers.

The well-proven \acrfull{DSSD} technology is the basic choice for the SVD. Section~\ref{sec:si_det} describes the optimization of sensor parameters for our application: shape, size, thickness, and readout strip pitch. The inner layer 3 has DSSD rectangular sensors with smaller size and pitch; the outer layers 4, 5 and 6 have larger rectangular sensors, and trapezoidal sensors for the lamp-shade part, necessitating dedicated R\&D and design. 

One of the major challenges in the design is the need to cope with the increased particle rates and hence with the readout channels occupancy, especially in the inner layers. The front-end readout \apv\ \acrshort{ASIC}~\cite{French:2001xb}, originally developed for the \acrshort{CMS} experiment, has the required properties of reduced signal shaping time, good time resolution, on-chip integrated pipeline for dead-time free readout, and radiation tolerance. 

The use of \apv\ and its implications for the project are described in section~\ref{sec:C2_ondet_elec}, dedicated to the on-detector front-end electronics. Faster shaping time comes with larger noise: to keep it at an acceptable level, readout chips are placed as close to the sensors as possible. Novel solutions to achieve this goal, while keeping a low material budget, are implemented by a ``chip-on-sensor'' concept. Thinned-down readout chips are mounted on a flex circuit placed on top of the sensors, separated by a layer of foam for thermal and electrical isolation; flexible fan-outs are wrapped around the edge of the sensor to reach the strips on the bottom side. This scheme is named \gls{origamischeme} with reference to the folding action. 

The mechanical integration of individual detectors and their thermal control are also challenging. Each detector layer is organized in mechanically and electrically independent subsets called \emph{\gls{ladder}s}, resembling the staves of a barrel. Although cold operation of the sensors, which operate around room temperature, is not required, active  cooling based on dual-phase \COtwo\ is needed to remove the heat produced by the readout electronics, leading to large temperature variations  (up to \qty{40}{\degc}) along the ladders and when the electronics is turned off. The employed materials and  the mechanical design are carefully optimized to avoid thermal stresses on the sensors and on the mechanical support. 

 The seven ladders of layer 3 are shorter and made of only two sensors, that can be read out by chips mounted on two traditional \emph{hybrid boards} located outside the active volume, at the forward and backward ends respectively. Layers 4, 5, 6 are made of 10, 12, and 16 longer ladders, each containing 3, 4, and 5 sensors, respectively,  for a total of 172 individual sensors. Only the first and last sensors in these ladders can be read out by the hybrid board arrangement; the central sensors require \glspl{origamiboard} for the readout scheme mentioned above. In all cases flexible \emph{pitch adapters} are used to connect sensor strips to \apv\ channels. 

The original solutions for these complex ladders, including insulating foam, DSSDs and support ribs are described in section~\ref{sec:modules}. For assembling the ladders in layers 4 to 6 the relevant sub-parts of each ladder are the slanted \emph{forward section} with one trapezoidal DSSD, and the remaining \emph{barrel part}, made of rectangular sensors. 

Figure~\ref{fig:C2_schematic_layout} gives an overview of the SVD layout with its organization in \gls{layer}s, \gls{ladder}s, and sensors.  The definition of the Belle II reference frame and of local frames for the individual sensors are recalled at the end of the present section. 
Sensors at different $z$ position along the ladders, forward (FW), central, backward (BW), are labeled with increasing numbers starting with one for the FW sensor. 
In the rest of the paper each sensor is indicated by its position in the layer, ladder ($\phi$), and $z$ position along the ladder with the following convention: L3.2.1 stands for the position in layer 3, ladder 2, sensor 1 (FW) along the ladder, as shown in figure~\ref{fig:C2_schematic_layout}. An `x' replacing a number indicates the entirety of sensors for that position; for instance L5.x.1 means all the sensors in layer 5, any $\phi$ position, $z$ position 1 (that is, all FW sensors in layer 5).

In total, the SVD has an active DSSD area of about \qty{1.1}{\m\squared} and features 223744 channels read out by a total of 1748 \apv\ chips. The material budget per DSSD layer corresponds to approximately 0.7\% of the radiation length. The detector is designed to withstand an integrated radiation dose of 10~Mrad before significant performance degradation.

The entire SVD is longitudinally divided in two azimuthal halves, which are separately assembled before being mounted on the beam pipe and the PXD layers. The ladders of each half are supported and held in position at their two ends by joining the ladder aluminum \emph{\gls{end-mount}s} to stainless-steel (half) \emph{\gls{end-ring}s} precisely glued on forward and backward carbon-fiber (half) \emph{\gls{end-cone}s}. The innovative solutions adopted to meet the demanding requirements of position precision and stability are discussed in section~\ref{sec:mech}. The same section also describes the \COtwo\ cooling system designed to remove the heat generated by the front-end chips, using meander-shaped pipes in thermal contact with front-end \acrshort{ASIC}s, connected at their ends to cooling channels integrated in the end-rings. 

Finally, the integration of the \apv\ readout in the Belle II data acquisition system required a very significant investment in design, prototyping, programming and tests. The overall back-end electronics design, and the floating power supplies allowing a flexible scheme for detector bias and front-end powering, are described in section~\ref{sec:C2_offdet_elec}. A critical aspect for the detector performance is the electrical grounding scheme, described in section~\ref{sec:C2_grounding}. The overall SVD design was validated by several cycles of prototypes and beam tests, summarized in section~\ref{sec:C2_beamtest}.

\begin{figure}[hbt]
    \centering
    \includegraphics[width=\textwidth]{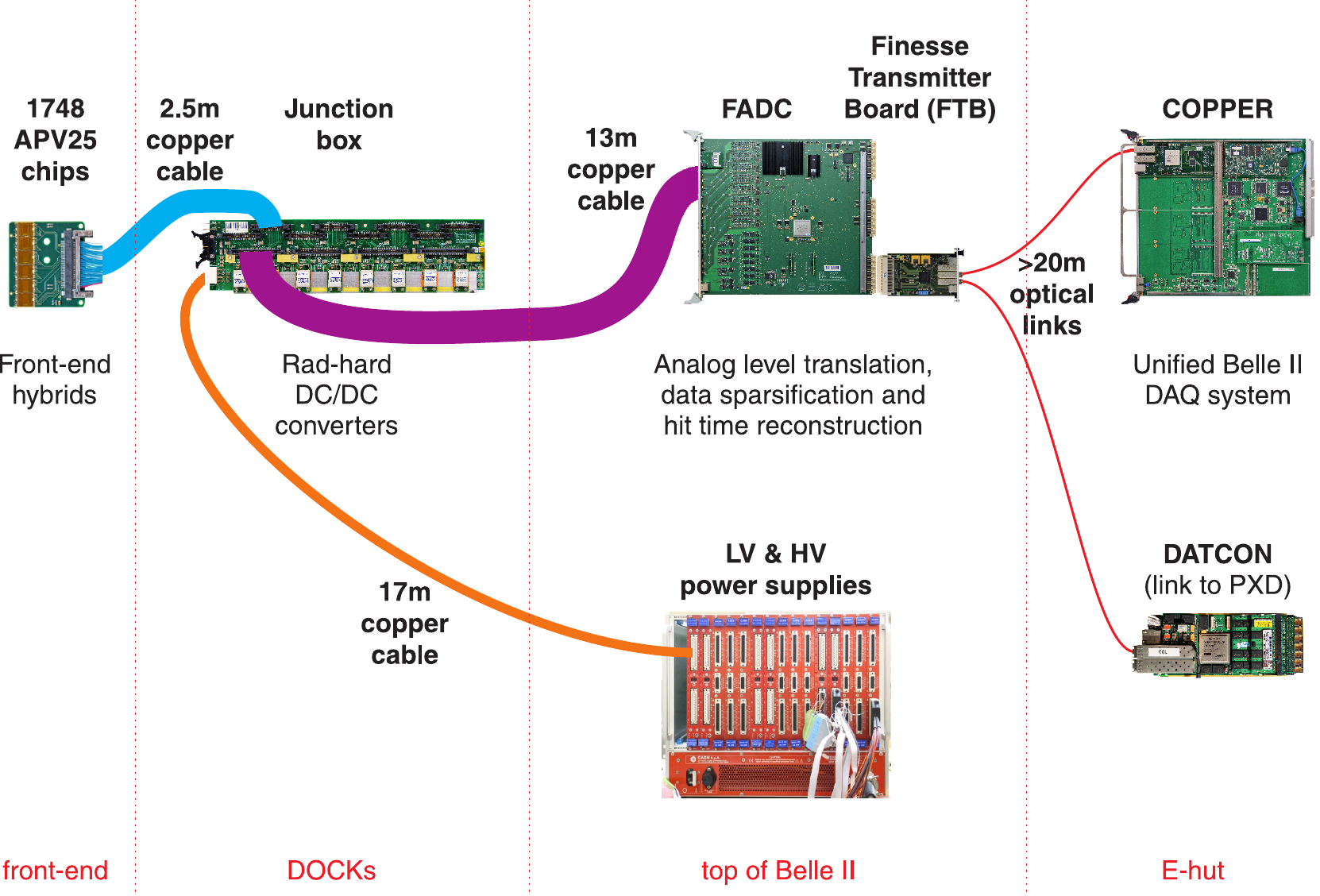}
    \caption{An overview of the elements of the Belle II SVD readout chain.}
    \label{fig:C2_svd_readoutchain}
\end{figure}

The main elements of the SVD readout chain are shown in figure~\ref{fig:C2_svd_readoutchain}. Both the hybrid boards and the origami flexes hosting \apv\ chips are connected by \qty{2.5}{\m} copper cables to \emph{junction boards} located in the forward and backward areas of Belle II, in front of the CDC end-wall, dedicated to interconnections, called \gls{DOCK}s. These junction boards also contain the radiation-hard voltage regulators, which provide the supply voltages needed by the \apv\ ASICs. Analog data are sent from the junction boards to the \emph{\acrfull{FADC}} boards located on top of the Belle II detector. In addition to analog level translation, the \acrshort{FADC}s perform first level data processing and formatting. Digital data are handed over to the \emph{\acrfull{FTB}} boards, which distribute it through optical links to the \emph{\acrfull{COPPER}} boards~\cite{Higuchi:2005yg} of the unified Belle II data acquisition system. Data is also sent to the \emph{\acrfull{DATCON}} boards~\cite{Bernlochner:2017mtp} of the PXD readout system, which use it to project candidate tracks and find regions of interest (\gls{ROIgls}s) in the PXD.

\begin{figure}[hbt]
    \centering
    \includegraphics[width=0.6\textwidth]{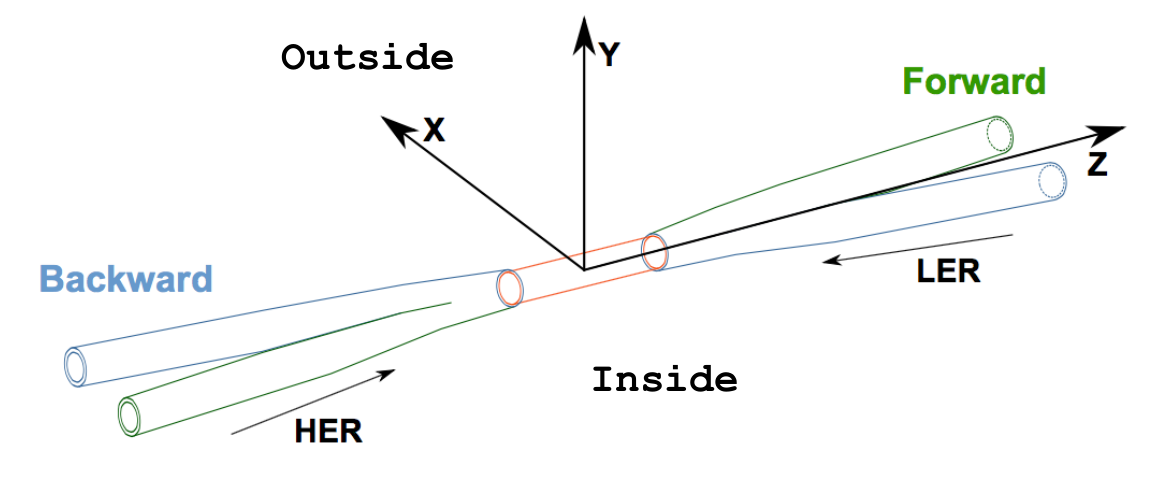}
    \caption{The Belle II reference frame. Electrons circulate in the High Energy Ring (HER), while positron in the Low Energy Ring (LER).}
    \label{fig:C2_Belle2_coordinates}
\end{figure}

 The right-handed global Belle II reference frame, used in figure~\ref{fig:C2_schematic_layout} and repeatedly throughout this report is shown in figure~\ref{fig:C2_Belle2_coordinates}: the origin is located at the nominal \acrfull{IP}; the $z$-axis is directed along the bisector of the angle between the two SuperKEKB beam pipes in the horizontal plane, pointing in the direction of the Lorentz boost (forward direction determined by the  high-energy electron beam); the $x$-axis points towards the outside of the SuperKEKB rings, the $y$-axis points vertically upwards.

The SVD covers all azimuth angles $\phi$ and a polar angle~$\theta$ ranging from 17$^\circ$ to 150$^\circ$. Moving from the in- to the outside, the SVD layers 3, 4, 5 and 6 are located at radii $R$ (measured from the $z$~axis) of \qtylist{39; 80; 105; 135}{\mm}. The DSSD readout strips on the $N$- and $P$-sides, as determined by the silicon diode configuration, read the $z$ and $R\phi$~coordinates, respectively.

\begin{figure}[hbt]
    \centering
    \includegraphics[width=0.5\textwidth]{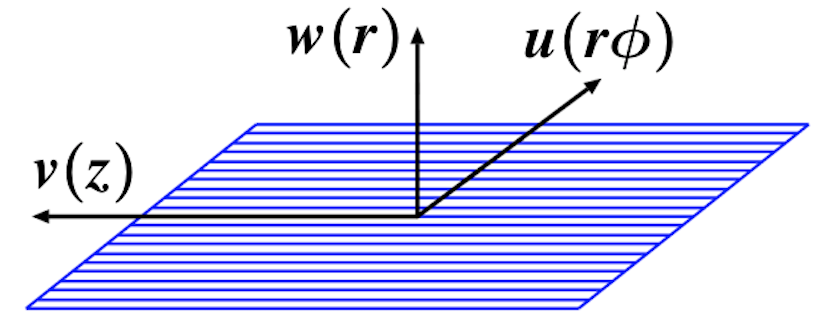}
    \caption{Local coordinates for each DSSD.}
    \label{fig:C2_local_coordinates}
\end{figure}

A local right-handed reference frame is also defined for each DSSD, with a $u$-axis in the sensor plane, pointing in the $R\phi$ direction, a $v$-axis parallel to the global $z$-axis, and the $w$-axis perpendicular to the sensor surface (figure~\ref{fig:C2_local_coordinates}). These local frames are used in the detector alignment as well as in the space point and track reconstruction. In different contexts, the detector sides are also identified as $v/N$-side and $u/P$-side.

\lvltwosection{Silicon Sensors} {\label{sec:si_det}}

%\emph{Writing assignment: Thomas (accepted)}

The design of the silicon sensors is driven by the fact that multiple scattering is a key factor for the Belle II track reconstruction. Thus, the largest available sensors ought to be used to minimize dead material and the design has to be double-sided to achieve a low material budget. At the time of SVD design, typical wafer processing facilities were able to produce six-inch wafers, resulting in a maximum sensor size of approximately \qty{12}{\cm} length and a width of \qty{6}{\cm}. 
Rectangular barrel sensors were fabricated by Hamamatsu Photonics (HPK)~\citeCompany{hpk}, while the production of  trapezoidal sensors for the forward part was carried out by Micron Semiconductor~\citeCompany{micron}.

The double-sided silicon-strip detectors from both companies are manufactured starting from an $N$-type bulk with high resistivity and a thickness of about \qty{300}{\um}, which corresponds to the standard material available in industry.  It also matches the design of the \apv\ front-end ASIC~\cite{French:2001xb} which is designed for capturing the signal charge from \qty{300}{\um} thick silicon sensors.The sensing strips are implanted in the $N$-type bulk either with acceptors or donors, depending on the sensor side. The acceptor-implanted side is the junction side, or ``$P$-side", while the other side is called the ohmic side, or ``$N$-side".
The width of the implants is \qty{14}{\um}  (\qty{30}{\um}) on the $P$-side ($N$-side) of the barrel sensors and  \qty{15}{\um}  (\qty{30}{\um})  on the $P$-side ($N$-side) of the trapezoidal sensors. The implanted strips are connected to a common bias rail using integrated bias resistors on each side with nominal values of \qty{10}{\Mohm}. 
The silicon oxide contains fixed positive oxide charges, which attract electrons and thus create a conductive layer in the silicon shorting all $N$-type doped areas. For that reason, the $N$-strips on the junction side are isolated by a dedicated $P$-implantation surrounding the strips. This $P$-implantation is called $P$-stop.

To improve spatial resolution a floating strip is placed between two readout strips on both $P$- and $N$-sides. The charge induced in the floating strip is shared by the neighboring strips and the effective strip pitch is reduced to half of the readout pitch.

The readout metal strips are placed on top of the implanted strips, separated by a dielectric silicon oxide, hence the readout strips are AC-coupled. The width of the metallization is  \qty{20}{\um}  (\qty{40}{\um})  on the $P$-side ($N$-side) of the barrel sensors and  \qty{25}{\um}  (\qty{40}{\um})  on the $P$-side ($N$-side) of the trapezoidal sensors. 
The capacitance of the AC coupling is 15 and \qty{30}{\pF/\cm} on the $P$- and $N-$side, more than one order of magnitude  larger than the typical interstrip capacitances, that are below \qty{1}{\pF/\cm}. 
The breakdown voltage of the dielectric is typically well above \qty{100}{\V}, with all the AC capacitors tested up to \qty{20}{\V}. These figures are adequate considering that the readout ASICs are operated above the detector bias voltage (the floating power supply scheme, section~\ref{sec:C2_power_supplies}). 
By using this biasing scheme, as explained in section~\ref{sec:C2_power_supplies}, the potential across the AC decoupling capacitor is below  \qty{1}{\V}, thus minimizing the probability of breakdown of the AC coupling. When a broken AC decoupling capacitor between the strip implant and the metal readout electrode is present, called a \gls{pinhole}, the corresponding pre-amplifier channel does not operate correctly, although some compensation are possible as explained in section~\ref{sec:C3_APVDAQ}. Pinholes can be created during sensor production or in the module assembly phases. A few additional pinholes, due to weaker capacitors, can also be generated during beam operation in case of large bursts of radiation delivered when the bias voltage is applied, as explained in  section~\ref{sec:C6_rad_spikes}. 

Table~\ref{tab:sensor_geo} shows the geometrical characteristics of the SVD sensors. All the sensor types carry the same kind of alignment marks (large F-marks in the corners) which can be detected automatically during the assembly process. The designs details of the forward and barrel sensors are described in the next two sections.

%%
% FF: in table below, removed the redundant word sensor, and introduced "rectangular", harmonizing with CH3.

\begin{table}[hbt]
    \centering
    \begin{tabular}{lccc}
        \toprule
        & Small rectangular & Large rectangular & Trapezoidal\\
        \midrule
        Readout strips $P$-side & 768 & 768 & 768\\
        Readout strips $N$-side & 768 & 512 & 512\\
        Readout pitch $P$-side (\unit{\um}) & 50 & 75 & $50-75$\\
        Readout pitch $N$-side (\unit{\um}) & 160 & 240 & 240\\
        Sensor active area (\unit{\mm\squared}) & $122.90\times 38.55$ & $122.90\times 57.72$ & $122.76\times (38.42-57.59)$\\
        Sensor thickness (\unit{\um}) & 320 & 320 & 300\\
        Manufacturer & Hamamatsu & Hamamatsu & Micron\\
        \bottomrule
    \end{tabular}
    \caption{Geometrical details of the double-sided silicon strip sensor types used in the SVD. All sensors have one intermediate floating strip between two readout strips.} \label{tab:sensor_geo}
\end{table} 

\lvlthreesection{Forward sensors} {\label{sec:si_det_Micron}}

\begin{figure}[hbt]
\begin{center}
\includegraphics[width = 0.8 \textwidth]{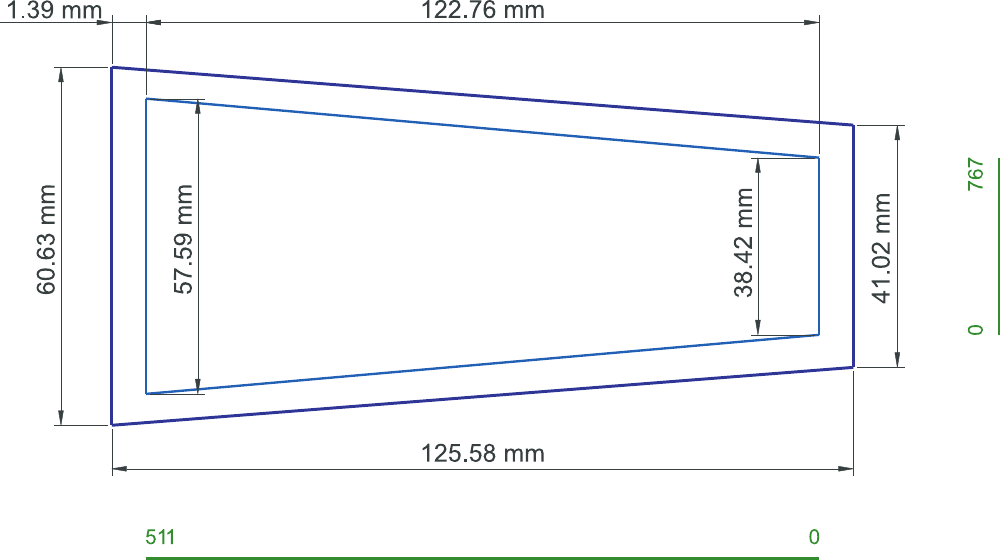}
\caption{Dimensions of the active area of trapezoidal sensors in millimeter. Not shown are the multi-guard ring and the edge ring structures about \qty{1.4}{\mm} from the edge of the active area. The strip numbering for both $P$- and $N$-sides are indicated.}
\label{fig:DSSD_Micron}
\end{center}
\end{figure}

Trapezoidal, \SI{300}{\um} thick sensors are used in the forward, lamp-shade part of the detector. Geometrical parameters are detailed in table~\ref{tab:sensor_geo} and figure~\ref{fig:DSSD_Micron}, while electrical parameters are given in table~\ref{tab:forward_sensor_electrical}. 
The strips of the $P$-side (junction side) feature a variable pitch, giving them a fan shape. The strips of the $N$-side (ohmic side) are parallel to each other and perpendicular to the central strip of the $P$-side.
Electrical strip separation on the ohmic side is achieved by $P$-stop blocking technique, where the so-called ``atoll" variant was chosen, maximizing the signal-to-noise performance both before and after irradiation~\cite{Valentan:2013bua}. 

\begin{table}[hbt]
  \centering
	\caption{Electrical parameters of the trapezoidal sensor~\cite{Abe:2010gxa}}
	\begin{tabular}{lc}
		\toprule
		Quantity                    & Value                \\
		\midrule
		Substrate doping           & $N$-type     \\
		Substrate resistivity & \qty{8}{\kilo\ohm\cm} \\
		Full depletion voltage & 30--\qty{40}{\V} \\
		%Crystal orientation         & $\langle 100\rangle $      \\
		%Full depletion voltage 		& $V_\mathrm{FD}=40$\;V (typ.), 70\;V (max.)              \\
		%Operation voltage           & $V_\mathrm{FD} \ldots 2 \times V_\mathrm{FD}$ \\
		Breakdown voltage           &  > \qty{100}{\volt}      \\
		Polysilicon bias resistor   & 10 $\pm$ \qty{5}{\Mohm}  \\
		%Interstrip resistance ($P$-side)   &  100\;M\ohm (min.), 1\;G\ohm (typ.)  \\
		%Interstrip resistance, ($N$-side)  &  10\;M\ohm (min.), 100\;M\ohm (typ.)  \\
		Coupling capacitance & \qty{100}{\pF}  (typ.) \\
		Dark current & \qty{2}{\uA} (typ.), \qty{10}{\uA}  (max.) \\
		\bottomrule
 	\end{tabular}
	\label{tab:forward_sensor_electrical}
\end{table}

To achieve and maintain a good HV stability and high breakdown voltage, the sensors are designed with metal strips extending a few micrometers beyond the width of the implant strips, causing some of the field lines to end in the aluminum rather than in the implant. This reduces the electric field strength at the edges of the strip implants and takes advantage of the much higher breakdown voltage of the silicon oxide.

An additional measure to improve the reliability is the so called \textit{edge-ring}. This is a heavily $N$ type doped zone close to the cut region of the sensor which  keeps the space charge region away from the imperfections created by the dicing. The edge ring is implanted on both sides and left floating. The edge rings on both sides  has the $N$-side potential, which confines the bias voltage drop to a defined area.

\begin{figure}[hbt]
\begin{center}
\includegraphics[width = 0.8 \textwidth]{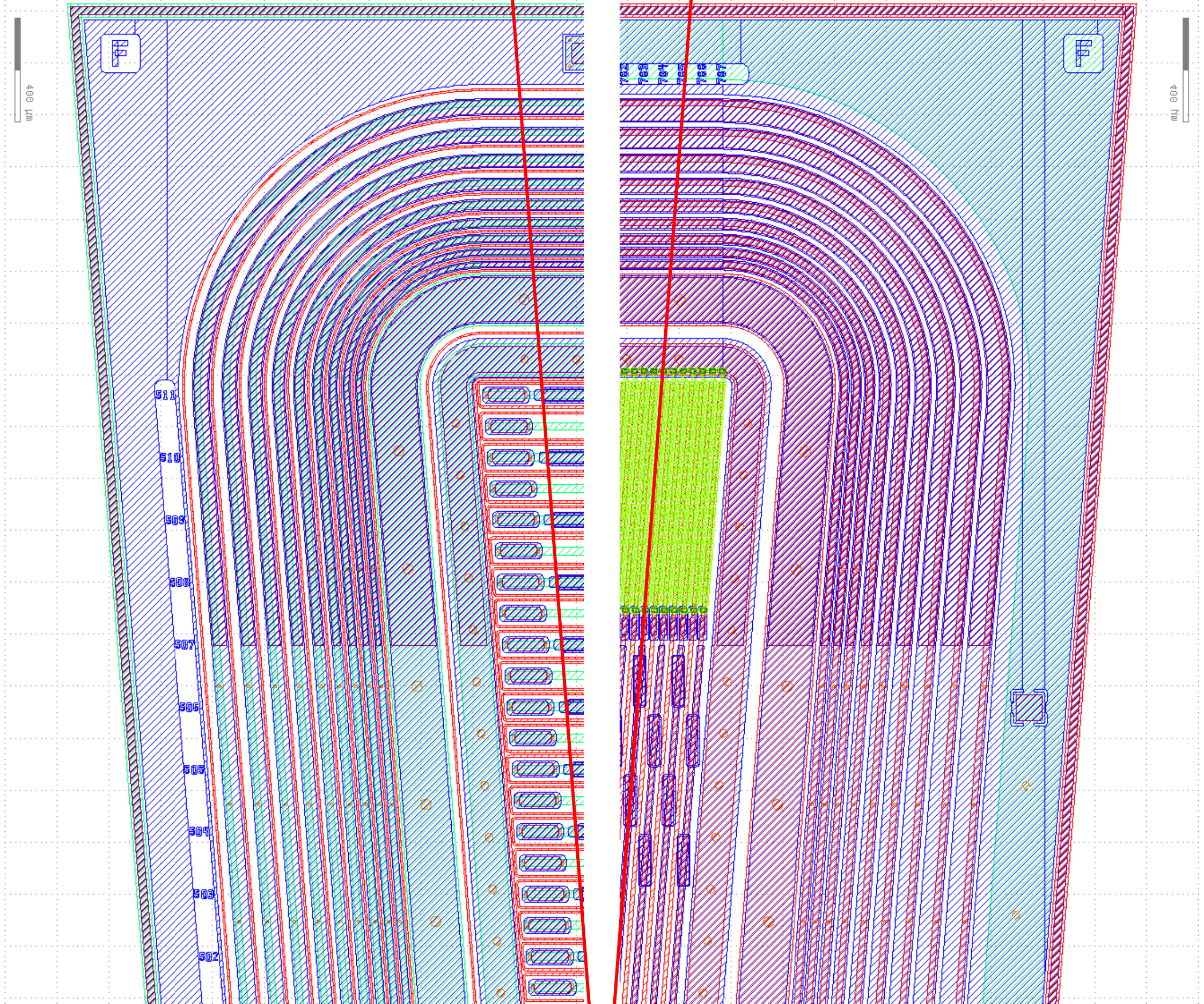}
\caption{Multi-guard ring structure of the Micron sensors for both the $N$-side (left) and the $P$-side (right). The small ruler in the top corners corresponds to \qty{400}{\um}} \label{fig:DSSD_Micron_multiguard}
\end{center}
\end{figure}

In the area between the bias and the edge ring a multi-guard ring structure is realized comprising nine individual floating guard rings. These rings are used to shape the electric field and constitute a special feature of the sensors produced by Micron. A picture of this region is shown in figure~\ref{fig:DSSD_Micron_multiguard}.

\begin{figure}[hbt]
\begin{center}
\includegraphics[width = 0.8 \textwidth]{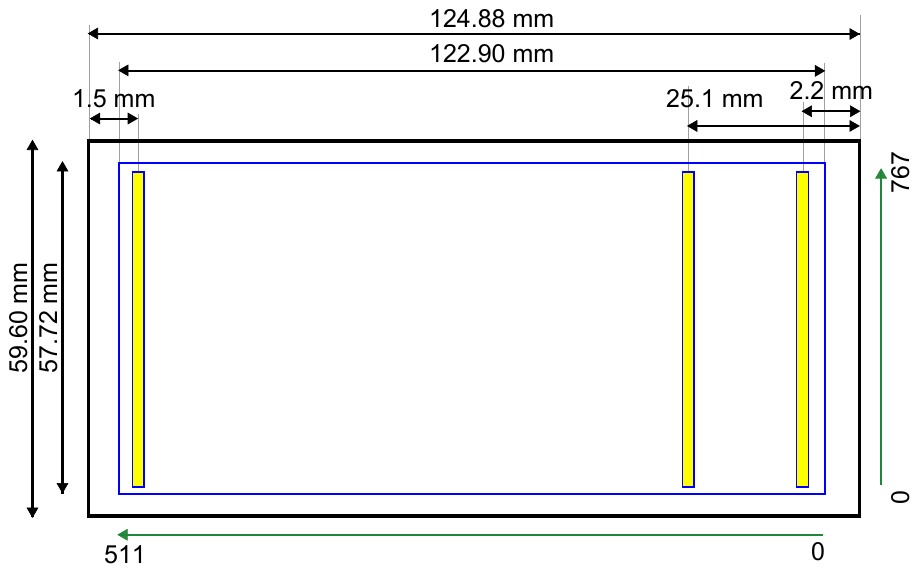}
\caption{Dimensions of the layer 4 to 6 rectangular sensors. The region outside the active area is less than 1 mm wide and implements the edge ring structure. The location of the $P$-side bonding pads rows is indicated by the yellow regions.}
\label{fig:DSSD_F2}
\end{center}
\end{figure}

\lvlthreesection{Barrel sensors} {\label{sec:si_det_HPK}}
Geometrical and electrical parameters are detailed in figure~\ref{fig:DSSD_F2} and tables~\ref{tab:sensor_geo} and \ref{tab:barrel_sensor_electrical}. The barrel sensors are produced from \SI{320}{\um} thick, \SI{15}{\cm} diameter wafers. 
As the silicon wafers become curved in the semiconductor processes, this is the smallest thickness that achieves the flatness required for the production at HPK.
The sensors are made from $N$-type floating-zone silicon wafers with \SI{6}{\kohm\cm} resistivity and the
expected full depletion voltage is about \SI{65}{\V}.
As for the trapezoidal Micron sensors, electrical strip separation on the ohmic side is achieved by an atoll $P$-stop blocking structure. 
The aluminum trace above the implant is slightly wider than the implant to mitigate the border electric field, which causes a micro-discharge resulting in an increase of the strip current.

\begin{table}[hbt]
  \centering
	\caption{Electrical parameters of the barrel sensors.}
	\begin{tabular}{lc}
    \toprule
	Quantity & Value\\
    \midrule
	Substrate doping & $N$-type\\
	Substrate resistivity & \qty{6}{\kilo\ohm\cm} \\
	Full depletion voltage & 50--\qty{60}{\volt} \\
    Breakdown voltage & >\qty{200}{\V} \\
    Polysilicon bias resistor & \qty{4}{\Mohm}  (min.), \qty{10}{\Mohm} (typ.)\\
    Coupling capacitance & >  \qty{100}{\pF} ($P$-side and $N$-side)\\
    %Coupling capacitor duration & > \qty{20}{\V}\\
    Dark current & < \qty{5}{\uA}  at $V_\mathrm{bias}=\qty{120}{\V}$ \\
    \bottomrule
\end{tabular}
\label{tab:barrel_sensor_electrical}
\end{table}

\begin{figure}[hbt]
\begin{center}
\includegraphics[width = 0.9 \textwidth]{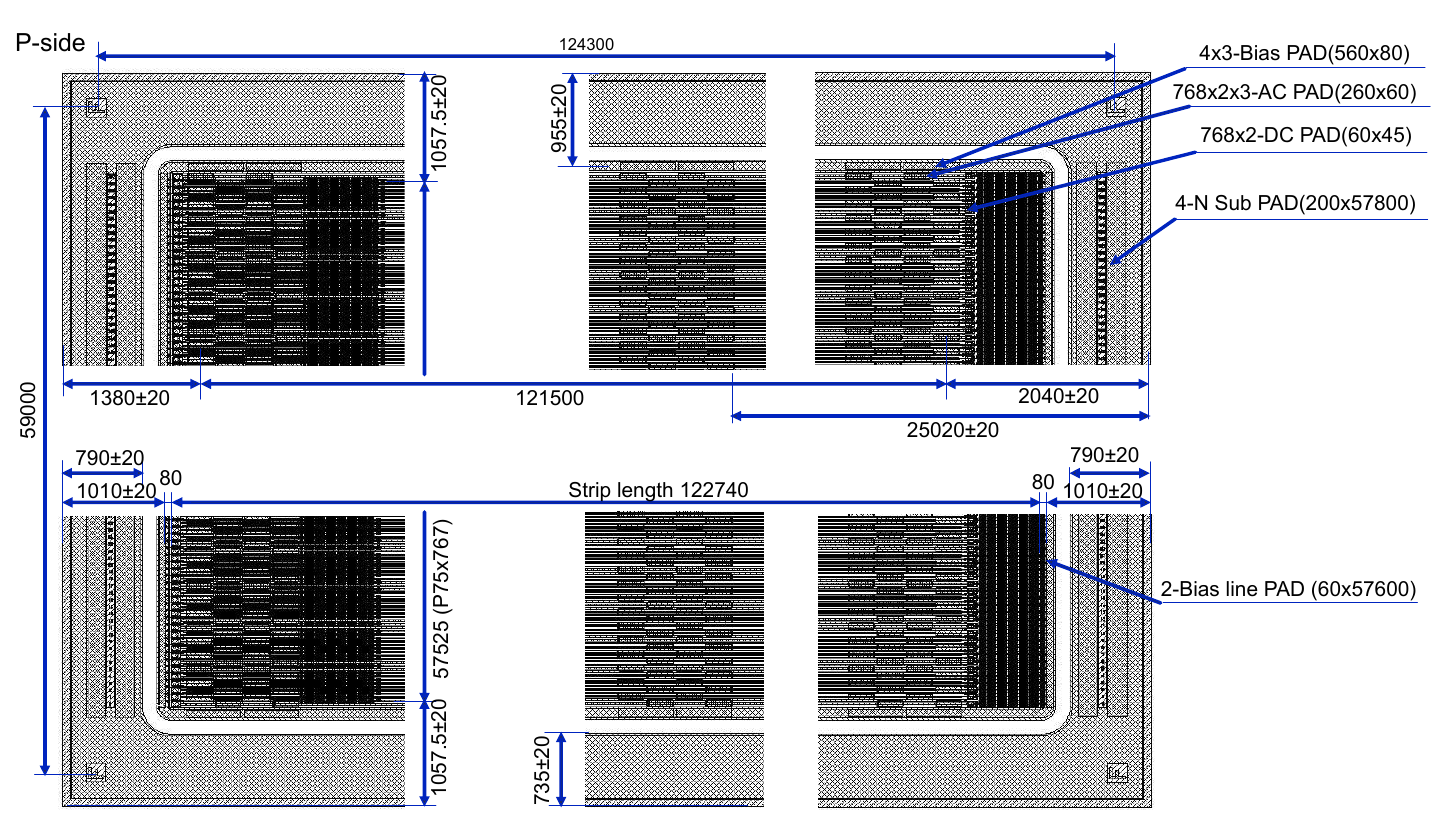}
\caption{Detailed drawing of the $P$-side of a large rectangular sensor.} \label{fig:DSSD_HPK}
\end{center}
\end{figure}

The resistance of the polysilicon bias resistor is typically \qty{10}{\Mohm} to keep the 
parallel noise negligible even given the short-shaping time of the AVP25 front-end. 
  The bonding pads and polysilicon resistors are located inside the sensitive region as otherwise they would couple electrically to the undepleted silicon bulk just below the 
dielectric and cause large additional capacitance. 
In the barrel sensor the dead area outside active region is less than \SI{1}{\mm} wide (figure~\ref{fig:DSSD_HPK}). In this area a heavily $N$-doped edge ring is implemented to prevent the extension of the depleted region to the cut edge. No multi-guard ring structures are present.

Two types of barrel sensors, one for layer~3 (\qty{124.88}{\mm} \mul ~\qty{40.43}{\mm}) and one for layers~4 to 6 (\qty{124.88}{\mm} \mul ~\qty{59.60}{\mm}) are produced (table~\ref{tab:sensor_geo}).
In addition to the wire bonding pads at the sensor edge, bonding pads are added in the middle of the sensor for the origami connection to the readout chips-on-sensor (figure \ref{fig:DSSD_F2}).

%The nominal full depletion and breakdown voltages are \SI{65}{\V} and \SI{180}{\V}.

\lvlthreesection{Radiation tolerance}
\label{sec:C2_rad_tolerance}

At the time of the \belletwo\ TDR a preliminary estimate of the expected dose in the innermost SVD layer was based on the extrapolation from the dose received in the Belle SVD sensors, of about \qty{90}{krad/ab^{-1}}. Assuming \qty{50}{ab^{-1}}  integrated luminosity for the projected lifetime of \belletwo, the radiation dose of the innermost SVD layer was then expected to be 4.5~Mrad.

To include some safety margin, a total dose of 10~Mrad was then initially set as requirement for the \belletwo\ SVD sensors. 

More detailed analyses of the beam background expected in the \belletwo\ SVD, with a dedicated full simulation, were performed in the following years and those results,  summarized in section~\ref{sec:C6_background_studies}, confirmed the previous assumption for DSSD radiation hardness requirements were adequate.  From these background studies it was calculated that at the design luminosity of \qty{8e35}{\per\square\cm\per\s}  the expected average background level in the layer 3 sensors corresponded to 1--3\% strip occupancy, 0.1--0.3~Mrad/yr integrated dose, and an equivalent neutron fluence of 0.2--\qty{0.6e12}{n\ensuremath{_\mathrm{eq}}/\cm\squared/yr}. The interval quoted corresponds to the values from the initial MC background simulation result and from a more refined estimate, that uses background measured on data,  up to 2020 run period,  to re-scale the initial MC prediction. 

Even with this more accurate background estimate, the 10~Mrad radiation hardness requirement, initially set for the SVD sensors, together with a requirement to an equivalent neutron fluence about 
\qty{e13}{n\ensuremath{_\mathrm{eq}}/\cm\squared} can be then considered adequate for 10 years of operation at design luminosity. The radiation hardness of DSSD sensors up to these levels were already proven in the past. 

As an example, the experience from the \babar\ Silicon Vertex Tracker (SVT) DSSD sensors,  produced by Micron with a design similar to the one used for our SVD sensors can be considered. The \babar\ SVT was exposed to a radiation field similar to the one expected in the SVD. The SVT sensors were successfully operated for several years in \babar\ reaching a total integrated dose of about
4.5~Mrad~\cite{TheBABAR:2013jta}. In addition during the lifetime of the \babar\ experiment several irradiation campaigns were performed on the DSSD sensors to futher study the effect of bulk damage, that was not negligible in that background environment. During these tests, described in \cite{Rashevskaya:2002nd} and \cite{Bettarini:2005im}, dedicated test structures, single DSSD sensors, and an entire module,  with its sensors connected to the readout electronics, were irradiated with a 0.9 GeV electron beam, up to about 9 Mrad, corresponding to an equivalent neutron fluence of about \qty{2.7e13}{n\ensuremath{_\mathrm{eq}}/\cm\squared}.  
The measured electrical properties of the sensor and the charge collection efficiency of the full module  proved that the DSSD detectors can be successfully operated without any problem up to the final irradiation step of 9~Mrad. No issues were found  even after type-inversion, reached at about 2--3~Mrad for that sensor with initial depletion voltage of \qty{25}{\V}.  After the final irradiation step a modest decrease of the charge collection efficiency, below 10\%, was reported, together with an increase in the sensor leakage current of about  \qty{2}{\uA/Mrad/\cm\squared} at  \qty{23}{\degreeCelsius}, similar to the observations in the BaBar sensors of about  \qty{1}{\uA/Mrad/\cm\squared} at  \qty{20}{\degreeCelsius}.

Dedicated irradiation campaigns were also performed on the Belle II SVD rectangular and trapezoidal  DSSD sensors, assembled with their readout chips in the origami and rigid hybrids. The modules were exposed to
9.5~Mrad of $^{60}$Co $\gamma$~radiation at SCK-CEN in Mol (Belgium) over one week~\cite{Valentan:2013uha}. 
Before and after this irradiation campaign
the system was tested at a CERN test beam line and the response to minimum ionizing particles was measured. 
Although a moderate increase of the sensor leakage current was observed,
the high signal-to-noise performance was maintained, and no degradation of resolution or the tracking performance was observed. 
In the beam test the system was operated with \COtwo\  cooling and the stability of the origami structure at low temperatures was also confirmed.

\lvltwosection{On-detector Electronics} {\label{sec:C2_ondet_elec}}
Following the design principle outlined at the beginning of this chapter, the chip-on-sensor concept was developed, where the readout chips, thinned down to \qty{100}{\um} to minimize their contribution to the material budget, are mounted on a flex circuit board that is placed on top of the sensors, separated by a layer of foam for thermal and electrical isolation. Moreover, flexible fanouts are wrapped around the edge of the sensor to reach the strips on the bottom side. In that way, it is possible to align all readout chips in a single row, cooled by a single pipe flowing two-phase \COtwo . As a reference to that ``folding'' concept and the country where it is used, that scheme was named ``\gls{origamischeme}".

\begin{figure}[hbt]
 	\centering
 	\includegraphics[width=\textwidth]{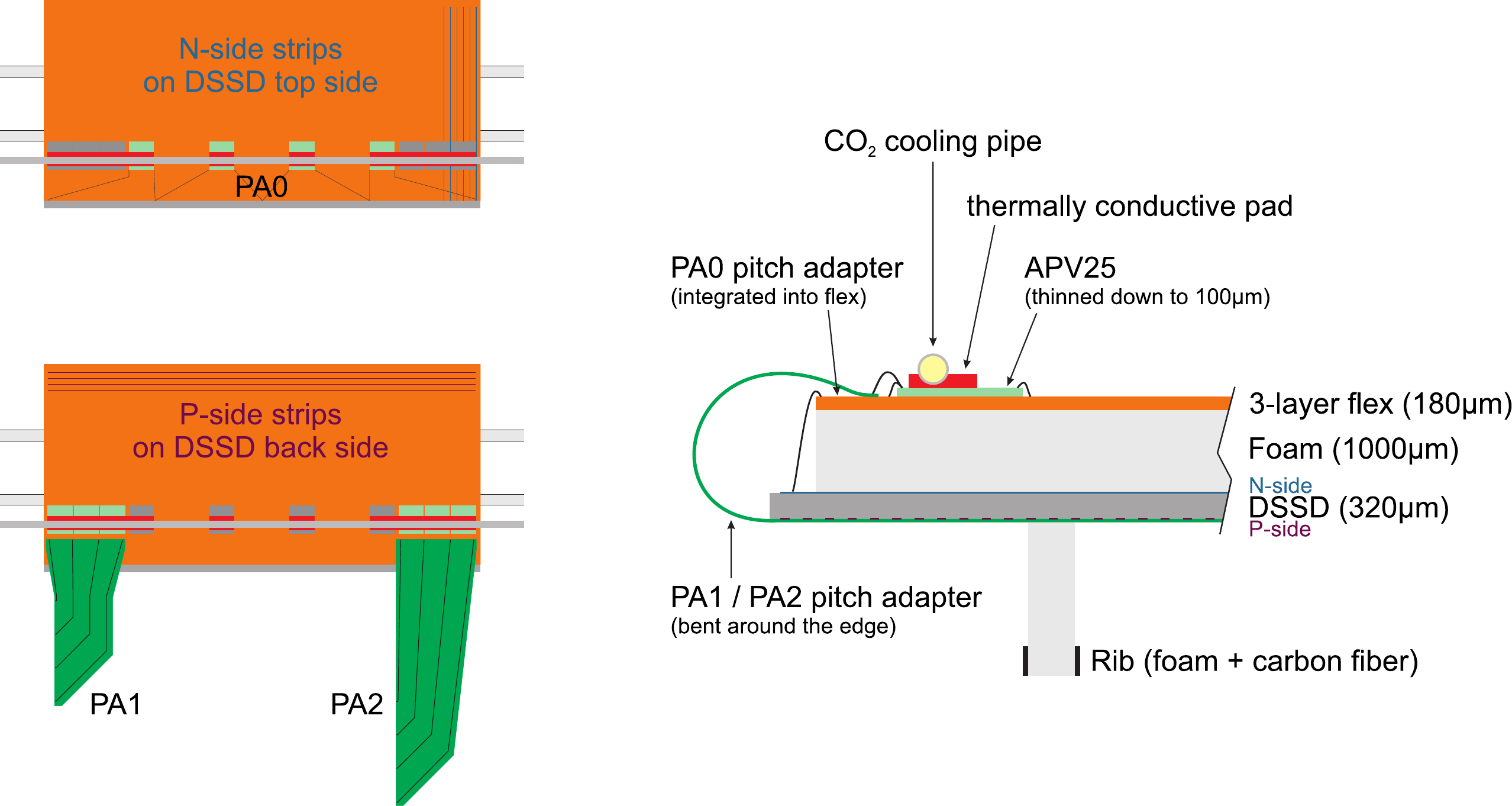}
 	\caption{Structure of the  \gls{origamischeme} . On the left, two views of the sensor are shown, both from the N-side. In the top sketch,  the N-side strips are shown, connected to the pitch adapter PA0 and then to the \apv\ chips. In the bottom sketch, the P-side strips are shown (as if the sensor was transparent), along with the PA1 and PA2 pitch adapters in a flat position. These pitch adapters are folded around the edge of the sensor, as shown on the right in a cross sectional view, so that the P-side strips can be connected to the \apv\ chips. }
 	\label{fig:C2_origami_sketch}
 \end{figure}

\begin{figure}[hbt]
  \centering
  \includegraphics[width=.7\textwidth]{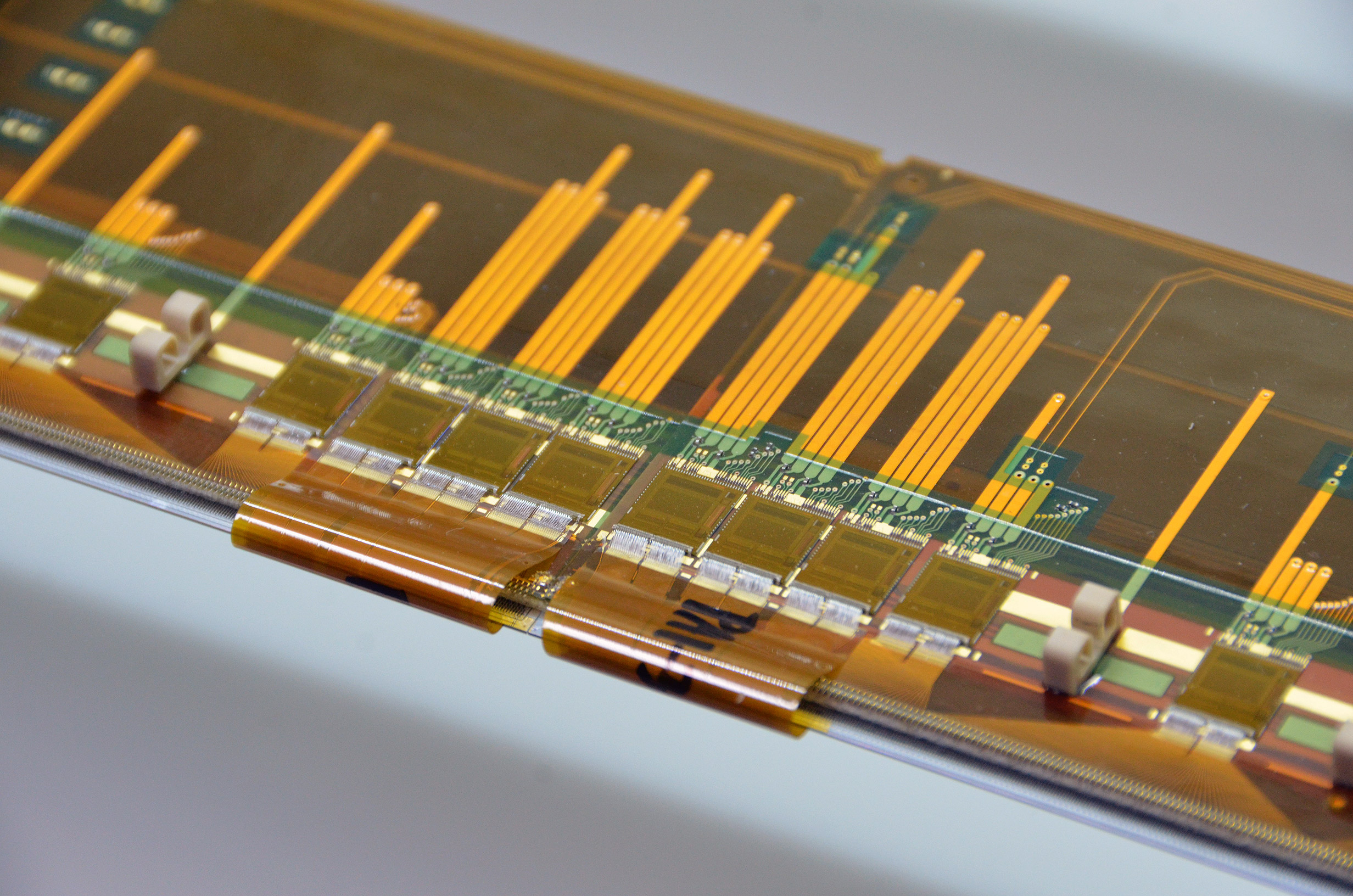}
  \caption{Photo showing a completed \gls{origamimodule} with $P$-side pitch adapters wrapped around the edge of the sensor.}
  \label{fig:C2_origami_photo}
\end{figure}

\lvlthreesection{Origami scheme}
Figure~\ref{fig:C2_origami_sketch} shows the structure of the origami scheme. The \apv\ chips are aligned along Z, in the active region. The u/P strips are readout with two flexible pitch adapters (PA1 and PA2) that bend around the edge of the sensor, while the v/N strips are connected through a short pitch adapter (PA0). The PA1 and PA2 pitch adapters bend over the wirebonds of the sensor to the PA0, making the assembly quite challenging. 
A photograph of a completed origami module is shown in figure~\ref{fig:C2_origami_photo}.

\lvlthreesection{\apv\ front-end chip}\label{sec:C2_apv25}
The \apv~\cite{French:2001xb} is a low-noise charge sensitive amplifier chip in
250~nm CMOS technology. Its final version, APV25S1, was released in the year 2000 and about 70,000 devices were installed in the CMS Tracker. The \apv\ has a shaping time of \qty{50}{\ns} (adjustable), a 192-cell deep
analog pipeline for each of its 128 channels, and a nominal operating clock of \qty{40}{\MHz}. The chip is known to withstand more than \qty{100}{Mrad} of total ionizing dose~\cite{NoahMessomo:2002vto}.

\begin{figure}[hbt]
    \centering
    \includegraphics[width=.7\textwidth]{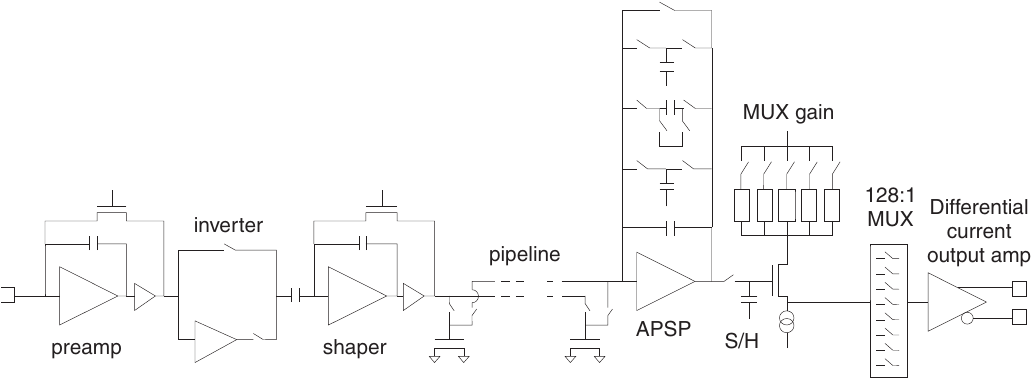}
    \caption{Building blocks of one of the 128 channels of the \apv\ front-end readout chip.}
    \label{fig:C2_svd_apv25}
\end{figure}

The internal structure of the \apv\ is shown in
figure~\ref{fig:C2_svd_apv25}. The usual preamp/shaper architecture can be seen on
the left, where an optional inverter is placed between these blocks. As the
supply rails are limited to $2.5\,\rm V$ in the \qty{0.25}{\um} technology,
this switch was introduced to optimize the dynamic range depending on the
polarity of the signal, i.e., reading out positive ($P$-side, inverter on) or
negative ($N$-side, inverter off) detector currents. Thus, a linear range of
approximately $-2$ to $+7\,\rm MIPs$ (referred to a standard
\qty{300}{\um}-thick sensor) can be achieved for both polarities.

The shaper output is written to a pipeline of 192 cells at the clock frequency.
In fact, the pipeline is implemented as a ring buffer memory with an additional
\acrshort{FIFO} of 32 words to label memory addresses that are requested for output by
pending triggers. Until those data are read out, the tagged cells are skipped
in the write cycle. Consequently, the available pipeline depth can vary between
160 and 192 cells, depending on the number of pending triggers. By
multiplication with the clock period, this translates to the maximum trigger
latency time.

After the pipeline, the \apv\ has an analog pulse shape processor (APSP).  This
is, in fact, a switched capacitor filter that can perform a
so-called \textit{deconvolution}~\cite{Bingefors:1992zr} or, in a different configuration,
simply pass on the pipeline contents. Subsequently, the amplitude can be adjusted to some extent in the so-called MUX gain stage. Finally, the strip data are multiplexed
through three hierarchical stages and sent to the differential-current-mode
output.

Various bias voltages and currents as well as general parameters of the \apv\
can be configured through its \acrshort{I2C} interface. Clock and trigger signals
are received by dedicated differential inputs. The trigger line also accepts
special 3-bit symbols, namely 100=trigger, 110=calibration request, and 101=soft
reset. Consequently, two triggers must be at least 3 clock cycles apart;
otherwise, they would be misinterpreted as a special symbol.

In the Belle II SVD setup, all
\apv\ chips used in the origami chip-on-sensor scheme are thinned down to \qty{100}{\um}, while conventional hybrids located at
the edges of each ladder, which reside outside the sensitive volume, use standard chips with a nominal thickness of \qty{325}{\um}. The \apv\ clock frequency in Belle~II is \qty{31.805}{\MHz}, corresponding to 1/16 of the SuperKEKB RF clock of \qty{508.887}{\MHz}. A trigger latency of 159 clock cycles or approximately \qty{5}{\us}  -- the maximum possible without reducing efficiency for high trigger rates -- is used.

The deconvolution requires clock-synchronous collisions, which is not applicable to Belle~II and therefore  the non-processing \textit{peak} mode is used or, more precisely, the \textit{multi-peak} mode, where three samples are recorded for each trigger. By duplicating the original trigger signal after three clock cycles, six consecutive samples are actually recorded.

A set of three or six samples along the shaped waveform allows to reconstruct both peak amplitude and hit time, provided that the timing is correct and the maximum sample is not at the edge. Especially, the peak time is important to distinguish real hits from off-time background. Initially, only 6-sample data are recorded, but a mixed 3/6 operation is in preparation to save both bandwidth and dead time at high luminosity operation. In that mixed mode, three samples are taken when precise trigger timing is available, while six samples are recorded in other cases -- dynamically switching event-by-event, depending on the trigger source. 
Simulation has shown that this mixture can reduce the dead time fraction~\cite{Nakao:2012aea} from more than 3\% (for 6-samples only) down to less than 1\% at the maximum trigger rate of $30\,\rm kHz$, although background rejection is negatively affected by a deterioration of the time separation between signal and background hits.

\begin{figure}[hbt]
    \centering
    \includegraphics[width=.7\textwidth]{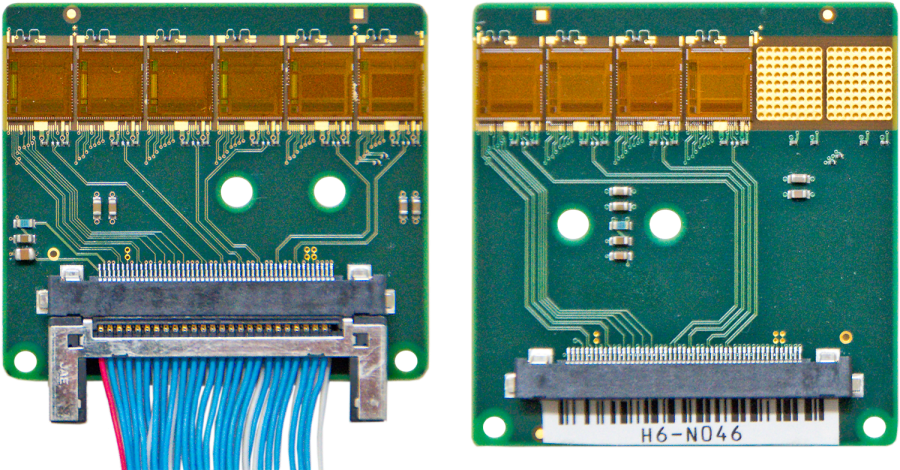}
    \caption{The hybrid boards used in layers 4/5/6 (left: forward $P$-side board with 6~APVs, and right: backward $N$-side board with 4~APVs).}
    \label{fig:C2_hybrids}
\end{figure}

\begin{figure}[hbt]
	\centering
	\includegraphics[width=\textwidth]{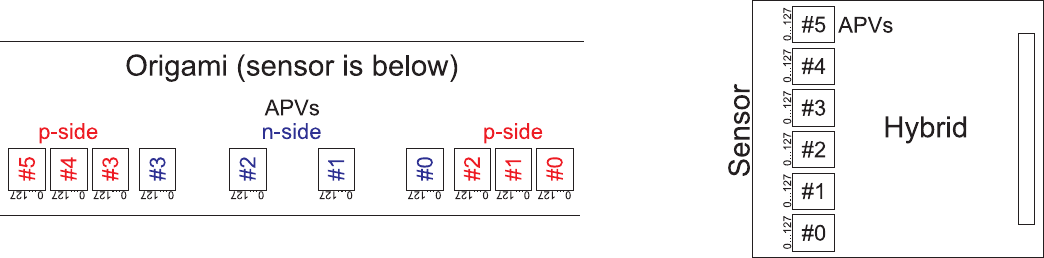}
	\caption{Numbering scheme of \apv\ chips on origami boards (left) and conventional hybrids boards (right). Chips on the P and N-side are numbered independently. The small numbers near the chip edge are the \apv\ channel numbers.}
	\label{fig:C2_origami_hybrid_numbering}
\end{figure}

\lvlthreesection{Hybrid boards}
\label{sec:C2_hybridboards}
The sensors at the edge of each ladder, i.e.\ the so-called \textit{forward} and \textit{backward} sensors, are read out from the ends in a conventional way, using hybrid boards made of glass-reinforced epoxy laminate material (FR4), as shown in figure~\ref{fig:C2_hybrids}. Slightly different variants of those boards exist for $P$- and $N$-sides, forward and backward as well as layer~3. Generally, $P$-side hybrids host six \apv\ chips, while $N$-side types have only four, except for layer~3 which features six chips on both sides. The numbering scheme of the \apv\ chips on the hybrid boards is shown in figure~\ref{fig:C2_origami_hybrid_numbering}.

For ease of construction, each hybrid board is single-sided, meaning that components and \apv\ chips are all attached onto the top side, while the bottom side remains empty. At a later stage, namely the construction of forward/backward modules and L3 ladders, pairs of $P$- and $N$-type hybrid boards are glued together back-to-back with a thin layer of thermally conductive, but electrically isolating, composite sheet of silicone rubber and fiberglass (``Sil-Pad'') in between. Finally, a ground wire is fed through matching vias in the hybrid boards and soldered onto both boards. This ground wire serves as a reference for the return lines of the HV bias supplies and is later attached to the metal structure of the ladder.

Each hybrid board is equipped with a 51 pin connector out of which 50 contacts are actually used~\citeCompany{ref:JAE}.
%n FI-R51S-HF connector by JAE, which has 51~contacts, but only 50 are actually used.
All analog and digital signals, low voltage (LV) power, and high voltage (HV) bias are sent over this connector and the attached halogen-free fine-pitch ($0.635\,\rm mm$) flat cable with twisted pairs.
%hybrid cable (3M 79992-25P-270A), a halogen-free fine-pitch ($0.635\,\rm mm$) flat cable with twisted pairs.
Only in case of L3, where space is too tight for such a connector, the hybrid cables are directly soldered to the hybrid boards. All hybrid cables have the same length of $249\,\rm cm$ to ensure synchronous delivery of clock and trigger signals. Any cable slack is stored near the Junction Boards.

\lvlthreesection{Origami flexible boards}
\label{sec:C2_origamiboards}
For sensors in the central part (one, two or three depending on the layer), the readout circuits are realized with flexible PCBs called \glspl{origamiboard}.
%and the \apv\ chips were thinned down to \qty{100}{\um}.
The origami boards are a complex stack of conductive and insulating layers with a maximum thickness of \qty{282}{\um}: 3 layers of \qty{9}{\um}-thick copper laminated on a \qty{25}{\um}-thick polyimide substrate; epoxy and insulating layers; polyimide coverlays to protect the circuitry.
Unlike the hybrids, the origami boards serve both $P$- and $N$-sides and therefore have two connectors at the far end which are stacked on top of the hybrid boards at the time of module assembly.
The connector region is stiffened by an H-shaped piece of FR-4 glued around the connector to prevent damage at the time of plugging/unplugging.

\begin{figure}[hbt]
    \centering
    \includegraphics[width=\textwidth]{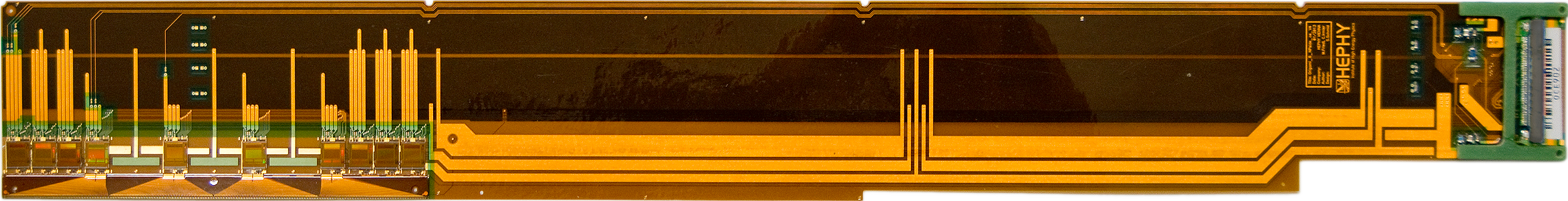}
    \caption{Fully loaded Origami\_CE flexible PCB with dimensions $450.71 \times 57.39\,\rm mm^2$. The 6 P-side and 4 N-side \apv\ are located on the far left of the circuit.}
    \label{fig:C2_origami}
\end{figure}

There are three different versions of \glspl{origamiboard}, serving the (up to three) barrel sensors, labeled according to their Z location: "-Z" or "+Z" and "CE" for central. 
Two of them are read out on the backward side (``Origami\_-Z'' and ``Origami\_CE''), while one flex needed only in L6 is routed over the slanted trapezoidal sensor (``Origami\_+Z''). Figure~\ref{fig:C2_origami} shows a fully loaded \gls{origamiboard} of the longest type (``Origami\_CE'').

On the \gls{origamiboard}, all \apv\ chips are aligned, such that they can be cooled using a single tube flowing dual-phase \COtwo . The four chips which read out the $N$-side strips (facing the flex) are located in the center of each sensor, being complemented by two groups of three chips on each side, which read out the $P$-side strips to be reached by flexible pitch adapters bent around the edge (shown in figure~\ref{fig:C2_origami_sketch}). The numbering scheme of the \apv\ chips on the \gls{origamiboard} is shown in figure~\ref{fig:C2_origami_hybrid_numbering}.

\begin{figure}[hbt]
    \centering
    \includegraphics[width=\textwidth]{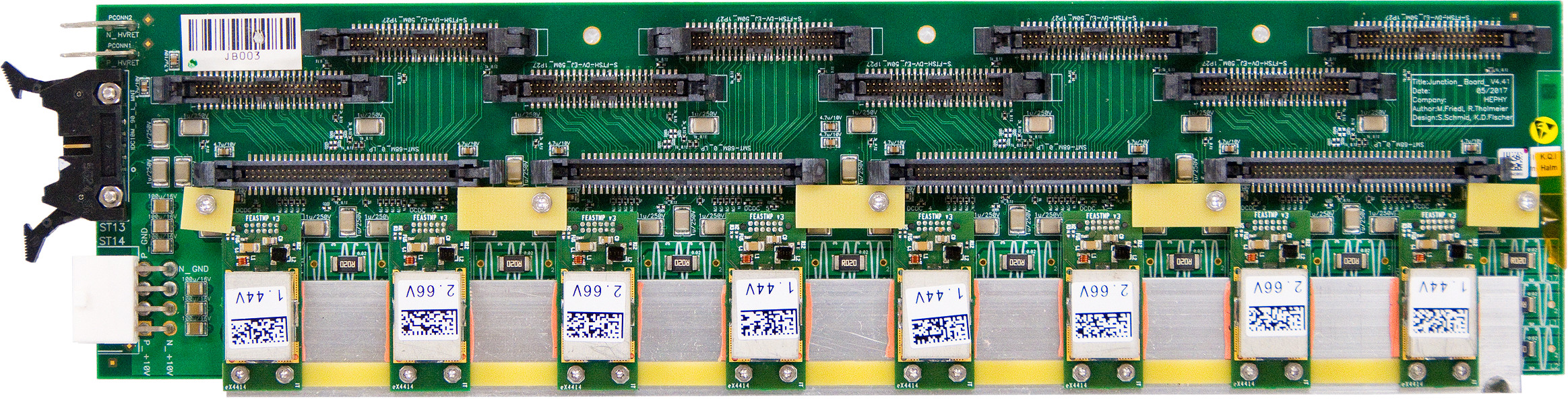}
    \caption{Junction Board with eight DC/DC converter modules.}
    \label{fig:C2_junction_board}
\end{figure}

\lvlthreesection{Junction boards}
\label{sec:C2_junctionboards}
As shown in figure~\ref{fig:C2_svd_readoutchain}, the junction boards serve two purposes: 1) to connect the hybrid cables from the front-end to the data cables to the Flash ADCs (\acrshort{FADC}s) and to the power cables from power supplies, and 2) to reduce the power supply voltages to a level accepted by the \apv\ chips.
The latter task is accomplished by DC/DC converter modules plugged onto the Junction Boards. Figure~\ref{fig:C2_junction_board} shows one junction board, which serves four DSSDs that are powered, biased, and read out on both $P$- and $N$-sides. 
In consequence, there are eight (staggered) connectors for hybrid/origami cables, which are merged onto four broader connectors that carry the data of one DSSD each to the \acrshort{FADC}s. LV and HV power are supplied from the two connectors on the left side.

The circuitry on the junction board ties together the negative side of the LV supplies to the respective HV bias potential on both $P$- and $N$-sides with an adjustable offset voltage \vsep\ (section \ref{sec:C3_APVDAQ}). 
This scheme implies that the front-end low voltages and thus also the output of the readout chips are not at ground level. The FADC boards provide electrical isolation of the chip signals to the ADC.

The outer walls of the CDC are located about \qty{1}{\m} in the backward direction and \qty{1.6}{\m} in the forward direction from the interaction point. Those regions host the \gls{DOCK} boxes of PXD and SVD. One box is used for \COtwo\  cooling on each side.
For the SVD, each \gls{DOCK} box contains six junction boards and has a water cooling pipe at its bottom. Connectors for environmental sensors like the thermal NTC sensors, the diamond radiation sensors and optical fibers (section~\ref{sec:environment}) are mounted onto some of the \gls{DOCK} boxes. 

Since the \gls{DOCK}s are located in a moderate radiation zone (tens of krad) as well as in the magnetic field (\qty{1.5}{\tesla}), specific design considerations must be applied. 
As the \apv\ analog output can drive the signal through up to \qty{25}{\m} long Cu cables, the junction boards do not require repeater functionality, which simplifies the radiation hard design.

Local DC/DC voltage conversion was adopted to reduce the current and the voltage drop on the long cables, thus protecting the front-end chips from potentially dangerous transient over-voltages occurring in conjunction with sudden load variations. It was considered that the enhanced safety was worth the additional  complexity in the junction boards design. 
Although off-the-shelf commercial components operating  in a radiation environment and inside a magnetic field were not available, 
the FEASTMP DC/DC converter modules~\cite{Faccio:2014iya}, developed by CERN for the use inside LHC experiments with tolerance up to \qty{100}{Mrad} and \qty{4}{\tesla}, meets the design requirements. 

Each Junction Board holds a set of four converters, one each for the two voltage levels (\qty{2.5}{\V} and \qty{1.25}{\V}) on the $P$- and $N$-sides. In order to account for a moderate voltage drop along the hybrid cables and \gls{origamiboard} traces, the converters generate nominal output voltages of \qty{2.66}{\V} and \qty{1.44}{\V}, respectively, from the input of up to \qty{12}{\V}. In addition to that primary set, there is a second, redundant arrangement of DC/DC converters,  where each device can be individually activated instead of the primary converter.  
All analog output voltages are monitored as well as the ``Power-Good'' output of each converter, so that DC/DC converter failures
can readily be diagnosed.

\lvltwosection{Detector Module and Ladder Design} {\label{sec:modules}}
%\emph{Writing assignment: Christian (accepted)}

\begin{figure}[hbt]
    \centering
    \includegraphics[width=\textwidth]{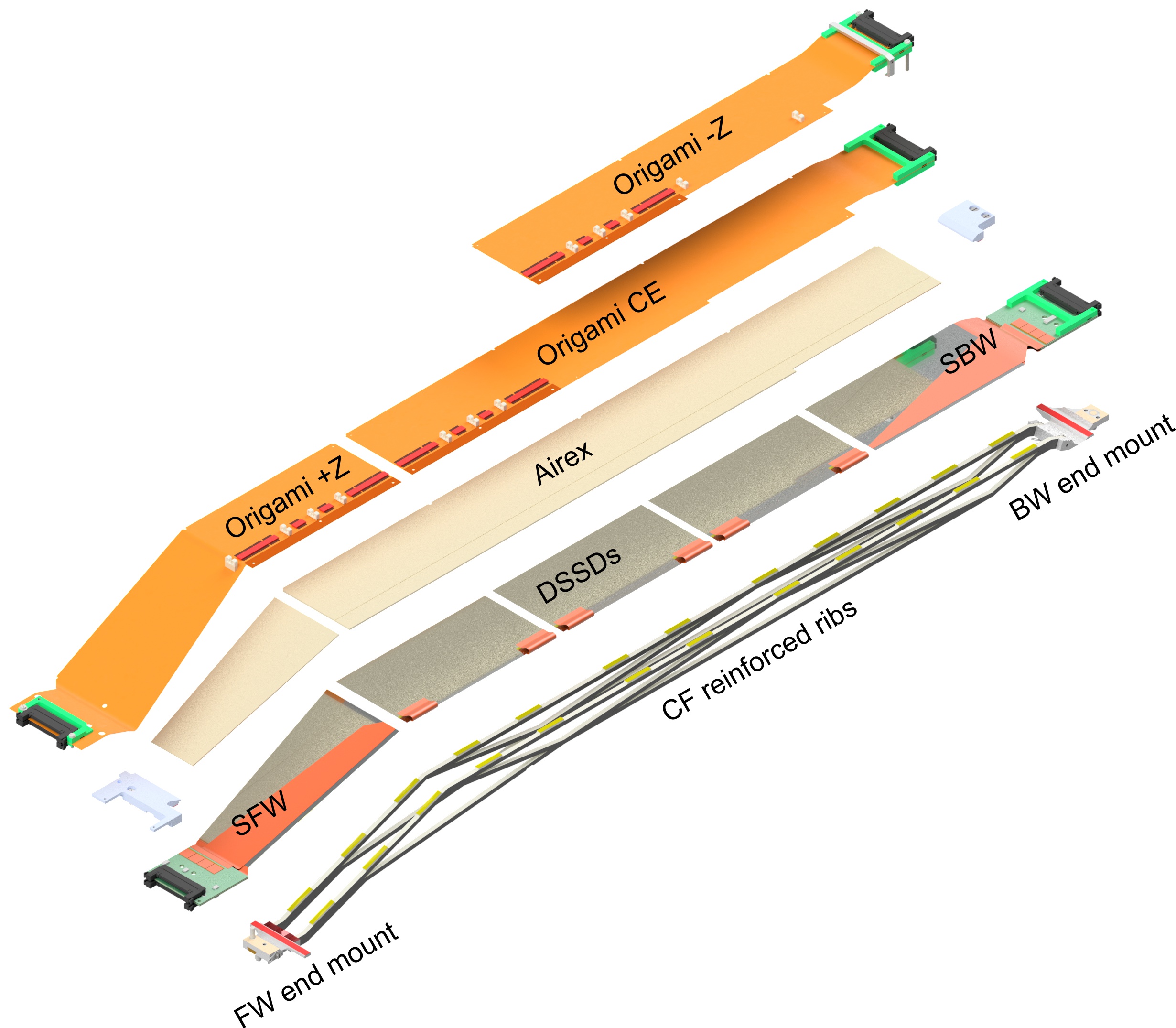}
    \caption{Exploded view of a layer 6 ladder. From bottom to top: rib sub-assembly, four rectangular and one trapezoidal DSSDs, pitch adapters (red), hybrid boards at both ends of the ladder, AIREX\textsuperscript{\textregistered} foam, three versions of origami flexes.}
    \label{fig:C2_l6_exploded_view}
\end{figure}

The ladder design 
aims at keeping the material budget inside the active detector volume as low as possible (figure~\ref{fig:C2_l6_exploded_view}). 
Two carbon fiber (CF) reinforced ribs with a core made of very lightweight foam were chosen as the mechanical support structure, onto which the sensors 
are glued with epoxy. At their ends, the ribs are glued into two aluminum blocks (\glspl{end-mount}), which are fixed to the SVD support structure. Since the readout chips dissipate heat and are actively cooled, the temperature along the ladder can vary from room temperature down to the coolant temperature (\qty{-20}{\degc}), requiring that the ladder support structure must be capable of  compensating the mechanical stress induced by thermal expansion. 
Hence the backward fixation is designed as a solid bearing, while the forward end-mount integrates a sliding mechanism. The mechanical support and cooling are discussed in section~\ref{sec:mech}.

\lvlthreesection{Target mechanical and electrical requirements}
\label{sec:C2_mech_and_elect_requirements}
The ultimate tracking accuracy is determined by the precision alignment obtained in software using actual particles (section~\ref{sec:C7_alignment}). 
The requirements for the mechanical precision are rather 
driven by the mechanical tolerances of the support structures and the narrow space between adjacent ladders. In the mechanical design a minimum clearance of about \qty{1}{\mm} between adjacent components is adopted.  
In order to have sufficient margin, and taking into account that the achievable precision decreases with the length of the ladders, a target tolerance in the $x$ and $y$ directions of $\pm \qty{100}{\um}$ for L3, L4 and L5 and 
$\pm \qty{150}{\um}$ for L6 is defined. For the $z$ coordinate a slightly relaxed tolerance of $\pm \qty{150}{\um}$ in L3, $\pm \qty{200}{\um}$ in L4 and L5, and $\pm \qty{250}{\um}$ for L6 is adopted. These tolerances refer to any reference point on any sensor with respect to its nominal position, measured while the ladder is still mounted on its assembly support.

In the floating powering concept (section~\ref{sec:C2_offdet_elec})  
the ground levels of the \apv\ chips are bound to +HV on the $N$-side and $-$HV on the $P$-side, respectively. It is thus important to ensure that all sensors and hybrid boards are well isolated from each other and also fully disconnected from conductive components of the support structure.  
On the other hand, a common grounding scheme is needed in order to reduce susceptibility to noise of the system.
Therefore, two ground points are defined at the forward and backward end-rings, respectively, where all electrically conductive components of the ladders as well as the HV return lines of the hybrid boards and origami boards  are connected.

\begin{figure}[hbt]
    \centering
    \includegraphics[width=\textwidth]{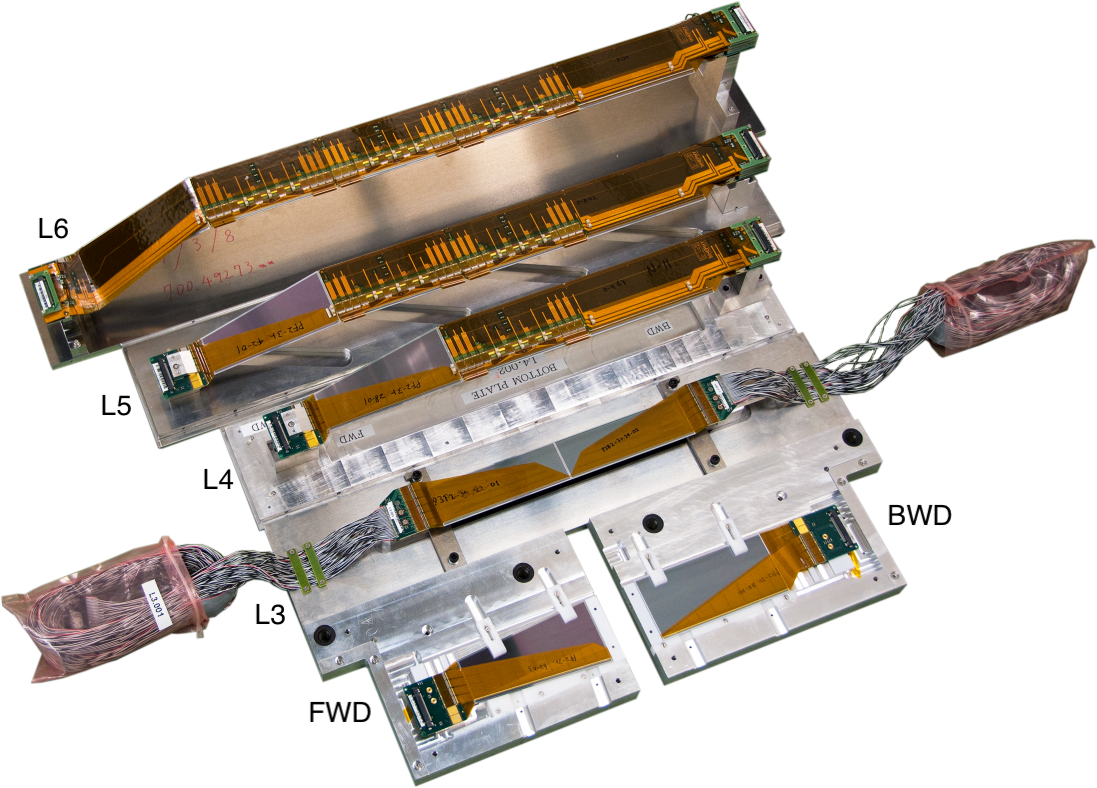}
    \caption{From top to bottom: Ladders of L6, L5, L4 and L3 placed on their transport frames, in the very front the forward and backward sub-assemblies are shown on their transport containers.}
    \label{fig:C2_ladder_design}
\end{figure}

\lvlthreesection{Ladder design}
\label{sec:C2_Ladder_Design}
The basic element of the SVD, a so-called ladder, consists of a mechanical support onto which the sensor modules are glued (figure~\ref{fig:C2_l6_exploded_view}). 
The ladder designs of layers 4, 5 and 6 (L4, L5 and L6) are very similar and 
can be divided into the barrel part with rectangular DSSDs and the slanted forward section with one trapezoidal DSSD. The ladders vary only in the slant angle of the trapezoidal sensor and the number of rectangular sensors in the barrel part. 
In contrast,  layer 3 (L3) ladders consist of two small rectangular sensors and do not have a slanted section. Another unique feature of the L3 assembly is that  the sensors face radially inwards while they face outwards in the other SVD layers. A photograph of all four kinds of ladders is shown in figure~\ref{fig:C2_ladder_design}. 

%- Components of ladders (PA, ribs, etc.) \par
\paragraph{Mechanical ladder support structure.}
\begin{figure}[hbt]
    \centering
    \includegraphics[width=\textwidth]{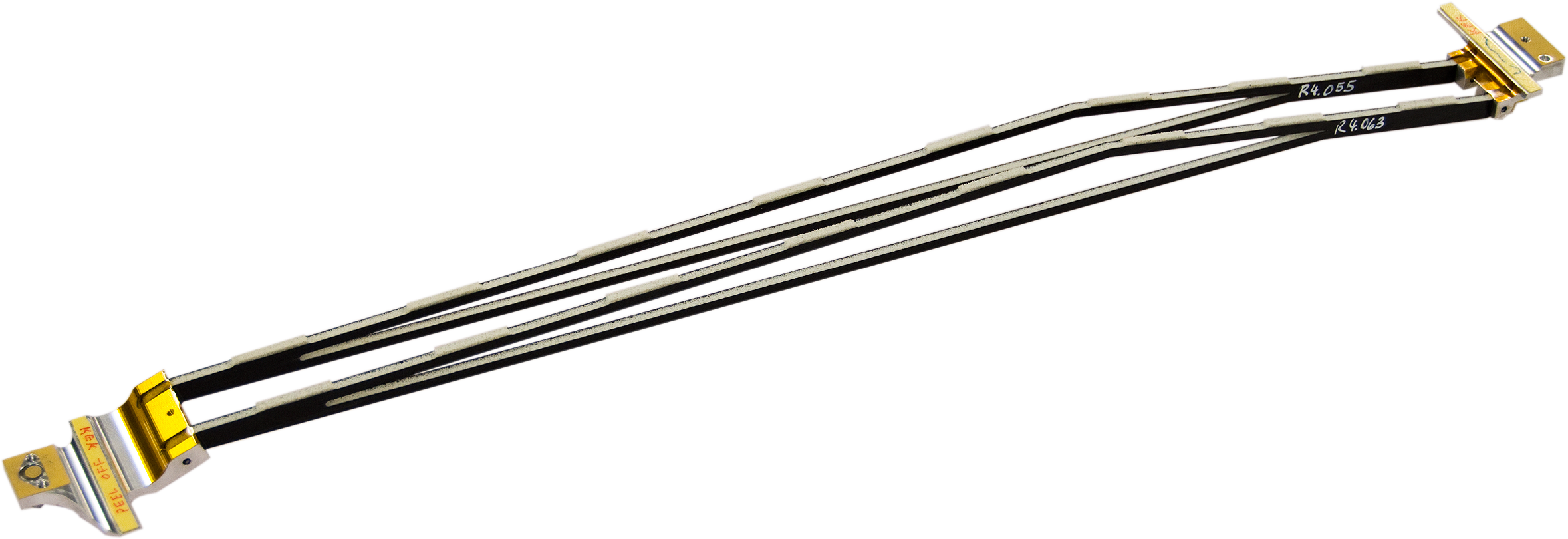}
    \caption{Mechanical support structure of a layer 4 ladder, consisting of a backward end-mount, 2 CF reinforced AIREX\textsuperscript{\textregistered} ribs, a forward end-mount with sliding mechanism and two ``Kokeshi'' pins (refer to the text for a description of these elements).}
    \label{fig:C2_rsa}
\end{figure}
The mechanical support structure of the ladders consists of two ribs made of 
\qty{3}{\mm} thick AIREX\textsuperscript{\textregistered} R82.60, a commercially available polymeric foam with outstanding dielectric properties, excellent stiffness to weight ratio, low water absorption, and high durability in particular in radiation environments~\citeCompany{airex}. The foam core is reinforced by a \qty{105}{\um}  thick carbon fiber (CF) sheet on either side. In L3 a simple bar-shaped rib design is sufficient to achieve the required mechanical stiffness, while for the longer outer layers a truss design is adopted for the ribs (figure~\ref{fig:C2_rsa}).

At their ends, the two ribs are glued into aluminum blocks called \glspl{end-mount}, forming the \acrfull{RSA}.
In L4, L5 and L6 precisely milled pins are inserted into holes located underneath the hybrid boards in the center of the end-mounts. With these pins the ladders are plugged into the precision holes of the end-rings and finally fixed with set screws. These set screws are designed to pull down the end-mounts onto the end-ring surface by a V-shaped notch. Since the shape of the pins looks like a traditional Japanese ``Kokeshi'' doll, these pins were named ``\glspl{kokeshi}" (see figure~\ref{fig:C2_kokeshi}). The fixation of the backward end-mount is designed as a rigid joint and the center of its \gls{kokeshi} defines the mechanical origin of a ladder. On the forward side there is a sliding mechanism in the  end-mounts to compensate for thermal expansion of the ladders:
a prism made of brass is held by the kokeshi-pin, while the end-mount is pushed onto it by spring-loaded stainless steel balls and can slide along the beam axis ($z$) within a range of more than $\pm 1\,\rm mm$. 

Due to geometric constraints, the aluminum blocks of layer 3 have a different design and are longer than those of the outer layers. They extend from the hybrids at both ends of the ladders and are called bridges. 
L3 ladders are fixed with normal M3 screws, without using kokeshi-pins. At the forward bridge an oblong hole allows compensation of thermal expansion.

\paragraph{Pitch adapters.}
Pitch adapters are flexible printed circuits used to connect DSSDs with \apv\ chips. They are produced in different shapes and lengths, depending on the sensor type, side, and position in the SVD layout.  
The electrical connection between the chips and the pitch adapters, as well as between pitch adapters and the sensor strips, is established by ultrasonic wedge wire bonding with \qty{25}{\um}-thick aluminum wire. The main requirements for the pitch adapters are thus mechanical flexibility, flatness, fine pitch, good capability of performing wirebonding on the pads, and electrical quality.
Mechanical flexibility is particularly important for the \gls{origamischeme} and a single copper layer design is chosen to enable bending without peeling off after gluing.
The base material is a 25~\um\ polyimide film with 5~\um\  copper. A coverlay of a 12~\um\  polyimide film with a 15~\um\  glue layer is added on both top and bottom of the base material. 
The bottom coverlay is required to ensure the flatness required for good gluing quality.
\begin{figure}[hbt]
	\centering
    \includegraphics[height=10cm]{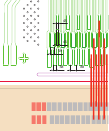}
	\caption{Bonding pad arrangement in three rows for the pitch adapters (top), used to achieve a small bonding pitch (\qty{55}{\um}) while keeping a sufficient bonding pad width (\qty{45}{\um}). The corresponding corner of the APV25 chip is shown below, together with a few wire bond connections in red. (In reality, APV25 and pitch adapter are farther apart.)}	
	\label{fig:C2_pa_three_rows}
\end{figure}
The bonding pad pitch is smallest on the APV~side and amounts to only 
44~\um. To achieve a pad pitch with enough bonding pad width for good wire bonding, the bonding pads are arranged in three rows as shown in figure~\ref{fig:C2_pa_three_rows}.
The minimum pattern width and pitch are 20~\um\ and 15~\um\ on the design, respectively.
For good wire bonding quality, a minimum bonding pad width of 30~\um\ is required, and a thin displaced gold plating is applied onto a nickel plating with a thickness of 3 to 4~\um.

\begin{table}[hbt]
    \centering
    \begin{tabular}{lcl}
	        \toprule
	        Name & Layer & Location\\
	        \midrule
	        P3F1, P3F2 & 3 & Forward sensor, $P$- (1) and $N$-side (2)\\
	        P3B1, P3B2 & 3 & Backward sensor, $P$- (1) and $N$-side (2)\\
	        PF1, PF2 & $4-6$ & Forward sensor, $P$- (1) and $N$-side (2)\\
	        PB1, PB2 & $4-6$ & Backward sensor, $P$- (1) and $N$-side (2)\\
	        PA0 & $4-6$ & $N$-side, on origami boards\\
	        PA1, PA2 & $4-6$ & Wrapped from the origami $P$ to the $N$-side\\
	        \bottomrule
	    \end{tabular}
    \caption{List of the eleven different types of flexible polyimide circuits, the pitch adapters, used in the layer $3-6$ ladders.}
    \label{tab:C2_pitch_adapters}
\end{table}

The name and functions of the eleven different pitch adapter geometries are summarized in  (table~\ref{tab:C2_pitch_adapters}). Their specific use and locations is described in the discussion of the ladder design and construction. 

\begin{figure}[!ht]
 \centering
 \includegraphics[width=\textwidth]{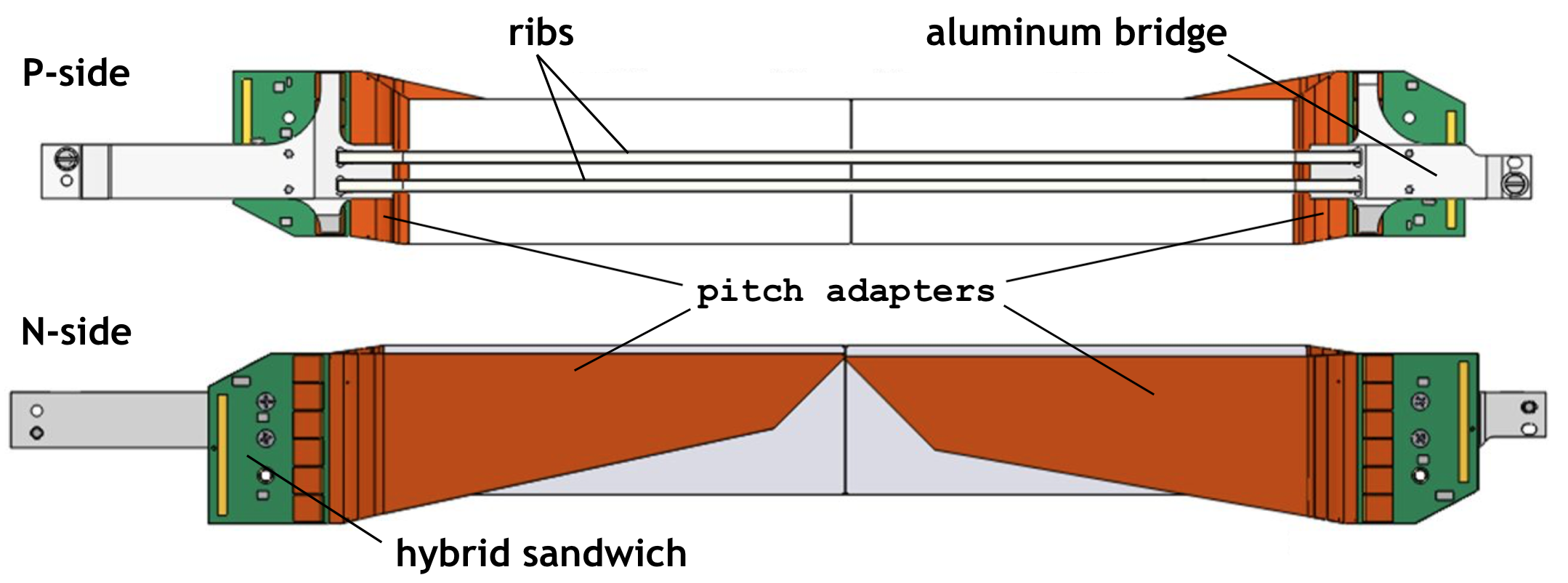}
 \caption{Schematic of a completed Layer 3 ladder, u/P side (top) and v/N side (bottom). The orange shapes are the pitch adapters. }
 \label{fig:C3_L3_ladder}
\end{figure}

\paragraph{Layer 3.}
Layer 3 ladders consist of two small DSSDs, glued on the mechanical support structure, made by two aluminum bridges and two support structures called ``ribs'', made by a 1 mm AIREX\textsuperscript{\textregistered} sheet sandwiched by 0.1 mm carbon fiber reinforced plastic (\acrshort{CFRP}) plates. 
The \apv\ readout chips are located on hybrid boards at both ends of the ladder. There are separate boards for the $N$- and $P$-side chips on each side, glued together with an electrically isolating but thermally conductive foil (Sil-Pad\textsuperscript{\textregistered} 800~\citeCompany{silpad}) in between. These hybrid sandwiches are then screwed onto the aluminum bridges.
The L3 pitch adapters are glued at one end to the hybrid boards in front of the readout chips and at the other end directly to the sensor surface. The heat produced by the \apv\ chips is drained through the bridges, which are screw-mounted to the \COtwo\  cooled end-rings. To improve the thermal contact between the $P$-side chips and the bridges, a $\rm 0.5\,mm$ thick soft and thermally conductive pad (Keratherm Softtherm 86/235~\citeCompany{keratherm}) is inserted. The \apv\ chips on the $N$-side hybrid board are indirectly cooled through the PCB.
A schematic of the full Layer 3 ladder, with the ribs and the bridges highlighted on the u/P side, is shown in figure~\ref{fig:C3_L3_ladder}.

%
% This figure was originally in Chapter 3
%

\paragraph{Layer 4, 5 and 6.}
Since the design of the ladders of L4, L5 and L6 is quite similar, the L6 design is described as an example. It comprises five DSSDs and is the most complex SVD ladder design. An exploded view of an L6 ladder is shown in figure~\ref{fig:C2_l6_exploded_view}.

At both ends of the ladder there are the \acrfull{SFW} and \acrfull{SBW}. Those are very similar to the sensor modules of the L3 ladders and consist of one DSSD, two pitch-adapters and two hybrid boards, each. The hybrid boards (section~\ref{sec:C2_hybridboards}) are conventional PCBs with four \apv\ chips on the $N$-side and six chips on the $P$-side, respectively. Two flexible polyimide pitch adapters are used to connect the sensor strips to the \apv\ inputs. 
As in L3 these pitch adapters are glued onto the hybrid boards and the sensors and connected by wires bonds. 
These hybrid boards are screwed to the forward and backward end mounts, which are located outside the sensitive area of SVD.

In the central section of the ladder, the origami chip-on-sensor scheme is used. Three rectangular DSSDs are glued onto the CF-reinforced ribs with epoxy glue. A sheet of \qty{1}{\mm} thick AIREX\textsuperscript{\textregistered} R82.60 foam~\citeCompany{airex} is glued on top of all sensors, including the forward and backward DSSDs. This AIREX\textsuperscript{\textregistered} sheet acts as a thermal and electrical isolation layer between the DSSDs and the readout electronics. In order to compensate the steps along the surface of the DSSDs, caused by the pitch adapters glued on the forward and backward sensors, the foam is cut from plates of \qty{1.2}{\mm} thick AIREX\textsuperscript{\textregistered} R82.60 and then thermally formed to the final shape and thickness. The readout of the central sensors is done via thin flex PCBs, the \glspl{origamiboard} , introduced in section~\ref{sec:C2_origamiboards}, which are glued on top of the AIREX\textsuperscript{\textregistered} sheet. 
The \glspl{origamiboard} carry only the \apv\ chips and a few capacitors and resistors in the acceptance region. The connectors and larger electronic components are all located at the end of the ladder and stacked on top of the hybrid sandwiches of the forward or backward modules, respectively. Therefore, the Origami\_-Z is glued on top of the tail section of the Origami\_CE flex. The Origami\_+Z flex is attached to the second forward sensor and is routed to the forward end of the ladder.
%In order to minimize the material budget, the \apv\ chips are thinned to \qty{100}{\um}. 
All the \apv\  chips of a ladder are arranged in a row in order to allow attachment of a single thin walled pipe for cooling. To hold the cooling pipe, small clips made from PEEK~\citeCompany{ref:PEEK}, a high performance plastic material, are glued in between the \apv\ chips.

\lvltwosection{Mechanical Support and Cooling} {\label{sec:mech}}
% \editor{Markus}

\begin{figure}[hbt]
\centering
\includegraphics[width=.9\textwidth]{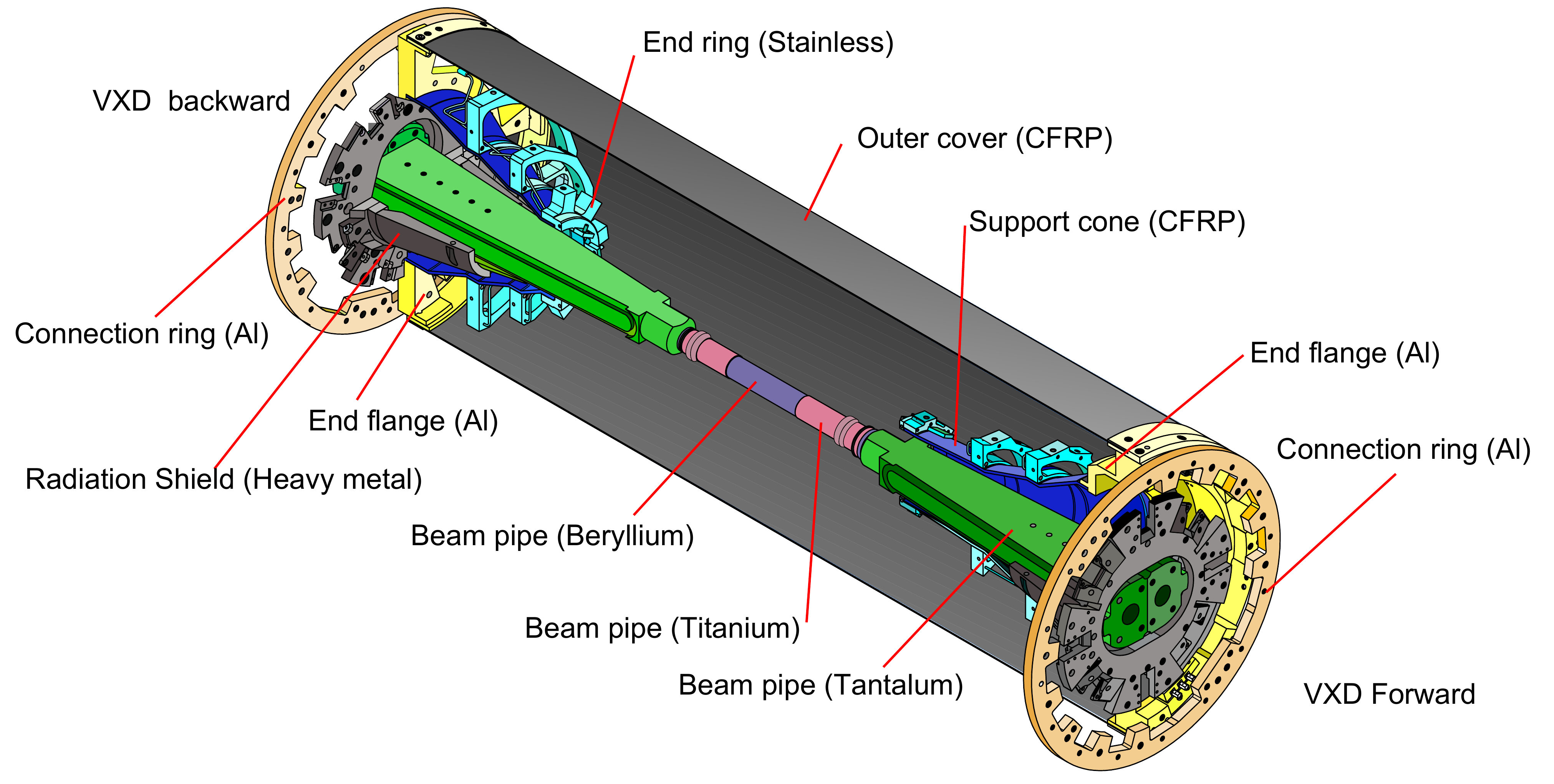}
\caption{Main components of the VXD structure. The SVD and PXD sensors are not shown.}
\label{TT:Endring_F1}
\end{figure}

A schematic view of the VXD mechanical structure is shown in figure \ref{TT:Endring_F1}. 
The rigid VXD cylindrical structure is made of 1 mm thick \acrfull{CFRP} cylinder (outer cover), and the end-flanges made of aluminum on each end.
The structure is divided into two parts along the $yz$~plane (naturally defining the halfs as $+x$ and $-x$) for installation of the SVD sensors. The end-flanges are divided into two halves combined by the connection rings in the forward and backward directions. The VXD components are connected to the end-flanges.
The radiation shield made of tungsten alloy and the tantalum part of the beam pipe weigh \qty{40}{\kg} and \qty{20}{\kg}, respectively, and are the main source of mechanical stress to the VXD structure. 

Inside the cylindrical structure, the end-rings, on which the ladders are mounted, are glued to the support cones, made from \qty{3}{\mm} thick CFRP, which are connected to the end-flanges. The beam pipe is supported near the center of gravity of the forward and backward tantalum chamber to reduce stress to the beryllium pipe.
The CFRP used for the outer cover and support cones~\citeCompany{TT:CFRPCOVERS} has tensile elasticity \qty{212}{\GPa}, shearing strength \qty{42}{\MPa}, and thermal expansion coefficient \qty{3}{ppm/\degreeCelsius}.

\begin{figure}[hbt]
    \centering
    \includegraphics[width=.9\textwidth]{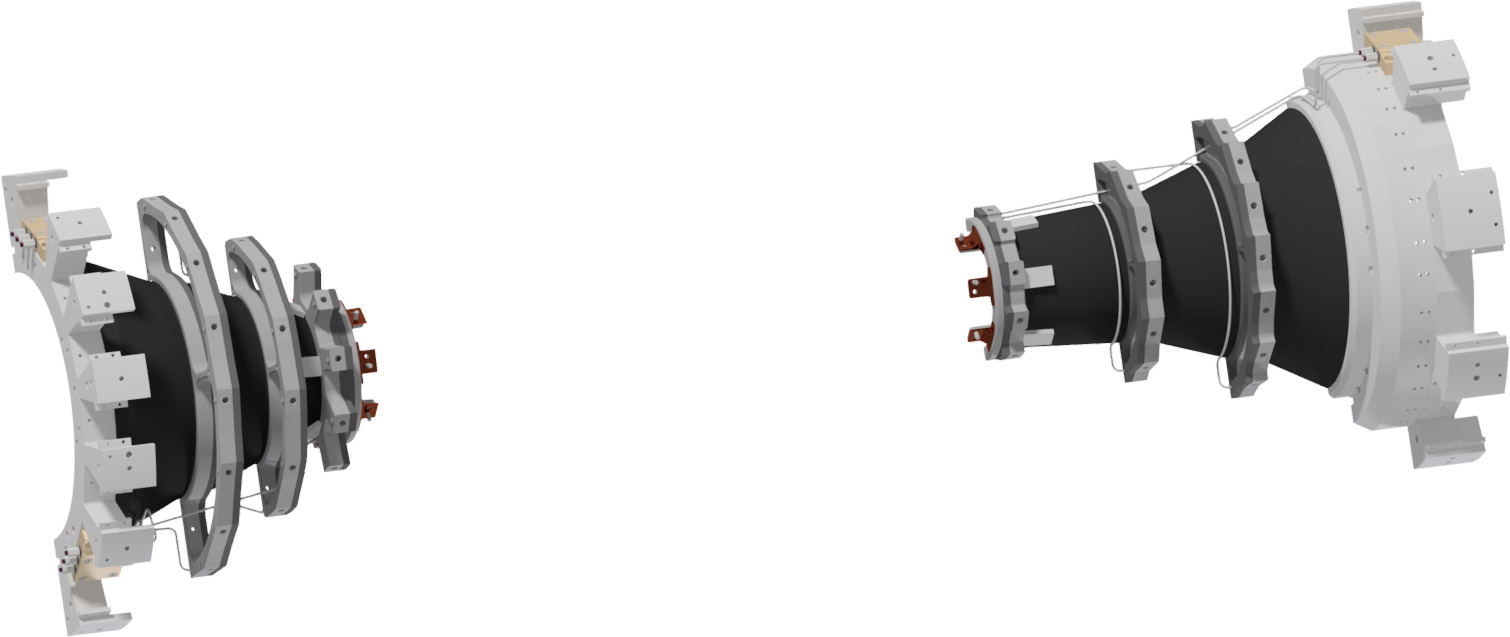}
    \caption{Support structure of the $+x$ half of the SVD, composed of carbon fiber cones, stainless steel end-rings, and aluminum end-flanges. Cooling pipes are interfaced through ceramic isolators (in beige plastic holders) at the end-flanges. The BWD (left) and FWD (right) parts are later connected by the outer-cover (not shown).}
    \label{fig:C2_support_structure}
\end{figure}

\begin{figure}[hbt]
    \centering
    \includegraphics[height=4.6cm]{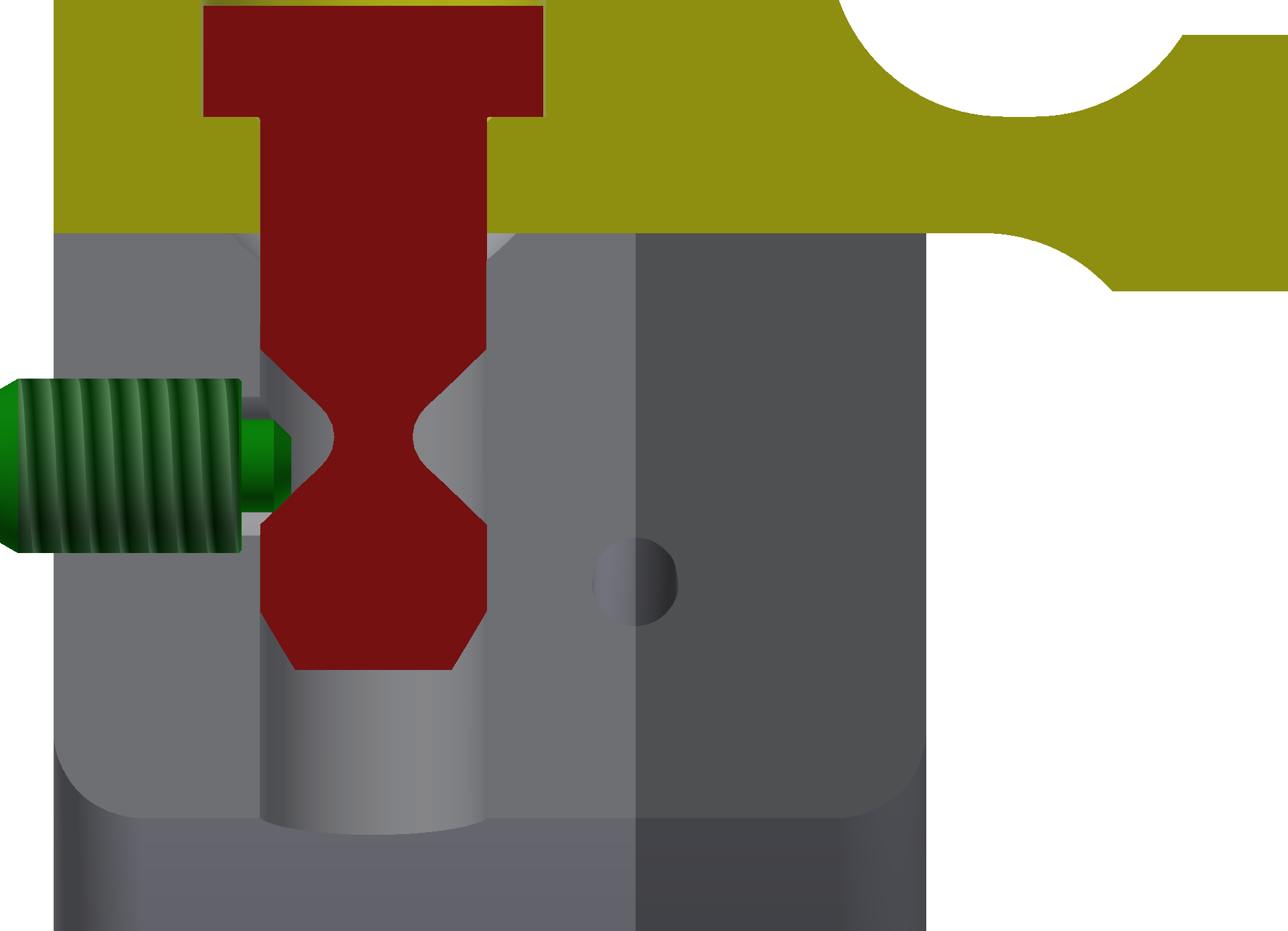} \quad \quad \quad \includegraphics[height=4.6cm]{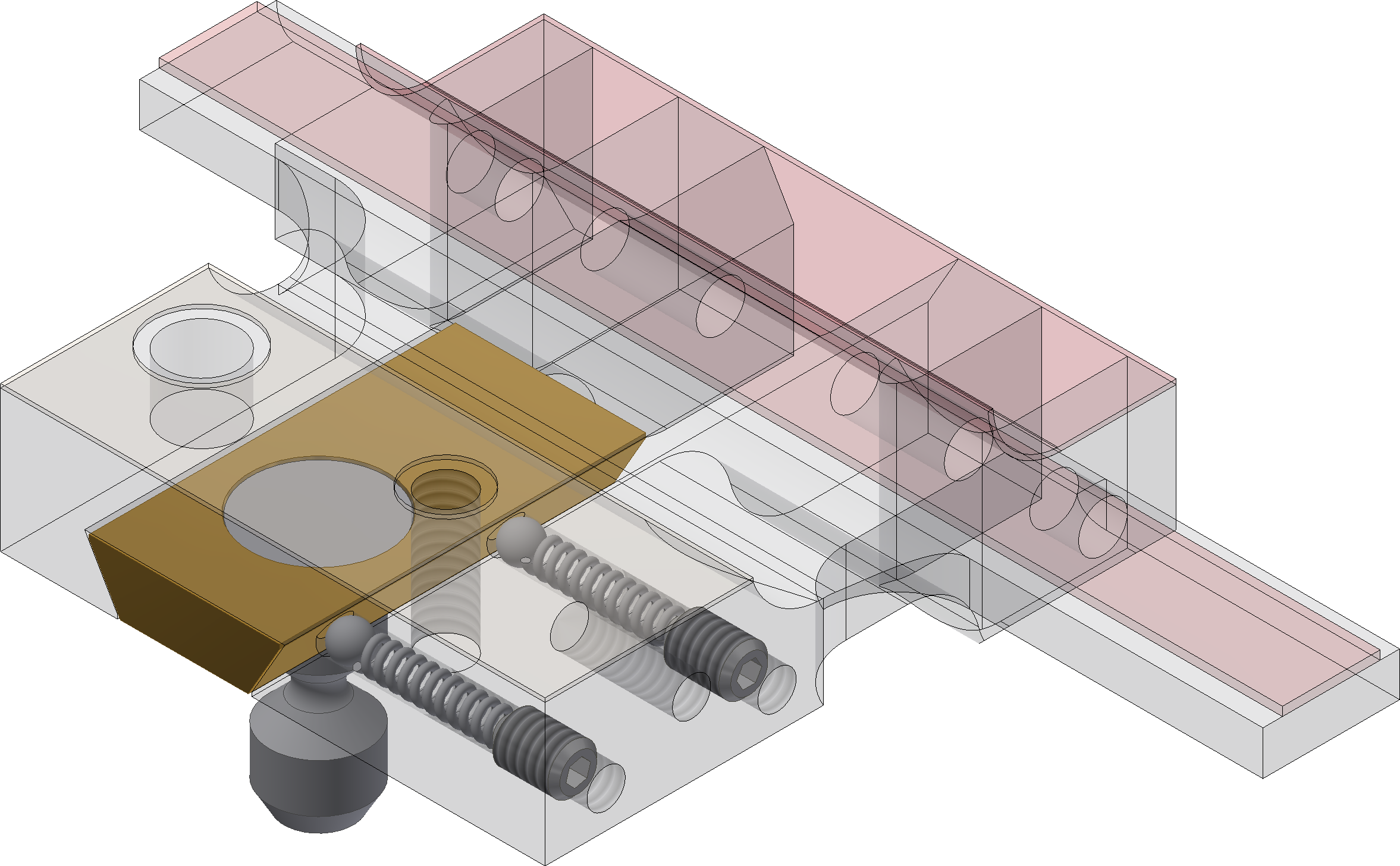}
    \caption{Left: cross section of the BWD ladder mount mechanism: end-mount (yellow) and ``Kokeshi'' pin (red) belong to the ladder, while end-ring (grey) and set screw (green) are support parts. The embedded cooling channel between the two diffusion-welded parts of the end-ring is clearly visible. Right: On the FWD side, a spring-loaded sliding mechanism allows relative motion along the beam axis ($z$) while maintaining a thermal contact.}
    \label{fig:C2_kokeshi}
\end{figure}

The ladders are mounted onto the end-rings by precision ``Kokeshi'' pins on each ladder that correspond to precision holes in the end-rings. All the support structures described in this section are located outside of the active volume and material budget is of less concern. 
Cooling channels integrated into the end-rings carry dual-phase \COtwo\  that is used to remove the heat dissipated by the \apv\ chips on the hybrid boards. 
Figure~\ref{fig:C2_support_structure} shows one half of the support structures mentioned above. The two halves are populated with ladders individually and later combined around the PXD. 

\begin{figure}[hbt]
    \centering
    \includegraphics[width=.7\textwidth]{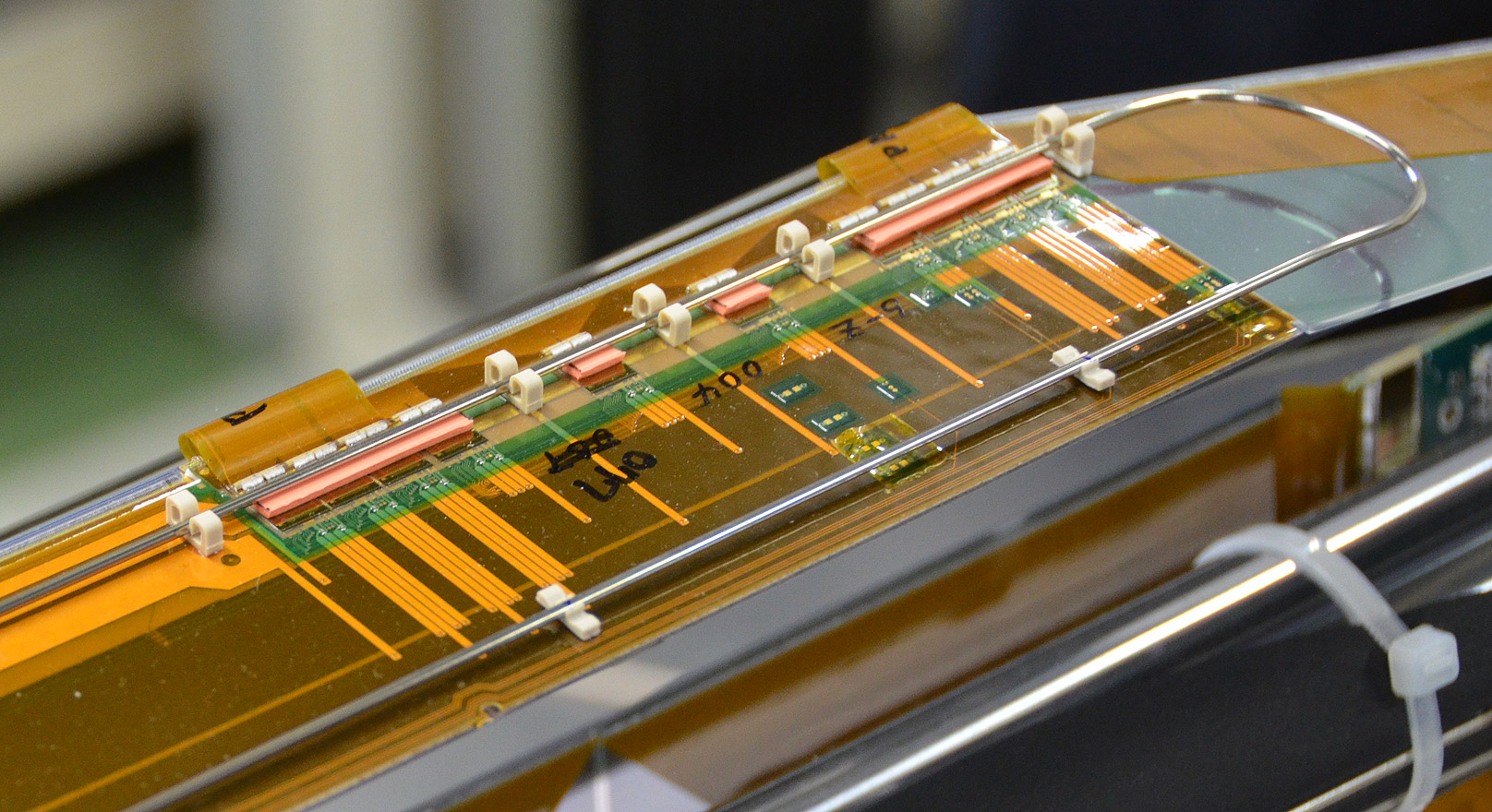}
    \caption{The \gls{origamipipe} is held by PEEK clips and heat is transferred from the \apv\ chips to the pipe through heat conductive soft pads. Due to the odd number of ladders, the pipe needs to be returned on the last ladder of L4.}
    \label{fig:C2_origamipipe}
\end{figure}

The \apv\ chips of the \glspl{origamiboard}  are cooled by a meander-shaped pipe, called \gls{origamipipe}, which is attached onto a half-layer once all its ladders are mounted (figure~\ref{fig:C2_origamipipe}). 
The circuits of layers 4 and 5 are connected in series, while layer 6 has its own circuit. The full list of cooling circuits is given in table~\ref{tab:C2_cooling_circuits}. A special miniature connector, originally developed for the CMS Pixel Detector and named ``Streuli'' after its inventor, is used for the interconnections of cooling pipes in the active region, where space constraints prohibit the use of bulky commercial products. Ceramic insulators are located in dedicated pockets of each end-flange and electrically separate the cooling pipes inside the SVD volume, which are grounded to the end-flange, from the outside world.

% cooling circuits

\begin{table}[hbt]
\begin{center}
\caption{List of SVD cooling circuits. There is a small asymmetry between $+x$ and $-x$ because of the different number of L3 ladders (4 and 3, respectively). A nominal power consumption of $0.4\rm\, W$ is assumed here for each \apv\ chip.}
\begin{tabular}{llrr}
\toprule
Half & Name & \apv\ chips & Power [W] \\
\midrule
\multirow{4}{*}{$+x$} & BWD\_end-ring & 238 & 95.2 \\
%\cline{2-4}
 & FWD\_end-ring & 238 & 95.2 \\
%\cline{2-4}
 & L45\_Origami & 170 & 68.0 \\
%\cline{2-4}
 & L6\_Origami & 240 & 96.0 \\
\midrule
\multirow{4}{*}{$-x$} & BWD\_end-ring & 226 & 90.4 \\
%\cline{2-4}
 & FWD\_end-ring & 226 & 90.4 \\
%\cline{2-4}
 & L45\_Origami & 170 & 68.0 \\
%\cline{2-4}
 & L6\_Origami & 240 & 96.0 \\
\midrule    
Total & & 1748 & 699.2 \\
\bottomrule
\end{tabular}
\label{tab:C2_cooling_circuits}
\end{center}
\end{table}

\lvlthreesection{Cones and end-rings}
% cones: carbon fiber
% end-rings: stainless steel, diffusion welded with internal cooling pipe
\label{sec:C2_cones}
The carbon fiber cones are shaped to fit the boundary of the PXD cable space underneath. Due to the lantern-like structure of the SVD, the FWD cone is significantly longer than its BWD counterpart. In total, there are four half-cones, one each for $+x$ and $-x$ halves as well as BWD and FWD, respectively.

For the end-rings, stainless steel 316L (1.4404)~\citeCompany{SS316L} was chosen because of its  relatively low \acrshort{CTE} (\qty{16}{ppm/\degreeCelsius}), high machinability, and small magnetic permeability that make it suitable to be used in the \qty{1.5}{\tesla} magnetic field of Belle II. The actual mount points for ladders are precision holes in the end-rings which receive the ``\Gls{kokeshi}'' of the ladders in layers 4, 5 and 6. The set screw presses onto the V-shaped notch of those pins and push the ladder end-mounts against the end-ring surface, as shown in figure~\ref{fig:C2_kokeshi}. The end-rings are made from two diffusion-welded stainless steel pieces with an internal cooling channel to remove the heat from the ladder end-mounts. The end-rings of layers 5 and 6 have large openings to accommodate the hybrid cables of the inner layers. Layers 3 and 4 are mounted onto a common end-ring. In contrast to the ``Kokeshi'' scheme, L3 ladders use straight precision pins and are screwed from top. Copper inserts are placed at the L3 mount points to supplement the limited thermal conductivity of stainless steel.

A ground braid connects each end-ring to the end-flange which serves as the local grounding point. Two NTC temperature probes are attached to each end-ring, and three diamond radiation sensors are glued onto each cone between the L3/4 and L5 end-rings. There are two nitrogen inlets (for creating the dry detector volume) and one suction pipe (for dew point monitoring) on each half on the FWD side, and one nitrogen inlet and one suction pipe on the $-x$ BWD half. The environmental and radiation monitors are described in chapter~\ref{sec:environment}.

\lvlthreesection{Origami pipes}
% stainless steel, 1.6mm outer diameter, 0.1mm wall thickness, (un)mountable with ladders installed
% clips (photo), Keratherm
Once all ladders of a half-layer (4/5/6) are mounted, a cooling pipe (\qty{1.6}{\mm} outer diameter, \qty{0.1}{\mm} wall thickness) with a meander-like shape following the surface of the ladders is attached to each ladder using custom-made clips made of PEEK. Soft and thermally conductive pads are placed between \apv\ chips and the pipe to efficiently transfer the heat from a flat surface to the round pipe. The clips allow to remove the pipe at a later stage and access the individual ladders if needed. 

Half-layers 5 and 6 have an even number of ladders, such that the inlet and outlet of the pipes are at the BWD end of the straight sections of first and last ladder. L4 however has an odd number, such that the pipe needs to be routed back to BWD on the last ladder, as shown in figure~\ref{fig:C2_origamipipe}.

\lvlthreesection{Outer-cover}
% carbon fiber, covered with aluminum foil (conductive shield)
% ensures mechanical relation between BWD and FWD

\begin{figure}[htbp]
\begin{center}
\includegraphics[width = 0.7 \columnwidth]{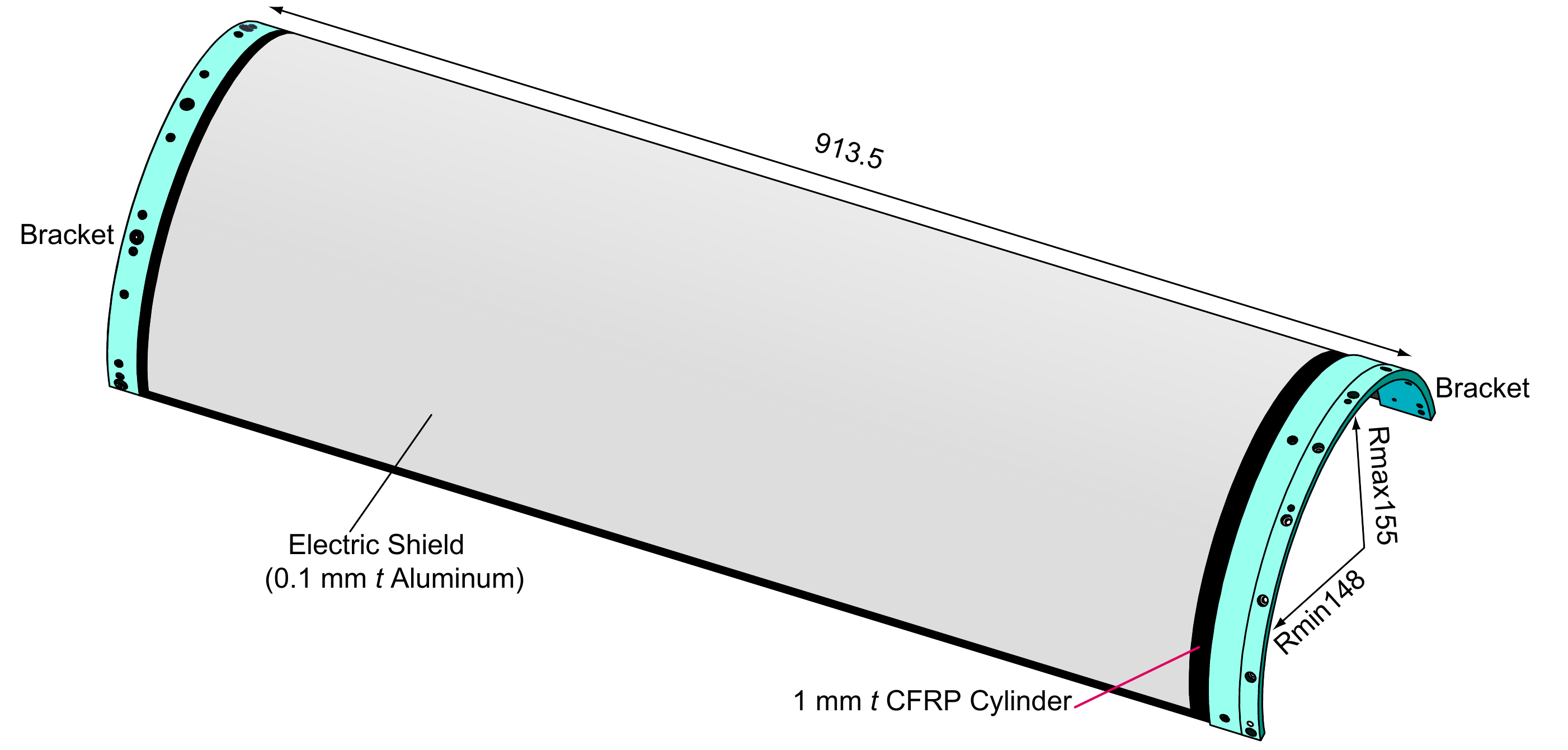}
\caption{The structure of the VXD outer cover. All dimensions are in mm.}
\label{TT:Outercover_F2}
\end{center}
\end{figure}

The detailed structure of the outer cover is shown in Figure \ref{TT:Outercover_F2}.
The two halves of the outer cover are identical, \qty{898}{\mm} long with a \qty{152.8}{\mm} radius.
The brackets are glued to the outer cover at both ends, so the total length of the outer cover becomes \qty{913.5}{\mm}. The brackets interface the outer cover to the end-flanges to keep the cylindrical shape of the outer cover under the strong stress due to the weight of the VXD.
To ensure a proper ground connection between the components of the read-out electronics, a \qty{0.1}{\mm} thick aluminum sheet is glued on the outer surface of the cylinder and electrically connected to the brackets, to establish an electrical connection with the forward and backward end-flanges.

Because of the dual-phase \COtwo\ cooling system, the temperature can vary from \qtyrange[retain-explicit-plus]{-20}{+20}{\degreeCelsius} 
(room temperature) within the volume of the VXD. 
The CFRP is a suitable material to reduce the thermal stress to the VXD sensors.

A finite element analysis using Ansys~\citeCompany{ref:ansys} was performed, assuming the VXD is supported at the forward and backward end-flanges. 
It was found that the maximum deviation of the outer cover is as small as 15~\um\ near the center, while the maximum stress on the outer cover is \qty{2.1}{\MPa} at the interface to the end-flange.
For the assembly of the VXD structure, the Stycast 2850FT epoxy adhesive with the 23LV catalyst~\citeCompany{TT:Stycaset}, originally developed for  cryogenic applications, was used. The adhesive has a large elastic modulus (\qty{6700}{\MPa}) and small CTE (\qty{39}{ppm/\degreeCelsius}).
To confirm the reliability of the adhesive, five thermal cycles between \qty{-200}{\degreeCelsius} (liquid nitrogen) and room temperature were performed. 
The glue samples were also irradiated  up to a dose of 10~Mrad with a $^{60}$Co $\gamma$ source and subjected to a shear strength test.
The shear strength measured after the irradiation was larger than \qty{7}{\MPa}, more than a factor 3 above the calculated maximum stress. 
The adhesive and the material were therefore judged strong enough to keep the structure robust over the entire lifetime of the \belletwo{} experiment.

\lvlthreesection{Cooling plant and distribution}
% Common with PXD, IBBelle location, transfer line, manifolds, flex lines, CO2 DOCK box
\label{sec:C2_coolling_plant}
\editor{T. Tsuboyama}

The power consumption of 1748 \apv\ chips amounts to 700 W (table~\ref{tab:C2_cooling_circuits}). 
As the neutron flux in the SVD region is expected to be about \qty{e13}{n\ensuremath{_\mathrm{eq}}/\cm\squared} after 10 years of operation at the design luminosity  (section~\ref{sec:C6_background_studies}), the bulk damage to the DSSDs is limited and  the SVD can be operated at room temperature.   
The two-phase \COtwo\  cooling system operated at $-20$ \textdegree C cools the \apv\ chips located in the active detector region. The following considerations lead to the choice of a \COtwo\ cooling system:
\begin{itemize}
\item  The liquid \COtwo\  has a large latent heat. When \qty{1}{\g} of liquid \COtwo\  evaporates at \qty{-20}{\degreeCelsius}, 
it absorbs \qty{300}{\J} of heat. Theoretically \qty{700}{\W} of heat can be removed with a flow of \qty{3}{\g/\s}. 
In case of the water cooling, to limit the temperature rise to \textDelta T= \qty{10}{\degreeCelsius} , a flow\qty{17}{\cm\cubed/\s}   of water is needed. 
\item The viscosity of the liquid \COtwo\ is small and allows the use of small-diameter tubes for coolant circulation.
For the SVD, the inner tube diameter is only \qty{1.5}{\mm}. The longest tube inside the VXD region is \qty{1}{\m} long, removing heat from 240 APV chips. 
\item A wall thickness of \qty{0.05}{\mm} provides sufficient pressure margin. As a result, the material budget for the cooling inside the Belle~II acceptance is reduced to 0.003\% radiation length per layer, which corresponds to about 1/7 of that of the water cooling~\cite{Abe:2010gxa}. % TDR
\end{itemize}
The liquid \COtwo\ plant, IBBelle~\cite{Verlaat:2017yzg}, supplies cooled 
liquid \COtwo\ to the VXD system (SVD and PXD) up to a rate of \qty{30}{\g/\s}. 
has been operated stably by the KEK-MPI collaboration. 
A detailed description of this cooling system is given in section~\ref{sec:C3_cooling_plant}.

\begin{figure}[hbt]
    \centering
    \includegraphics[width=.7\textwidth]{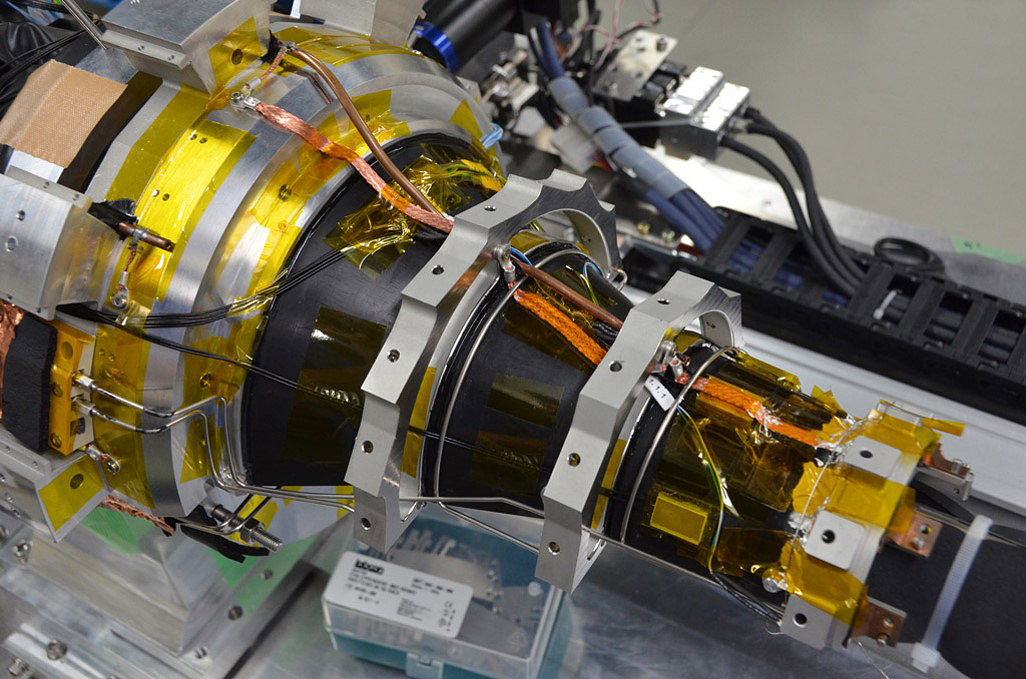}
    \caption{The $+x$ FWD half-cone assembly on the ladder mount table with diamonds, NTCs, grounding braids, $\rm N_2$ and sniffing pipes, cooling pipes and ceramic insulators.}
    \label{fig:C2_fwd_cone}
\end{figure}

\lvlthreesection{Mechanical integration}
% ladder mounting: +x, collision test, L3, L4, Origami4, L5, Origami5, L6, Origami6, -x (repeat)
% wedding with PXD
% installation
% mechanical precision after ladder mounting?
The ladder mounting procedure for each half consists of numerous steps, and generally works from inside out, starting with an empty set of half-cone assemblies mounted on a dummy beam pipe. The cones are equipped with environmental sensors, grounding, and cooling infrastructure (figure~\ref{fig:C2_fwd_cone}).

Before the first actual ladder is mounted, mechanical interference tests are performed between PXD and L3 as well as all neighboring ladders (within and between adjacent layers) using geometrically accurate dummies.
Due to the overlapping windmill-like pattern of the sensors, the ladder mounting must be done in the direction of ascending $\phi$ angle.

For layer 3, the procedure is completed after mounting the 4 ($+x$ half) or 3 ($-x$ half) ladders. All outer layers (L4/5/6) require attachment of a pre-bent \gls{origamipipe} after placing all ladders of one layer. Each ladder is electrically tested after mounting (without cooling) and each half-layer is tested (with cooling) after the corresponding \gls{origamipipe} is mounted. This is done inside a dry box covered with infrared transparent foil, so that the actual cooling contact between the pipe and all \apv\ chips can be confirmed with a thermal camera. On the ladder mount table in Tsukuba hall B1, the cooling was done with an open-loop \COtwo\ system.

\begin{figure}[hbt]
    \centering
    \includegraphics[width=.9\textwidth]{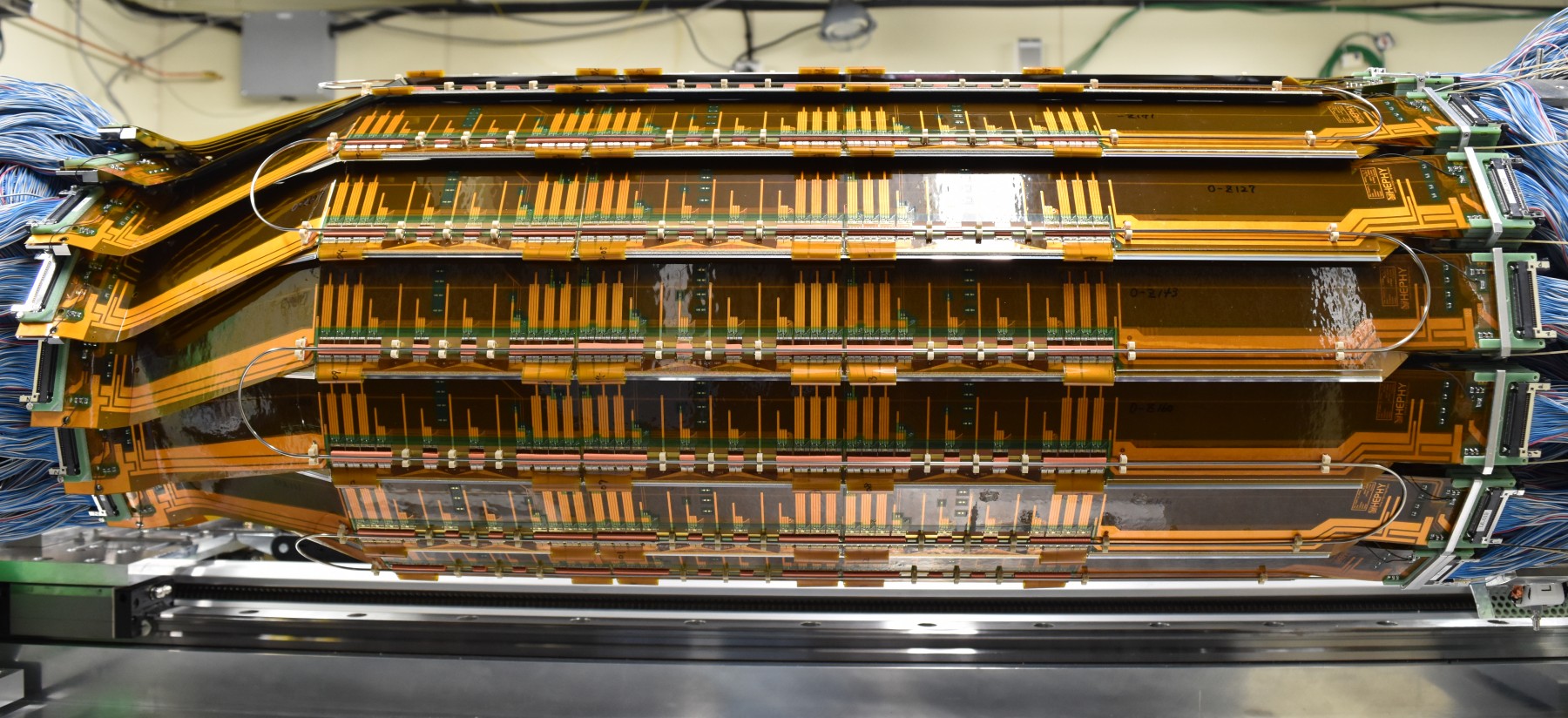}
    \caption{Fully completed $+x$ half of the SVD. The visible ladders as well as their \gls{origamipipe}  belong to the outermost layer 6.}
    \label{fig:C2_svd_plusx}
\end{figure}

After completing a half (see figure~\ref{fig:C2_svd_plusx}), it is tested as a whole, followed by attaching the outer-cover and its transfer to the storage area, in order to make space for the other half on the ladder mount table. Later, the two halves are placed into a dry and dark box for cosmic testing before being attached onto beam pipe and PXD. After a combined cold testing, the completed VXD is installed in Belle~II.

\lvltwosection{Off-Detector Electronics} {\label{sec:C2_offdet_elec}}
%\editor{Markus}
%\lvlthreesection{Design choices and Belle II DAQ}

The general outline of off-detector electronics is summarised in figure~\ref{fig:C2_svd_readoutchain} and consists of Junction Boards, the \acrfull{FADC} boards, the \acrfull{FTB} boards, the \acrfull{COPPER} boards, and the \acrfull{DATCON} readout modules. Their functions and characteristics are described here in more detail.
Due to the lack of active electronics on the Junction Boards, the \apv\ chips have to drive their output all the way up to the \acrshort{FADC} boards and need to be properly terminated there. Signal distortions due to the limited bandwidth of these $15.5\,\rm m$ of twisted pair cables need to be addressed as well as potential reflections. Moreover, the signals are at positive and negative bias potentials and need to be translated to ground levels before they can be digitized.

As the analog circuitry needs a considerable amount of space, a 9U VME system was chosen which features large boards ($400 \times 366.7\,\rm mm^2$).
The ``intelligence'' of the system is packed into a powerful FPGA as the central device on each \acrshort{FADC} board. It receives the digitized data streams from the front-end and sends the processed data out on a parallel bus leading to the optical ``Belle2link'' (Xilinx RocketIO over fiber) that transmits data to the unified \acrshort{COPPER} Belle II DAQ modules~\cite{Higuchi:2005yg}. Moreover, it has an interface to the VME bus and a Gigabit Ethernet port. A second optical link is used to stream the same data to the \acrshort{DATCON} units of the PXD, which are used for online tracking in order to find regions of interest (RoI) on the pixel planes and reduce the PXD data transmission rate.

\begin{figure}[hbt]
    \centering
    \includegraphics[width=.7\textwidth]{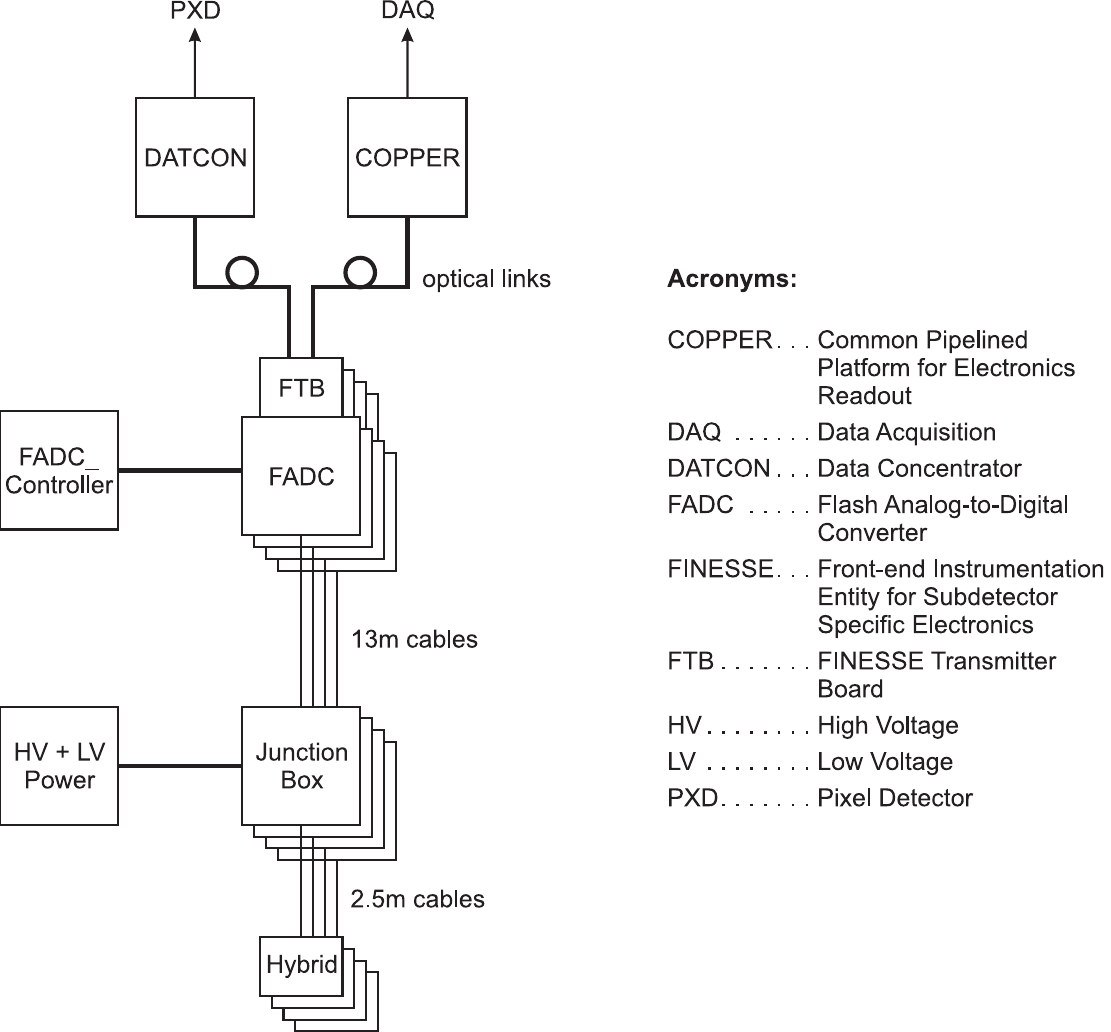}
    \caption{Schematic readout chain of the Belle II SVD~\cite{Friedl:2013gta}.}
    \label{fig:C2_fadc_system}
\end{figure}

The \acrshort{FADC} system (figure~\ref{fig:C2_fadc_system}) is distributed over four 9U VME crates, two each on FWD and BWD sides, located on top of Belle II. It consists of one FADC\_Controller board that interfaces to four Buffer boards, one in each crate, and a total of 52~FADC boards, each of which sends out data to the central data acquisition (DAQ) and \acrshort{DATCON} through an attached \acrshort{FTB}. The communication within each crate takes place over a customized Backplane between Buffer and \acrshort{FADC} boards. An asymmetric distribution of {FADC} boards (32 on BWD and 20 on FWD) reflects the fact that the majority of cables exit the VXD on the BWD side. One Junction Board is attached to one {FADC} board, except for L3 where the output of each Junction Board is shared between two {FADC} boards in order to reduce the bandwidth burden for the upstream DAQ system.

\begin{figure}[hbt]
    \centering
    \includegraphics[width=\textwidth]{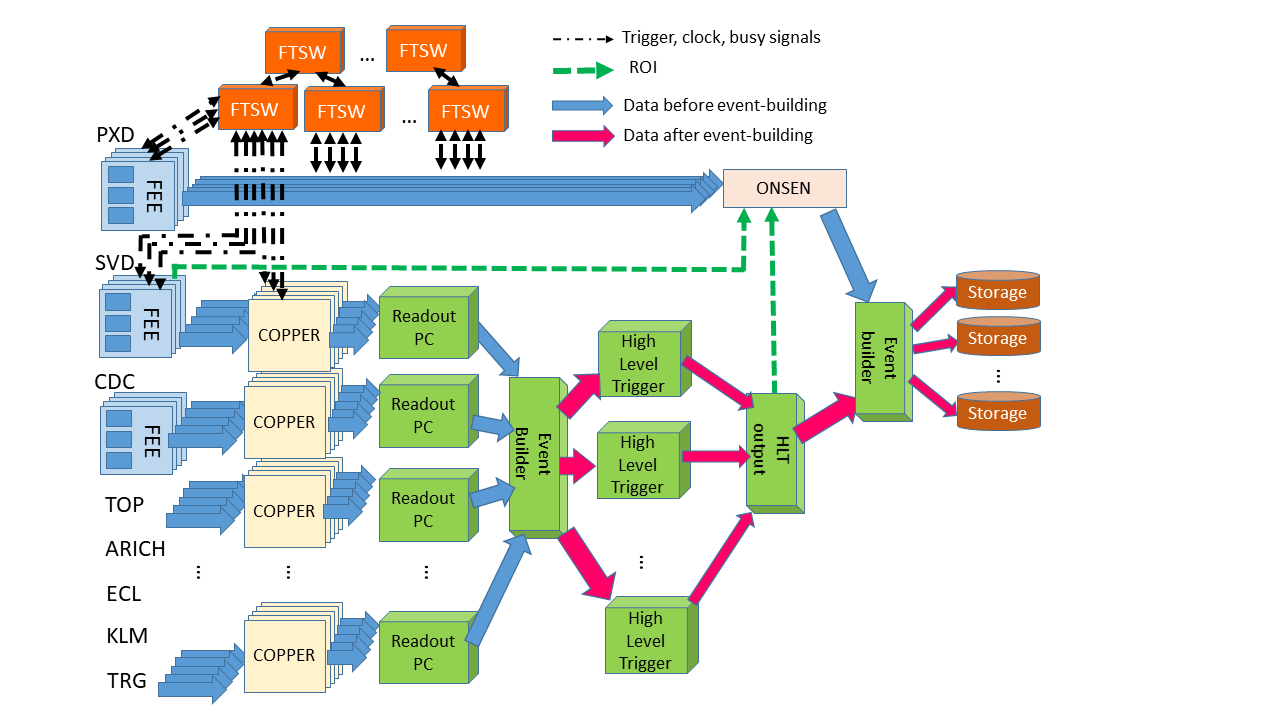}
    \caption{Overall Belle II DAQ scheme.}
    \label{fig:C2_daq_yamada}
\end{figure}

The unified DAQ~\cite{Yamada:2015xjy} consists of 9U VME modules named \acrshort{COPPER}, which collect the data from all subdetectors through the optical links mentioned above, as shown in figure~\ref{fig:C2_daq_yamada}. Those data are then read out by PCs, combined in the Event Builder and processed by a computer farm known as \acrfull{HLT}, which performs tracking and reconstruction and finally selects events to be written to disk. Clock, trigger and related information are distributed to the FADC\_Controller and each \acrshort{FTB} board by the \acrfull{FTSW}~\cite{Nakao:2012aea}.

\begin{figure}[hbt]
    \centering
    \includegraphics[width=.9\textwidth]{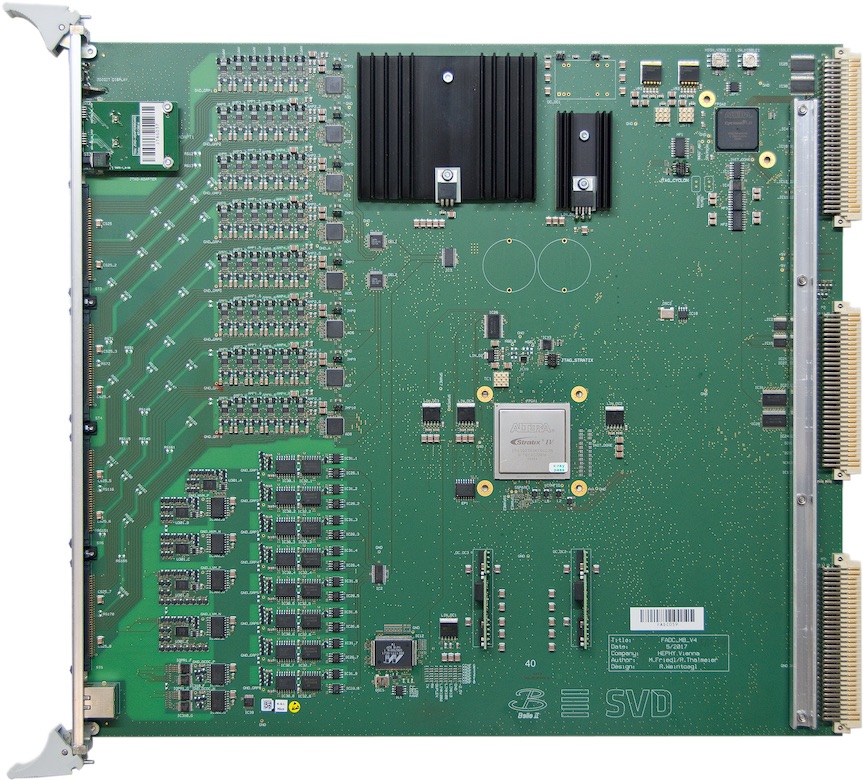} % Note: changed to jpg to reduce file size. FF
    \caption{FADC board (9U VME).}
    \label{fig:C2_fadc}
\end{figure}

\lvlthreesection{FADC system}
\label{sec:C2_FADC_system}
The \acrshort{FADC} system has one FADC\_Controller which interfaces to 52~\acrshort{FADC} boards, distributed over four crates, through one buffer board per crate. The controller receives its input (such as a trigger) either over VME (in case of local runs) or from the FTSW system (global runs). Some monitoring and also spy data can be read from the \acrshort{FADC} boards through the VME bus. The primary data output is through a parallel interface to \acrshort{FTB} and from there optically to DAQ (and \acrshort{DATCON} in parallel). The communication between the single controller and each \acrshort{FADC} board is performed through Buffer boards and custom Backplanes to which every FADC (figure~\ref{fig:C2_fadc}) is connected. Each \acrshort{FADC} also features a Gigabit Ethernet (GbE) port, which is intended for fast local data taking.

Each 9U VME crate has three 6U slots on the left side (VME controller, empty, Buffer), followed by 18 full-sized slots (\acrshort{FADC}s). The Backplane is attached on the rear of the J2 connectors and runs from slot~3 (Buffer) to 21. It utilizes pins for internal communication which are not used by the VME bus. Broadcasts (from the FADC\_Controller to the FADCs) are sent over differential LVDS lines, and a few open collector lines are available for communication in the opposite direction.

In general, one Junction Board is connected to one \acrshort{FADC} board, and thus the latter receives data from up to four $P$- and four $N$-type hybrid boards, corresponding to up to 48~channels. 
To avoid readout bandwidth limitations in the {FADC} to \acrshort{COPPER} connection, the readout of Junction Boards which serve L3 hybrids are split between 2~{FADC} boards each.
Therefore, the SVD system has 48~Junction Boards in total, but 52~{FADC} boards.

\begin{figure}[hbt]
    \centering
    \includegraphics[width=.9\textwidth]{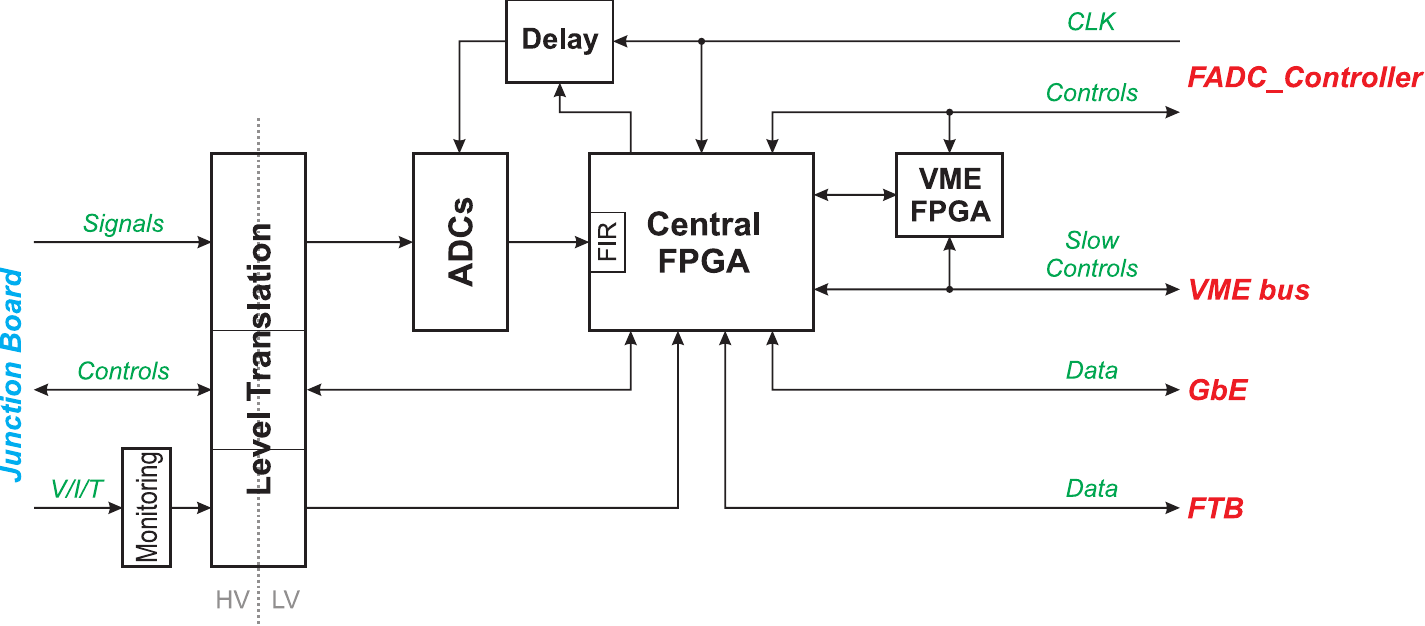}
    \caption{Schematic block diagram of the FADC~\cite{Friedl:2013nea}.}
    \label{fig:C2_fadc_blocks}
\end{figure}

Figure~\ref{fig:C2_fadc_blocks} shows the functional blocks of an \acrshort{FADC} board. Data processing mainly happens 
within the firmware of the central FPGA, an Altera (now Intel) Stratix~IV~GX. It is supported by another simpler FPGA (Altera Cyclone~II) that can be used to flash the firmware over VME onto the main device.

The signals to/from the Junction Boards are shown to the left in figure~\ref{fig:C2_fadc_blocks}: clock ($31.805\,\rm MHz$), trigger and $\rm I^2C$ control signals are sent to the \apv\ front-end chips, while their analog output is received as well as HV bias current, LV and Junction Board temperature measurements. Moreover, Enable and Power Good signals are propagated to/from the DC/DC converters on the Junction Boards. All those signals are related to the \apv\ supply voltages, which reside on top of the negative ($P$-side) and positive ($N$-side) HV bias levels and therefore need to be translated to ground levels on the \acrshort{FADC}. This is done using capacitive coupling for the signals (essentially a high-pass filter) and commercial digital insulators (based on integrated HF transformers) for the digital signals. After level translation, the analog signals are digitized with adjustable clock phase and sent to the central FPGA.

\begin{figure}[hbt]
    \centering
    \includegraphics[width=.9\textwidth]{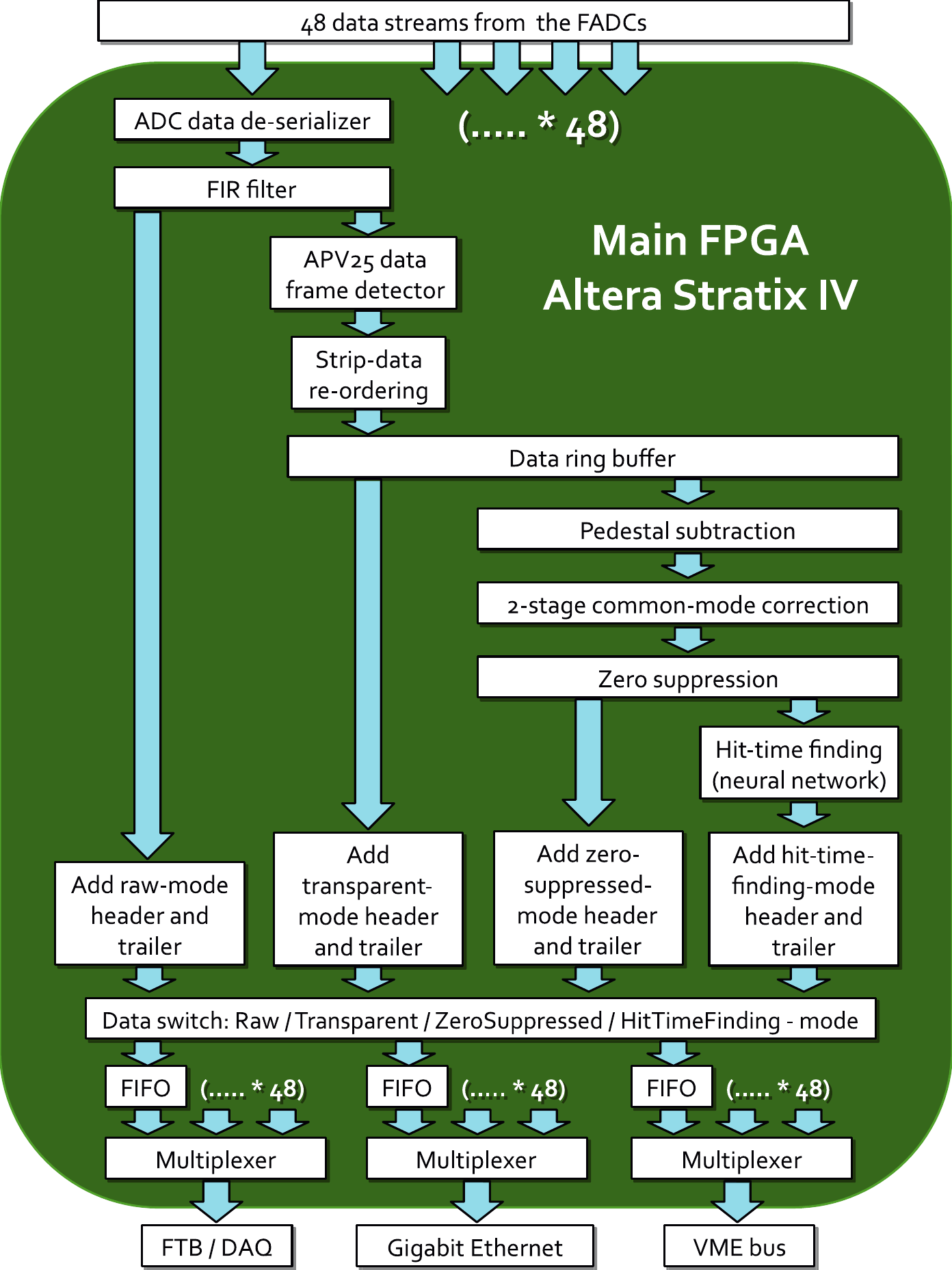}
    \caption{Functional blocks of the data processing in the \acrshort{FADC} firmware. Depending on the mode setting, data is processed according to one of the four columns shown (Raw, transparent, zero-suppressed or hit-time-finding). Adapted from~\cite{Belle-IISVD:2017pzz}.}
    \label{fig:C2_fadc_firmware_blocks}
\end{figure}

Inside the FPGA firmware (shown schematically in figure~\ref{fig:C2_fadc_firmware_blocks}), each 10-bit ADC input data stream is first de-serialized and subjected to a digital \acrfull{FIR} filter with eight coefficients. This effectively compensates for the nonlinear transfer function of the long cables (due to bandwidth limitation) and removes moderate reflections (due to imperfect termination). In {\it raw mode}, these data are sent to the output. Otherwise, \apv\ data frames are detected in the data stream and extracted, followed by re-ordering of the strip data -- which is necessary because the order is twisted due to the 3-level output multiplexer of the \apv\ chip. Only the strip data, but without further processing, are sent out in {\it transparent mode}. In {\it zero-suppressed} and {\it hit-time-finding modes}, the pre-loaded pedestals of each strip are subtracted. 

Common mode noise is removed by subtracting the average amplitude of strips from each individual strip value (``Common Mode Correction'', CMC). This CMC can be performed either in groups of 32 (default), 64 or 128~strips, i.e.\ in quarters, half or full chips, and it is performed in two stages. In the first round, masked strips (typically noisy or known bad ones) are excluded to avoid abnormal distortion of the average value. In the second pass, also strips with a signal above the hit cut threshold (normally three times the RMS noise of each strip) are excluded in addition to masked strips for obtaining an unbiased average. During the CMC calculation, the strip values are stored in 10~bits, with a range of $-512$ to 511.
Finally, the data are zero suppressed, discarding all strip data below a certain threshold (by default three times the strip noise). The results are presented as unsigned 8~bit numbers (range $0 \ldots 255$).

In terms of monitoring, the \acrshort{FADC} measures the values of each of the eight sensor bias currents and the temperatures of the two NTC probes on the Junction Board it is connected to. Moreover, it probes the \apv\ supply voltage levels ($2.5\,\rm V$ and $1.25\,\rm V$ nominal) near the \apv\ chips on each hybrid board using sense lines. All that metering is performed with slow 24bit ADCs which are read out over VME at a rate of approximately \qty{1}{\Hz}.

The processed data of the \acrshort{FADC} are passed on to the \acrshort{FTB} boards attached to the J3 connector at the rear of each \acrshort{FADC}. A busy handshake is implemented towards the DAQ chain to avoid transient data loss. In case of persistent busy, this can lead to back pressure that eventually blocks further triggers because ultimately the \apv\ internal buffers  become full.

Special runs are used to determine operational parameters such as the optimum clock phase for each ADC or FIR filter coefficients. They are very similar to the ones used in the common APVDAQ test system and are described in Sec.~\ref{sec:C3_APVDAQ} and~\ref{sec:C4_local_runs}.

\lvlthreesection{FTB boards}
\editor{Katsuro}

\begin{figure}[ht]
    \centering
    \includegraphics[width=.7\textwidth,angle=180]{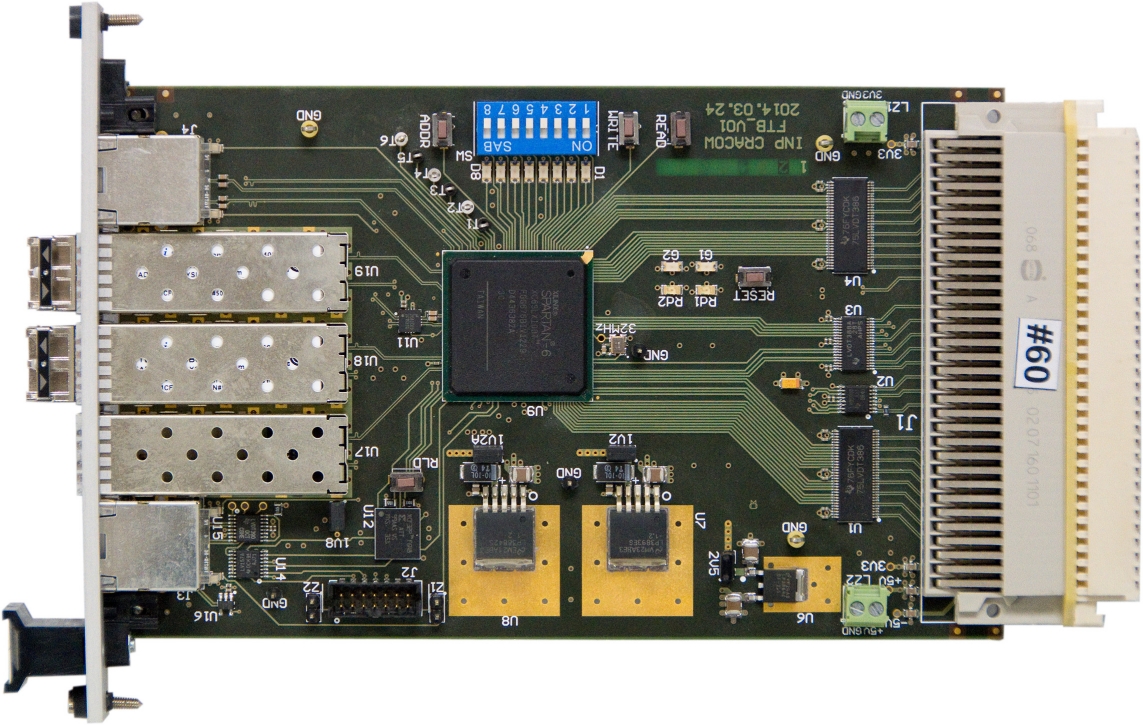}
    \caption{FTB board.}
    \label{fig:C2_FTB}
\end{figure}

The \acrshort{FTB} is a 1U board that transmits the \acrshort{FADC} data to the \acrshort{COPPER} board
and \acrshort{DATCON} module (figure~\ref{fig:C2_FTB}). The left side of the board is connected to the J3 connector on the rear side of the \acrshort{FADC} board, and
the 32-bits data stream is received %with \qty{31.8}{\MHz}
by the Xilinx
Spartan-6 FPGA (XC6SLX100T).
In the FPGA, the data are duplicated and sent to the \acrshort{COPPER} board and to the \acrshort{DATCON} module outputs. 
At each data output in the FPGA, the data are serialized using
8b/10b encoding and transmitted using a Gigabit %GTP
transceiver in the Spartan-6 FPGA.
For the serial-data-transmission to the \acrshort{COPPER} board,
a dedicated protocol called {\it belle2link} is used. 
The serialized data rate of belle2link is 1.27~Gbps, which is fast enough
that all the \acrshort{FADC} data can be sent without buffering in the \acrshort{FTB} board.
The belle2link protocol also allows remote register access to the FTB
board through the \acrshort{COPPER} board.
For the serial-data-transmission to the \acrshort{DATCON} board, the Xilinx Aurora module is used for the data transmission. 
For the error detection, the CRC16 checksum of the \acrshort{FADC} data is checked,
and also the \acrshort{FTB} board calculates another CRC16 checksum
for another check downstream.
The \acrshort{FTB} board also receives the 127~MHz Belle~II DAQ clock and level-1
triggers from the \acrshort{FTSW} board through the RJ45 connector.

\begin{figure}[hbt]
    \centering
    \includegraphics[width=\textwidth]{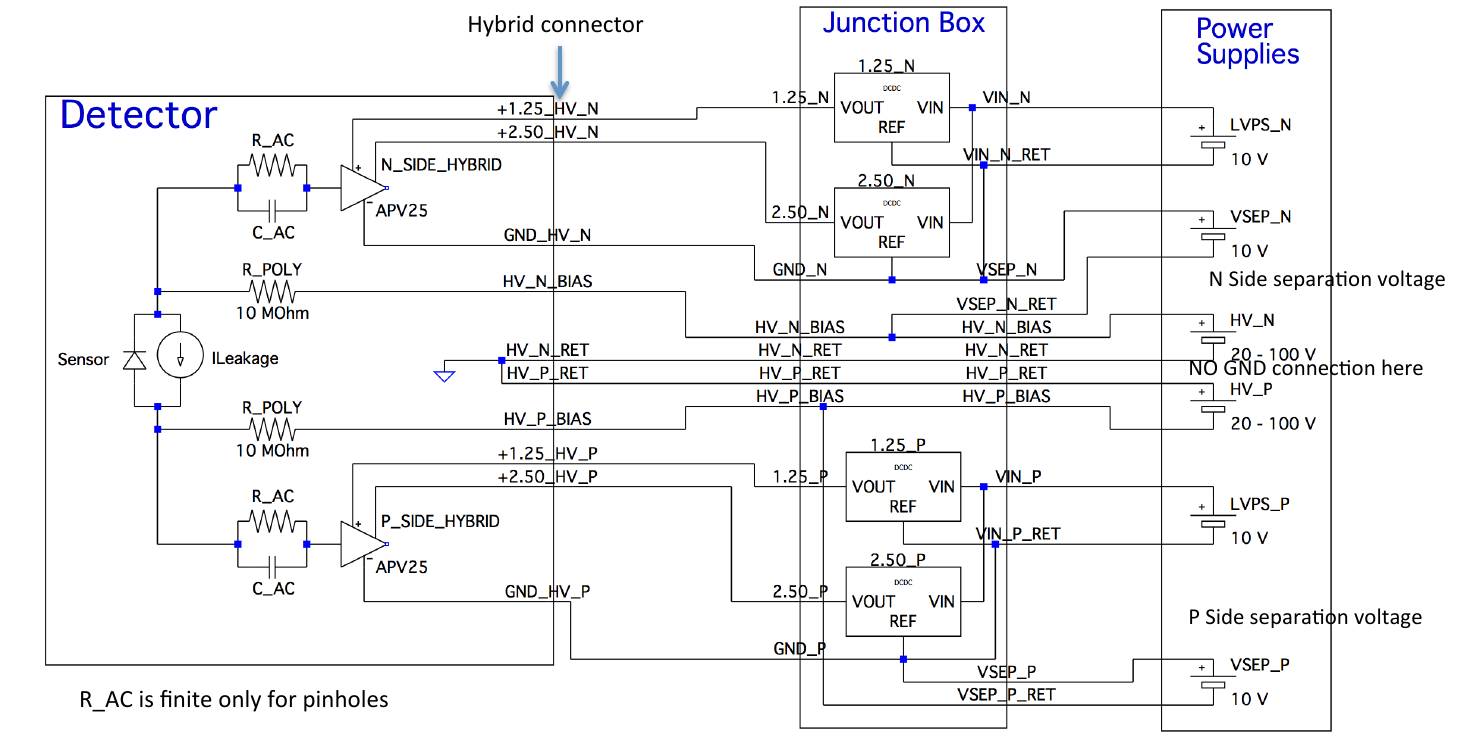}
    \caption{Power supply schematics.}
    \label{fig:C2_PS_schematics}
\end{figure}

\lvlthreesection{Power supplies}
\label{sec:C2_power_supplies}
%\editor{Francesco. Note: text taken from specs, to be adapted}
The power supply design is based on fully floating low and high voltage sources with a common ground connection made on the detector to avoid ground loops.
The \apv\ readout ICs require two regulated positive voltages, \qty{1.25}{\V} and \qty{2.5}{\V}, while the sensors require a bias voltage, called HV, ranging from \qty{40}{\V} to \qty{200}{\V}. 
Bias to the sensors is provided by two independent \qty{100}{\V} floating voltage supplies on the two sides: HV\_P with negative polarity on the $P$-side, and HV\_N with positive polarity on the $N$-side. 
The two HV supply return lines are connected to ground at the detector side, not at the power supply side, as shown in the simplified schematic of the power supply connections in figure~\ref{fig:C2_PS_schematics}. 
The hybrids on which the \apv\ chips are mounted are referenced to the HV potential of the sensor side ($P$ or $N$) to which they are connected, thus limiting to a few volts the potential difference between the input of the \apv\ chip and the sensor bias applied to the strip implant, i.e.\ the voltage across the strip coupling capacitor.
The two \apv\ supply voltages are generated by radiation-hard DC/DC converters installed on the Junction Boards (section~\ref{sec:C2_junctionboards}) located inside the detector. These converters are powered by a regulated floating \acrfull{LVPS}, that should be adjustable between 5 and 12V to optimize the DC/DC converter efficiency.

Since the input to the \apv\ chip stays at approximately \qty{0.8}{\V},  in addition to the main power supply system described above, an additional floating voltage source is required, called \acrfull{VSEP}, slightly offsetting the \apv\ reference w.r.t. the sensor bias voltage. 
The \vsep\ is used to adjust the \apv\ input voltage to prevent current flowing into the pre-amplifier in the case of broken AC coupling capacitors on the sensors (pinhole compensation). 
The \vsep\ polarity can be reversed to allow full diagnostic on the faulty capacitors (see section~\ref{sec:C3_APVDAQ} for details). This polarity inversion is realized with a relay in a separate hardware component, the Power Distribution Panel (not shown in figure) which also allows the matching of connectors between the cables from the detectors and the power supply front panels. 

For reasons of cost and dimensions of the cable plant, hybrids are connected in parallel in groups, and powered from the same LV source, with at most four hybrids in every group. Sensors in two hybrid groups (at most eight hybrids) share the same bias HV. Hybrid groups are defined following certain requirements: same readout direction (forward or backward); same sensor side ($P$ or $N$); same detector layer; same sensor type (wedge and rectangular). These considerations lead to a total number of 96 hybrid groups, 60 in the forward direction and 36 in the backward direction, and therefore 96 low voltage channels, 48 HV channels, and 48 \vsep\ channels.  
Considering the power consumption of the \apv\ chips, the efficiency of the DC/DC converters, and the expected leakage current of the sensors, the LVPS channels should be able to deliver a power of at least \qty{20}{\W} between \qty{5}{\V} and \qty{10}{\V}, while the HV channels should supply \qty{1}{\mA} at \qty{100}{\V}. The \vsep\ channels are required to deliver only a few volts, with virtually zero current. To allow sufficient setting accuracy, the actual \vsep\ voltage is obtained through a resistive divider ($\times 10$) from the power supply voltage.

\lvlthreesection{Power and data cables}
\label{sec:C2_cables}
%\editor{Markus}
% table with types, numbers, length
% LV has sense lines
% HV cable has HVp, HVn, Vsepp, Vsepn
% data cables are twisted flat
% include information on double (triple) set of cables
Three types of cables lead to the Junction Boards: LV power (Leoni~\citeCompany{leoni} LEHC 002054, $17\,\rm m$), HV power (3M HF659/10, $17\,\rm m$), and data (Amphenol~\citeCompany{amphenol} 125-3097-998, $13\,\rm m$). The first one contains four conductors with a cross-section of $1.5\,\rm mm^2$ each, plus two shielded pairs, which serve as sense wires, while the cable is originally meant for a fieldbus installation, delivering power and data. The HV cable is actually a 10-conductor flat cable rolled up and shielded. It delivers the bias voltages and the offset voltages between HV and LV (\vsep) for both $P$- and $N$-sides, thus utilizing 8 of the available wires. Thin conductors (28 AWG) are sufficient here because the currents are tiny ($\ll 1\,\rm mA$).

In order to avoid any impedance disruption, the data cable is of a similar type as the hybrid cable (see section~\ref{sec:C2_hybridboards}), but with 68 instead of 50 conductors (because two hybrid data are merged into one cable at the Junction Board).
Unlike the hybrid cable, it is a flat cable (with twisted pairs), which cannot be laid in curves. Thus, to lay this cable on the outside of the CDC end-walls, a machine was built 
that removes the plastic cover on the outside of the flat cable by heat, makes the cable round and wraps Kapton tape around it. The first $6\,\rm m$ of every data cable are converted into a round cable in this way in order to cover the entire CDC region even for the longest cable path. All cables are flame retardant and free of halogens. 

\lvlthreesection{Grounding}
\label{sec:C2_grounding}

\begin{figure}[hbt]
    \centering
    \includegraphics[width=\textwidth]{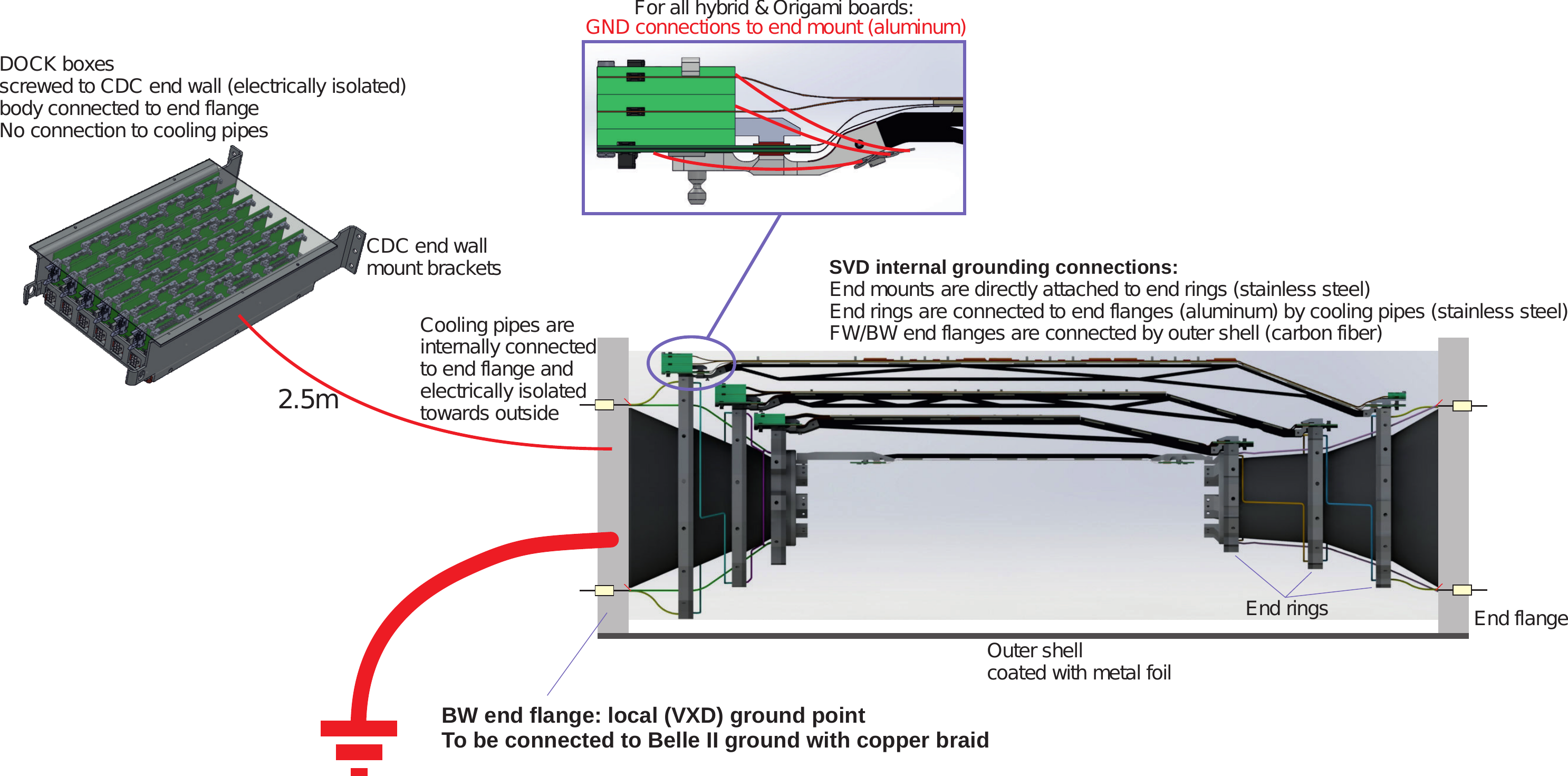}
    \caption{SVD grounding scheme: the BWD end-flange acts as the central grounding point for the SVD. The FWD side end-flange is connected to ground through the metal coated outer-cover.}
    \label{fig:C2_svd_grounding_scheme}
\end{figure}

As shown in figure~\ref{fig:C2_svd_grounding_scheme}, the local grounding point of the SVD is the end-flange of the BWD side. It is connected to the metal structure of Belle~II with a thick grounding cable. The FWD side end-flange is connected to the local ground point through the aluminum foil on the outer-cover. Each end-ring is connected to the end-flange with a copper braid, and the same applies to each \gls{origamipipe}. All cooling pipes are electrically separated from the outside world by ceramic insulators located at the end-flanges.

Each ladder is electrically and thermally coupled to end-rings on both BWD and FWD sides, and ground wires lead from the end-mount to each hybrid and \gls{origamiboard}, where they are connected to the HV return lines of both the $P$- and $N$-side bias voltages.

The whole SVD is inserted into the CDC, but electrically isolated from CDC, PXD and also the beam pipe. Similarly, the DOCK boxes are electrically decoupled from the CDC, but connected to the corresponding end-flange with individual ground cables. Therefore, all on- and off-detector electronics are connected (in a star topology) to the central SVD ground point.

\lvltwosection{Beam Test Validation of Design}
\label{sec:C2_beamtest}
%\emph{Writing assignment: Christoph}

A number of test beam experiments were performed to develop and validate the various parts of the SVD~system. In Table~\ref{tab:C2_testbeam} the main experiments performed since 2004, the beams used, and the goal of the test are listed.
\begin{table}[hbt]
    \centering
    \begin{tabular}{llp{4cm}p{5cm}}
    \toprule
    Date & Site & Device under test & Comments\\
    \midrule
    Aug 13--18, 2004 & CERN SPS T7 & Striplet (single-sided) & \\
    Apr 6--11, 2005 & KEK PS & Striplet & UV module, doubled-sided readout\\
    Aug 10--12, 2005 & PSI piE1 & Striplet, MWPC & \\
    Aug 26--29, 2006 & PSI piM1 & Flex & \\
    Nov 2007 & KEK Fuji hall & Striplet, JP-single, Flex & SVD2.5 readout system (NECO, MAMBO, REBO)\\
    Jun 2008 & CERN SPS H6 & JP-ganged, JP-single, Micron, Striplet, Flex & \\
    Nov 2008 & KEK Fuji & JP-ganged, JP-single, Micron, Striplet, Flex & \\
    Aug 2009 & CERN SPS H6 & Origami 4", Micron & First test of origami concept\\
    Oct 2010 & CERN SPS H6 & Origami 6" module, Wedge module, Micron baby	& Including irradiation at SKC-CEN Brigitte~\cite{SCKCEN:Rita:Brigitte}, first full size origami module with B2 HPK sensor\\
    Oct 2011 & CERN SPS H6 & 3x origami 6" modules & First Origami flex with high density JAE connector~\citeCompany{ref:JAE}\\
    Oct 2012 & CERN SPS H6 & 2-DSSD origami, 2x origami, Wedge & Including irradiation at SKC-CEN Rita~\cite{SCKCEN:Rita:Brigitte}, \COtwo\ cooling\\
    Jan 2014 & DESY TB24 & L3 module, 3x origami 6" (L4-L6) & Together with PXD, installation started in Dec 2013, final \acrshort{FADC} readout\\
    Nov 13-24, 2014 & CERN SPS	& L3 module, 3x origami 6" (L4-L6), SFW993, SBW993 & First test with CAEN power supply\\
    May/Jun 2015 & CERN SPS H6 & L5.903 + \acrshort{FADC} readout	& With EMC tests in Zaragoza, first completed L5 class B ladder\\
    Apr 2016 & DESY TB24 & Phase 2 cartridge (first iteration) & \\
    May 2017 & DESY TB24 & Phase 2 cartridge &  \\
    \bottomrule
    \end{tabular}
    \caption{Belle II SVD test beam experiments since 2004. The ``JP" modules where an early design described in \cite{Abe:2010gxa}}
    \label{tab:C2_testbeam}
\end{table}

 %25 pages 

\lvlonesection{Detector Construction}
\label{sec:C3_construction}
\editor{Francesco, approx 25 pages}

The validation of the detector design by a series of tests on prototypes (section~\ref{sec:C2_beamtest}) is the starting point for the \acrshort{SVD} construction, presented here. This process involved several institutions and companies, and required a complex organization as well as a careful and continuous quality control to ensure the integrity of the final detector. The timeline spanning several years, the logistics across several sites tracked in a construction database, the optimization and standardization of procedures and tests for quality control and assurance are outlined in section~\ref{sec:C3_overview}. 

The silicon sensors were produced by two manufacturers, Hamamatsu Photonics K.K.~\citeCompany{hpk} and Micron Semiconductor Ltd.~\citeCompany{micron}. 
Section~\ref{sec:C3_sensors} recalls the sensor specifications and describes the tests and full characterization performed on each device. The resulting quality classification was used to select the sensors for the assembly of ladders, installed in the experiment or kept as spares.

The procurement of the components of the on-detector electronics (\apv, pitch adapters, hybrid and origami boards, junction boards: section~\ref{sec:C3_electronics}) required similar attention to details, and remedial actions in a few problematic cases.

The assembly of sensors and front-end electronics in modules and \gls{ladder}s for different detector \gls{layer}s had to proceed in parallel at different sites, under the responsibility of several institutions. 
As described in section~\ref{sec:C3_module}, this process required the development of common assembly techniques and procedures, standardising  assembly jigs, as well as alignment, gluing, wire bonding, and electrical test procedures, including their quality assurance, in order to reach the required mechanical tolerances and detector performance.

The implementation of the innovative mechanical design is described in section~\ref{sec:C3_mechanical}: support \gls{end-cone}s, \gls{end-ring}s and an outer cover form the rigid support structure of the detector, with an embedded system of channels and pipes for the circulation of the dual-phase CO$_{2}$ coolant.  Section~\ref{sec:C3_offdet_elec1} describes the fabrication and tests of the off-detector electronics: \acrshort{FADC} and power supplies.

The final step of the detector construction, the so called ``\gls{ladder} mount", was performed in a dedicated clean room at KEK, with the contribution of experts from several institutions, as detailed in section~\ref{sec:C3_ladder_mount}. A rotating mounting support was used to precisely position and mount the ladders shipped to KEK from the different construction sites, starting with the inner layer, for each of the two detector halves.  Cooling pipes were installed with a special positioning device after the completion of each half-layer. Temperature monitoring sensors were also installed at this time. Complete functional tests on each ladder confirmed its performance before and after installation.  

During the several years of SVD construction, distributed over several laboratories,  many challenges appeared along the way, requiring considerable effort to be overcome. Section~\ref{sec:C3_issues_during_svd_consruction} gives a short report on these issues and on the lessons learned during the construction.

\begin{figure}[phbt]
    \centering
    \includegraphics[angle=90,height=1.3\textwidth]{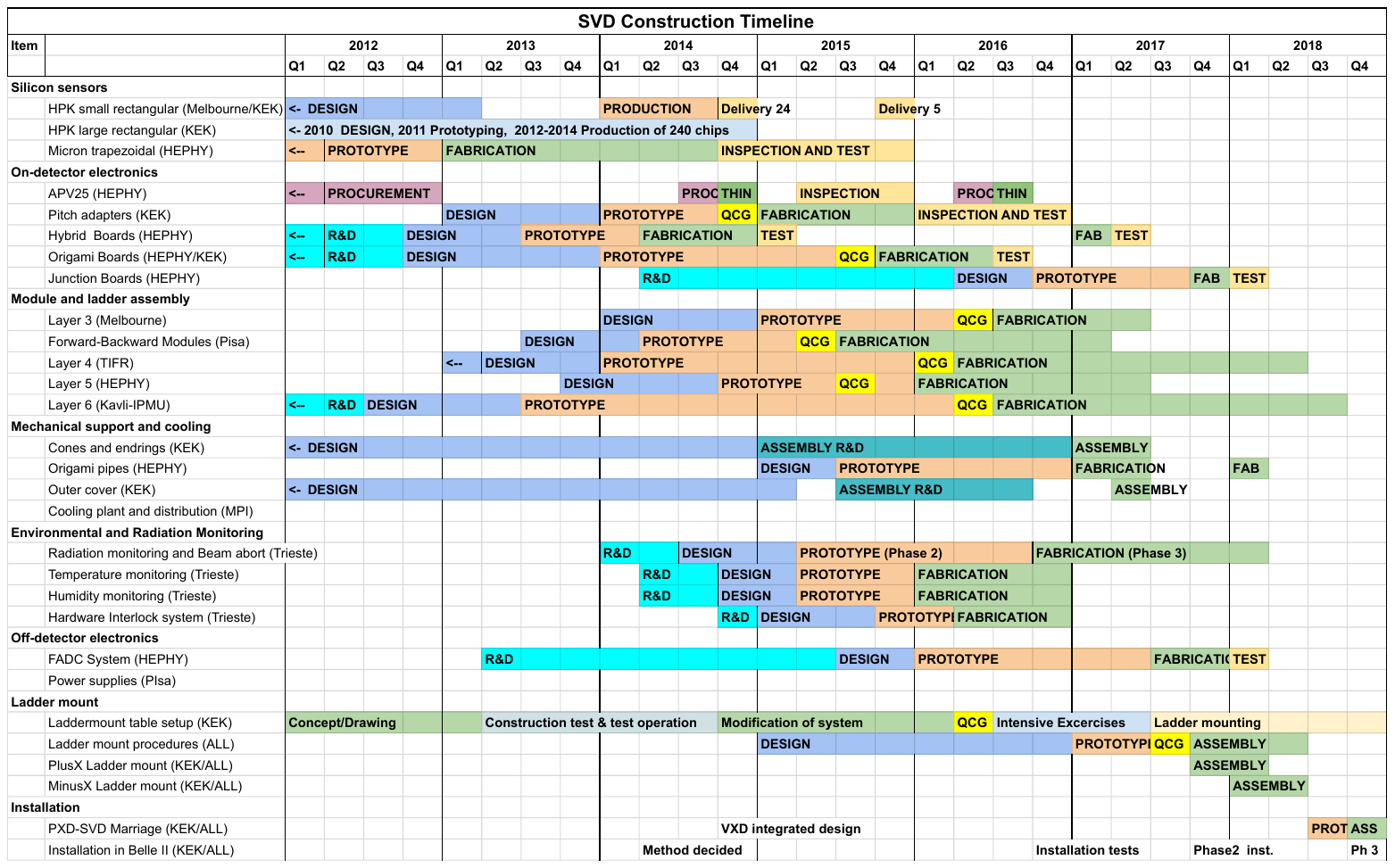}
    \caption{Timeline and major responsibilities of the production of the SVD components and assembly. For Layer 4 (TIFR operating at Kavli-IPMU) the timeline includes a regular break of about three weeks during the prototyping and fabrication phase owing to the travel scheduling of the team. }
    \label{fig:C3_timeline}
\end{figure}

%========================
\lvltwosection{Organization and Quality Control}
\label{sec:C3_overview}
\editor{Francesco  - 2 pages}
%\lvlthreesection{Overview of the construction process}
%\paragraph{Organization of SVD construction}
The SVD is composed of a large number of different components, described in section~\ref{sec:C2_design}, that were designed, fabricated, and tested at different places and times. The integration, quality assurance and control of the entire system was challenging, requiring careful coordination and a dedicated \acrfull{QCG}, as discussed in section~\ref{sec:C3_QCG}. The SVD construction effort required several years. Figure~\ref{fig:C3_timeline} gives an overview of the timeline for the R\&D, fabrication, and construction of the various components.

An effort was made to minimize the number of different module designs and standardize the production, so that components could be reused as much as possible in different assemblies. 
Individual detector modules were assembled by gluing together a sensor, pitch adapters and hybrid or origami boards, which are equipped with \apv\ readout chips; subsequently the electrical interconnections are made using ultrasonic wire-bonding techniques. Four basic detector module designs exist: Layer 3 modules;
forward modules used in Layers 4, 5 and 6;
backward modules used in Layers 4, 5 and 6;
origami modules used in Layers 4, 5 and 6 (although the origami design depends upon the location). 

Detector modules were then assembled into ladders using a hierarchical structure so that rework and reuse is possible until the last gluing steps. Completed and qualified ladders were shipped to KEK in preparation for the final ladder mount phase. 
In parallel, the mechanical support structure, the cooling circuits, the environmental and radiation monitoring sensors (described in section~\ref{sec:environment}), and the off-detector electronics were prepared. 

The ladders were then mounted on the two half-SVD mechanical supports, installing the required environmental sensors and origami cooling pipes. Each completed detector half was then commissioned and stored until the coupling with the PXD detector prior to the installation in \belletwo.

%%\begin{itemize}
%    \item Complexity of the SVD construction and distributed nature
%    \item Summary of distinct construction components and reference to %Fig.~\ref{fig:C3_timeline}
%\end{itemize}

%===========================================
\lvlthreesection{Construction sites}
\label{sec:C3_constructionsites}

The SVD construction was carried out by an international team distributed across several institutions and sites. A high-level summary of the construction responsibilities for the various components is shown in figure~\ref{fig:C3_timeline}.
Regarding the actual assembly and mounting of the detector ladders, it was carried out at five major sites, as shown in table~\ref{tab:C3:construction_sites}. It should be noted that export regulations forbid the shipping of the radiation-tolerant \apv\ chips to India. This is a key reason why the TIFR-Mumbai team, who had the responsibility of L4 ladder assembly, relocated their operation to the Kavli-IPMU clean room, that was expanded to host them. 
 
%\begin{itemize}
%%    \item Description of the different sites and responsibilities, with reference to Tab.~ \ref{tab:C3:construction_sites}
%    \item Workflow of construction and shipment issues. Prepare figure showing flow of material in Fig.~\ref{fig:C3_material_flow}
%\end{itemize}

\begin{table}[hbt]
    \centering
    \begin{tabular}{p{0.3\linewidth} p{0.6\linewidth}}
\toprule	
        Site & Activities \\
\midrule	
        INFN -- Pisa  & Assembly of forward-backward modules \\
        University of Melbourne & Assembly of L3 ladders \\
%%        Tata Institute of Fundamental Research NOTE: Too long
        TIFR -- Mumbai$^{(*)}$ & Assembly of L4 ladders \\
%%        Institute for High Energy Physics ( -- NOTE: Too long
        HEPHY -- Vienna & Assembly of L5 ladders \\
        Kavli IPMU -- Tokyo & Assembly of L6 modules \\
        KEK & Assembly of mechanical structure, ladder mount, commissioning and installation \\
\bottomrule
    \end{tabular}
    \caption{SVD assembly and construction sites. }
    \label{tab:C3:construction_sites}
    $^{(*)}$operating at Kavli IPMU -- Kashiwanoha, University of Tokyo 
\end{table}

The assembly of ladders for each SVD layer was assigned to a different site, so that the site could specialize and limit the number of different components to process. As a consequence, components had to travel between the construction sites, that resulted in a rather complex logistical organization.

\lvlthreesection{Quality control and assurance}
\label{sec:C3_QCG}

The completed SVD must be of high quality for \belletwo\ to obtain its anticipated physics performance. It is therefore of the utmost importance that all the components and ladder assembly operations maintain the highest possible quality standard. 
This is essential at all steps of the process, particularly because the limited number of spare components makes the replacement of faulty or under-performing ladders difficult and expensive. 
Given the distributed nature of the assembly of the SVD ladders and the many groups involved, it was not straightforward to guarantee that the quality control and assurance procedures were similar at all sites.

It was therefore decided to constitute the \acrshort{QCG} with the charge of defining and documenting all the assembly and quality control procedures, as well as verifying that the different sites were prepared for the work they had to carry out.
The \acrshort{QCG} was formed by site representatives and some additional expert members from the SVD group. 
The QCG carried out numerous site visits to verify the conditions at the sites, and to ensure maximal cross-communication between the experts at different sites. Besides regular meetings, the QCG held a number of focused reviews that resulted in written reports, which was very important with respect to identifying the points that required action, such as the safety and detailed logging of operations, the reproducibility of the gluing or bonding steps, or the storage and analysis of the electrical measurements.

%\label{sec:qcgrades}
Components were graded into classes according to their quality.
Class A and B are fully working and usable components, with class B identifying lower quality components, for instance with more than 1\% of defective sensor strips. Class C and D are mechanical components used for both assembly-procedure optimization and training. Class C are non-working real components that can therefore be used exactly as the real components in the assembly procedures, including wirebonding. Class D are instead substitute components, with the same mechanical dimension, for instance made of aluminum instead of silicon. Assembly operations were designed and optimized starting from class D components, progressively transferring the experience gained to class C, B, and finally class A components assemblies, to avoid wasting good components during the optimization stage.

Over the years, the QCG covered many topics, ranging from very detailed operations, like gluing or wirebonding procedures, to broader activities like the ladder mount or installation activities. The group's composition was adapted to the topic being considered at a given time, including also experts from other \belletwo{} subsystems. The most relevant reviews are, in chronological order:

\begin{itemize}
	\item Site qualification - verifying that the sites have the proper infrastructure and personnel.
	\item Class B assembly qualification - review the assembly procedures on electrically working but reduced quality modules, produced with some class B components. Passing this review was a pre-requisite to use class A components. 
	\item Class A assembly qualification - review the assembly procedures on an installation quality class A module.
	\item Class A monitoring - review over time the assembly procedures and the quality of the modules.
	\item Ladder mount operation - review the suitability and safety of the ladder mount operation.
	\item SVD commissioning - review the safety and appropriateness of the SVD commissioning.
	\item SVD operation - review the safety and organization of the SVD operation.
	\item Environmental and radiation monitoring - review the environmental and radiation monitoring systems.
\end{itemize}

Detailed assembly manuals were prepared for each assembly site, specifying all site-specific details, while following the common procedures.  They included a standardized process flow diagram, describing the sequence of all assembly steps, and a set of detailed work instructions for each task. The completeness and correctness of the manuals was reviewed during the site qualification reviews. 

The main function of the QCG group was to provide a unified view of the construction process, enforcing the use of a common workflow, procedures, electrical test setup, and quality assurance methodology.

\lvlthreesection{Logistics for the mass-production}
\label{sec:logistics}
%{\it HEPHY DB} --> Christian?\\

The complexity of the SVD modules and ladders, as well as the distributed nature of the assembly process, led to a complex logistic organization that required a unified tracking tool. 
To keep track of the actual location, assembly status and quality of all these individual components, as well as the assemblies made from them, a construction database named HephyDB~\cite{Wurkner:2015}, a web application based on the CakePHP~\citeCompany{cakephp} framework and utilizing the MySQL~\citeCompany{mysql} database engine, was developed and routinely used at all assembly sites. 

The basic element in the database is an item, assigned to an item type, which categorizes the individual items according to their features and functionality. The item type also defines whether the item is a simple component consisting of a single piece 
(such as amplifier chips and sensors) or a composite item built from other items 
(such as electronic boards, ladder sub-assemblies and so on). For each composite item type a parts list is defined, which has to be filled with corresponding items when the component is created. Thus, it is easy to trace the items used for any given assembly.
For the assembly of more complex items, like ladders, in addition to the parts list, a checklist was defined, specifying the sequence of assembly steps and all the required measurements and tests to be performed during the assembly process. 
The checklist allowed the assembly to be monitored and ensured continuous quality control during the production of detector components.
For each item and assembly HephyDB provided a feature to upload files like pictures and documents as well as to store measurement results, which can be displayed in an online data browser. Finally, the database was also used to track the shipment of the components and assemblies between the assembly sites and KEK, so that the actual location of every component at any given time could be traced.

\lvlthreesection{Common electrical test setup}
\label{sec:C3_APVDAQ}
%\editor{Markus}
% general APVDAQ description, locations
% 4 channels, so 2 boards for p-side (and 2 for N-side of L3 - Melbourne and KEK)
% HV scan, ADC Delay Scan, FIR Run, Pedestal/noise, IntCal, Vsep scan
% source/laser scan (depending on location, Pisa and Melbourne: laser, HEPHY and IPMU: source, KEK: neither)

The Belle~II SVD front-end electronics were tested with a unified setup called the ``APVDAQ system''~\cite{Irmler:2008zz} at various stages of construction, i.e.\ from bare hybrids and origami boards, over BWD and FWD sub-assemblies, to full ladders of all flavors. The purpose of these tests was to verify the overall electrical functionality as well as to diagnose specific issues like \gls{pinhole} strip defects of the silicon sensor. The results were recorded after each test, and the defects compared to the corresponding results from the manufacturer's sensor testing, so that any new defects introduced in the assembly process could be readily identified.
The same system was replicated at each assembly site and at KEK so that measurements at different sites could be directly compared. 

\begin{figure}[hbt]
    \centering
    \includegraphics[width=13cm]{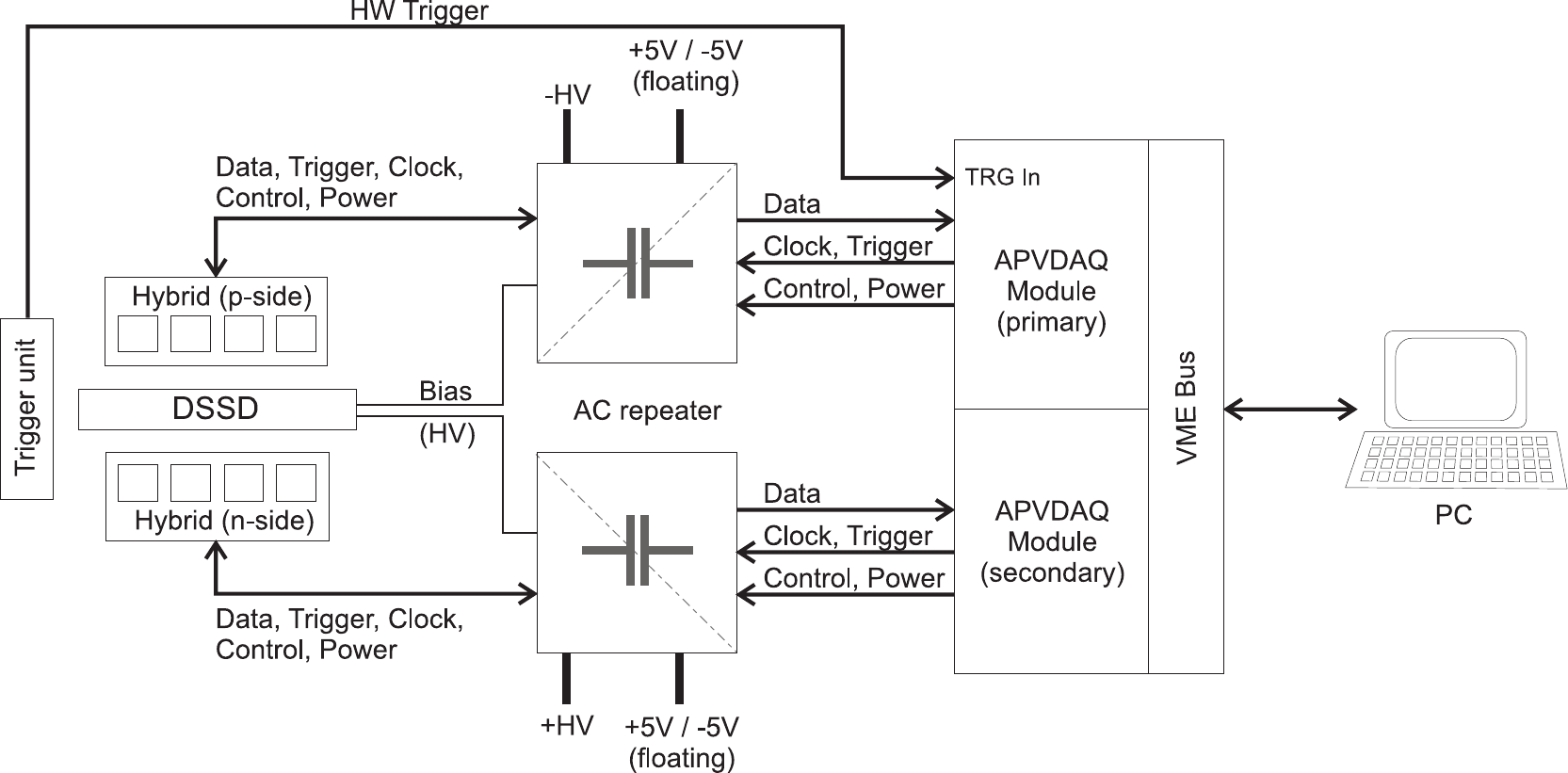}
    \caption{A schematic view of the APVDAQ system in a standard configuration reading out a double-sided silicon sensor.}
    \label{fig:C3_apvdaq_system}
\end{figure}

The APVDAQ system is a modular VME-based set of electronics modules developed and built by HEPHY Vienna to read out \apv\ chips connected to silicon sensors. It was designed for laboratory and beam tests and consists of 6U VME modules connected to repeater boards that can be up to \qty{30}{\m} away. 
The repeater boards provide signal amplification and implement AC coupling for both control signals and readout data, allowing the  system to work with double-sided silicon sensors (DSSDs), where front-end supply voltages are tied to the bias potentials. Furthermore, the APVDAQ system accepts a hardware trigger signal, typically generated by a scintillator attached to a photomultiplier when used with real particles, or by an external clock when used with a laser. Figure~\ref{fig:C3_apvdaq_system} shows the elements of the APVDAQ system when reading out a DSSD with four \apv\ chips on each side.

%%
% FF: it seems too detailed for this paper.
%

The APVDAQ system was adapted to the requirements of the SVD construction, and in particular a small \acrshort{DAC} board was added to each sensor side to create a programmable offset voltage in the range $\pm 5~\mathrm{V}$ between the sensor bias and the APV25 ground potential called \vsep, which is a powerful tool to detect broken strip coupling capacitors.

The results of each run are saved to disk in various formats, including an xml file suitable for subsequent uploading to the construction database (see~\ref{sec:logistics}). 
Several run types listed in table~\ref{tab:C3_apvdaqruntypes} are defined, and normally executed in that sequence. 
Since the analog signals from the APV25 are sampled at a reasonably high frequency (\qty{40}{\MHz} in the APVDAQ setup, while \qty{31.8}{\MHz} is used in \belletwo) a careful adjustment of the ADC clock phase is required (ADC delay scan). In addition, a digital \acrfull{FIR} filter is applied to reduce signal distortion and reflection effects. The coefficients for the FIR filter are optimized in a dedicated run (FIR calculation). 
The pedestal and noise of individual strips are calculated in the Pedestal Run, while the gain is measured by injecting a known charge through test capacitors connected to each channel during the Internal Calibration (IntCal) Scan. 
In the IntCal vs.\ \vsep\ Scan, the response of each channel is measured varying  \vsep\ from \qty{-5}{\V} to  \qty[retain-explicit-plus]{+5}{\V}, allowing a powerful defect-finding analysis, as described below. Finally, the Normal Runs are used to record actual sensor signals, that can be generated with a laser, a radioactive source, or a particle beam. 
For modules with a sensor attached, the sensor current-voltage characteristic is measured by the IV scan, varying the voltage between the N- and P-side from 0 to  \qty{100}{\V}. 
Based on the APVDAQ system experience, the same run types were later implemented in the Belle~II SVD readout system (see~\ref{sec:C4_local_runs}).

\begin{table}[hbt]
\begin{center}
\caption{Run types of the APVDAQ system.}
\begin{tabular}{ll}
\toprule
Run type & Purpose \\
\midrule
ADC Delay Scan & Find the best clock phase for each ADC \\
FIR Calculation & Obtain FIR filter coefficients for each input \\
Pedestal Run  & Measure pedestals and noise of each strip \\
Internal Calibration Scan & Obtain the gain of each strip \\
IntCal vs.\ \vsep\ Scan & Detect pinholes and other strip defects \\
Normal Run & Record radioactive source or laser hits \\
\bottomrule
\end{tabular}
\label{tab:C3_apvdaqruntypes}
\end{center}
\end{table}

\begin{figure}[hbt]
    \centering
    \subfigure[Strip with pinhole. Signal vs. time\label{fig:aDFexamples1_a}]{\includegraphics[trim=0 0 0 0.6cm, clip=true, width=.48\columnwidth,keepaspectratio, width=.48\columnwidth, keepaspectratio]{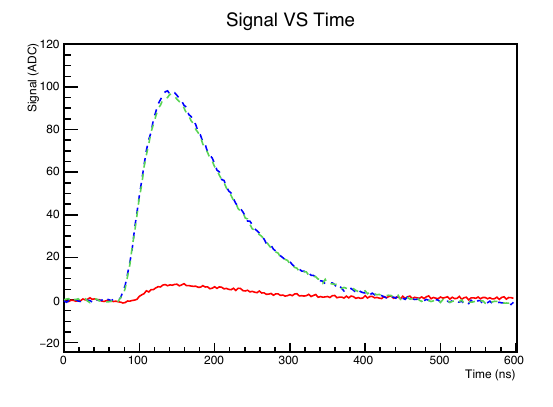}}
    \subfigure[Shorted strip. Signal vs. time\label{fig:aDFexamples1_b}]{\includegraphics[trim=0 0 0 0.6cm, clip=true, width=.48\columnwidth,keepaspectratio, width=.48\columnwidth, keepaspectratio]{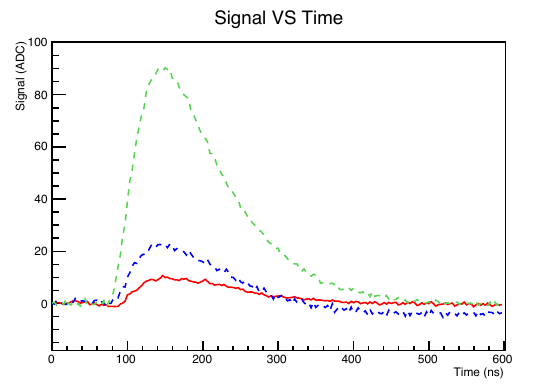}}\\
    \subfigure[Strip with pinhole. Signal vs. \vsep. \label{fig:aDFexamples1_c}]{\includegraphics[trim=0 0 0 0.6cm, clip=true, width=.48\columnwidth,keepaspectratio, width=.48\columnwidth, keepaspectratio]{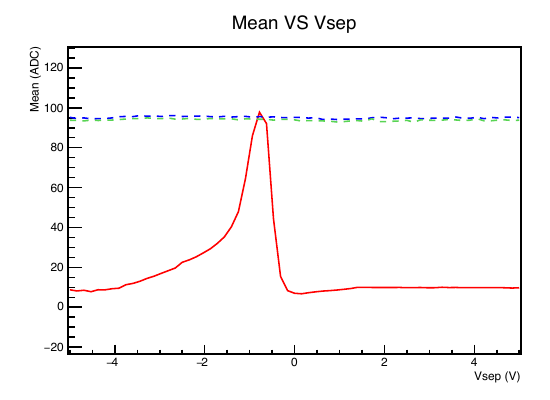}}
    \subfigure[Shorted strip. Signal vs. \vsep.\label{fig:aDFexamples1_d}]{\includegraphics[trim=0 0 0 0.6cm, clip=true, width=.48\columnwidth,keepaspectratio, width=.48\columnwidth, keepaspectratio]{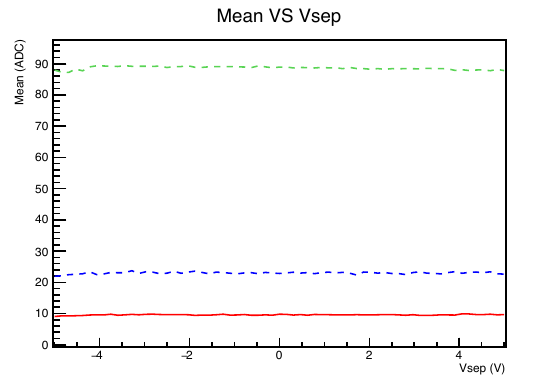}}
    \caption{Examples of plots analyzed by {\it aDefectFinder} for different types of defects. The defective strip is shown in red, while the adjacent strips are show in green and blue.
    The upper row shows the analog output of the APV25 vs time, without applying any \vsep{} voltage,  the lower row shows the height of the peak of the APV25 output, used to estimate the channel gain,  as a function of \vsep.
    The first column, (a) and (c), shows an example of a pinhole response, with gain becoming equal to normal strips when \vsep{=-0.8V}). The second column, (b) and (d), shows a shorted strip, with low gain for two adjacent channels not depending on the applied \vsep. 
    }
    \label{fig:aDFexamples1}
\end{figure}

\begin{figure}[hbt]
    \centering
    \subfigure[Open channel. Signal vs. time\label{fig:aDFexamples2_a}]{\includegraphics[trim=0 0 0 0.6cm, clip=true, width=.48\columnwidth,keepaspectratio, width=.48\columnwidth, keepaspectratio]{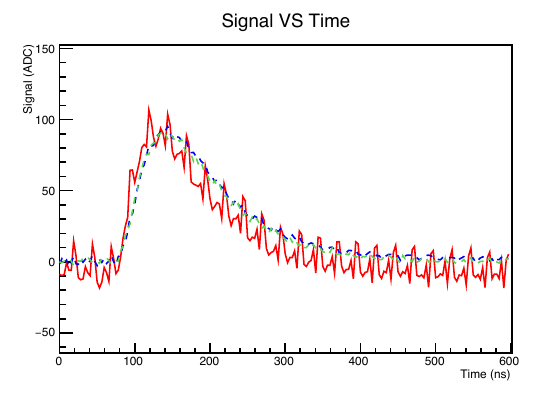}}
    \subfigure[Noisy strip. Signal vs. time\label{fig:aDFexamples2_b}]{\includegraphics[trim=0 0 0 0.6cm, clip=true, width=.48\columnwidth, keepaspectratio]{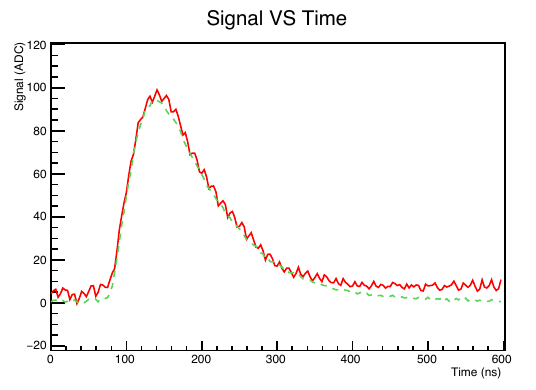}}\\
    \subfigure[Open channel. Signal peak vs. \vsep.\label{fig:aDFexamples2_c}]{\includegraphics[trim=0 0 0 0.6cm, clip=true, width=.48\columnwidth, keepaspectratio]{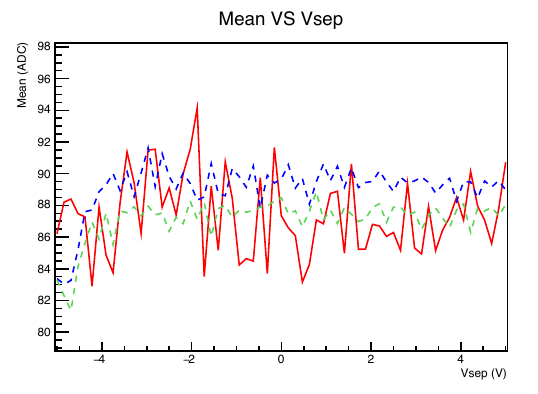}}
    \subfigure[Noisy strip. Signal peak vs. \vsep.\label{fig:aDFexamples2_d}]{\includegraphics[trim=0 0 0 0.6cm, clip=true, width=.48\columnwidth, keepaspectratio]{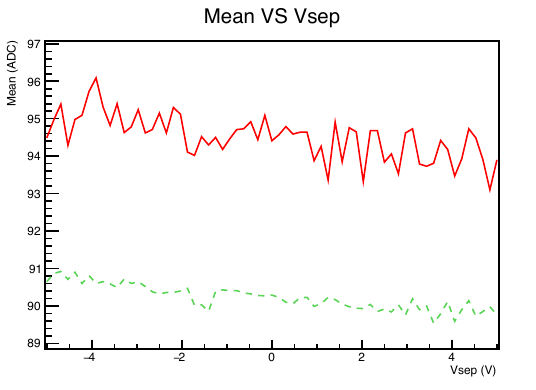}}
    \caption{Examples of plots analyzed by {\it aDefectFinder} for different types of defects. The defective strip is shown in red, while the adjacent strips are show in green and blue.
    The upper row shows the analog output of the APV25 vs time, without applying any \vsep{} voltage,  the lower row shows the height of the peak of the APV25 output, used to estimate the channel gain,  as a function of \vsep.
    In the first column, (a) and (c), the response of an open channel is shown, with a large noise due to the common mode subtraction (see text for more details). The second column, (b) and (d), shows a regularly connected noisy strip.}
    \label{fig:aDFexamples2}
\end{figure}

The results of the tests performed with the APVDAQ or FADC system (see section~\ref{sec:C2_FADC_system}) were analyzed by a dedicated software package called {\it aDefectFinder} developed by INFN-Pisa. 
The {\it aDefectFinder} package analyses the output of the electrical tests in order to identify the defective strips, checking the noise, the time and amplitude of the \apv\ waveform in response to the injected signal, and the response from the  \vsep{} scan output. 
Examples of defects and their response during the electrical tests are shown in figures~\ref{fig:aDFexamples1} and~\ref{fig:aDFexamples2}. 
The possible defects and their appearance during electrical tests are the following:

\begin{enumerate}
    \item Pinhole: the AC decoupling capacitor between the implant and the aluminum of a readout strip is broken. Since the \apv\ input potential is at \qty[retain-explicit-plus]{+0.8}{\V} with respect to the strip implant voltage, a pinhole leads to a non-negligible current flowing into the \apv\ preamplifier, saturating it; the \apv\ channel exhibits low gain and low noise that prevents it from working properly, as shown in figure~\ref{fig:aDFexamples1_a}.
    In most of the cases the pinhole can be recovered applying a \vsep=\qty{-0.8}{\V} to remove the voltage difference between the strip implant and the \apv\ input. 
    This compensation effect is evident in figure~\ref{fig:aDFexamples1_c}, where the measured \apv\ amplitude output signal, in response to the charge injection in the internal calibration circuit, is shown as a function of the \vsep{} voltage.  
    In very few cases on the u/P side, the AC oxide layer is broken to the N-type substrate, instead of to the implanted P strip. These are called ``\glspl{badpinhole}'' or ``bad pinholes'' because of their detrimental effect on the entire sensor.
    Because to the reverse bias, the natural potential in the N substrate in that region is about \qty[retain-explicit-plus]{+2}{\V}, so the direct connection to the \apv\ input with a substrate pinhole causes not only the saturation of the corresponding \apv\ channel, but also a huge sensor current, that can only be compensated with \vsep{}>\qty[retain-explicit-plus]{+2}{\V}.   Since a single \vsep{} setting is in common to many sensors, it was decided to compensate the majority of standard pinholes with \vsep=\qty{-0.8}{\V}, and to disconnect the very few substrate pinholes.
    \item Short: two or more strips are shorted together, therefore the signal collected on one strip is shared with one or more other strips. During the calibration run, the charge injected in one \apv\ channel is also shared with the adjacent channel, resulting in a lower gain for both adjacent channels, as shown in figures~\ref{fig:aDFexamples1_b} and \ref{fig:aDFexamples1_d}. 
    \item Open: the signal collected on the strip does not reach the \apv\ channel input. Since the \apv\ channel is not connected to the sensor strip, its noise is very low. But since in the electrical test a common mode correction is applied (see section~\ref{sec:C4_local_runs}), a single disconnected channel appears to have an extremely high noise corresponding to the common mode noise of the connected channels, as shown in figure~\ref{fig:aDFexamples2_a} and \ref{fig:aDFexamples2_c}.
    \item Noisy strip: the noise of the strip is higher than that of regular strips, although not as high as that if a short, as shown in figures~\ref{fig:aDFexamples2_b} and \ref{fig:aDFexamples2_d}.

\end{enumerate}

If any of the criteria tuned to identify the defects described above are matched, then the program classifies the defective strip accordingly. 
First it checks if the strip is a pinhole, by checking if the gain of the \apv\ channel changes for different \vsep{} values, as shown in figure~\ref{fig:aDFexamples1_c}.  
If the strip is not a pinhole, but has a low gain for all the  \vsep{} values, then it is classified as a short. If the noise is very large and above a given threshold it is classified as open, while if the noise is high,  but below the threshold optimized to identify the opens, it is classified as a noisy strip.  Specific values of the selection and classification criteria change with sensor types and locations and a detailed quantitative description is beyond the scope of this paper. It should be noted that in some cases strips exhibit a borderline behaviour, which makes the classification uncertain.  
A strip is also classified as a defect if, in spite of showing a normal electrical behaviour, during the laser or source scan it records a number of hits  much lower (typically less than 50\%) than the average number of hits observed in the neighbouring strips. The threshold for this defect classification is adjusted depending on the specific conditions at the different construction sites. 

%=========================================================

\lvltwosection{Silicon Sensor Fabrication and Test} {\label{sec:si_det1}}
\label{sec:C3_sensors}
%\editor{- 3 pages}

The silicon sensors, whose specifications can be found in table~\ref{tab:sensor_geo}, were tested by the manufacturers as well as by the SVD group in different ways. The small and large rectangular sensors produced by HPK were fully characterized by the vendor, with only verification sample-testing performed by SVD groups. For the trapezoidal sensors produced by Micron, on the other hand, only basic tests were performed by the manufacturer, while the full characterization of every sensor was performed by the SVD groups.
% Markus: I wrote that for the Micron section but then realized it also concerns the HPK tests in the very same way, so it should be in a common section preceding those sections.
The tests performed on all DSSDs can be grouped into two categories: 

\paragraph{Global measurements.}Overall current and capacitance measurements (IV and CV curves) are used to judge the overall health of the sensor and to measure the depletion and breakdown voltages; 
\paragraph{Strip scans.} AC (metal) and DC (implant) strip scans are performed on both P- and N-side to characterize each individual strip. The DC strip parameters, related to the quality of the wafer, implants, and bias resistors, are the strip leakage current, strip insulation with respect to its neighbors, and the poly-silicon resistance. The AC parameters are the coupling capacitance and the capacitor leakage current, related to defective AC coupling capacitors (pinholes) or to other defects in the metal layer, such as strip shorts and interruptions.

\lvlthreesection{Rectangular sensors}
%\editor{Tsuboyama} 

The HPK rectangular sensor production was carried out using 15 cm wafers over the course of several years.
% The specifications (shapes, number of strips, and arrangement of bonding pads for the origami readout) of 
% Large (120~mm~$\times$~60~mm) DSSD,
% Small (120~mm~$\times$~40~mm) DSSD
% and Trapezoidal (120~mm~$\times$~(40 to 60)~mm) DSSD sensors were already determined.
% The Large sensor was chosen for the first production.
The first prototypes of the large rectangular sensors were delivered in March 2010 and used for design verification and ladder assembly procedures development.
Following the full characterization of the prototype sensors~\cite{DOLEJSCHI:2832}, in summer 2010 an origami flex module was assembled and successfully tested on a minimum ionizing particles beam by the HEPHY group, allowing the start of the DSSD mass production.

The full characterization of the DSSDs, which presents distinctive quality control challenges, was performed at HPK contacting the wafers only on the front side. It included global measurements and complete DC and AC strip scans on the P-side, where sensor biasing was provided through a front-side N-type bulk contact realized on the sensor P-side periphery, effectively contacting the back side.This technique is not viable on the N-side, since it is not possible to realize a contact from the N-side surface to the opposite side P strips. In any case on the N-side  only AC strip scans, not requiring sensor biasing, are required, since the N strips individual currents are not important for detector characterization. 
The initial diamond-saw wafer dicing method was replaced in 2012 by laser dicing, leading to improvements in the quality of the sensor edges and in the bias IV characteristics.
The production of the large sensors was completed in 2013, while the small sensors were delivered in 2014 and 2015.

%==========
\lvlthreesection{Trapezoidal sensors}
\label{sec:C3_micron_dssd}
%\editor{Markus (, Lorenzo) } 
%Timeline of fabrication\par
%Tests performed\par
%Overview of the sensor quality\par

% 60 good sensors were ordered in Jan 2013 (38 installed, 44 required including spares)
% Micron delivered 81 sensors (+5 mechanical samples) -- from August 2013 until September 2014
% Due to manpower/scheduling issues only quick test (IV) @ Micron, but full strip scan in Vienna and/or Trieste
% (Longterm stability tests performed on all (?) Micron sensors using Vienna climate chamber)
% Tests: 23 in Vienna (only), 43 in Trieste (only), 15 first in Vienna and later in Trieste
% Sensor quality results: 52 grade A;  19 grade B;  10 grade C
% Installed --> SVD+X: 18 A, 1 B;   SVD-X: 14 A, 5 B

In January 2013, a contract was made with Micron Semiconductor (UK) for the fabrication of trapezoidal sensors to be installed in the forward modules of layers 4,5, and 6. The first sensors were delivered in August 2013 and the original plan was to have the fabrication completed by the end of that year. However, it was realized that sensor characterization was a bottleneck in the production at Micron. To avoid excessive delays, it was decided to shift the time-consuming full strip scans to the clean rooms of HEPHY Vienna and INFN Trieste. A full back side contact technique was used in these laboratories, allowing complete DC and AC strip scans on both sides of the sensors.
The final batch of sensors was delivered in September 2014.
%  Numbers should be in the summary table. 
%% accounting to a total of 81 functional devices plus five mechanical samples.

%the procurement of at least 60 usable sensors. As one trapezoidal sensor is required for each ladder in layers 4/5/6, a total amount of 38 such DSSDs is installed, while a few more were meant for spare ladders. The final batch of sensors was delivered in September 2014, which is considerably more than originally planned and thus gave us a comfortable margin of contingency.

%Despite the strip scan mitigation plan, there was an additional manpower issue in Vienna, such that some of the sensors that were initially tested (partially) at HEPHY were later fully re-tested in Trieste. Finally, o
%Out of the total amount of 81 DSSDs, 23 were tested in Vienna, 43 in Trieste and additional 15 were first characterized in Vienna and later in Trieste. The final result grades (see~\ref{sec:C3_QCG}) are listed in Tab.~\ref{tab:C3_micron_dssd}.

\lvlthreesection{Quality assurance of sensors}
Similar quality assurance procedures were applied to both HPK and Micron sensors, 
although they had some evolution during the several years of production.
Specifications included limits on the depletion voltage ($<\qty{120}{\V}$), 
on the leakage current  ($<\qty{10}{\uA}$), and on the bias resistor  ($> \qty{4}{\Mohm}$), 
as well as on the percentage of defective strips as identified by the AC and DC strip scans.

All coupling capacitors were tested at \qty{20}{\V}. The average fraction of the AC failures (pinholes) was about 0.1\% (0.03\%) in the P-side (N-side) for the HPK sensors, and about 0.4\% (0.2\%) in the P-side (N-side) for the Micron sensors.
%The average fraction of the AC failures (pinholes) is 0.1~\% in the p-side and 0.07~\% in the N-side, roughly proportional to the area of individual strips. 
The fraction of DC failures (shorts and interruptions) was in many cases as large as 5\%, but it was found that most of these strips behaved normally after the sensor was connected to the readout system. Therefore, the impact of the DC defects on the sensor quality was calculated as an effective defect fraction, obtained dividing the actual defective strip fraction by a failure-mode-dependent factor (up to 10) estimated from the sensors that had been connected to the readout system. 
Sensor quality classification was done using the sum of the effective DC defect fraction summed to the full AC failure fraction. Sensors with an overall fraction of defective strips less than  1\% were graded as class A.
The rest of sensors were graded as class B and used for the sensor and assembly studies, as well as spare ladder production.

%Overview of the sensor quality

An overall summary of sensor production, their use in completed ladders, and their installation in the experiment is shown in table~\ref{tab:C3_sum_dssd}. It should be noted that although the overall number of class A sensors would have been sufficient to avoid installing lower quality sensors, the complexity of the assembly procedures, a few incidents and unforeseen challenges (like the peel-off one described in Sec.~\ref{sec:C3_fwbw_modules}) caused losses of modules or portions of ladders, forcing the use of a few class B sensors in the installed ladders. 

\begin{table}[hbt]
\begin{center}
\caption{Summary of the DSSD production quantities and their usage in the assembly process.}
\begin{tabular}{clccc}
\toprule
Class & Usage & HPK Small & HPK Large & Micron \\
\midrule
\multirow{3}{*}{A} & Total & 26 & 188 & 52 \\
& Used in ladders  & 24 & 170  & 48  \\
& Installed in SVD & 14 & 114  & 32 \\
\midrule
\multirow{3}{*}{B} & Total & 2 & 27 & 35 \\
& Used in ladders  & 2 & 17  & 12  \\
& Installed in SVD & -  & 6  & 6 \\
\bottomrule
\end{tabular}
\label{tab:C3_sum_dssd}
\end{center}
\end{table}

%The SVD+X half (mounted first) contains 18+1 ladders with Micron sensors of grades A+B, while the SVD-X half (mounted later) has 14+5, respectively.

%========================
\lvltwosection{On-detector Electronics Fabrication and Test} {\label{sec:ondet_elec1}}
\label{sec:C3_electronics}
%\editor{Markus, Richard} - 3 pages
The front-end electronics were built in a modular way, such that readout units (hybrid and origami boards) could be tested before being integrated with silicon sensors, thus avoiding loss of expensive good-quality items by attaching defective electronics. 
Known good dies from the \apv\  production were assembled onto hybrid and origami boards. The connection from the \apv\  to the sensor strips is realized with pitch adapter flexible circuits that come in several different designs as discussed in section~\ref{sec:C2_Ladder_Design}.

\lvlthreesection{\apv\  fabrication and thinning}
\label{sec:C3_apv25}

% Markus

The \apv\  chips were produced in \qty{0.25}{\um}  CMOS technology on 8" wafers with a thickness of $325\pm \qty{25}{\um}$  by an IBM foundry and subsequently wafer tested using an automated probe card setup at Imperial College, London. The overall production yield (fraction of good chips) was 88\%~\cite{Raymond:2005qa}, translating to more than 300 known good dies per wafer. About 75,000 chips were installed in the CMS Tracker, while almost twice that amount of good chips were produced in total, making some excess quantities  available for other experiments like Belle~II.

Chips mounted on the origami boards are located in the detector active region and therefore require thinning down to \qty{100}{\um}, while thick chips are mounted on the hybrid boards.
Since wafer level thinning is a well established procedure,  several wafers were thinned and subsequently diced. The chips were mounted on the origami boards and connected to the first batch of PA0 pitch adapters,  which turned out to be defective (see section~\ref{sec:C3_origami}).
When the problem was discovered, all the remaining \apv\  wafers with standard thickness were already diced, and the replacement of the defective origami boards required a non-standard thinning procedure applied to individual chips. In the first attempt it was observed that microscopic cracks appeared at the edges of many chips, sometimes causing malfunction. 
In the second attempt with a different company, the single-die thinning was successful, as verified by optical inspection and probe card testing, yielding a sufficient number of chips to populate a second set of origami boards. The numbers of the purchased and used \apv\  chips are summarized in table~\ref{tab:C3_APVnumbers}.

\begin{table}[hbt]
\begin{center}
\caption{Summary numbers for the \apv\  production. Thick chips are $325\pm \qty{25}{\um}$ thick, while thin chips were thinned down to \qty{100}{\um}.}
\begin{tabular}{lrrr}
\toprule
 & Thin & Thick & Total \\
\midrule
Purchased & & & $\approx 7100$ \\
Used on modules & 3330 & 2172 & 5502 \\
Installed in \belletwo  & 820 & 928 & 1748 \\
\bottomrule
\end{tabular}
\label{tab:C3_APVnumbers}
\end{center}
\end{table}

\lvlthreesection{Pitch adapter fabrication and test}
\label{sec:C3_pitch_adapters}

A contract for the final fabrication of the pitch adapters was awarded in 2015.
%% Note: Variable Energy is a consultant company, so should not be mentioned here. made with Variable Energy in 2015. 
%The pitch adapter flexs are produced by Tokai Denshi. Tokai Denshi is no longer active. 
The quality of the produced pitch adapters was ensured by performing three main checks:
\begin{itemize}
	\item  Visual inspection at the pitch adapter flex production company, Tokai Denshi~\citeCompany{TokaiDenshi}, to check the quality of materials, pattern defects, and dimensional tolerances.
	\item Electrical measurement of all lines on the pitch adapters. Quality control procedures require all lines to have low resistance, no interruption, and no short circuit to adjacent lines. The resistance of each line and the resistance between neighboring lines were measured to identify open lines and shorts. The electrical measurements were performed by Daiei Electronics~\citeCompany{Daiei}.
	\item Dimensional verification of the bonding pad width to ensure bondability. The widths of each pad in the innermost row at the \apv\ side, which is the narrowest in the design, were measured at IPMU with an optical \acrfull{CMM} and were required to be larger than \qty{30}{\um}. Additional visual inspection was performed at each assembly site.
\end{itemize}
%\rem{still missing PA0 numbers }
The number of produced pitch adapter circuits is summarized in table~\ref{tab:C3_pitchadapter}.

% May 21, 2021 K. Hara
% type  Produced installed
% PA1  % 200 % 82
% PA2  % 200 % 82
% PF1  % 100 % 38      
% PF2  % 100 % 38
% PB1  % 100 % 38 
% PB2  % 100 % 38
% P3F1 % 30  % 7
% P3F2 % 30  % 7
% P3B1 % 30  % 7
% P3B2 % 30  % 7
% PA0  % 270 % 7  
% number of PA0 passed electrical test in 6 batches 90+14+6+26+21+113=270 
\begin{table}[hbt]
\begin{center}
\caption{Approximate numbers of produced pitch adapter boards for the different uses.}
\begin{tabular}{lcccc}
\toprule
  & Layer 3& FWD/BWD modules & \multicolumn{2}{c}{Origami}  \\
  & P3F1/2, P3B1/2  & PF1/2, PB1/2  & PA1/2 & PA0   \\
\midrule
Purchased             & 120 & 400 & 400 & 270 \\
Installed in Belle II & 28  & 152 & 164 & 82  \\
\bottomrule
\end{tabular}
\label{tab:C3_pitchadapter}
\end{center}
\end{table}

\lvlthreesection{Hybrid board fabrication and test}
\label{sec:C3_hybrid_boards}
% Markus

% PCBs produced in industry
% components placed with automatic pick & place machine, then vapor phase soldered at HEPHY
% conductive glue applied by (adapted) CMM, APV25 chips placed on pick & place machine (+manual alignment)
% tested with APVDAQ

The hybrid boards are PCBs composed of six layers with an overall thickness of \qty{0.7}{\mm} and were manufactured in industry with gold plated surfaces to enable aluminum wire bonding to the \apv\  chips. The boards were electrically tested for open- and short-circuits with a needle probe tester at the manufacturer~\citeCompany{ref:LEITON}, with 16 hybrid boards combined on one single panel.

The assembly of electronic components was done at HEPHY Vienna. First, solder paste was applied using a manual stencil dispenser with the proper solder mask corresponding to the PCB panel. The passive components (resistors, capacitors and the connector) were then positioned with an automatic pick and place machine, followed by vapor phase soldering. After that, conductive glue was dispensed at the locations of \apv\  chips using an adapter on the CMM. Then, \apv\  chips were positioned using again the pick and place machine, with some additional manual alignment. After curing of the glue, the individual hybrid boards on a panel were separated and all \apv\  chips on each hybrid were electrically connected to the PCB by aluminum wire bonding. For L3 hybrid boards, the hybrid cable was directly soldered onto each board in the last step, using a shield to protect the \apv\  chips and their wire bonds. All hybrid boards were electrically tested using the APVDAQ setup 
(see section~\ref{sec:C3_APVDAQ}). 

The numbers of produced hybrid boards are summarized in table~\ref{tab:C3_hybridtests}. In this case, class A refers to a completely flawless hybrid board, whereas class B indicates fully functional boards with some minor defects, such as: some \apv\  chips are slightly displaced from their nominal position; one of the (up to four) redundant wire bonds is missing; the amount of conductive glue is slightly extending beyond the normal area, without causing unwanted connections. Class C indicates not-functional, mechanical-quality boards that can be used for testing assembly procedures. 
%, i.e.\ a condition that reduces the -- generally very comfortable -- margin a bit but by itself is no reason for malfunction. S
Additional dummy boards were produced initially for training purposes, but those are not included in this list. A total of 180 hybrid boards are installed in the Belle~II SVD (among which are 161 of class A and 19 of class B) -- each of the 45 ladders holds a pair of hybrid boards glued back-to-back on either end.

% details (maybe we don't need that detail level, just sums for A/B/C):
% H3P:  A/B/C = 27/9/2
% H3N:  A/B/C = 28/6/8
% H6+P: A/B/C = 59/16/1
% H6+N: A/B/C = 74/3/5
% H6-P: A/B/C = 56/13/2
% H6-N: A/B/C = 66/4/6

\begin{table}[hbt]
\begin{center}
\caption{Summary of the hybrid board production.}
\begin{tabular}{lr}
\toprule
Grade & Number \\
\midrule
A & 310 \\
B & 51 \\
C & 24 \\
\midrule
Total & 385 \\
\bottomrule
\end{tabular}
\label{tab:C3_hybridtests}
\end{center}
\end{table}

\lvlthreesection{Origami board fabrication and test}
\label{sec:C3_origami}
% Koji/Toru

The first mass production of origami boards started in 2013, after full verification of a test module. About 150 such boards were produced in the first batch. 
After mounting \apv\  chips and passive components, the boards were tested performing calibration with test pulses. 
However, a major problem was identified only after the connection to the sensor, with many channels showing no signal at all.
Using a microscope for visual inspection, many tiny cracks of the copper lines on the PA0s, i.e.\ the flexes connecting the sensor N side strips to the \apv\  (see section~\ref{sec:C2_Ladder_Design}), were found, localised near the bonding pads at the sensor side,  where the curvature of the copper lines is largest. 
All the inspected origami boards (about 20) had cracks affecting  about 20\% of the copper lines, mostly on the bottom layer of two PA0 layers. 
Due to the very high crack fraction, and the unfeasibility of any repair work, a second mass production of PA0 pitch adapters and origami boards was required, as well as a new thinning campaign for the \apv\  chips, as described in section~\ref{sec:C3_apv25}. 

After detailed investigation of the  production procedure several weak points were identified:  the design of the copper lines, the cutting method of PA0s, and the lack of individual lines electrical testing. In addition, the glue used between the two PA0 layers was changed from acrylic to epoxy type out of concerns of radiation tolerance, leading to an increase in stiffness. 

The problematic points were solved by improving the production and testing process as well as the sign-off procedures used to define reviewed and approved specifications:
the design was improved to be more tolerant to mechanical stress; the original acrylic glue, more flexible and sufficiently radiation tolerant, was used; a  laser cutting method was introduced; electrical measurements of each individual line resistance and isolation from the neighboring lines were performed; fully documented specifications including all the design details, tolerances, and testing procedures were prepared, formally signed-off, and forwarded to the manufacturer.

The final mass production started in 2015 at Taiyo Industrial~\citeCompany{Taiyo}.
Electrical tests were performed by Taiyo for origamis and Daiei Electronics~\citeCompany{Daiei} for PA0s, selecting only circuits with no failures. Gluing of the PA0 on the origami, the assembly with surface mount passive components and \apv\ chips, as well as the wirebonding of the chips digital and power lines, were performed in companies, with frequent visual inspection steps, while the final inspection and electrical test on the assembled origami boards were performed at KEK.

During the first batch of mass production it was discovered that blisters were produced on origami flexes in the automated reflow soldering process. 
Although there was suspicion that the problem could be due to an excessive water content between the various origami layers, a workable solution could not be found, and hand soldering was adopted instead of reflow soldering, to avoid the blisters.

Table~\ref{tab:C3_Origamitests} summarizes the mass production numbers. 
Origami boards that passed all the inspections and electrical tests with no bad channels are selected to be class A for the ladder assembly. Of these boards 82 are installed in the experiment.
Class B identifies origami boards with defects observed in the visual inspection or some minor non-compliance of electrical specifications. 
Class C boards have significant failures, such as dead \apv\  chips or assembly defects, and are used only for mechanical tests. 

% type :  A / B / C / other
% O-Z  : 54 / 8 / 4 / 2 
% OCE  : 39 / 6 / 4 / 0 
% O+Z  : 23 / 4 / 1 / 1 
% 3 "others" above are flagged as "destroyed" on db,
% but included in class C for simplicity in the table below
% O-Z111 (small blister by hand soldering of GND cable)
% O-Z140 (assembly terminated due to partial peeling off of gold plating)
% O+Z115 (solder fully covers a PA0 pad)

\begin{table}[hbt]
\begin{center}
\caption{Summary of the origami board production.}
\begin{tabular}{lr}
\toprule
Grade & Number \\
\midrule
A & 116 \\
B & 18 \\
C & 12 \\
\midrule
Total & 146 \\
\bottomrule
\end{tabular}
\label{tab:C3_Origamitests}
\end{center}
\end{table}

\lvlthreesection{Junction board fabrication and test}
% Richard/Markus

\begin{figure}[hbt]
	\centering
		\includegraphics[height=6.5cm]{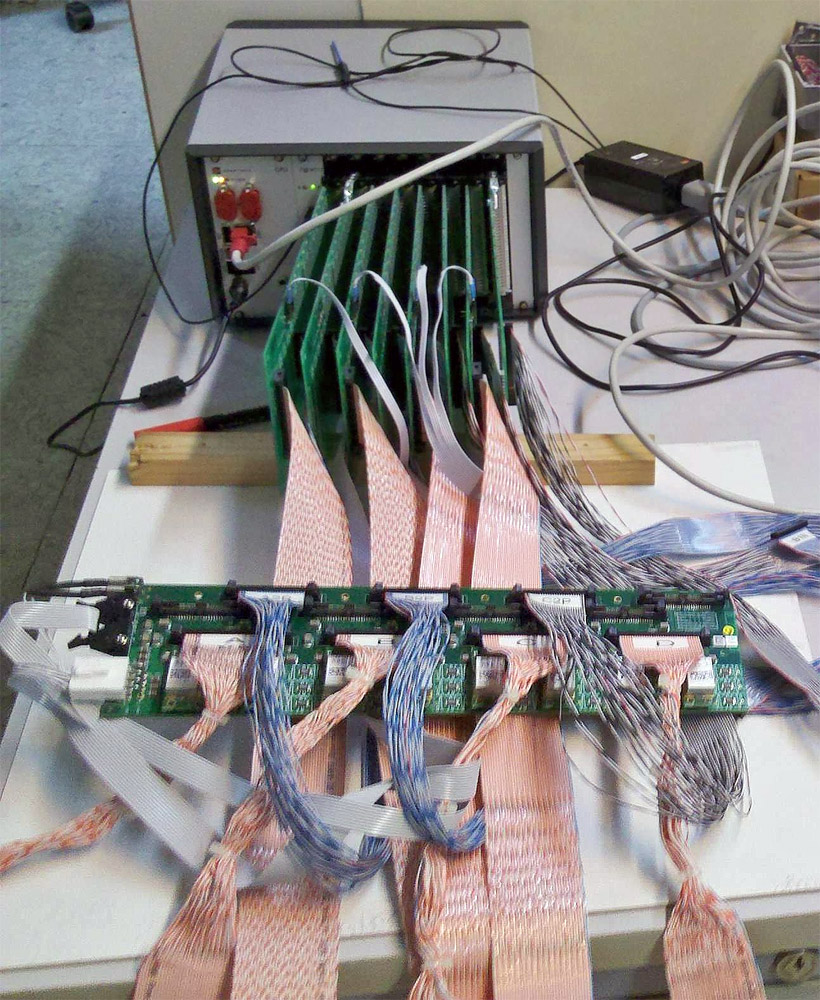}\hfill
		\includegraphics[height=6.5cm]{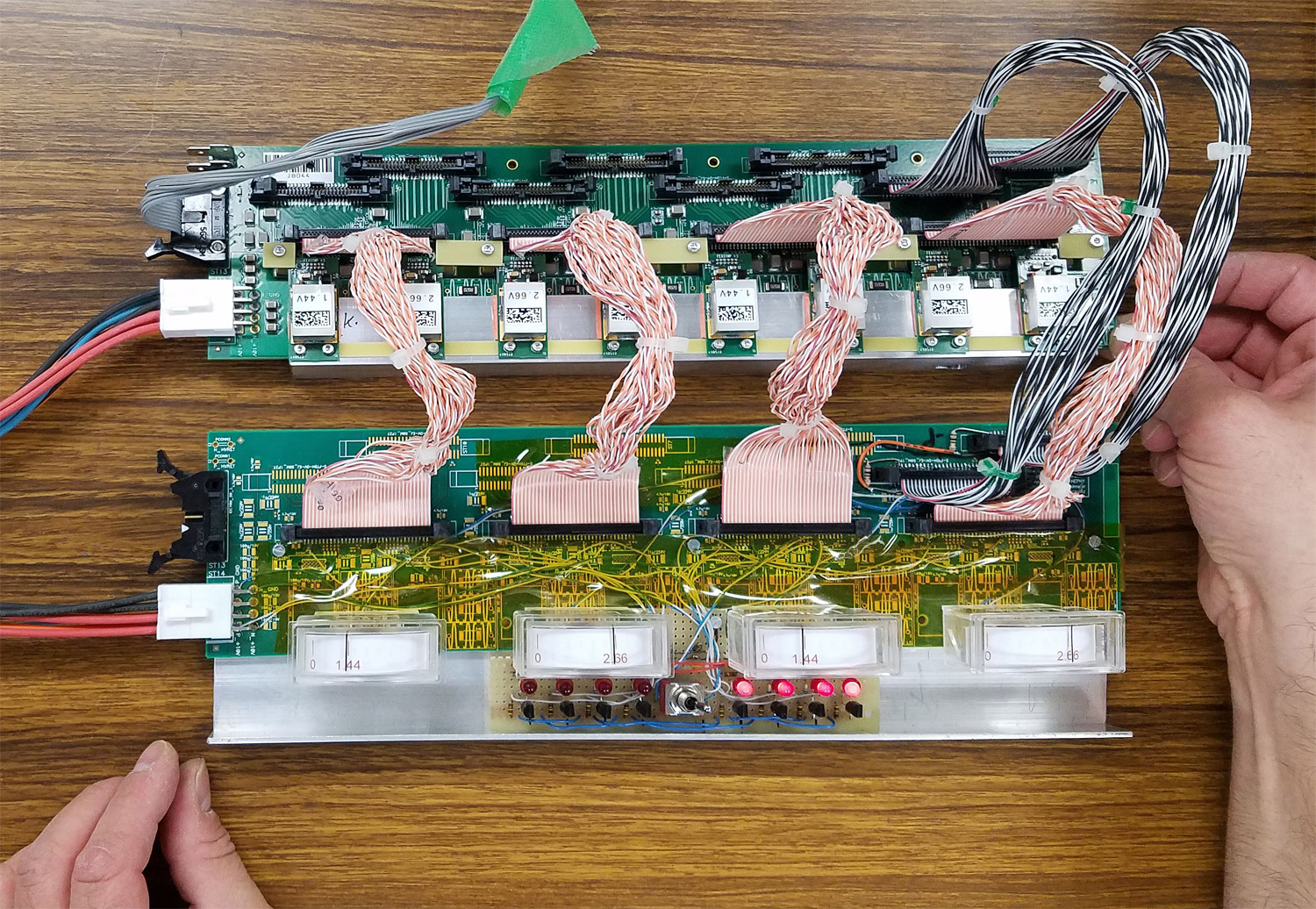}
	\caption{Left: junction board (bottom) under test with the cable tester (top). Right: Voltage display (bottom) for a junction board under test (top). The switch on the bottom allowed selecting between the default and redundant converter for a complete test.}	
	\label{fig:C3_jb_test}
\end{figure}

The junction board \acrshort{PCB}s, which arrange the \apv\  signals and produce the 1.25~V and 2.5~V supplies for the \apv\  by using air-core DC/DC converters~\cite{Faccio:2014iya}, were manufactured and assembled in industry, where they were also subjected to optical inspection, but no functional test. At HEPHY Vienna, they were equipped with the mounting/cooling bracket and the DC/DC converter modules. Then, each board was tested using a commercial cable tester (Adaptronic NT-210, see left side of Fig.~\ref{fig:C3_jb_test}). This device measures resistances and capacitances as well as diode characteristics between any combination of its 512 I/O pins and thus in two passes compares a whole junction board connected to it to a previously recorded reference board. 

Before their assembly on the junction boards, each DC/DC converter module was characterized individually by recording two IV curves with a dummy load using a Source Measure Unit (SMU, Keithley 2410), once with its ENABLE signal on and once off, in order to verify their healthiness. After installing them on the junction board, the individual output voltages (nominally \qty{2.66}{\V} and \qty{1.44}{\V}) were measured using a dedicated metering board that can switch between the default and redundant converters (see right side of Fig.~\ref{fig:C3_jb_test}), allowing a complete test of the system and verifying that the correct voltage variants were installed.

%=======================================================
\lvltwosection{Module and Ladder Assembly} {\label{sec:modules1}}
\label{sec:C3_module}
% \editor{Higuchi} - 4pages
Modules and ladders were produced with responsibilities shared among SVD collaborating institutions: \acrfull{SFW}s and \acrfull{SBW}s by INFN Pisa, Layer 3 ladders by the University of Melbourne, Layer 4 ladders by TIFR-Mumbai, Layer 5 ladders by HEPHY-Vienna. The Layer 6 ladders were produced by a consortium comprising Kavli IPMU, University of Tokyo, Science University of Tokyo, Niigata University, Nippon Dental University, Tohoku University, KEK, Kyungpook National University, and Seoul National University.  
The TIFR-Mumbai team shared the clean room at Kavli IPMU because of restrictions  on the export of radiation hard components to India.

The modules and ladders were assembled using the same techniques across all the production sites.  In particular, since Layer 4, 5 and 6 ladders share the same ladder structure they were fabricated following very similar assembly procedures.
In this section,  techniques and procedures common to all assembly sites are first presented, followed by the module and ladder production procedures for the Layer 3 ladders, the forward and backward modules, and the Layer 4, 5 and 6 ladders.

\lvlthreesection{Common techniques and procedures}

\paragraph{Clean room setup.}
All the module and ladder assembly procedures were carried out in a controlled environment to avoid contamination and improve the reproducibility of the operations. The assembly areas were required to  maintain an ISO8 cleanliness level, equivalent to class 100000 of US Fed.  Std. 209E~\citeCompany{ref:cleanroom}.
The environment of the clean rooms was controlled and/or monitored to ensure constant conditions, needed to ensure dimendional stability of the parts and reproducible glue deposition and curing characteristics. Typical values were  \qty[uncertainty-mode=separate]{23\pm 1}{\degreeCelsius}  for the temperature and  $(50\pm 3)$\% for the relative humidity.
Clean rooms were equipped with site-specific anti-static protection systems including anti-static floor, mats, bracelets, boots; proper grounding of the operators and equipment was routinely verified.

\paragraph{Assembly jigs and alignment.}
A large number of assembly jigs were developed in the different sites and were either fabricated in the local workshops or outsourced to precision mechanical machining companies. Depending on the use and required precision, aluminum or stainless steel was used. All the surfaces that would have been in contact with the sensors or other delicate components were fabricated in Delrin\textsuperscript{\textregistered}~\citeCompany{delrin} or 
Teflon\textsuperscript{\texttrademark}~\citeCompany{teflon} to prevent scratches. Inside most of the jigs, vacuum channels or holes were realised to allow vacuum chucking of the sensors.
The jigs were all qualified with a touch-probe CMM to ensure compliance with the required tolerances, which were set between 10 and \qty{100}{\um}, depending on the mechanical elements to be held and on the length of the jigs.
Precision coupling between different jigs was realized using commercial pins and bushings or precision holes. Particular care was devoted to avoid any interference between the jigs and the module wire-bondings, especially on the side of the module that is lying on the jig, which is not visible and cannot be visually checked.
Many assembly steps were carried out under a CMM that allowed a precision alignment of the components and the measurement of their position. Micrometer screw gauges were used in many locations to adjust the position of the various components. Silicon sensors have an ``F-shaped'' alignment mark in each corner which allows automatic detection and position measurement by the CMM.  For the positioning of the sensors, a dedicated XYZ$\Theta$-stage, allowing a 4-axis adjustment of the sensor position under the CMM, was developed and produced in several copies for the different ladder assembly sites.

\paragraph{Component gluing.}
The different components of the modules and ladders are for the most part glued to each other.
The Araldite 2011$^{\circledR}$~\citeCompany{ref:araldite} epoxy glue, whose radiation hardness had been confirmed by the LHC experiments~\cite{Guarino:531818}, is used throughout the assembly process.
The position and spread of glue deposited below the pitch adapters are very tightly controlled using optimized glue deposition patterns, to ensure a sufficient glue spread below the pitch adapter that makes a solid foundation for wire-bonding pads. Under-flown glue could leave an empty region under wire-bonding pads, while over-flown glue from the pitch adapter could spoil bonding pads near the pitch-adapter edge. Both conditions could result in wire-bonding failure. The typical glue thickness under the pitch adapters was \qty{50}{\um}, controlled by the position of the gluing jigs with a precision of  \qtyrange{5}{10}{\um}.
Each assembly site developed a special setup to control the glue position and spread. 
%Details of the equipment used at different sites is available in table~\ref{tab:C3_sum_equipment}.
%The environment of the clean room was controlled and/or monitored to ensure constant conditions during glue curing: typical values were  \qty[uncertainty-mode=separate]{23\pm 1}{\degreeCelsius}  for the temperature and  $(50\pm 3)$\% for the relative humidity. 
The typical glue curing time was 24 hours.

\begin{figure}[!ht]
 \centering
 \includegraphics[width=.8\columnwidth,keepaspectratio]{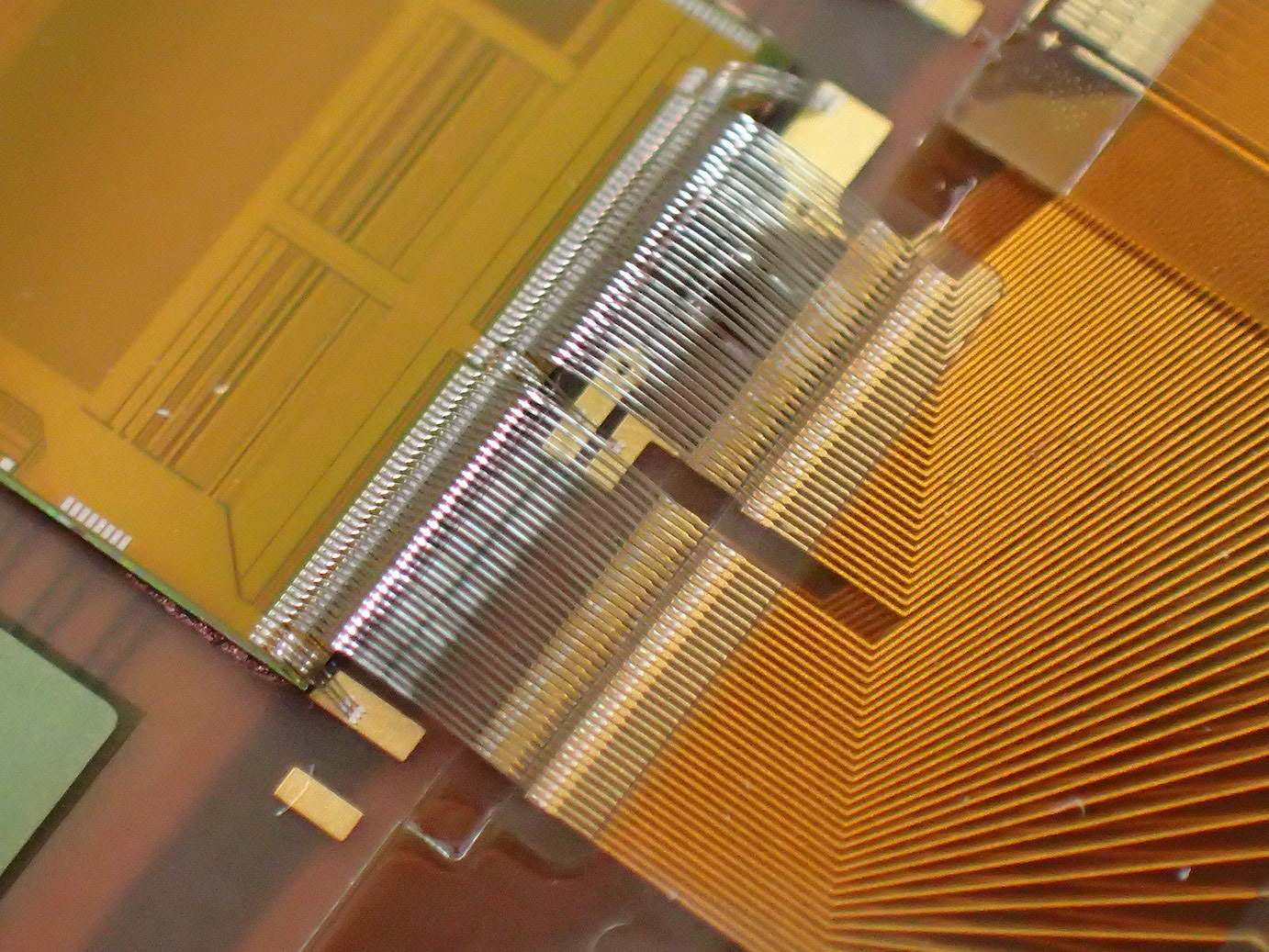}
 \caption{Photo of the wire bonding between a pitch adapter (PA0) and an \apv\  chip.}
 \label{fig:C3_Wire_Bonding}
\end{figure}

\paragraph{Wire bonding.}
The sensor strips, the pitch adapters, and the \apv\  chips are electrically connected with \qty{25}{\um} diameter aluminum wires using the ultrasonic wedge bonding technology. An example of wire bonding is shown in figure~\ref{fig:C3_Wire_Bonding}.

The requirements are set to a bonding efficiency greater than $99\%$ and a pull force greater than \qty{5}{gf} (gram force), as measured with a destructive pull tester.  To satisfy the requirement, each assembly site optimized the parameters of its wire-bonding machine, such as the ultrasonic bonding power and time, the wedge pressure on the pads, the shape and height of the wire loop.
Table~\ref{tab:C3_Nwirebonds} summarizes the total number of wire-bonds performed for the various layers at the different sites, considering all connections between sensor, pitch adapter and \apv\  chips ($128\times 2$ per chip). The wire-bonds for the connections of the chip bias (18 per chip considering redundancy), signal, clock and control lines (48 per chip) were made at the hybrid or origami production sites, and correspond to an additional 26\% with respect to the numbers in table~\ref{tab:C3_Nwirebonds}.

\begin{table}[!htp]\centering
\caption{Number of wire-bonds for the installed modules for the readout channels. Considering all the actually produced modules a factor 1.5 to 2 should be considered.} \label{tab:C3_Nwirebonds}
\begin{tabular}{lrrrr|r}\toprule
Site &L3 &L4 &L5 &L6 &Total \\
\midrule
Melbourne &43008 & & & &43008 \\
Pisa & &51200 &61440 &81920 &194560 \\
TIFR & &25600 & & &25600 \\
HEPHY & & &61440 & &61440 \\
IPMU & & & &122880 &122880 \\
\midrule
Total &43008 &76800 &122880 &204800 &447488 \\
\bottomrule
\end{tabular}
\end{table}

\paragraph{Electrical test.}
Ladders and modules were electrically tested at almost every stage of the assembly process, as well as before and after their transfer to KEK, using the common APVDAQ readout system described in section~\ref{sec:C3_APVDAQ}.
The APVDAQ test stand was complemented with a system capable of injecting charge in the sensor, equipped with stepping motors to scan the entire sensor surface:  in Pisa and Melbourne an infrared light laser emitter was used, while in HEPHY and IPMU a radioactive $\beta$ source was used.
The response of individual strips was analyzed using the {\it aDefectFinder} software package, described in section~\ref{sec:C3_APVDAQ}, to classify defects and compare them with the list of known defects from the sensors manufacturer or from previous testing steps. Such frequent and systematic testing allowed the early detection and often repair of defects, and prevented the use of problematic modules in the successive stages of the assembly process.

\paragraph{Procedure documentation and tracking.}
An extensive set of manuals was prepared, reviewed by the \acrshort{QCG}, and made available to all collaborators. Each step in the procedures was described in the manual, with an associated checklist to be followed. The checklist was also implemented in the construction database and all the relevant information, such as optical inspection pictures, mechanical measurements, and results of electrical tests, were inserted in the database, so that any piece of information on the module parts could be recovered easily.

%========================================
\paragraph{Equipment at assembly sites.}
Although the assembly sites had the same basic capabilities, the specific equipment differed from site to site, leading to slight differences in the procedures that had to be fully understood and documented.  A summary of the equipment used in the assembly sites is shown in table~\ref{tab:C3_sum_equipment}.

\begin{table}[hbt]
\begin{center}
\caption{Summary of the equipment used at different sites.}

\begin{tabularx}{1.0\textwidth} { 
   >{\raggedright\arraybackslash}X 
   >{\raggedright\arraybackslash}X 
   >{\raggedright\arraybackslash}X 
   >{\raggedright\arraybackslash}X 
   >{\raggedright\arraybackslash}X  }

%\begin{tabular}{p{0.1\textwidth}p{0.2\textwidth}p{0.2\textwidth}p{0.2\textwidth}p{0.2\textwidth}}
\toprule
Site  & CMM & Glue robot & Glue dispensing  & Wire Bonding \\
\midrule
Melbourne & Mitutoyo QV-PRO302 & SONY CAST Pro II & EFD 1500 XL & Hesse \& Knipps BJ715M  \\[5pt]
Pisa  & Mitutoyo BHN506 & I\&J Fisnar 750 & EFD  & K\&S 8090  \\[5pt]
TIFR  ($^*$) & Mitutoyo QV-X606P1L-C &  Musashi ShotMaster 3 & Musashi  ML-5000XII & F\&K Delvotec 6400 \\[5pt]
HEPHY  & Mitutoyo Euro-C776 & Customized Mitutoyo CMM &  Loctite 97006  & F\&K Delvotec 64000 G5 \\[5pt]
IPMU  & Mitutoyo QV-X606P1L-C & SONY CAST Pro; JANOME SCARA & Musashi  ML-5000XII  & Cho-on-pa Kogyo REBO-7   \\[5pt]
\bottomrule
% \end{tabular}\\
\end{tabularx}\\
($^*$) some of the equipment is shared between TIFR and IPMU.
\label{tab:C3_sum_equipment}
\end{center}
\end{table}

% \lvlthreesection{General scheme for module and ladder assembly}
\lvlthreesection{Layer 3 ladder assembly}
\label{sec:C3_layer3_assembly}
%\editor{Francesco, Antonio}
Layer 3 ladders are made by two small rectangular sensors connected to \apv\  chips through pitch adapters. A description of Layer 3 ladders is given in section \ref{sec:C2_Ladder_Design}. The following procedure describes the main steps of Layer 3 ladder assembly. \par

At first, the Layer 3 hybrid boards of u/P and v/N sides are glued together to build the Layer 3 hybrid sandwich: the P-side board is positioned on the hybrid gluing jig, where it is glued with the N-side hybrid board using two countersunk screws to find the alignment. After 24 hours of glue curing, the hybrid sandwich is optically inspected and electrically tested. Since for Layer 3 hybrid boards no connector is used and cables are soldered, handling the layer 3 hybrid sandwich requires extreme care to avoid tensions caused by the permanently attached cables. 

\begin{figure}[!ht]
 \centering
 \includegraphics[width=\textwidth]{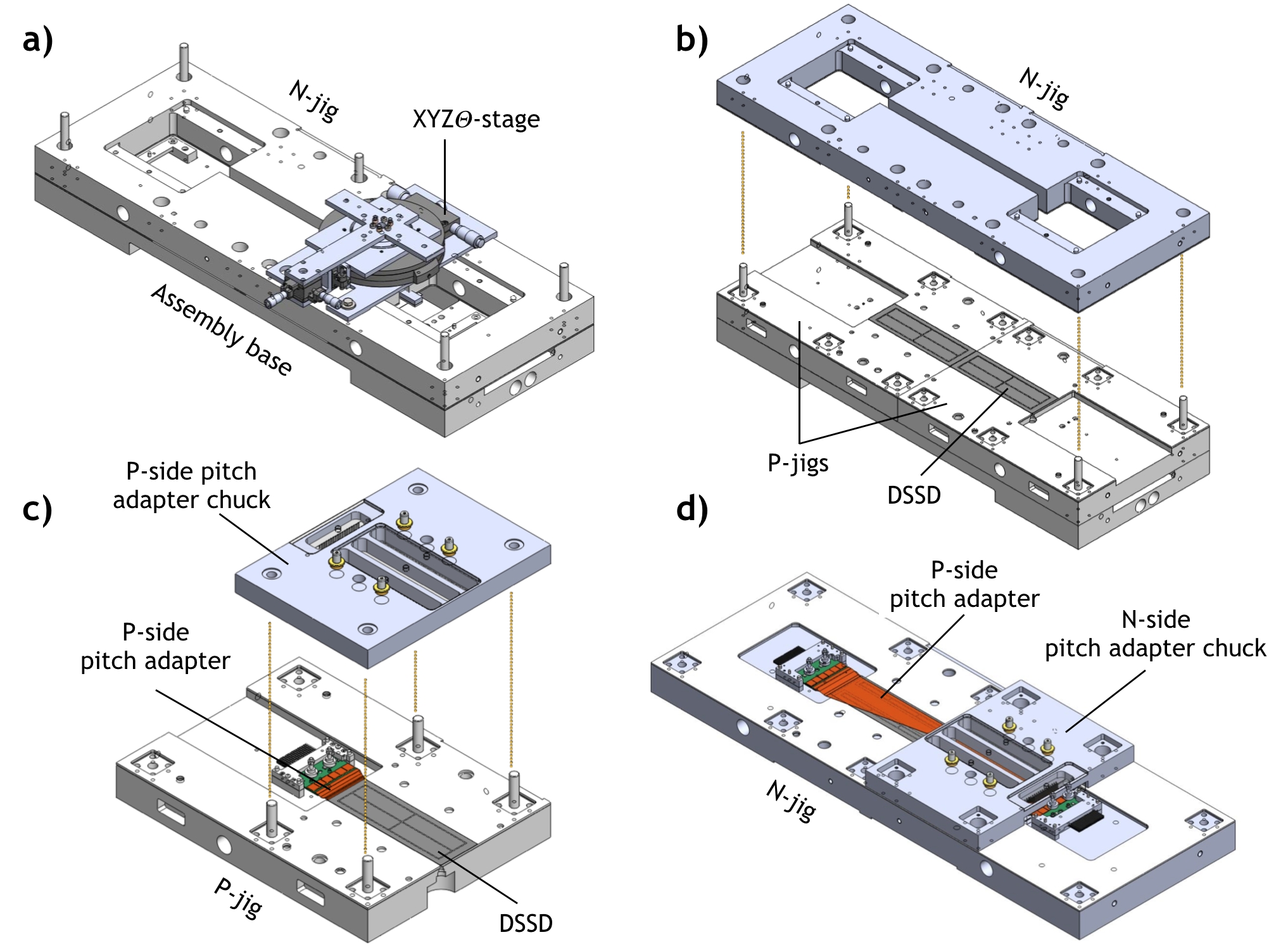}
 \caption{Some steps of Layer 3 ladder assembly: a) alignment of DSSD using XYZ$\Theta$-stage; b) positioning silicon sensors on the P-jigs; c) gluing of P-side pitch adapter between DSSD and hybrid sandwich; d) gluing of N-side pitch adapter.}
 \label{fig:C3_L3_assembly}
\end{figure}

The assembly of a Layer 3 ladder requires wire-bonding and other operations to be performed on both sides of the sensors, hence N-jigs and P-jigs were produced. They can be coupled through precision pins, allowing picking-up a sensor and turning it upside down without losing the alignment. The jigs are divided in two halves handling the forward and backward sensors of the ladder.
The two silicon sensors are positioned on the N-jig, then the XYZ$\Theta$-stage is used to align the sensors under the CMM, using the F-marks of sensor corners as a reference, as shown in figure~\ref{fig:C3_L3_assembly} a). The main reference system is determined using reference points of the assembly base, which is a support structure where the N-jig is positioned. After the alignment, the N-jig is lifted, the two P-jigs are positioned on the assembly base, then the N-jig is lowered again on top of the P-jigs, where the sensors are transferred, with their u/P side up. Figure~\ref{fig:C3_L3_assembly} b) shows the N-jig lifted up after sensors have been transferred to P-jigs. For each P-jig, the P-side pitch adapter is glued on the sensor and the hybrid sandwich. P-side pitch adapters are positioned on a dedicated jig, and glue is dispensed using a robot. Figure~\ref{fig:C3_L3_assembly} c) shows the P-jig with the hybrid sandwich, the sensor and the u/P side pitch adapter already glued, and the P-side pitch adapter chuck used to perform the gluing. 
After glue curing and optical inspection, P-side wire bonding and electrical test is performed. Each sensor is then picked up with one half of the N-jig, and is turned upside-down to prepare it for the N-side pitch adapter gluing, which is performed after properly adjusting  the height of the hybrid sandwich. Figure~\ref{fig:C3_L3_assembly} d) shows the N-jig with one v/N side pitch adapter glued, and with the other under the N-side pitch adapter chuck, during gluing operation. After glue curing, the two halves of the ladder are optically inspected, wire bonded on the N-side and electrically tested. At this stage, two half-ladders are ready for the assembly of a full Layer 3 ladder.\par

The two half-ladders are positioned P-side up on the basement-jig. Forward and backward aluminum bridges are positioned and fixed on the corresponding hybrid sandwiches. The ribs are aligned and glue is dispensed on the stand-offs and in the rib slots of the forward and backward bridges. The ribs are then lowered on the P-jig until the stand-offs are at $\sim \qty{100}{\um}$ from the sensor and the two ends are inserted in the respective slots of the bridges. After 24 hours of glue curing, the ladder is completed. It is optically inspected and electrically tested, then it is finally stored in the transportation box and prepared for the shipment to KEK.

\lvlthreesection{Forward - backward modules}
\label{sec:C3_fwbw_modules}
%\editor{Stefano}

The following procedure applies to both forward and backward modules. Two different gluing jigs are needed for the P-side and N-side gluing of the pitch adapters. Each jig consists of: a tower with a vacuum chuck holding the silicon sensor on its Teflon surface; a second tower where the double-sided hybrid board is fixed by screws; a transparent vacuum chuck holding the pitch adapter positioned on top of the two towers. 
All the operations are done under a CMM. All the manipulation and transfer of the modules are done with several different types of vacuum chucks specifically designed to match the relevant dimensions and avoid damage to wirebonds. 
\par
The silicon sensor is placed P-side up, aligned by a mechanical stop on its tower, and the reference system is defined using the CMM software. The hybrid sandwich is then placed on its tower and aligned to the reference system. The pitch adapter is positioned on its transparent vacuum chuck exploiting two reference holes, and horizontally aligned to the detector-hybrid pair by means of micrometer screws. The pitch adapter chuck can then be removed, turned upside down, and placed under a glue deposition robot, keeping the mechanical reference to the detector-hybrid pair. After the glue-line deposition on the pitch adapter, the chuck is returned to its position. The joint is realized by raising the sensor and hybrid towers to the nominal height, filling with glue the \qty{50}{\um} gap between the pitch adapter and the detector-hybrid plane. The transparency of the chuck allows to see the spreading of the glue, both on the detector and the hybrid sides. An optical survey after glue curing verifies that the glue line has spread uniformly below all the bonding pads of the pitch adapter.

The module is transferred to a P-side multi-purpose chuck, which can be coupled with precision pins to an N-side multi-purpose chuck to turn the module upside-down. Vacuum is used in order to hold safely the assembly. After flipping, the module is transferred to the N-side jig, where a second pitch adapter gluing operation is repeated for the N-side. After the final survey of the glue spread, the module is wire-bonded, using dedicated P-side and N-side wire-bonding jigs. 
The module is then transferred to the multi-purpose chuck, which allows both vacuum chucking and mechanical clamping of the module and is used for storage, electrical tests, and shipping.  
A full electrical characterization is performed (see section~\ref{sec:C3_APVDAQ}): I-V curve; pedestal and  internal calibration run; VSep scan. To test the complete functionality of the module a scan under an infrared laser is performed. The modules are then  shipped to the various assembly sites for Layer 4, 5 and 6 ladder production.

The production consisted of 60 SFW + 56 SBW modules and was completed in 2017. With respect to the nominal 47 pairs needed to assemble all ladders, some extra spare modules were produced to cope with the possible failures during the construction of the ladders. All the built modules showed good quality:
\begin{itemize}
\item 
the achieved mechanical precision of sensor position was better than \qty{100}{\um};
\item very few defects (<0.1\%) were added during assembly with respect to the initial sensor defects (at the level of $\simeq 1\%$ or better);
\item the typical signal-to-noise ratio was measured to be $\sim$24 for the u/P side and $\sim$30 for the v/N side for APV25 calibration pulses.
\end{itemize}

About 10\% of the SFWs developed a high bias current after the connection to \apv\  chips, which  was traced to the presence of substrate pinholes, discussed in section~\ref{sec:C3_APVDAQ}. 
Disconnecting the faulty strip from the amplifier input (by removing the wire bond), the modules could be recovered losing only one strip.

When the module production was already well advanced, in some of the completed Layer 4 and Layer 6 ladders the P-side pitch adapter started to detach from the forward sensor. The root cause of this failure, called "peel-off issue", was traced to a limited overlap between the forward sensor and the pitch adapter (only 1.3~mm), leading to a weak glue joint. When the SFW is rotated to its nominal angle in the ladder assembly, this glue joint is subject to a large mechanical stress that in some cases caused the detachment of the P-side pitch adapter. Although the joint robustness was tested in Layer 5 ladder assembly and no peel-off happened, it turned out that the mechanical conditions in Layer 4 and Layer 6 were significantly more severe. A delicate emergency reinforcement procedure was applied to all SFWs, consisting in the insertion of a thin glue-covered Mylar\textsuperscript{\textregistered}~\citeCompany{mylar} strip between the P and N-side pitch adapters near the sensor edge, resulting in a much stronger joint.  

\begin{figure}[!ht]
 \centering
 \includegraphics[width=0.9\columnwidth,keepaspectratio]{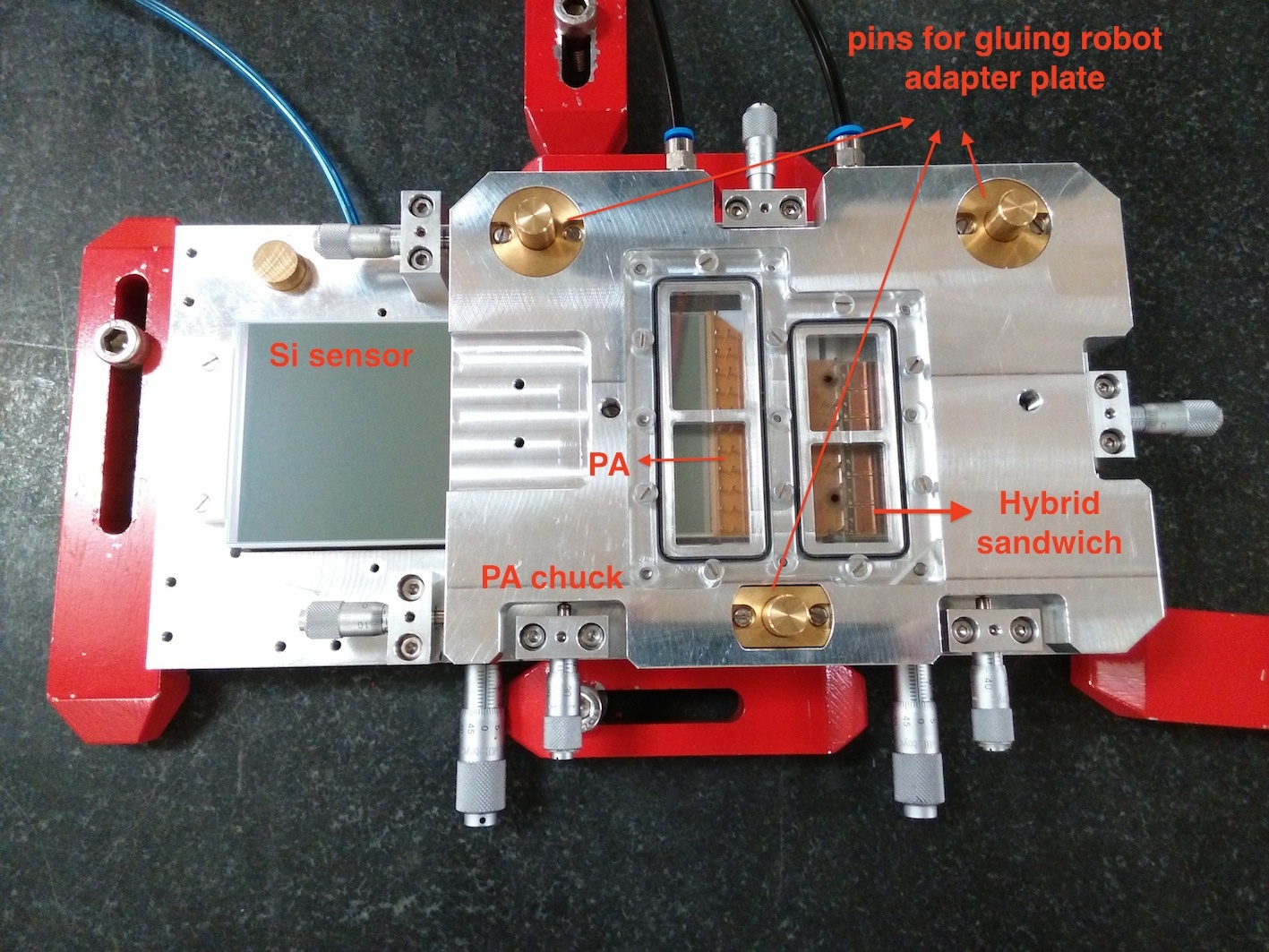}
 \caption{Backward module P-side gluing jig.}
 \label{fig:C3_BW_P_gluing}
\end{figure}

%==========================================
% plot of final mech. precision of all L5 ladders
% plot with total number of defects per ladder

%==========================================
\lvlthreesection{Layer 4, 5, and 6 ladder assembly}

\begin{figure}[!ht]
 \centering
\includegraphics[width=0.8\columnwidth,keepaspectratio]{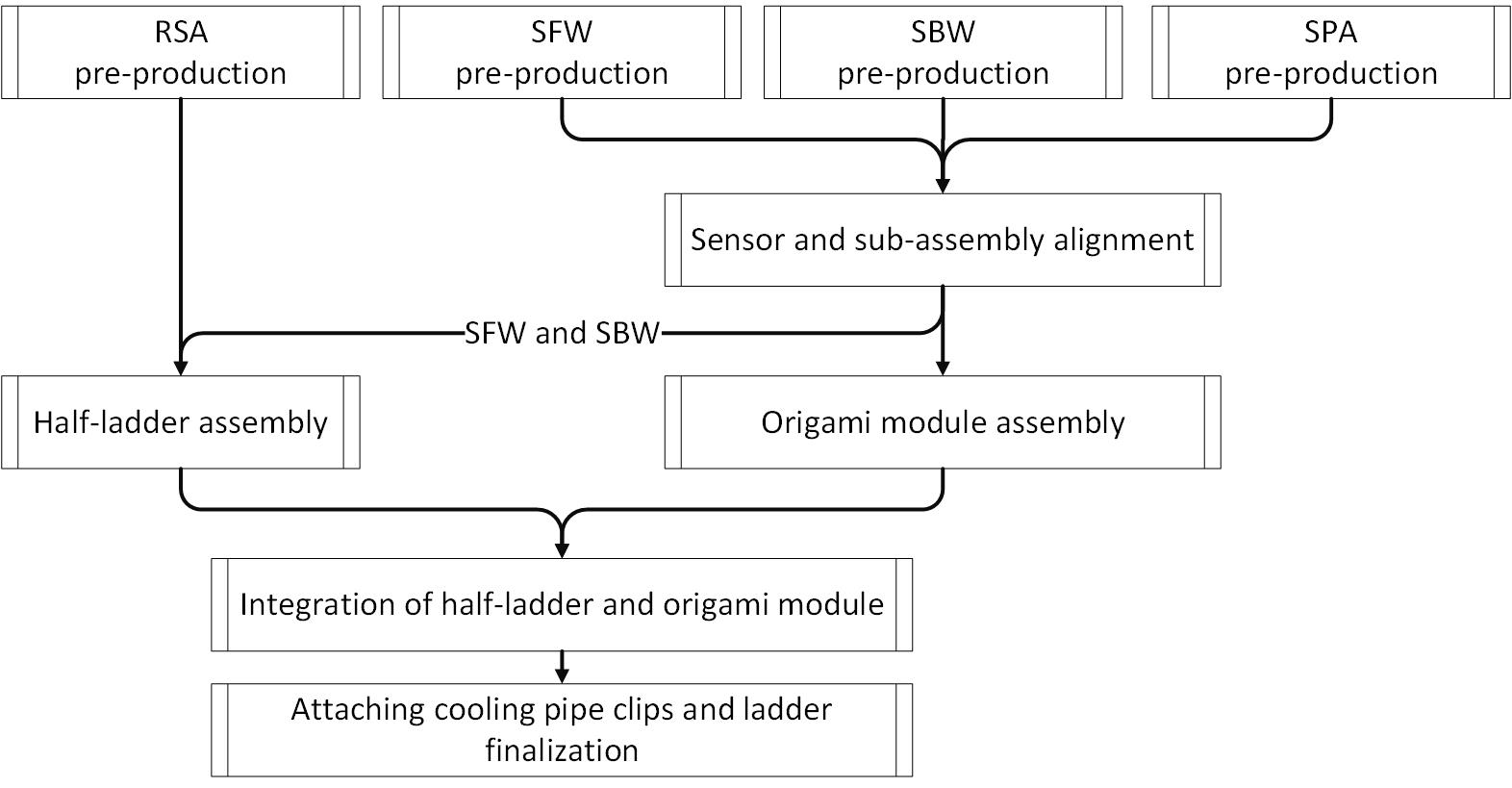}
 \caption{Process flow of main steps of the layer 4, 5, and 6 ladder assembly procedure. Adapted from~\cite{Irmler:2016qyg}. }
 \label{fig:C3_ladder_assembly}
\end{figure}

The procedure for the assembly of Layer 4, 5 and 6 ladders is complex and requires a large number of steps, which are summarized in figure~\ref{fig:C3_ladder_assembly}. 
The main points, which are common to the three layers, are described in this section in some detail to allow an understanding of the entire process. 

Layer 4, 5 and 6 ladder are composed of a forward module, one origami module with one to three sensors, and a backward module connected together by the rib sub-assembly, as shown schematically in figure~\ref{fig:C2_l6_exploded_view}. 
The main difference in the assembly procedures between the layers  is due to the different number of sensors in the origami module, as shown in table~\ref{tab:C3_OrigamiUsedInLadder}. 

\begin{table}[htbp]
\begin{center}
\caption{SFW/SBW and origami modules used in the ladders of different layers.}
\label{tab:C3_OrigamiUsedInLadder}.
\begin{tabular}{rccccc}
\toprule
Ladder & SFW module &Origami\_+Z & Origami\_CE & Origami\_-Z & SBW module\\
\midrule
L4 & \checkmark &   &   & \checkmark & \checkmark\\
L5 & \checkmark &   & \checkmark & \checkmark & \checkmark\\
L6 & \checkmark & \checkmark & \checkmark & \checkmark & \checkmark\\
\bottomrule
\end{tabular}
\end{center}
\end{table}

The Layer 4, 5 and 6 full ladder assembly procedure can be separated into the following six steps:
\begin{enumerate}
    \item Pre-production of  \acrfull{SPA}.
    \item Pre-production of \acrfull{RSA}.
    \item Sensor alignment of one to three SPAs, one SBW, and one SFW.
    \item Half-ladder assembly.
    \item Origami module assembly.
    \item Integration of half-ladder and origami module.
\end{enumerate}

In the following, these six steps are described in some detail, but without trying to account for slight differences of procedures at the various production sites.
\begin{enumerate}[leftmargin=*]
    \item {Pre-production of \acrfull{SPA}}\par
Two pitch adapters, PA1 and PA2, are glued onto the P-side of the large rectangular silicon sensors, and then electrically connected to the sensor strips by wire-bonding. This sub-assembly is called SPA and is produced in advance, to save time in the ladder assembly process. The SPA structure is shown in figure~\ref{fig:C3_YO_SPAandRSA}.
The SPA is assembled with the following steps:
\begin{enumerate*}
    \item The DSSD is placed on the DSSD-jig P-side up. The jig has a Delrin\textsuperscript{\textregistered}  %~\citeCompany{delrin} 
    plate with a vacuum circuit inside, allowing the sensor to be held by vacuum chucking.
    \item PA1 and PA2 are placed, with their backside up, on the PA-jig using two alignment pins, and held by vacuum chucking. Glue is deposited on the backside of PA1 and PA2 using a programmable glue dispensing robot.
    \item  The PA-jig is flipped and positioned on top of the DSSD-jig, using two guide shafts to find the correct alignment between the parts.
    \item After a few minutes, the vacuum of the PA-jig is released and the PA-jig is lifted, leaving PA1 and PA2 on the surface of the sensor. A visual check is made to spot any possible glue overflow on the pitch adapters bonding pads.
\end{enumerate*}
After waiting 24 hours for glue curing, the wire-bonding is performed. The SPA is then stored into a Gel-Box\textsuperscript{\texttrademark}~\citeCompany{gelbox} container, which is kept in a desiccator until it is used for the ladder assembly.

\begin{figure}[htbt]
 \centering
 \includegraphics[width=.9\columnwidth,keepaspectratio]{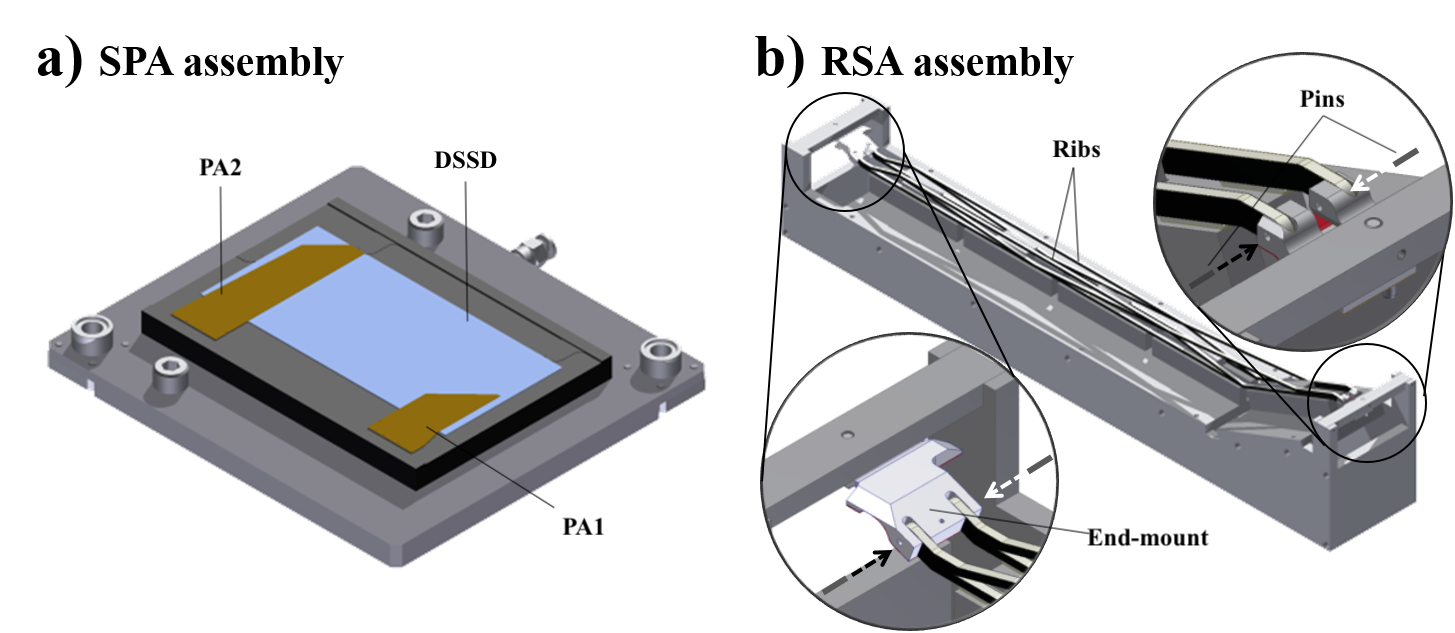}
 \caption{a) SPA assembly and components. b) RSA assembly and components.}
 \label{fig:C3_YO_SPAandRSA}
\end{figure}

\item {Pre-production of \acrfull{RSA}.}\par
The rib sub-assembly is the support structure of a ladder. Like the SPA, it can be produced in advance, in order to speed up the ladder assembly process.
It is composed of two truss-shaped ribs, made by a 1 mm thick AIREX\textsuperscript{\textregistered} sheet sandwiched by two 0.1 mm thick \acrshort{CFRP} plates, a forward and a backward end-mount. The RSAs are different for each layer, while the end-mounts are all the same and are the mechanical interface between the ladder and the end-rings. The RSA assembly is shown in figure~\ref{fig:C3_YO_SPAandRSA}.
A dedicated jig for each layer is required to produce the RSA, but the procedure is exactly the same: 
%The main steps are summarized below and are shown in Figure~\ref{fig:C3_YO_RSAassembly}.\\
%(a) Two 3 mm $\phi$ holes and several pairs of height blocks define the position of the mount blocks and the position of ribs with respect to the mount blocks.\\
%(b) The mount blocks are positioned using the 3 mm $\phi$ holes as a reference. \\
%(c) Ribs are placed in the slots of the mount block. The ribs stand-offs, where the silicon sensors will be glued, are placed onto the height blocks to find the correct height. 
\begin{enumerate*}[leftmargin=*]
    \item Both end-mounts are positioned on the jig, using a pin to find the correct alignment.
    \item Ribs are placed in the corresponding slots of the end-mount and aligned on the plane defined by the jig. 
    \item Glue is deposited in the slots of the end-mounts, and a stainless steel pin is inserted to reinforce the glue joint.
\end{enumerate*}

\item {Sensor alignment of SPA(s), SBW and SFW.}\par
The alignment of SPA(s), SBW and SFW is the first step of the ladder assembly. The precision of the alignment determines the mechanical quality of the full ladder. The alignment is performed under the CMM, using the XYZ$\Theta$-stage to position the sensors with a \qty{10}{\um} precision with respect to the nominal position of the sensors fiducial F-marks.  Figure~\ref{fig:C3_YO_DSSDalignment2} shows the SPA(s), SBW and SFW alignment and pickup procedure:
\begin{enumerate*}[leftmargin=*]
    \item The SBW/SFW and SPA(s) are placed on the assembly bench using jigs and procedures designed to avoid collisions. The assembly bench is moved under the CMM, where each sensor is precisely aligned using the XYZ$\Theta$-stage, which allows to move the sensor in two directions and to rotate it around the z-axis.
    \item After the alignment, the SBW and SFW are picked up with two jigs and are prepared for the half-ladder production (described in step 4), while the SPA(s) stay on the assembly bench for the origami module assembly (described in step 5).
\end{enumerate*}

\begin{figure}[htbt]
 \centering
 \includegraphics[width=.8\columnwidth,keepaspectratio]{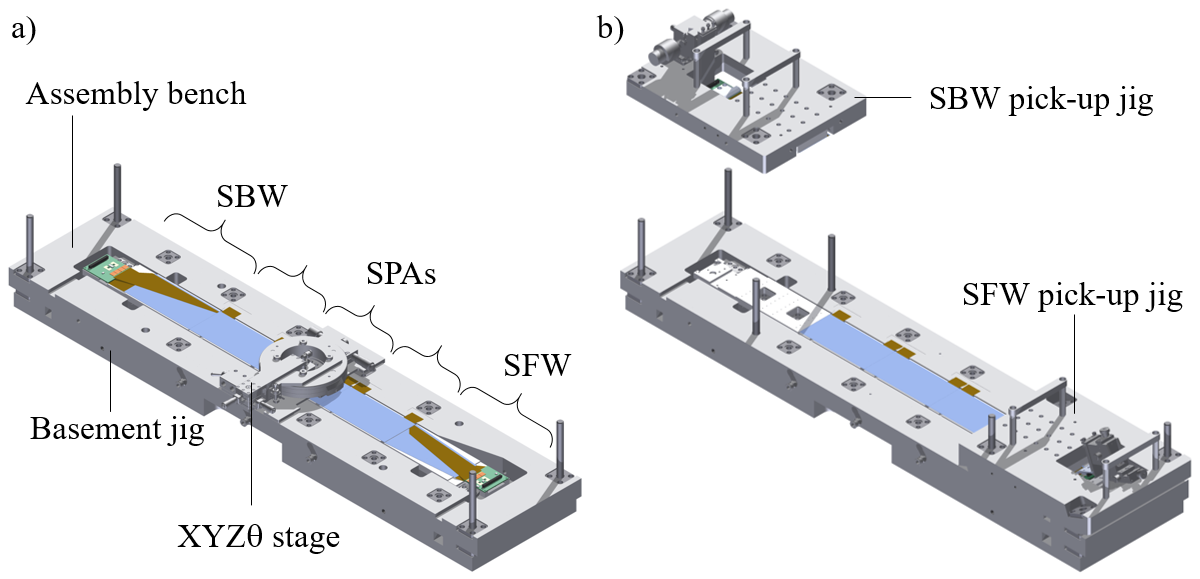}
 \caption{a) SPA(s), SBW and SFW placed on the assembly bench and aligned under the CMM using by the XYZ$\Theta$-stage. b) SBW and SFW  picked up for half-ladder assembly.}
 \label{fig:C3_YO_DSSDalignment2}
\end{figure}

\item {Half-ladder assembly.}\par
The SBW and the SFW  are glued on the RSA to make a half-ladder.
The steps of the half-ladder assembly are shown in Figure~\ref{fig:C3_YO_FWDBWDassembly2}:
\begin{enumerate*}[leftmargin=*]
    \item The assembly bench is lifted up and put aside. The rib-jig is placed on the basement-jig. The RSA is positioned on the rib-jig. Then height blocks and height shafts are placed.
    \item Glue is manually dispensed on the ribs, where the SBW and the SFW are glued. The SFW hybrid board is bent to its design angle with respect to the sensor surface and held in this position with a clamp. Then the pick-up jigs holding the SBW and the SFW are placed on top of the rib-jig, to glue the sensors to the ribs. The pick-up jigs are coupled to the basement-jig with precision pins, so that the initial alignment is kept through the assembly steps.
\end{enumerate*}
After glue curing, the SBW and SFW hybrid boards are fixed to the RSA end-mounts, then the vacuum is released and the SBW and SFW pick-up jigs are removed. The half-ladder is optically inspected and electrically tested, then it is stored until its integration with the origami module.

\begin{figure}[!ht]
 \centering
 \includegraphics[width=.8\columnwidth,keepaspectratio]{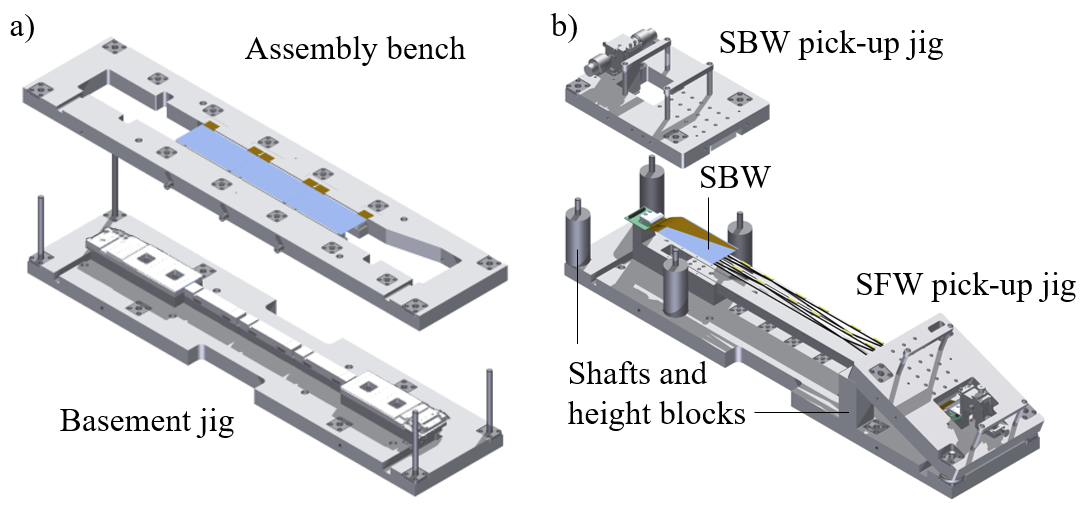}
 \caption{a) assembly bench removed from the basement jig. b) SBW/SFW glued on the RSA.}
 \label{fig:C3_YO_FWDBWDassembly2}
\end{figure}

\item  {Origami module assembly.}\par
The origami module assembly is shown in figure~\ref{fig:C3_YO_OrigamiAssembly2}.
\begin{enumerate*}[leftmargin=*]
    \item The assembly bench with SPAs is placed back on the basement-jig.
    \item Glue is dispensed on the sensors with a programmable robot, then an AIREX\textsuperscript{\textregistered} sheet is glued on the sensors, using a dedicated jig to find the correct alignment. The origami boards are glued on top of the AIREX\textsuperscript{\textregistered} sheet, again using dedicated jigs to align them to the sensors.
    \item Wire-bonding of the N-side is performed between each sensor and the corresponding origami board, as well as between the PA0 and the \apv\  chips.
    \item After an optical inspection of the N-side wire-bondings, both PA1 and PA2 of each sensor are wrapped and glued on top of the origami board.
    \item After glue curing, the final step is the wire-bonding between wrapped pitch adapters and P-side \apv\ chips.
    \item The origami module is optically inspected and electrically tested before proceeding to the integration with the half-ladder.\\
\end{enumerate*}

\begin{figure}[htbt]
 \centering
 \includegraphics[width=1.\columnwidth,keepaspectratio]{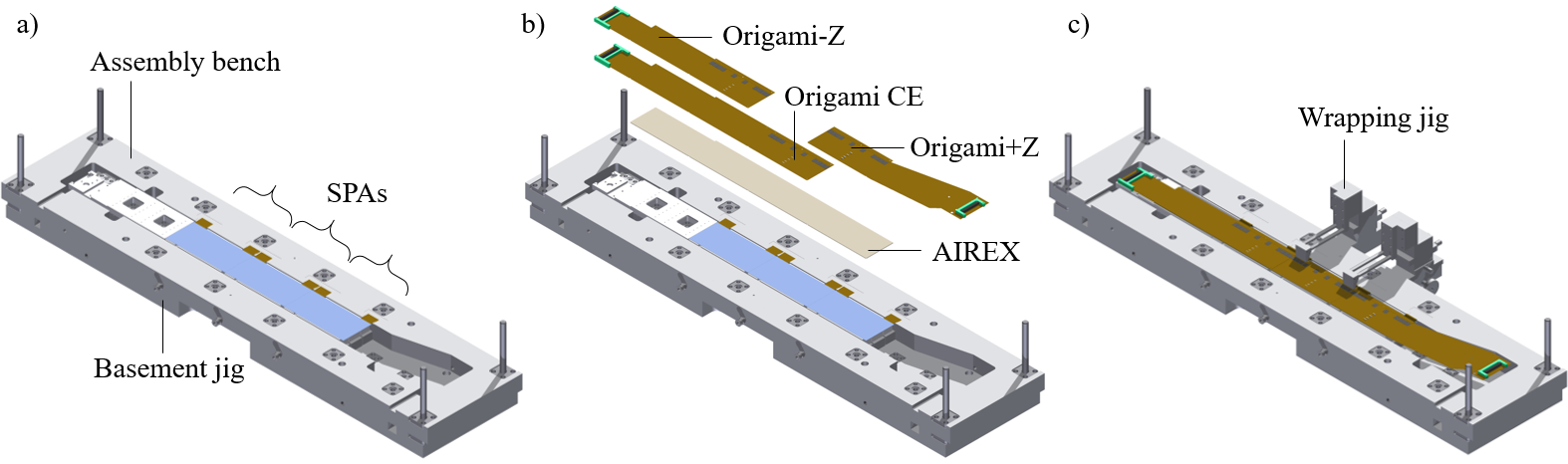}
 \caption{a) The assembly bench positioned on the basement jig. b) Origami boards gluing. c) PA1/PA2 wrapping from DSSD P-side to the top of the origami boards.}
 \label{fig:C3_YO_OrigamiAssembly2}
\end{figure}

\item {Integration of half-ladder and origami module.}\par
The final step, shown in Figure~\ref{fig:C3_YO_LadderAssembly2}, is the integration of the half-ladder and the origami module.
\begin{enumerate*}[leftmargin=*]

    \item The half-ladder on the rib-jig is placed on the basement jig and glue is dispensed on the ribs. The assembly bench is then placed on two electrically controlled coupled vertical stages and slowly lowered until the gap between the sensors of the origami module and the ribs is around \qty{100}{\um}. After glue curing, the tails of the origami boards are fixed on the SBW and SFW hybrid boards.
    \item The ladder is picked up from the assembly bench. A mechanical survey is done under the CMM to assess the mechanical quality of the ladder, then an optical inspection and a full electrical test are performed before storing the ladder in its container. The ladder is then shipped to KEK to be mounted on the SVD support structure.
\end{enumerate*}

\begin{figure}[htbt]
 \centering
 \includegraphics[width=.8\columnwidth,keepaspectratio]{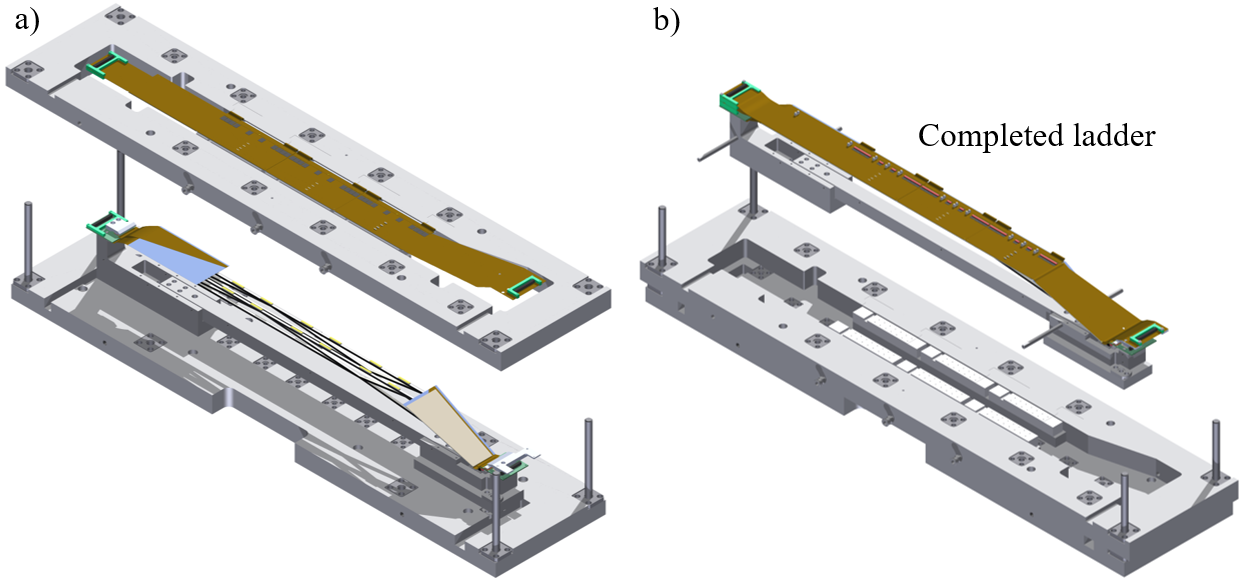}
 \caption{a) Integration of half-ladder and origami module. b) Completed ladder.}
 \label{fig:C3_YO_LadderAssembly2}
\end{figure}

\end{enumerate}
%===============================

\lvlthreesection{Summary of module and ladder production}

\begin{table}[hbt]
\begin{center}
\caption{Summary of the ladder production quantities and their usage in the installed SVD.}
\begin{tabular}{clcccc}
\toprule
Class & Usage & L3 & L4 & L5 & L6 \\
\midrule
\multirow{3}{*}{A} 
& Installed in SVD & 7 & 10  & 12 & 16 \\
& Spares & 5 & 2  & 3 & 3 \\
& Total & 12 & 12 & 15 & 19 \\
\midrule
{B} & Total & 2 & 2 & 2 & 2 \\
% N/A & Not completed & - & 2 & -  & 5 \\
\bottomrule
\end{tabular}
\label{tab:C3_sum_ladders}
\end{center}
\end{table}

The total numbers of produced ladders are summarized in table~\ref{tab:C3_sum_ladders}.  They include the class B ladder used in the \phasetwo\ commissioning cartridge described in section~\ref{sec:phase2SVDSetup}
Each completed ladder was mechanically surveyed. The position of the fiducial F-marks of the silicon sensors was measured and compared to the nominal position, in a local reference frame defined for the ladder assembly, where the x-y plane corresponds to the sensor surface, with the origin in center of the backward kokeshi pin, and the z-axis is perpendicular to the sensor surface. 
\begin{table}[hbt]
\begin{center}
\caption{Summary of the ladder mechanical precision. Average and maximum deviations, in \unit{\um}, from the nominal position of the sensor fiducial F-marks in the completed class A ladders.}
\begin{tabular}{lcccccc}
\toprule
Layer & \multicolumn{2}{c}{$\Delta$x} & \multicolumn{2}{c}{$\Delta$y} & \multicolumn{2}{c}{$\Delta$z} \\
 & Aver. & Max. & Aver. & Max. & Aver. & Max. \\
\midrule
L3 & 13 & 42 & 28 & 78 & 80 & 173 \\ %updated
L4 & 55 & 327 & 37 & 145 & 67 & 478 \\ %updated  
L5 & 69 & 254 & 65 & 317 & 79 & 342 \\  %updated
L6 & 40 & 233 & 45 & 199 & 62 & 234 \\ %updated
\bottomrule
\end{tabular}
\label{tab:C3_ladder_mech_prec}
\end{center}
\end{table}

% CI:  The result for L5 is the following:
% You can find the calculation and the results in the green boxes of the tabs Summary x, Summary y and Summary z of the attached spreadsheet.
% For the calculation, I only considered the class A ladders, which are L5.002 – L5.016.
% Please ignore columns W to AA, those does not consider the absolute values.

% I also did the same math for L4 class A ladders in L4 Measurement Overview.xlsx with the following results:
% % avg.       max.
% L3 delta x                          13          42
% L3 delta y                          28          78
% L3 delta z                          80          173
% avg.       max.
% L4 delta x                          56          327
% L4 delta y                          37          145
% L4 delta z                          67          478
% avg.       max.
% L5 delta x                          69          254
% L5 delta y                          65          317
% L5 delta z                          79          342
% 
% avg.       max.
% L6 delta x                          40          233
% L6 delta y                          45          199
% L6 delta z                          62          234

A summary of the ladders mechanical precision is reported in table~\ref{tab:C3_ladder_mech_prec}, where the average and maximum deviations for each axis are reported. Comparing the average deviations with the target tolerances presented in section~\ref{sec:C2_mech_and_elect_requirements}, it can be observed that the overall achieved precision is well within the requirements for all axes. The maximum deviations are higher than the required tolerances, but the occurrences of these deviations are very rare exceptions that do not affect the mechanical quality of the ladders. Furthermore, as long as the ladders are mechanically stable, the final position resolution is determined by the alignment with tracks, as discussed in section~\ref{sec:C7_alignment}.

%========================
\lvltwosection{Mechanical Support and Cooling Fabrication} {\label{sec:mech1}}
\label{sec:C3_mechanical}

% Maybe insert a SHORT INTRO, but not clear what to say. 

The overall mechanical structure of the SVD plays an essential role in guaranteeing the mechanical precision and stability of the ladder location and ultimately the detector performance. The cooling circuits are fabricated inside the mechanical structure in the case of the end-rings, while are applied on the already mounted ladders in the case of the origami pipes.  The support cones and the outer cover were fabricated and assembled by KEK. The origami pipes were fabricated by HEPHY, while the pipe assembly technique and jigs were developed by Pisa. 

%\editor{Francesco- 2 pages}
\lvlthreesection{Support cones, outer cover and end-rings}
%\editor{Toru Tsuboyama}
% NOTE: new commands should go in the file svd-sym.tex

\begin{figure}[htbp]
\begin{center}
\includegraphics[width = 0.8 \textwidth]{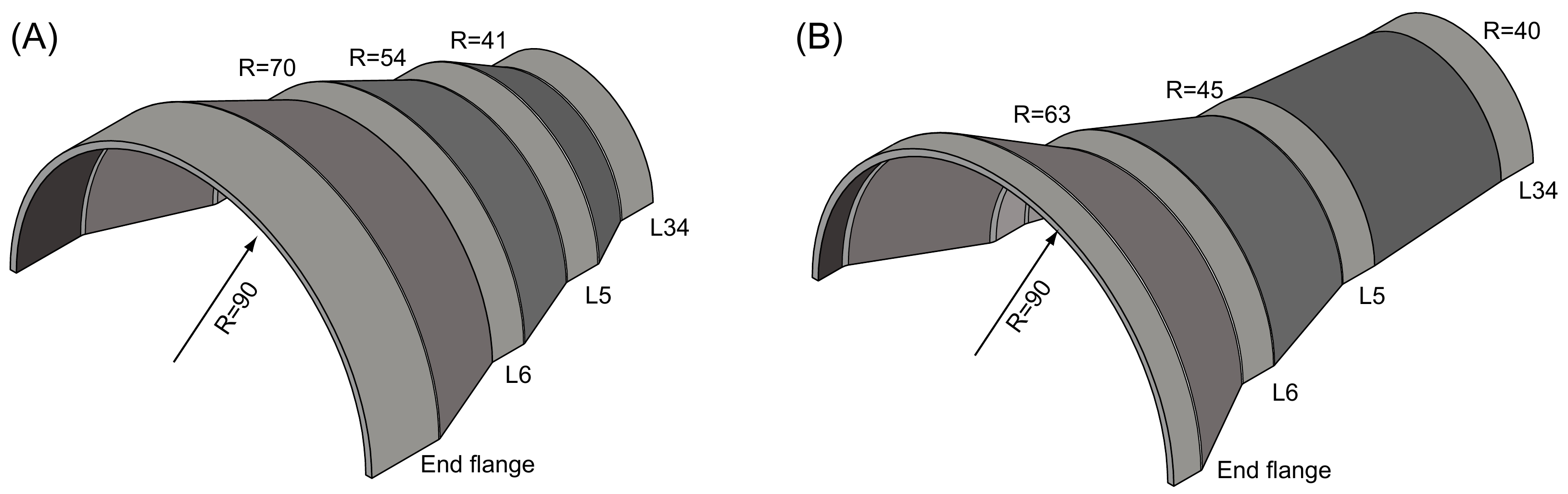}
\caption{The CFRP support cone for (A) backward (B) forward region.}
\label{TT:SUpportCones_F3}
\end{center}
\end{figure}

The support cone design is shown in Figure \ref{TT:SUpportCones_F3}.
The cylindrical regions are for the gluing of the end-flanges and the end-rings. The \acrshort{CFRP} cones are machined to allow 0.1-0.3~mm gap for the glue.

\begin{figure}[htbp]
\begin{center}
\includegraphics[width = 0.95 \columnwidth]{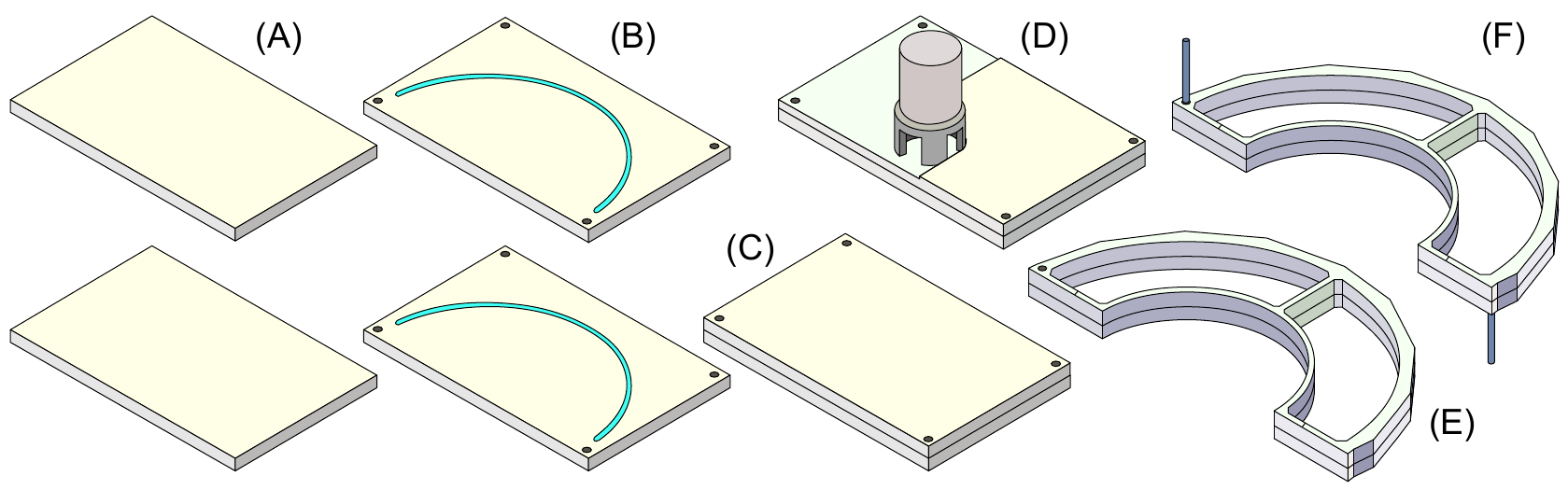}
\caption{The production of the end-ring with diffusion bonding.
(A) The surface of the stainless plates are flattened and polished.
(B) Grooves are milled on the surface of both plates.
(C) The two plates are stacked face to face and diffusion bonding is performed at high temperature and pressure. Metal crystals are developed at the border resulting in a strong connection between the surfaces.
(D) The plates are thinned to their design thickness.
(E) Machining to the final shape.
(F) \COtwo\ tubes are brazed in a vacuum oven. 
}
\label{TT:Welcon_F4}
\end{center}
\end{figure}

The end-mounts of the SVD ladders are fixed to the mounting surfaces of the end-rings.
The heat produced by the readout chips flows through the end-mount to 
the end-ring and is removed by the \COtwo\ coolant flowing in the channel embedded in the end-ring. The channels are made by using the diffusion bonding technique, illustrated in  
figure~\ref{TT:Welcon_F4}. 

\begin{figure}[htbp]
\begin{center}
\includegraphics[width = 0.85 \columnwidth ]{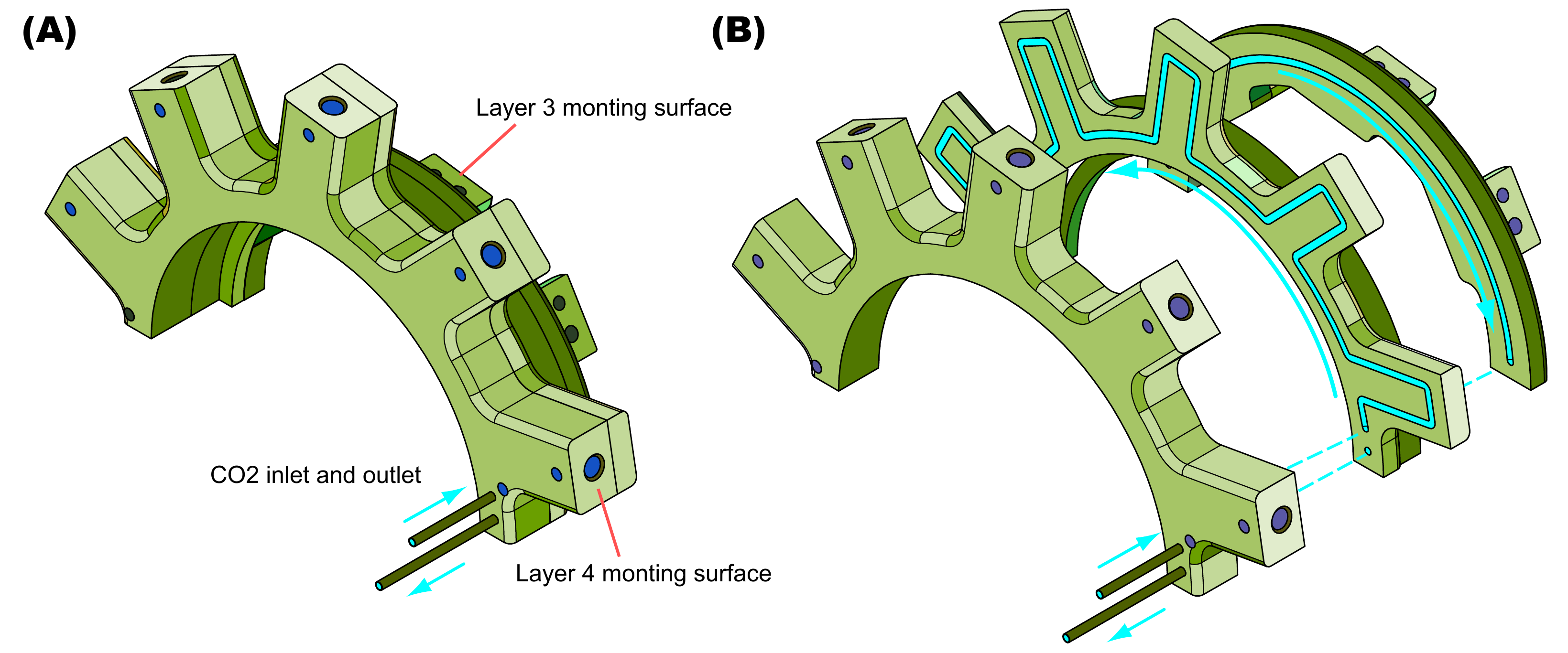}
\caption{The end-rings for the Layer 3 and Layer 4 in the backward direction. (A) The end-ring in its final shape, after bonding;
(B) exploded view of the individual layers before bonding, showing the \COtwo\ channels.}
\label{TT:L34Endring_F5}
\end{center}
\end{figure}

Because of the slanted sensor in Layer 4, the supporting points in Layer 3 and Layer 4 are so closely spaced that the layer 3 and layer 4 end-rings are combined into one body, as shown in figure~\ref{TT:L34Endring_F5}. 
% A ``three-stage'' diffusion bonding was performed to realize two embedded \COtwo\ channels. 

\begin{figure}[htbp]
\begin{center}
\includegraphics[width = 0.85 \columnwidth ]{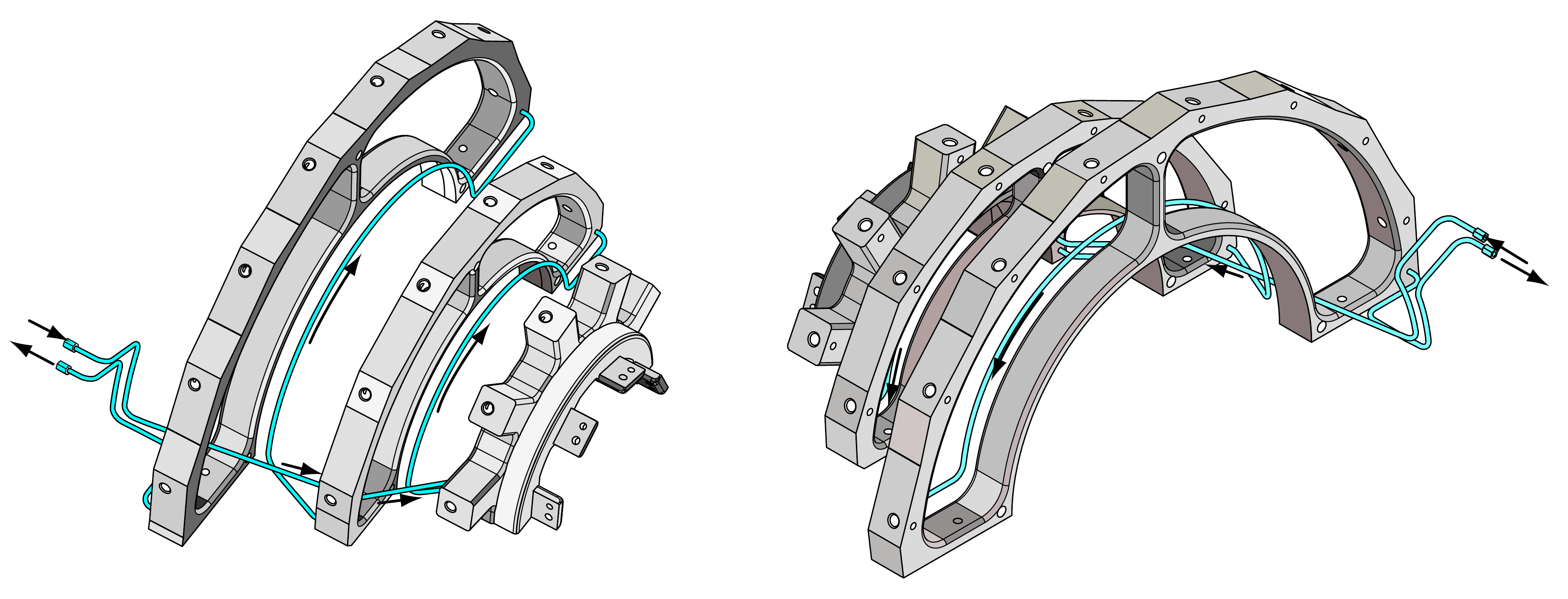}
\caption{The brazing of end-rings and \COtwo\ cooling tubes shown  in the front and rear views.}
\label{TT:Weldingpipes_F6}
\end{center}
\end{figure}

Figure \ref{TT:Weldingpipes_F6} shows the brazing scheme of the \COtwo\ tubes and the end-rings in the backward region. 
Since the stainless steel tube has an outer diameter of 1.7mm and a wall thickness of only 0.1mm, a special brazing jig is prepared to tightly fix the end-rings at their final distance. 
The full VXD assembly procedure is designed to avoid stress on the cooling tubes,  maintaining the relative position of the end-rings until they are glued to the support cones.

\begin{figure} [htbp]
\begin{center}
\includegraphics[width=0.6\columnwidth ]{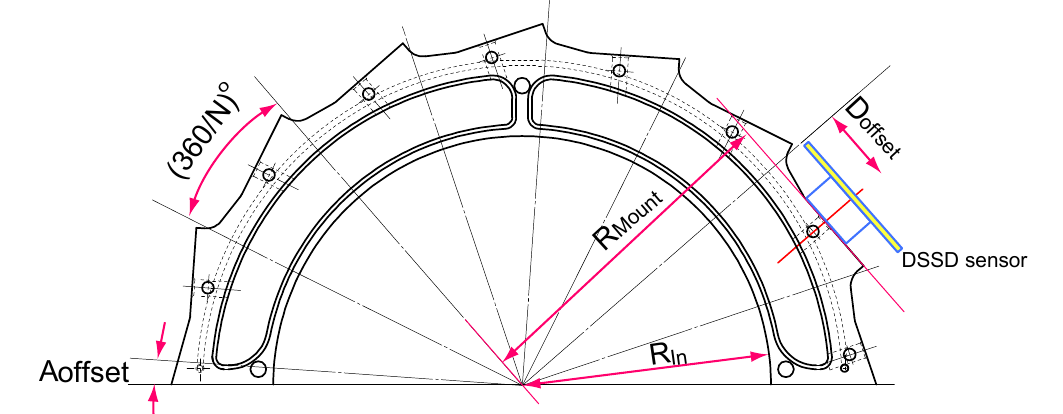}
\caption{Parameters describing the shape of the end-ring. $N$ denotes the number of ladders in a layer.}
\label{TT:EndringParam_F7}
\end{center}
\end{figure}

The design parameters of the end-rings are shown in figure~\ref{TT:EndringParam_F7} and Table \ref{TT:Endring_T1}.
To allow the overlaps of sensors in each layer, the ladders are arranged in a wind-mill like shape.
Parameter $D_\mathrm{offset}$ is the offset of the support point of the ladder from the symmetric position. 
An angular offset $A_\mathrm{offset}$ is introduced to optimize the clearance between sensors in neighboring layers.
The end-rings and support cones are designed for a   \qty{0.2}{\mm}  gluing gap, ensured by the end-ring gluing jig that positions the parts at the proper distance during the gluing procedure.

\begin{table}[htbp]
\begin{center}
\caption{Production parameters of the end-rings. Note that L3 is mounted on the same ring as L4 and has therefore the same R$_\mathrm{in}$.}
\label{TT:Endring_T1}.
\begin{tabular}{rccccc}
\toprule
Layer &  3  & 4 & 5 & 6 & Unit\\
\midrule
N$_\mathrm{ladders}$ & 7 & 10 & 12 & 16 \\
R$_\mathrm{in}$ (B)  &   & 41.2 & 54.2 & 70.2 &  \unit{\mm}\\
R$_\mathrm{in}$ (F)  &   & 40.2 & 45.2 & 63.2 &  \unit{\mm}\\
R$_\mathrm{mount}$ (B)  & 39.0 & 67.1 & 89.3 & 120.3 &  \unit{\mm}\\
R$_\mathrm{mount}$ (F)   & 38.0 & 48.8 & 66.3 & 85.3 &  \unit{\mm}\\
D$_\mathrm{offset}$&  4.16  & 10.63 & 10.97 & 19.1 &  \unit{\mm}\\
A$_\mathrm{offset}$&  5 & 10 & 8 &  4 & \unit{\degrees} \\
\bottomrule
\end{tabular}
\end{center}
\end{table}

\lvlthreesection{Origami pipes}
% Thanks to Markus to provide such info. 
\begin{figure}[!ht]
 \centering
 \includegraphics[width=.7\columnwidth,keepaspectratio]{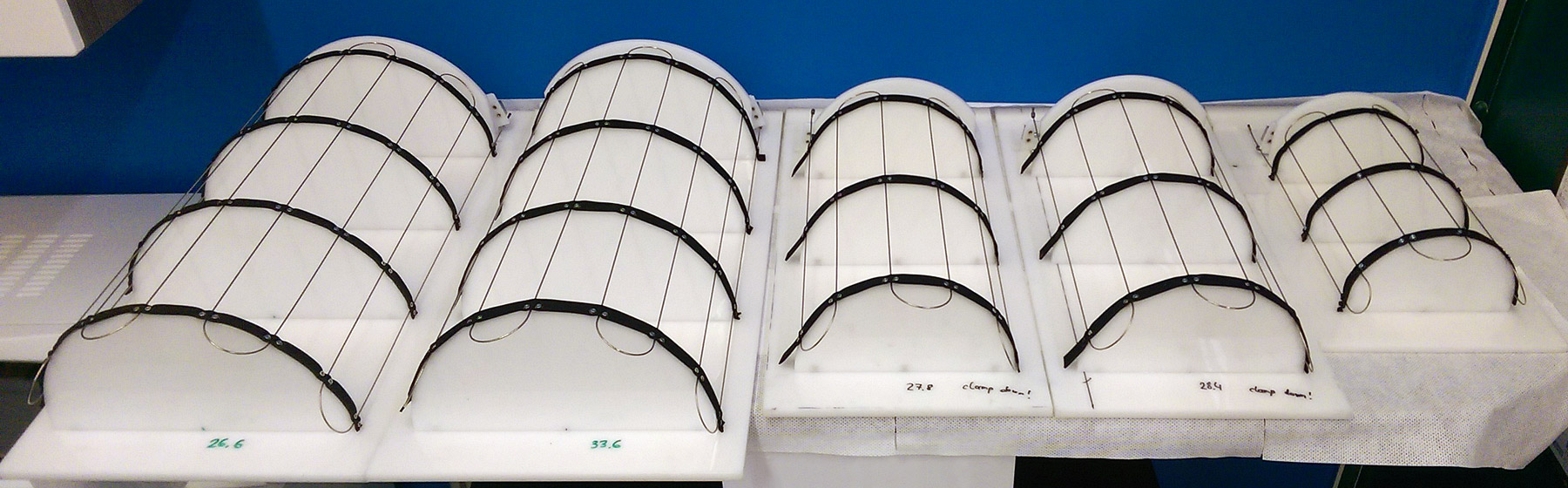}
 \caption{Finished Layer 4, 5 and 6 (from right to left) pipes on their storing and shipping masks.}
 \label{fig:C3_origami_pipe}
\end{figure}

The origami cooling pipes are made of  stainless steel SUS304 (1.4301)~\citeCompany{SUS304} with a \qty{1.6}{\mm} outer diameter and a \qty{0.1}{\mm}  wall thickness. Starting from \qty{6}{\m}  long straight pipes, they are bent in a 3D fashion by using a dedicated jig for each left and right bending. Three different sizes of cooling pipes were prepared for Layer 4, 5 and 6. After bending, metal-seal \qty{300}{bar} stainless steel connectors are brazed at each end. 
Figure~\ref{fig:C3_origami_pipe} shows the bent pipes on their storage containers, also used for shipment. 
Before the shipment to KEK, each pipe was tested for tightness by immersing in a water bath and applying a pressure of 200 bars for 15 minutes. No leaks were ever detected.
A good thermal contact between the pipe and the \apv\ chips is realized by interposing a \qty{1.0}{\mm} or \qty{1.5}{\mm}-thick (depending on the location) layer of Keratherm~\citeCompany{keratherm},
a soft foam with good thermal conductivity.

%%%% Include part on cooling plant. Maybe move to chapter 4
%%\subfile{03_03-01_Cooling_Plant}

%========================
\lvltwosection{Off-Detector Electronics Fabrication and Test}
\label{sec:C3_offdet_elec1}
%\editor{Markus}
SVD-specific off-detector electronics consist of the FADC data acquisition system and the power supplies. They were fabricated through a combination of external companies and in-house personnel, under the responsibility of HEPHY-Vienna for the FADC and INFN-Pisa for the power supplies.  

\lvlthreesection{FADC system fabrication and test}
\editor{Richard}

The \acrshort{FADC} boards were developed by HEPHY-Vienna, and assembled commercially.  The FADC boards were optically tested at this company, but not electrically. This is the reason why a test system had to be implemented to check all the components, interconnections, and external connections on these boards~\cite{Thalmeier:2020nrl}.
%======

The testing procedure performed is partially automated. It can be performed by following a well-defined step-by-step instruction manual, without detailed knowledge of the hardware or firmware.
Some hardware-testing equipment is connected to specific ports of the device under test, while some test firmware is programmed into the \acrshort{FPGA}, controlled by a PC via the VME crate. The operator can see if the test is passed and study the plots taken during the test to identify faulty or marginal components. All measured values including the graphs are recorded for later reference.
The first tests are thermal inspection, using an infrared camera, and power consumption, measured by the internal current monitors of the VME crate.

%\begin{figure}[!hbp]
\begin{figure}[!hbp]
	\centering
	\includegraphics[width=13cm]{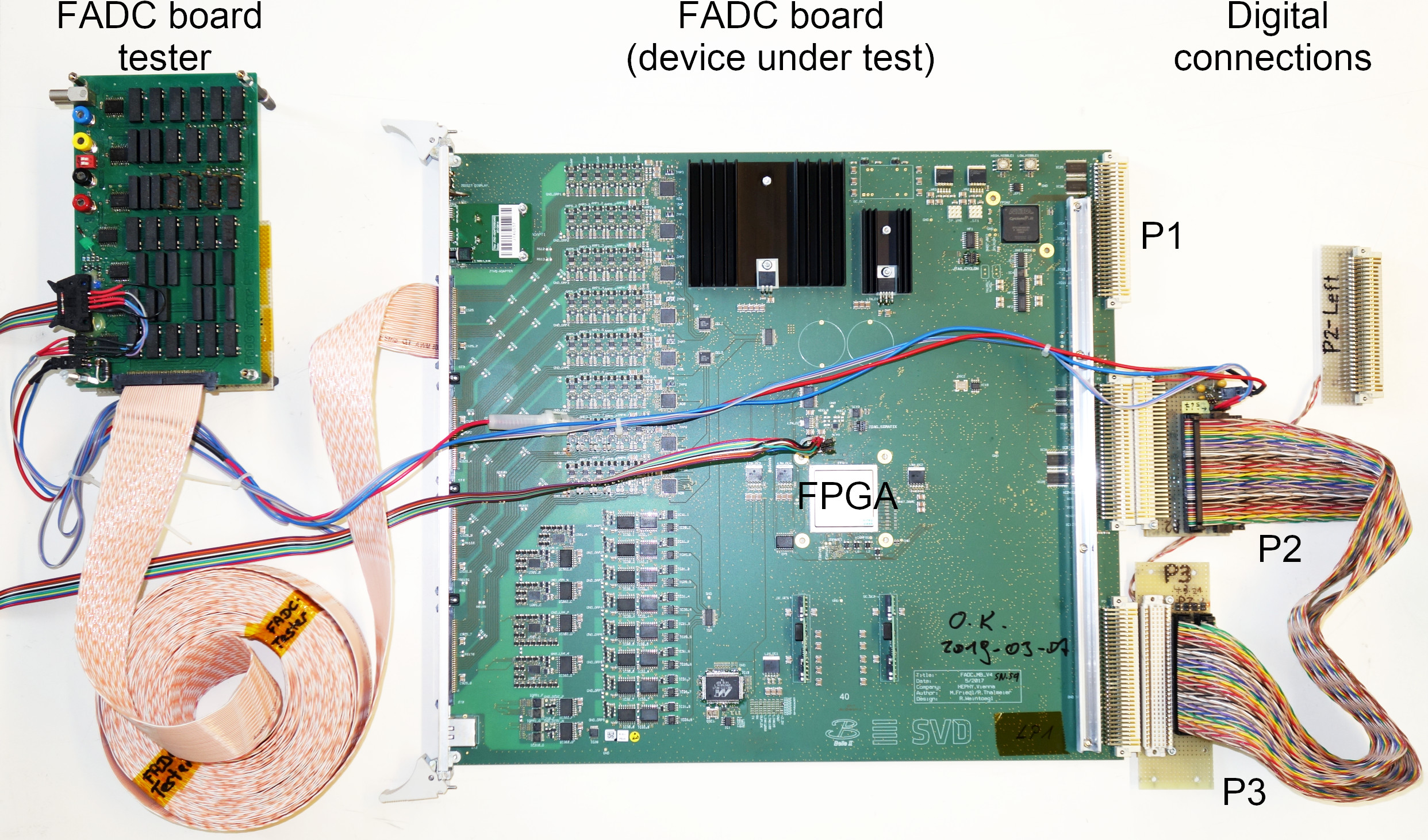}
	\caption{FADC board testing setup. In reality, this ensemble is in a VME crate with a controller, a customized backplane for P2 and other peripherals.}
	\label{fig:C3-RT-FADC-tester-blockdiagram}
\end{figure}

A special hardware test set, shown in figure~\ref{fig:C3-RT-FADC-tester-blockdiagram}, was developed for the FADC boards together with firmware for their onboard FPGAs and software for the PC controlling the VME bus. 

%\paragraph{FADC boards, digital connections}
\label{FADCdigitaltests}
%====

To test each external digital connection for conductivity and isolation against the others, the firmware sends 1024 signal pulses to one single output at different frequencies up to 40 MHz depending on the type of the interface hardware. Independent counters enumerate the signal changes on all inputs individually at the same time. The results are then read out and interpreted by the software. If the input connected to the active output counted any other number than 1024, or if any other input counted anything else but zero, the test has failed, and the faulty line is reported. The same is then repeated for all other lines on these two connectors successively, and also for some digital signals of the front connectors like $\rm I^2 C$ lines or the power-good and enable signals of the DC/DC converters on the junction boards. This digital test only needs a few seconds per board.

%\paragraph{FADC boards, high-speed analog data}
\label{FADCanalogtests}

\begin{figure}
	\centering
	\includegraphics[width=8cm]{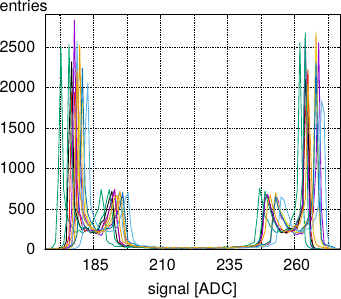}
	\caption{Statistical distributions of the ADC output data with uncorrelated square-wave input from the FADC tester board. Eight individual channels are shown in different colors. See chapter~\ref{FADCanalogtests} for explanation.}
	\label{fig:C3-RT-FADC-tester-digitallevels-histograms}
\end{figure}

The analog inputs of the FADC boards for the detector signals are tested in a different way. 
The testing hardware is connected to one of the four front connectors of the FADC board by a 68-conductor twisted-pair cable (figure~\ref{fig:C3-RT-FADC-tester-blockdiagram}, lower left corner). 
The main FPGA on the FADC board uses a dedicated testing firmware to generate several square-wave differential signals e.g. on the CLK and TRG lines, which are sent through the cable to the FADC tester board. 
There the signals are converted to analog levels using two voltage dividers, of which one or both can be selected through reed relays. 
%These switches are operated by $\rm I^2 C$ chips (PCF8574D), which are controlled by the firmware in the FPGA on the FADC board and accessed by software in a PC communicating with the FPGA over a VME bus. 
The square-wave output signals of the voltage dividers then can be connected to any of the analog inputs of the current connector on the FADC board via a relay matrix controlled by the FPGA. 
These signals are distorted and slurred a little by the cable which has the same length as the signal cable in the real system, in order to get characteristic signal shapes. 
It is important to note that the frequency of those square-wave signals and the sampling clock of the ADCs on the FADC board are completely uncorrelated in this test. In that way, histograms of the ADC levels are recorded to obtain statistical distributions (figure~\ref{fig:C3-RT-FADC-tester-digitallevels-histograms}) by performing \mbox{(pseudo-)} random equivalent-time sampling, as used in most spectrum analyzer devices. If the input were an ideal square wave,  only  two sharp peaks in the histogram for low and high states would be visible. However, as the bandwidth is limited and there is also some ringing, entries exist between (and beyond) those ideal states.
These histograms are then automatically examined by the software, checking if they fit into envelope templates generated from known good boards (pass) or not (fail).
Since the FADC board tester can verify one of the four front connectors of an FADC board at a time, it is connected to each of the four ports successively by the operator as instructed by the software.

%\paragraph{Remaining interface connections and components of the FADC boards}

The remaining lines on the front connectors are analog signal inputs for slow control measurements like voltage and temperature monitors. They are  automatically verified by feeding them with voltages generated by different voltage dividers also switched by reed relays on the FADC tester board. They are converted to 24-bit digital data by sigma-delta ADCs on the FADC boards which the software reads and compares to the nominal levels. The JTAG connectors are tested using a USB blaster to flash and verify the EEPROMs of the two FPGAs (which are additionally verified by rebooting in the procedure). The onboard JTAG flasher is checked by flashing the Stratix FPGA via the Cyclone FPGA, and the Gigabit Ethernet by sending and receiving packets to and from the PC using an Ethernet cable. On-board components like the delay chips are tested using sweep signals while recording histograms that are automatically analyzed afterwards.

%\paragraph{DAQ and integration tests}
\label{chapter-daq-tests}

In the next test phase one FADC after the other gets connected to a junction board and to a spare front-end detector module. The electrical isolation of the sensor bias voltage (``HV'') on the FADC boards, the cables, and the junction boards are tested using a $\pm$200V power supply with current measurement, and all channels of the ladder are read out by the DAQ PC. This procedure (but without the HV power supply) is repeated later for each channel every time before the final detector gets connected to the system.

%\paragraph{Results, status and conclusion}

During the initial FADC tests at HEPHY Vienna in the spring of 2018, some component assembly errors were detected, like a couple of so-called tombstone SMD resistors and capacitors (where only one side of the component is soldered properly and the other one is up in the air), a few resistors soldered in wrong positions, and several SMD connectors where some pins were without solder. One clock distributor IC was rotated by 180 degrees, and one FPGA had badly soldered BGA pins. These failures were successfully repaired, and all 60 boards behaved well afterwards.
On the junction boards, some minor errors were found, for example a loose SMD resistor jammed in between two connector contacts, and they were corrected on-the-fly.

No damage was found on the retest of the individual components after the shipment to KEK, and also the DAQ tests of the individual boards performed as expected.
The whole readout system was successfully tested during the commissioning of the two separate SVD halves and later on after the coupling with PXD, when the full VXD was assembled.

\lvlthreesection{Power supplies fabrication and test}
% \editor{Francesco}
The power supply system was based on commercial components and fabricated by CAEN SpA~\citeCompany{ref:caen}. High density fully floating switching power supply sources are used, allowing the installation of the full power supply system in just two standard rack-mount 19" crates (model SY4527).
The selected modules are specified in table~\ref{tab:C3_caenps}. 

\begin{table}[hbt]
\begin{center}
\caption{Installed power supply modules types. Technical documentation on the modules is available on the CAEN website~\protect \citeCompany{ref:caen}. }
\begin{tabular}{llllll}
\toprule
Type & Model & Specifications & Channels/ & Installed  & Total  \\
     &       &                & module    &  modules   & channels\\
\midrule
LV     & A2519A  & 5--\qty{15}{\V} / \qty{5}{\A} (< \qty{50}{\W}) & 8  & 12  & 96\\
HV     & A1519B  & \qty{250}{\V} / \qty{1}{\mA}          & 12 & 4  & 48\\
\vsep  & A1510   & \qty{100}{\V} / \qty{10}{\mA}         & 12 & 4  & 48\\
\bottomrule
\end{tabular}
\label{tab:C3_caenps}
\end{center}
\end{table}

Pre-production boards showed a large noise when channels were connected in series as foreseen in the schematics. The problem was promptly corrected by the company with additional filtering. All the boards were tested with a resistive load mimicking the load in the experiment, measuring the noise spectrum. In addition, all channels were verified on a test SVD module, performing a noise run with the full FADC system. The noise is fully comparable with what was observed with linear power supplies.
%, i.e. around 2.5 ADC counts for good channels. 

%========================
\lvltwosection{Ladder Mount and Mechanical Survey}
\label{sec:C3_ladder_mount}
\editor{Francesco, Antonio}

The assembly of the two SVD half-shells is performed in a dedicated clean room at KEK. In this section, a description of the workflow is given, describing the clean room and the setup of the ladder mount table, the procedure to mount a single ladder and to attach the cooling pipe on layers 4, 5 and 6 half-shells, the mechanical survey performed after a layer half-shell is completed. After the whole half-shell is assembled, it is stored on a dedicated table, where it is kept for the commissioning period and until the SVD is combined with PXD to assemble the full VXD.

\lvlthreesection{Clean room setup and ladder mount table}
\editor{Katsuro, Koji, Antonio}
The clean room is divided into two areas. The larger area, \qtyproduct[product-units=power]{4.5 x 4.5}{\m} , is used for the ladder mount and the tests, with the ladder mount table placed in the center of the area. The smaller area,  \qtyproduct[product-units=power]{3.0 x 4.0}{\m}, is used for the inspection of the ladders just before mounting, and for the mount tool preparation and storage.
The cleanliness of the room is maintained at the required ISO8 level (equivalent to class 100000 of US Fed. Std. 209E) by three \acrshort{HEPA} fan filters.
The floor of the room is covered by an anti-static mat connected to the ground.
The temperature, humidity, cleanliness, $\rm O_2$ and \COtwo\ densities in the clean room are monitored and recorded continuously. $\rm O_2$ and \COtwo\  sensors are installed in the clean room, to monitor their levels especially during the tests of the cooling system, operated with liquid \COtwo\ .

\begin{figure}[!ht]
 \centering
 \includegraphics[width=.8\columnwidth,keepaspectratio]{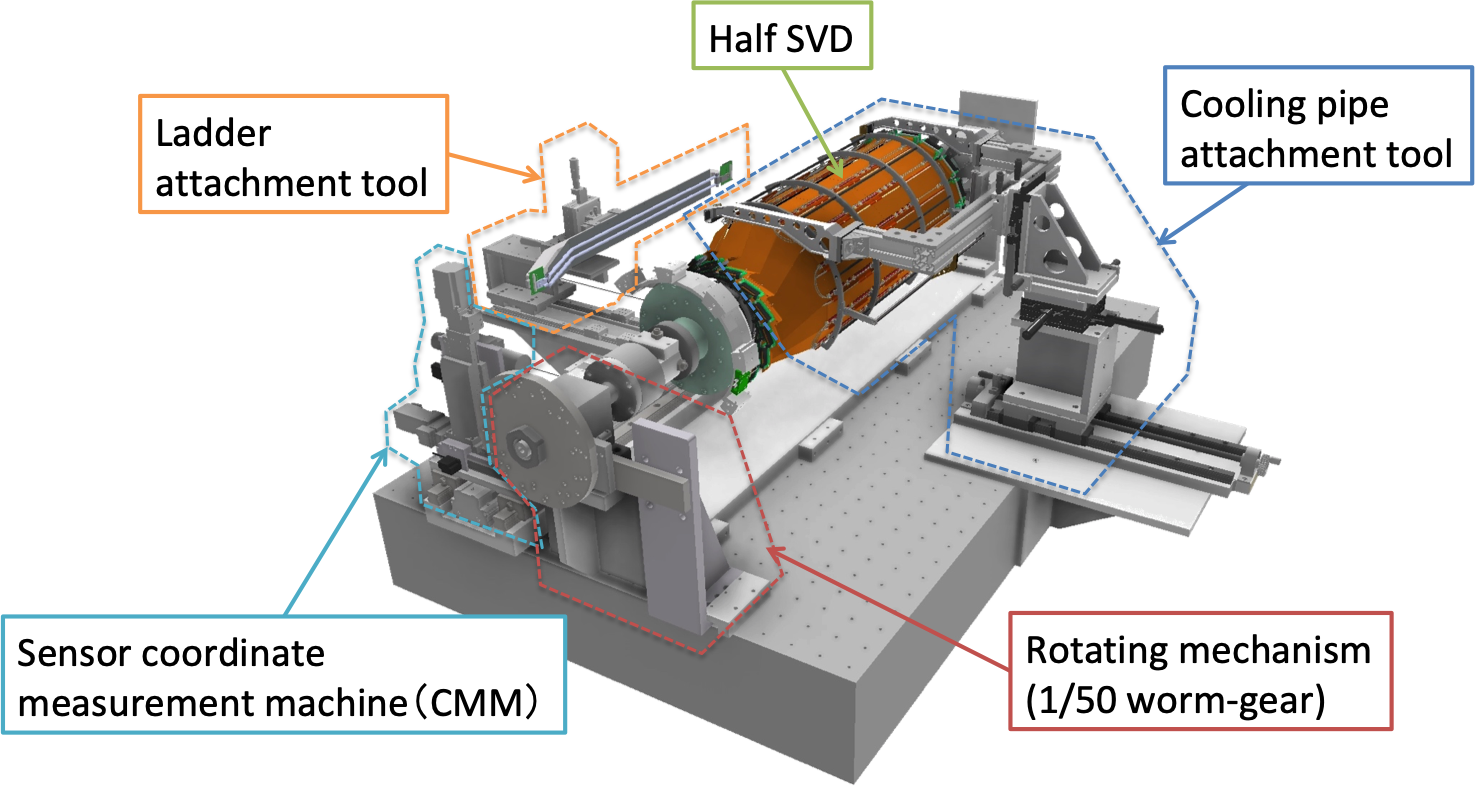}
 \caption{A CAD drawing of the Ladder mount table.}
 \label{fig:C3_08-02_LadderMountTable_CAD}
\end{figure}

The ladder mount table, shown in figure~\ref{fig:C3_08-02_LadderMountTable_CAD}, is designed to position the support structure that is used to assemble the SVD. The support structure, on which the end-rings are mounted, is equipped with a worm-gear that can rotate around the z-axis to allow mounting all SVD ladders from the same side of the table, with the procedure described in section~\ref{sec:C3_single_ladder_mount}. The worm-gear has a reduction factor of 50 to allow the setting of the angle of the end-rings with a precision of 0.001 degrees. The table accommodates a Coordinate Measurement Machine (\acrshort{CMM}) system used to perform mechanical measurements before and after the ladder mount, as described in section~\ref{sec:C3_mechanical_survey}. The CMM is located on the same side where the ladder mount attachment tool is positioned. On the opposite side of the table, the structure for the cooling pipe attachment tool is located. During the ladder mount period, this structure holds a cover that protects the already installed ladders, while after layer 4, 5 and 6 half-shells are completed, the cover is temporarily removed to perform the cooling pipe attachment, described in section~\ref{sec:C3_cooling_pipes_mount}.

After the cooling pipe of each layer is mounted, a cooling test is performed to check that there are no leaks in the cooling circuit and to verify the thermal contact between the cooling pipe and the \apv\ chips using an infrared camera. Given that the half-shell cannot be moved from the table at this point, the test is performed on the ladder mount table, using a custom design open circuit \COtwo\ cooling system connected to the half-shell cooling pipe through temporary  connections. After checking the connections for tightness, a small box is mounted around the SVD half-shell to create a dry environment with a low dew-point to avoid water condensation when the  cooling test is performed at \qty{-20}{\degreeCelsius}. The volume is flushed with nitrogen for the whole duration of the cooling test. 

%%%%%%%%%% SINGLE LADDER MOUNT PROCEDURES %%%%%%%%%%
\lvlthreesection{Single ladder mount procedures}
\label{sec:C3_single_ladder_mount}
\editor{Katsuro,Antonio}

Some preparatory steps are needed on the ladder mount table as well as on the ladder to be mounted before the installation of each ladder. 

\paragraph{Ladder preparation}
Four plexiglass covers of different lengths and radii were prepared, one for each layer, to be installed just before the assembly of each half layer, to protect the ladders already installed. The protection covers are sustained by the same support structure used for the cooling pipe attachment, described in Section~\ref{sec:C3_cooling_pipes_mount}. Ladders are installed from one side of the ladder mount table, moving horizontally towards the end-rings. The protection covers are designed to leave enough space for the ladder mounting, without being moved. On the table, the end-rings are rotated so that the planes of the two end-ring supports on which the ladder is mounted are perpendicular to the surface of the table. A measurement of the \Gls{kokeshi} hole and of four points for each end-ring support is performed with the 3D CMM system, to verify the position and orientation of the plane on which the ladder is installed.
The tolerance on the planes orientation with respect to the nominal angle is set at $\pm$0.15 degrees for both $\theta$ and $\phi$ angles. It turned out that this tight tolerance could not be reached for L3 rings after gluing the Layer 3 supports to the Layer 4 ring. Nonetheless, such a reduced tolerance on Layer 3 planes is acceptable because the larger gaps between Layer 3 modules requires less demanding accuracy to avoid mechanical interference.

Meanwhile, the ladder is optically inspected in the dedicated smaller area of the clean room. The goal of the inspection is to verify that the ladder has no mechanical issues: after removing the top and side walls of the ladder container, all wire-bonds, pitch adapters gluing points and \COtwo\  clips are checked. If the optical inspection is successful, Layer 4, 5 and 6 ladders, still fixed on the container supports, are brought to the Keratherm\citeCompany{keratherm} attachment station, while Layer 3 ladders are ready to be mounted. \par

\begin{figure}[!ht]
 \centering
 \includegraphics[width=\textwidth]{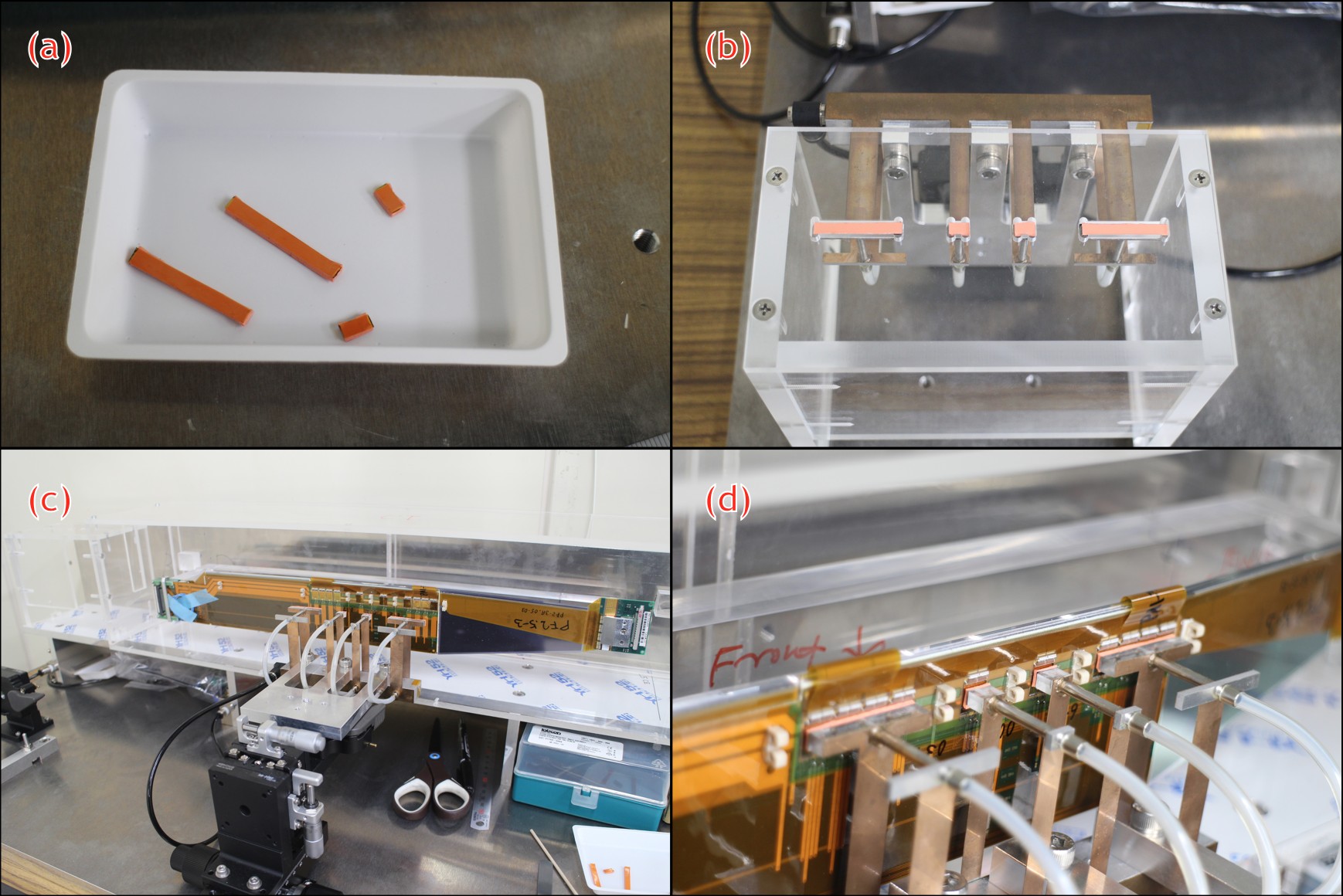}
 \caption{Main steps of Keratherm attachment on an origami board of a Layer 4 ladder. (a) the Keratherm strip is cut into four pieces; (b) the pieces are positioned and aligned on the Keratherm-jig; (c) the Keratherm-jig is positioned on the movable stage that is used to align the strips with the ladder; (d) Keratherm strips are attached to the \apv\  chips of an origami board.}
 \label{fig:C3_keratherm_attachment}
\end{figure}

The Keratherm, used to ensure proper thermal contact with a cooling surface, has to be positioned on the surface of \apv\  chips of the origami boards on which the cooling pipe is mounted. Extreme care must be taken during the installation of Keratherm strips on top of \apv\  chips of the origami boards because they are placed very close to the \apv\ chips wire-bonds, so any mistake in the operation can lead to wire-bonding damage, that could compromise the functionality of \apv\  chips in the ladder. To properly attach Keratherm, a specific procedure was established: four Keratherm strips are cut, as in Figure~\ref{fig:C3_keratherm_attachment} (a), and are then positioned on a jig that holds them by vacuum chucking. The ladder is positioned as in Figure~\ref{fig:C3_keratherm_attachment} (c), tilting the ladder container so that the top surface of the ladder is facing the operator. Keratherm strips are aligned on the Keratherm-jig, and the protective film is removed to expose the adhesive layer. The Keratherm-jig is then positioned on a movable stage with four degrees of freedom (X-Y-Z-$\phi$). The alignment of Keratherm strips with \apv\  chips is verified with the help of a magnifying glass, then the movable stage is brought forward until Keratherm strips touch all \apv\  chips. Finally, the vacuum is switched off, the Keratherm jig is removed and the plastic film on the other surface of Keratherm strips is removed. The procedure is repeated for all origami boards of the ladder. 
Once Keratherm strips have been attached, all wire-bondings of the ladder are optically inspected again to verify that no damage has occurred during Keratherm attachment. After this last optical inspection, the ladder is ready to be installed on the end-rings.

\paragraph{Ladder installation}
For Layers 4, 5 and 6 ladders, hybrid cables are tested, assigned to each connector of the ladder and labeled on both sides, to easily recognize which cable is attached to which connector. For Layer 4 ladders, cables must be fed through Layer 5-6 end-rings prior to the ladder installation. The last step of the preparation for Layers 4, 5 and 6 ladder, before their installation, is the verification of \acrfull{FOS} (see Section~\ref{sec:temp_mon}) insertion, to be sure that the fibre can be properly inserted in the dedicated channel of the AIREX\textsuperscript{\textregistered} layer.\par
\begin{figure}[!ht]
 \centering
 \includegraphics[width=\textwidth]{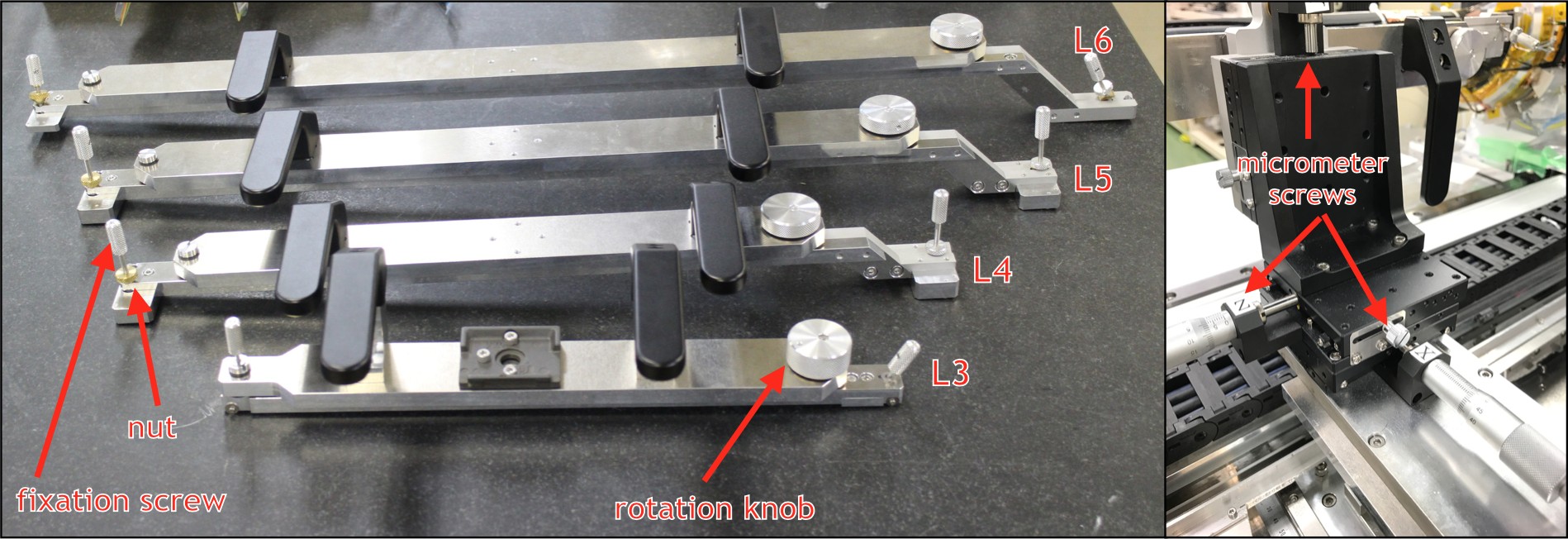}
 \caption{On the left side, the ladder mount tools are shown. This tools are used to grab the ladders, allowing the operator to engage them on the ladder mount stage. See text for a step by step description of the procedures.  On the right side, the ladder mount stage.}
 \label{fig:C3_LM-tools}
\end{figure}
\begin{figure}[!ht]
 \centering
 \includegraphics[width=\textwidth]{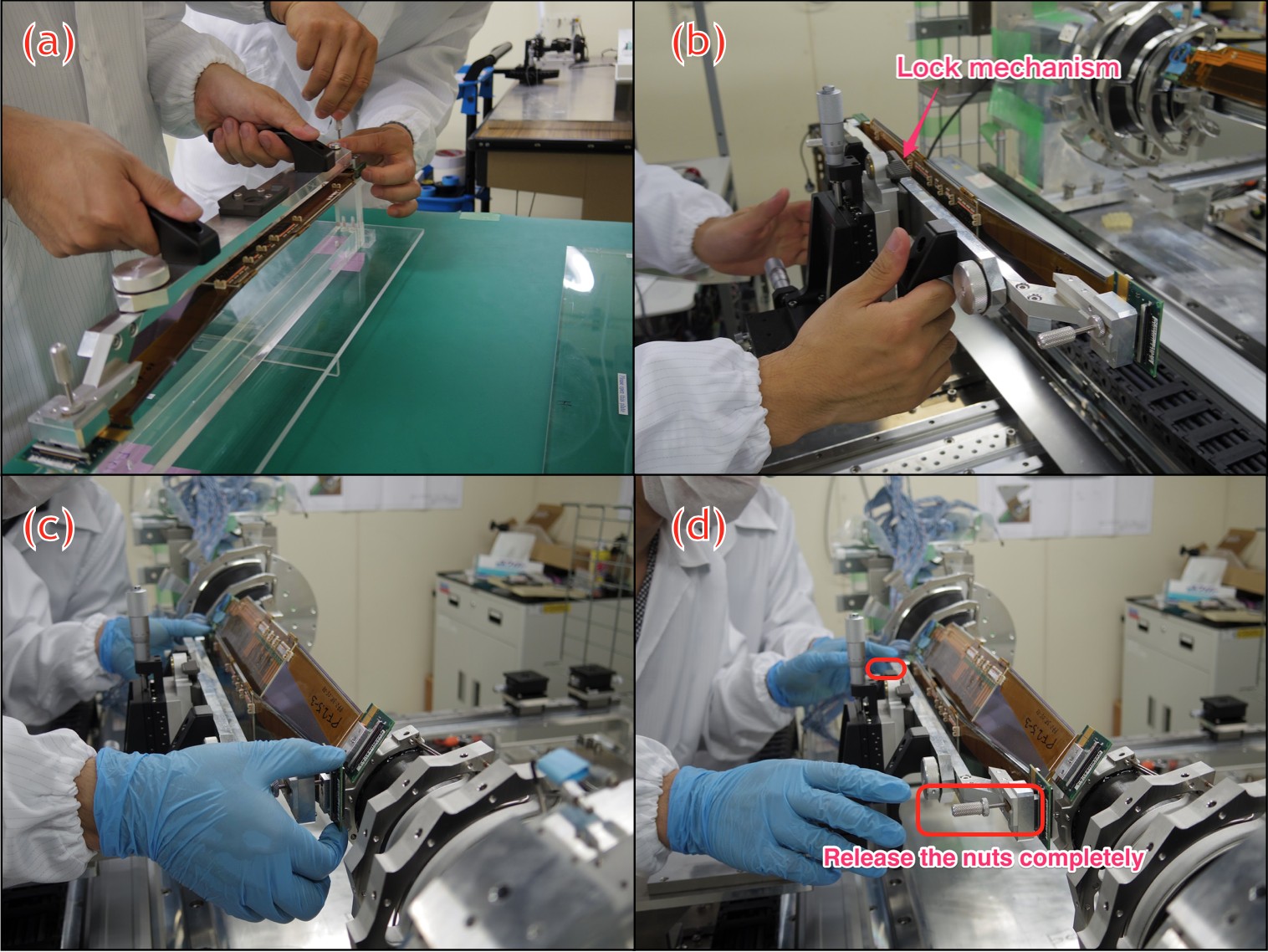}
 \caption{Key steps of the ladder mount procedure. (a) the ladder mount tool is attached on the ladder using the fixing screws and the nuts; (b) the ladder mount tool is fixed on the ladder mount stage; (c) after the alignment of kokeshi-pin heads with the corresponding holes, the ladder is inserted in the end-rings; (d) the ladder is released from the ladder mount tool. Note that in the actual ladder mount procedures gloves were worn at all times to avoid sensor contamination. Some pictures were taken in early stages of the procedures development, with mechanical samples that required less attention.}
 \label{fig:C3_LM-steps}
\end{figure}

The ladder container, without all side walls, is positioned on a table beside the ladder mount table. A ladder mount tool, shown on the left side of Figure~\ref{fig:C3_LM-tools}, is used to grab the ladder: made of aluminum, the tool has two screws and two corresponding nuts at the extremities that are used to fix the ladder on the FWD and BWD ends, two handles used by the operator to grab the tool, a knob connected to a gear that allows a rotation of the ladder around the X-axis, and a quick-release plate that is used to fix the tool on the ladder mount stage. The operator brings the ladder mount tool on top of the ladder, then two more operators fix the tool to each end of the ladder, engaging the screw and fixing the nut (figure~\ref{fig:C3_LM-steps} (a)). Now the pick-up screws that hold the ladder to two vertical supports of the container are released, so that the ladder can be extracted from the container. The main operator starts lifting the ladder, while another operator guides the extraction of kokeshi-pins from the container vertical supports. Once the ladder is extracted, the main operator brings and fixes the ladder mount tool on the ladder mount stage engaging the locking mechanism of the quick-release plate (figure~\ref{fig:C3_LM-steps} (b)). The ladder mount stage, shown on the right side of Figure~\ref{fig:C3_LM-tools}, is equipped with three micrometer screws used to move the ladder along the three axes. The whole stage block is mounted on rails on which it can slide towards the ladder mount table, with a vacuum lock mechanism that must be released by pressing a pedal. After all these operations, hybrid cables are temporarily connected to the ladder and an electrical test is performed to verify that the ladder has not been damaged during the operation. \par

After the electrical test, the FWD hybrid cables are detached and the ladder is slowly moved towards the end-rings, sliding the ladder mount stage on its rails, while the distance to the end-rings is continuously monitored with the help of a ruler mounted on the rails. The stage is stopped when the ladder is 5 mm far from the end-rings, then the stage is locked on the rails. Any further movement is controlled by the micrometer screws on the ladder mount stage. A helper on the backward side takes care of pulling the still connected BWD hybrid cables while the ladder is moved toward the end-rings, to avoid any entanglement with the end-rings and interference with any other previously mounted ladder. The main operator turns the X-axis micrometer screw bringing the BWD kokeshi-pin surface at a distance of approximately 1~mm from the end-ring; then turns the Z-axis micrometer screw to bring the ladder in the proper position in Z. The BWD kokeshi-pin head is aligned with the corresponding hole of the end-ring acting on Z and Y axes micrometer screws. Once the BWD kokeshi-pin is aligned, the position of the FWD one is adjusted using the knob on the ladder mount tool holding the ladder: the rotation axis of the ladder mount tool is close to the BWD kokeshi-pin location, so that only the FWD kokeshi-pin location changes while turning the knob. A fine adjustment of the alignment between kokeshi-pin heads and end-rings holes may be done if necessary, then the ladder is moved towards the end-rings using the X-axis micrometer screws so that the chamfered part of the kokeshi-pin heads are engaged in the end-rings holes. The main operator grabs the ladder from the FWD and BWD extremities, while two helpers on both sides release the nuts so that the ladder is released from the ladder mount tool (figure~\ref{fig:C3_LM-steps} (c)). The fixing screws of the tool, however, are still fixed to the ladder, so the ladder cannot fall even if it slips from the grab of the operator. The main operator inserts the ladder fully into the end-rings, then the helpers fix the set screw using a torque controlled screw driver set to 50 N$\cdot$cm. The fixation screws of the ladder mount tool are released (figure~\ref{fig:C3_LM-steps} (d)), the ladder mount stage is brought back to its original position and the ladder mount tool is finally removed from the stage. \par

FWD hybrid cables are connected, then all cables on both sides are properly routed through the assigned slots of the end-flange. For Layers 4, 5 and 6 ladders, the FOS is inserted in the corresponding channel in the AIREX\textsuperscript{\textregistered} sheet and the clamp is fixed with a set screw to the fiberglass surrounding the connectors on the BWD side. Finally, pictures of the installed ladder are taken and the reference points of each sensor are measured with the 3D coordinate measurement system. A full electrical test of the installed ladder, including a readout test of the FOS, follows. If the test is successful, the end-rings are rotated and prepared for the next ladder. \par
After the installation of all ladders of each SVD half-shell, the cooling pipe attachment and its test are done for Layers 4, 5 and 6 as described in the next section.

\begin{figure}[!ht]
 \centering
 \includegraphics[width=.8\columnwidth,keepaspectratio]{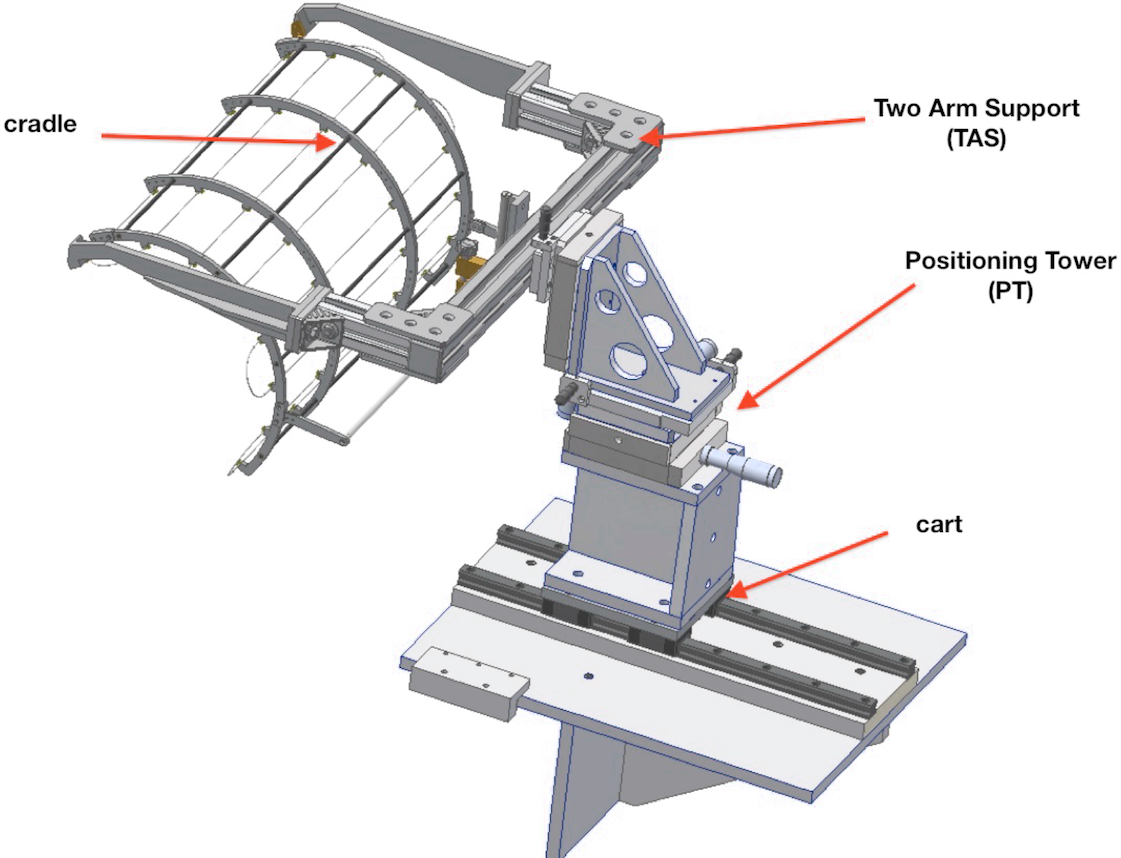}
 \caption{Drawing of the mechanical system used to mount the cooling pipes.}
 \label{fig:C3_mechanical_system_for_origami_pipe}
\end{figure}

\begin{figure}[!ht]
 \centering
 \includegraphics[width=.8\columnwidth,keepaspectratio]{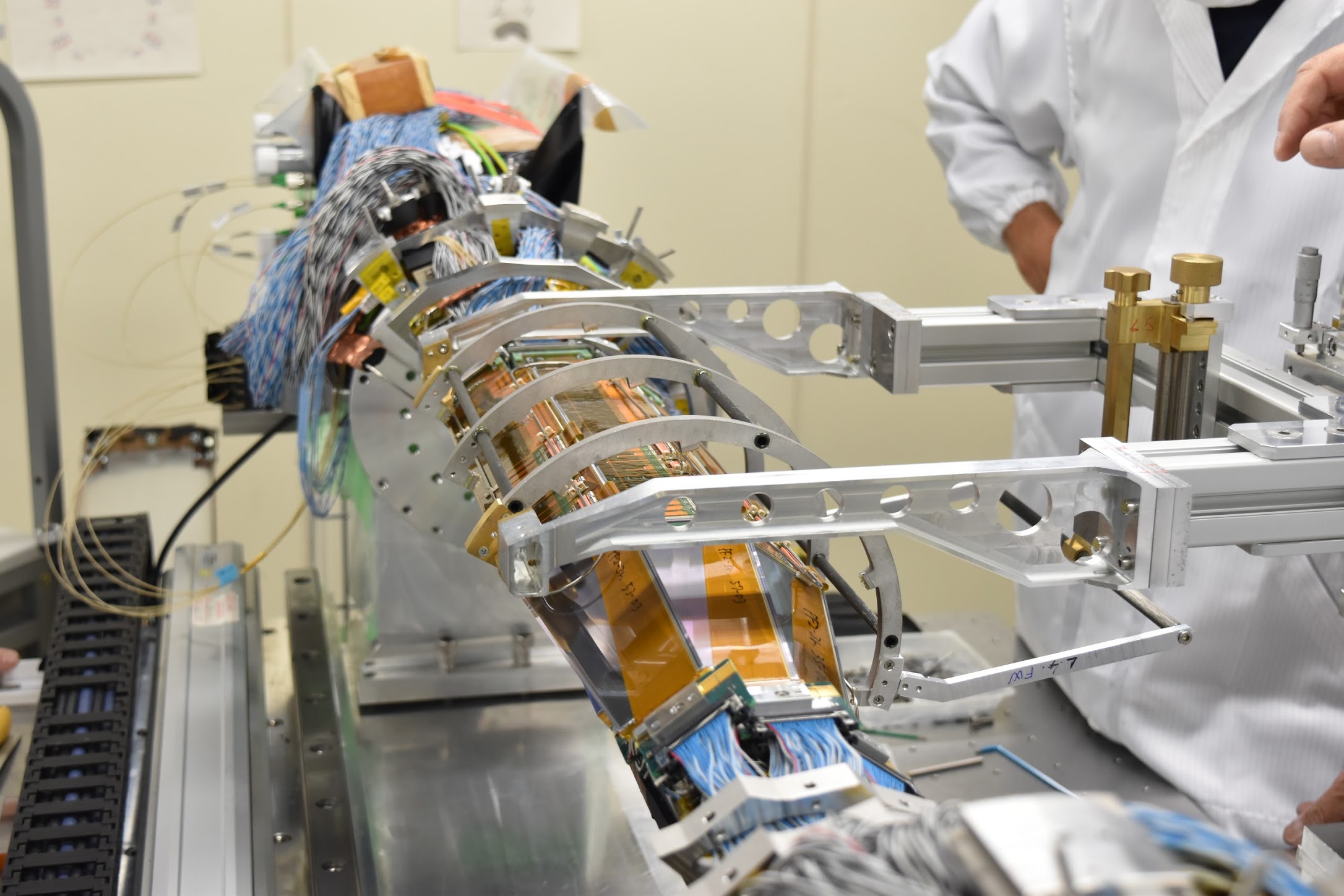}
 \caption{Layer 4 cooling pipe attachment: after positioning the cradle on the Layer 4 half-shell, the pipe is inserted into the \COtwo\ clamps of the ladders.}
 \label{fig:C3_L4_cooling_pipe_attachment}
\end{figure}

\lvlthreesection{Cooling pipe mounting procedures}
\label{sec:C3_cooling_pipes_mount}
\editor{Stefano, Antonio}
In order to mount the cooling pipes on a completed half SVD Layer 4, 5 and 6, a mechanical system, shown in figure~\ref{fig:C3_mechanical_system_for_origami_pipe}, was developed.
The system is positioned on the ladder mount table, on the opposite side with respect the ladder mount tool. On a thick base plate, fixed to the ladder mount table, two rails allow a coarse translation of the positioning tower (\acrshort{PT}) toward the SVD half. Micrometer screws are used to drive the fine movements along the x, y and z axes, together with a rotational stage. The two-arms support (\acrshort{TAS}) holds a cradle, on which the cooling pipe is fixed through plastic clamps. The TAS can be rotated to align the horizontal plane of the two arms with the plane of the ladder mount table. The cradle can be wrapped around the half SVD layer by means of a coarse mechanism and a fine micrometer screw.
In the following, general  procedures that are common to all layers, are  summarised.
\begin{enumerate}[leftmargin=*]
\item{Mechanical System setup.}\par
The cart is in its starting position, far from the SVD. The PT-TAS system is rotated by 180$^o$, pointing outside of the table. The cradle is detached from the TAS.
\item {Pipe pick up.}\par
The cradle is positioned on the cooling-pipe container, where the cooling pipe is fixed with some screws. The cooling pipe is released from the container, one straight line at a time, and inserted in the corresponding clamps of the cradle. Once the cooling pipe is fully attached, the cradle is positioned on the TAS, engaging the cantilever parts of the TAS. The PT-TAS-cradle system is then rotated toward the SVD.
\item {Approach the SVD.}\par
The cart moves the PT-TAS cradle system towards the SVD, reaching a pre-determined safe position, using the SVD end-rings as a reference. This is the starting point of finer adjustments. Acting on x and y-axes micrometer screws, the cradle is moved so that the first line of the cooling pipe reaches the mounting position. If necessary, the cradle can be rotated in order to align the pipe with the ladder at the top. All movements performed by micrometer screws are carefully checked by three watchers, who make sure that the cooling pipe nor the cradle touch the SVD ladders.
The cradle is moved with the micrometer screw around the axis of the top cooling line, until it reaches its final position, where all the straight  lines of the cooling pipe are at a radial distance of 1.5~mm from the clamps on the ladders.
\item {Cooling pipe attachment.}\par
Starting from the first line on top, the cooling pipe is released from the cradle clamps and engaged onto the ladder \COtwo\ clamps using the mounting tool, which is held by hand by an experienced operator. The mounting tool is at first engaged in the two rings of the ladder \COtwo\ clamp, then the cooling pipe is gently pushed down. 
This procedure is repeated for every clip on the ladders, following the path along the cooling pipe. Figure~\ref{fig:C3_L4_cooling_pipe_attachment} shows the stage at which the cooling pipe is attached to the Layer 4 ladders, with the cradle still in its final position. 
Finally, after the cooling pipe is mounted on the SVD, the cradle is brought to its "open" position again and the PT-TAS-cradle group is moved away from the SVD using the cart.
\end{enumerate}


The procedures were tested and tuned with mechanical prototype ladders for each layer, resulting in a cooling pipe mounting time of a few hours. Only in one case a connection of the cooling pipe was found damaged, and a detachment of the installed pipe was needed in order to replace it with a spare cooling pipe. No damage was observed on the ladders after each cooling pipe installation.

\begin{figure}
    \centering
    \includegraphics[width=0.9 \columnwidth,keepaspectratio]{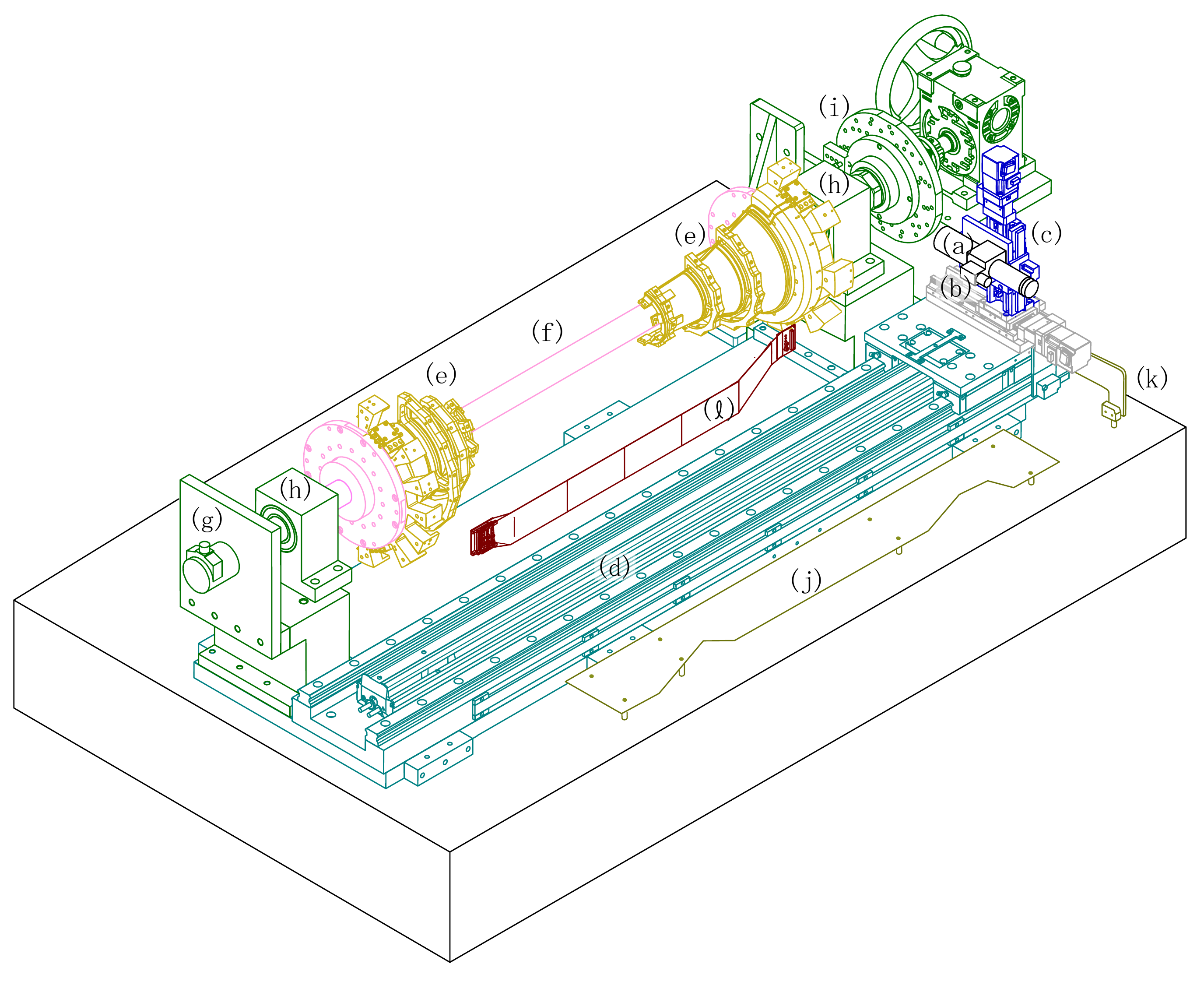}
       \caption{The 3D coordinate measurement machine system on the ladder mount table. 
        The ladder mount arm is not shown here. 
        (a) The CCD camera with working length of 55~mm,
        (b) (c) (d) $X$, $Y$, $Z$  linear rails with position encoder,
        (e) End-rings,
        (f) Support rod,
        (g) Angle encoder,
        (h) Bearings,
        (i) Index plate and the worm gear for the rotation of (f),
        (j) (k)  Limiter plate and switch for crash prevention, and,
        ($\ell$) A ladder.
         }
    \label{fig:C3_TT_3Dfig1}
\end{figure}

\lvlthreesection{Mechanical contactless survey system and results}
\label{sec:C3_mechanical_survey}
\editor{Tsuboyama}
% Markus (originally was in chapter 2, but better fits here)
%Tsuboyama uncommented on 11 Sept 2020.
The mechanical precision of both end-rings and installed ladders is verified on the ladder mount table using a built-in CMM positioned on the side of the table where ladders are mounted, as shown in figure~\ref{fig:C3_TT_3Dfig1}. The position of the precision holes for the kokeshi-pins has to be within \qty{150}{\um} of the nominal position and the angular deviation must be less than 1.5~mrad. The position of the sensors fiducial F-marks on each mounted ladder has to be within \qty{400}{\um} of its nominal position, with a maximum rotational error of 2.5~mrad. These specifications are determined mainly by the safety requirement of avoiding interference between ladders. 
 
%\lvlfoursection{The principle of operation}
\paragraph{The principle of operation}
The CMM system consists of a \acrshort{CCD} microscope camera (Schott SOD-10X Telecentric lens) with working distance of 55~mm and focusing depth of \qty{17}{\um}, mounted on three-axes linear rails.
Each axis is equipped with an absolute position encoder (Heidenhain LIP481, LIP581 absolute linear encoder) of $<$ \qty{0.1}{\um} precision. The end-rings are fixed to a rotating axis bar. The distance between the lens and the object is determined by using the CCD image. The rotation angle is measured with an angle encoder (Heidenhain ROC413 absolute angle encoder) with a resolution of 0.05\degree. The resolution of the coordinate measurement of the camera is better than \qty{1}{\um} in all directions.

To cover the whole SVD structure, the length of the $Z$ linear rail is 1 meter long.
The overall measurement ambiguity of the system, mainly due to uncertainties in the mechanical setting and alignment, is estimated to be \qty{50}{\um}.

When the position of an object is measured, the camera is first moved to the nominal position of the target by the CMM control program. Then the measurement operator adjusts the $X$, $Y$ and $Z$ coordinates by watching the CCD image on the PC screen.
The final adjustment below \qty{100}{\um} is done by a focus finding program.
The resolution of the automated focusing is better than \qty{1}{\um}, so the uncertainty due to different operators is avoided. After the final adjustment, the coordinates and the focus image are immediately stored to the PC.

 To avoid accidents due to operation errors or hardware malfunctions, the CMM movement was constrained both by software limits and by a hardware safety system based on limit switches. 
%
%\lvlfoursection{Measurement procedure}
\paragraph{Measurement procedure and results}
Before a ladder is mounted, the rotation angle of the end-ring is set within 0.001\textdegree{} with respect to the nominal angle, so that the kokeshi-pins of the ladder are inserted straight into the mounting holes of the end-rings. Eight points of the end-ring surface and the center of the mounting hole are measured.

After the ladder is mounted, the coordinates of the four F-marks of each sensor are measured. The measured coordinates are compared with the nominal position to judge if the ladder is properly mounted on the support system. If the deviation is above the \qty{400}{\um} limit, the ladder mount procedure is examined, and if necessary the ladder is un-mounted. For the Layers 4, 5 and 6 ladders, the measurement is repeated after the \COtwo\ pipe is mounted, to check if the layer structure remains within the required tolerance.

The measured coordinates are compared to the measurements taken after the ladder assembly process, to understand if the ladder was deformed during the mounting process. Only in one case one layer 3 ladder was found deformed outside the defined tolerances, and it was replaced with another ladder. For all other ladders, no significant differences were found comparing the measurements before and after the ladder mount.

The measurements performed after the cooling pipe attachment show a slight deformation of the ladders due to the intrinsic force exerted by the cooling pipe. The biggest deformation is observed on the slanted forward sensors and is of the order of a few hundreds of micron, still within the acceptable range and with no danger of interference with ladders of the other layers.

The coordinate measurements data are the starting point for the alignment of the SVD with respect to the other \belletwo{} sub-detectors. The final position of the sensors in the \belletwo{} detector are determined with real particle tracks from cosmic ray or beam collisions by using an alignment software, described in section~\ref{sec:C7_alignment}.

%%%%%%%%%% SVD HALF-SHELL STORAGE %%%%%%%%%%
\lvlthreesection{SVD half-shell storage}
\label{sec:C3_svd_half_shell_storage}
\editor{Antonio}
After its completion, the first SVD half-shell had to be removed from the ladder mount table and stored safely in order to assemble the second half-shell. In addition, between the completion of both halves and the full VXD assembly, the commissioning of the detector had to be performed. For this purpose, two similar movable tables and two so called ``pick-up tools'' were used to store the two SVD half-shells. \par

Two 1 mm CFRP outer covers are designed to surround the two SVD half-shells, to protect the whole VerteX Detector (VXD) once it is fully assembled. Once an SVD half-shell is complete, the first step before dismounting it from the ladder mount table is the installation of the outer cover. On the inner surface of each cover, an optical fiber temperature sensor, similar to the one described in Section~\ref{sec:temp_mon}, is attached. The outer cover is placed on top of the SVD half-shell and taper pins of 2.5~mm diameter are inserted in the corresponding holes to accurately align the end-flange and the outer cover. Screws are then used to fix the cover to the end-flange. Taper pins are finally extracted from their location. \par

\begin{figure}[hbt]
    \centering
    \includegraphics[width=14cm]{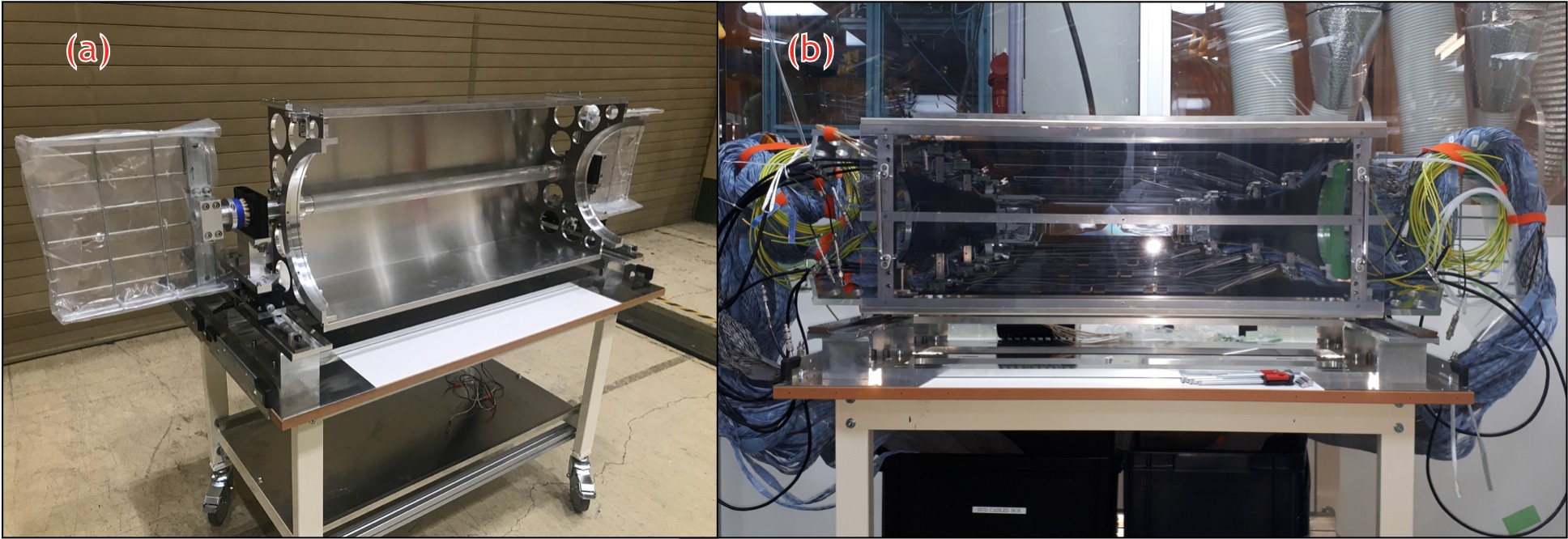}
    \caption{(a) The bare SVD pick-up tool on the storage table. (b) SVD half-shell on the storage table held by the pick-up tool, with the front cover installed.}
    \label{fig:C3_pick-up_tool}
\end{figure}

The pick-up tool structure is shown in Figure~\ref{fig:C3_pick-up_tool}. It is positioned on a storage table equipped with wheels, most of it is made of aluminum, on the two side supports two rails with braking mechanisms are mounted, on which the tool can slide toward the SVD half-shell. The side supports are also equipped with six ball plungers that allow the movement of the whole tool between the tables. The storage table is brought beside the ladder mount table and raised up so that the surfaces of the two tables are at the same height. The whole pick-up tool structure is moved on the ball plungers and fixed on the ladder mount table, then the pick-up tool approaches the SVD half-shell until the two curved ends of the tool engage the SVD end-flanges on both sides. The SVD half-shell is fixed to the pick-up tool, then all cables are fixed to some supports installed on the pick-up tool, and finally the SVD is released from the support of the ladder mount table. Once the SVD half-shell is fully detached from the ladder mount table, the pick-up tool is moved back on the storage table, where it is fixed with screws. In the bottom part of the table, two containers are placed to store the cables. In order to protect the inner part of the SVD half-shell, which is exposed at this point, a protection cover is installed on the side. The table was stored into a box and constantly flushed with dry air until the commissioning started.

%============================   LESSONS   ======================
\lvltwosection{Issues and Lessons Learned During the SVD Construction}
\label{sec:C3_issues_during_svd_consruction}
%\editor{Francesco}
The SVD construction was a complex and challenging task carried out over the course of several years. 
As in all large projects, many surprises, problems and difficulties appeared and were overcome, sometimes with considerable effort. 
In this section  these issues and the relevant lessons learned during the construction process are briefly reported. 

\lvlthreesection{Quality control of components}
In several cases some components were not of sufficient quality, because of either insufficient design margins, or lack of proper quality control procedures. This required extended quality control procedures, or the fabrication of additional components, sometimes with design modifications possibly realized at very short notice.  

For instance, initial pitch adapter productions had issues with the pad size and material control, leading to great difficulties in the wire bonding process and even resulting in the creation of cracks in the metal traces causing electrical connection failure. These problems were discovered in some of the first prototypes allowing an optimization of the pitch adapter design and production process.

Another example was the appearance of cracks on the \apv\  chips after the thinning and dicing process. In spite of the positive electrical characterization of the devices, there was the concern that mechanical stress induced especially by large temperature variation could cause the cracks to extend leading to failures. A manual chip inspection program was hastily setup to identify the cracked dies and reject them.

On more than one occasion unexpected features of components were observed, e.g. various traces of material on top of the sensors, leading to investigations and analyses to determine whether the features were the indication of some problem, or rather a normal outcome of the production process. 

The lesson learned from these problems is that quality control and assurance must be carried out and documented in the most systematic and extensive fashion possible; experience indicates that what is not checked is likely to become problematic, which can be particularly damaging in  systems where replacing single failing components is hard or impossible. 

\lvlthreesection{Assembly and integration issues}
Ideally all the assembly procedures and final integration should be fully defined and tested with real components before production starts. In practice, this is often impossible and one has to test the procedures only piece-wise, maybe with surrogate components, opening the possibility that bad surprises appear in the middle of the production process, requiring retro-fitting and adaptation. 

One important and potentially very damaging issue appeared during the construction of the ladders: the pitch adapters of some forward modules started to detach from the sensor after the assembly of the ladder. Although the pitch adapter gluing procedures had been tested and verified, it turned out that in layer 4 and layer 6 ladders the pitch adapter bending radius is significantly smaller, leading to much higher stress on the glue joint. An emergency retrofitting program was launched on all forward modules, including the ones already installed on ladders, with the development of a reinforcement procedure which was quite challenging and risky. In this case the problem was generated by the combination of a marginal design (very small glue area between pitch adapter and sensor) and the lack of a proper prior analysis and verification of what would be the most extreme usage situation of the components.

Another example of problems discovered only very late was related to the fixing of layer 3 modules to the support cones. The screw initially foreseen in the design had the potential of causing a short on the hybrid board since the ground plane was very close to the screw hole. An insulation system was developed and implemented. This example shows how crucial it is to perform final and complete integration tests with final components to spot design imperfections early on. 

Ensuring the complete reproducibility of the assembly procedures has proven to be extremely challenging, since both the laboratory conditions and the personnel undergo changes that may result in small difference in the procedures, which should be avoided as much as possible. The Quality Control and Assurance Group described in section~\ref{sec:C3_QCG} was instrumental in ensuring that all the minute details where specified in the manuals and that the instructions where consistently and uniformly followed at all sites. 

\lvlthreesection{Project management}

Running a large construction project requires considerable effort and must be properly organized and staffed. In general, very marginal contingency was used in planning the SVD construction activities, resulting in an often slipping schedule. Expert person-power was limited and there were many cases of single points of failure, namely operations that could be performed only by a single person.
Preliminary risk assessment should be given serious consideration, to facilitate a timely and correct completion of the project. Ensuring a correct design requires great attention to detail, as well as a continuous internal and external review process. For instance, the initial  design of the SVD had some serious mechanical incompatibilities. As another example, the Layer 3 cooling design was discovered to be insufficient very late in the process, thanks to simulations and a test mockup, leading to a last minute retrofit to avoid excessive temperature gradients. 
As a general observation, spares were insufficient at all levels, from the individual component to the complete ladder. Although the additional cost and effort needed for spares is sometimes difficult to justify, having a significant number of spares makes the project much more resilient to problems, such that it is more likely to be completed in time.

 % 25 pages

\lvlonesection{Environmental and Radiation Monitoring}
{\label{sec:environment}}
%% Here the introductory text (Track 1)

Safe operation of the vertex detector (\acrshort{SVD} and \acrshort{PXD}) requires continuous and reliable monitoring of its environmental parameters, in particular radiation dose rates, temperature, and humidity. If their values happen to be outside accepted limits, an interlock system, independent of the computer network and of the slow control software, must be able to promptly shut down the power supplies as well as to notify the central interlock system of the experiment.
Environmental and radiation monitoring were introduced at a late stage in the \acrshort{VXD} design. The solutions successfully adopted are briefly described here, including the requirements, the design and construction at INFN Trieste, and the installation at KEK.

The radiation from beam losses is monitored by a system based on diamond detectors, that also generates beam-abort requests to SuperKEKB in case of excessive beam losses; this function is essential to prevent severe local radiation damage to VXD parts. This system is described in section~\ref{sec:radmon_ba}.

Temperature is monitored by \acrfull{NTC} thermistors at the inlets and outlets of the SVD cooling pipes and by Bragg-type \acrfullpl{FOS} close to the SVD front-end \acrshort{ASIC}s, as described in section~\ref{sec:temp_mon}.

The dew point of the residual humidity in the dry nitrogen circulating in the VXD volume is kept under control via ``sniffing pipes'' connected to external humidity sensors. As explained in section~\ref{sec:humid_mon}, this is essential to prevent condensation or ice formation on parts of the SVD cooling system, based on dual-phase CO$_{2}$ circulating in thin pipes (section~\ref{sec:mech1}).

Section~\ref{sec:hard_interlock} describes the \acrfull{PLC}-based hardwired interlock system, which protects both SVD and PXD by disabling their power supplies when local and general \belletwo\ environmental conditions exceed acceptable limits.

Finally, section~\ref{sec:C3_cooling_plant} briefly describes the working principles and design of the \COtwo-based evaporative cooling plant. 

\lvltwosection{Radiation Monitoring and Beam Abort}
{\label{sec:radmon_ba}}

The radiation monitor and beam abort system is designed to protect VXD sensors and front-end electronics from damaging beam losses and to continuously monitor the radiation dose rates in the interaction region. Detailed descriptions of the design, construction, calibration, and performance of this system are given in references~\cite{Bassi:2021dno,Bacher:2021frk}.

\lvlthreesection{Requirements}

According to extrapolations from the former B-factories, KEKB and PEP-II, and simulations of the expected beam backgrounds at SuperKEKB, the system was expected to measure dose rates from a few \unit{\micro rad/\s}  to several \unit{\kilo rad/\s}, depending on the beam currents and accelerator conditions.

The SuperKEKB beam-abort process, triggered by a large increase in beam losses and radiation dose rates, is constrained by the \qty{10}{\us} revolution time of electrons and positrons in the SuperKEKB rings. The beam abort system should react in a few \unit{\us}, before a full revolution of the beams. Localised radiation damage in SVD microstrip detectors may occur when a large radiation dose, in excess of about $1$~rad, is delivered in a time interval of the order of \qty{1}{\ms} (see discussion in section~\ref{sec:C6_rad_spikes}). 
For the dose-rate monitoring, archival, and integration, a recording rate of \qty{10}{\Hz}  is required.

A configuration of radiation detectors distributed on the beam pipe and SVD support cones was initially specified; a request followed to instrument also the bellows, flexible connections between the vacuum beam pipes of the two rings, and the beam pipe section inside the VXD, located close to the \acrfull{QCS} magnets. The present configuration includes a total of $28$ diamond detectors: eight on the beam pipe, twelve on the SVD support cones, eight on the bellows, as shown in figure~\ref{fig:C3_diamonds_installation}.

\lvlthreesection{Detectors design and construction}

Artificially grown single-crystal \acrshort{CVD} (sCVD) diamond sensors are chosen for their fast and temperature-independent response as well as their radiation resistance.
Electrons and holes produced by ionisation in the diamond bulk drift to respective electrodes located on opposite faces, when a bias voltage is applied to them. The induced current in the external circuit is proportional to the dose rate. 

\begin{figure}
	\centering
		\includegraphics[width=7.5cm]{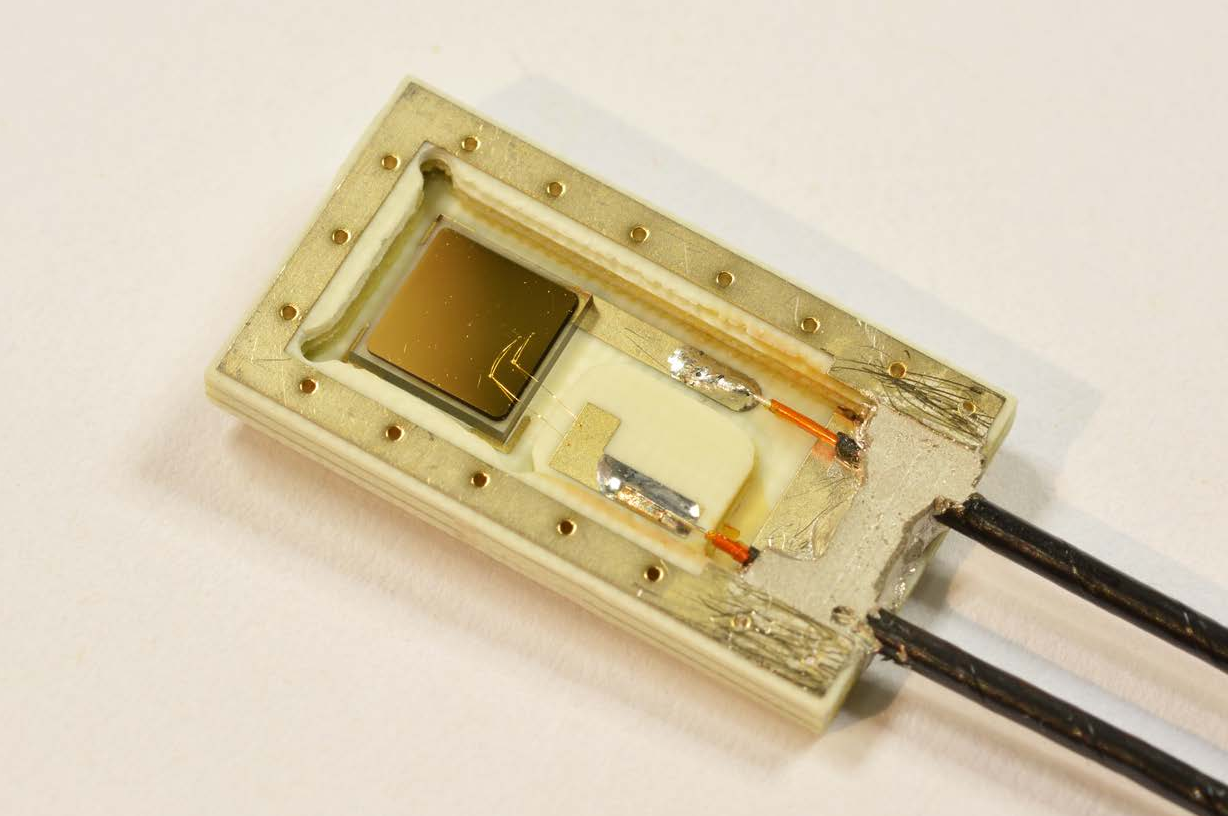}
	\caption{A diamond sensor packaged into a detector unit, as described in the text.} 
	\label{fig:C3_diamonds_package}
\end{figure}

The electronic-grade sCVD sensors~\citeCompany{ref_e6} have lateral dimensions $(4.5 \times 4.5)$~mm$^{2}$ 
and $0.50$~mm thickness; two $(4 \times 4)$~mm$^{2}$ ($100$~nm Ti + $120$~nm Pt + $250$~nm Au) electrodes are deposited~\citeCompany{ref_cividec} on the two opposite faces.
Each diamond sensor is mounted on a ceramic-like~\citeCompany{ref_rogers} printed-circuit board, providing back screening and electrical contacts, as shown in Figure~\ref{fig:C3_diamonds_package}. The diamond sensor is glued to a square pad, which provides connection of the back-side electrode to a soldering pad. The front-side electrode is connected to the second soldering pad by ball-bonded gold wires.  The inner conductors of two miniature coaxial cables, each $2.5$~m long, are soldered to the printed board pads, while conductive glue is used to mechanically fix the cables and establish a connection of their outer conductors with the outer shielding of the printed circuit, completed by a thin Aluminum cover on the front side. The cable length of $2.5$~m is fixed by the space constraints within the \acrshort{VXD} set-up.

Each detector is characterised and calibrated as detailed in reference~\cite{Bassi:2021dno}.
The quality of the detector assembly is assessed by measurements of dark current $I$ as a function of bias voltage $V$ ($I-V$). In all cases, the measured dark current is less than a few pA at $500$~V and not more than about $1$~pA at $100$~V.
The uniformity of the transport properties of the charge carriers and of the electric field in the diamond bulk are validated using the \acrfull{TCT}~\cite{Pernegger:2005}, based on the localised injection of a controlled amount of electrons and holes at a very small depth, close to one of the electrodes, by monochromatic $\alpha$-particles from a $^{241}$Am source.
For each detector the radiation-induced current-voltage relation is established with a constant electron flux from a radioactive $^{90}$Sr $\beta$-source. At least one bias polarity shows a clear plateau and stable response, leading to the choice of the operating bias voltage at $100$~V for the optimal polarity.

Finally the measured current is related to the radiation dose-rate and an individual calibration factor is determined for each detector. The flux of ionizing electrons from the almost point-like radioactive $^{90}$Sr $\beta$-source is controlled by varying the distance between the source and detector. 
The measured current $I$ in the detector polarisation circuit is compared with the dose rate $dD/dt$ predicted by a detailed simulation of the entire set-up; a reference silicon diode is used to constrain the source activity and reduce the systematic uncertainties in the simulation.

The measured conversion coefficients $k$ in the linear relationship  $dD/dt = k I$ correspond to about $35$~(mrad/s)/nA on average.
They can differ from the average by up to approximately $\pm 50$\% for different detectors. The uncertainty of about $8$\% in their determination is predominantly systematic~\cite{Bassi:2021dno}. Measurements with a strong $^{60}$Co $\gamma$ source confirm a linearity in the response to the dose rate over at least three orders of magnitude.

\lvlthreesection{Electronics, monitoring and beam abort}

Groups of four diamond detectors are controlled and read out by a purpose-designed \acrfull{DCU}. The DCU digital core is a board hosting an \acrshort{FPGA}, that receives commands via an Ethernet interface, drives four \acrshort{HV} modules independently through a \acrshort{DAC}, and accepts input data from front-end modules providing amplification and analog-to-digital conversion of the input signals. The DCU is also able to deliver VXD abort requests for the \acrshort{HER} and \acrshort{LER} beams, and receives the SuperKEKB abort signals. The current signals from diamond detectors are amplified by trans-impedance amplifiers and digitised by 16-bit \acrshort{ADC}s at $50$~Msamples/s. The FPGA firmware stores sums of 125 ADC samples every 2.5~$\mu$s at $400$~kHz in a large revolving buffer memory. 

Further sums of five million ADC values can be read out at $10$~Hz. The online control software converts these values from ADC to dose-rate units after subtracting pedestals; these data are archived and used for continuous monitoring. From the archived values, accompanied by time stamps, integrated doses are computed. 

Three current-measurement ranges can be selected: $36$~nA, $9$~$\mu$A, and $4.5$~mA. The first range allows precise monitoring of relatively small beam losses, while the third range avoids saturation in the detection of large radiation spikes, and is used for beam abort. The intermediate range may be used in the future, with increasing dose rates.

For each diamond detector, two moving sums of $400$~kHz data are computed by the DCU firmware. The moving sums are updated at each \qty{2.5}{\us} cycle, by subtracting the oldest added value and adding the newest one. The two moving sums, representing integrated doses, are compared with programmable thresholds. The resulting eight logical signals in a DCU can be combined using programmable logic to produce an abort request, if the DCU is enabled to do so. In the present configuration of seven DCUs connected to the 28 detectors, only one DCU, controlling four detectors located on the beam pipe, is enabled to deliver abort requests. 

Communication with the SuperKEKB beam abort system is based on four signals. The diamond system delivers two abort requests, separately for the electron \acrfull{HER} and the positron \acrfull{LER}. When the abort kicker magnets complete an abort that either originated from this system or from other sources, two "SuperKEKB abort" timing signals (HER and/or LER) are broadcast and received by the DCUs. These signals stop the buffer memory writing. The $400$~kHz data can then be read out and used for a "post-abort" analysis of the beam losses preceding the abort. 

\begin{figure}
	\centering
		\includegraphics[width=14cm]{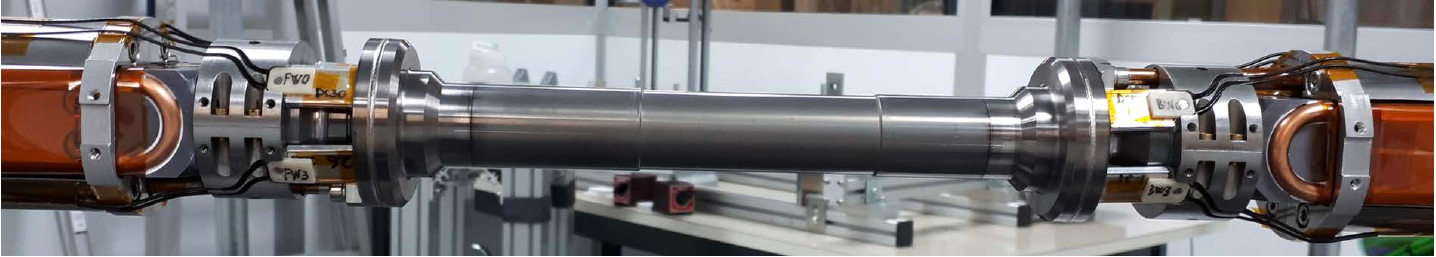} \\
		\includegraphics[width=8.8cm]{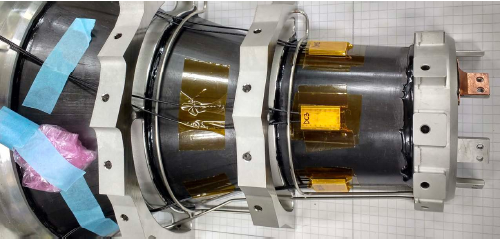}
		\includegraphics[width=5.4cm]{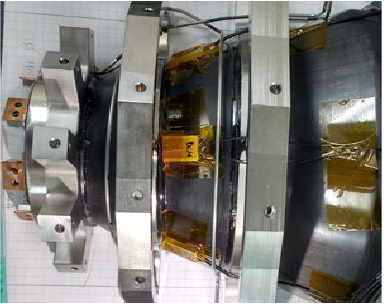} \\
		\includegraphics[width=8cm]{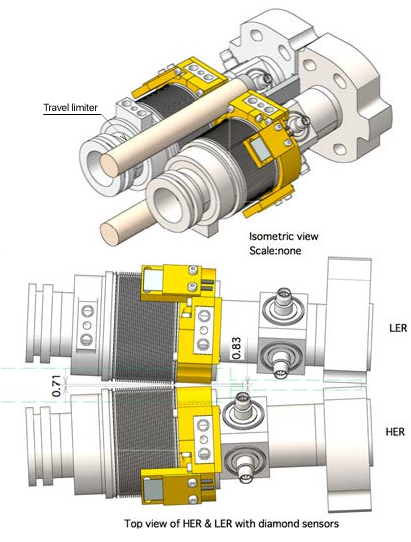}
	\caption{Installation of diamond detectors: (upper) on the beam pipe; (middle) on SVD cones, forward and backward; (lower) on the beam pipe bellows.}	
	\label{fig:C3_diamonds_installation}
\end{figure}

\lvlthreesection{Installation}

Four diamond detectors and prototype electronics were installed prior to the accelerator commissioning in 2017 (Phase 1), as part of the \acrshort{BEAST} II detectors dedicated to background studies~\cite{Lewis:2018ayu}. During a pilot run in 2018 (Phase 2) the final version of the electronics was tested with eight detectors in their final position on the beam pipe, as discussed in section~\ref{sec:phase2SVDSetup}. An additional 20 diamond detectors were delivered and installed at KEK in 2017 on the beam pipe (4 + 4), the SVD cones (6 + 6), while the eight detectors of Phase 2 were recycled to instrument the forward and backward beam-pipe bellows, as shown in Figure~\ref{fig:C3_diamonds_installation}. The $2.5$~m short, thin coaxial cables are linked at the \gls{DOCK}s (section~\ref{sec:C2_junctionboards}) to longer and thicker double-shielded ($25 - 30$~m) coaxial cables, connected to seven \acrshort{DCU}s in the Belle II electronics hut.

\lvltwosection{Temperature Monitoring}
{\label{sec:temp_mon}}
%\editor{Livio}

The power dissipation by the \acrshort{PXD} \acrshort{DEPFET} sensors and their front-end electronics amounts to about $18$~W per module, $360$~W in total for the $20$ PXD modules. The APV25 front-end chips of the SVD dissipate about $700$~W in total during full operation. 
This power is removed by a cooling system based on heat exchange with a dual-phase CO$_{2}$ fluid at a temperature of about \qty{-20}{\degreeCelsius}, circulating in thin pipes with good thermal contact with the front-end readout chips, as described in section~\ref{sec:mech1}. 

Temperature monitoring is therefore important to ensure proper operation of the cooling system, with a required absolute accuracy of \qty{1}{\degc} and a resolution of \qty{0.1}{\degc}. It is composed of two subsystems. The first system measures input and output temperature of the CO$_{2}$ cooling pipes and is based on \acrshort{NTC} thermistors.
%~\citeCompany{ref:murata_thermistors} 
The second system (\acrshort{FOS}) monitors the temperature of SVD \glspl{ladder}, with distributed measurement points close to the front-end chips, based on Bragg-type sensors.

\begin{figure}[htbp]
	\begin{center}
		\includegraphics[width=0.8\textwidth]{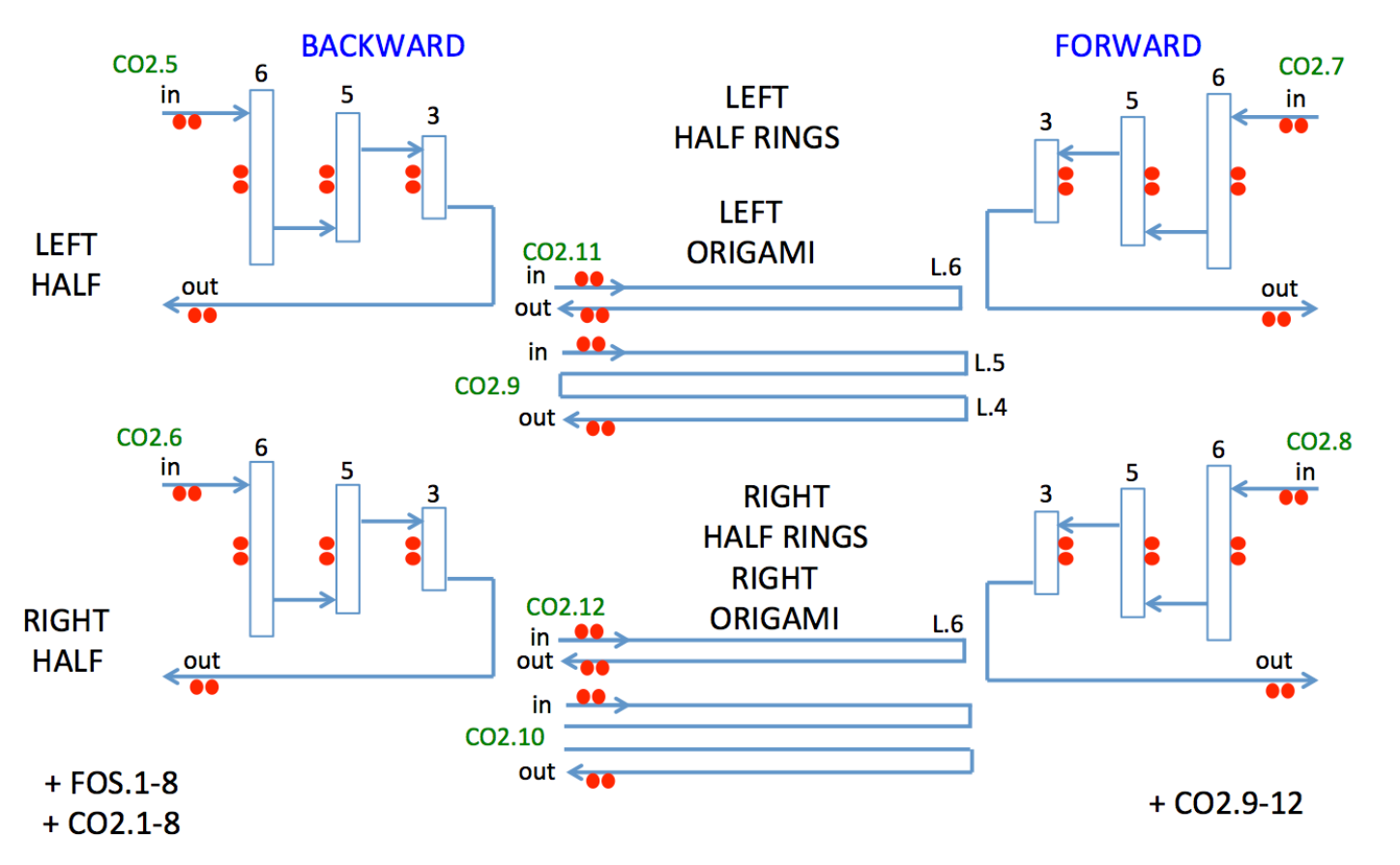} \\
		\caption{Physical layout of the \acrshort{NTC} system:  the red dot pairs represent NTC sensors placed on half \gls{end-ring}s and on the inlets and outlets of cooling pipes. }
		\label{fig:C2_temp_mon_NTC_layout}
	\end{center}
\end{figure}

\lvlthreesection{NTC system design}
For the first task $12$ \acrshort{NTC} thermistors
%~\citeCompany{ref:murata_thermistors} 
are mounted on the outer surface of the SVD support \gls{end-ring}s (one sensor on each of the six half-rings) and $16$ sensors on the cooling pipes (one for each inlet and outlet). Doubling the number of sensors for redundancy at each measurement spot brings the total number to $56$ sensors in the SVD volume. Additional sensors are installed on the CO$_{2}$ transfer lines, in the docks chiller cooling system, and at cross-reference points with the \acrshort{FOS} sensors, for a total of $84$ sensors. The location of the sensors is shown in figure~\ref{fig:C2_temp_mon_NTC_layout}.

\begin{figure}[htbp]
	\begin{center}
		\includegraphics[width=0.8\textwidth]{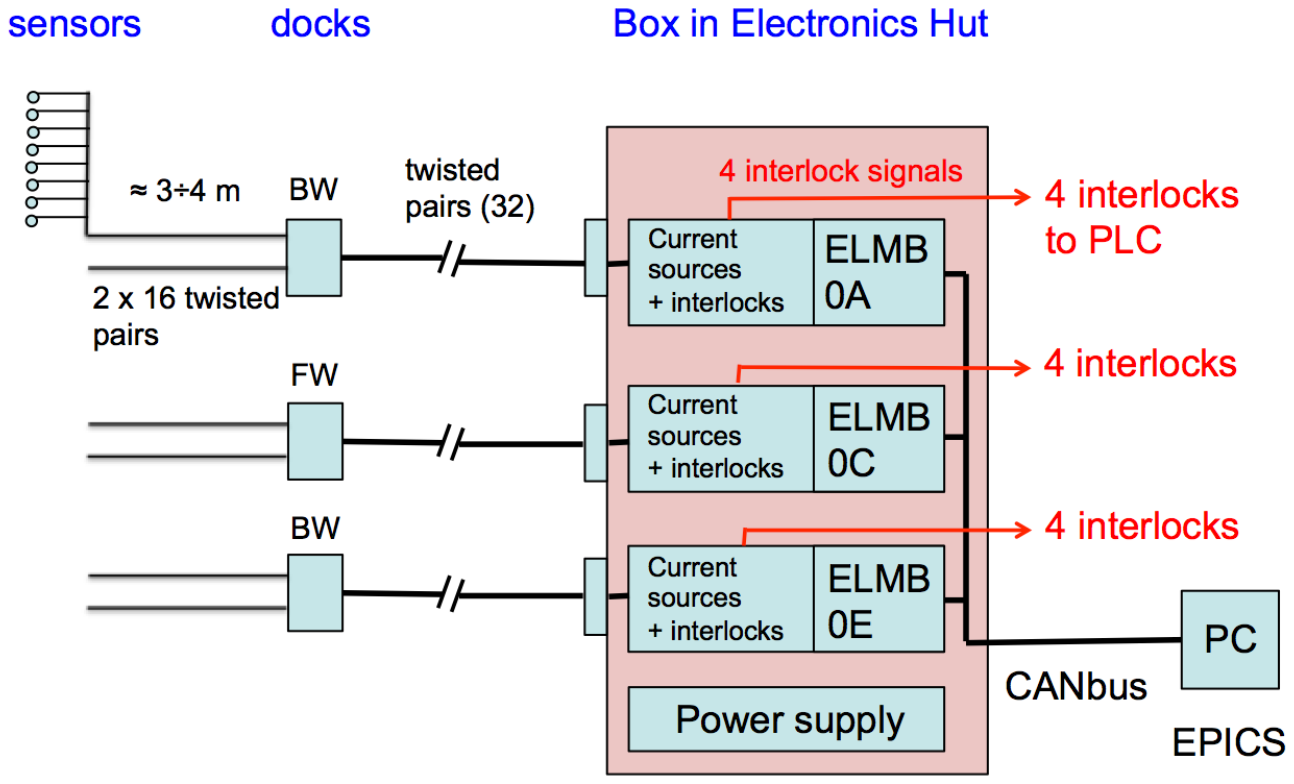}
		\caption{Block diagram of the readout of the \acrshort{NTC} system: one readout unit hosts three motherboards with one \acrshort{ELMB} card and four interlock lines each, as explained in the text (see also section~\ref{sec:hard_interlock}).}
		\label{fig:C2_temp_mon_NTC_connection}
	\end{center}
\end{figure}

\acrshort{NTC} thermistors have accuracy better than \qty{1}{\degreeCelsius}. They require only two wires per sensor, since their resistance (typically $100$~k$\Omega$) can be chosen to be much higher than that of connecting cables. The readout of NTC sensors is performed by a custom-designed system, built by INFN Trieste. It is based on the \acrfull{ELMB}, designed and used in large numbers at CERN~\cite{Hallgren:530675}.
Each ELMB hosts a CANbus-connected~\citeCompany{CANbus} processor and an ADC multiplexed to 64 input channels. To optimize accuracy, half of them are connected to reference resistors; the remaining 32 input channels measure the voltage drop across the NTC thermistors. Each ELMB is mounted on a motherboard providing the current sources and hosting  comparators with trimmer-adjustable thresholds, OR-ed in groups of eight, which provide four fast interlock signal outputs for $32$ input channels. These signals activate the hardware interlock system (section~\ref{sec:hard_interlock}) when the preset threshold temperature is exceeded. Three motherboards with their ELMBs are mounted together with power supplies in one unit, providing $96$ input channels, and $12$ interlock outputs in total. The main features of the NTC readout system are shown in figure~\ref{fig:C2_temp_mon_NTC_connection}.

\lvlthreesection{FOS system design}

The \acrshort{FOS} system measures the temperature of the SVD \gls{ladder}s, close to the APV25 chips. For \gls{layer}s from $4$ to $6$, optical fibres are inserted in channels prepared in the Airex foam, with several sensors along the ladder in positions roughly corresponding to the presence of the front-end ICs on the \glspl{origamiboard}  (section~\ref{sec:C2_origamiboards}). Each sensor on a fibre is realized by a \acrfull{FBG} with a characteristic reflection wavelength, mechanically shifted by temperature-sensitive acrylate coating.
%~\cite{ref:C2_temp_mon_Bragg}
The temperature is derived by the measurement of shifts in reflection wavelength, with an accuracy better than \qty{0.6}{\degreeCelsius}. 

In total $260$ sensors are distributed on $44$ fibres with nominal wavelength at \qty{20}{\degreeCelsius} between $1522$ and $1573$~nm, distributed between the PXD ($12$ sensors), SVD layers and outer cover ($248$ sensors in total). For all contiguous \acrshort{FBG} sensors, the separation in nominal wavelength is in steps of $3$~nm. For cross-calibration, a subset of eight fibres have an additional sensor located outside the Airex foam channel, thermally coupled with an NTC thermistor. 

\begin{figure}[htbp]
\begin{center}
\includegraphics[width=10cm]{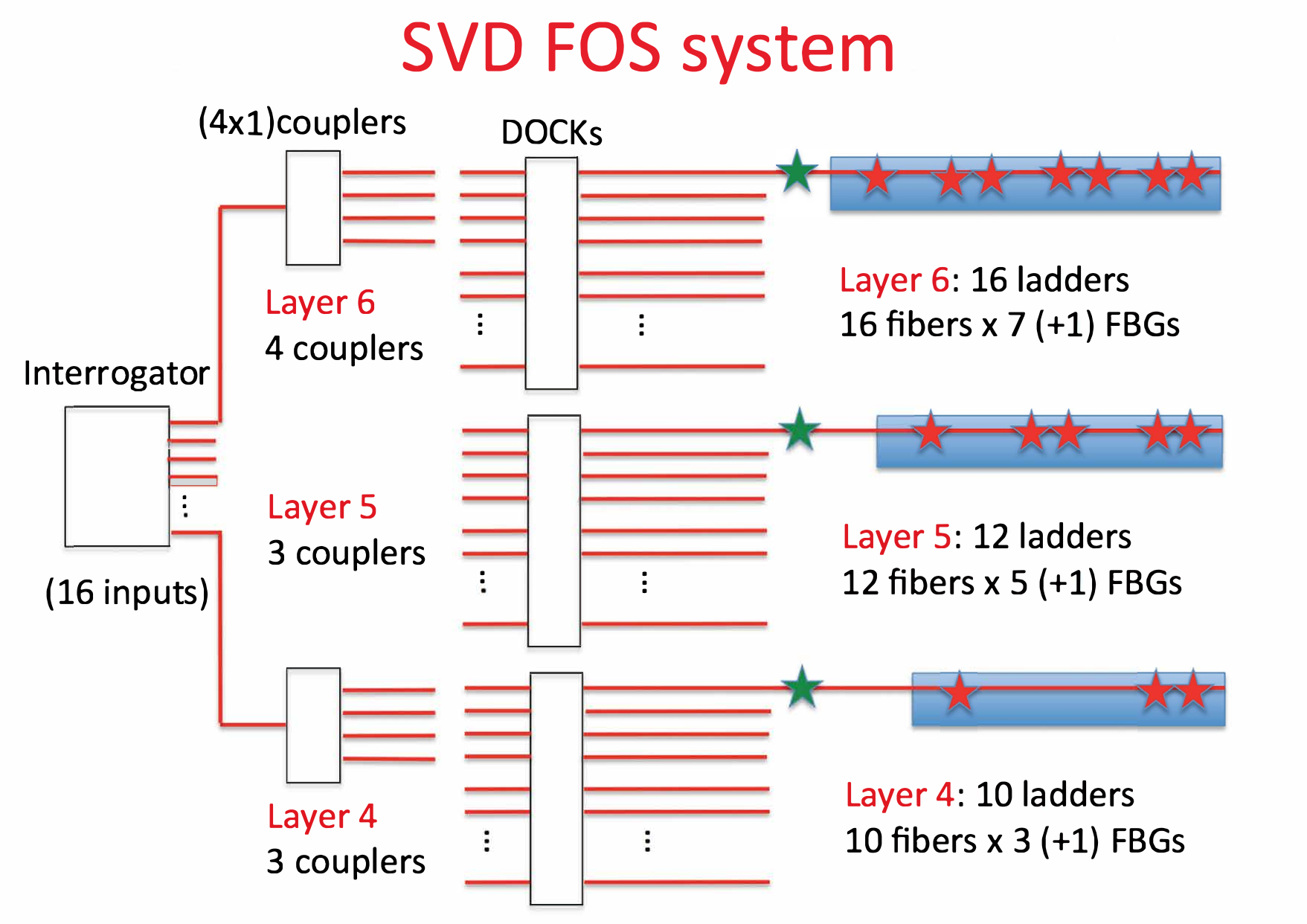} \\
\caption{Readout scheme of \acrshort{FBG} sensors for SVD: each \gls{ladder} in \gls{layer}s 4, 5, and 6 hosts one fibre with the indicated number of sensors. Groups of four fibres are connected via ($4 \times 1$) couplers to one of the $16$ inputs of the interrogator.}
\label{fig:C2_FBG_sensors_readout}
\end{center}
\end{figure}

The readout of \acrshort{FBG} sensors is performed by an ``optical sensing interrogator''~\citeCompany{ref_FOS_interrogator}, featuring a high-power, low-noise swept-wavelength laser. The interrogator has $16$ input channels and an internal $4 \times 4$ multiplexer; external multiplexing is provided by $16$ $1 \times 4$ splitter/coupler modules. The association of fibres with external multiplexers and interrogator input channels is optimized to exclude overlapping nominal \acrshort{FBG} wavelenghts at the same input. The overall readout scheme is sketched in Figure~\ref{fig:C2_FBG_sensors_readout}.

For calibrations and debugging the Micron Optics ENLIGHT sensing analysis software~\citeCompany{ref:Enlight}
provides a single suite of tools for data acquisition, computation and analysis of optical sensor networks.
%~\cite{ref:Interrogator}
A custom Micron Optics protocol via Ethernet is used for readout by the SVD run and slow control software (section~\ref{sec:C4_OnlineSW}).

\lvlthreesection{NTC system construction, calibration and installation}

The \acrshort{NTC} thermistors, with \qty{10}{\kohm} resistance at \qty{20}{\degreeCelsius}, are radiation-resistant and halogen-free~\citeCompany{ref_NTC_data_sheet}. Their calibration is obtained by measuring their resistance at  \qtylist{-30; 10; 50}{\degreeCelsius}, and fitting the Steinhart-Hart coefficients~\citeCompany{ref:Steinhart}. The obtained accuracy, better than $\pm$ \qty{1}{\degreeCelsius}, is sufficient for our purpose. 
At each step of SVD cooling pipes mounting procedure (section~\ref{sec:C3_cooling_pipes_mount}), the relevant thermistors and their cables are checked to ensure they are well connected and properly working. Halogen-free flat twisted-pair cables (SVD cables: section~\ref{sec:C2_cables}) were used to connect the sensors to intermediate patch panels at the VXD \gls{DOCK}s and finally to the readout unit in the electronics hut (\gls{e-hut}). Two spare readout units were built and used in parallel during construction and commissioning: one for the ladder-mount setup and the other for the SVD commissioning setup.

\lvlthreesection{FOS calibration and installation}

Each \acrshort{FOS} was calibrated in a climatic chamber by determining the coefficients of a polynomial fit to the measured temperature-wavelength relation.

During the SVD \gls{ladder} mount for \gls{layer}s 4, 5 and 6 (section~\ref{sec:C3_single_ladder_mount}) one fiber was inserted in each ladder
%(Figure~\ref{fig:C3_NTC_FOS}), 
and checked for proper functioning. Particular care was required to handle the fibers and store their excess length on rolls next to the detector, when they were temporarily connected to the interrogation device, which was located outside the clean room. All tests were repeated during the assembly and commissioning of the two SVD halves. After the installation of SVD in the final position, the fibers were laid down in cable trenches and connected to the multiplexers and the interrogator in the \gls{e-hut}.

\lvltwosection{Humidity Monitoring}
{\label{sec:humid_mon}}
%\editor{Livio}

\begin{figure}[htbp]
\begin{center}
\includegraphics[width=7cm]{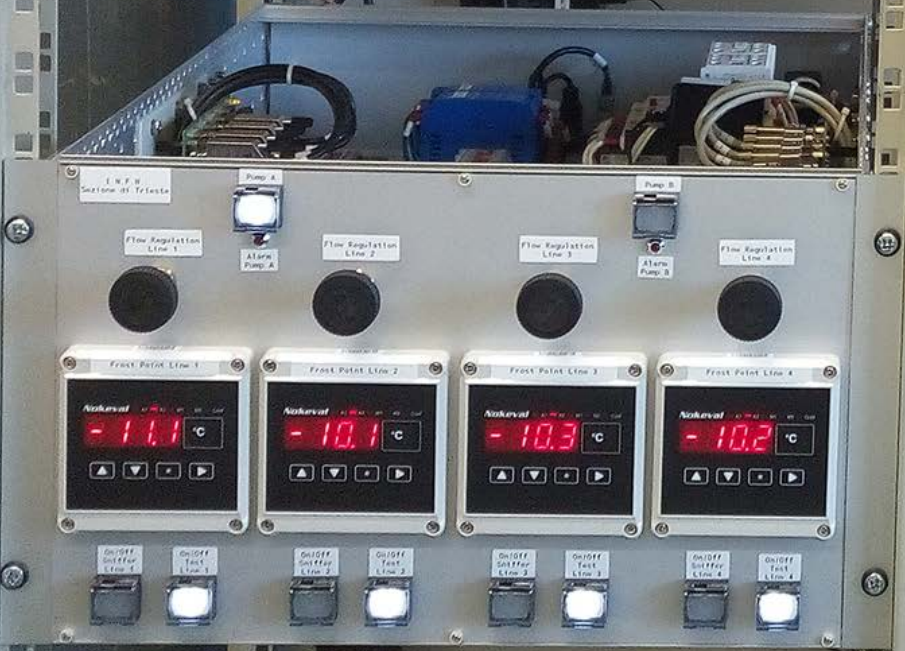}
\includegraphics[width=7cm]{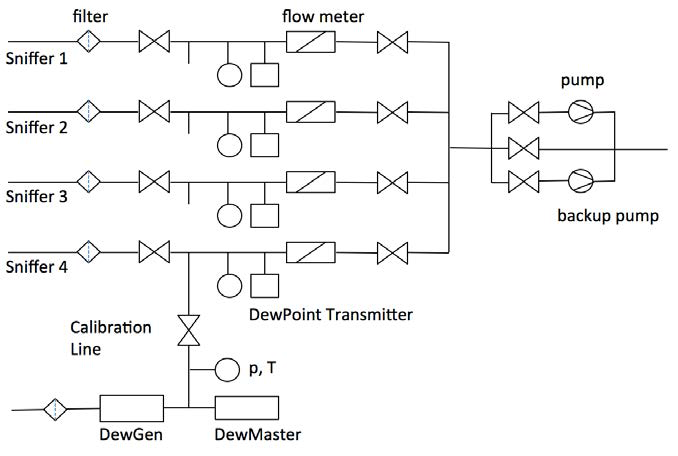}
\caption{Simplified block diagram of the VXD humidity monitoring system, showing the four external units, each equipped with a dew point transmitter, a mass flow meter and a pressure transmitter.}
\label{fig:C2_humidity_sniffers}
\end{center}
\end{figure}

The whole volume of the \acrshort{VXD} must be kept at low humidity by a flux of dry Nitrogen, to prevent water vapour condensation and ice formation on the cooling pipes and other cold components. The atmosphere inside the VXD volume needs to be kept at a dew point below \qty{-30}{\degc}, whereas the measured dew point temperature is about \qty{-80}{\degc}. The miniature humidity sensors available on the market are not sufficiently radiation hard and their replacement during operation is not feasible. Therefore to monitor the dew point, samples of the circulating gas are extracted from the dry volume and channeled to humidity monitor units located in the \gls{e-hut}, consisting of a dew point transmitter, a mass flow meter and a pressure transmitter, as shown in Figure~\ref{fig:C2_humidity_sniffers}.

\lvlthreesection{Design}
The system collects six 
%6/4mm \textcolor{red}{(steel/ copper)} 
input pipes from the forward and backward sides of the cold and warm dry volumes of the VXD. 

Four of these lines are each connected to one humidity measurement unit. Two electro-valves allow the connection of each fully equipped line to the VXD dry volumes or to a calibration gas source. The four lines merge into a vacuum pump allowing the gas suction from the detector dry volumes. The suction flow of each line can be precisely set by means of manual flow regulators using the mass flow meter information. An additional vacuum pump grants redundancy: interchange of pumps is controlled by automatic motor protection switches. An alarm LED in the front panel warns if a pump switching occurs.

The two remaining lines, not connected to the suction pump, are connected to a pressure transmitter, to monitor the dry volume pressure. 

\lvlthreesection{Construction and installation}

The humidity monitoring system was designed, built, and tested at INFN Trieste. It was then shipped to KEK and installed in a rack in the electronics hut (\gls{e-hut}). The required pipes were installed between VXD dry volume and the e-hut.
%Six copper pipes were routed from the VXD dry volumes to the \gls{e-hut}, and connected to the monitoring system. Four pipes were connected to the humidity sniffers, to monitor the dew point. Two pipes were used to measure the pressure inside the dry volume. 
The readout electronics, integrated in the system, was interfaced to the SVD environmental monitor server via USB connections. The system was fully tested and commissioned before the beginning of the physics run.

The system was originally equipped with miniature dew point transmitters Vaisala DMT143~\citeCompany{ref:vaisala}.
This device is able to measure the dew point in the range  \qtyrange[retain-explicit-plus]{-80}{+20}{\degreeCelsius} with an accuracy in air and nitrogen of $\pm \qty{2}{\degreeCelsius}$; it was well matched to the expected a dew point of about \qtyrange{-60}{-70}{\degreeCelsius}, but the level of humidity turned out to be lower, with a dew point below \qty{-80}{\degreeCelsius}. As a consequence, after a few months of operation the sensors frequently went into a self-calibration state, resulting in error conditions, and were subsequently substituted  with miniature dew point transmitters Vaisala DMT152~\citeCompany{ref:vaisala},
with a measurement range  \qtyrange[retain-explicit-plus]{-100}{0}{\degreeCelsius}   and the same accuracy. These sensors need to be re-calibrated every two years. Short period re-calibration is performed by a built-in software stored in the sensors internal memory. 

The DMT152 provides both analogue ($4-20$~mA signal) and digital (RS485) readout. The most important information, the dew point readings, are sent to both front panel displays and serial-over-USB lines. The analog current outputs of the Vaisala sensors are connected to loop-powered seven-segment displays of type Nokeval 302~\citeCompany{ref:nokeval}, located in the front panel. Those displays compare the measured values to a programmable range and provide two relay outputs for upper and lower alarms, respectively, whenever the measured value is out of range. The alarm relays are connected to the VXD hardware interlock system. As DMT143/DMT152  communicate digitally via the RS485 bus, a 4-ports \acrshort{FTDI} USB-COM485-Plus4 interface was used to connect the sensors to the readout server via USB interface.

In addition, a similar, extended system,  
with increased redundancy for safe long-term operation, was designed and built at INFN Trieste and is ready for installation. It features one additional spare measurement line, for a total of five. Each line has redundant dew point measurements with two different sensors: an Alpha Moisture Systems Model AMT~\citeCompany{ref:alpha_moisture} (\qtyrange[retain-explicit-plus]{-120}{+20}{\degreeCelsius} range) and a 
Michell Systems Easydew~\citeCompany{ref:michell}
(\qtyrange[retain-explicit-plus]{-110}{+20}{\degreeCelsius}). Mass-flow meters and pressure sensors are the same type as in the other system. Each line is managed by a dedicated processor board, with both USB and Ethernet readout for an improved interfacing. The system will be installed at the first shutdown opportunity. 

\lvltwosection{Hardwired Interlock System}
{\label{sec:hard_interlock}}
%\editor{Livio}

A fast but orderly shutdown of the PXD and SVD power supplies should occur on a number of critical conditions, triggered by hardwired signals, independent of the slow control network and software. 

\lvlthreesection{Design}

The \acrfull{VLHI} must be both reliable and flexible, to accommodate different digital and analog input signals, and evolving interlock conditions. An optimal solution, adopted by several experiments, is based on industry-standard programmable logic controllers (\acrshort{PLC}s), that are reliable, programmable, and easily expandable with both digital and analog input/output modules. For the VXD, the requirement of \acrshort{EPICS} compatibility led to the choice of a Schneider M340 \acrshort{PLC}~\citeCompany{ref:schneider_plc}. 
The VLHI receives input signals and alarms from::
\begin{itemize}
    \item $12$ groups of NTC temperature sensors;
    \item $4$ dew point sensors and their flowmeters and pressure sensors;
    \item the water cooling system of the dock chiller (water leaks, water flow); 
    \item the CO$_{2}$ cooling system;
    \item the central \belletwo\ interlock system (solenoid, water leak, \gls{e-hut} power, environmental monitors).
\end{itemize}
The firmware of the PLC implements a "heartbeat counter", which is incremented continuously. This counter has to be reset by the slow control software within a given maximum period. Exceeding this timeout is counted as a software error and an interlock is triggered.

The following interlock output signals are triggered due to a combination of the input signals:
\begin{itemize}
    \item SVD low voltage (LV) and high voltage (HV) power supplies;
    \item PXD power supplies;
    \item dock chiller and CO$_{2}$ cooling system; 
    \item central \belletwo\ interlock system.
\end{itemize}

\lvlthreesection{Construction and installation}

The \acrshort{VLHI} hardware was assembled and the PLC programmed at INFN Trieste. A test box was used to simulate all the input signals with the agreed protocols and their electrical properties. The \acrshort{PLC} program was thoroughly tested and debugged in the laboratory, before the shipment to KEK. All tests were then repeated for each input and output, in the VXD commissioning phase before the start of physics data taking.

%% Move the cooling plant here.

%\subfile{03_03-01_Cooling_Plant}
%%%% COOLING PLANT SECTION
%
% moved to separate file to facilitate inclusion in different chapters
%
\lvltwosection{Cooling Plant and \texorpdfstring{\COtwo}{CO2} Distribution}

%\lvlthreesection{Cooling plant and \COtwo\ distribution}
\label{sec:C3_cooling_plant}
% \editor{Toru Tsuboyama}
%\section{\COtwo\ cooling system and its operation}

\begin{figure}[htbp]
\begin{center}
\includegraphics[width=0.975 \linewidth]{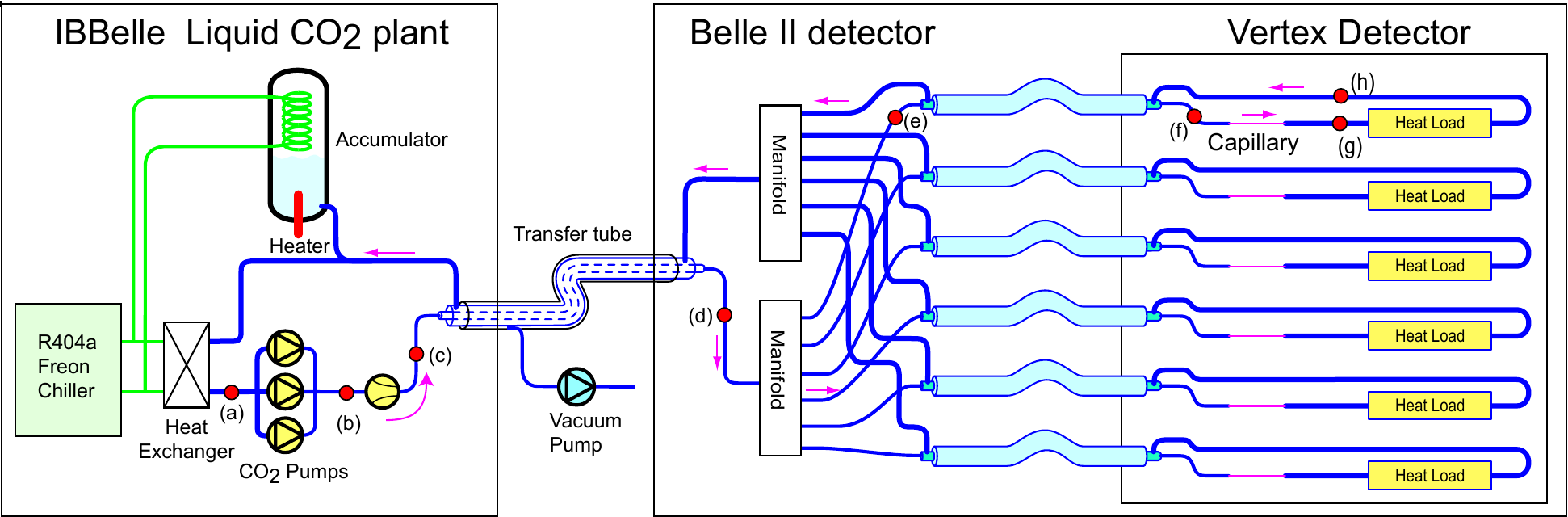} % New version with letters
\caption{A schematic of IBBelle and \belletwo{} VXD cooling system. The approximate length of the pipes is \qty{34}{\m} from IBBelle to the distribution manifold and  \qty{9}{\m} from the  manifold to the VXD.}
\label{fig:C3-TT-CO2-fig1}
\end{center}
\end{figure}

\begin{figure}[htbp]
\begin{center}
\includegraphics[width=0.9 \linewidth]{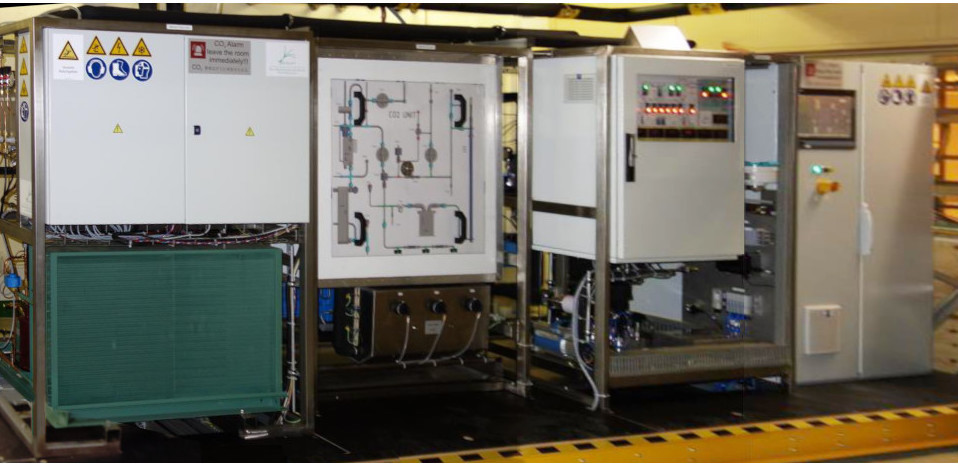}
\caption{The IBBelle cooling plant. The four chassis from left to right are for the R404A refrigerator, the \COtwo\ pumps and tubing, the accumulator, and the control unit. 
In total, the cooling plant occupies a \qtyproduct[product-units = power]{6 x 1.5 x  2 }{\m} volume. }
\label{fig:C3-TT-CO2-fig1b}
\end{center}
\end{figure}

%The SVD electronics produces about $700$~W that must be removed from the detector volume. In the initial conceptual design, water cooling was foreseen, since there the sensors do not require cold operation. Given the more stringent requirements coming from the PXD,  a two-phase \COtwo\ cooling system was finally chosen to remove heat produced in  the VXD, providing advantages in terms of pipe dimension, material budget, cooling efficiency. 
A schematic of the cooling system of \belletwo{} VXD, serving both the PXD and SVD (whose design and motivations are discussed in section~\ref{sec:C2_coolling_plant}), is illustrated in figure~\ref{fig:C3-TT-CO2-fig1}.
The liquid \COtwo\ plant, IBBelle, was originally designed for the IBL pixel detector for the ATLAS experiment \cite{Verlaat:2017yzg}. A photo of the plant is shown in figure~\ref{fig:C3-TT-CO2-fig1b}. 

IBBelle is based on the 2PACL (2-Phase Accumulator Controlled Loop)~\cite{2PACL:1, 2PACL:2} concept, and consists of a chiller, a heat exchanger, liquid \COtwo\ pumps,
and an accumulator. The R404A refrigerant, which can be operated below \qty{-50}{\degreeCelsius}, is chosen for the chiller. The refrigerant flows to the heat exchanger as well as to the cooling coil in the accumulator.
The liquid \COtwo\ is pressed by the membrane pumps and supplied to the \belletwo{} detector through a \COtwo\ transfer tube. 
After absorbing heat at the detector, the \COtwo\ returns from the \belletwo{} detector and goes through the heat exchanger again to start a new cycle.
The accumulator is a stainless-steel reservoir tank of \COtwo; it has a volume of \qty{50}{\liter} and includes the cooling coils and electric heaters. 
The cooling and heating power determines the temperature, pressure and amount of liquid \COtwo\ in the accumulator. Consequently, the accumulator operation determines the temperature of \COtwo\ at the heat load. It also determines the stability of the whole system. Therefore, the accumulator requires a dedicated control and monitoring system.

The connection between IBBelle and the \belletwo{} detector is done by using a triple wall coaxial transfer tube. The \COtwo\ from IBBelle goes to the inner tube and the returning \COtwo\ flows through the intermediate tube. The outer tube is evacuated for thermal insulation. On the detector side, the \COtwo\ is distributed by using a manifold. There are 12 cooling circuits: eight for the SVD and four for the PXD. Before the VXD heat load a capillary is inserted, reducing the pressure  of the liquid, as discussed in the next section. The \COtwo\ after the heat load flows back to the \COtwo\ plant through the vacuum isolated transfer tubes up to the manifold, where the individual lines are joined and then further to the accumulator of the IBBelle plant.

In order to avoid water condensation, the regions that face the cooled \COtwo\ are sealed with insulation foam and filled with nitrogen gas coming from the cold evaporator tank. The dew point is normally kept at about \qty{-80}{\degreeCelsius}. 

\begin{figure}[htbp]
\begin{center}
\includegraphics[width=0.9 \linewidth]{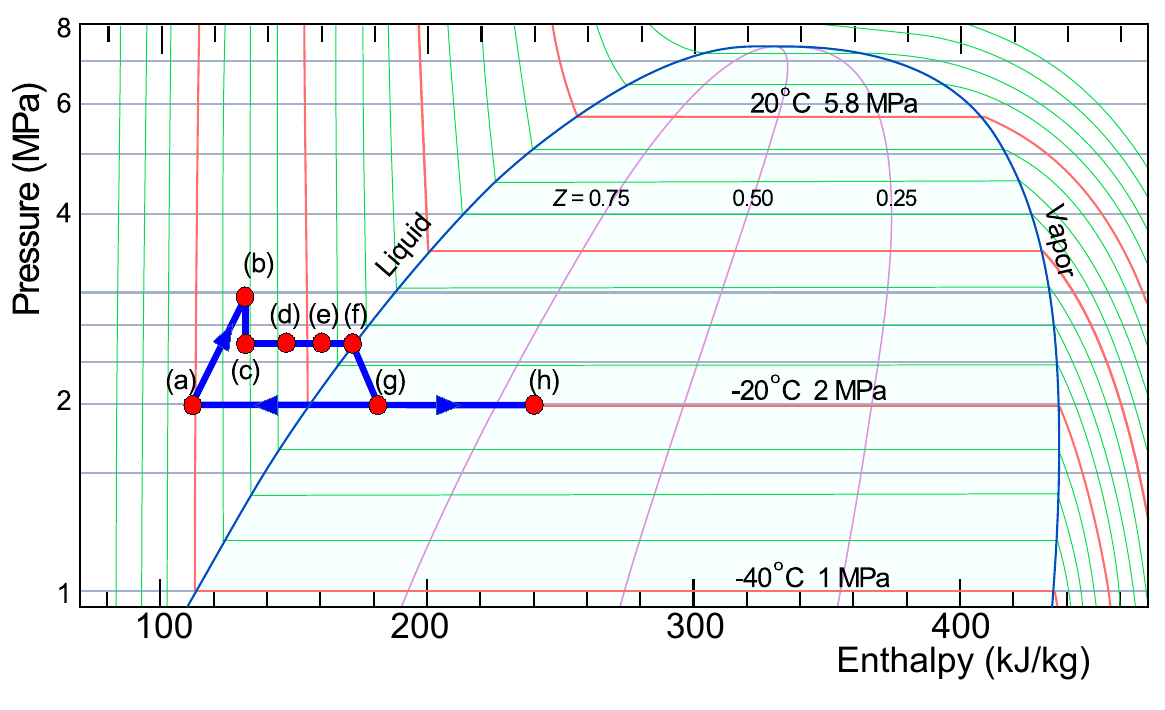} % new version removing point (i)
\caption{The phase diagram of \COtwo. The horizontal axis shows the enthalpy, the vertical axis shows the pressure. The red and green lines represent  constant-temperature transformations. The purple lines correspond to specific values of $Z$, the fraction of liquid component in the dual-phase mode.}
\label{fig:C3-TT-CO2-fig2}
\end{center}
\end{figure}

\paragraph{Operation of the \COtwo\ system}
Before the cooling operation, the system is evacuated and then filled with liquid \COtwo\ at room temperature. 
At \qty{30}{\degreeCelsius} the pressure of \COtwo\ reaches \qty{7}{\MPa}. With a 1.5 safety factor, the design pressure of the system is \qty{11}{\MPa} and the acceptance tests were performed with even higher pressure (\qty{16}{\MPa}). The cold operation starts when the R404A chiller is activated.
Figure \ref{fig:C3-TT-CO2-fig2} shows a pressure-enthalpy diagram of the \COtwo. 
The region between the liquid and vapor curves is the dual-phase region, where heat can be removed from the heat load at constant temperature and pressure. 
% FF: I think  the critical temperature and triple point are not really relevant here.
% The critical temperature is $T_{\rm{C}}$=31.1\degrees C. 
%Above $T_{\rm{C}}$, the liquid and vapor states are indistinguishable. The triple point is at -51\degrees C and 0.5~MPa. Below the triple point pressure, the \COtwo\ gas makes a phase transition directly to solid \COtwo. 

The blue line with the red points shows the IBBelle cycle after the operation is stabilized, which can be briefly described as follows, 
with reference to the points in figures~\ref{fig:C3-TT-CO2-fig1} and \ref{fig:C3-TT-CO2-fig2}.\\
\inlinelist{a} The liquid \COtwo\ is first cooled with the refrigerant from the chiller, typically to \qty{-40}{\degreeCelsius}.\\
\inlinelist{b} The \COtwo\ is pressurized by the \COtwo\ pumps. \\
\inlinelist{c} The pressure of the \COtwo\ is reduced by an orifice located between points $(b)$ and $(c)$.
The volume of the \COtwo\ changes very little and the enthalpy is kept constant in the process.\\
\inlinelist{d} \inlinelist{e} \inlinelist{f}
The \COtwo\ flows in the transfer tube with constant pressure. The enthalpy and temperature are increasing because of friction with the tube.
The \COtwo\ lines are branched with the manifold $(d)$. From the manifold to the VXD, triple wall coaxial tubes with smaller diameter are used. 
While the \COtwo\ flows through the transfer tube, the returning \COtwo\ removes the heat of the going \COtwo\ to minimize the increase of enthalpy. \\
\inlinelist{g}
A capillary tube is inserted just before the heat load, between points $(f)$ and $(g)$, to reduce the \COtwo\ pressure and temperature to the target  values of \qty{-20}{\degreeCelsius}  and  \qty{2}{\MPa}. The pressure is determined by the return line connected to the accumulator.\\
\inlinelist{h}
The coolant removes heat from the detector and the enthalpy increases, as well as the fraction of vapor, that increases to about 30\% as shown in figure~\ref{fig:C3-TT-CO2-fig2} by the purple lines.
The vapor fraction should be kept below 50\% for an efficient heat removal. The returning \COtwo\ flowing through the outer part of the triple wall coaxial transfer tube gives enthalpy to the \COtwo\ flowing in the opposite direction in the inner part.\\
%
%\inlinelist{i}
%At this point the fraction of the liquid component is almost zero and the cooling capability is lost.
After the \COtwo\ returns to the \COtwo\ plant, it goes through the heat exchanger to be cooled down before a new cycle.

The \COtwo\ circulation system is connected to the accumulator. The refrigerant flow rate and the heater power are controlled to stabilize the operation and total amount of \COtwo\ in the whole system.

\lvlonesection{Detector Control and Data Acquisition}
\editor{Giulia - total of < 10-11 pages}
{\label{sec:software_online}}
%\rem{Introduction to be adjusted to the new structure}

The detector control and data acquisition software is responsible to ensure proper operation of the \belletwo\ SVD during physics data taking and local calibration runs, and acts as an interface between the SVD hardware, the \belletwo\ run and slow control, the SuperKEKB control system as well as the operators. It provides software modules to configure, control, and monitor the power supplies, the front- and back-end data acquisition hardware, the environmental monitoring sensors, the detector safety and interlock unit, and the radiation monitoring system. 

In addition to these run and slow control tasks, it allows dedicated calibration runs to be performed, which are necessary to determine operational parameters like pedestals, noise levels, amplifier gain and calibration constants, which are then used in the back-end electronics for data processing in order to reduce the amount of data sent to \belletwo\ data acquisition. These parameters are also uploaded to the Conditions Database and finally used by the reconstruction software (see section~\ref{sec:software_offline})  to decode and analyse the data.

Section~\ref{sec:C4_OnlineSW} describes the grouping and functions of the SVD control software modules, integrated in the \belletwo\ framework and synchronized by state machines. Efficient and error-free operation of the detector relies on well designed operator interfaces.

The readout chain and data stream are described in section~\ref{sec:online_daq}, following the path from the SVD front-end APV25 chips to the back-end as well as to the \belletwo\ central DAQ. The characteristics of local runs and details of data processing are also reported.

\lvltwosection{Detector Control Software}
\label{sec:C4_OnlineSW}
\editor{Christian}
%\lvlthreesection{Overview}
In the \belletwo\ run and slow control systems the subdetectors are integrated into two frameworks, the custom Network Shared Memory~2 (NSM2)~\cite{Nakao:842639}, and the Experimental Physics and Industrial Control System (EPICS)~\citeCompany{epics}. NSM2 is used for the global \belletwo\ run and slow control (\belletwo\ \acrshort{RC}) as well as for the outer \belletwo\ subdetectors and the central data acquisition system (\belletwo\ DAQ). The VXD and in particular most modules of the SVD run and slow control system (SVD RC/SC) are implemented in EPICS. Only a few DAQ related modules of SVD RC/SC are within the NSM2 regime. The interface between these two platforms is realized by an NSM2-to-EPICS gateway, which translates process information and control requests between them.  
In EPICS, software modules are implemented as so-called input/output controllers (\acrshort{IOC}s), which either interface to hardware components or fulfil dedicated software tasks, such as data analysis or the computation of process parameters. Every IOC stores its data in process variables (PVs), which are shared among other IOCs via an Ethernet network. Each PV in the network has a unique name and can be accessed by other clients via a sequence of UDP and TCP requests. Thus, EPICS allows for a simple and standardized data exchange between the individual software modules and moreover provides an interface to standardized services like an archiver, an alarm system, a message logger and so on.

\begin{figure}[hbt]
    \centering
    \includegraphics[width=1\columnwidth,keepaspectratio]{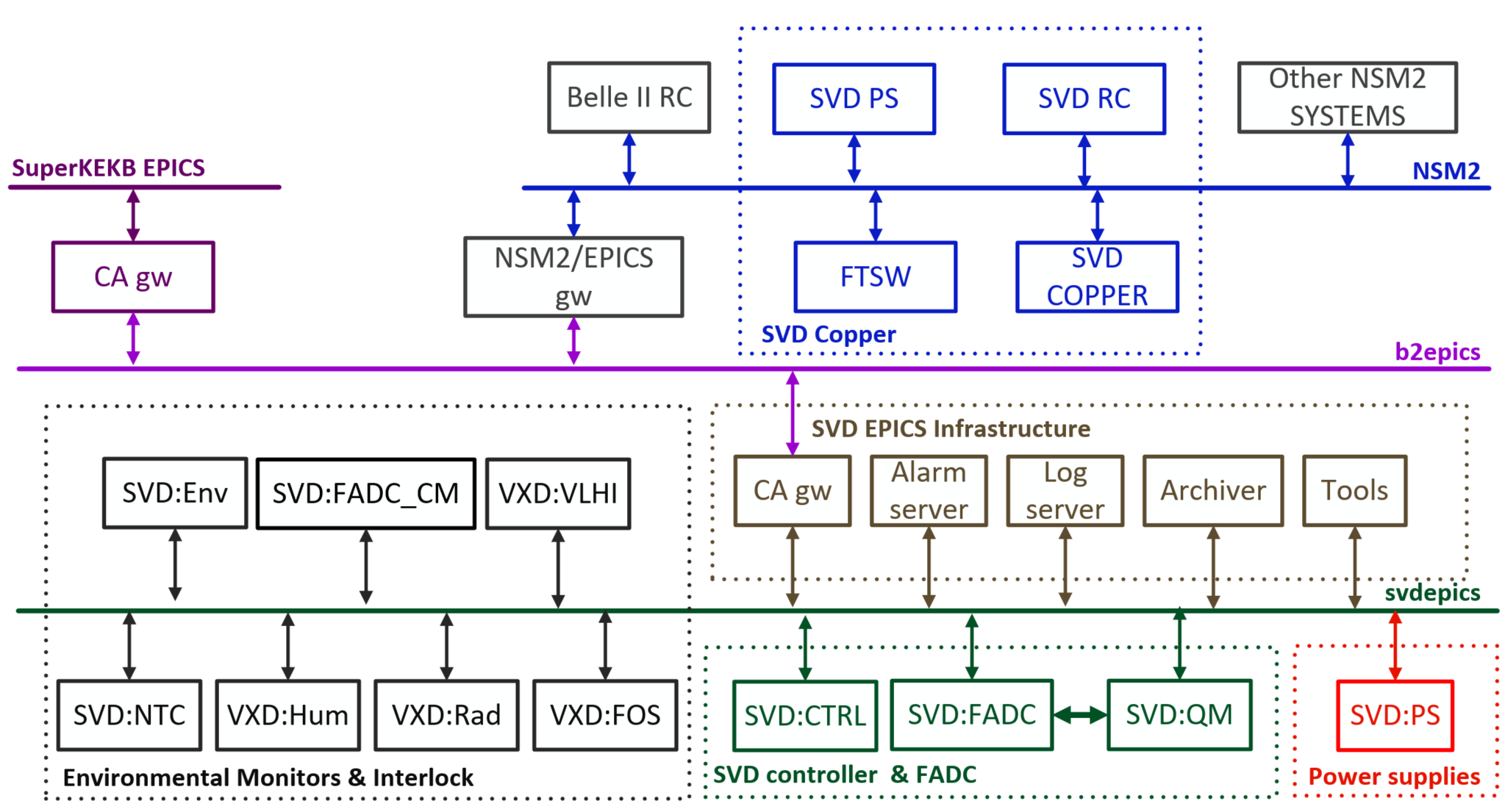}
    \caption{The block diagram of SVD I/O controllers.}
    \label{fig:C4_SC_diagramm}
\end{figure}

\lvlthreesection{SVD online software modules}
%\lvlthreesection{I/O Controllers}
\label{sec:C4_OnlineSW_modules}
The \belletwo\ SVD online software covers all software modules required to power, configure, control, and monitor the detector. As shown in figure~\ref{fig:C4_SC_diagramm}, the SVD online software is divided into four groups of I/O controllers according to their functionality:
\begin{itemize}
    \item SVD Controller \& FADC;
    \item Power Supplies;
    \item Environmental Monitors \& Interlock;
    \item SVD EPICS Infrastructure.
\end{itemize}

The {\bf SVD controller \& FADC} group contains all core IOCs required to operate the SVD and control data taking. Those are the SVD run controller (\ioct{SVD:CTRL}), the IOC to configure, control, and monitor the FADC system described in section~\ref{sec:C2_offdet_elec} (\ioct{SVD:FADC}) and the online low-level data quality monitoring (\ioct{SVD:QM}). \ioct{SVD:FADC} is also responsible to acquire spy data (as described in~\ref{sec:C2_FADC_system}) during physics runs, which are forwarded to \ioct{SVD:QM} where they are processed and evaluated. Since the amount of these data is rather large, a direct TCP socket connection is established between \ioct{SVD:FADC} and \ioct{SVD:QM} to avoid large data traffic over the SVD EPICS network. The control of the low- and high-voltage power supplies is performed by the \ioct{SVD:PS} IOC, which is the only member of the {\bf power supplies} group in the block diagram. \ioct{SVD:PS} directly communicates with the CAEN power supplies ensuring that the correct voltages are set and all currents are within limits.

The {\bf Environmental Monitors \& Interlock} group comprises all IOCs used to configure, monitor, and acquire data from the environmental monitoring sub-systems, like temperature, humidity, and radiation dose monitoring. The detailed functionality of these IOCs is listed below.
\begin{itemize}
	\item \ioct{SVD:Env}: master ENV IOC that is used to collect the state of the subordinated ENV system and propagates errors and software interlock to \ioct{SVD:PS} and \ioct{SVD:CTRL}; 
	\item \ioct{SVD:NTC}:	monitors the temperatures of cooling pipes and end rings, measured by NTC resistors;
	\item \ioct{VXD:FOS}:	measures the temperatures of SVD and PXD ladders with fiber optic sensors (FOS);
	\item \ioct{VXD:Hum}:	dew point monitoring of the VXD volume;
	\item \ioct{VXD:Rad}: 	radiation dose monitoring with diamond sensors;
	\item \ioct{VXD:VLHI}:	monitoring of the state of the VXD hardwired interlock system (VLHI).
\end{itemize}
Most of the ENV IOCs have prefix VXD, which means that they are responsible for systems jointly used for SVD and PXD. Those with prefix SVD are responsible for sub-systems dedicated to SVD alone.
In the fourth group, the {\bf SVD EPICS Infrastructure} group, all common services like the archiver, the alarm system, the message logger and the gateway between the SVD EPICS network and the \belletwo\ EPICS network are collected. So far, available standard EPICS tools are used for those systems.
%The core IOCs required to operate the SVD and control data taking are grouped in 

\begin{figure}[hbt]
    \centering
    \includegraphics[width=.7\columnwidth,keepaspectratio]{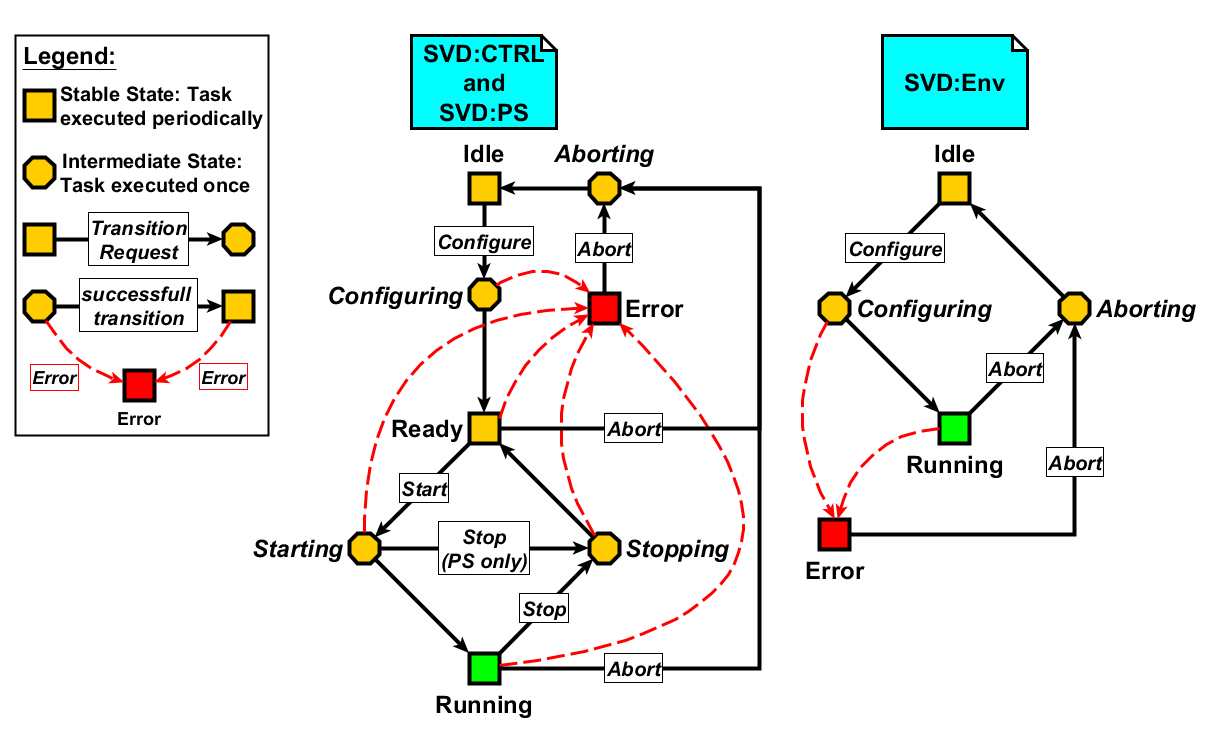}
    \caption{State Machines of SVD Run Control~\cite{irmler2020vci}.}
    \label{fig:C4_state_machine}
\end{figure}

\lvlthreesection{State machines}
The IOCs of the SVD run and slow control are synchronized by a finite state machine (FSM), with four static states (\textit{Idle, Ready, Running,} and \textit{Error}) and four transitional states (\textit{Configuring, Starting, Stopping,} and \textit{Aborting}), which are adopted during the transition between the static states. The state diagram of the FSM is shown in  figure~\ref{fig:C4_state_machine}. From each state there are typically very few possible requests to initiate the transition to the next state. In \textit{Idle} the IOC does not have any information about the hardware of the sub-system, and waits for a \textit{Configure} request to switch to the transitional state \textit{Configuring} as well as to execute all tasks required to initialize and configure the sub-system. If this is completed successfully, the FSM changes to the state \textit{Ready}, in which the system is configured and waits for the \textit{Start} request in order to initiate the transition from \textit{Ready} to \textit{Running}, in which the system is ready for data taking. It stays there until either a \textit{Stop} or \textit{Abort} request is received. The exact meaning of the states for the individual run control IOCs is listed in table~\ref{tab:C4_fsm_states}. 
Whenever an error occurs, the FSM goes into \textit{Error} state, where it remains until the confirmation by the operator. In any state only valid requests are accepted and executed, any invalid request, e.g. \textit{Start} in state \textit{Idle}, is ignored. 

\begin{table}[hbt]
	\begin{center}
		\begin{tabular}{lccc}
			\toprule
			\small \textbf{State} & \small \textbf{\ioct{SVD:PS}} & \small \textbf{\ioct{SVD:FADC}} & \small \textbf{\ioct{SVD:QM}}\\
			\midrule
			\small \textbf{Idle} & \small LV off, HV off & \small off & \small off\\
%			\hline
			\small \textbf{Ready} & \small LV on, HV on (stand-by voltage) & \small configured & \small configured\\
%			\hline
			\small \textbf{Running} & \small LV on, HV on (peak voltage) & \small running, & \small analyis\\
			\small  & \small & \small data taking & \small ongoing\\
			\bottomrule
		\end{tabular}
		\caption{Meaning of states for each run control IOC.}
		\label{tab:C4_fsm_states}
	\end{center}
\end{table}

For the environmental monitoring IOCs a simpler FSM, as shown in the right diagram of figure~\ref{fig:C4_state_machine}, consisting of two stable (\textit{Idle, Running}) and two transitional states (\textit{Configuring, Aborting}), is chosen. The reason is that these IOCs are mainly used for monitoring purposes,  staying in the \textit{Running} state most of the time, and are not directly controlled by the \belletwo\ RC. They provide data even when the \belletwo\ detector and thus also the SVD is off.
On the other hand, the SVD run control IOCs are controlled by the \belletwo\ \acrshort{RC} during global data taking and thus have to follow the run states of the whole \belletwo\ detector. 

The interfacing between \belletwo\ RC and the SVD slow control is realized via a set of dedicated PVs. To initiate a state transition, the \belletwo\ RC sets the so-called request PV to the requested transition, e.g. \textit{Configure}. The SVD IOC reads \textit{Request} PV and sets its state PV to the corresponding state, e.g. \textit{Configuring}. Once the transition is completed the request PV is set to \textit{Processed}, which indicates that the IOC has finished the transition and reached a new stable state. If the transition is successful, the state PV is set to the new state, e.g. \textit{Ready}. In the case of a failure, the state PV is either set to \textit{Error}, if the error is unrecoverable, or back to its previous state, if a safe fallback is possible. The \belletwo\ RC regularly polls these two PVs in order to verify that the SVD slow control has finished the transition and in which state it finally arrived.

\lvlthreesection{Operator interfaces}
%\textcolor{red}{TO BE WRITTEN}
%OPIs decoupled from IOC, usage CSS, screenshot ctrl OPI and one ENV OPI

The operator interfaces (\acrshort{OPI}) of the SVD run and slow control are implemented in Control System Studio (CSS) \citeCompany{css}. CSS is a collection of tools to monitor and operate large-scale control systems. It directly interacts with the underlying EPICS network to read and set the PV values. Apart from graphical OPIs it provides a data browser, which allows to retrieve from an archiver and generate history plots of each archived PV value.
\begin{figure}[hbt]
    \centering
    \includegraphics[width=.9\columnwidth,keepaspectratio]{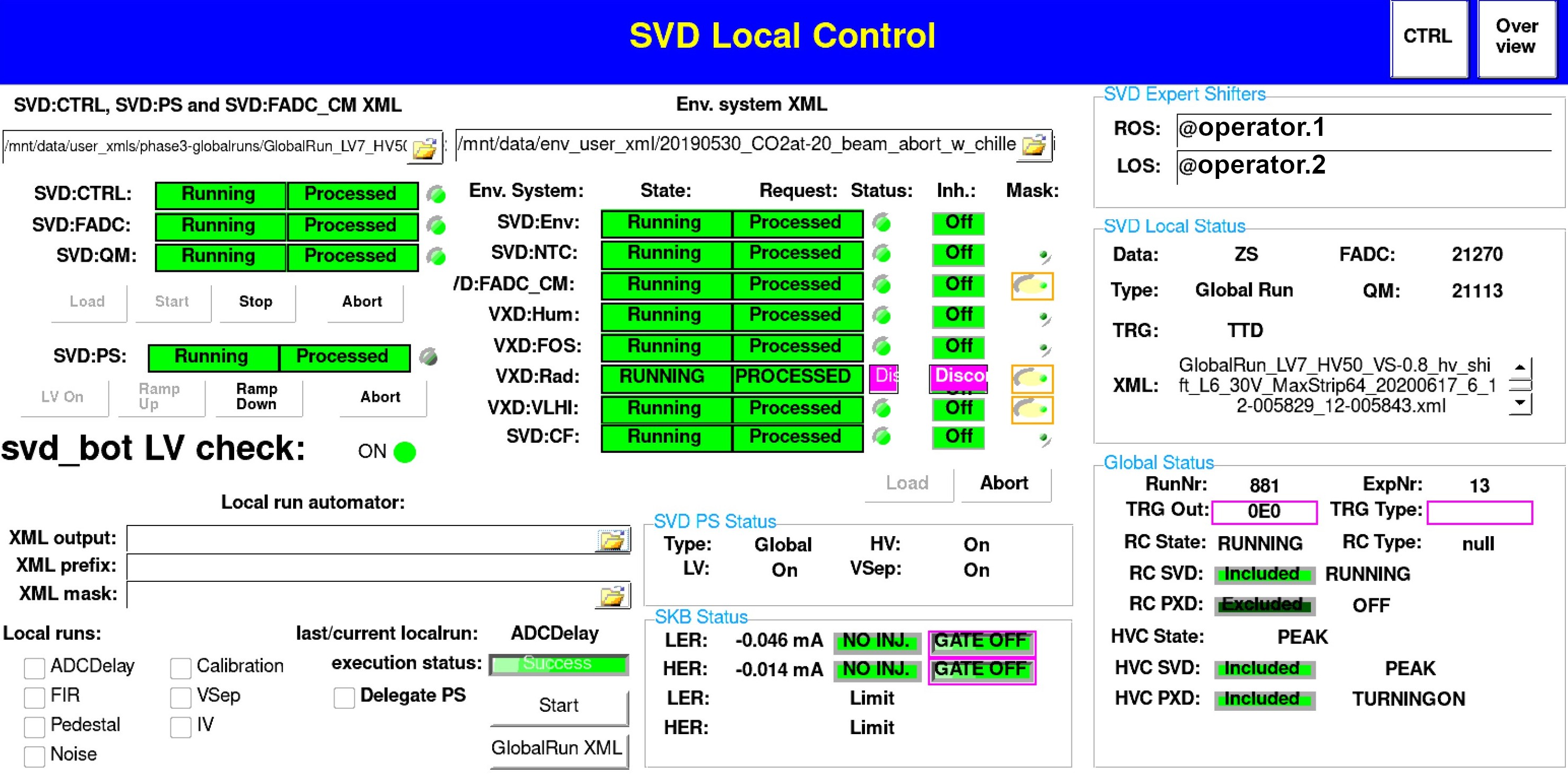}
    \caption{SVD control operator interface.}
    \label{fig:C4_ctrl_opi}
\end{figure}
The main OPIs of the SVD RC/SC are the SVD overview OPI and the SVD local control OPI (figure~\ref{fig:C4_ctrl_opi}). The first one displays the general state of all SVD sub-systems, each SVD ladder, the power supplies, the VME crates, as well as general informations from \belletwo\ RC like experiment and run numbers, the actual trigger configuration and so on. This OPI is foreseen as an entry point, allowing the operator to navigate to other OPIs with more detailed information of each sub-system. 
In addition to these SVD RC/SC user interfaces, there is a dedicated OPI for each environmental monitoring sub-system, which shows the actual temperature (figure~\ref{fig:C4_fos_opi}), humidity, and radiation values as well as the condition of the interlock system.
\begin{figure}[hbt]
    \centering
    \includegraphics[width=.5\columnwidth,keepaspectratio]{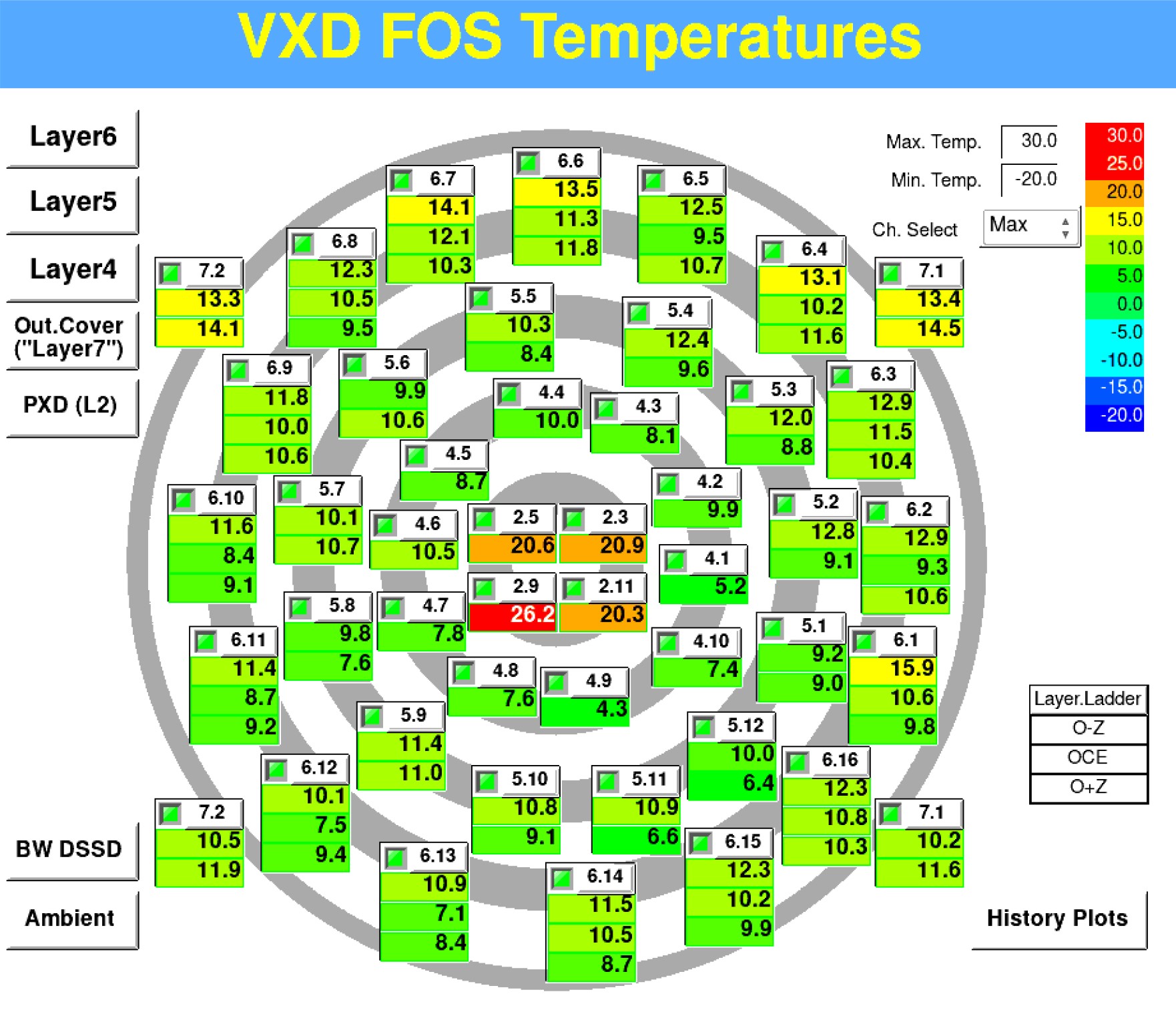}
    \caption{VXD FOS temperatures OPI.}
    \label{fig:C4_fos_opi}
\end{figure}

\begin{figure}[tbph!]
	\centering
	\includegraphics[width=1\linewidth]{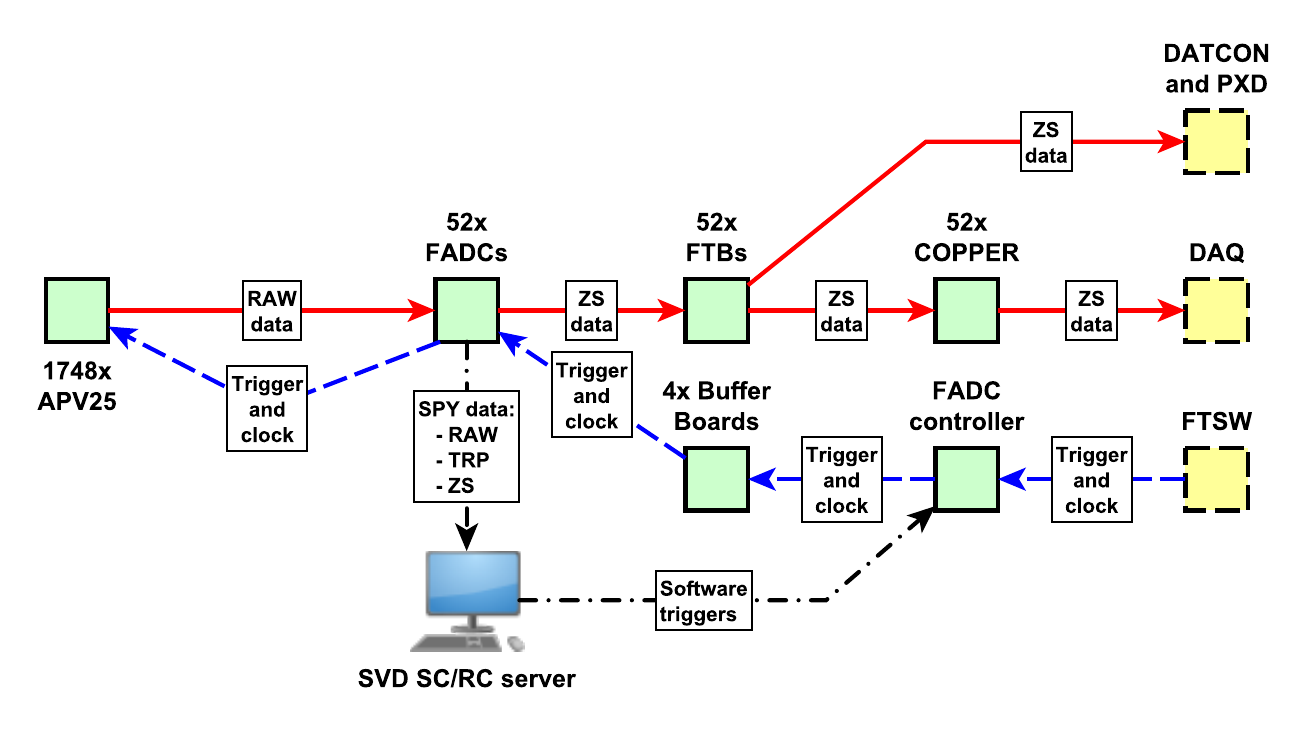}
	\caption{Data stream within the SVD FADC system.}
	\label{fig:C4_daq_local_data_stream}
\end{figure}

\lvltwosection{Online DAQ Software and Data Processing}
\label{sec:online_daq}
%(2/3 of the online part)
The SVD readout chain is shown in figure~\ref{fig:C4_daq_local_data_stream}. 
Its key components are the FADC\_Controller board, the FADC boards, and the frontend APV25 readout chips, see sections~\ref{sec:C2_apv25} and \ref{sec:C2_offdet_elec}. The FADC\_Controller board is the main interface to the FTSW, which forwards the \belletwo\ clock and trigger to each subdetector. The incoming clock received by the FADC\_Controller is scaled down from $\sim 127$ MHz (one fourth of the \acrshort{RF} clock) to $\sim 31.8$ MHz, the SVD system frequency that drives the FADC boards and APV25 chips. The FADC\_Controller moreover converts incoming trigger signals to the APV25 trigger format. Both signals are in turn distributed with 4 Buffer boards over 52 FADC boards to the 1748 APV25 chips.

In idle mode, the APV25 outputs the so-called tick mark every 35 %corr MF (it is 35, not 32) 
clock cycles. Upon receiving a trigger the next tick mark is replaced with a data frame composed of:
\begin{itemize}
	\item three header bits: used to identify the start of the data frame;
	\item eight pipeline address bits: unique address of the internal analog APV25 data buffer, that is incremented every clock cycle;
	\item an error bit: indicating either an latency or buffer full error;
	\item 128 strip data samples in a multiplexer (MUX) output order.
\end{itemize}
The end of the output is indicated by either a tick mark or the three header bits of the next frame. The number of APV25 data frames per \belletwo\ trigger depends on the configured APV25 operation mode as well as the generated trigger output of the FADC\_Controller. During a typical run, the APV25 chips are operated in the multi-peak mode generating three frames per trigger. To acquire 6 frame data the FADC\_Controller board generates two consecutive APV25 triggers for each incoming \belletwo\ trigger.
When 6 or 3 data frames are acquired, the acquisition mode is referred to as 6- or 3-sample mode, respectively. As anticipated at the end of section~\ref{sec:C2_apv25}, only  6-sample event data has been acquired in the initial phase of the experiment.
During specific local runs, the APV25 chips are operated in a single peak mode with one data frame output per trigger.

Besides the main data stream chain that provides the data to \belletwo\, there is an additional {\it spy} channel that allocates a fraction of the data to a dedicated local readout server running the slow control system. Details on how the spy data are acquired and what information is extracted from them can be found in section~\ref{sec:C4_local_runs}, while section~\ref{sec:C4_data_processing} describes the processing of the data that are handled to \belletwo\ DAQ.

\lvlthreesection{Local data acquisition and local runs}
\label{sec:C4_local_runs}
The FADC system also provides an additional side channel, the so-called \texttt{SPY} channel, that reads out a fraction of events from the main data stream. The \texttt{SPY} data are transferred over four \texttt{VME} connections to a dedicated local readout server running the slow control system, as discussed in section~\ref{sec:C4_OnlineSW}. The \ioct{SVD:QM} and \ioct{SVD:FADC} perform configuration and calibration of the FADC system. During the local calibration runs the \belletwo\ trigger input as well as the FTB data output are masked. The system is operated in a stand-alone mode. The \ioct{SVD:FADC} issues the trigger generation on the FADC\_Controller board over \texttt{VME}. The \texttt{SPY} data format returned by each FADC board can be set to either one of two data types:
\begin{itemize}
	\item Raw data (\texttt{RAW}): raw APV25 output without any processing; 
	\item Transparent data (\texttt{TRP}): strip signals, applying the FIR correction, the data frame detection, and reordering the strip signal from MUX to the APV25 channel order.
\end{itemize}

The two data types above are mainly used for debugging and local runs.
The \texttt{SPY} data in turn are forwarded to \ioct{SVD:QM} over a TCP connection for processing and evaluation of the calibration. The resulting hardware parameters are propagated back to the \ioct{SVD:FADC} over a configuration file. The same file is in turn pushed to the offline Conditions Database, housing the parameters required for offline analysis. 

Several types of local runs, described below, are implemented and routinely performed during detector operation as discussed in section~\ref{sec:C6_calibration}.
\begin{itemize}
	\item The \textbf{ADC delay scan} evaluates the sampling delay of the ADC chip. Therefore, the height of the APV25 tick mark is scanned while varying the delay in increments of one 64th of the clock cycle ($\sim0.5$ ns). An example of the ADC delay scan is shown in figure~\ref{fig:C4_delay_curves}.
	\begin{figure}[htbp]
		\centering
		\includegraphics[width=.85\linewidth]{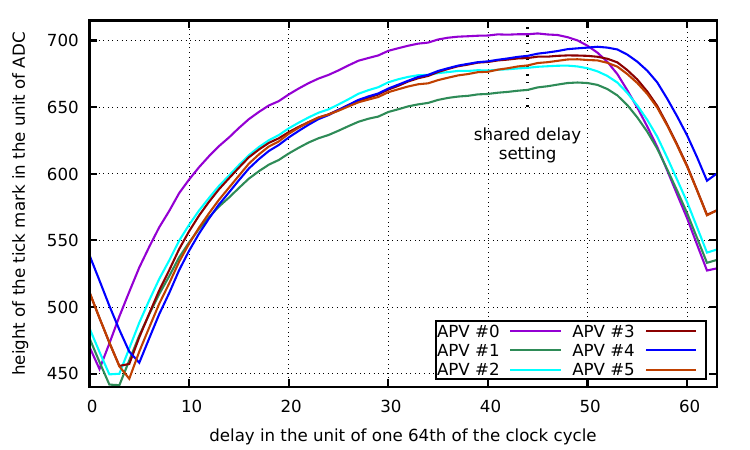}
		\caption{Tick mark vs delay of APV25s connected to the same delay channel, the vertical black dotted line marks the evaluated delay setting.}
		\label{fig:C4_delay_curves}
	\end{figure}
	
	The optimal delay setting of each individual APV25 depends on the layout of the front-end electronics. Since groups of six APV25 chips are connected to one ADC chip, a shared delay setting, dotted line, is selected with following conditions:
	\begin{itemize}
	    \item the common delay is at least 4 ns away from the falling edges of each chip, solid lines;
	    \item the signals of the tick marks are maximized within the limits of the previous condition.
	\end{itemize}
	
	\item The \textbf{FIR scan} evaluates the \acrshort{FIR} filter coefficients used to compensate the signal dispersion of the data cables. This dispersion is originated in the transfer function of the data cables between front-end electronics and FADC boards and the huge amplitude of digital data at the beginning of the APV25 data frame. 
	The extent of the distortion can be observed in the first seven baseline samples after a tick mark in figure~\ref{fig:C4_tickmark_distortion}.
	\begin{figure}[htbp]
		\centering
		\includegraphics[width=.85\linewidth]{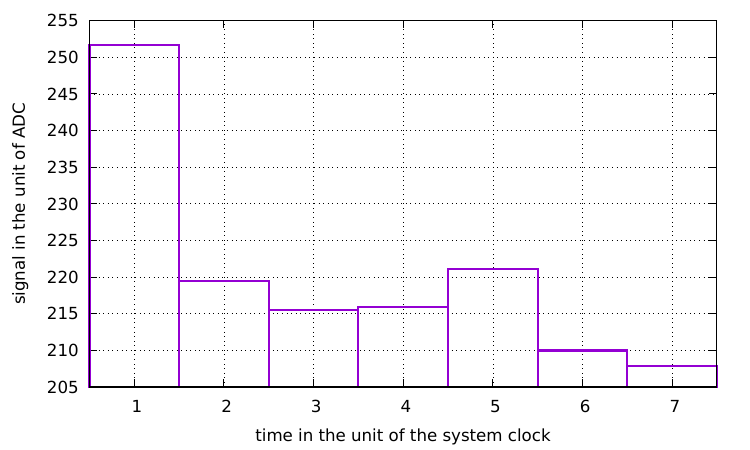}
		\caption{Baseline distortion, after a tick mark with a signal height of 705 ADC (which corresponds to
sample number 0 not drawn here).}
		\label{fig:C4_tickmark_distortion}
	\end{figure}
	The tick mark resembles an impulse response, where the following seven samples are affected by the dispersion. Thus eight FIR coefficients determined by the following equation
	\begin{equation}
		\begin{matrix}
			\begin{pmatrix}
				1 \\ 0 \\ 0 \\ 0 \\ 0 \\ 0 \\ 0 \\ 0
			\end{pmatrix}
		\end{matrix} = 
		\begin{matrix}
		\begin{pmatrix}
			d_0 & 0 & 0 & 0 & 0 & 0 & 0 & 0 \\ 
			d_0 & d_1 & 0 & 0 & 0 & 0 & 0 & 0 \\
			d_0 & d_1 & d_2 & 0 & 0 & 0 & 0 & 0 \\ 
			d_0 & d_1 & d_2 & d_3 & 0 & 0 & 0 & 0 \\ 
			d_0 & d_1 & d_2 & d_3 & d_4 & 0 & 0 & 0 \\
			d_0 & d_1 & d_2 & d_3 & d_4 & d_5 & 0 & 0 \\
			d_0 & d_1 & d_2 & d_3 & d_4 & d_5 & d_6 & 0 \\
			d_0 & d_1 & d_2 & d_3 & d_4 & d_5 & d_6 & d_7 \\
		\end{pmatrix}
		\end{matrix} \cdot
		\begin{matrix}
			\begin{pmatrix}
			k_7 \\ k_6 \\ k_5 \\ k_4 \\ k_3 \\ k_2 \\ k_1 \\ k_0
			\end{pmatrix}
		\end{matrix}
	\end{equation}
	were chosen, with $d_t$ representing the tick mark ($t=0$), and seven following baseline samples after offset subtraction and normalization.
	%% correction: Hao 20210907
	%% The $k_t$ is the FIR coefficient at sample $t$. The incoming samples $s^*_t$ are corrected with the relation $s=\sum^{T=7}_{t=0}k_t\cdot s^*_{T-t}$.
	The $k_t$ is the FIR coefficient at sample $t$. The incoming samples $s^*_t$ are corrected with the relation $s=\sum^{7}_{t=0}k_t\cdot s^*_{7-t}$, with s representing the corrected sample.
	\item The \textbf{pedestal run} samples the average of signals without the sensor hit, called \gls{pedestal} in the rest of the paper. During this stage the FIR filter coefficients are uploaded to the FADC. The frame detection and strip signal extraction are performed by the slow control software.
	\item The \textbf{noise run} evaluates the strip \gls{noise} as root mean square (\texttt{RMS}) of the each strip signal. After the strip samples have been extracted and the pedestal subtracted, a simple \gls{common-mode noise} correction (\acrshort{CMC}) is applied. To that end, the strips connected to a single APV25 chip are divided in groups of 16, \textbf{32} (default setting), 64 or 128 channels. The common mode within a group is approximated with the truncated average over 50\% of strip signals. Channels with known defects, and 50\% of the strips with highest and smallest signals are excluded from the evaluation. Finally, the noise of every channel is computed as \texttt{RMS}. As an example the strip noise of a sensor in the innermost layer is shown in figure~\ref{fig:C4_noise}.
	\begin{figure}[htbp]
		\centering
		\includegraphics[width=.85\linewidth]{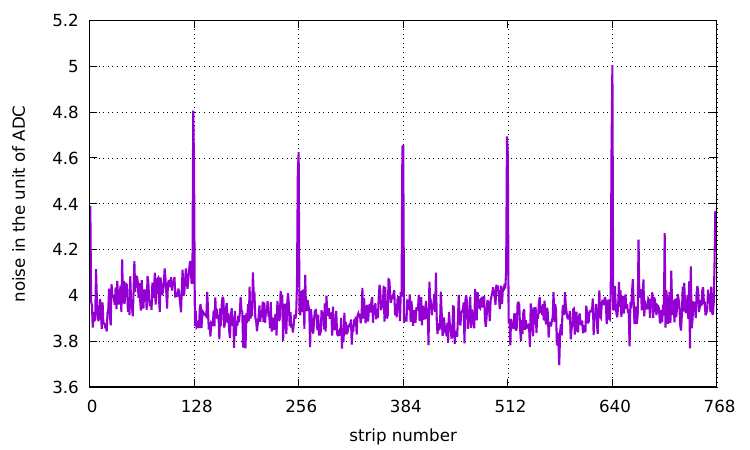}
		\caption{Strip noise of a layer 3 u/P side sensor. The spikes at every 128 strips (APV25 borders) are caused by crosstalk, originating from the power inputs of the chips and are further discussed in section~\ref{sec:C6_observations}.}
		\label{fig:C4_noise}
	\end{figure}
	\item The \textbf{calibration run} provides an estimate of the signal generated by a Minimum Ionizing Particle \acrshort{MIP} in units of ADC counts. The APV25 chip has a so-called internal calibration circuit, which injects configurable amount of charge into the preamplifier of each channel. The absolute calibration of the injected pulse, corresponding to an MIP signal, was performed with testbeam data. Given the spread of about 15\% in the relative gain among different APV25 chips due to the significant processing variations in the values of the small on-chip test capacitors~\cite{French:2001xb}, this absolute calibration performed with only a few chips from the testbeam data is accurate at the 15\% level.     
	The timing of the charge injection is configurable in the step of one eighths of the system clock tick. Figure~\ref{fig:C4_shaper_pulse}  shows the analog outputs of all the 128 strips connected to an APV25 chip as a function of time, in response to the calibration injection pulse.
	\begin{figure}[htbp]
		\centering
		\includegraphics[width=.85\linewidth]{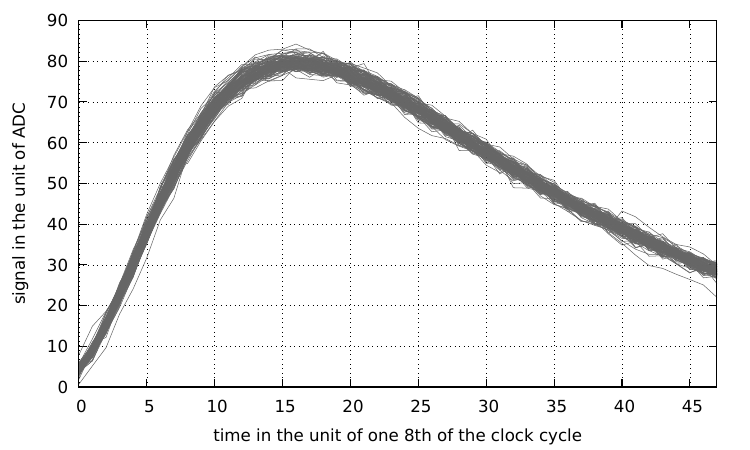}
		\caption{Analog outputs of 128 strips connected to an APV25 chip.}
		\label{fig:C4_shaper_pulse}
	\end{figure}
	The maxima correlating to the deposited charge are evaluated with an third-order %"3rd order" 
	polynomial fit over the full width half maximum. The two output parameters of the calibration run are the CalPeak and CalTMax ($t_{\rm peak}$), the height and peaking time of the signal corresponding to an MIP. 
	
	These parameters from local runs (i.e. the noise, the gain from calPeak, and $t_{\rm peak}$) are uploaded on the Conditions Database (see section ~\ref{sec:C4_cdb}) and used to calibrate the strip charge and time, as explained later in section~\ref{sec:C4_strip_reconstruction}. 
	\item The $\mathbf{V_{SEP}}$ \textbf{scan} is used to detect pinholes and determine a \vsep\ setting to be set during physics runs, to compensate for pinholes. The maxima of the injected voltage over the internal calibration circuit are sampled at various \vsep\ settings, see also section~\ref{sec:C3_APVDAQ}.
	\item The \textbf{IV} scan produces a current vs. voltage plot for each sensor. Since radiation damage is expected to cause an increase in leakage current, the IV curve measurements are used to monitor degradation of the sensor performance with time.
\end{itemize}

\lvlthreesection{Data processing}
{\label{sec:C4_data_processing}}
%\editor{Hao?, Richard?, Katsuro? - written by Giulia}
A single FADC board processes the output of up to 48 connected APV25 chips. The following sequence of operations on the input data is performed in each FADC board:
\begin{enumerate}
    \item de-serialization and digitization of the analog APV25 signal with signed 10-bit precision;
	\item signal conditioning: removal of reflections and compensation of nonlinear transfer function by a FIR filter;
	\item re-ordering of the strip data;
	\item pedestal subtraction: the strip pedestal, determined in a dedicated local run and pre-loaded to the FADC, is subtracted from the raw signal;
	\item common mode correction: the first pass consists in the subtraction of the average amplitude of strips (excluding masked strips) from each individual strip value. The average is computed in groups of 32 consecutive strips. The second pass consists in the subtraction of the average amplitude of strips (excluding masked strips and strips with signal above three times the noise) from each individual strip value. The average is computed in groups of 32 consecutive strips;
	\item online zero suppression: keep only strips with at least one sample above $n$ times the noise, whereas $n$ is an integer coefficient factor, programmed to the FADC boards during configuration. Since no clustering is performed in the firmware, the same cut is applied online for all strips, currently set to three for \belletwo\ physics runs.  Different cuts are instead applied in the offline cluster reconstruction  for the cluster seed and for neighbouring strips (\ref{sec:C4_clustering}).  
	
	%For \belletwo\ physics runs, the factor $n$ is currently set to three.
	\item strips samples are finally presented as unsigned eight bit numbers ($0...255$).
\end{enumerate}
The zero suppressed data are forwarded to the corresponding FTBs (see figure~\ref{fig:C4_daq_local_data_stream}) that connect to both the PXD system and the \belletwo\ DAQ. In particular, in the \belletwo\ DAQ chain, the COPPER boards that receive the data from FTB, encapsulate them in objects that can be read by the \belletwo\ software~\cite{basf2}, so that they can be processed by the {\it offline} software.
The data contain several additional information besides the strip data (FADC board number, APV25 chip number, APV25 channel, 3 or 6 digitized samples) including:
\begin{itemize}
    \item trigger number;
    \item trigger type (e.g.: subdetector that triggered the event, random trigger, delayed physics trigger, ...);
    \item acquisition mode (3- or 6-acquired sample per trigger);
    \item several possible errors detected in the processing up to FTB.
\end{itemize}

Beside the zero-suppression acquisition mode used in normal runs and described above, there are two more acquisition modes that are used for specific studies or calibrations. In the raw  mode
 data are sent to the output after the FIR correction; in the transparent mode data are sent to the output after the re-ordering of the strip data.

 % 10  --> 25 ?

\lvlonesection{Detector Commissioning and Installation}
\label{sec:C5_commissioning}
%\editor{Katsuro about 10 pages in total.}

A well-planned and thoroughly executed commissioning of the detector is a key factor in its successful and efficient operation at the beginning of the physics data taking.

The initial commissioning of SuperKEKB with collisions was carried out from April to July 2018, before the installation of the \acrshort{VXD}. During this period, called \phasetwo, the inner volume of the \belletwo\ detector was instrumented with detectors dedicated to the characterization of beam-related backgrounds. As described in section~\ref{sec:phase2_commissioning}, final \acrshort{PXD} and \acrshort{SVD} ladders, one per layer, were positioned at the design radii in the horizontal plane of the intersecting beams, where backgrounds were expected to be larger. These  measurements were essential to provide a realistic extrapolation of the level of backgrounds from  \phasetwo\ to \phasethree\  and ensure the safety of the full VXD, which was a pre-condition for the detector installation. In addition, it confirmed the performance of the PXD and SVD sensors and their readout electronics in the SuperKEKB environment before the VXD installation. 

In parallel with Phase 2, the complete SVD, fully equipped with electronics and environmental monitors, was operated and tested with cosmic muons for about two months. This commissioning phase, summarized in section~\ref{sec:stand_alone_commissioning}, allowed a thorough verification of all hardware and software components, including cooling, calibration procedures, track reconstruction, and efficiency mapping.

The final installation and commissioning steps are described in section~\ref{sec:installation}. In October 2018, the two halves of the SVD were mounted around the beam-pipe - PXD assembly in a clean room. The procedure required a careful alignment of the SVD. The VXD could take calibration and cosmics data in these conditons for about one month and was fully characterized before its insertion in the \acrshort{CDC} of the \belletwo\ detector. Special procedures and mechanical devices were developed for the insertion process, made difficult by tight tolerances and limited space for cables.
The insertion of the VXD was carried out on 21 Novenber 2018, and followed by the cabling and piping until the middle of December 2018.

\lvltwosection{\phasetwo\ SVD Commissioning}
\label{sec:phase2_commissioning}
%%%
%{\it about 3 pages.}
%%%

Before the final VXD could be installed into the Belle II detector system, a commissioning operation of SuperKEKB was carried out from April to July 2018.
This campaign of the SuperKEKB operation without the VXD is called ``\phasetwo'', while the operation after the VXD installation is called ``\phasethree''.
The \phasetwo\ operation was a major milestone for SuperKEKB to confirm the feasibility of the nano-beam scheme~\cite{Abe:2010gxa} as well as to understand the beam background on various Belle~II subdetectors. 
During the \phasetwo\ operation, SuperKEKB succeeded to squeeze the beams down to $\beta_{y}^{\star}=3$~mm with beam currents of more than 800~mA for the both LER and HER rings.
The peak luminosity achieved in \phasetwo\ was  \qty{5.55e33}{\per\square\cm\per\s}.

Although the final SVD was not installed in \phasetwo, this operation provided  a good opportunity to test the sensors and the developed FADC system under the operating conditions of the experiment as a \phasetwo\ SVD commissioning.
For this purpose, one SVD ladder per layer was installed in \phasetwo. 

\lvlthreesection{Commissioning goals}
\editor{Katsuro}

The main goal of the \phasetwo\ commissioning was to evaluate the beam background condition in the VXD volume before its installation.

Being very close to the interaction point, high background levels could cause radiation damage to the SVD sensors, as well as an increased hit occupancy, which could degrade tracking performance and increase data rates in the DAQ.
The most restrictive limit was related to the degradation in the tracking performance.
Background levels in the SVD are measured by the hit occupancy, and the equivalent limit in the Layer~3 hit occupancy is about 3\%.
%This limit doesn't take into account beam background rejection by the hit time selection of the collision events, which will improve the limit by a factor of about 2.
Another task of the \phasetwo\ commissioning was to check the performance and stability of the SVD.
The \phasetwo\ commissioning was the first test in the real Belle~II environment where any noise interaction to other subdetector systems could occur.
Another important purpose of \phasetwo\ was to test the VXD installation, using the developed procedures and tools.

\lvlthreesection{\phasetwo\ SVD setup}
%\editor{Katsuro}
\label{sec:phase2SVDSetup}

\begin{figure}[!ht]
 \centering
 \includegraphics[width=.9\columnwidth,keepaspectratio]{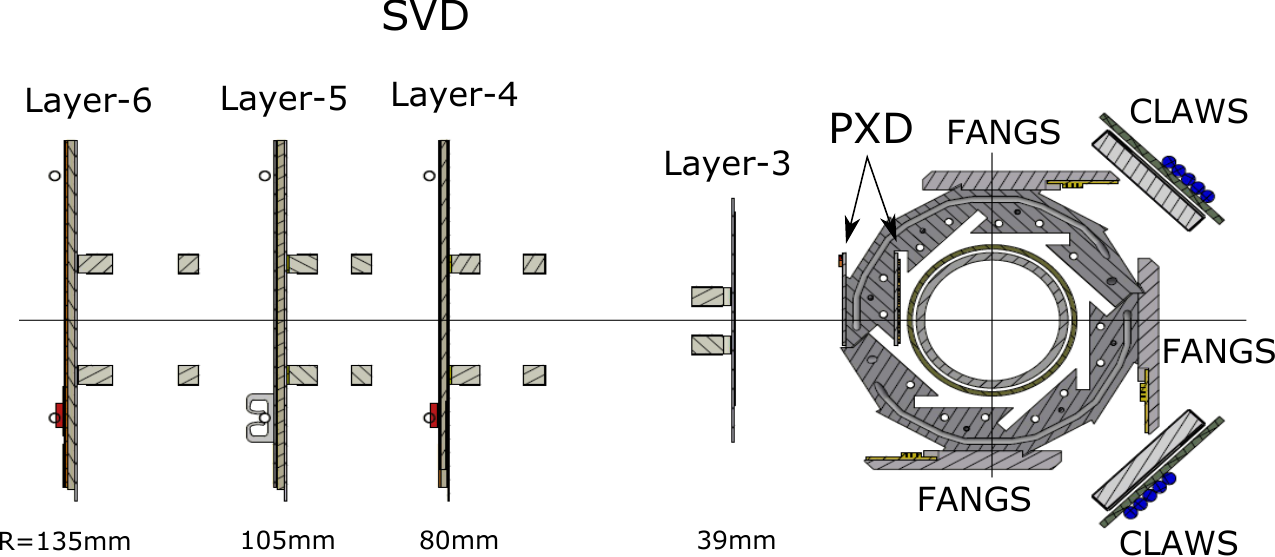}
 \caption{A cross sectional view of the \phasetwo\ VXD setup, which contains PXD ladders, SVD ladders, FANGS sensors, CLAWS sensors, and PLUME sensors. The PLUME sensors are not shown here.
 The radial distances of the SVD sensors from the IP are written below the sensors.}
 \label{fig:C5-01_Phase2VXD}
\end{figure}

\begin{figure}[!ht]
 \centering
 \includegraphics[width=.8\columnwidth,keepaspectratio]{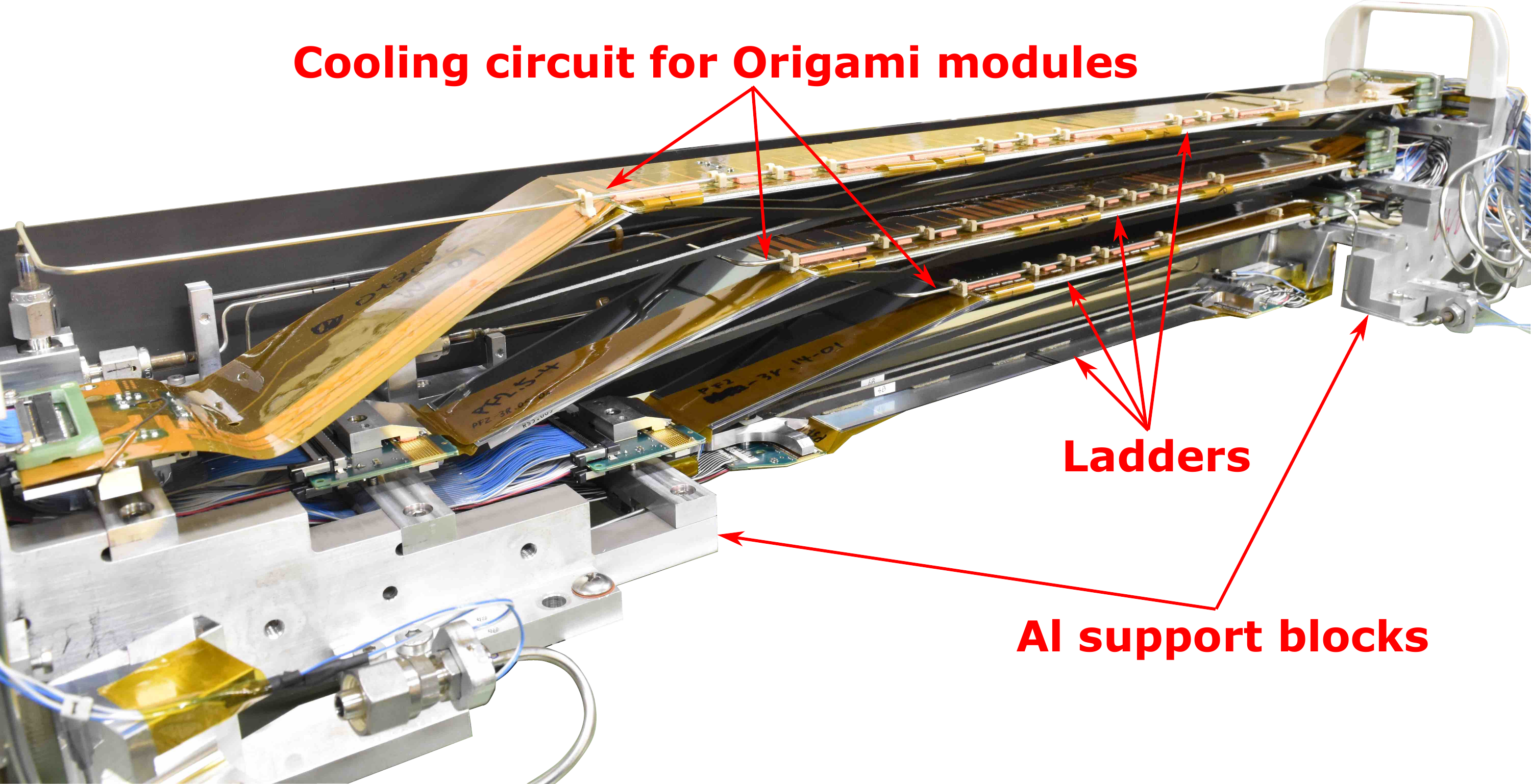}
 \caption{The SVD cartridge setup for the \phasetwo\ commissioning.}
 \label{fig:C5-01_SVDCartridge}
\end{figure}

%(Concept of the \phasetwo\ SVD setup. Concept of the SVD cartridge.)\\
For the \phasetwo\ commissioning, a special VXD which had one PXD or SVD ladder per layer, with a detector configuration shown in figure~\ref{fig:C5-01_Phase2VXD} was designed to study the detector performance in each layer. The FANGS, CLAWS, and PLUME sensors constituted the BEAST II system for background measurement~\cite{Liptak:2021tog}.
For the ease of the installation of those four SVD ladders, an ``SVD cartridge'' accommodating the four ladders, mechanical supports, and cooling services was developed, as shown in figure~\ref{fig:C5-01_SVDCartridge}.
These ladders had the same longitudinal direction parallel to the beam direction, and the sensor planes were facing the interaction point as shown in figure~\ref{fig:C5-01_Phase2VXD}.
The cartridge was mounted on the beam pipe flanges at the outside direction of the SuperKEKB main ring (the $+X$ direction in the Belle~II coordinate system).
The radial distances of the sensors from the \acrshort{IP} were the same as the \phasethree\ design, i.e. $R=$~39~mm, 80~mm, 105~mm, and 135~mm.
The outside direction was chosen because beam-background simulations~\cite{HTanigawa:2018MThesis} indicate that the outside has the largest background contribution from the storage beams.
There were three CO$_2$ circuits in the cartridge which absorbed heat from the APV25 chips on the DSSD modules, in which flowed the dual-phase CO$_2$ provided by IBBelle.
One long circuit was attached on the Origami modules of Layer 4, 5, and 6.
The remaining two were attached on the forward and backward aluminum support blocks that cooled the chips on the Forward and Backward modules via the aluminum mount blocks of the ladders.
Two CFRP plates of 1~mm thickness were located in parallel on both sides of the ladders to support the cartridge structure.
As seen in figure~\ref{fig:C5-01_SVDCartridge}, all the four ladders, support aluminum blocks, and cooling circuits were accommodated in the SVD cartridge, whose installation was easily done by just attaching it on the VXD support structure. 
The region between the forward and backward aluminum blocks was the physics acceptance where only light materials could be located.

\lvlthreesection{Results and conclusions}
\editor{Katsuro}

\begin{figure}[!ht]
 \centering
 \includegraphics[width=.6\columnwidth,keepaspectratio]{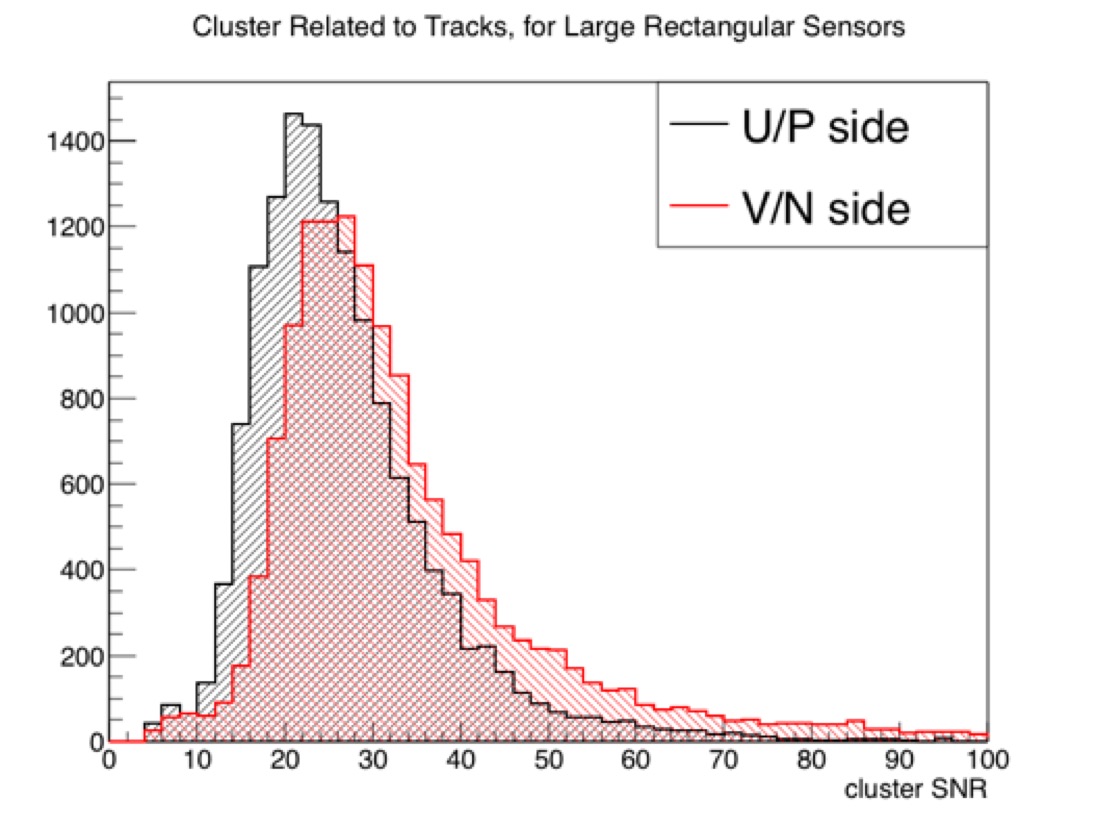}
 \caption{Cluster Signal-to-Noise Ratio (\snrCL) distribution on the large rectangular sensors in the \phasetwo\ Layer~4, 5, 6 ladders in Exp.7 Run.783.
 Black and red histograms correspond to the clusters on the u/P side and v/N side, respectively.}
 \label{fig:C5-01_Phase2_SNR}
\end{figure}

During the \phasetwo\ operation, several beam background studies were performed to understand the background composition and compare it with the \acrshort{MC} expectation.
The studies provided correction factors to extrapolate MC simulation results to the final SuperKEKB machine condition.
In \phasetwo, the measured occupancy on the innermost layer, Layer 3, was less than 0.4~\%, which is well below the 3\% limit.
The expected occupancy in early \phasethree\ was similar to the one measured in \phasetwo\, since beam currents and machine optics were expected to be similar.
From this observation and studies, the background conditions  for early \phasethree\ were considered safe for the installation of the full vertex detector.

Analyzing the data taken during the \phasetwo\ SuperKEKB operation, the performance of the SVD ladders was evaluated.
The noise and gain of all sensors were found to be stable and no visible change was observed.
Figure~\ref{fig:C5-01_Phase2_SNR} shows distributions of the cluster  \acrfull{SNR} (see the definition in section~\ref{sec:C4_clustering}) for the large rectangular sensors.
The most-probable-values of the distributions were over 20.
These high \snrCL values resulted from the low noise characteristics due to
the chip-on-sensor concept described in section~\ref{sec:C2_ondet_elec}.
Efficiencies on the Layer 3 forward sensor
of $99.08\pm 0.02$~\% and $99.44\pm 0.02$~\% were measured for the u/P side and v/N side, respectively.
From these measurements, the excellent performance of the SVD ladders was confirmed.

\lvltwosection{Stand-alone Full SVD Commissioning}
\label{sec:stand_alone_commissioning}
%%%
%{\it about 3 pages.}
%%%

Two SVD halves for \phasethree\ were fully assembled in the first basement floor of the KEK Tsukuba experimental hall and then moved to a storage box in the fourth basement floor (B4) in February 2018 for the first half and in July 2018 for the second half.
In the storage box, before the coupling with PXD in October 2018, a stand-alone commissioning test for the SVD halves was performed using cosmic muons.
This commissioning was performed in parallel with the \phasetwo\ commissioning, which is described in section~\ref{sec:phase2_commissioning}.

%%%
%%% https://confluence.desy.de/display/BI/SVD+phase+3+commissioning+at+B4 is a good reference to complete this page.
\lvlthreesection{Targets of stand-alone SVD commissioning}
\editor{Katsuro}

The first target in this commissioning test was to check the healthiness of all the DSSD sensors, front-end electronics, and environmental monitoring sensors (NTC and FOS, see section~\ref{sec:temp_mon}) on the SVD halves, because this was the last performance test before coupling PXD and SVD.
Another target was to confirm the tracking capability of the SVD analysis software measuring cosmic muons, since it was the first opportunity to detect the particle tracks with the real geometrical configuration of the multiple layers.

\begin{figure}[!ht]
 \centering
 \includegraphics[width=\columnwidth,keepaspectratio]{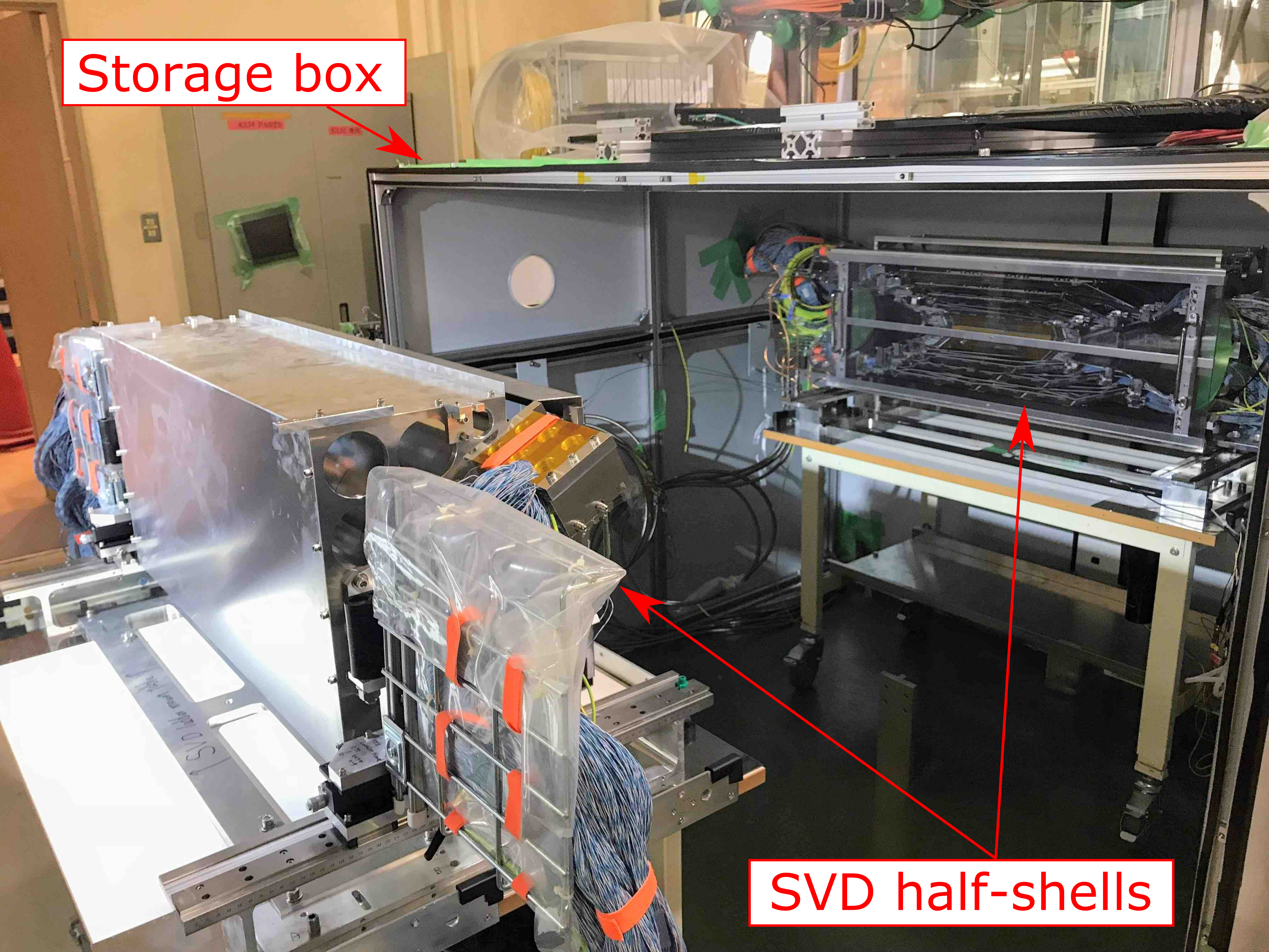}
 \caption{Stand-alone SVD commissioning setup. The two SVD halves were put in the storage box (in the photo, the front panel is open), and fixed on the floor with anchors as a protection against earthquakes.
 After closing the front panel of the box, they were tested in the box.
 }
 \label{fig:C5-02_SVDB4CommissioningSetup}
\end{figure}

Figure~\ref{fig:C5-02_SVDB4CommissioningSetup} shows the setup of the
stand-alone commissioning test.
For the commissioning, the two SVD halves held by the pick-up tool (see section~\ref{sec:C3_svd_half_shell_storage}) were moved into the storage box built with aluminum frames and panels.
The box offered a protection for the detector, but was also a dark environment that allowed applying high voltage to the sensors.
It was also flushed with nitrogen to offer a dry environment with a dew point lower than \qty{-40}{\degreeCelsius}, necessary to operate the cooling system which uses CO$_2$ at \qty{-20}{\degreeCelsius}.
Scintillation counters were set on the roof of the box and under the storage tables to provide triggers for the cosmic rays.
During the two months commissioning, detailed calibrations were performed for the first time with the full SVD ON and with the final configuration of the CAEN power supply system and the FADC readout system.

\begin{figure}[!ht]
 \centering
 \includegraphics[width=.4\columnwidth,keepaspectratio]{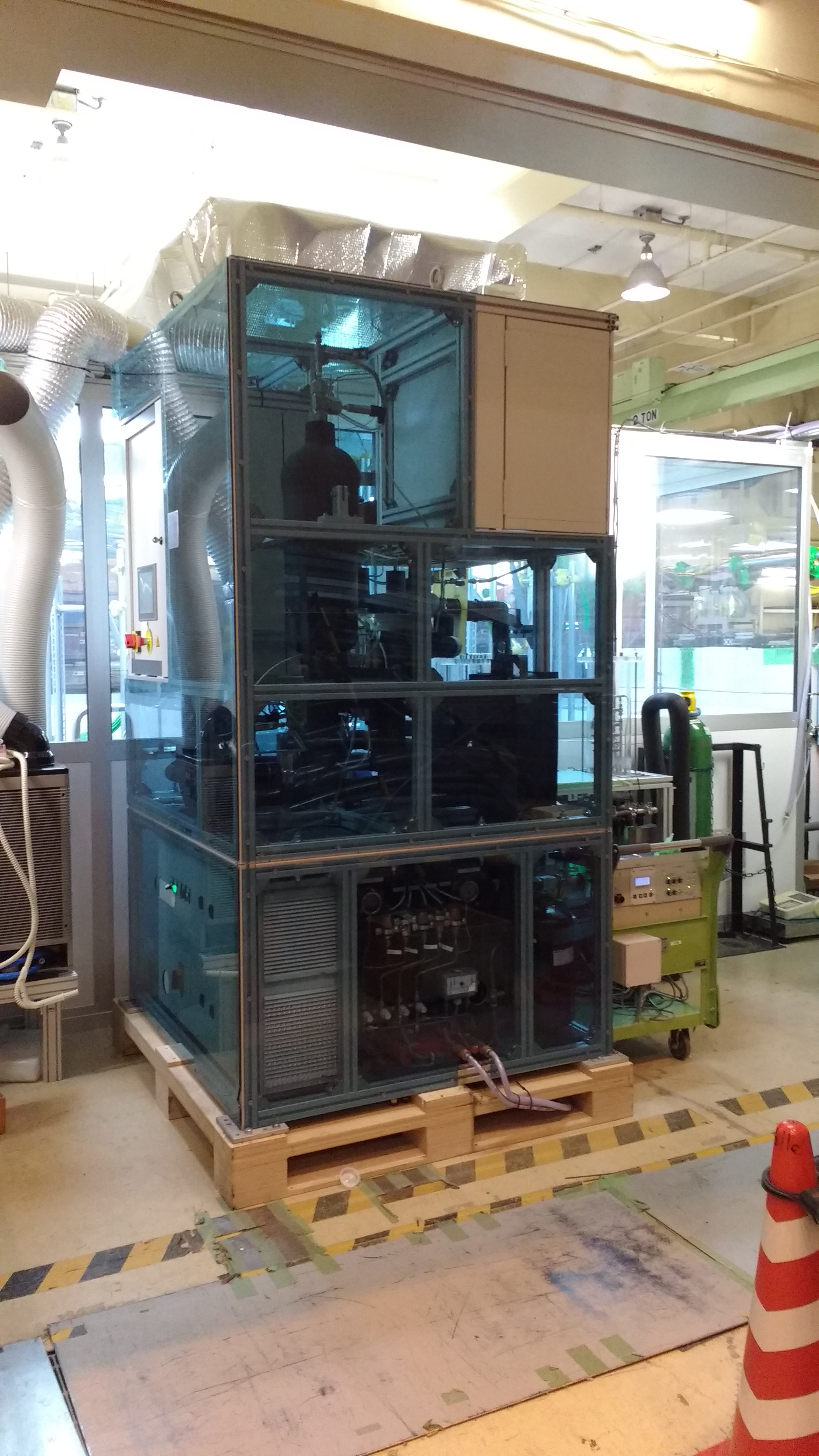}
 \caption{MARCO cooling system installed in the Tsukuba hall B4 floor.}
 \label{fig:C5-02_marco_B4}
\end{figure}

%\textcolor{red}{Explanation of the cooling system: MARCO (Multi-purpose Apparatus for Research on CO2 cooling)}.
For the cooling of the SVD during commissioning, the MARCO (Multi-purpose Apparatus for Research on CO$_2$ cooling) cooling system was utilized.
MARCO is a CO$_{2}$ cooling unit based on the 2PACL (2-Phase Accumulator Controlled Loop)~\cite{2PACL:1, 2PACL:2} concept.
The cooling capacity of MARCO is in the order of \qty{1}{\kW}, and the operational temperature ranges from the room temperature to \qty{-40}{\degreeCelsius}.
MARCO was designed for non-expert users with all its necessary operations being automated with a PLC.
The CO$_{2}$ cooling design of MARCO was utilized also for IBBelle and the ATLAS IBL cooling.
Figure~\ref{fig:C5-02_marco_B4} shows MARCO installed in the Tsukuba hall B4 floor for the commissioning.

In the commissioning test, the final system of FADC and FTB boards, that were described in section~\ref{sec:C2_offdet_elec}, were used for the readout of the detector signals.
Also, the final environmental system to monitor the temperature of the detector and the CO$_{2}$ cooling tubes, that was described in section~\ref{sec:temp_mon}, was used.
The functionality of the equipment and the firmware and software was confirmed.

\lvlthreesection{Results and conclusions}
\editor{Katsuro}

The SVD halves ran stably during the commissioning and about $3\times 10^{7}$ cosmic-muon events were recorded, confirming that all the sensors were working properly. 
%No change in noise and defect of strips was observed.
All the noise and calibration constants were stable over the entire two-month period.
The noises in the sensors were consistent with our expectations which took into account the various noise contributions significantly depending on the sensor side and location in the ladder.
A summary of the noise contributions for the various sensors is shown in section~\ref{sec:C7_performance}.
The number of defective strips, which had anomalous values of the noise, gain, or pedestals, was about 1\% per sensor.
This number was driven by the pinholes which were already present in the produced silicon sensors, and only a few additional bad channels, like short- or open-circuits, introduced during the ladder assembly.

It was also possible to observe and fully characterize a number of peculiar effects, for example the sensor turn ON effect, or cross talk effects, which were visible for the first time during commissioning in the cold and dry environment.
Although they do not affect the performance, they deserve attention to operate SVD in a proper way.
Details of these effects are described in section~\ref{sec:C6_operation}.

The flow of the CO$_{2}$ coolant in the cooling circuits was also confirmed by checking temperatures of the cooling channels measured with the NTC.
Only one NTC sensor was found dysfunctional, however not affecting the ability of the environment monitor and control thanks to the redundancy of the sensors.
The functionality of other NTC and all FOS for the temperature measurement was confirmed. 

\begin{figure}[!ht]
 \centering
 \includegraphics[width=.6\columnwidth,keepaspectratio]{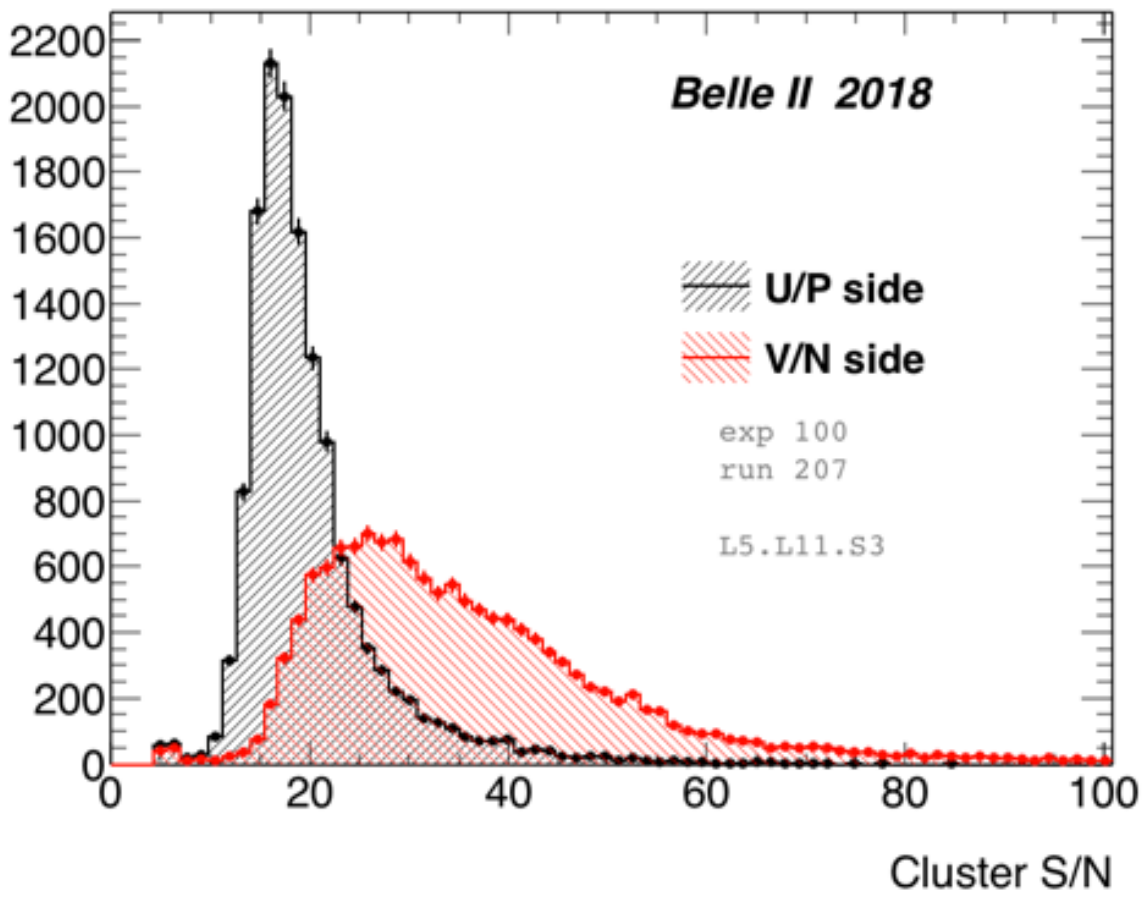}
 \caption{\snrCL distribution of a Layer 5 ORIGAMI $-$Z sensor measured in
 the stand-alone SVD commissioning with cosmic rays.}
 \label{fig:C5-02_SVDB4Commissioning_SNR}
\end{figure}

Tracks of the cosmic muons are reconstructed from the hits on the sensors.
Using the clusters associated to the reconstructed tracks, the \snrCL and hit efficiency of SVD sensors are measured.
Figure~\ref{fig:C5-02_SVDB4Commissioning_SNR} shows the \snrCL distributions measured in the stand-alone commissioning.
Depending on the sensor location, the incident angle distribution of the cosmic muons is found to be different, and this affects the signal charge and the cluster size of the hit on the sensor.
The \snrCL distributions change according to the sensor location and side.
By considering the effect of the incident angle, it was confirmed that the measured \snrCL and efficiency are consistent with that observed during the beam tests and \phasetwo\ commissioning.

\lvltwosection{Installation}
%%%
\label{sec:installation}
%{\it about 4 p\\ages.}
%%%

As the final step of the SVD construction, it needed to be coupled to the PXD to build the full VXD.
Another commissioning phase was carried out for the full VXD, as a final performance check before its installation.
Once the performance was confirmed, the VXD was finally installed in the Belle~II detector.
In the following subsections, those steps until the VXD installation are explained.

\lvlthreesection{Coupling to PXD}
\editor{Katsuro}

%%% Brief explanation of the SVD and PXD combining. %%%
The coupling of the SVD halves around the PXD to obtain the full VXD was performed at the beginning of October 2018.
Before the coupling, which took place on a granite table in a clean room built on the B4 floor, the interaction point section of the beam pipe (IP beam pipe) was assembled with heavy metal shields,  and the PXD, fabricated in two halves in several German sites and  transported to KEK, was  mounted around the IP beam pipe.
In the coupling the two SVD halves were attached around the PXD from both sides.
The same pick-up tool used for the SVD pick-up procedure (see section~\ref{sec:C3_svd_half_shell_storage}) as well as for the storage of the two SVD half shells, was used in the coupling operations.

\begin{figure}[!ht]
 \centering
 \includegraphics[width=\columnwidth,keepaspectratio]{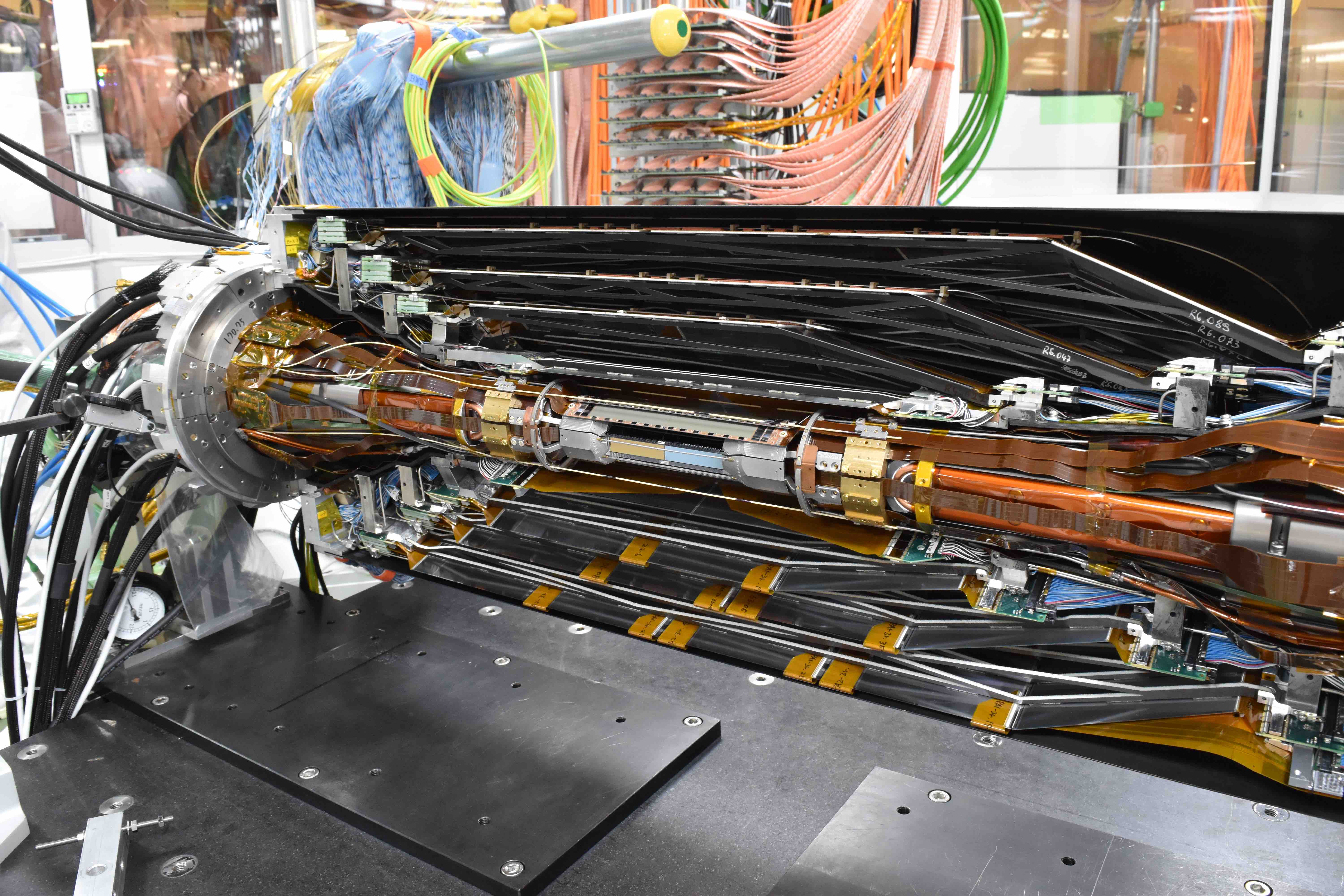}
 \caption{A picture of the first SVD half already coupled to the PXD.}
 \label{fig:C5-03_1stSVDHalf}
\end{figure}

\begin{figure}[!ht]
	\centering	\includegraphics[width=\columnwidth,keepaspectratio]{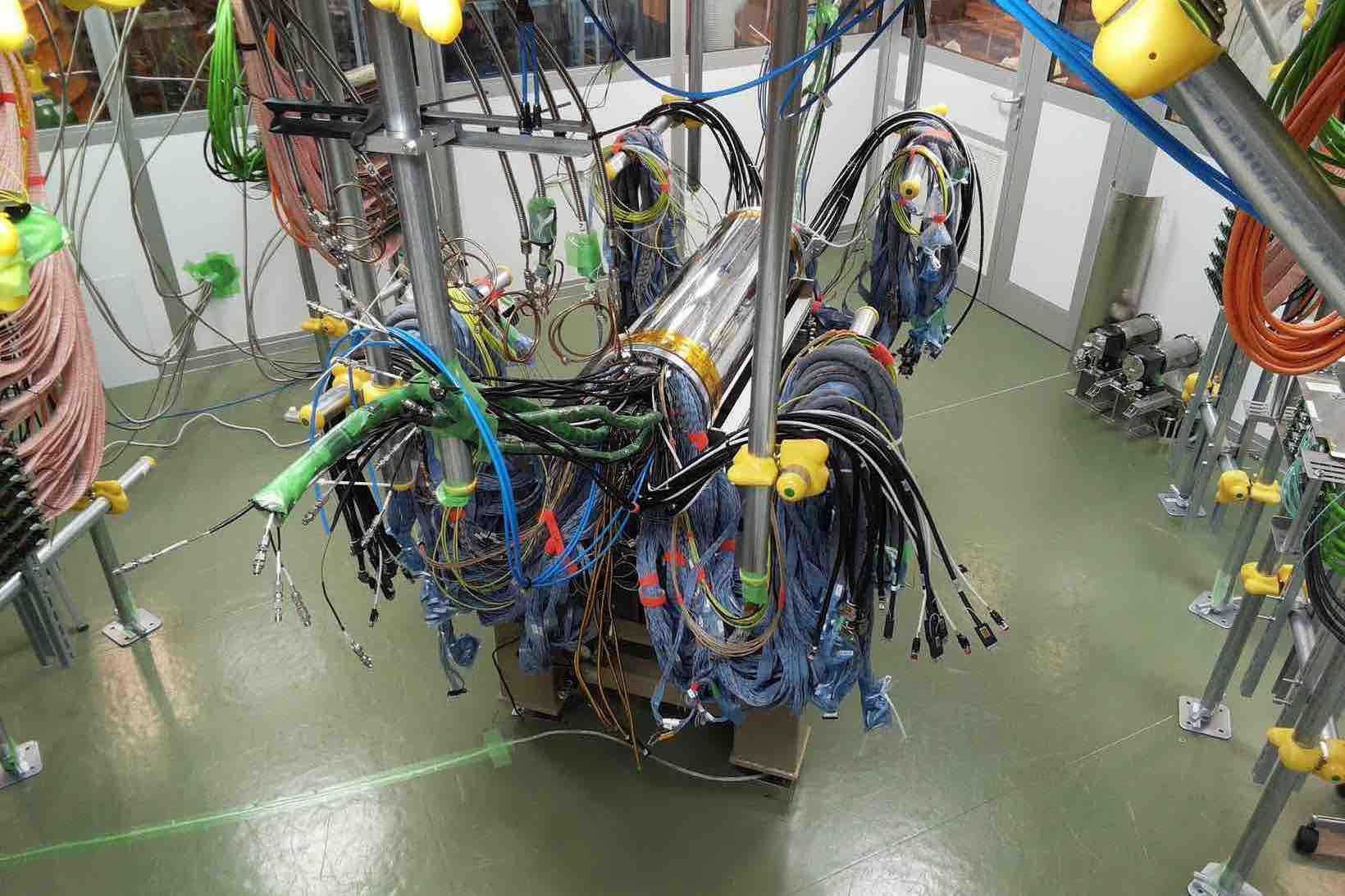}
	\caption{A picture of the fully assembled VXD in the B4 clean room.}
	\label{fig:C5-03_SVDAndPXD}
\end{figure}

%%% Careful point in the combining. and alignment function %%%
During the attachment, an accurate position control of the moving SVD half was important to prevent the PXD and SVD from touching each other, since the design clearance between them was only 1.5 mm.
The pick-up tool allowed us to control the vertical position and rotation angle of the SVD half, and they were precisely aligned by using alignment gauges.
Also, guide pins on the beam pipe flanges prevented any accidental displacement of the SVD, which could result in a mechanical interference with the PXD.
With the alignment mechanism and the guide pins, coupling to the PXD was performed successfully.
Figures~\ref{fig:C5-03_1stSVDHalf} and \ref{fig:C5-03_SVDAndPXD} show pictures of the first SVD half coupled to the PXD and the fully assembled VXD.
The coupling of the SVD with the PXD took one day for each half, while two days were required afterwards to perform precise alignment of the two halves and connect all additional mechanical components and services.

\lvlthreesection{Combined PXD-SVD commissioning}
\editor{Katsuro}

After the coupling of the SVD to the PXD was completed, commissioning of the fully assembled VXD was performed for about a month starting from the middle of October 2018.
In the combined PXD-SVD commissioning, key performances of the SVD were checked as the final step before the VXD installation.
At first, the healthiness of all SVD components was checked in order to ensure that no damage was caused to the detector during the coupling with the PXD.
It was confirmed that the full detector can be read out without any issue, and observed noise, gain, and pinholes were consistent with the results before the coupling.
Another important task was to confirm the moderate temperature of the detector using FOS and NTC when sensors were biased, \apv\ chips were powered on and configured, and CO$_2$ cooling was active.
This test was important because it was the first temperature evaluation inside the closed SVD volume with the full heat dissipation from \apv\ chips (in total, about 700~W) and the heat absorption of the CO$_{2}$ cooling using the real setup. 
The origami modules showed temperatures from about 4~\si{\degreeCelsius} to 16~\si{\degreeCelsius}, confirming that the produced heat was removed effectively and there was no overheating of the detector volume.
The hardware, software, and firmware were also successfully tested with the full-scale FADC system.

\begin{figure}[!ht]
 \centering
 \includegraphics[width=.6\columnwidth,keepaspectratio]{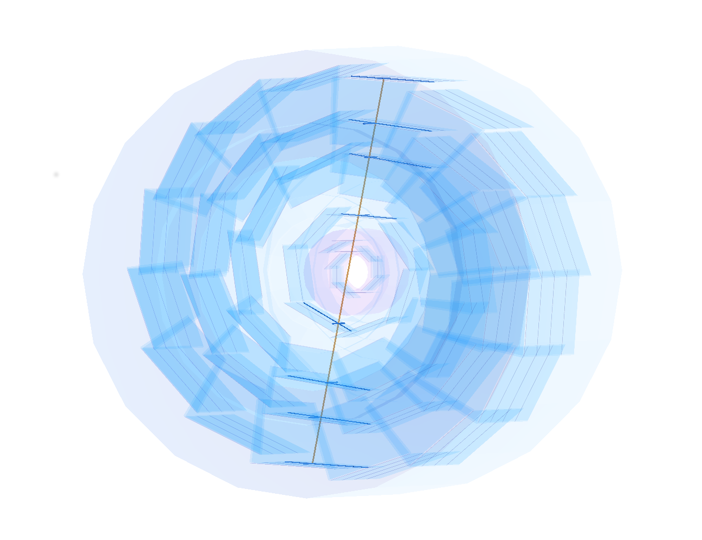}
 \caption{An event display of a cosmic muon track in the combined PXD-SVD  commissioning. 
 It shows inner two layers of PXD sensors (the second layer is incomplete) and outer four layers of SVD sensors.
 Hit strips in the SVD are indicated by blue lines, and a reconstructed track from these hits is shown as a brown line.}
 \label{fig:C5-03_SVDPXDCosmicRay}
\end{figure}

As for the stand-alone SVD commissioning, scintillation counters on top and bottom of the VXD were installed to provide triggers for cosmic data taking.
Figure~\ref{fig:C5-03_SVDPXDCosmicRay} shows one event display of the cosmic event reconstructed with the SVD reconstruction software, that is described in section~\ref{sec:software_offline}.
Using the cosmic muon hits, the reasonable distributions of the energy loss and the performance of the alignment software were obtained.

\lvlthreesection{Installation in Belle II}
\editor{Shuji Tanaka}

\begin{figure}[!ht]
	\centering
		\includegraphics[width=\columnwidth,keepaspectratio]{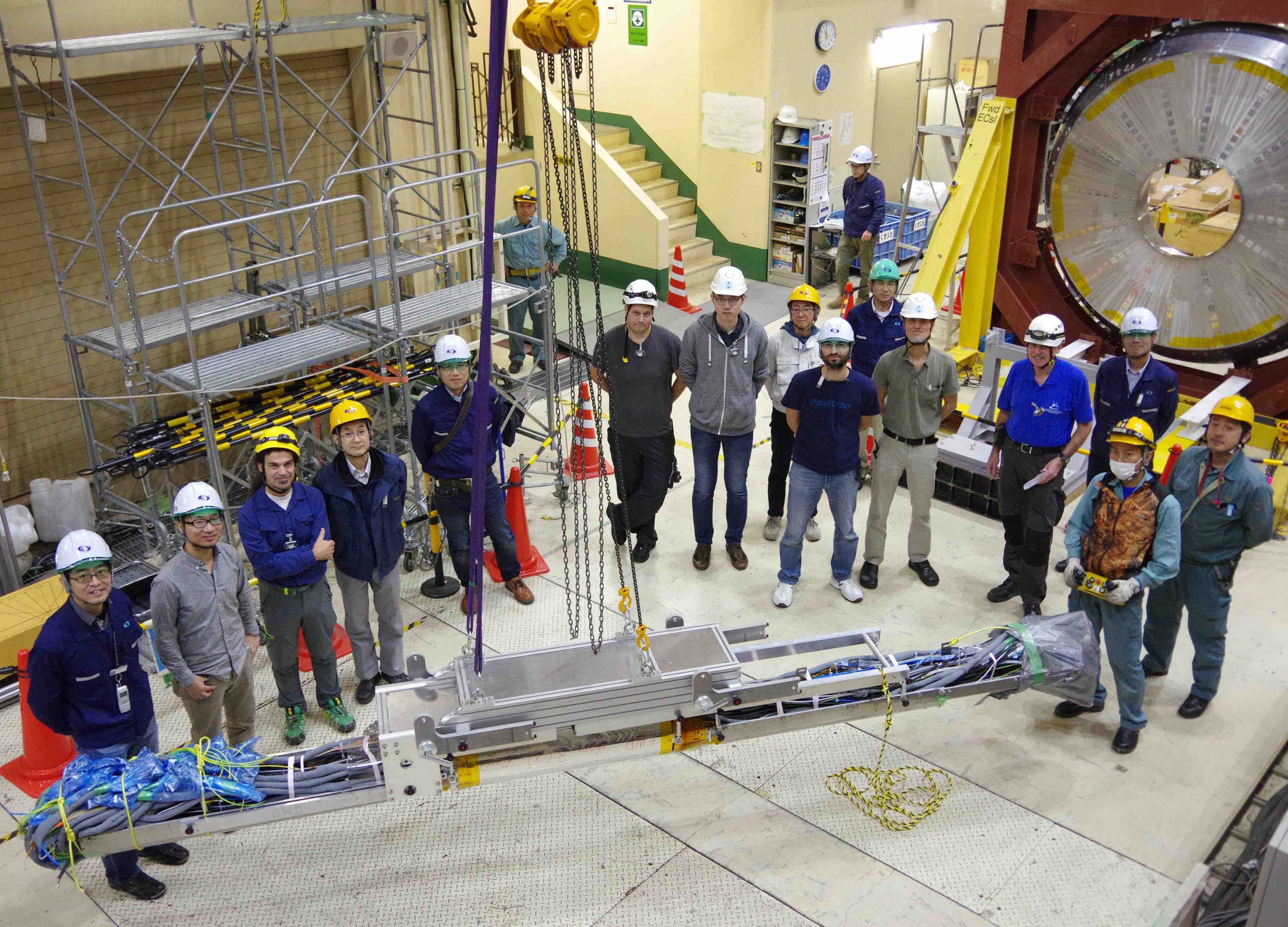}
	\caption{Photo taken during the VXD installation in the \belletwo\ detector: lift-up of the VXD with a crane. The cables and pipes were fixed in the cables trays on both sides.}
		\label{fig:C5-03_VXD_Crane}
\end{figure}

\begin{figure}[!ht]
	\centering
	\includegraphics[width=\columnwidth,keepaspectratio]{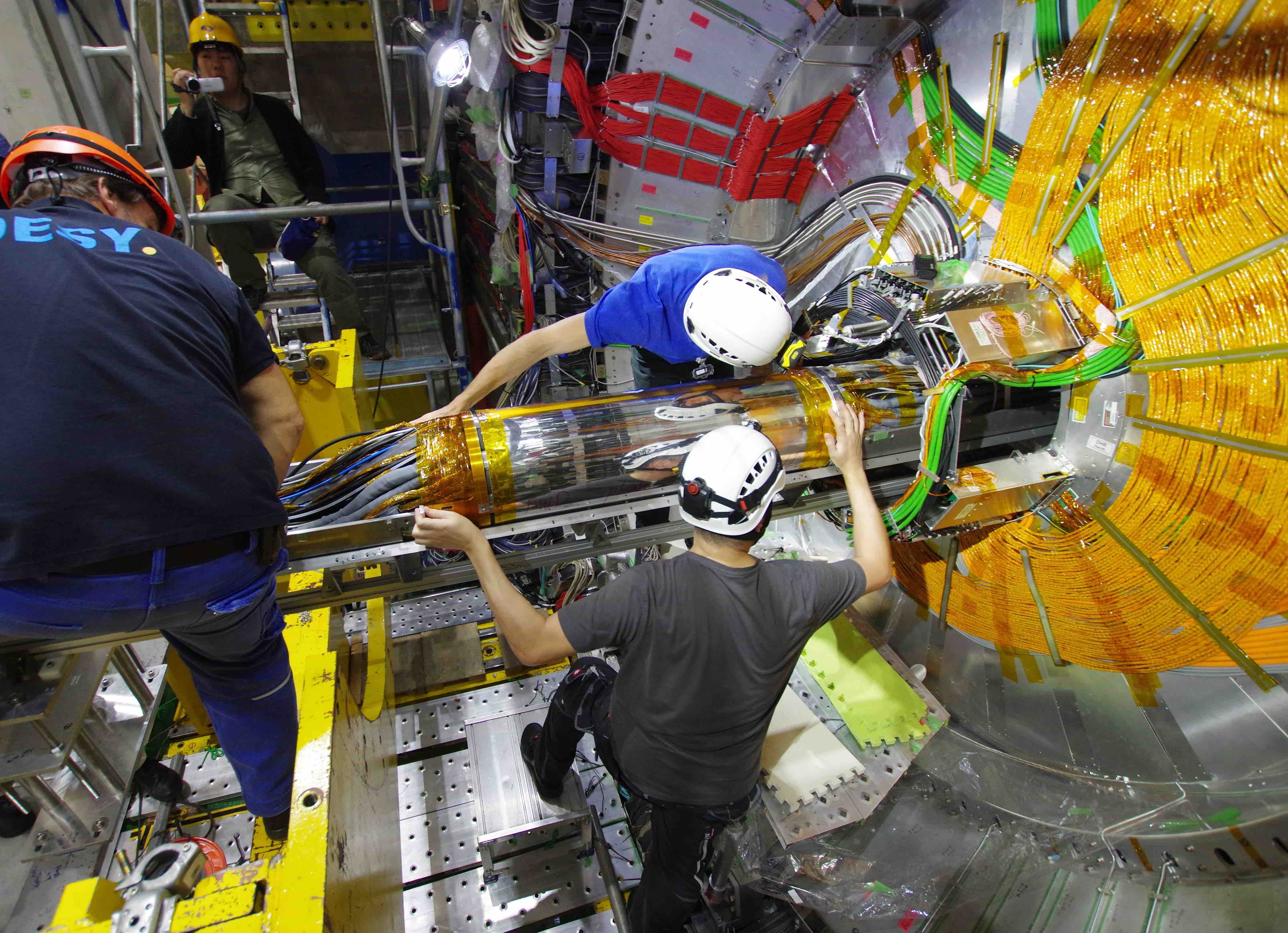}
	\caption{Photo taken during the VXD installation in the Belle~II detector:  installation of the VXD in the Belle~II detector. The slide guide installed in the \belletwo\ detector using the slide guide installed in the forward part of \belletwo.}
%	\label{fig:C5-03_VXD}
	\label{fig:C5-03_VXD_Installation}
\end{figure}

After the VXD commissioning, the bellows pipes were attached to both ends of the IP beam pipes to connect them to the final focus beam pipes.
On the bellows pipes, the diamond sensors were installed to monitor the radiation dose around the pipes (see section~\ref{sec:radmon_ba}).

During the VXD installation preparation, the cables and pipes of the PXD and SVD were laid straight and fixed in cable trays connected to the VXD.
During the transportation, the VXD and cable trays were supported by the crane tool, and the combined structure was lifted by the crane as shown in figure~\ref{fig:C5-03_VXD_Crane}.
The VXD and cable trays were lowered and put on the slide guide, that was then installed inside the Belle~II detector and extended up to the forward side of the QCS.
The slide guide was designed to safely insert the VXD into the dedicated location at the center of the CDC cylinder.
The outer radius of VXD is 155~mm while the inner radius of the CDC is 160~mm, therefore there is a 5~mm clearance in the design.
After the crane tool was detached, the VXD and cable trays were slid on the guide by hand to insert them into the CDC, as shown in figure~\ref{fig:C5-03_VXD_Installation}.

The VXD was fixed on the CDC with alignment pins.
The positions of the alignment pins on the VXD were adjusted in advance based on the results of a position survey, in which the CDC position with respect to the accelerator devices was measured with a precision of about \qty{100}{\um}.
With this alignment, the VXD and IP beam pipe were located at the correct position.
After the fixation of the VXD, the slide guide was disassembled.
The part inside the Belle~II detector was removed by pulling it out from the gap between the VXD and CDC.

The transportation and insertion of the VXD were completed on 21 November, 2018.
After the insertion of the VXD, the cables and pipes from the PXD and SVD were laid and connected to the junction boxes, which were then connected to the readout electronics, the power supplies, and the CO$_{2}$ cooling plant IBBelle.
The cabling and piping works were carried on until the middle of December, 
due to the large amount of cables and the limited service space for cables and pipes.
As a last commissioning step, the installed VXD was tested again using cosmic rays.
The results of this last test were found to be consistent with the ones taken before the installation, and confirmed an excellent performance of the SVD.

 % 10 pages

\lvlonesection{Detector Operation}
\label{sec:C6_operation}
\editor{Antonio}

Careful operation is essential to efficiently record high-quality data for subsequent analysis, as well as to identify performance trends in the evolving high luminosity conditions of   SuperKEKB.
Section~\ref{sec:C6_periods_shifts} gives a summary of the data taking periods in 2019 and 2020, followed by a description of the organizational efforts to progressively simplify the tasks of operators, and to relieve part of the burden on KEK residents, by allowing remote operation from the collaborating institutes scattered around the globe.

A relevant part of the detector operation is the periodic recalibration during local runs (section~\ref{sec:C6_calibration}), to provide updates of the calibration constants (pedestals, noise, gain), used for zero-suppression and offline analysis, and of the maps of defective channels.
The experience accumulated in monitoring the evolution of calibration constants and defective channels is summarized in section~\ref{sec:C6_stability}. A few issues on specific sensors and on the power supply system, encountered during two years of operations, are also briefly mentioned.

A great deal of attention is devoted (section~\ref{sec:C6_radiation}) to monitor the accumulated radiation dose, by observing the evolution of noise and leakage current in the \acrshort{DSSD} sensors. A specific tool is developed to monitor the depletion voltage of the sensors, while local calibration runs are used to search for additional defective channels after high beam-loss events: the limited number of new defects confirms the robustness of the \acrshort{SVD} sensors.

Further observations on the behaviour of the detector components are detailed in section~\ref{sec:C6_observations}, in particular cross-talk effects and baseline shifts depending on the trigger rate. In section~\ref{sec:C6_apv_trigger_veto} the implementation of a firmware feature to minimise the dead time in case of high occupancy is described.

Section~\ref{sec:C6_blm_env_intlks} is dedicated to the operation of the beam loss monitor and beam abort system, based on diamond detectors; a brief outline of the experience on temperature and humidity monitors is also given, followed by some remarks on the hardware interlock operation.
Finally, section~\ref{sec:C6_background_studies} presents some results on machine induced-backgrounds, both in single-beam configuration and with colliding beams. These preliminary studies are essential for the extrapolation of the background levels to future higher-luminosity conditions of SuperKEKB.

%%%%%%%%%%%%%%%%%% Operational overview %%%%%%%%%%%%%%%%%%
\lvltwosection{Data Taking Periods and Shift Organization} 
% - 2 pages
\label{sec:C6_periods_shifts}
The full \acrshort{VXD} system was placed inside the \belletwo{} detector in November 2018. After a series of tests made to verify that the system was working properly, \phasethree{} operations started on January 21, 2019. Each year was divided into three running periods. The runs considered in this chapter are the following:
\begin{itemize}
    \item \textbf{2019a}: January 21, 2019 - March 31, 2019
    \item \textbf{2019b}: April 1, 2019 - July 1, 2019
    \item \textbf{2019c}: October 15, 2019 - December 12, 2019
    \item \textbf{2020a}: February 25, 2020 - March 31, 2020
    \item \textbf{2020b}: April 1, 2020 - July 1, 2020
\end{itemize}
The first six weeks of operation of the full \belletwo{} detector in the 2019a run were devoted to cosmic ray runs, used to test the full \acrshort{DAQ} chain and the detector performance. Machine operations started on March 11, 2019 with single beams circulating in the machine, achieving the milestone of first collisions in \phasethree{} on March 25. Unfortunately, a fire in a building near the Linac complex stopped machine operations for four weeks. After the restart, injection studies allowed operation of both beams with continuous injection from May 14, 2019.  
%In this mode, the charge lost by each bunch is continuously restored, leading to higher background originating from the individual bunches affected by the trickle charge until they stabilize. To cope with this higher background, a trigger veto is applied for small time window around the injected bunch for a number of turns. 
The machine ran stably with these conditions until the end of 2019b run. No other major issues occurred during the 2019c run. Some 2019 beam time was devoted to machine tuning and studies, during which the \belletwo{} detector was turned off, but the whole DAQ chain was kept running in order to test its stability. Moreover, background studies were performed periodically, and in particular every time the machine optics was changed, to understand and mitigate the background induced by the accelerator on the detector. Results for some of these studies are discussed in section~\ref{sec:C6_background_studies}. \par
To manage SVD operations in \phasethree, the role of an ``operation coordinator'' was established.
%For 2019a and 2019b periods (first runs of \phasethree) it was decided to have two operation coordinators to take care of organizational and technical aspects of the SVD operation.
A shift system, that included local and remote shifters, was developed in order to safely operate the SVD detector. During 2019 runs, the local shifters ensured constant presence at KEK, being "on-call" in case there were problems with the SVD that required actions from an expert.  The remote shifters were in charge of monitoring the detector and the environmental variables, as well as of checking data quality using \acrfull{DQM} plots.
In 2020, with the spread of the COVID-19 pandemic and travel restrictions, the shift system was modified adding an expert remote shifter to operate the SVD detector remotely, with the local shifter role being limited to a few non-ordinary operations that required physical presence in the experimental hall.

%%%%%%%%%%%%%%%%%% Detector Calibration %%%%%%%%%%%%%%%%%%
\lvltwosection{Detector Calibration} % - 2 pages
\label{sec:C6_calibration}
To efficiently operate the detector, it is important to monitor and constantly update the calibration constants, like pedestal, noise and gain, since they are used both for the online zero suppression during data taking, and in offline reconstruction for the signal energy conversion and hit time reconstruction as explained in detail in section~\ref{sec:C4_strip_reconstruction}. 
For the online zero suppression the strip signal, pedestal subtracted, is compared with a threshold selecting only strips that have a signal higher than three times their noise (that is the strip Signal-to-Noise Ratio is greater than 3). During reconstruction the strip signal is then converted  to charge  using the strip gain measured with the internal APV25 calibration circuit.

Periodic calibration runs are also useful to monitor defective channels, i.e. those with anomalous calibration constants, that need to be masked out to avoid adverse impacts on the detector performance. Periodic measurements of the characteristic IV curve of all the sensors were also performed to evaluate effects of radiation damage. All types of calibration runs, here referred to as ``local runs'', are described in section \ref{sec:C4_local_runs}.

Different sets of local runs were then defined to monitor all important parameters and they were taken with different frequency during data taking. This frequency was optimized based on some initial experience, which showed that in general the calibration constants were very stable; these constants were found to slowly evolve with radiation damage or due to temperature changes related to different detector settings that could change the power consumption. \par

Local runs were arranged in three different groups, depending on the time available as well as the calibration constants to be monitored. Local run groups are defined as follows:
\begin{itemize}
    \item \textbf{Short}: pedestal, noise ($\sim$20 min). The short local run is used when there is a short no-beam time, to monitor noise and pedestals of every channel in the SVD. This is particularly important since the signal of each channel depends on pedestal and noise values, so even a small change in one of these parameters can affect the signal of each channel.
    \item \textbf{Intermediate}: \acrshort{ADC} delay scan, \acrshort{FIR}, noise, pedestal, calibration ($\sim$55 min). This kind of local run can be done when there is a longer period without beams, or during maintenance days (every two weeks). In the intermediate set of local runs, the ADC delay and the FIR are also evaluated to check their stability over time, while the gain of each channel is evaluated with the calibration.
    \item \textbf{Complete}: ADC delay scan, FIR, noise, pedestal, calibration, \vsep{} scan, IV scan ($\sim$150 min). In addition to all runs of the intermediate set of local runs, \vsep{} and IV scans are performed. The \vsep{} scan is used to find defective strips, pinholes in particular, while the IV scan is used to evaluate the characteristic IV curve of each sensor. Typically, the complete set of local runs is taken every two weeks during maintenance days, or during the downtime after a severe beam abort due to large radiation spikes, that can potentially create damage in the SVD sensors by adding new pinholes.
\end{itemize}
%Pedestals and noise are used online to evaluate the signal from each channel, while CalPeak and CalTmax \textbf{(reference to be added)} are used in the offline analysis, hence a periodical monitoring of these constants is essential to preserve SVD performance. 
The information on calibration constants is included in the configuration file used for data taking, which is generated using calibration constants measured during a specific local run. When a new local run is performed, calibration constants are compared with the ones used previously for data taking. If the differences in the calibration constants are significant, a new configuration file is generated using the calibration constants of the last local run.\par

%%%%%%%%%%%%%%%%%% Stability %%%%%%%%%%%%%%%%%%
\lvltwosection{Stability} % - 6 pages
\label{sec:C6_stability}

During the first two years of operation, a number of small issues and effects on sensors were also observed, and are reported in this section, together with the evolution of calibration constants, which were monitored taking local runs frequently.

%%%%%%%%%%%%%%%%%% Evolution of calibration constants and defective strips %%%%%%%%%%%%%%%%%%
\lvlthreesection{Evolution of calibration constants and defective strips}
\label{sec:C6_evolution_of_constants_and_strips}

Before the accelerator started operating, calibration constants were very stable, while after the start of beam operations an evolution related to radiation damage effects was observed. The largest effect was visible in Layer 3 sensors, which are the most exposed to machine induced background. 

\begin{figure}[hbt]
    \centering
    \includegraphics[width=15cm]{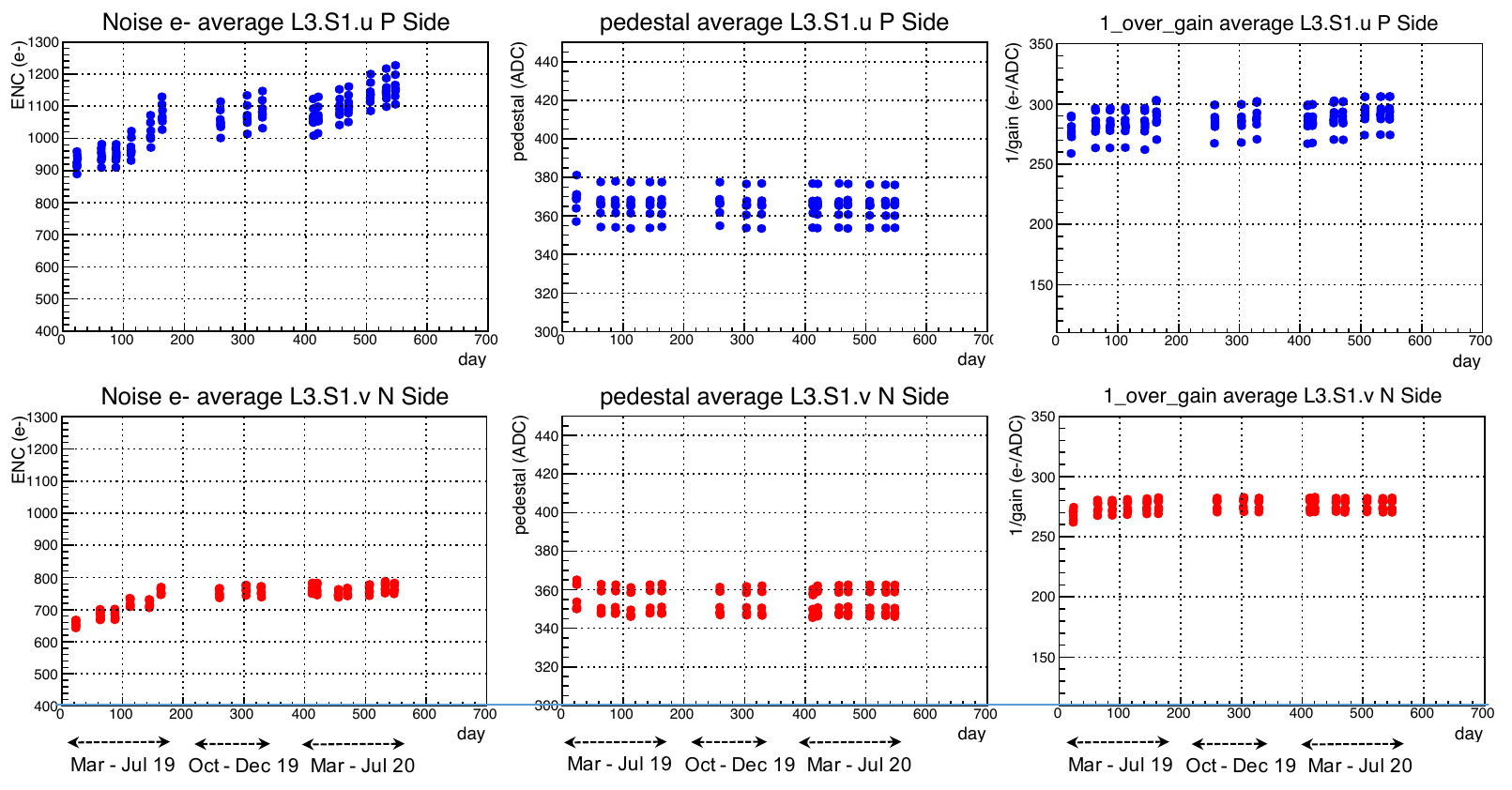}
    \caption{Temporal evolution of calibration constant for Layer 3 FW sensors. Top plots refer to u/P side, bottom plots refer to v/N side. From left to right, the plots show the evolution of noise, pedestals and 1/gain.}
    \label{fig:C6_noise_ped_gain_L3_S1}
\end{figure}
%- plot of noise/pedestal average vs time \par

The evolution of calibration constants from March 2019 to July 2020 is shown in figure~\ref{fig:C6_noise_ped_gain_L3_S1} for Layer 3 sensors in the forward region. Large changes were observed on noise levels due to radiation damage: in Layer 3 mid plane, the most exposed to radiation, a noise increase up to 15\% and 25\% was observed on v/N and u/P sides, respectively, with some saturation effect already visible. In the external layers exposed to lower dose the increase in noise levels was around 5-10\% only until July 2020. More details on the evolution of the noise are shown in section~\ref{sec:C6_radiation_damage_noise}. Pedestals initially changed by about 2-4\% and then stabilized; a small reduction of the gain around 5\% was also measured.

The number of defective SVD channels, i.e. strips masked on the basis of their anomalous values of noise, pedestal, and gain from local runs, was monitored and was very stable during all 2019 runs and the first two runs of 2020. The total number of masked strips was about 1\% on each sensor's side since the beginning of operations in March 2019. The number of masked strips is mainly due to the number of defective strips in the produced ladders, related to sensor defects, which had already been identified during ladder production. A few additional defects, mainly short- and open-circuits, were introduced during ladder assembly. There are also a few “hot strips” that developed during data taking; these strips become very noisy after some time, producing a significant increase in the sensor \gls{occupancy} as well as distorting the distribution of the relevant quantities for the sensor performance monitoring. Hot strips need to be masked out so that they do not affect data quality monitoring. In total, only 16 new hot strips developed in the entire SVD in 2019, and 3 in 2020, corresponding to 0.008\% of the total SVD strips.  The overall number of masked strips (including defective and hot strips) remained basically constant throughout the run: the total fraction of masked strips at the end of 2019c run was 0.96\%, with 1.00\% of strips masked on the u/P side and 0.86\% of strips masked on the v/N side.
Pinholes are not masked, because they are normally compensated by the \vsep{} = $-0.8$ V (see \ref{sec:C3_APVDAQ} for the \vsep{} definition).
\par

%%%%%%%%%%%%%%%%%% Issues on specific sensors %%%%%%%%%%%%%%%%%%
\lvlthreesection{Issues on specific sensors}\label{sec:C6_sensors_issues}
During the 2019 run, some peculiar behaviour of a few sensors was observed, and is briefly described in this section. It should be noted that most of these effects did not cause any performance deterioration, except for one single chip in Layer 3 that was excluded from data taking (causing an efficiency loss in the corresponding sensor) for a few months in spring 2019 and which was later recovered during the summer shutdown.

\textbf{Connection issue} - From March 10, 2019, one APV25 chip of sensor L3.2.2.n was responsible for some SVD errors during data taking. At first, it was thought that ADC delay scan or FIR parameters were changed or wrongly evaluated during the last local runs, causing the errors. Other sets of local runs were taken to verify this hypothesis, but the errors were still issued even during local-run data taking. The only way to proceed with data taking without issuing errors was to disable the problematic APV25 chip, losing a fraction equal to 1/84 of the Layer 3 v/N side acceptance. \par
When the access to FADC crates was granted during a maintenance day, attempts to reconnect data cables on the FADC board front panel and to use different FADC channels were made without success. Once access to dock boxes was granted at the end of 2019b run, data cables were disconnected and reconnected on the junction boards. Tests performed after the reconnection showed a recovery of the DC/DC converter voltages, a good signal output coming from the APV25 chip previously disabled, and good noise levels. The issue was considered solved and the full SVD was again available for data taking. However, at the beginning of 2020a run, DC/DC converter voltages corresponding to the same APV25 chip decreased to a level lower than normal, indicating a remaining potential issue with the connection. The lower DC/DC voltages did not affect the performance of the APV25 chip, which was subsequently included in data taking. \par

\begin{figure}[hbt]
    \centering
    \includegraphics[width=15cm]{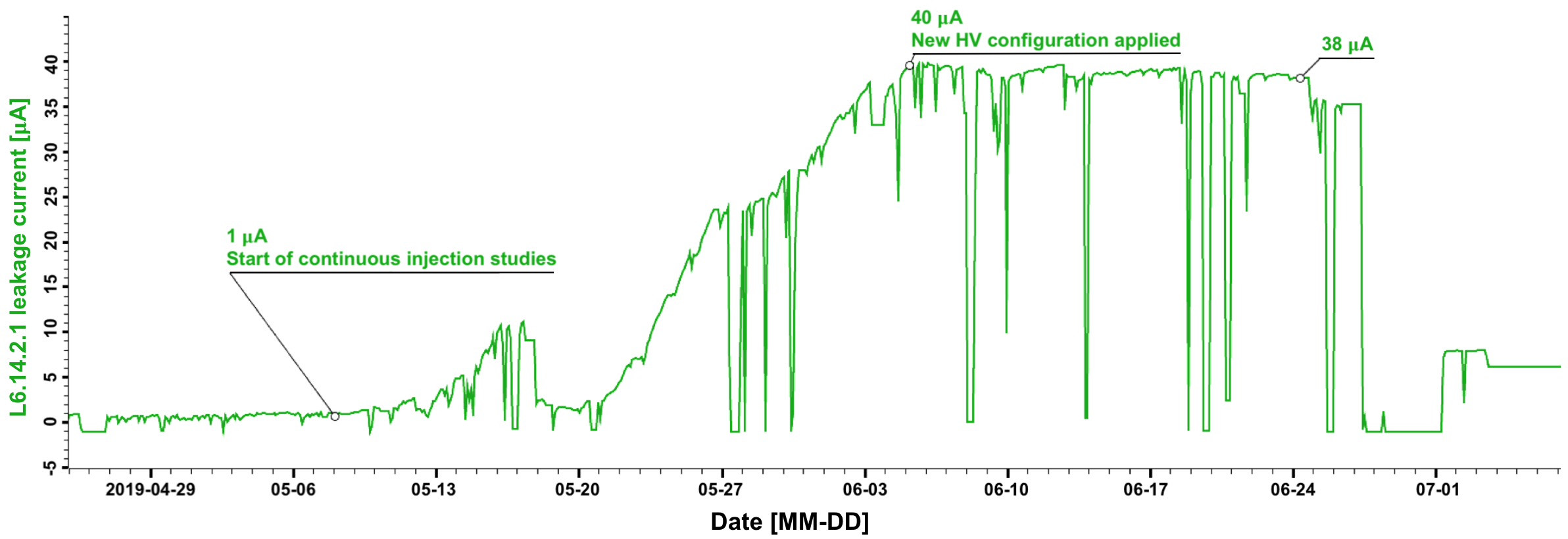}
    \caption{L6.14.2 v/N side leakage current evolution from the start of continuous injection studies to the end of the run.}
    \label{fig:C6_L6-14_current_evolution}
\end{figure}

\vspace{5mm}
\textbf{High sensor current} - When the machine started using continuous injection, which entails keeping the beam currents almost constant by performing top-up injections for both beams, an anomalous rise of the leakage current of sensor L6.14.2 v/N side was observed. The rise was seen only when beams were continuously operated and the bias voltage was applied to the sensors, while no increase was visible with the bias voltage applied but no beams in the machine. The increase in the leakage current was around  \qty{4}{\uA}  per day, as shown in figure~\ref{fig:C6_L6-14_current_evolution}. After one week, SVD was unbiased for some days, while the machine was operating with high currents, and this caused a sudden decrease of the L6.14.2.n leakage current. As soon as the detector was biased again, the increase restarted at the same rate as before, reaching  \qty{40}{\uA}  after two weeks of operation. \par

The hypothesis to explain this behaviour was that some positive charge had accumulated on the u/P side of sensor L6.14.2 due to both ionization created by beam background radiation and the presence of an electric field in air between the v/N side of Layer 5 ladders and the u/P side of Layer 6 ladders. The positive charge created on top of the u/P side can increase the electric field at the junction side, close to a sensor defect, and could increase the leakage current due to generation from impact ionization at that position. The hypothesis was consistent with the observation that when beams were stopped the increase in the leakage current also stopped, confirming the relation with beams. A similar effect was also observed in the past in the BaBar silicon sensors~\cite{TheBABAR:2013jta}. %(Nuclear Instruments and Methods in Physics Research A 729 (2013) 615–701, already in bib)
In addition, a few additional small groups of noisy strips on the two sides of the sensor were observed during local runs taken with high leakage current. On the u/P side the new noisy strips developed close to a preexisting pinhole, probably due to a local defect on the sensor that prevented the healthy growth of the oxide. This was an additional indication that the high current could be localized in that part of the sensor, at the intersection of the two groups of noisy strips on either sides of the sensor, close to a preexisting defect, with the effect being constantly aggravated by an accumulation of positive charges on the u/P side surface of the sensor.

To verify the hypothesis of higher leakage current due to impact ionization, the sensor leakage current was measured at different temperatures (changing the set point of the cooling plant by 5 and \qty{10}{\degreeCelsius}). All sensors in the same Layer 6 ladder doubled the leakage current with a temperature change of \qty{7}{\degreeCelsius} , as was expected for a leakage current due to thermal generation in the bulk or surface. On the problematic sensor, the current increased by only 10\% with the higher temperature setting, supporting the hypothesis of generation due to impact ionization. \par
\begin{figure}[hbt]
    \centering
    \includegraphics[width=12cm]{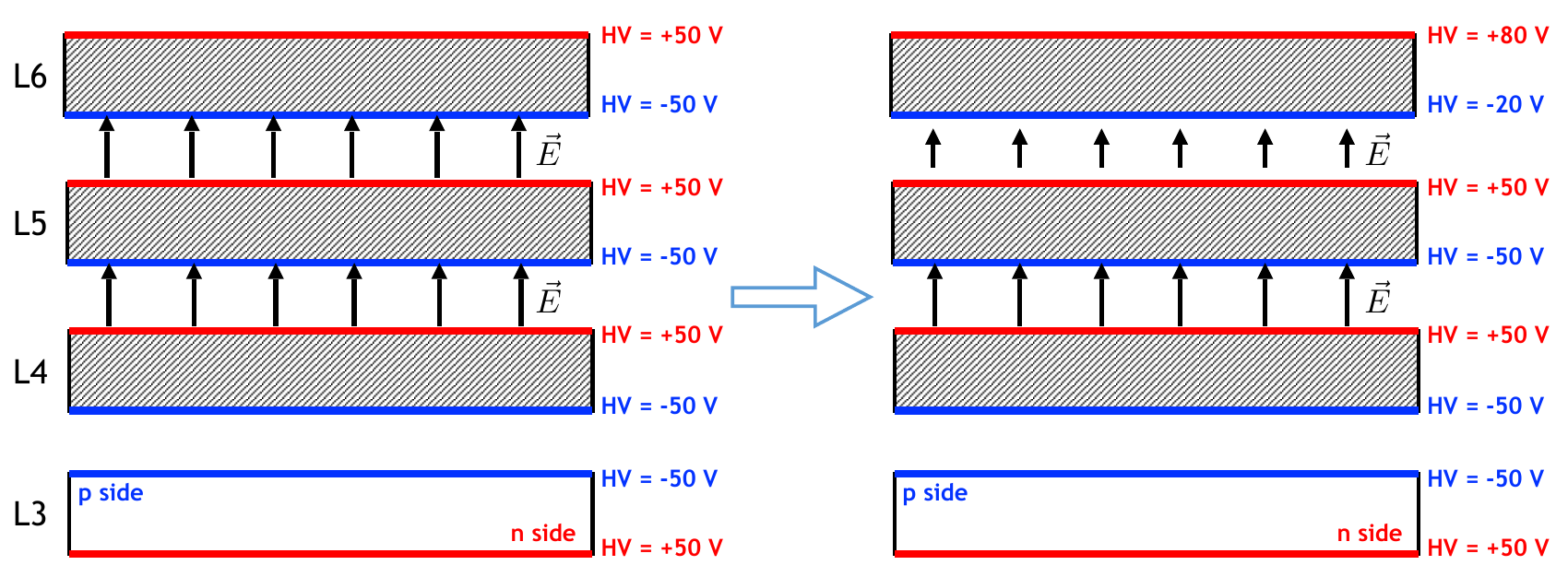}
    \caption{High voltage configuration between the four layers of the SVD. On the left side the original configuration is shown, with symmetrical bias voltages on P and N sides for all layers. On the right side, the modified configuration is shown, with Layer 6 sensors' voltages changed so that the electric field between the v/N side of Layer 5 and the u/P side of Layer 6 is reduced to 70~V~\cite{Rizzo:2021pgu}.}
    \label{fig:C6_HV_configuration}
\end{figure}
In order to mitigate the problem, which could lead to the sensor reaching its breakdown point, the electric field in air between the v/N side of Layer 5 and the u/P side of Layer 6 was reduced by applying different absolute values of high voltage in Layer 6: keeping $\Delta V = \qty{10}{\V} $ between the u/P and v/N sides of L6 ladders, but reducing the electric field between the v/N side of Layer 5 and the u/P side of Layer 6. A schematic view of the solution is shown in figure~\ref{fig:C6_HV_configuration}.

The new \acrfull{HV} configuration was applied in June 2019 with an HV shift of 30~V, which reduced the voltage between the v/N side of Layer 5 and the u/P side of Layer 6 to 70~V. The decrease in the electric field between Layer 5 and Layer 6 ladders was sufficient to arrest the increase in the sensor leakage current, that remained stable around \qty{38}{\uA}  until the end of the 2019b run. During the summer 2019, the current recovered to very low values, as expected due to the discharge, and started to rise with a smaller slope during the 2019c beam operation. The current reached a saturation value around  \qty{42}{\uA}  and remained stable until the end of the 2019c run. The HV shift adopted as a countermeasure is still in place, effectively mitigating this effect. No other sensors developed the same symptoms. \par

\vspace{5mm}
\textbf{Leakage current instability} - In May 2019, sensor L6.8.3 developed an instability in the leakage current, with spontaneous jumps that were not correlated to  particular interventions on either the detector or the machine. The jumps in the leakage current were symmetrical between P and N sides, with the highest recorded value around $16-17~\mu\mathrm{A}$. Some tests were made to understand the behaviour, but none of them gave a clear explanation of the phenomenon, although the IV curve showed some soft breakdown toward 100V. After some time, the leakage current went back to a low and stable value, within $5~\mu\mathrm{A}$, still being above the average with occasional small fluctuations. No particular countermeasures were taken against this issue, since its origin was not clearly understood, and the effect did not cause sensor performance degradation. \par

%%%%%%%%%%%%%%%%%% Turn ON effect of Micron sensors %%%%%%%%%%%%%%%%%%
\lvlthreesection{Turn-on effect of Micron sensors}
Already during commissioning, an effect was observed on some of the slanted sensors in the forward region: immediately  after turning ON the high voltage, the v/N side sensor noise and therefore occupancy increased with time, while the leakage current decreased with time, with a time constant up to 12 hours. Due to this effect, a local run taken right after turning ON the high voltage would underestimate the noise level.
The effect could be explained assuming that after the sensor bias is turned ON, positive
charges start to accumulate on the passivation layer of the N side. This effect can increase the extension of the electron accumulation layer,  always present on the N side at the Si-SiO$_2$ interface, close to the v/N side strip. The accumulation of negative charges extends the width of the v/N side implanted strip, therefore increasing  the inter-strip capacitance, which results in the observed increase in noise and occupancy on the v/N side. This mechanism also explains the reduction in the leakage current: electron accumulation reduces the extension of the surface depletion region that contributes to the surface leakage current, which results in a smaller current.
This effect was already observed~\cite{Chilingarov:2006sv}: the steady state conditions in the e$^-$ accumulation layer at the Si-SiO$_2$ interface is slowly reached, due to the movement of charge on top of the passivation layer, with a time constant that depends on humidity and bias history. This effect in our double sided sensors can influence both inter-strip capacitance and leakage current, resulting in the observed anticorrelation.
\begin{figure}[hbt]
    \centering
    \includegraphics[width=15cm]{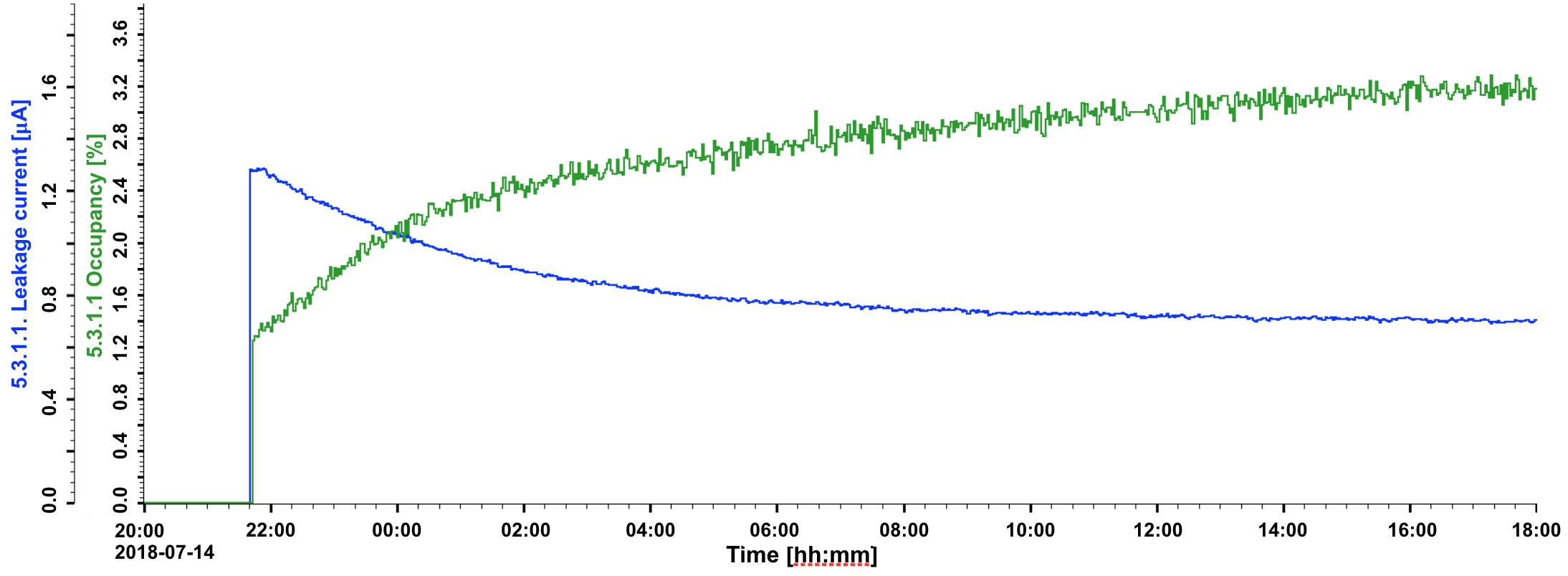}
    \caption{L5.3.1 v/N side leakage current (blue) and occupancy (green) after turning ON the sensor bias.}
    \label{fig:C6_turn_on_effect}
\end{figure} 
The behaviour is shown in figure~\ref{fig:C6_turn_on_effect}, where leakage current and occupancy are shown for the v/N side of L5.3.1 sensor: right after the high voltage is turned ON, the leakage current and occupancy start changing until, after some hours, the sensors settle to a condition where leakage current and occupancy have stable  values similar to other sensors. There is no specific cure for the observed issue, but it is important to avoid taking local runs right after turning ON HV, especially if HV was OFF for a long time, because in this way the noise would be underestimated, with the consequence of a high occupancy observed once these sensors reach the operating condition. \par

%%%%%%%%%%%%%%%%%% Power supply related issues %%%%%%%%%%%%%%%%%%
\lvlthreesection{Power supply related issues}
Two issues regarding power supplies arose in 2019.
The first issue was observed during the 2019a and 2019b runs, when a few channels of some \acrfull{LVPS} boards had a failure of a \acrshort{MOSFET}, and were temporarily removed from the experiment to replace the defective transistor. The availability of spare boards (25\% of those installed) allowed the experiment to keep running during the repair of problematic channels, reducing the unavailability of the SVD just to the time needed to diagnose the issue and replace the board, around two to three hours. As a precaution, at the end of the 2019c run, in all LV boards a systematic replacement of the problematic transistors with ones that have a higher maximum rating was made to avoid similar failures in the future. \par

A second issue was found that was related to the high voltage boards. It was observed, when setting HV = 0~V, that the leakage current of some sensors increased up to a few hundreds of \unit{\uA}. After tests performed with a high voltage spare board and a spare Layer 5 ladder, it turned out that under particular conditions the internal regulation circuit of the HV board caused an oscillation on the output voltage that could have resulted in a wrong sign current on the sensors. To avoid this possibility, a modification on the regulation circuit was made on one spare HV board, that was first tested with the spare Layer 5 ladder, then installed in the experiment for a couple of months to verify its functionality. After verifying that the modification had no drawbacks, at the end of the 2019c run the modification was systematically applied to all HV boards. \par

Both modification campaigns were performed by specialized CAEN~\citeCompany{ref:caen} technicians. Modified boards were first tested by technicians, tested with the spare system and then reinstalled in the experiment. All the 15 LV boards and 5 HV boards were successfully modified.

%%%%%%%%%%%%%%%%%% Radiation damage %%%%%%%%%%%%%%%%%%
\lvltwosection{Radiation Damage}
\editor{Antonio}
\label{sec:C6_radiation}

\begin{figure}[hbt]
    \centering
    \includegraphics[width=12cm]{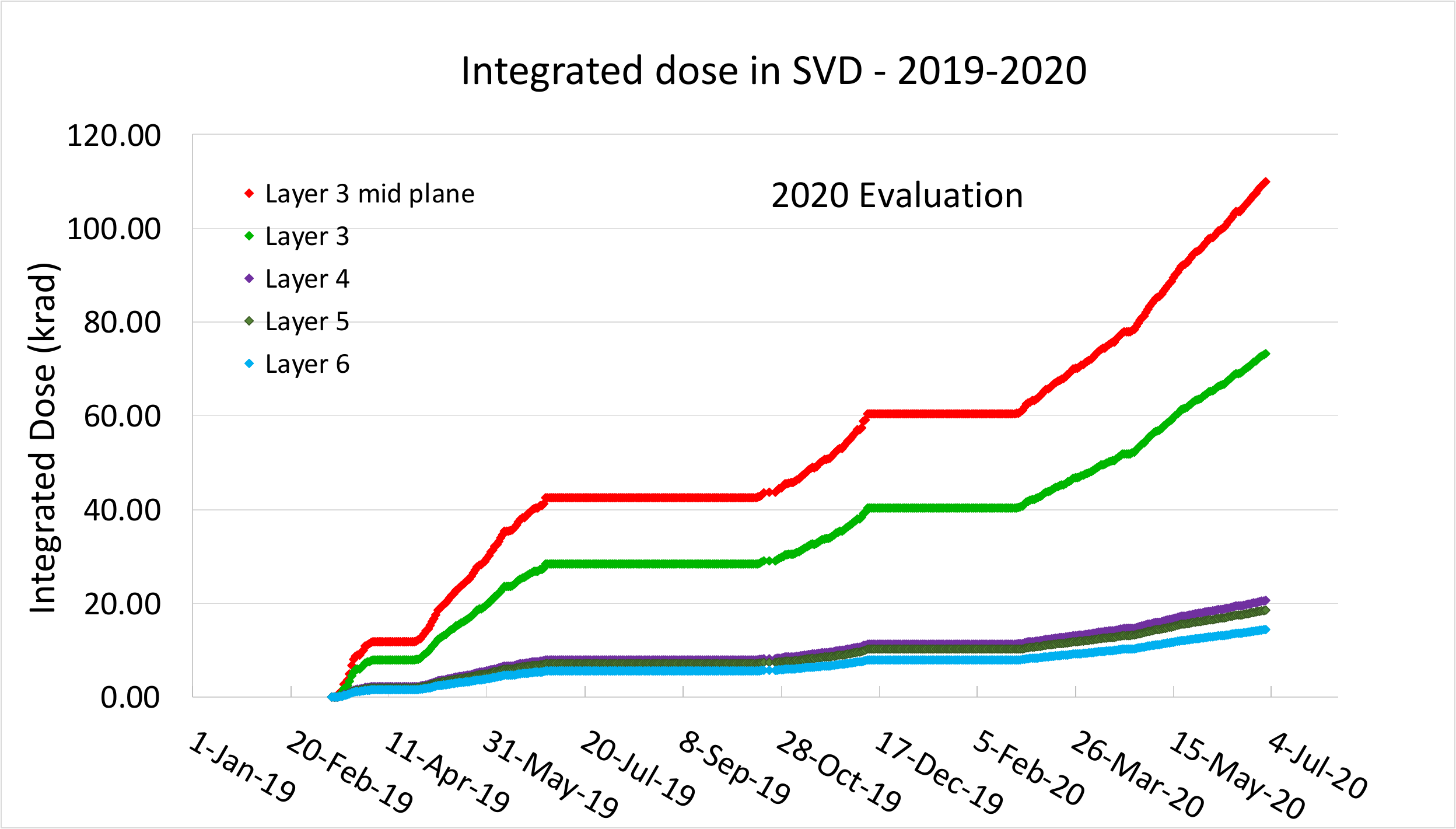}
    \caption{Integrated dose estimated for all SVD Layers as evaluated in 2020. See references in section~\ref{sec:conclusion} for updated radiation study results. Adapted from~\cite{Belle-IISVD:2021uar}.}
    \label{fig:C6_integrated_dose}
\end{figure}

Exploiting the good correlation between the SVD occupancy and the diamond sensors' instantaneous dose rate, it was possible to estimate the integrated dose  in the different SVD layers, which is shown in figure~\ref{fig:C6_integrated_dose} as evaluated in 2020. The ladders that received the highest dose are those in Layer 3 that lay on the horizontal plane, as expected. 
The method  has large uncertainty associated, and a more accurate analysis is under development,  but  this initial  estimate was very useful to correlate the effect of radiation damage seen in SVD with the expectations.

The 1~MeV equivalent neutron fluence in Layer 3 mid-plane sensors corresponding to the integrated dose was also estimated, using two methods. 
In the first method, also used for the discussion in section~\ref{sec:C2_rad_tolerance}, the full background simulation is performed to estimate, in the same location, the ratio between the integrated dose in SVD and the equivalent neutron fluence, which can be considered a reasonable estimate, although the absolute value of background from simulation is known to be affected by large uncertainties.
Since irradiation results~\cite{Ritter:2013cja} indicate that the energy-dependent damage factors used in the simulation to calculate the 1~MeV equivalent neutron fluence may be overestimated, an independent  calculation was also performed. The particle fluence, extracted from the measured occupancy, was translated to the 1~MeV equivalent neutron fluence using the \acrfull{NIEL} model and a damage factor of 0.01, measured in~\cite{Ritter:2013cja} for 10~MeV electrons. 
The dose to fluence factor in Layer 3 mid-plane, calculated from the average of the two methods described, is \qty{1.5e12}{n\ensuremath{_\mathrm{eq}}/\cm\squared} per Mrad.

%The expected 1~MeV equivalent neutron fluence corresponding to the, extracted using a data-driven scaling of background simulations, is also shown. The expected 1~MeV equivalent neutron fluence in Layer 3 mid-plane sensors was also estimated independently

\lvlthreesection{Noise}
\label{sec:C6_radiation_damage_noise}
As already shown in section~\ref{sec:C6_evolution_of_constants_and_strips}, after beam operations started, noise levels increased due to initial radiation damage on the sensor surface, with different effects depending on the sensor position.

\begin{figure}[hbt]
    \centering
\includegraphics[width=7cm]{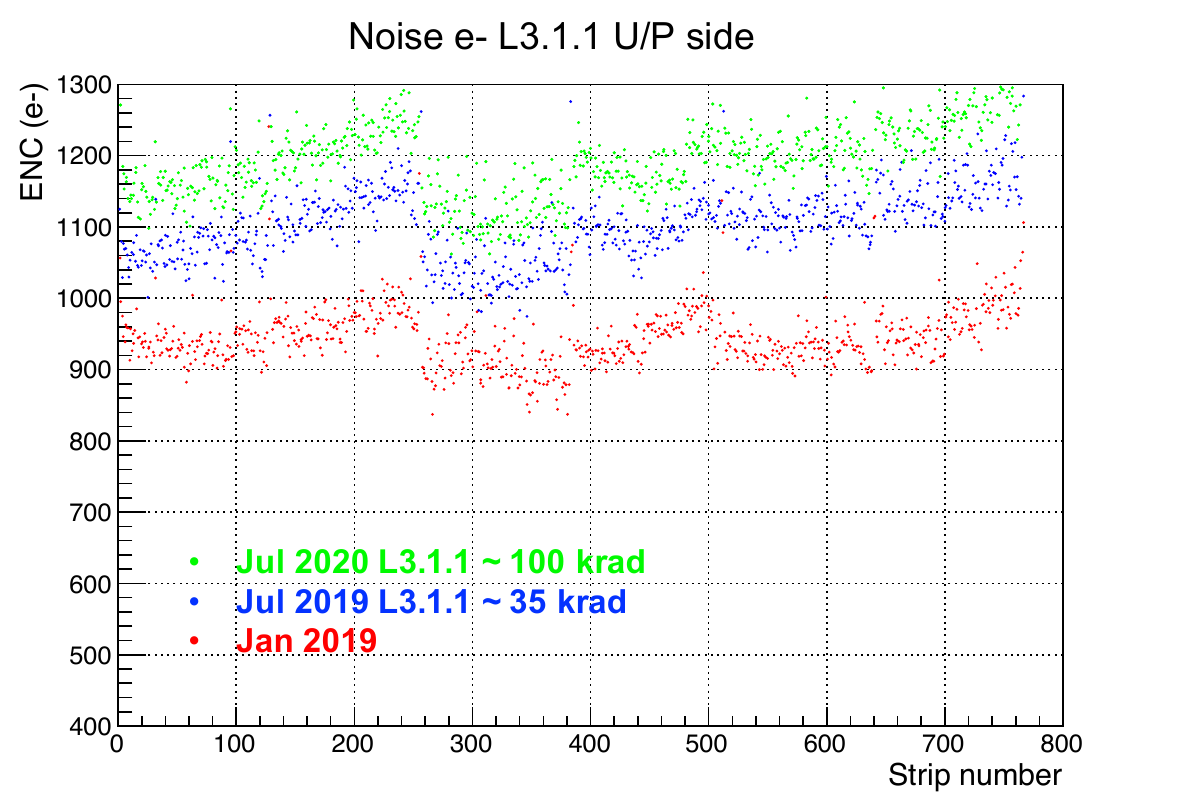}
\includegraphics[width=7cm]{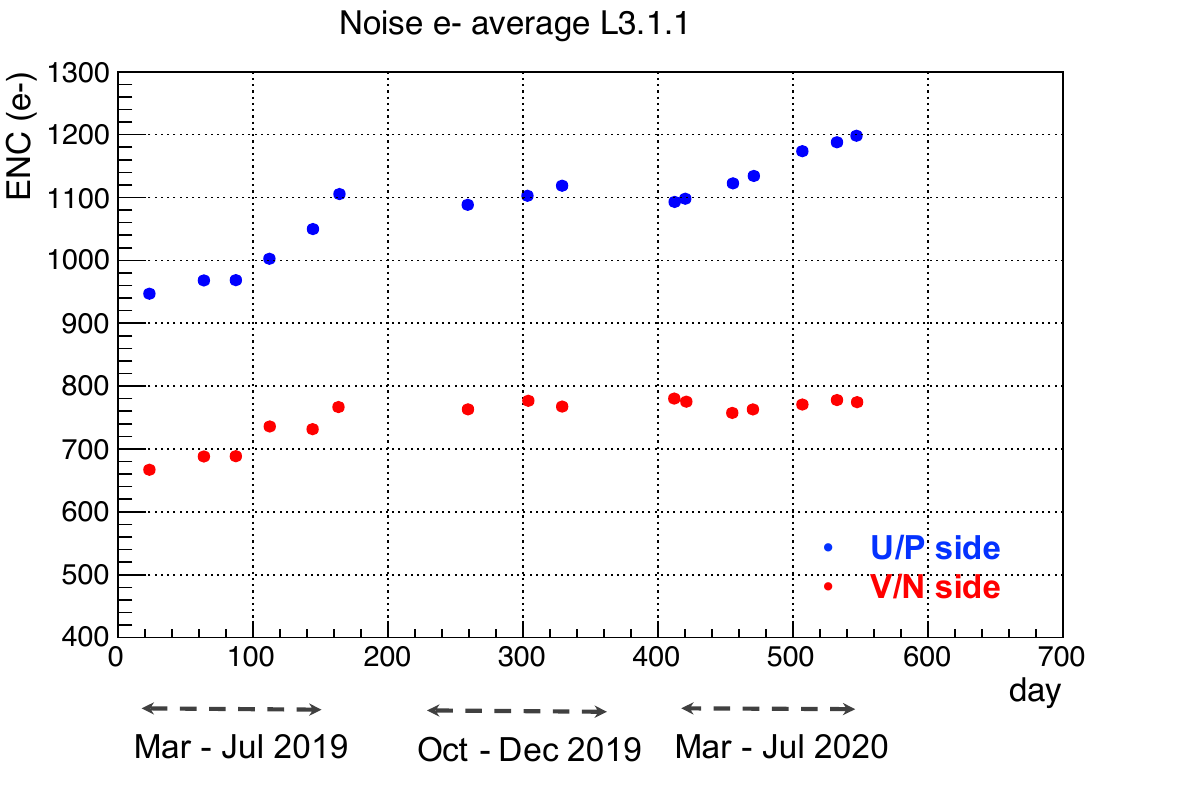}
    \caption{Noise evolution of sensor L3.1.1 u/P side between January 2019 (before starting operations) and July 2020. On the left, the change in the noise levels for the u/P side strips for a Layer 3 mid plane sensor is shown; on the right, the evolution of sensor average noise level vs time, from January 2019 to July 2020, is shown. Adapted from~\cite{Belle-IISVD:2021uar}.}
    \label{fig:C6_noise_evolution}
\end{figure}

An example is given in figure~\ref{fig:C6_noise_evolution}: on the left the change in the noise levels for the u/P side strips for a Layer 3 mid-plane sensor is shown; on the right, the evolution of sensor average noise level vs. time, from January 2019 until July 2020, is reported.
As expected, the noise started to rise due to the radiation damage effect: on the sensor surface the higher fixed oxide charge, induced by irradiation, increases the inter-strip capacitance and therefore the strip noise. After the first months, in July 2019, the noise had increased by about 15\% on both sides. On the v/N side the increase in noise was already almost saturated at that point, while on the u/P side the noise continued to rise during the 2020 runs, but with a lower slope, reaching about 25\% increase in the considered Layer 3 mid-plane sensors. The noise increase is also expected to saturate in the u/P side, since the increase in the fixed oxide charge also saturates. In Layers 4, 5 and 6, exposed to lower dose, the increase in noise was at the level of 5-10\% only, as of July 2020. 
%The noise levels increased by $\sim$10-12\%. A similar comparison was done at the end of 2019c run, but in this period the change in noise levels was lower than during 2019b. This result is consistent with expectations due to initial bulk damage due to incident radiation absorbed by silicon sensors. The change in noise levels caused a small change in the signal-to-noise ratio, which anyway didn't affect the detector performance.

\begin{figure}[hbt]
    \centering
\includegraphics[width=14cm]{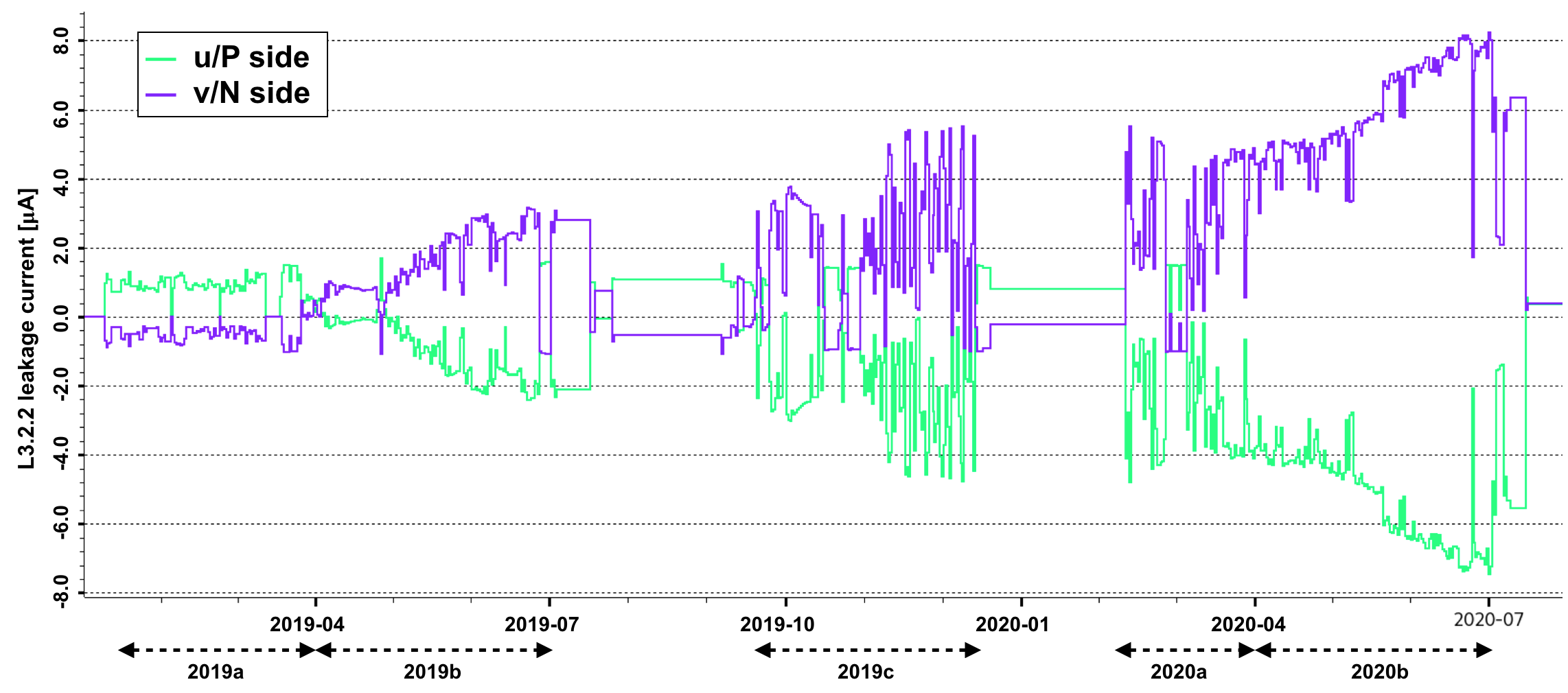}
    \caption{Evolution of leakage currents of one backward Layer 3 sensor up to July 2020.}
    \label{fig:C6_leakage_evolution}
\end{figure}

\lvlthreesection{Leakage currents}
Radiation damage is also responsible for the increase in the sensor leakage current. The effect is more visible in Layer 3 sensors, where the radiation absorbed is higher.
Figure~\ref{fig:C6_leakage_evolution} shows the evolution of the leakage current of one backward Layer 3 sensor up to July 2020. 
The general trend shows a clear correlation of the increase in leakage current with beam operations. Some \gls{annealing} effects are also visible when the detector was operated with no beams. In the first months of 2020, before the start of the operation with beams, a reduction of the leakage current in small periods is visible in the plot and it is correlated with the operation at lower temperature, which coincided with the \acrshort{PXD} being off.

It has to be underlined that during the 2020 run period there were a few changes in the temperature close to Layer 3, by around 2-\qty{3}{\degc}, related to changes in the power settings of the PXD, which sometimes caused significant changes in the Layer 3 sensor current.
Although these changes were compensated for during data taking by promptly updating the calibration constants, the detailed analysis of the leakage current increase and its correlation with dose is more complicated because of the temperature dependence.

\begin{figure}[hbt]
    \centering
\includegraphics[width=8cm]{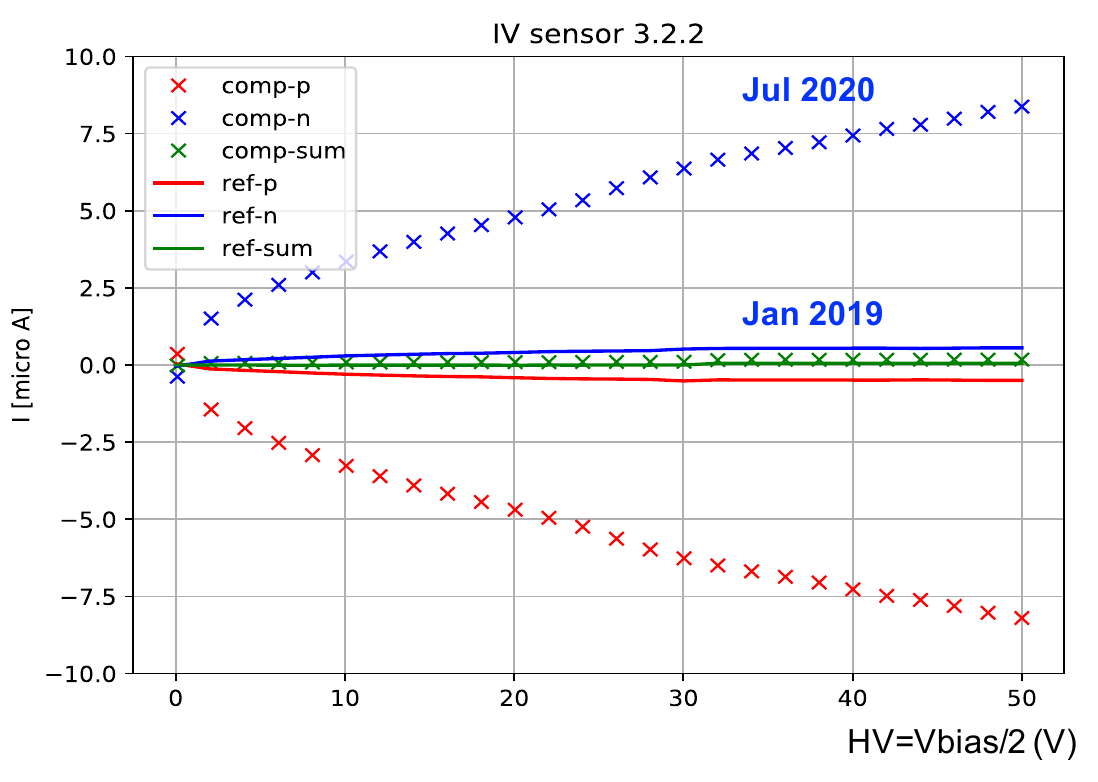}
    \caption{Comparison among IV measurements for one of the Layer 3 sensors in the mid plane (L3.2.2) taken in Jan 2019 (line) and July 2020 (crosses).}
    \label{fig:C6_IV_current_L3_2_2}
\end{figure}

The leakage current increase for one of the Layer 3 sensors in the mid plane (L3.2.2) is visible in figure~\ref{fig:C6_IV_current_L3_2_2}, where the comparison of two IV measurements taken in January 2019 (before operations with beams started) and in July 2020 is shown.
After exposure to around 100~krad integrated dose, the increase in leakage current for the mid-plane sensors was of about  \qty{8}{\uA}, both due to surface and bulk damage, although it is difficult to distinguish between the two sources.  

The observed increase is in reasonable agreement with the expectation based on the increase in leakage current measured in similar sensors of the BaBar experiment~\cite{TheBABAR:2013jta} of about \qty{1}{\uA/Mrad/\cm\squared} at  \qty{20}{\degreeCelsius}.
The leakage current increase measured in the BaBar sensors, and confirmed with irradiation campaigns to be mainly due to bulk damage from low-energy electrons is  a good estimate of the effect  expected in the SVD, since the radiation fields in BaBar and \belletwo{} vertex detectors are expected to be similar. 
With an estimated integrated dose of about 100~krad this increase rate corresponds to an \qty{8}{\uA}  current increase in Layer 3 mid-plane sensors (\qty{47.5}{\cm\squared} area), assuming a running temperature of  \qty{26}{\degreeCelsius}.  
%During the BaBar experiment a detailed irradiation campaign with 0.9 GeV e- was performed on some sensors and test structures to prove the origin of the effects seen on leakage current and depletion voltage in the BaBar SVT sensors were due to bulk damage, caused by relatively low energy e- \cite{TheBABAR:2013jta}). 

A calculation of the expected current increase from NIEL scaling using the estimated neutron fluence was also tried. Considering the current related damage rate $\alpha = \qty{4e-17}{\A/\cm} $ \cite{Lindstrom:421210} and the dose to fluence factor of \qty{1.5e12}{n\ensuremath{_\mathrm{eq}}/\cm\squared} per Mrad, the expected increase in leakage current at \qty{20}{\degreeCelsius}  in our \qty{320}{\um} thick sensors would be \qty{1.9}{\uA/Mrad/\cm\squared}. This is a factor two higher than the estimate calculated with the previous method based on BaBar data, but here no detailed estimation of the annealing effect with temperature and time is considered, and this can reduce the $\alpha$ parameter even by a factor two.

The SVD performance did not change and is not expected to deteriorate with the increase of leakage currents. In fact, even with an absorbed dose up to 10~Mrad, the strip noise contribution from leakage current, thanks to the very short APV25 shaping time, is expected to be very small with respect to the main capacitance contribution.

\lvlthreesection{Depletion voltage monitor}

An effective and relatively fast method to monitor possible changes in the depletion voltage due to bulk damage was developed. 
%which can vary the effective doping concentration in the coming years up to type inversion with equivalent neutron fluence above  a few $10^{12} \ \rm{n_{eq}/cm^2}$. 
The depletion voltage for the installed sensors is evaluated with a simple set of calibrations, measuring the noise as a function of the bias voltage applied. With this method there is no need to use beam time to monitor the depletion voltage, as in other techniques based on the reconstruction of the clusters with different bias voltages. In the $\rm p^+/n/n^+$ double-sided sensors used in SVD, the depletion region expands from the P-side, so the N-side strips become insulated only when the N-type bulk is fully depleted. When N-side strips are insulated, their noise drops to a minimum level, indicating that the full depletion voltage is reached.
With over-depletion voltage applied, the noise on the N-side continues to slightly decrease: the inter-strip capacitance, which is the dominant noise contribution, is proportional to the ratio between the strip implants width and the pitch. The over-depletion bias reduces the electron accumulation layer present on the N-side surface, which on N-type bulk sensors causes a reduction of the effective width of the N-strip implant.

%On n-type bulk sensors the effective width of the n-strip implant is  extended by the electron accumulation layer present on the n-side surface. Since over-depletion reduces the electron accumulation layer the noise is also reduced. 

\begin{figure}[hbt]
    \centering
\includegraphics[width=\columnwidth,keepaspectratio]{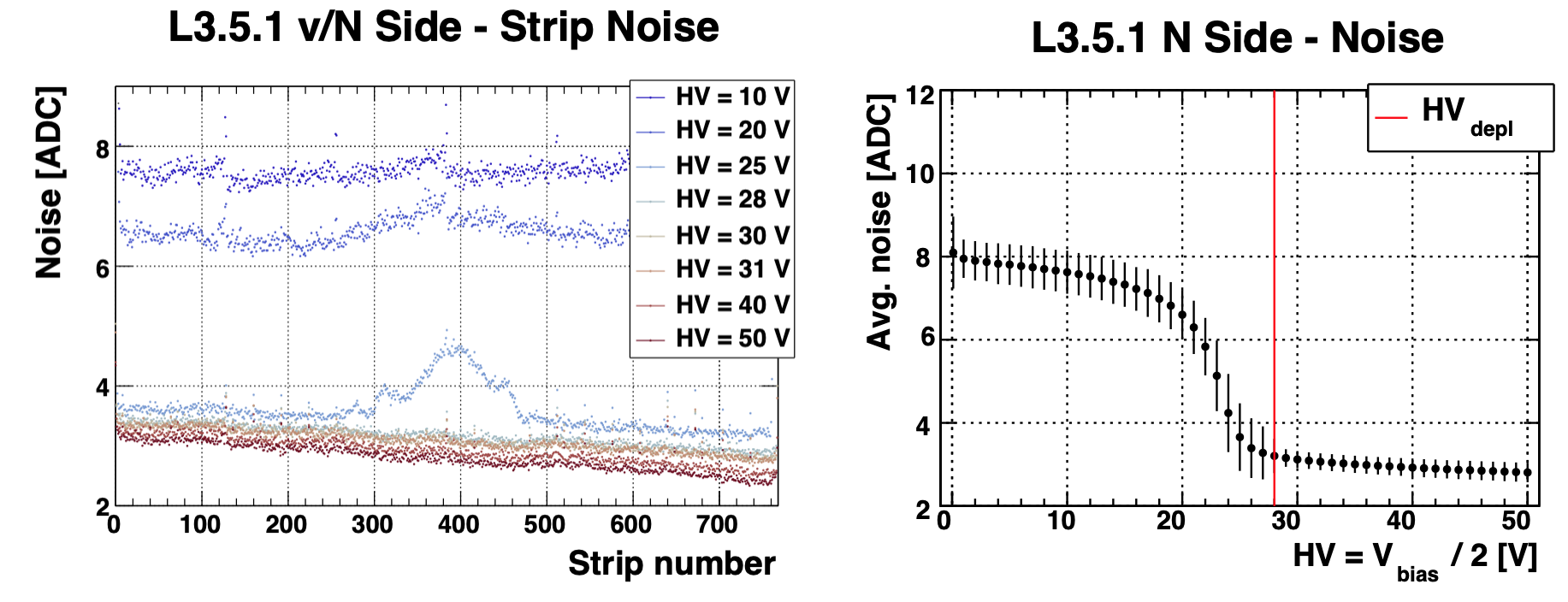}
  \caption{Example of the noise in a L3 sensor during the July 2020 HV scan, used to measure the depletion voltage. The indicated HV is half of the bias voltage. On the left, the N-side strip noise as a function of the strip number for increasing HV is shown. Red points correspond to full depletion. The noise dependence on the strip number is due to the variation in length (and capacitance) of the pitch adapter traces connecting the sensor strips to the \apv.  The bump around strip 400 is due to a slightly lower resistivity in the wafer center, with a locally higher depletion voltage.  On the right, the average N-side noise as a function of the applied HV, with the red vertical line indicating the drop at V$\rm_{depl}$= \qty{56}{\volt} ($2\times$\qty{28}{\volt}). Adapted from~\cite{Rizzo:2021pgu}. }
    \label{fig:hv_scan}
\end{figure}

The method was proven to work, with initial scans taken in the spring/summer 2020, when still no changes in depletion voltage were expected given the low integrated fluence of about 
\qty{1.5e11}{n\ensuremath{_\mathrm{eq}}/\cm\squared}. 
Results obtained are in good agreement with the depletion voltage measured directly on the sensor before their assembly. An example of the behaviour of the N-side strip noise as a function of the bias voltage is shown in figure~\ref{fig:hv_scan} for a Layer 3 sensor, showing a depletion voltage of \qty{56}{\V}. 
All the 172 installed sensors have an initial depletion voltage between 20 and \qty{60}{\V}.  

\lvlthreesection{Additional defects after large beam loss}
\label{sec:C6_rad_spikes}
Silicon sensors and electronics are subject to damage when a relatively large amount of radiation is absorbed in a very short time. In accelerators, this can happen when beams become unstable and in case of beam dust events.
On June 9, 2019, a quench of some superconducting magnets of the final-focus system was caused by a so called ``beam dust'' event. During events of this kind beams became unstable, causing a sudden jump in background levels on the detector. It was estimated that in one of these events a radiation dose above 3~rad was delivered on the \belletwo{} detector in $\sim$40~$\mu$s. After this event, local runs were taken from each \belletwo{} subdetector to evaluate any possible damage caused by the high radiation dose deposited on the detector in such a short time. For the SVD, a comparison between the two local runs taken before and after this event showed that a few new pinholes were created. 
This result was somehow expected as observed in other experiments~\cite{TheBABAR:2013jta}. 
In normal conditions the voltage drop across the AC decoupling capacitors is very small,  thanks to the floating power supply biasing scheme (section~\ref{sec:C2_power_supplies}).
Bursts of radiation releasing a large amount of charge in the silicon can induce a rapid discharge of the sensor capacitance, causing  the full bias voltage (100 V) to be applied across the AC decoupling capacitors. Although these capacitors have a typical breakdown voltage well above 100V, and all of them were tested up to 20 V, this effect could induce a breakdown of the oxide layer in the weaker capacitors, creating additional pinholes. The limited number of new defects confirmed the robustness of the AC coupling capacitors and of the sensors. 
%This result was somehow expected to occur, and the limited number of new defects confirmed the robustness of AC coupling capacitors of the sensors.
A  protocol was established in order to perform local runs and check the health of each \belletwo{} subdetector every time a similar event occurs. \par

%%%%%%%%%%%%%%%%%% Observations on data during operation %%%%%%%%%%%%%%%%%%
\lvltwosection{Observations on Data During Operation}
\label{sec:C6_observations}
\editor{Yuma}
Analyzing the ADC distributions after the pedestal and common-mode-noise subtraction during the operation, it was found that, under certain conditions, the distributions are shifted by a quantity that depends on the strip position. With further investigations, the sources of this shift were understood and classified into three cases: crosstalk from APV25 data outputs, crosstalk from APV25 power inputs, and baseline tilt at high trigger rate. Although such effects in the ADC distributions are not desirable, the effect on the performance is found to be negligible. In this section, the sources and the characteristics of these three cases are explained.

\begin{figure}[hbt]
    \centering
    \includegraphics[width=8cm]{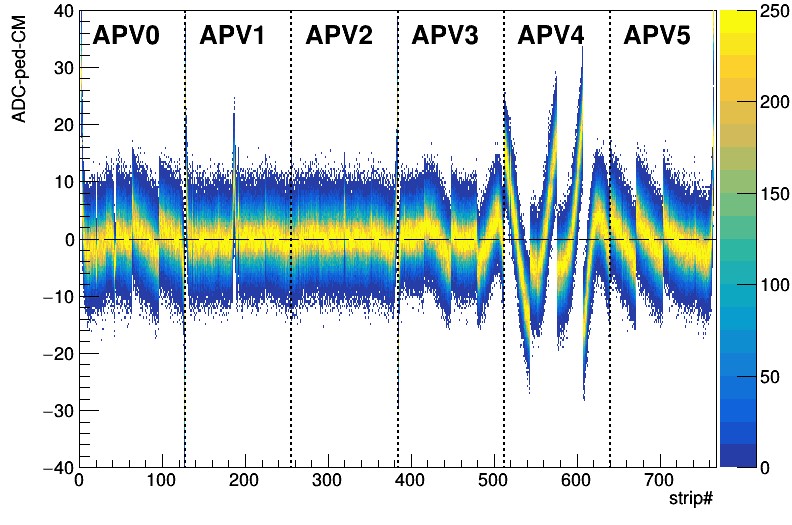}
    \caption{ADC distributions for strips in L4.1.2 u/P side, after the pedestal and \gls{common-mode noise} subtraction. Data are taken when the synchronization pulse is generated on the APV25 output lines. The ADC distributions are shifted due to the crosstalk from the synchronization pulse.}
    \label{fig:C6_crosstalk_noise}
\end{figure}

\lvlthreesection{Crosstalk from APV25 data outputs}\label{sec:crosstalk}
A part of the output lines of APV25 (see section~\ref{sec:C2_apv25}) on \gls{origamiboard} has capacitive couplings to the sensor strips, especially where they are aligned in parallel.
Thus, the APV25 output can create crosstalk signals on the corresponding strips through the couplings. These crosstalk signals become prominent when the APV25 output changes rapidly at the transitions between digital-low and digital-high signals.
This rapid transition happens when the APV25 transmits the tick mark for the synchronization (which is called `synchronization pulse') or the data header~\cite{French:2001xb}.
Figure~\ref{fig:C6_crosstalk_noise} shows the shifts of the ADC distributions observed at the synchronization pulse in a Layer 4 sensor. 
The most visible shifts are observed in APV4 (see \ref{sec:C2_origamiboards} for the APV25 numbering scheme on origami boards), where the output lines in the origami boards are parallel to the u/P side strips.

\begin{figure}[hbt]
    \centering
    \includegraphics[width=5.5cm]{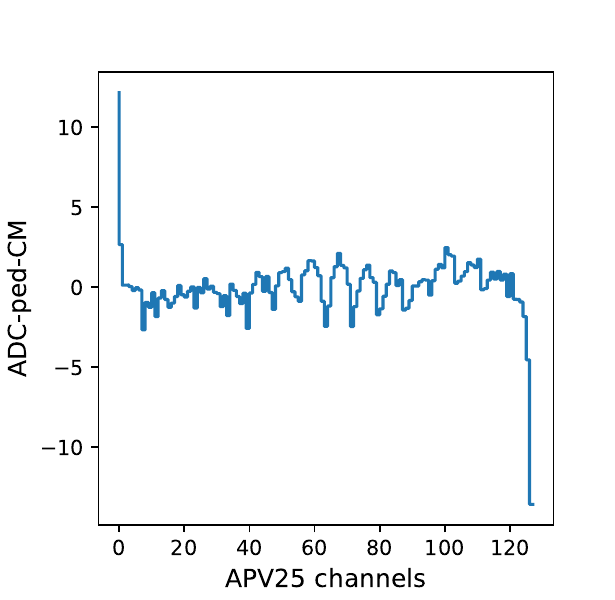}
    \includegraphics[width=7cm]{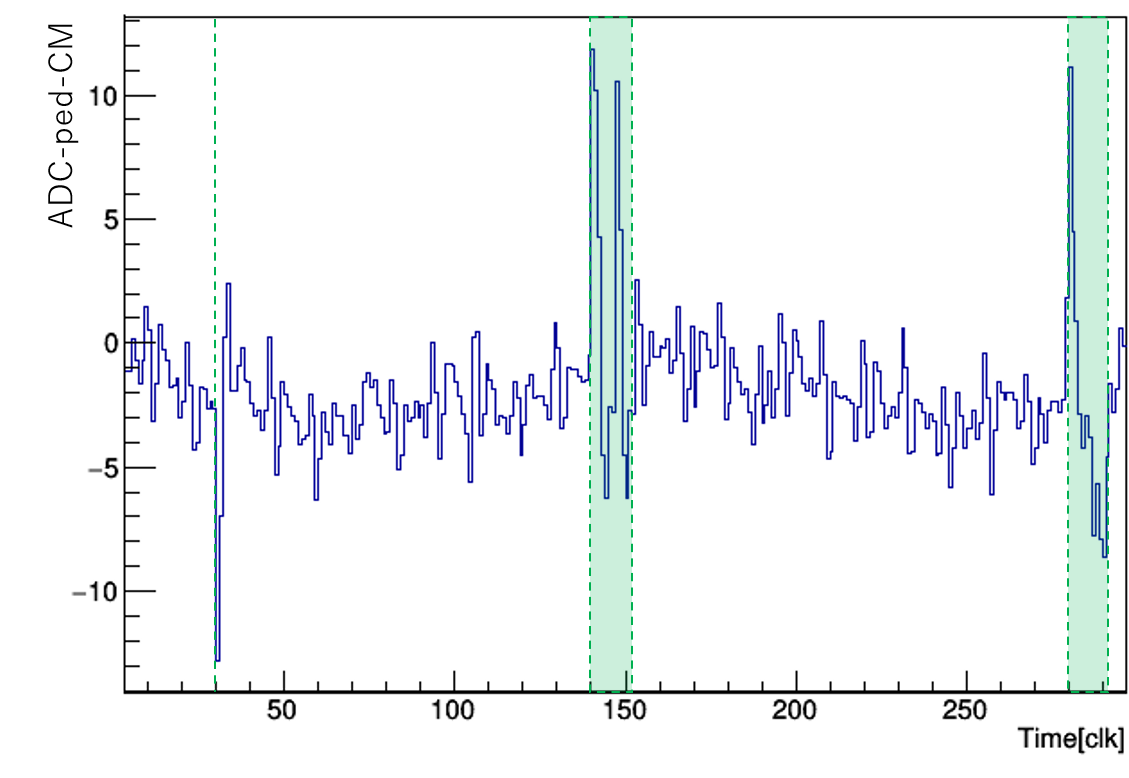}
    \caption{Left: mean values of the ADC distributions after the pedestal and common-mode-noise subtraction for the strips of one APV25 in L3.1.1 u/P side at the start of the APV25 data readout. The polarity of the spike is opposite between the two edges of the APV25 channels, reflecting the polarity of the power supply of $\pm$\qty{1.25}{\V}.
 Right: time dependence of the mean values at channel 0 in the same APV25 chip.
 The green shaded areas correspond to the data headers and the green dashed line is the timing of the retrieval of the data samples from the pipeline which happens 110 clock-cycles before the data header transmission.}
    \label{fig:C6_edge_effect}
\end{figure}

\lvlthreesection{Crosstalk from APV25 power inputs}
The wire-bonds and the bonding pads of the APV25 power inputs on the chip are adjacent to the strip-signal inputs of channels 0 and 127. Therefore, there are capacitive couplings between them.
The current in these power supply lines rapidly changes when APV25 processes data, as reported by the CMS collaboration~\cite{Hahn:2009rua}. Such rapid changes can cause crosstalk on channels 0,1 and 126,127 via the capacitive couplings.
The strip-number dependence of the ADC distribution shifts due to the crosstalk is shown in the left plot of figure~\ref{fig:C6_edge_effect}, in which a significant effect can be observed only on channels 0 and 127. The polarity of the spike is inverted between the two edges, reflecting the polarity of the power supply of +\qty{1.25}{\V} close to channel 0 and $-1.25$~V close to channel 127.
The time structure of the output in channel 0 when APV25 starts sending data is shown in the right plot of figure~\ref{fig:C6_edge_effect}. The structure is almost identical to that of the power supply current reported by the CMS collaboration; the ADC output also sharply increases at the start of data readout (i.e. at the data header) and decreases 110 clock-cycles before.
Due to this time structure, the ratio of the affected events is proportional to the trigger rate.

\begin{figure}[hbt]
    \centering
    \includegraphics[width=7cm]{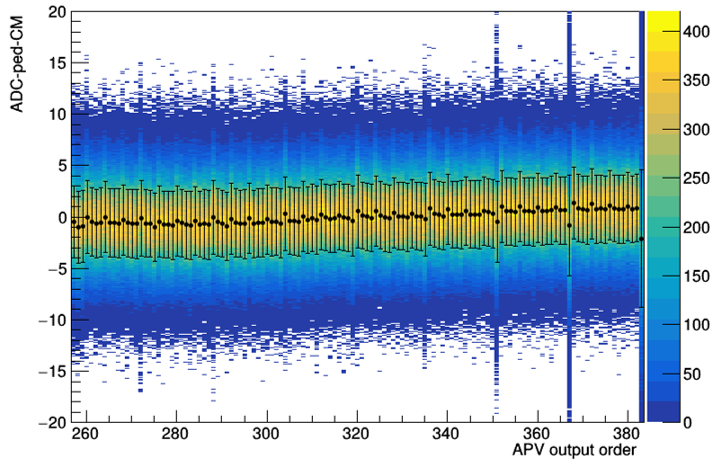}
    \includegraphics[width=7cm]{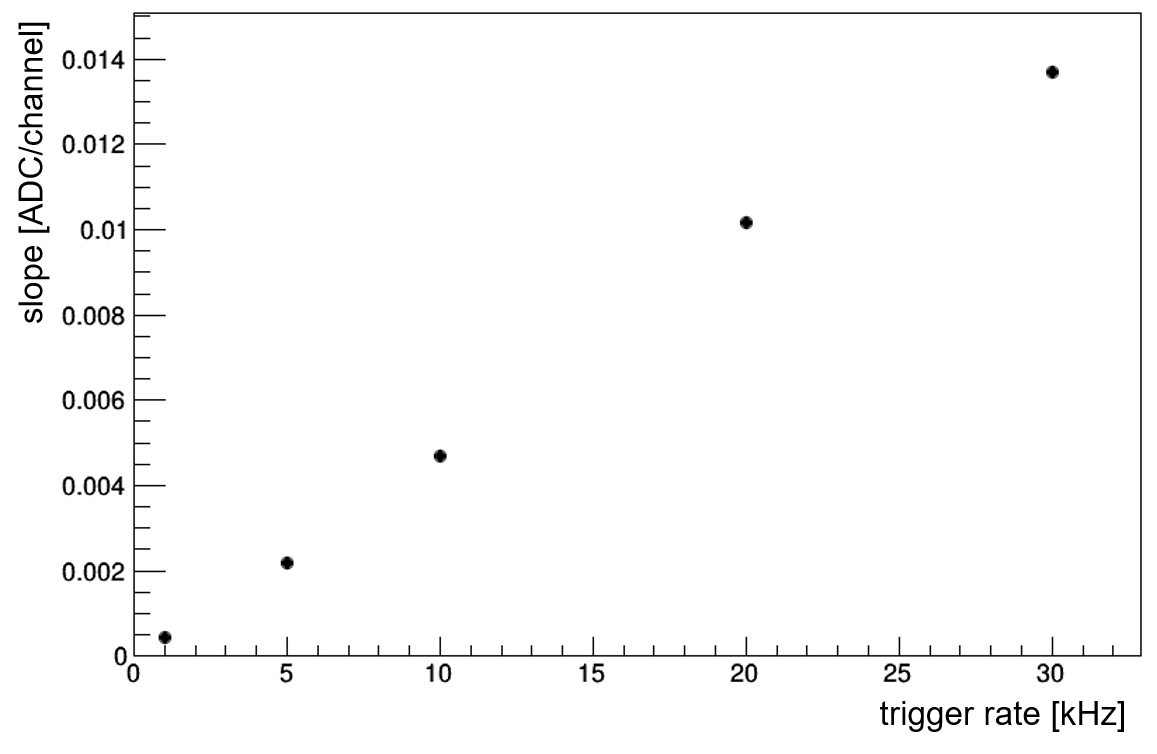}
    \caption{Left: ADC distributions for strips of one APV25 in L4.1.2 u/P side, after the pedestal and common-mode-noise subtraction. The data are taken with a 30 kHz Poisson trigger. The circle marker represents the mean value for each column, and the bar represents the standard deviation. The baseline tilts up to around $\pm$1~ADC. Right: the slope of the linear-fitting in the same APV25 chip for 1, 5, 10, 20,  \qty{30}{\kHz} trigger rate. The fitting uncertainties are smaller than the point size. There is a linear correlation between the slope and the trigger rate.}
    \label{fig:C6_baseline_tilt}
\end{figure}

\lvlthreesection{Baseline tilt at high trigger rate}
When the trigger rate approaches \qty{30}{\kHz}, another shift of the distribution is observed that corresponds to a linear increase with APV25 output order, as shown in the left plot of figure~\ref{fig:C6_baseline_tilt}.
% This shift comes from shifts in the ADC-zero level.
Since the data are read out only through AC-coupling in \acrshort{FADC}, the circuitry tries to balance out any excursions, so that the long-term average becomes equal to the DC operating point of each channel. This has some implications depending on the trigger rate.
% edited by MF upon request from Yuma:
At a low trigger rate, the digital baseline is by far the dominant signal, so it is almost equal to the operating point. In that case, an exponential decay in the analog strip signals is observed, because those are higher than the digital baseline. This is also the condition in which the pedestals are recorded.
However, at a high trigger rate, the actual strip data become the dominant signal, while the digital baseline largely disappears, and then the AC coupling mandates that the average strip signal is leveled out to the operating point. In that case, the strip signals are flat, and the exponential decay is observed only in the (rarely occurring) digital baseline.
This also implies that the pedestal subtraction overcompensates a slope that is not present anymore, and therefore introduces the observed rising slope.
%DC-component of the output such as the baseline of the analog strip data decreases exponentially to the %ADC-zero set at the average value of the signal output. 
%At a low trigger rate, the ADC-zero is almost equal to the digital baseline. The analog strip data show an exponential slope, which is compensated by pedestal subtraction with an exponential slope.
%However, at a high trigger rate, the dominant pattern is the analog strip data themselves and thus shows a smaller exponential slope.
%The pedestal subtraction overcompensates this slope and introduces another rising slope, as observed.
% As the trigger rate gets higher, this slope gets more tilted.
The tilted slope is linearly increasing with the trigger rate, as shown in the right part of figure~\ref{fig:C6_baseline_tilt}. The tilt cannot be mitigated by the current pedestal nor common-mode-noise subtraction.
The characteristic of this baseline tilt also appears in the occupancy.
If the baseline is above the average, the noise is more likely to exceed the signal threshold. Thus, the occupancy increases in half of the sensors, and decreases in the other sensors, but the effect is of the order of $10^{-4}$, so the performance is barely affected.

\begin{figure}[hbt]
    \centering
    \includegraphics[width=12cm]{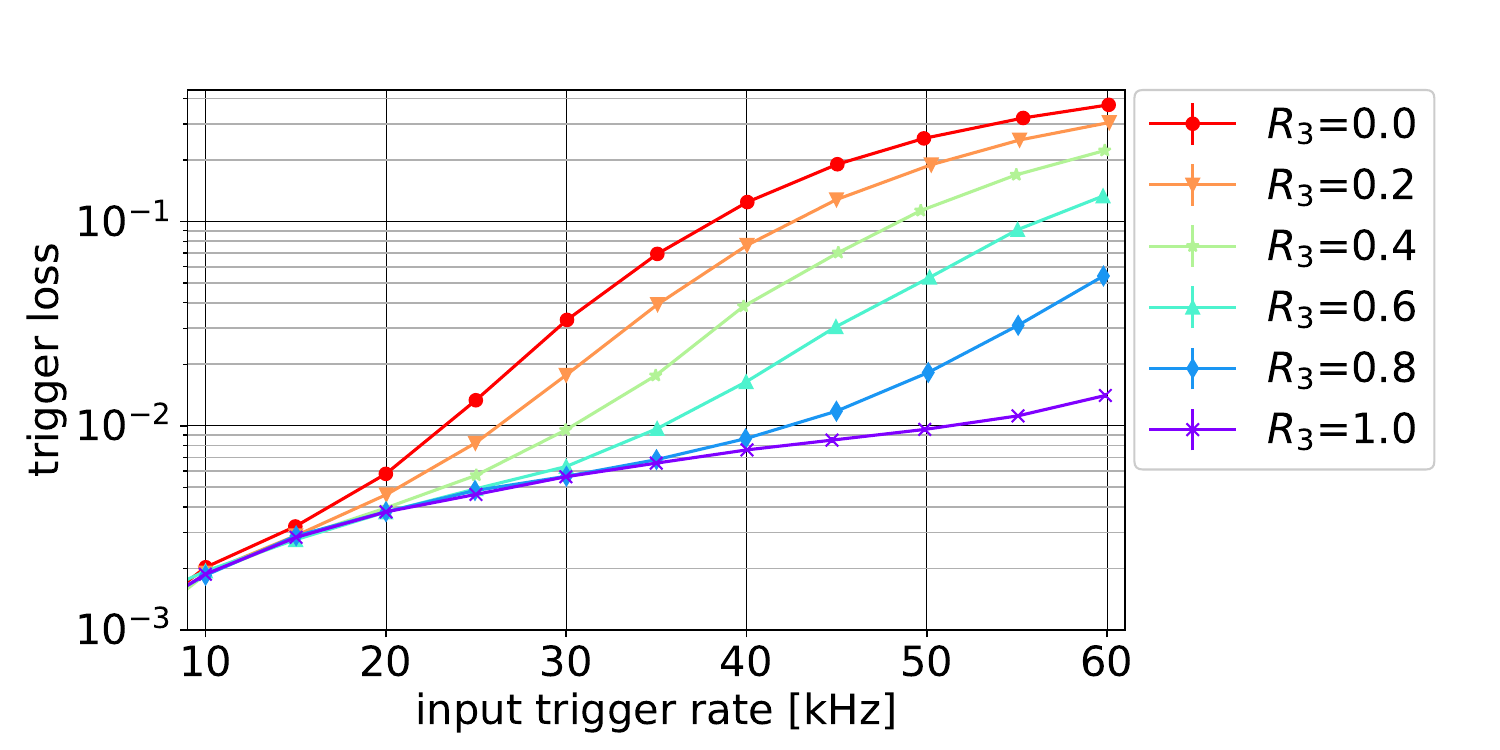}
    \caption{Dead time due to APV trigger veto versus input trigger rate in mixed 3/6-sample readout. The fraction of 3-sample triggers, $R_3$, is changed.}
    \label{fig:C6_APV_deadtime}
\end{figure}

%%%%%%%%%%%%%%%%%% APV trigger veto %%%%%%%%%%%%%%%%%%
\lvltwosection{APV Trigger Veto}
\label{sec:C6_apv_trigger_veto}
\editor{Hikaru}
The pipeline readout of the APV25 is implemented as a ring buffer and a \acrshort{FIFO} (see \ref{sec:C2_ondet_elec}). It is notable that the chip is not equipped with a mechanism to prevent the overflow of the FIFO. Without countermeasures, 6-sample readout at a trigger rate of \qty{30}{\kHz} causes a FIFO overflow error in a few milliseconds, which then requires resetting the chip. To overcome this error, the ``APV trigger veto'' module was developed and implemented in the firmware of the most upstream \acrfull{FTSW}~\cite{Nakao:2012aea}.
The module monitors trigger and reset signals that are eventually sent to the APV25 chips and calculates the future FIFO occupancy. The triggers are vetoed if they would overflow the FIFO. In this way, the system prevents the FIFO overflow with minimal dead time, rather than sending busy signals from front-end electronics. 
According to a simulation shown in figure~\ref{fig:C6_APV_deadtime}, the dead time due to the APV trigger veto is around 3\% for 6-sample readout of \qty{30}{\kHz} triggers following the Poisson distribution. The simplest alternative, vetoing until the FIFO gets empty after every trigger, would yield 44\% dead time. With the mixed 3/6-sample acquisition mode the dead time would be suppressed to less than 1\% if the fraction of 3-sample triggers is more than 0.4.

The trigger veto module was integrated to the DAQ since \phasetwo{} and is working as expected.

%%%%%%%%%%%%%%%%%% Beam loss monitor and aborts, environmental monitors and interlocks %%%%%%%%%%%%%%%%%%
\lvltwosection{Beam Loss Monitor and Aborts, Environmental Monitors and Interlocks}
\label{sec:C6_blm_env_intlks}
%\editor{Livio}
The monitoring of environmental parameters is a critical aspect of the operation and performance stability of SVD.  Some key aspects of the operational experience concerning the radiation monitor and beam abort, the temperature and humidity monitors, and the hardware interlock are summarized here. 
The details of these systems are described in \ref{sec:environment}.

After the installation and commissioning of the full system of 28 diamond sensors, in 2019 the DCU firmware was modified, to allow a faster abort cycle: \qty{2.5}{\us} instead of \qty{10}{\us}, as described in section~\ref{sec:radmon_ba}. As a result of the experience and improvements, the full system was recently operated with $24$ diamond detectors in monitor mode (in the most sensitive \qty{36}{\nA} range) and four beam-pipe diamond sensors dedicated to abort requests (in the \qty{4.5}{\mA} range), providing effective protection to both VXD and QCS magnets. The diamond system issued 305 beam-abort requests in 2019, and 655 beam-abort requests in the first semester of 2020. Two abort thresholds were recently applied in parallel:
\begin{itemize}
    \item ``very fast'': \qty{4}{mrad} integrated in a moving time-window of \qty{10}{\us}, updated and compared with threshold every \qty{2.5}{\us}; 
    \item ``fast'': \qty{40}{mrad} integrated in a moving time-window of \qty{1}{\ms}, with the same abort cycle.
\end{itemize}
The ``very fast'' threshold detects sudden large spikes in beam losses, while the ``fast'' threshold is meant to catch a slower building up of unwanted radiation doses. Most aborts were of the ``very fast'' type.

Data from the most sensitive ``range 0'' (section~\ref{sec:radmon_ba}) diamond sensors, read out at \qty{10}{\Hz}, have also been used to set up ``injection inhibit'' algorithms, to veto continuous injection when it results in large radiation doses. 

The \qty{10}{\Hz}  data are also used for the measurement of integrated radiation doses. The integrated dose is underestimated, if range 0 is saturated during the short injection time intervals. An upgrade program to improve the dynamic range of the diamond electronics and to avoid or mitigate such saturation effects is now considered.

The temperature monitoring by \acrshort{NTC} and \acrshort{FBG} on \acrshort{FOS} sensors (section~\ref{sec:temp_mon}) was in operation without failures since their installation. Archived monitor data provided useful correlations with the detector operation conditions. Hardware interlock signals on preset temperature thresholds behaved as expected.

The humidity monitoring system (section~\ref{sec:humid_mon}), after about two months of continuous operation in 2019, experienced some failures of Vaisala Dew Point Transmitter DMT143 sensors. The problem was traced to the fact that the frost point of the dry volume went lower than originally expected, typically down to about -\qty{85}{\degreeCelsius}, outside the measurement range of the instrument (-\qty{80}{\degreeCelsius} to +\qty{20}{\degreeCelsius}), causing very frequent self-calibrations and finally error conditions. The substitution of these sensors with Vaisala DMT152 (-\qty{100}{\degreeCelsius} to \qty{0}{\degreeCelsius}) fixed the problem.

Some hang-ups also occurred in the USB read-out connection: at some point no more data were coming from one of the four Vaisala sensors. As a result, the \acrfull{VLHI} system immediately shut down the VXD and set the IBBelle cooling plant (\ref{sec:C3_cooling_plant}) to warm operation (+\qty{15}{\degreeCelsius}), as expected when there is a failure of one of the environmental monitoring systems. The reason for the failure was found between the communication board of the humidity system and the corresponding \acrfull{IOC}: data corruption was interpreted by the IOC as a malfunction of the Vaisala sensor, causing the VXD interlock. Improving the code library used to manage the communication between the humidity system and the corresponding IOC made the communication more stable and avoided similar problems in the next runs. As noted in section~\ref{sec:humid_mon}, a new system, with different redundant humidity sensors and more robust Ethernet readout, was built at INFN Trieste and is ready for shipping and installation.

The VLHI hardware interlock system, described in section~\ref{sec:hard_interlock}, was in continuous operation without failures. It protected the VXD in a few events, due to alarms from the \COtwo{} cooling system or from water leaks sensors, detecting condensation of water on some cryogenic transfer lines. A complete spare system is ready for shipment and installation for the long-term operation of the experiment.

A small issue that caused unexpected VXD interlocks at the beginning of 2019a run was due to the chiller used for the dock-boxes cooling system. The chiller is connected to the VLHI system so that in case of failure the SVD is automatically shut down. At the beginning of 2019a run some interlocks were issued by the chiller line, although the chiller was working properly. Looking at the output of the low-pass filter module of the chiller that sends the signal to the VLHI hardware, spikes were observed when one of the chiller's relays switched. These spikes gave fake inputs to the VLHI system, that asserted the interlock at each time. The input was temporarily disabled and subsequently a software interlock was implemented on the water flow of the chiller. The possibility of filtering the short spikes coming from the chiller was considered and implemented in 2020, to re-enable the hardwired interlock of the chiller on the VLHI system. \par

%%%%%%%%%%%%%%%%%% Background studies %%%%%%%%%%%%%%%%%%
\lvltwosection{Background Studies}
\label{sec:C6_background_studies}
\editor{Antonio}

\begin{figure}[hbt]
\centering
\includegraphics[width=10cm]{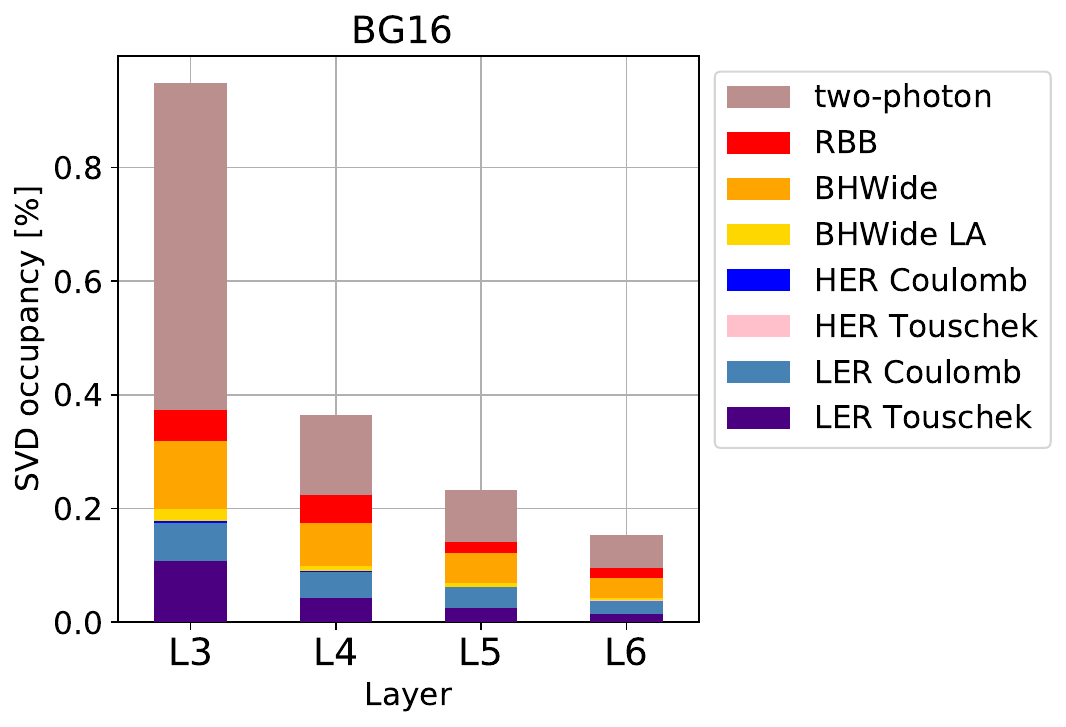}
\caption{Occupancy of all SVD Layers estimated by preliminary Monte Carlo simulations at design machine parameters, L= \qty{8e35}{\per\square\cm\per\s}. Different contributions from luminosity related background (two-photon processes and radiative Bhabha, which include ``RBB'', ``RHWide'' and ``BHWide LA'', referring to different photon emission angles) and single beam related background (beam-gas scattering and Touschek effect) are shown.}
\label{fig:C6_bg_estimation}
\end{figure}

The challenging design luminosity of SuperKEKB means that particular care must be taken of the machine induced background conditions, which are expected to be severe and can affect the lifetime of the detector and its performance.
The main background sources in SuperKEKB can be classified in two groups: single beam backgrounds (Touschek effect, beam-gas scattering, synchrotron radiation, injection background) and luminosity related backgrounds (radiative Bhabha scattering, two-photons processes). A brief description of these background sources and results of the first background studies performed in \phaseone{} are given in~\cite{Lewis:2018ayu}.
Before the start of \phasetwo, preliminary Monte Carlo simulations were used to estimate the expected background levels at design machine parameters. Figure~\ref{fig:C6_bg_estimation} shows the estimated SVD occupancy levels due to backgrounds, with the contributions of each background source. This initial background extrapolation at design luminosity gave on average for Layer 3 sensors an estimate of about 1\% occupancy, 0.1 Mrad/yr integrated dose, and an equivalent neutron fluence of about
\qty{0.2e12}{n\ensuremath{_\mathrm{eq}}/\cm\squared/yr}.
Considering the radiation hardness of the DSSD sensors,  conservatively set to 10 Mrad and about \qty{3e13}{n\ensuremath{_\mathrm{eq}}/\cm\squared} (as discussed in section~\ref{sec:C2_rad_tolerance}), these numbers give enough margin to safely operate the SVD even for 10 years at design luminosity. Also the extrapolated occupancy was well below the 3\% limit set to ensure good tracking performance, and it was considered promising for the operation of the VXD and the full \belletwo{} detector.

In order to understand the real background levels in the VXD volume, dedicated studies were conducted during \phasetwo, with the machine in its final configuration, the \belletwo{} detector in position, but with only one slice of the Vertex Detector. The remaining volume of the VXD was instrumented with the BEAST II detector system, designed to measure the radiation field in the inner detector region. The full description of the system and the results of the studies are available in \cite{Liptak:2021tog}.
After the VXD installation and the start of \phasethree, many studies were conducted during every data taking period by machine and detector groups to carefully evaluate background levels, in order to separate background components and to optimize collimators settings, to lower the backgrounds in the interaction region preserving the beam lifetime and injection efficiency \cite{Belle-IISVD:2020wwc}.

\begin{figure}[hbt]
\centering
\includegraphics[width=10cm]{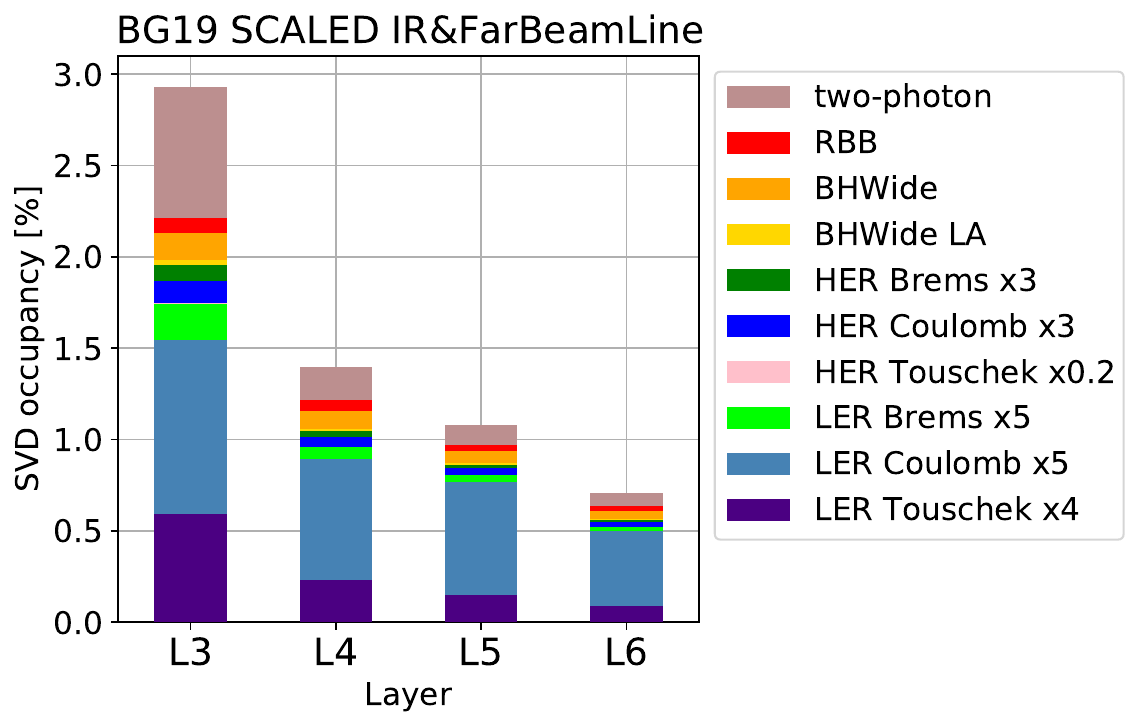}
\caption{Occupancy of all SVD Layers after re-scaling MC simulation using background studies measurements based on 2020 data set. Different contributions from luminosity related background (two-photon processes and radiative Bhabha, which include ``RBB'', ``RHWide'' and ``BHWide LA'', referring to different photon emission angles) and single beam related background (Bremsstrahlung, Coulomb scattering, and Touschek effect) are shown. Design machine parameters are considered, L=\qty{8e35}{\per\square\cm\per\s}.}
\label{fig:C6_bg_extrapolation_scaled}
\end{figure}

Background studies allowed a better understanding of background levels in the interaction region. Data from these studies and comparisons with simulations were used to re-scale MC backgrounds at design machine parameters to reflect the real background conditions in the interaction region. The re-scaled MC background levels,  according to recent background studies based on 2020 runs data set, are shown in figure~\ref{fig:C6_bg_extrapolation_scaled}, again with contributions of each background source.

A comparison with figure~\ref{fig:C6_bg_estimation}
shows that the SVD background levels in the re-scaled MC are much higher by almost a factor 3: for Layer 3 sensors about 3\% occupancy, 0.3 Mrad/yr integrated dose and an equivalent neutron fluence of about 
\qty{0.6e12}{n\ensuremath{_\mathrm{eq}}/\cm\squared/yr}   
are expected. This estimation, though still affected by a large uncertainty, is more reliable than the preliminary one shown in figure~\ref{fig:C6_bg_estimation}, because it is calculated using data taken with the actual machine and detector. 
In terms of integrated radiation damage, even with the re-scaled background estimate there are no concerns to safely operate  the SVD for several years at design luminosity. However, the estimated occupancy level is close to the 3\% limit for good tracking performance, and further mitigation strategies are now considered to increase the safety margin.
%but occupancy levels close to the limit for good performance of the tracking.

%With the running conditions and machine parameters used in 2019 and 2020, background levels were quite low ($<0.5$\%) and not affecting detector performances, but in the future, approaching design luminosity, the SVD is expected to hit performance limits. Some countermeasures were discussed in order to mitigate backgrounds: 
With the running conditions and machine parameters used in 2019 and 2020, background levels were quite low (< 0.4\% ) and did not affect detector performance. To mitigate the higher background levels expected at design luminosity, some countermeasures are  being evaluated and implemented. 
On the machine side, new beam collimators were installed, the design of additional ones is being considered, and a new beam pipe for the interaction region is being prepared, allowing the installation of better optimized heavy metal shields. On the SVD side, the 3-sample DAQ mode and the SVD hit time, described in section~\ref{sec:C7_hit_time_determination}, can be used to reduce the required bandwidth, maintaining the tracking efficiency in high background conditions.
More details of the strategy to cope with high background conditions are given in section~\ref{sec:conclusion}.

In conclusion, operations of the SVD was smooth in the first two years of the experiment. A constant periodic monitoring has shown the expected evolution of the relevant parameters of the silicon sensors, with a few issues that are under control, well understood, and not affecting performance.
 % 10 pages

\lvlonesection{Simulation and Reconstruction Software}
\editor{Giulia - total of <25 pages}
\label{sec:software_offline}
The SVD plays a key role in the \belletwo\ reconstruction, both for charged particle tracking, and for  low momentum particle identification.  Tracks reconstructed in the \acrshort{CDC} are first extrapolated to the SVD  and then to \acrshort{PXD}, to significantly improve the precision of the track parameters. The SVD is also used for stand-alone pattern recognition, specifically for low-transverse-momentum tracks (approximately below \qty{90}{\MeV/c}) that do not reach or are not reconstructed in the CDC because of their bending in the magnetic field. In addition, using the energy loss in the material, the SVD significantly improves the particle identification capability below \qty{1}{\GeV}. As an example, this capability is essential to reconstruct the slow pion of the \Dstarp\to \Dz \pips decay from \B decays, which is crucial for the \belletwo\ physics program.

The complete tracking chain is run online in the High Level Trigger (\acrshort{HLT}) to determine the regions of interest (\acrshort{ROI}) on the PXD sensors in which  \gls{hit}s from  tracks are expected, allowing to reduce by one order of magnitude the PXD data, which would otherwise violate storage and bandwidth limits. 
Hence, there are quite stringent limits on the execution time of the tracking chain (including SVD reconstruction) as well as on the quality of the tracking that must be achieved during data taking, i.e. before the dedicated calibration of the tracking detectors and algorithms that is usually done \textit{using} the data that have already been collected. In fact, a sub-optimal quality of the online tracking, especially the SVD part, may result in a permanent loss of PXD signal clusters with in a significant degradation of the track quality.

The high quality and precision needed for the \belletwo\ physics program require the SVD detector to be efficient and precise in the reconstruction of the \gls{hit}s produced by charged particle crossing the sensors. One of the main challenges of the SVD reconstruction is the reduction of the background hits produced by uninteresting electromagnetic processes, such as $\epem \to \epem\epem$, occurring hundreds of nanoseconds before the triggered events. 
Since the bunch crossing frequency is $\simeq 250$~MHz, and the tail of the response of the APV is order of 400~ns long, the SVD sees particles hitting it since roughly 100 bunch crossings before the triggered event. These off-time hits populate the detector, increasing significantly its \gls{occupancy} from $10^{-5}$ to a few percent, and making the pattern recognition much more challenging.
The measurement of the \gls{hit} time is particularly powerful to remove this background: a simple selection on the hit time allows to remove $\simeq 40\%$ of the background 1-D hits while keeping an efficiency above $99.9\%$. 

The main technical features of the SVD software are summarized in section~\ref{sec:C8_SVD_software}: integration in the \belletwo\ framework, reconstruction data flow, simulation workflow (section~\ref{sec:C4_simulation}), and their interactions with the data store and Conditions Database (section~\ref{sec:C4_cdb}).

The reconstruction (section~\ref{sec:C4_reconstruction})
starts with the unpacking of the raw data and the calibration of the strips (section~\ref{sec:C4_strip_reconstruction}), that are then combined in \gls{cluster}s, representing the 1-D \gls{hit} on the sensor plane. The computation of the cluster position and time is done using the strip time and charge. The cluster charge is calculated summing the strip charges and is then used for the particle identification. As described in section~\ref{sec:C4_clustering},
clusters on the two sides of the sensors are finally combined in \gls{spacepoint}s that represent the 3-D \gls{hit}s in space and are the key ingredient for the track finding algorithms. At this stage, clusters are rejected based on their hit time.\\
An essential part of the reconstruction performance is the calibration of the detector and of the reconstruction algorithms. The strip noise and gain calibrations are measured in local runs, and stored in the Conditions Database. The calibration of the hit time is done exploiting its correlation with the time of the event with respect to the trigger arrival (section~\ref{sec:C4_CoGCalibration}). Calibrations of each sensor side are also stored in the Conditions Database.

In order to monitor the quality of the data during data acquisition, the reconstruction of a fraction of the events is performed online on dedicated computer farms. The \acrfull{DQM} software 
(section~\ref{sec:C4_dqm}) produces a set of plots that are shown live in the control room and automatically change colors in case some anomaly is detected. This is essential to immediately identify and react if something is not working as expected.  A set of expert-mode plots are also available to identify  issues in the shortest time possible, to maximize data taking efficiency. 
%The SVD DQM plots available for the control room shifters include checks of: the data format, the average efficiency and occupancy of the sensors, and the correct location of the acquisition window with respect to the trigger signal arrival.

The accurate modeling and simulation of the interaction of charged particles with the silicon sensors, the APV response to the induced charge, and the DAQ chain is very important. In the first place, the track finding algorithms are calibrated and optimized using simulated data: a mis-modeling of the data would result in a degraded tracking performance and would contribute to increase the systematic uncertainty for the physics analysis.
Furthermore, the simulation provides an essential tool to understand the  features observed in the data, and to predict the reconstruction performance for different configurations of the detector and for varying external conditions, such as an increased background.

As outlined in section~\ref{sec:C4_simulation}, the simulation starts from generated MC particles, whose interaction with the material is simulated by Geant4~\cite{Agostinelli:2002hh}, and ends with a list of raw strips, equivalent to the output of the unpacker on the real data. The sensor response is obtained by  producing a set of small groups of ionization charges along the track, that are then drifted in a simplified electrical and magnetic field towards the electrodes on both sensor sides reaching the collection implants. The charge collected on the implants is coupled to the nearby floating and readout strips, to simulate the charge sharing. No weighting field is used to model signal induction during charge drift. 
The APV response is then simulated, including the strip gain, while the electronic noise is taken directly from random trigger data in run-dependent simulations. Neither  \gls{common-mode noise} correction, nor the \acrshort{FIR} are simulated.
The calibration of the charge sharing parameters as well as of the strip gain was performed using test-beam data.
This simplified model provides a fair agreement for cluster charge signal-to-noise ratio, but underestimates the size of the clusters with respect to data. A study is ongoing with a re-parameterization of the charge sharing variables and a re-calibration with data in order to improve the agreement in the cluster charge and cluster size. 
%and, more importantly, to increase the cluster size and achieve an agreement withing $2\%$ with data. %Since the study is not fully completed, results are not shown in this paper.

\lvltwosection{\belletwo\ SVD Software} 
\editor{Giulia}
\label{sec:C8_SVD_software}

The SVD Offline Software is fully integrated into the  \belletwo\ official software with a dedicated package in the \belletwo\ Software Analysis Framework (\acrshort{basf})~\cite{basf2}. The full \basf\ source code is publicly available in~\cite{basf2-zenodo}.

\begin{figure}[hbt]
    \centering
    \includegraphics[width=\textwidth,keepaspectratio]{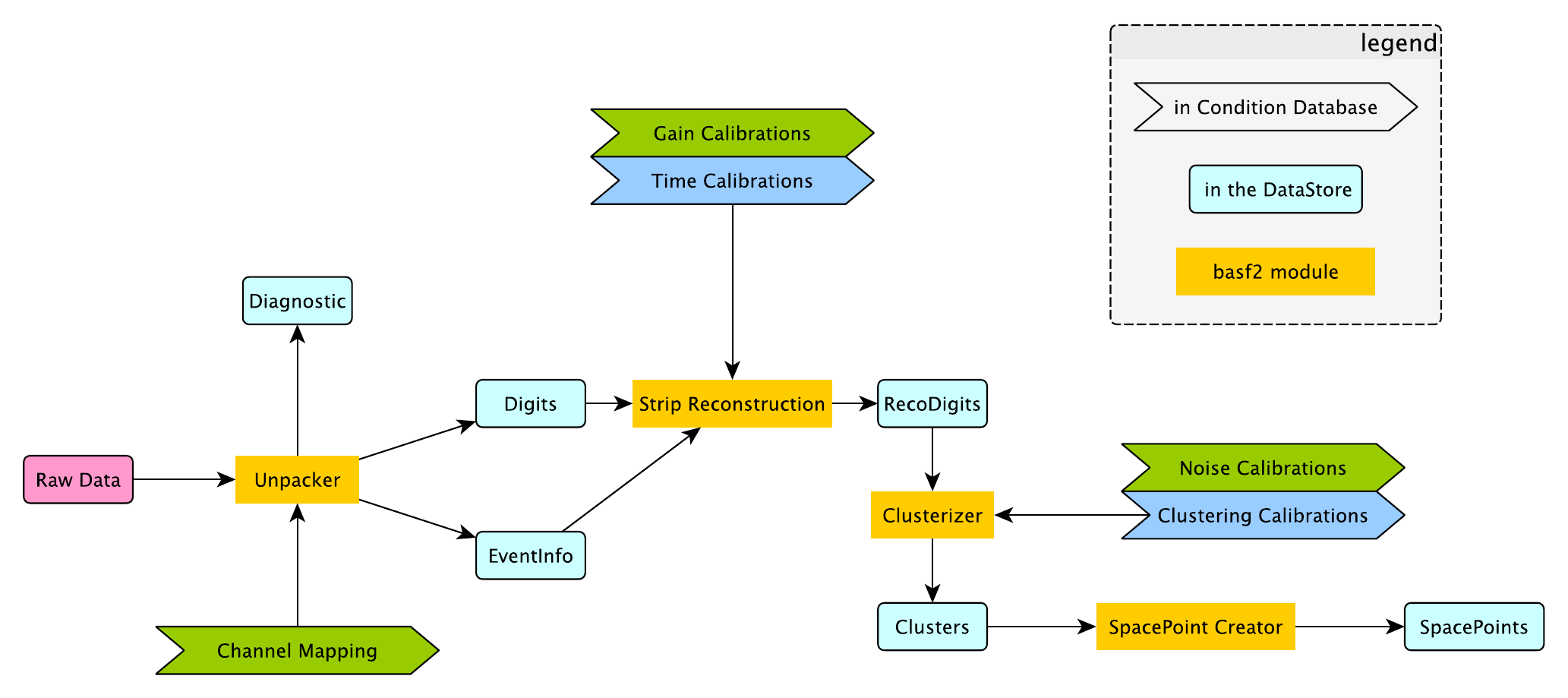}    
    \caption{SVD Offline Reconstruction data flow. Modules, inputs from the DataStore and inputs from the Conditions Database are shown. For the sake of clarity, the prefix SVD is removed. The conditions in green are uploaded on the database during data taking, while the ones in blue are computed offline.}    \label{fig:C4_offline_reconstruction_data_flow}
\end{figure}

The building blocks of the \basf\ processing chain are the {\it modules}, each having a particular task that goes from a simple task like reading the data from disk, to more complex ones like clusterization. A python steering file is used to define an ordered list of modules that are executed one after the other for each event, producing the required output. The input data of a module (e.g. list of strips of the event) are stored in a common storage, the DataStore. Each module has read and write access to the storage.
Run condition information are also needed for certain modules (e.g. strip noise calibration), and are stored the Conditions Database.
As an example, in figure~\ref{fig:C4_offline_reconstruction_data_flow} a schematic view of the data flow of the SVD offline reconstruction with indications of modules, objects read and written to the DataStore   and conditions taken from the Conditions Database is shown.

The SVD package in \basf\  contains the necessary software that allows to:
\begin{itemize}
    \item unpack real data and simulate the strip signals recorded by the sensor in response of a charged particle traversing it, processed by the APV25, and digitized in the \acrshort{FADC}s;
    \item reconstruct the hit time, position, and ionization charge from the digitized signal of the strips;
    \item calibrate the simulation and reconstruction algorithms, storing and retrieving the information from the Conditions Database;
    \item monitor the detector performance during data taking.
\end{itemize}{}
The simulation and calibration software run offline, while the reconstruction and monitoring software run offline for data reprocessing, and are also executed online on the High Level Trigger farm for event selection 
and on dedicated machines for the online data quality monitoring (ExpressReco). In particular, the SVD reconstruction (and tracking) runs on the \acrshort{HLT} to identify regions on the \acrshort{PXD} where the track has most probably crossed the sensor (\gls{ROIgls}) in order to reduce the data size of the PXD, both for bandwidth and storage limitations. The required data reduction factor is 10, with an efficiency which must be close to 1 in order not to loose the precious information of the PXD hit for offline reconstruction. Since SVD reconstruction runs online, there are additional requirements on the software, especially in terms of the execution time and the memory usage. Moreover, the online reconstruction needs to work reasonably well with calibrations (e.g. for clustering and hit time computation) determined on previously recorded data, while for the offline reconstruction the calibration is done using a subset of the runs that will be re-processed. 
%The online reconstruction for the \acrshort{DQM} is less critical as it is used to judge the quality of the data, and therefore it does not need to be excellent, but reasonable.
The tracking pattern recognition algorithms are trained on simulated events, therefore a good simulation of the SVD is fundamental to have a high tracking efficiency.

\lvlthreesection{Local coordinate systems} 
\label{sec:C4_coordinates}

Apart from the global coordinate system defined in Chapter 2, each SVD sensor has a local right-handed coordinate system ($u,v,w$) centered at its center, with $v$ in the direction of $z$, $w$ pointing away from the IP, and $u$ pointing along the global $r\phi$ direction.
The N strips measure the local $v$ coordinate, while the P strips measure the local $u$ coordinate. In the slanted sensors (forward sensors of layer 4,5,6), because the u/P strips are not parallel to each other but form a varying angle with the global $r\phi$ direction, the $u$ position of the hit can be determined only if its $v$ position is also known.

\lvlthreesection{Time reference frames} 
\label{sec:C4_timeref}
There are two main time reference systems: the SVD reference system with $t_{\rm SVD} = 0 $ at the time of the first sample of the APV, and the global \belletwo\ reference system with $t_{\rm glob} = 0$ at the trigger arrival time, simultaneous for all subdetectors.
The two systems are synchronized, but the APV25 clock frequency is 4 times smaller than the trigger clock. In order to switch from one reference frame to the other, the FADC registers the trigger signal arrival time within one APV clock, then stored it in the \gls{triggerbin} variable. For example, if the trigger signal arrives during the first quarter of the APV clock, the trigger bin is 0.
Therefore, the trigger bin allows us to translate the hit time measured in the SVD reference frame to the trigger reference frame, common to all detectors, and vice-versa. 

\lvlthreesection{SVD simulation} 
\label{sec:C4_simulation}
\editor{Peter Kv., Giulia} %, details of the content of the section have been moved in the Jira ticket B2SVD-204

\begin{figure}[hbt]
    \centering
    \includegraphics[width=.8\columnwidth,keepaspectratio]{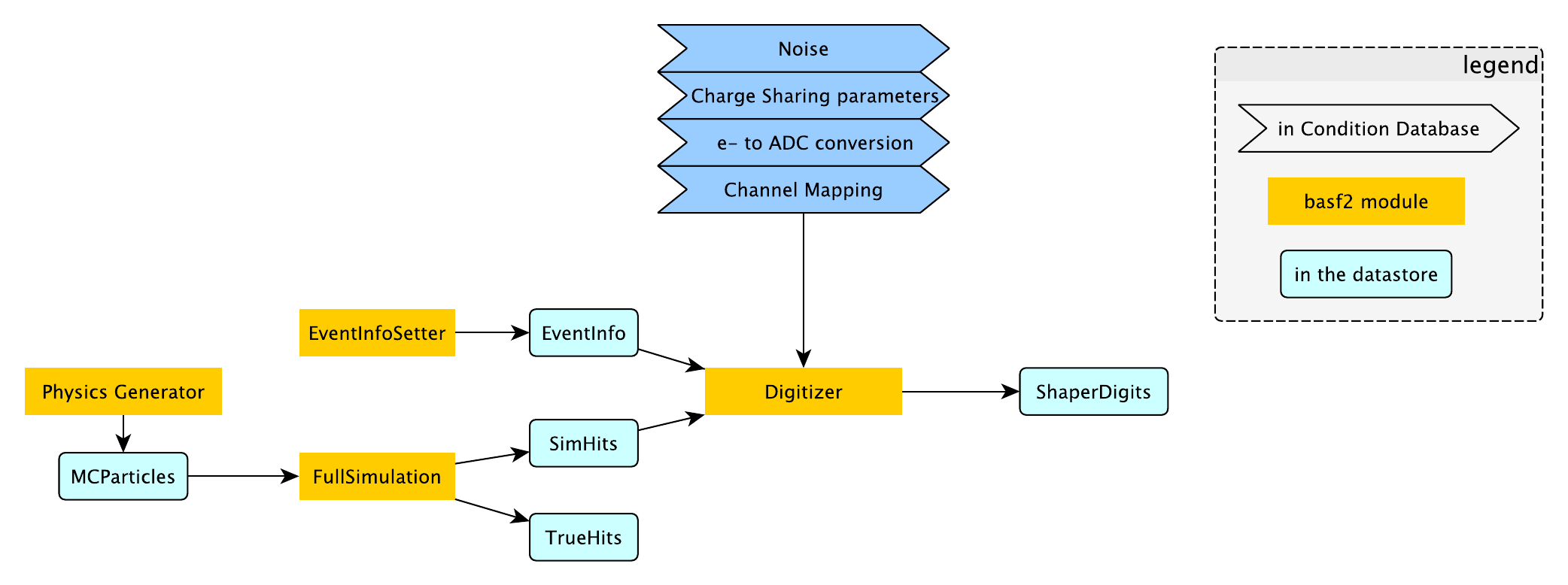}    
    \caption{SVD simulation workflow, showing \basf\ modules and objects in the DataStore that form the SVD simulation.}    \label{fig:C4_simulation_workflow}
\end{figure}

The SVD Simulation covers the part of \basf\ software that simulates SVD  response to particles  traversing the sensitive volumes of SVD sensors. 
In figure~\ref{fig:C4_simulation_workflow} a schematic view of the data flow of the SVD simulation, with indications of modules, objects written to the DataStore, and conditions taken from the Conditions Database is shown.

Technically, detector simulation is a three-step process, with \emph{physics event generators} simulating collisions in the accelerator and other relevant radiation-generating processes (step 1), and the Geant4~\cite{Agostinelli:2002hh} engine propagating the output of physics generators through the detector (step 2). The Geant4 tracker uses callbacks to individual subdetector routines to defer the calculation of detector response from energy depositions in the sensitive subdetector volumes to the dedicated subdetector software (step 3). The last process is called \emph{digitization} and is the main topic of this section. 

There are several physics event generators supported in \basf\ as part of its \emph{externals} software package, including \emph{EvtGen}~\cite{Lange:2001uf}, \emph{PYTHIA8}~\cite{pythia8}, and \emph{KKMC}~\cite{kkmc}.
All such generators produce a list of Monte Carlo (MC) Particles, \nc{MCParticle}.
A crucial part of \basf\ software is a geometric representation of the whole \belletwo\ detector and (parts of) the accelerator, as well as the related mechanical and civil structures. The geometry is implemented as a hierarchy of Geant4 volumes and stores its parameters in the \belletwo\ Conditions Database. The geometry is built subdetector-wise, with each subdetector defining its geometry builder class (\nc{GeoSVDCreator} for SVD) and its \nc{SensitiveDetector} Geant4 callbacks.

SVD simulation starts with a collection of \nc{MCParticle}'s tracks going through sensitive the regions of the SVD and ends with a collection of \nc{SVDShaperDigits}\ objects, which are the same objects used to encode the data on fired SVD strips during the real data acquisition process, thus producing data that imitate what is provided by the SVD hardware for reconstruction. 

\textbf{The SVD Sensitive Detector} class is the piece of software that gets a Geant4 (MC Particle) track from the Geant4 tracker and stores its energy deposition to be used by the \nc{SVDDigitizer}. For this, the software uses objects called \nc{SVDSimHits}, which represent segments of a piecewise-linear approximation to the MC Particle track through the sensor. They store a detailed profile of energy deposition along the segment. 
The \nc{SVDSensitiveDetector} also forms another collection of data objects called \nc{SVDTrueHits}. An \nc{SVDTrueHit} represents a single MC Particle passage through an SVD sensor. These objects facilitate a simple MC-truth-based tracking.

\textbf{Sensor Information.} Each silicon sensor has an associated \nc{SensorInfo} instance that stores sensor-specific parameters, including position, strip electric parameters (strip capacitances and electronic noise for the simulation) and material constants. The \nc{SensorInfo} class also has methods to calculate some simulation-related quantities for the sensor, such as the electric and magnetic fields, charge carrier mobilities, and drift velocities. The parameters and equations used are listed in table~\ref{tab:C4_SensorInfo-formulas}. 
%Once the Geant4 geometry is built at the beginning of a \basf\ run, the \nc{SVD::SensorInfo} objects for all SVD sensors are available via the \nc{VXD::GeoCache} singleton. 

%% -------------- Table: Simulation-related quantities in SVD::SensorInfo.--------------------- 
%% tab:C4_SensorInfo-formulas

\newcommand{\tvi}{\vrule height 17pt depth15pt width 0pt} 

\newcolumntype{F}[1]{>{\hfil$\displaystyle} p{#1} <{$\hfil}} 
\newcolumntype{L}{>{\centering\arraybackslash}m{3cm}}

\begin{table}
\begin{center}
\begin{small}

\begin{tabular}{l F{6.5cm} L}\\ \toprule %6cm
\textbf{Quantity}       & \text{\bfseries{Formula}}           & \textbf{Parameters}    \\ \midrule
\tvi Electric field     &  E(z) = \frac{2V_{\rm depl}}{d} \left(\frac{z - d/2}{d}\right) - \frac{V_{\rm bias} - V_{\rm depl}}{d} %E(z)
                        & \multicolumn{1}{m{4cm}}{
                           $V_{\rm depl} = 40\,V$ \newline 
                           $V_{\rm bias} = 100\,V$ \newline 
                           $d = 300-\qty{320}{\um}$ \newline 
                           }  \\ \midrule 
\tvi Carrier mobility~\cite{Canali:1975lgt}   
                        & \mu(E(z), T) = \frac{v_m/E_c}{\left(1 + (\frac{E}{E_c})^{\beta}\right)^{1/\beta}} &  \multicolumn{1}{m{4cm}}{Electrons: \newline 
                           $v_m = 1.53 \cdot T^{-0.87} \cdot 10^9\,cm/s$ \newline 
                           $E_c = 1.01 \cdot T^{+1.55}\,V/cm$ \newline 
                           $\beta = 2.57 \cdot T^{+0.66} \cdot 10^{-2}$ \newline 
                           Holes: \newline 
                           $v_m = 1.62 \cdot T^{-0.52} \cdot 10^8\,cm/s$ \newline 
                           $E_c = 1.24 \cdot T^{+1.68}\,V/cm$ \newline 
                           $\beta = 0.46 \cdot T^{+0.17} $
                           }  \\ \midrule % \mu(E(z), T) = 
\tvi Hall factor        &  r_H(T) = r_H^0 + r_H^T \cdot (T-273.15 K) 
& \multicolumn{1}{m{4cm}}{ Electrons: \newline 
                           $r_H^0 = 1.13 \newline r_H^T = 0.0008\,K^{-1}$ \newline
                           Holes: \newline
                           $r_H^0 = 0.72 \newline r_H^T = 0.0005\,K^{-1}$
                           }  \\ \midrule %r_H(T) =
%\tvi Drift velocity    & \boldsymbol{v}(\boldsymbol{E}, \boldsymbol{B}) =  \frac{\mu\boldsymbol{E} + \mu\mu_H \boldsymbol{E} \times \boldsymbol{B} + \mu \mu_H^2 \boldsymbol{B}(\boldsymbol{E} \cdot \boldsymbol{B})}{1 + \mu_H^2 \|\boldsymbol{B}\|^2}
\tvi Drift velocity    &  \boldsymbol{v}(\boldsymbol{E}, \boldsymbol{B}) = \frac{\mu\boldsymbol{E} + \mu\mu_H \boldsymbol{E} \times \boldsymbol{B} + \mu \mu_H^2 \boldsymbol{B}(\boldsymbol{E} \cdot \boldsymbol{B})}{1 + \mu_H^2 \|\boldsymbol{B}\|^2}
& \multicolumn{1}{m{4cm}}{  $\mu$ carrier mobility, \newline
                            $\mu_H = \mu \cdot r_H$ Hall mobility
                           }  \\ \bottomrule %\boldsymbol{v}(\boldsymbol{E}, \boldsymbol{B}) = 
\end{tabular}
\caption{Simulation-related quantities provided by \nc{SVD::SensorInfo} methods. The value of the temperature is $T = 300$ K.}
\label{tab:C4_SensorInfo-formulas}
\end{small}
\end{center}
\end{table}
%%-------------------- End of table definition -------------------------- 

\textbf{Digitization} is the process of calculating strip signals from energy depositions in the sensor.
The digitizer gets the information about SVD data acquisition and trigger arrival from the \nc{SVDEventInfo}\ object. The latter stores the DAQ mode (6- or 3-sample acquisition), the trigger type, and the \gls{triggerbin}.
%: the arrival time of the L1 trigger relative to the APV sampling clock.

The \nc{SVDSimHits} store detailed information about energy deposition used in the digitization  four steps:
\begin{enumerate}
    \item create charge carrier clouds based on the energy deposition pattern;
    \item allow charge carriers to drift and diffuse in the electric and magnetic field in sensor bulk, reaching the collection implants;
    \item calculate the signal that this induces on the sensor strips through capacitive coupling;
    \item process the raw strip signals by taking into account strip noise, charge sharing, and readout electronics.
\end{enumerate}
The \basf\ implementation of SVD digitization is implemented in the \nc{SVDDigitizer} class.  
Digitization starts with energy deposition profiles in individual \nc{SVDSimHits}, encoding energy deposition along a quasi-linear piece of a particle track through a silicon sensor. It also includes information about the time of MC Particle passage through the sensor. The piece of track is divided into several small pieces of \qty{5}{\um}, and the energy deposited along each piece is converted into free charge carriers -- electrons and holes -- by dividing the energy deposition by the conversion energy for silicon, $3.65$\,eV. The number of charge carriers is randomized by the Poisson smearing. In the following, these small clouds of electrons and holes are called \emph{chargelets}. A chargelet is characterized by its position within a sensor, the number of carriers it contains, and time of its creation. 

%\textbf{Drift and diffusion}
Chargelets drift in the sensor bulk under the combined effect of the electric and magnetic fields (including the Lorentz shift effect). Mobilities and velocities of charge carriers are calculated using the formulas in table~\ref{tab:C4_SensorInfo-formulas}. Diffusion is  also simulated by increasing the size of the chargelet cloud proportional to $\sqrt{Dt}$, where $D$ is the diffusion coefficient.

A simplified model that does not include the effect of  induced signal on the strips due to charge moving in the sensor (Ramo-Shockley theorem~\cite{1686997, doi:10.1063/1.1710367}) is used in the simulation. The signal is instead produced only when the charge is {\it collected} on the sensor surface and it is assigned to the nearest floating or readout strip. A single chargelet can produce signal on both a readout and a floating strip, assuming a Gaussian distribution of the carriers in the chargelet a fraction of the charge corresponding to the area of the Gaussian distribution is assigned to each strip. 

\begin{figure}[hbt]
    \centering  
        \includegraphics[width=10cm]{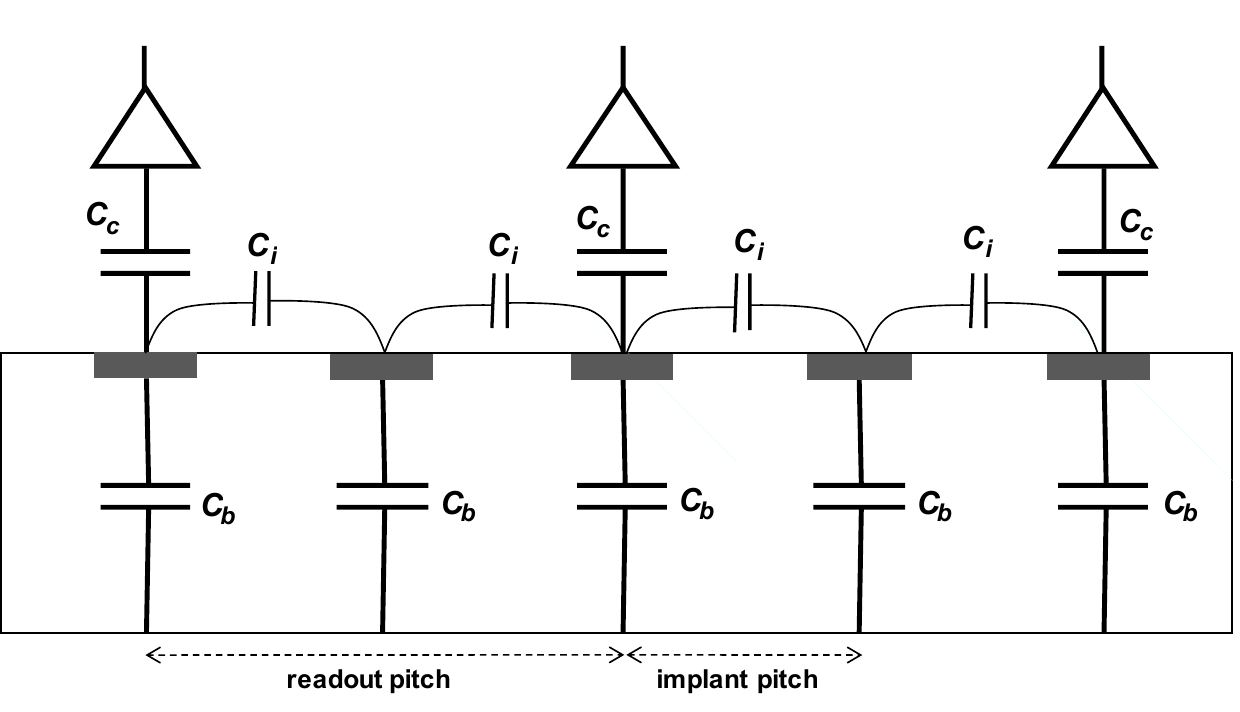}
    \caption{Sensor capacitance coupling scheme implemented in simulation.}
    %\rem{readout pitch/2 -> implant pitch}}
    \label{fig:C4_sensor_capacitance}
\end{figure}

After this step, charge sharing among the various implants is simulated using a simplified model based on the circuit shown in figure~\ref{fig:C4_sensor_capacitance}, that takes into account:
\begin{itemize}
\item $\mathbf{C_i}$: the interstrip capacitance between an implanted strip and its first neighbouring strip. The capacitance between one implant and the second neighboring strip is not considered, as it is expected to be negligible.  $C_i$ varies from  0.5 to 1 pF/cm depending on the different sensor geometry and sides.  These estimated values are affected by quite a large uncertainty, depending not only on details of the sensor design, like the ratio of the width of the implant over the implant pitch (half the readout pitch), but also on the positive oxide charge on the sensor surface. The latter can be different from sensor to sensor, and also varies with irradiation. 
\item $\mathbf{C_b}$: the capacitance to the back of the sensor. It can be calculated from the geometrical parameters of the sensor and it scales with the implanted pitch.  $C_b$ is about 0.1 pF/cm for u/P side, with smaller pitch, and about 0.4 pF/cm for v/N side with larger pitch.
\item $\mathbf{C_c}$: the decoupling capacitance to the APV input. It is more than an order of magnitude larger than $C_i$ and $C_b$, with measured values of 15 (30) pF/cm on the u/P (v/N) side. With this large value, the charge collected on a readout implant is efficiently collected on the corresponding APV input. 
%with only a small loss of about 2-5\%.
\end{itemize}

The coupling coefficients among implanted strips calculated considering the simplified network shown in figure~\ref{fig:C4_sensor_capacitance} are the following:
\begin{itemize}
    \item $k_1$, the fraction of the charge on a floating strip coupled to each neighboring readout implant: 
    $$k_1=\frac{C_i}{(2C_i+C_b)}$$
    %$$k_1=\frac{C_i}{C_{\rm tot-float}}=\frac{C_i}{(2C_i+C_b)}$$
    Ideally it should be 50\%, which is the case for small pitch, i.e. small $C_b$. In this case even when the charge is collected on a floating strip  a full collection of the signal on the two nearby readout strips is obtained. For large pitch, when $C_b$  becomes comparable to $C_i$, some fraction of the signal is instead lost. In the layer 4, 5, 6 sensors for the v/N side $k_1$ is  35-40\%, and a total charge loss due to this effect can be  20-30\%, as  observed in data.
    \item $k_2$, the fraction of the signal on a readout strip coupled to the next readout implant: %$$k_2=\frac{1}{2}\frac{C_i}{C_{\rm tot-readout}}=\frac{1}{2}\frac{C_i}{(C_i+C_b+C_c)}.$$
    $$k_2=\frac{1}{2}\frac{C_i}{(C_i+C_b+C_c)}.$$
    Since $C_c$ is much larger than the other capacitances this coupling is very small and the signal coupled from one readout strip to the next one is in general below threshold and then lost. This effect can give rise to an additional few percent charge loss, larger on the u/P side where $C_c$ is smaller compared to the v/N side. 
 
\end{itemize}
 
After charge sharing simulation, the charge of each readout strip is smeared with a  Gaussian distribution whose width includes  the Fano factor~\cite{Fano:1947zz}.

Finally, the response of the APV for each signal on each readout strip is simulated by the waveform of the response as a function of time. The start time of the waveform is computed as half the drift time of the chargelet to the sensor surface. For each strip the APV responses of each chargelet is summed, and electronic noise is added  on top of the signal waveform, obtaining the final APV response. This waveform is then digitized by sampling it with the APV clock frequency and converting its amplitude in 8-bit ADC counts with a conversion factor, \nc{ADUEquivalent}. The Online zero suppression is applied to limit the number of \nc{SVDShaperDigits} produced in output of the simulation. In particular, only \nc{SVDShaperDigits} with at least one sample larger than three times the noise of the strip are stored, as done in real data acquisition.

\textbf{Simulation calibration.}
There are several parameters in the simulation that need to be tuned separately for the two sensor sides of each sensor type (small rectangular, large rectangular, and trapezoidal or wedge):
\begin{itemize}
    \item noise superimposed to the APV waveform;
    \item \nc{ADUEquivalent}\ to convert electrons into ADC;
    \item the three capacitances for the charge sharing description. For $C_c$ and $C_b$ the design parameters provide already a good estimation, while $C_i$ is much more difficult to estimate from design parameters. Therefore, only the latter has been tuned on data.
\end{itemize}
%The simulation was initially tuned using 2017 test-beam data.

\begin{figure}[hbt]
    \centering  
        \includegraphics[width=.45\columnwidth,keepaspectratio]{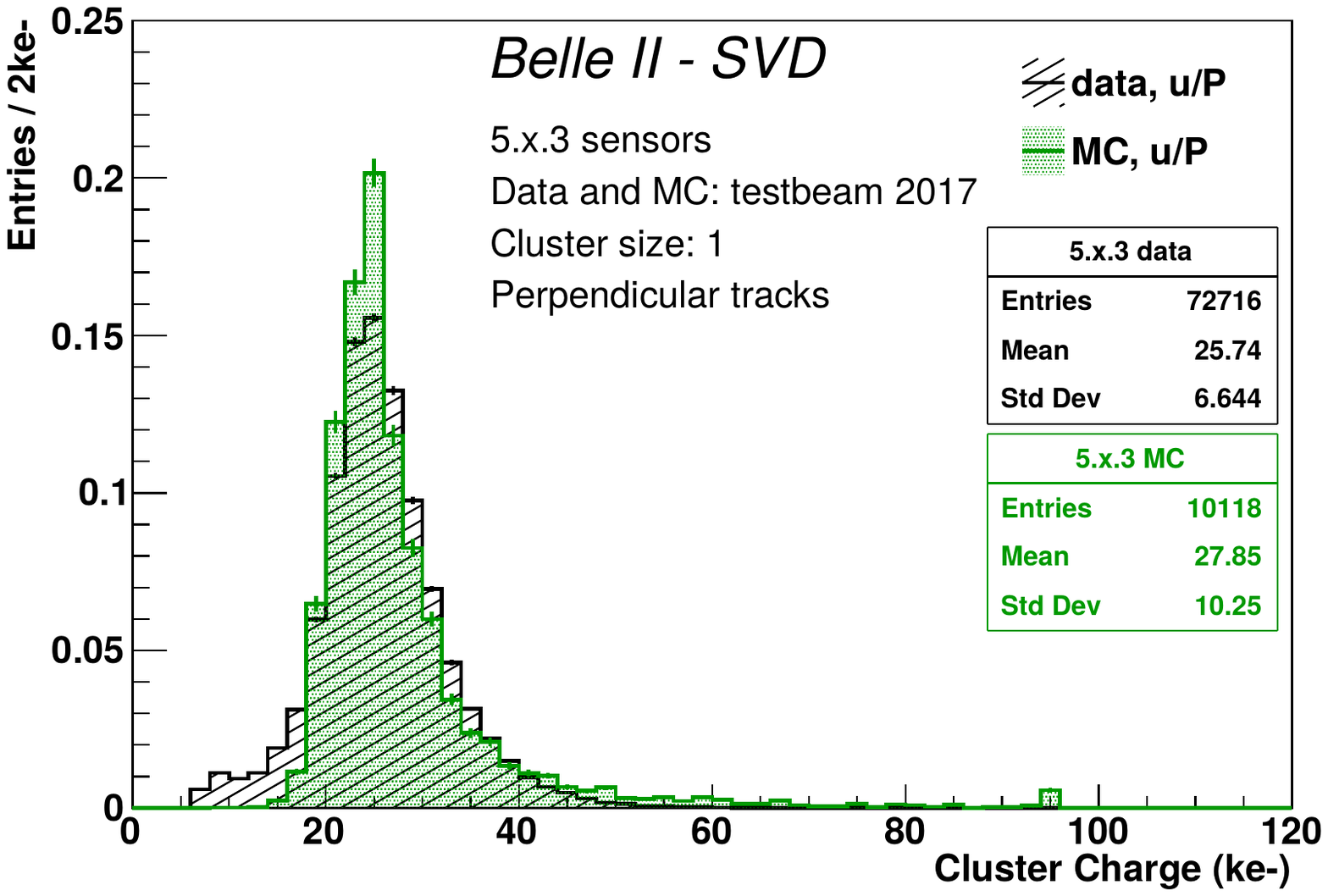}
        \includegraphics[width=.45\columnwidth,keepaspectratio]{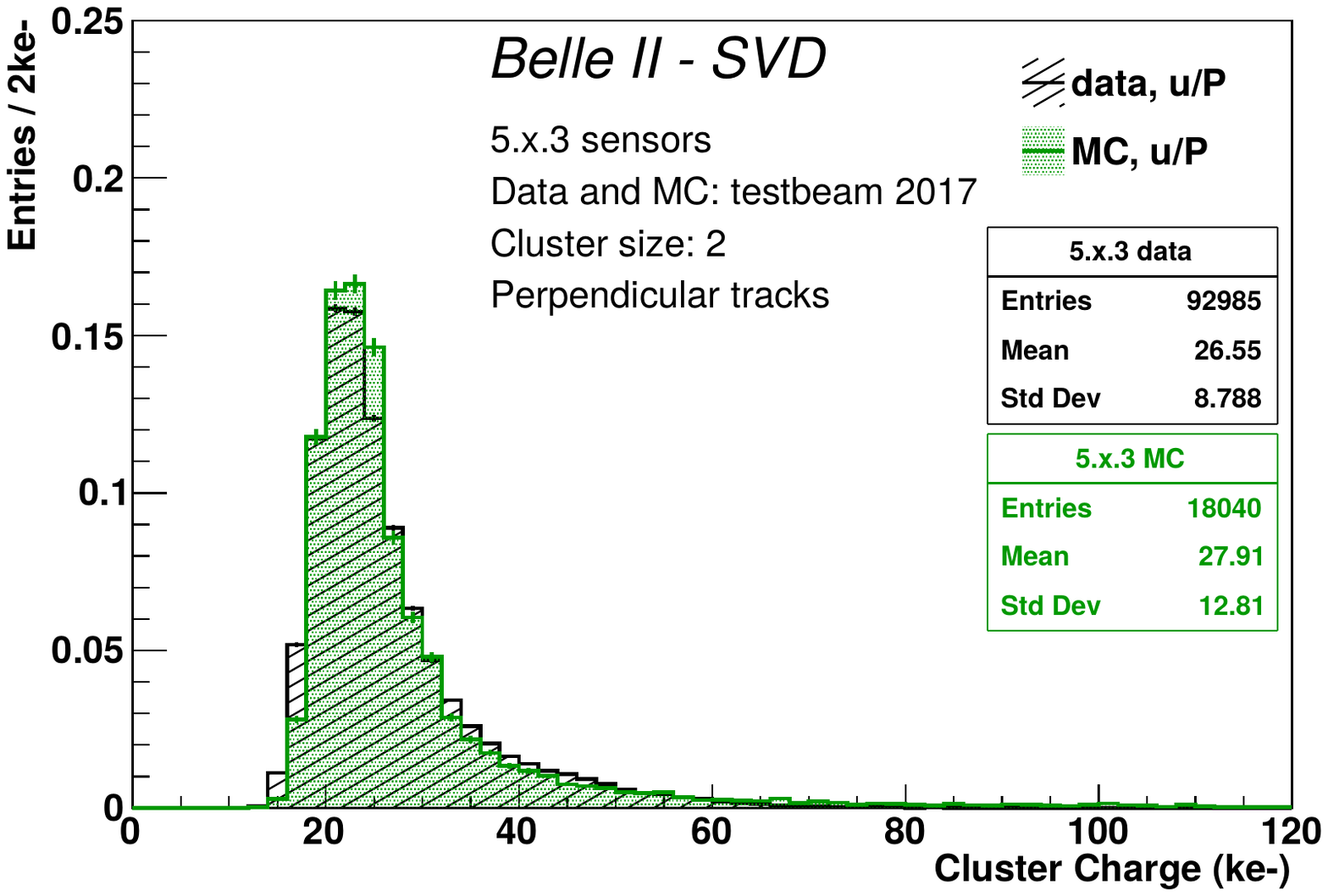}
        \includegraphics[width=.45\columnwidth,keepaspectratio]{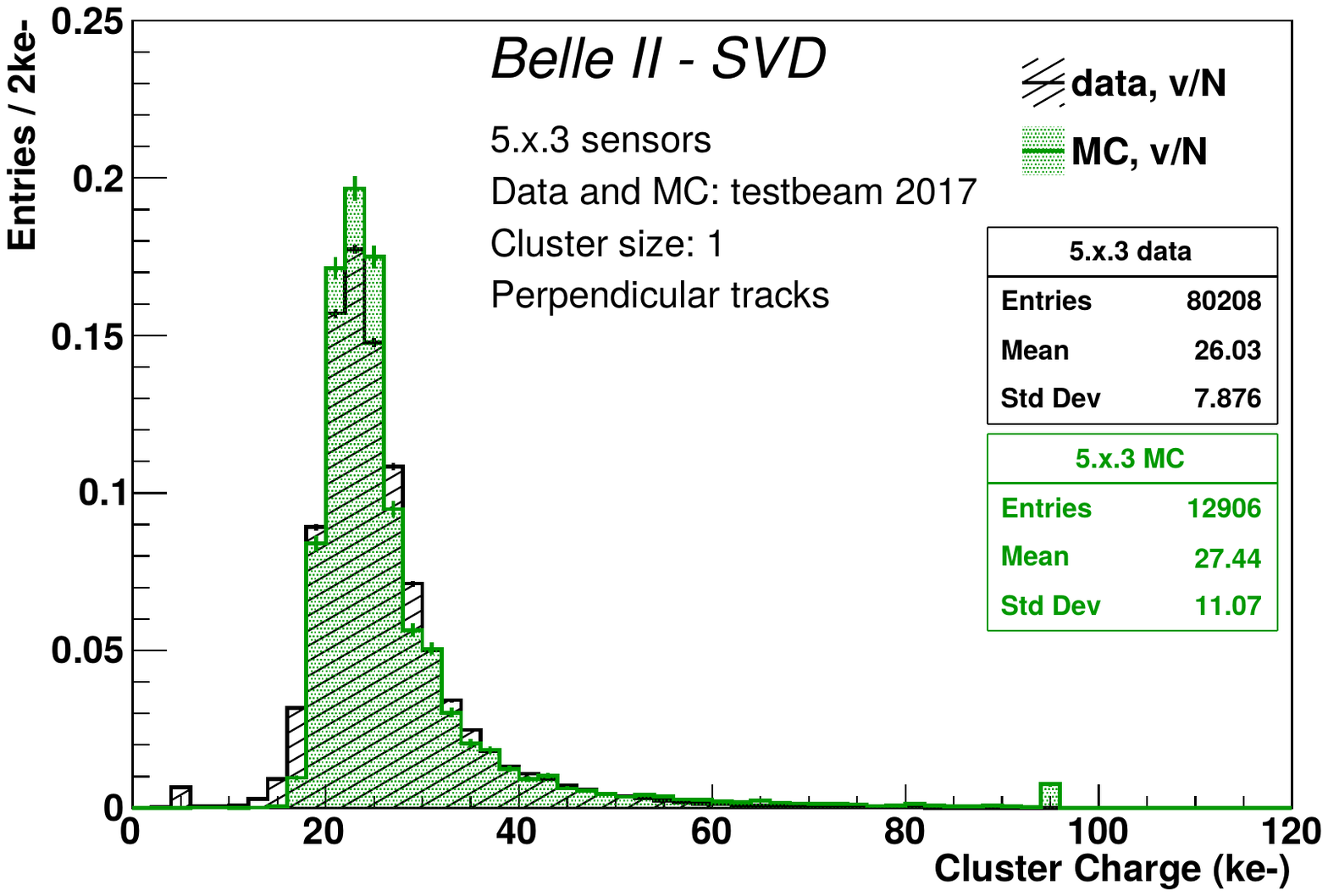}
        \includegraphics[width=.45\columnwidth,keepaspectratio]{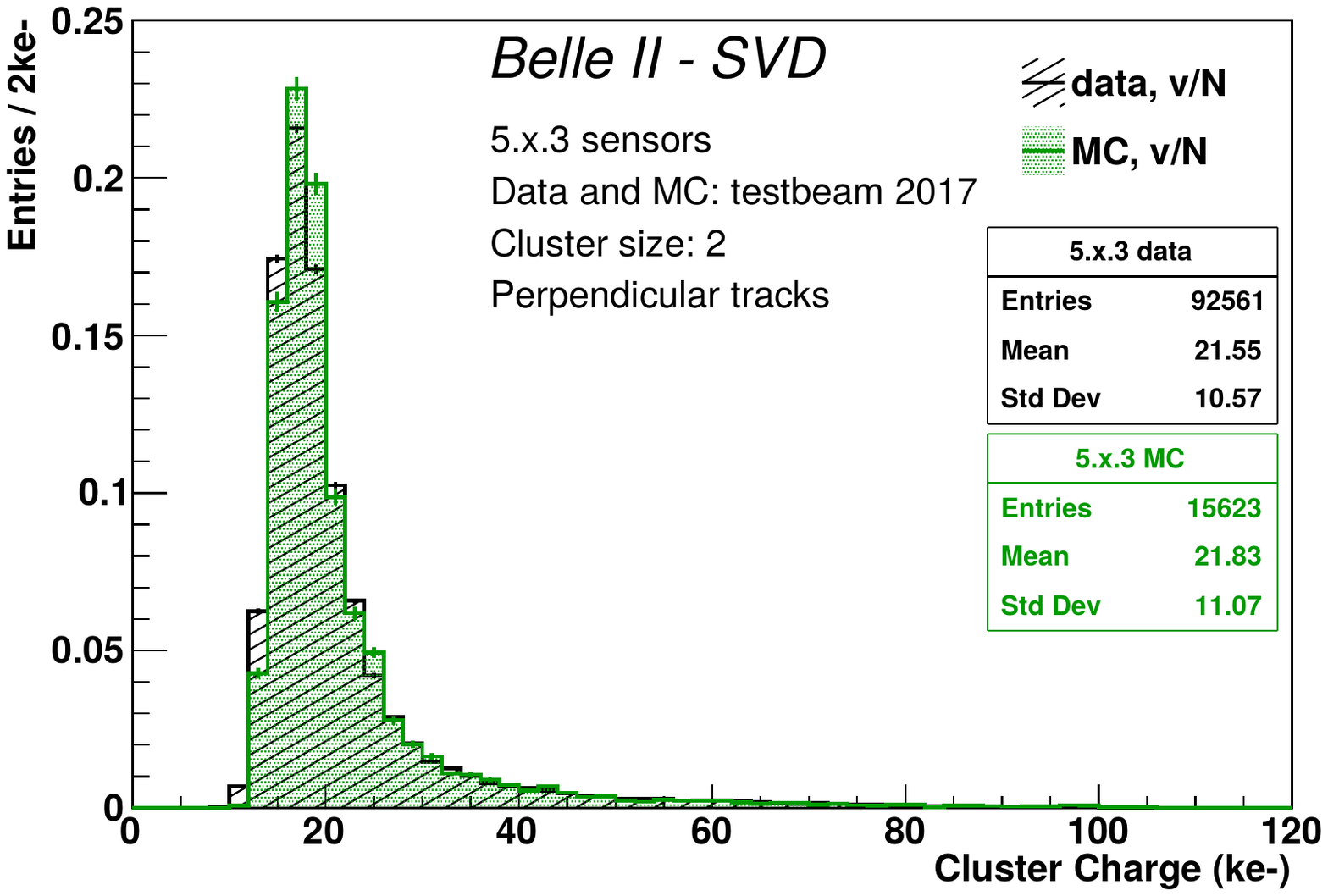}
    \caption{Comparison of cluster charge between test-beam data and simulation after tuning of the $C_i$ and  \nc{ADUEquivalent} values for Layer 5 sensors, the u/P side on the top row and the v/N side on the bottom row; for cluster size 1 on the left column and cluster size 2 on the right column. As described in section~\ref{sec:C2_design}, 
    L5.x.3 means all the sensor in layer 5, any $\phi$ position, $z$ position 3 (that is, all Origami\_-Z sensors in layer 5).    }
    \label{fig:C4_cl_charge_tbdata2017_simulation}
\end{figure}

The noise in electrons was estimated using the measured noise (in ADC) converted in electrons using the measured gain. 
For the tuning of $C_i$ and \nc{ADUEquivalent}, 2017 test-beam data (see table~\ref{tab:C2_testbeam}) was used. In the test-beam, performed with high energy electrons crossing the sensor with perpendicular incidence, most of the clusters have a cluster size 1 or 2, depending on the impact point of the tracks. Only a small amount of clusters have higher cluster size and are not used for this tuning. Due to the coupling coefficients $k_i$ described above, the total cluster charge for cluster size 1 and 2, $Q_{\rm cls1}$ and $Q_{\rm cls2}$ are sensitive to both parameters $C_i$ and \nc{ADUEquivalent}, since the other capacitances in the charge sharing model are assumed to be well known and fixed.  The ratio $R=Q_{\rm cls2}/Q_{\rm cls1}$ is instead only sensitive to $C_i$ and this sensitivity can be exploited to tune the interstrip capacitance $C_i$ with data in the following way. From the full simulation the dependence of the ratio $R$ from $C_i$ is extracted; then the value of $R$ measured on test-beam data is compared with simulation to extract $C_i$ for the various sensors types and sides. After this step, the total cluster charge in data and simulation is compared to extract the \nc{ADUEquivalent}\ value.  A quite good agreement for the total cluster charge distributions between simulation and test-beam data is achieved with this method, as shown in figure~\ref{fig:C4_cl_charge_tbdata2017_simulation} for a sensor in a Layer 5 ladder.  %(picture from Mateusz  presentation done on 7 03 2018)
For cluster size 2 the effect of charge loss of about 20\% on the v/N side with respect to the u/P side is clearly visible, which is related to the presence of a floating strip and larger pitch on the v/N side, as described above.
Although in test-beam data a fair agreement with the simulation was achieved, further tuning of the charge sharing model  is now ongoing to improve the matching of the cluster properties between simulation and collision data.

\textbf{Timing in simulation.}
In order to simulate trigger jitter, the time of all \nc{MCParticles} of an event are randomly  shifted using a Gaussian distribution centered at 0, and with a width of 10 ns; consequently, also the time of  \nc{SVDSimHits} are shifted. The simulation also handles the start of the APV sampling with respect to the arrival of the trigger. As explained in section~\ref{sec:C4_timeref}, in real data taking the quarter of the APV clock period in which the trigger arrived is stored as the trigger bin. In simulation the trigger bin is chosen as a random number between 0 and 3, and the beginning of the sampling is shifted by the trigger bin times the APV period divided by 4 ($\simeq$ \qty{7.86}{\ns}). 
%The trigger latency is tuned on data looking at the distance between the signal peak and the background bump, see figure~\ref{fig:C4_clusterTime} for simulation and figure~\ref{fig:C7_TimeMinusEventT0} for data.

\textbf{Run-dependent simulation.}
In addition to the run-independent simulation (same conditions for all events), the run-dependent simulation is produced, to properly take into account effects of the detector that vary from run to run.
Strips fired on noise and positive signals from the cross-talk effect described in section~\ref{sec:crosstalk} are automatically, and realistically, added to the run-dependent simulation by overlaying random-triggered real events to simulated signal events. The overlay allows to sum sample-by-sample the \nc{SVDShaperDigits} from simulation to the ones in real data.
The negative signals from cross-talk are not part of the simulation, but  studies on efficiency and cluster position resolution do not indicate that the performance is sensitive to these negative signals.
Finally,  all the strips and APV chips  disabled during data taking for the simulated run  are masked. This information is stored on the Conditions Database.

\textbf{Simulation performance.}% \textcolor{red}{Giuliana, I think we need to show here some data-MC comparison plots.}
\begin{figure}[hbt]
    \centering
     \includegraphics[width=.45\columnwidth,keepaspectratio]{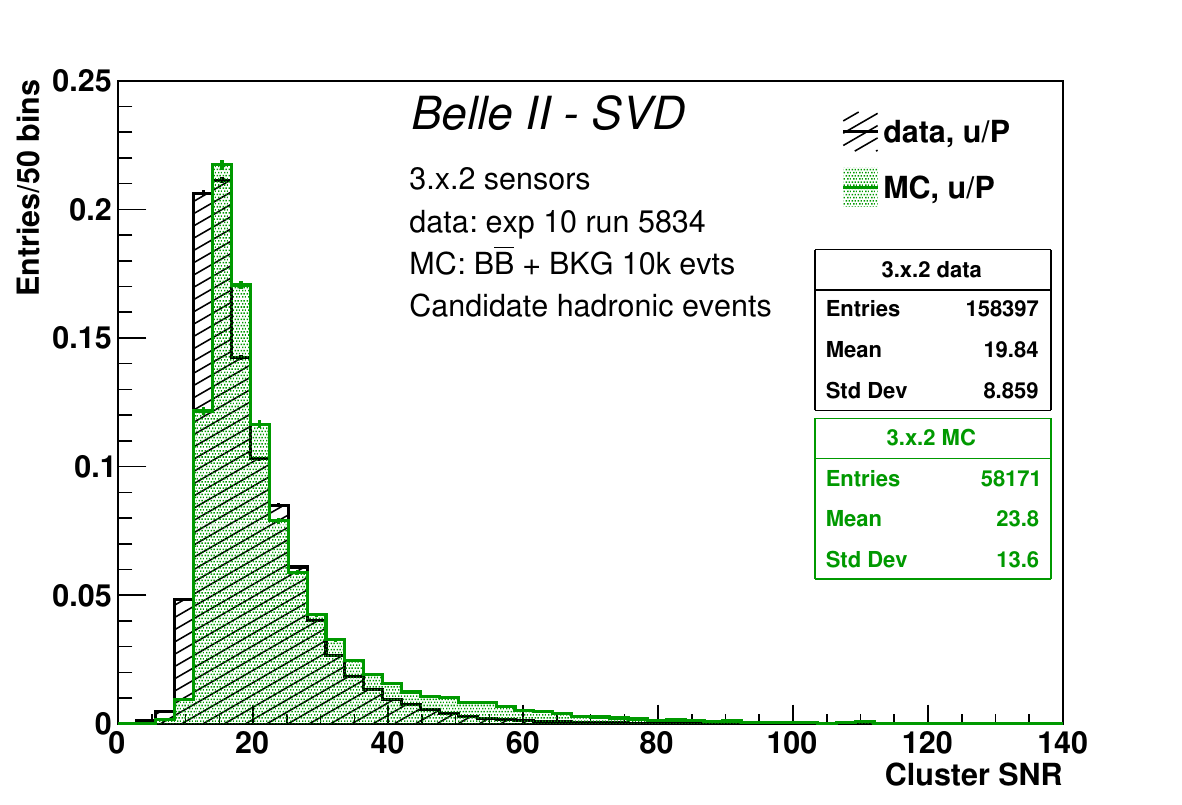} 
    \includegraphics[width=.45\columnwidth,keepaspectratio]{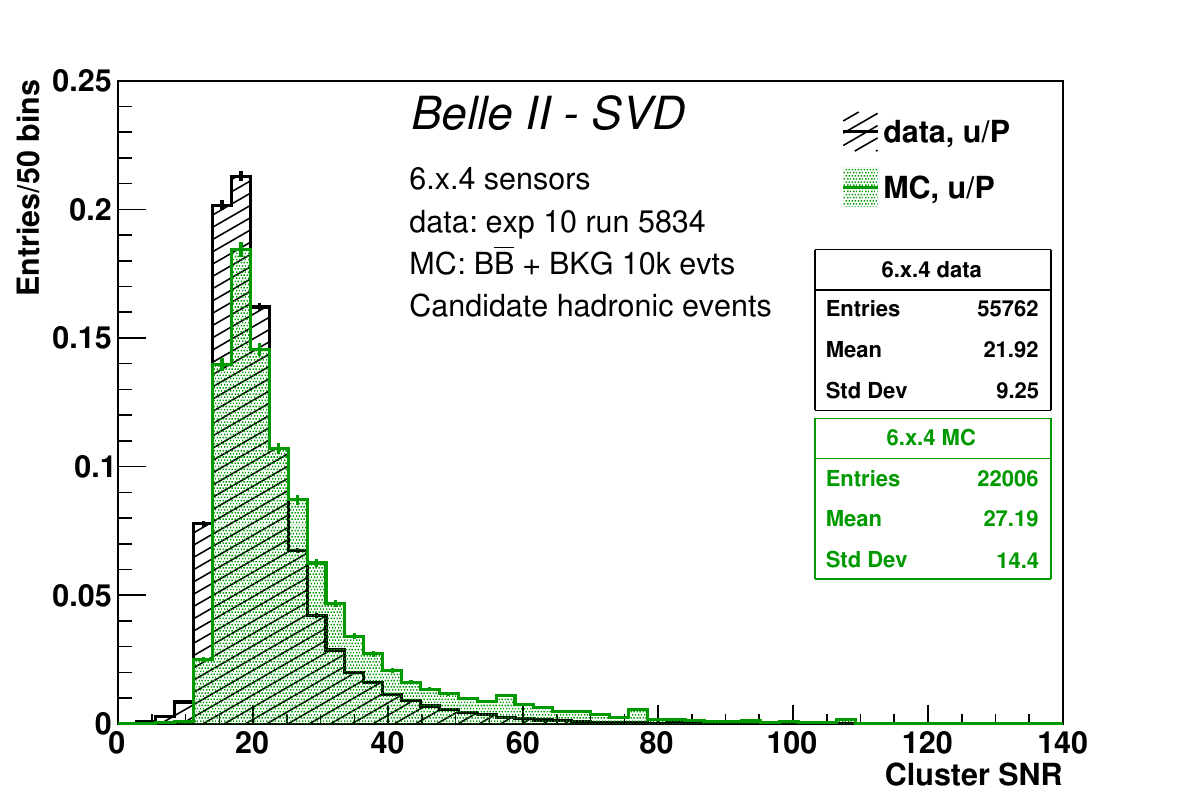} 
    \includegraphics[width=.45\columnwidth,keepaspectratio]{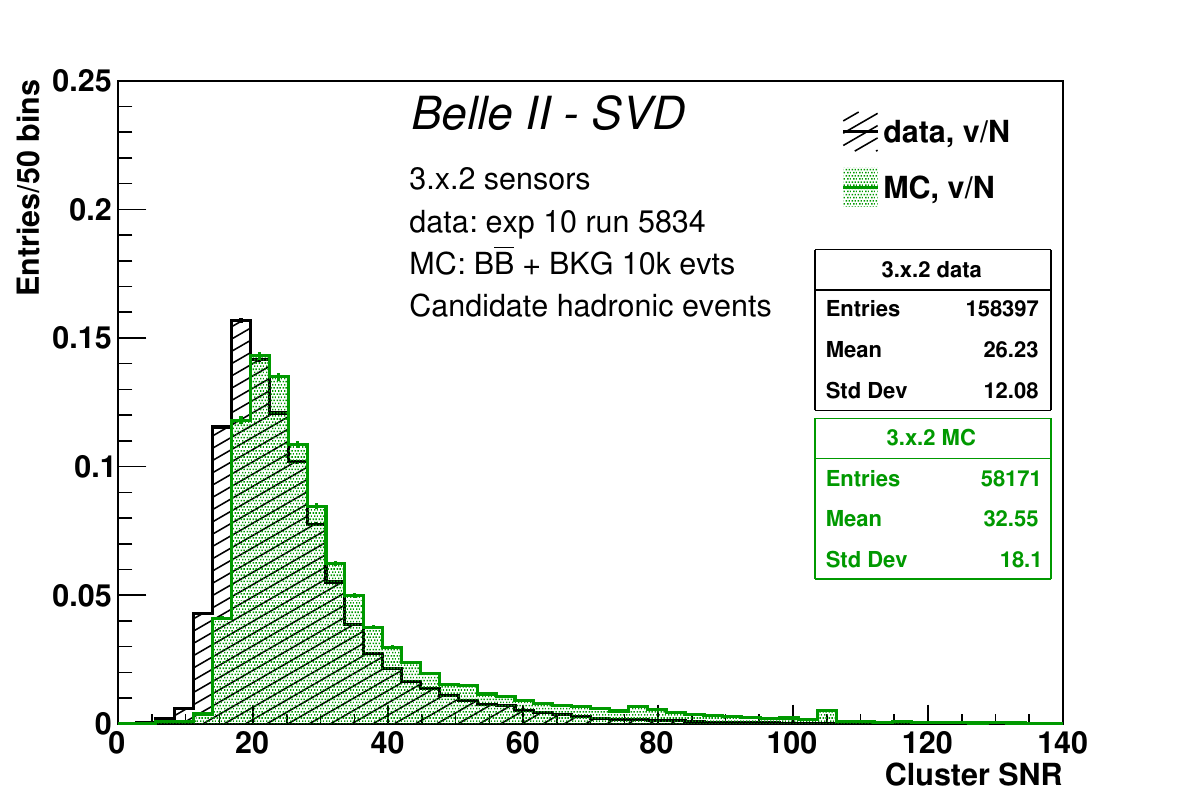} 
    \includegraphics[width=.45\columnwidth,keepaspectratio]{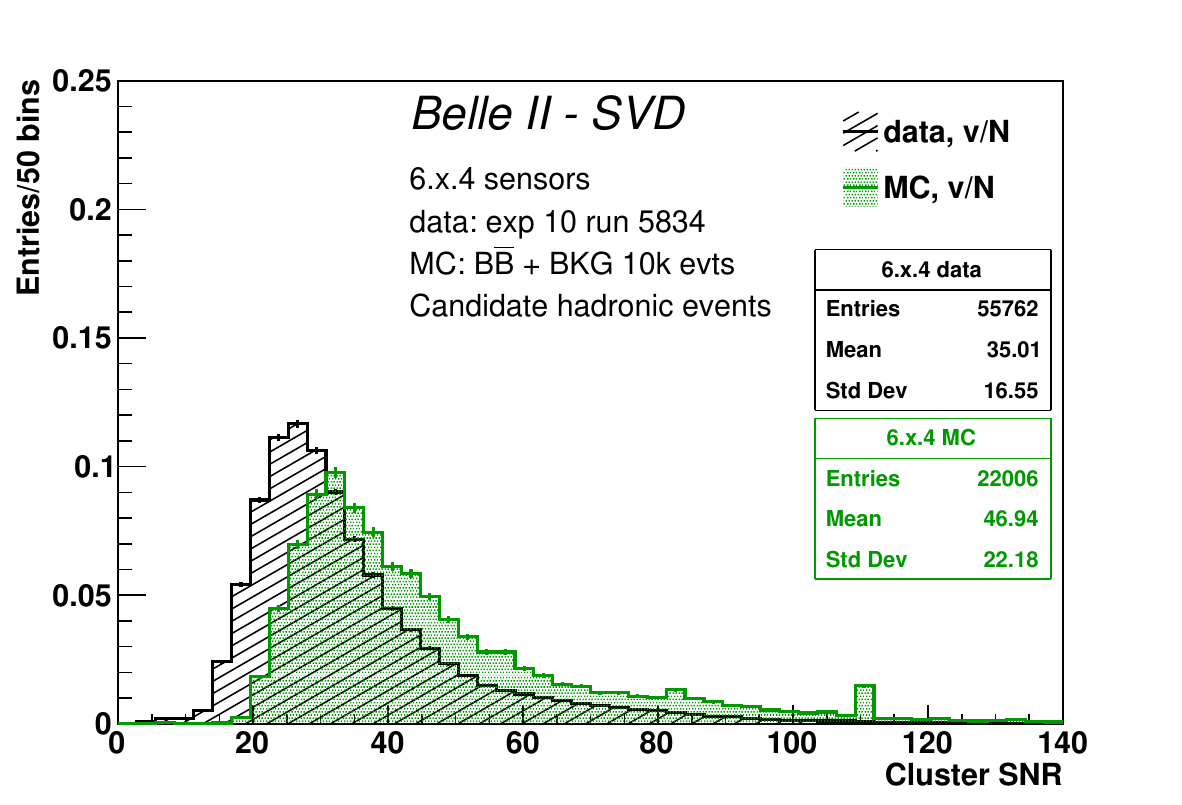} 
    \caption{SVD cluster \snrCL distributions for data (black) and simulated $B\bar{B}$ plus beam background (green) for the u/P side on the top row and the v/N side on the bottom row. L3 clusters on BW sensor on the left, and L6 cluster on one of the Origami sensors on the right.}
    \label{fig:C4_dataMC_clSNR}
    \end{figure}
The reader should note that, at this point of the chapter, the reconstruction has not been discussed yet, therefore this section contains references to later parts of the text, in particular to section~\ref{sec:C7_performance} for what concerns data.

\begin{figure}[hbt]
    \centering
    \includegraphics[width=.45\columnwidth,keepaspectratio]{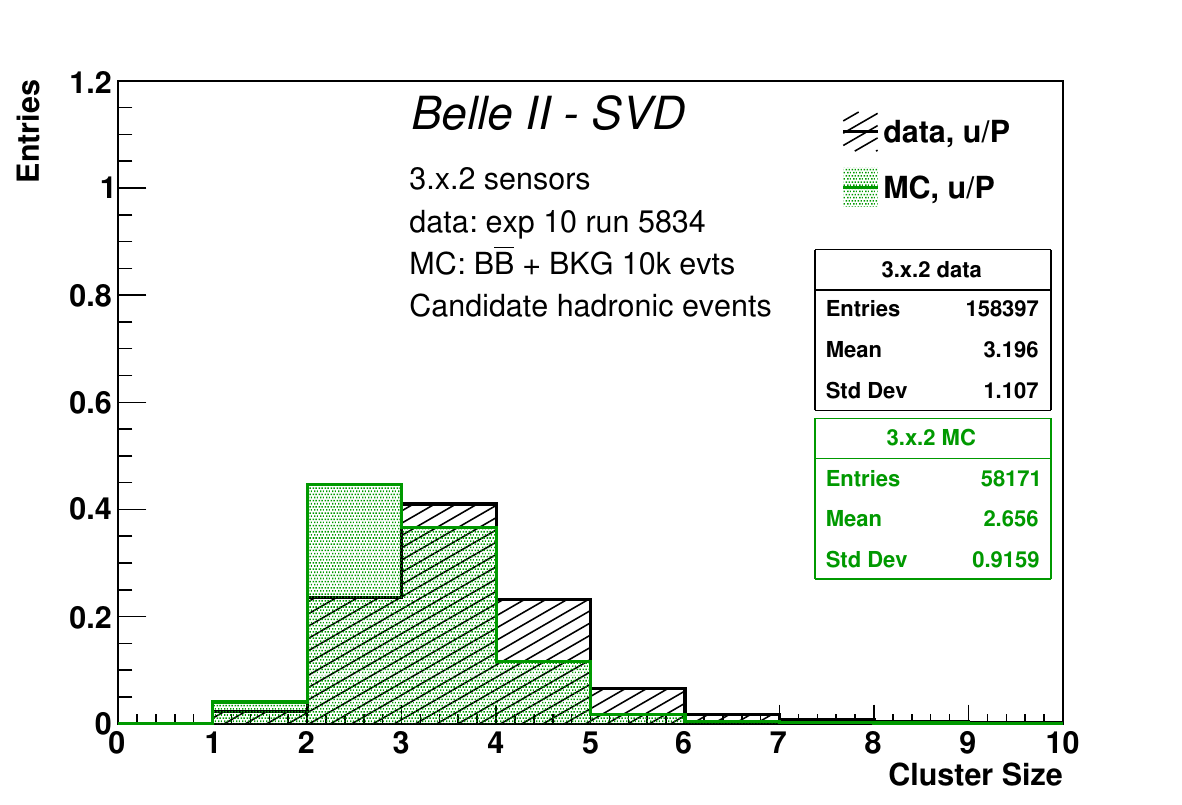} 
    \includegraphics[width=.45\columnwidth,keepaspectratio]{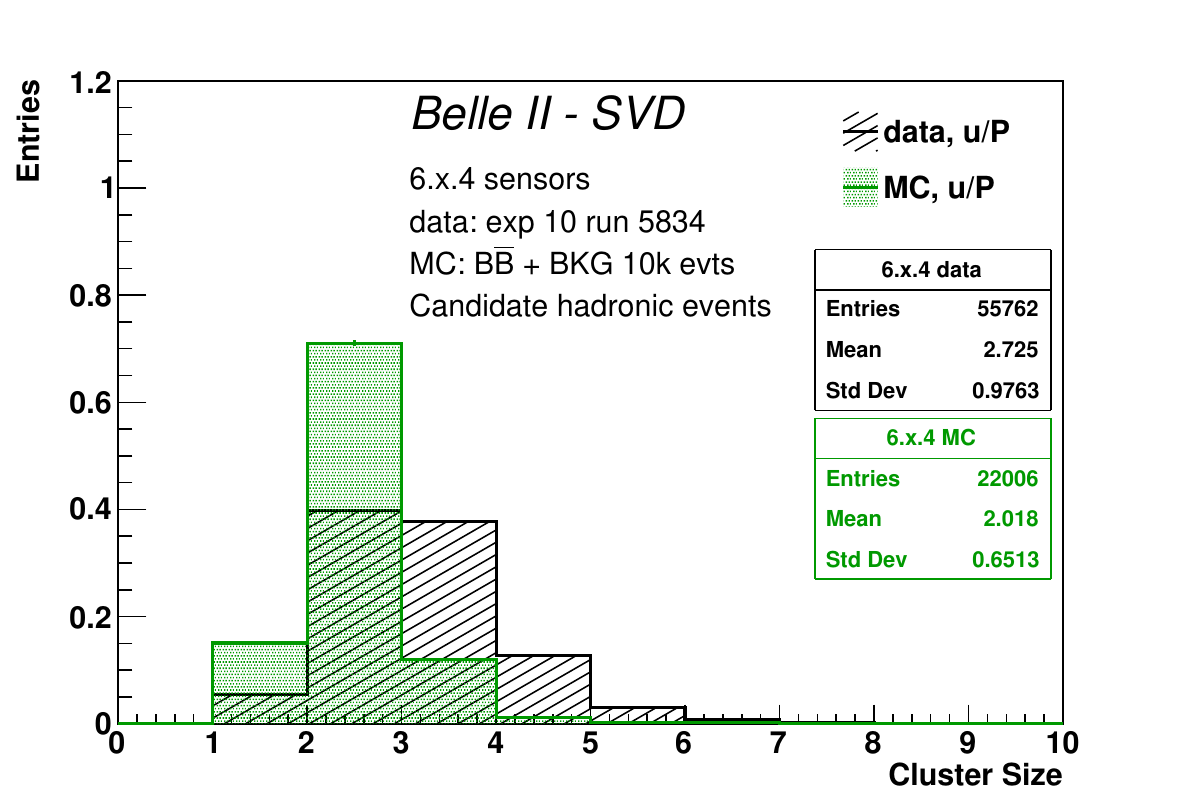} 
    \includegraphics[width=.45\columnwidth,keepaspectratio]{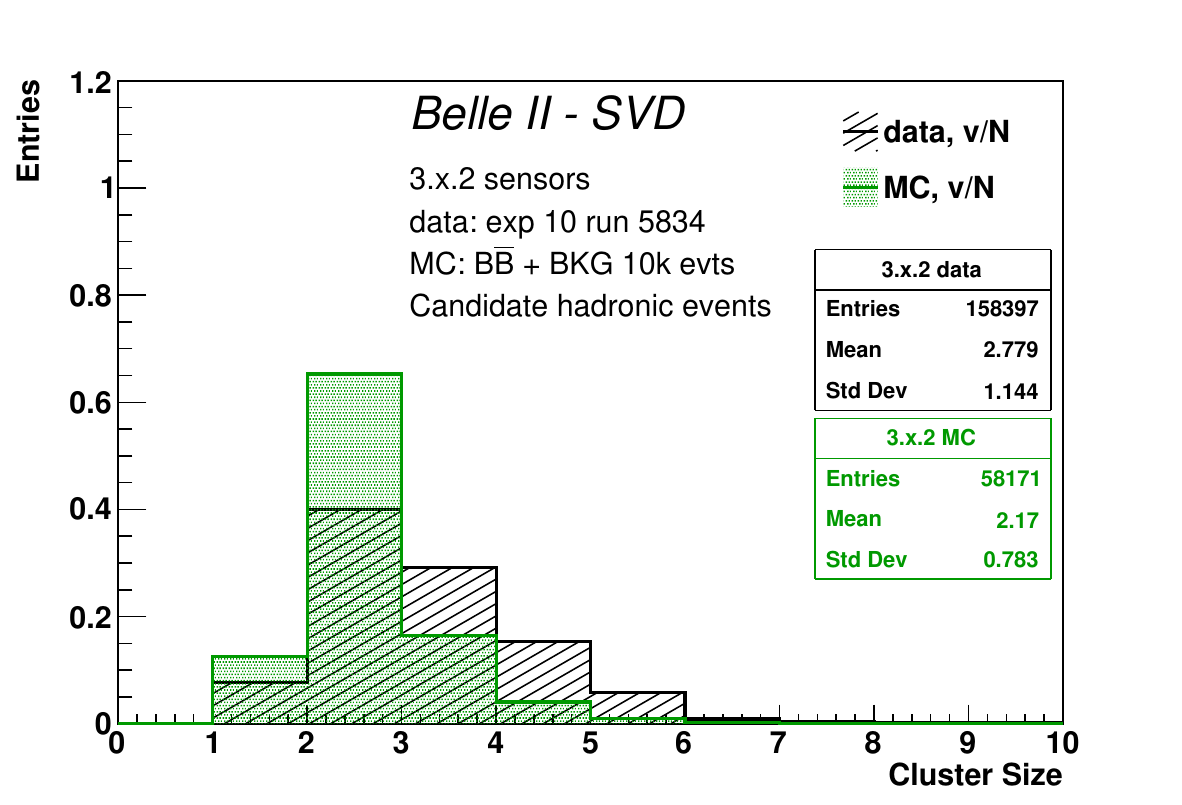} 
    \includegraphics[width=.45\columnwidth,keepaspectratio]{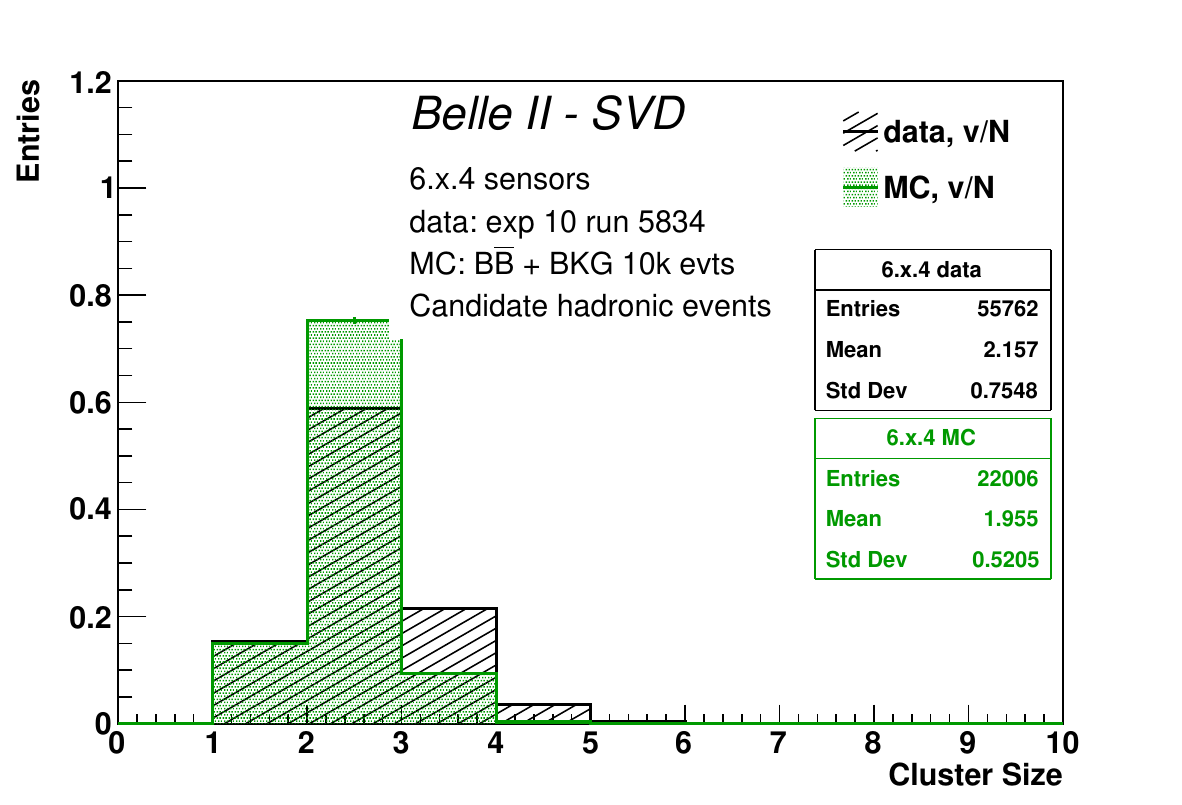} 
    \caption{SVD cluster size distributions for data (black) and simulated $B\bar{B}$ plus beam background (green) for the u/P side on the top row and the v/N side on the bottom row. L3 clusters on BW sensor on the left, and L6 clusters on one of the Origami sensors on the right.}
    \label{fig:C4_dataMC_size}
    \end{figure}

The simulation reasonably describes the cluster properties  for the collision data. A fair agreement was found for the cluster time, as can be seen comparing figure~\ref{fig:C4_clusterTime} for simulation and figure~\ref{fig:C7_TimeMinusEventT0} for data. The collected signal and the signal-to-noise ratio are also well simulated on the u/P side, while on the v/N side some data-MC differences up to 30\% were seen for the layers 4, 5, and 6 sensors with larger pitch, indicating  that still some tuning of the charge loss and sharing model is needed to improve the agreement. This can be seen in figure~\ref{fig:C4_dataMC_clSNR}. The cluster size is systematically smaller on simulation with respect to collision data, especially for the u/P sides, which have a smaller pitch, as shown in figure~\ref{fig:C4_dataMC_size}.

\begin{figure}[hbt]
    \centering
     \includegraphics[width=.45\columnwidth,keepaspectratio]{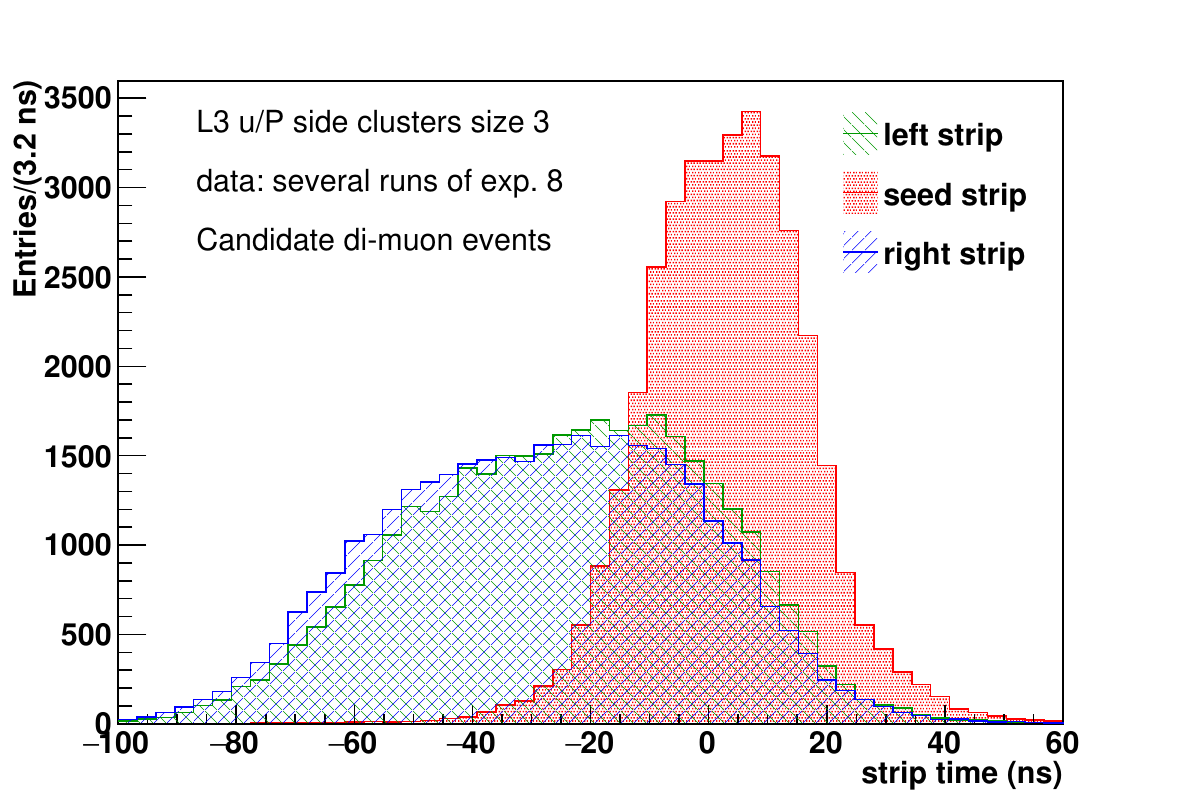}
     \includegraphics[width=.45\columnwidth,keepaspectratio]{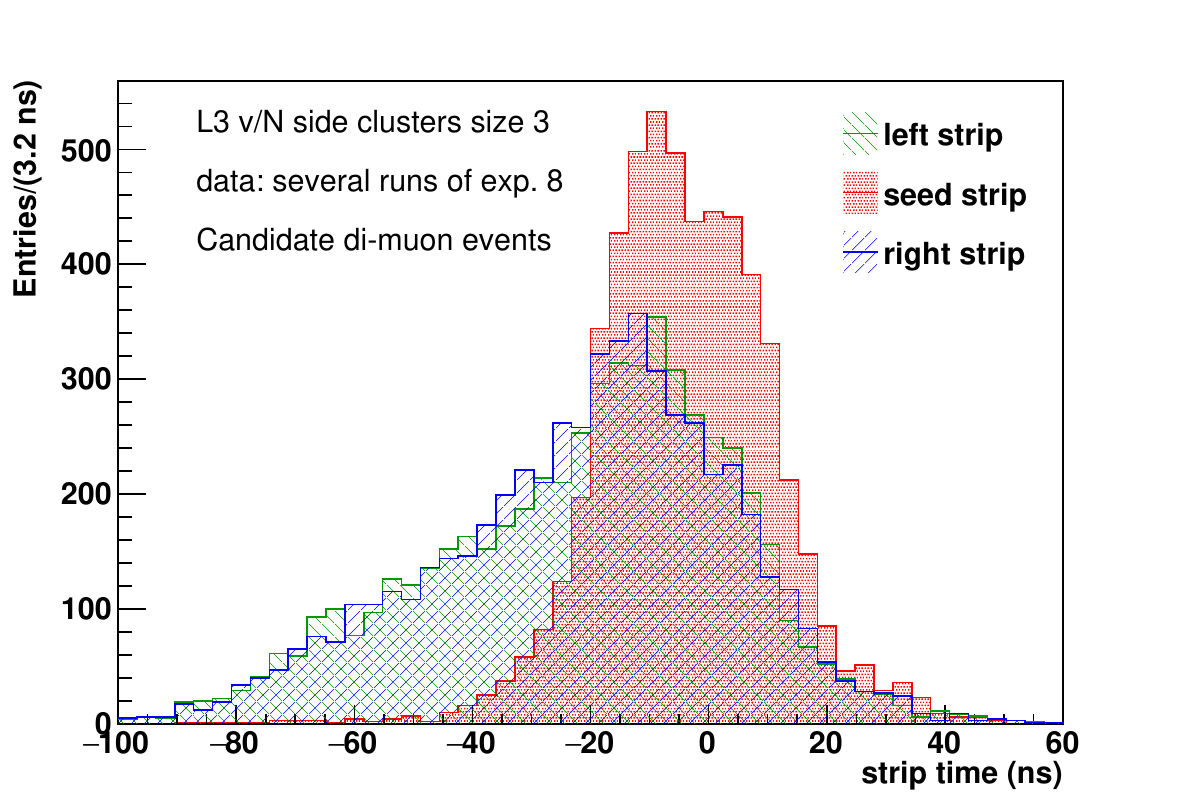}    
    \caption{Time distributions of the strips belonging to layer 3 clusters of size 3 for perpendicular tracks in di-muon events reconstructed in several runs of 2019 data. The time distribution of central strip is shown in red, while the overlaid time distributions of the left (right) edges strips is shown green (blue).}
    \label{fig:C4_stripTimes}
\end{figure}

The disagreement in cluster size is an indication that the simplified signal formation model implemented in the simulation is not good enough to reproduce the collision data.
This level of disagreement was not observed in test-beam data used to tune the simulation, in which there was a much better agreement in cluster size between data and simulation. This difference could be due to a different online zero-suppression criterion applied during test-beam ($\snrST > 5$) with respect to physics runs ($\snrST > 3$). A tighter zero-suppression criterion reduces the effect of the noise on the cluster size, but it degrades the cluster position resolution.
Beside this, there is another significant difference between data and simulation regarding the strip time. In data, the strip at the edge of a cluster of size 3 has a hit time significantly smaller than that of the central strip, as shown in figure~\ref{fig:C4_stripTimes}. The effect is more evident for the u/P side with smaller pitch, than the v/N side.
Both these effects are probably due to the fact that the signal induced by moving carriers (Ramo-Shockley theorem) is not simulated: the fast APV25 may catch the  signal induced by charges moving in the weighting field even if this signal integrates a null charge. 
Studies using an external software~\cite{weightfield2} confirm that the timing structure of the strip in the clusters is reproduced when the signal formation is simulated with the Ramo-Shockley theorem. Further studies and simulation tuning are now ongoing to improve the implemented model for signal formation in order to have a better agreement with data.

Finally, the cluster position resolution is optimistic in simulation if compared to data, as shown in figure~\ref{fig:C7_resolution}. As commented in section~\ref{sec:C7_resolution}, this effect is not a surprise given the disagreement in cluster size and the partial inaccuracy of the charge sharing model. Preliminary simulation studies extending the time structure of the strip signals as a way of obtaining a more accurate cluster position estimate do not seem to significantly deteriorate the simulated resolution to match the measured one.  However, these studies are not based on first principles, but on a phenomenological parametrization of what is observed in calibrations, and may therefore still miss the physics that would have an impact on the resolution.

\lvlthreesection{Belle~II conditions database}\label{sec:C4_cdb}
\editor{Giulia} %, details of the content of the section were moved in the Jira ticket B2SVD-205
The SVD detector status and calibrations are stored in the \belletwo\ Conditions Database~\cite{Wood:2019twj}. Conditions are stored in serialized ROOT objects called payloads.
Intervals of  Validity (IoVs) specify  the  time interval for  which  a  payload is valid. Time intervals are defined with run-level granularity. A given payload may have multiple IoVs assigned to it if it is valid for different periods of time.
A Global Tag is a list of IoVs and associated payloads, and it is used to select a complete set of conditions for a given processing or reprocessing effort.

The Online Global Tag contains all the conditions that are used during data taking by the High Level Trigger as well as for the online data-quality monitoring.
The SVD detector status and the results of the local calibrations, described in section~\ref{sec:C4_local_runs}, are promptly uploaded to the Online Global Tag. This allows an optimal data quality for the following runs that ensures an unbiased High Level Trigger selection.

Other calibrations cannot be performed directly during data taking and are performed on triggered events. This is for example the case of the hit time calibration described in section~\ref{sec:C4_CoGCalibration}. The results of these calibrations are uploaded to the Global Tags that are used for the regular reprocessing of the data. These reprocessing allow the best possible data quality for physics analyses.

\lvltwosection{Reconstruction Software}
\label{sec:C4_reconstruction}
\editor{Giulia}%, details of the content of the section were moved in the Jira ticket B2SVD-203; for this introductory part, see B2SVD-217}

The goal of the SVD reconstruction is to provide the 3D position, time, and ionization charge of the charged particles crossing the SVD sensors. This information is used to reconstruct the tracks and to provide information on the particle identification. The reconstruction of charged particles with low transverse momentum completely relies on SVD reconstruction, as the CDC-only tracking efficiency is $\sim80\%$ for $p_{\mathrm T} \simeq 100$ MeV/c.

The SVD reconstruction was developed to cope with an expected background hit rate of 1.5 MHz/cm$^2$ at the SuperKEKB design luminosity, $8\times 10^{35} {\rm cm}^{-2}{\rm s}^{-1}$, which corresponds to an occupancy of $\simeq 1.5\% $ in the innermost layer, layer 3. In order to reconstruct the signal tracks (the average number of tracks in an $\Upsilon(4S)$ event being 10) most of the hits from particles belonging to a different collision event, mostly due to machine background particles (off-time hits), should be identified and removed before the tracking stage. Therefore, beside the ability to correctly reconstruct the position of the hit on the sensor, the hit time must be estimated with a precision that allows to distinguish on-time from off-time hits.
As of 2020, given the low machine backgrounds, all reconstructed hits are provided to the tracking, including the off-time hits. In the future, if the machine background further deteriorates, the distinction between on- and off-time hits will become crucial for the tracking performance of the experiment. 

In the following, the algorithms that allow to reconstruct the clusters and space-points are described, and  the reconstruction performance on simulated events are briefly shown, while in chapter~\ref{sec:C7_performance} the reconstruction performance on data is discussed. Performance results refer to 6-sample acquired events both for data and simulation. From the point of view of reconstruction and hit time calibration, the simulated events are treated exactly in the same way as the data.

\lvlthreesection{Unpacking and strip reconstruction}
\label{sec:C4_strip_reconstruction}
\editor{Giulia}%, details of the content of the section were moved in the Jira ticket B2SVD-15}

As shown schematically in figure~\ref{fig:C4_offline_reconstruction_data_flow}, the raw data received from \belletwo\ DAQ are unpacked into
\begin{itemize}
    \item an array of digits (\nc{SVDShaperDigits}), one per each strip passing the online zero suppression, and consisting of the 6 digitized values for the sampled waveform;
    \item a single object that contains event-wise information \nc{SVDEventInfo}: 
    the data acquisition mode (6- or 3-sample acquisition), the trigger type, and the \gls{triggerbin}: the arrival time of the L1 trigger relative to the APV sampling clock.
    The trigger bin allows to move the hit time measured
 in the SVD reference frame to the trigger reference frame, common to all subdetectors, and vice-versa.
\end{itemize}

\begin{figure}
    \centering
    \includegraphics[width=.45\columnwidth,keepaspectratio]{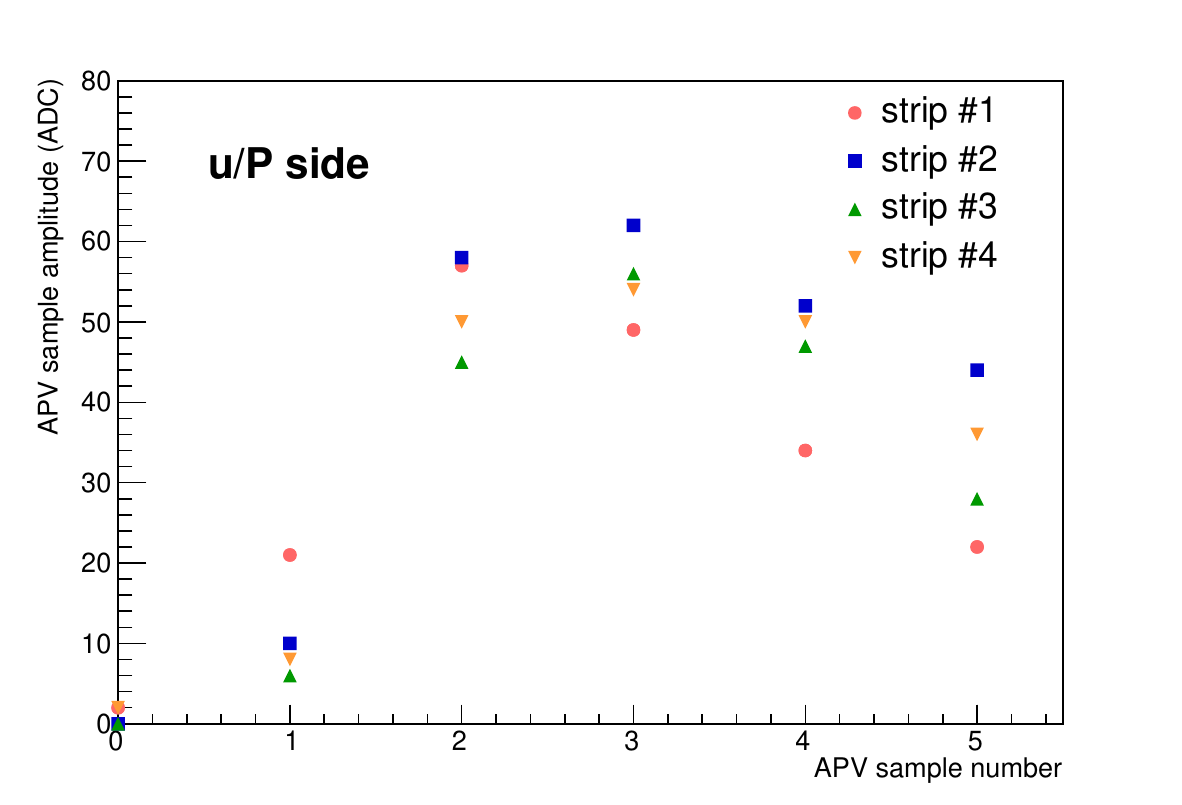}
    \includegraphics[width=.45\columnwidth,keepaspectratio]{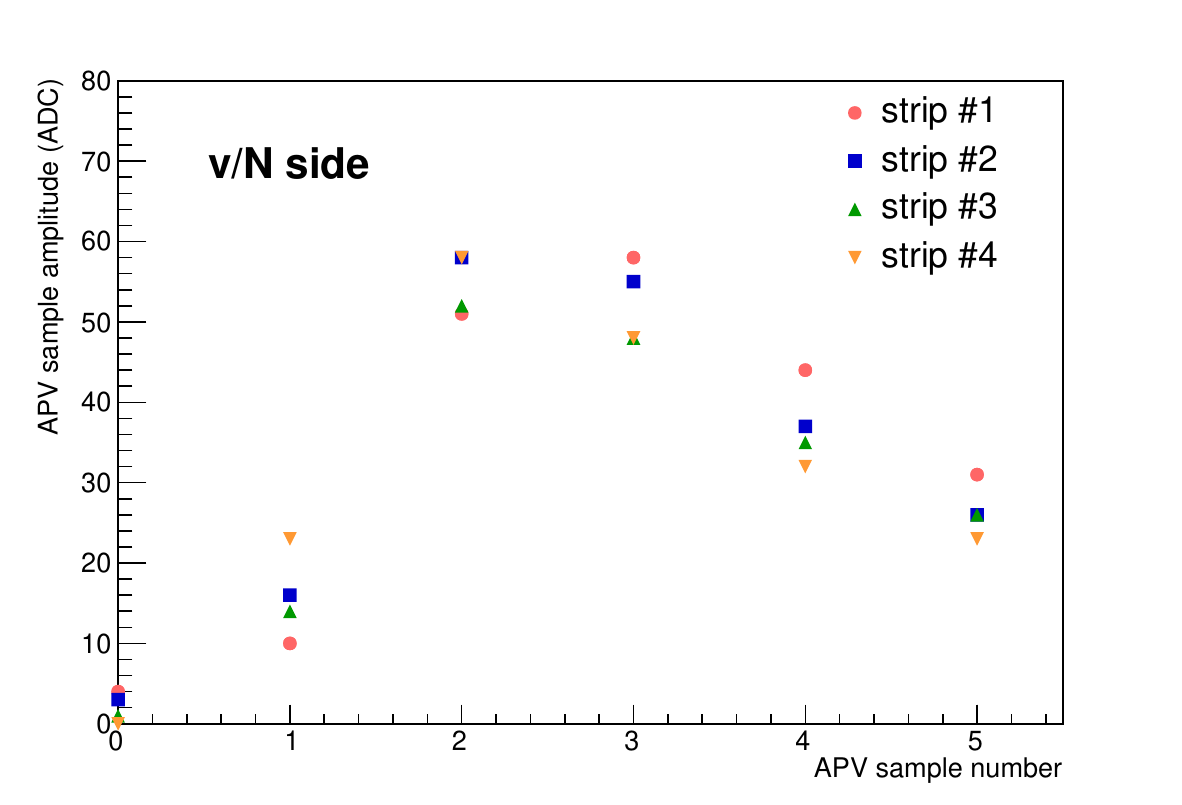}
    \caption{u/P side (left) and v/N side (right) digitized waveforms of the APV25 in data, shown for 4 random strips, in hadronic events collected in 2020. 
    	The noise ranges between 2.5 and 4 ADC on u/P and v/N side.}
    \label{fig:C4_shapers}
\end{figure}

The unpacker module checks the data format, and the consistency of  event-wise information across the FADC, FTB and APV. It also checks for errors from FADC, FTB or APV reported in the raw data.
Detected errors are written in a dedicated object, which is not stored for reconstruction  but  is used in the online Data Quality Monitor. If an error is detected, the unpacker does not produce any strip information in output, hence that event does not have SVD data available for reconstruction.
In figure~\ref{fig:C4_shapers} examples of the digitized waveforms of strips for both sides are shown.
%compared to the noiseless analog output of the APV recorded during a calibration.

{}

The \nc{SVDShaperDigits}\ are reconstructed by a dedicated module that computes the charge and the hit time of each strip, then stored in a dedicated array named \nc{SVDRecoDigits}.
The {\bf charge} of the strip corresponds to the highest of the six digitized samples, converted in electrons using the strip-gain as determined during the local calibration (section~\ref{sec:C4_local_runs}), and read from the Conditions Database. 
The raw {\bf hit time} ($t_{\rm raw}$) is determined as a weighted average of the sampling time ($t_i$) with the ADC count of the $i$-th sample {$A_i$} corrected by $t_{\rm peak}$ to remove differences in peaking times among the strips:
\begin{equation}
    t_{\rm raw} = \sum_{j=0}^{j=5} \frac{t_j \cdot A_j}{A_{\rm tot}} - t_{\rm peak},\quad {\rm with} \quad A_{\rm tot} = \sum_{j=0}^{j=5} A_j
\end{equation}
where $t_j = j \cdot 1/f_{\rm APV}$ and $f_{\rm APV} = 31.805$ MHz, and $t_{\rm peak}$ is determined for each strip from the local calibration and stored in the Conditions Database, see section~\ref{sec:C4_local_runs}.\\
As the last step of the strip reconstruction, the raw strip time is calibrated:
\begin{equation}
    t_{\rm strip} = f(t_{\rm raw})+ \delta t
\end{equation}{}
where $f(t_{\rm raw})$ is the calibration function and $\delta t$ is an event-wise time shift that allows to move the hit time from the SVD reference frame to the trigger time reference frame, as explained in section~\ref{sec:C4_CoGCalibration}. The calibration function is stored in the Conditions Database, together with the value of its parameters, that depend on each sensor side.

\lvlthreesection{Cluster reconstruction and SpacePoint creation}
\label{sec:C4_clustering}
\editor{Giulia}
The cluster is reconstructed from the strips passing the online zero suppression, which rejects the signal from a strip whose maximum signal height $S_i$ is less than three times the noise $N_i$ of the strip. Please note the difference between the strip \snrST, defined as the ratio
of the strip signal to its noise, and the cluster \snrCL, defined as ratio of the total charge of a cluster to its total noise (discused below). 

A bunch of strips, which comprises only adjacent strips with $\snrST > \snrADJ$, is regarded as a cluster if at least one of the strips (seed strip) in the bunch has $\snrST > \snrSEED$ and the bunch has $\snrCL > \snrCL^{\rm min}$.
%A collection of consecutive strips each one with a minimum signal of $\snrADJ\times N_i$ are considered a cluster if at least one strip has a minimum signal of $\snrSEED\times N_i$ and if the cluster has a minimum SNR, $\snrCL^{\rm min}$.
In the current reconstruction $\snrADJ=3$ (the same as the online zero-suppression criterion, therefore all strips available at the reconstruction level are also available for clustering), $\snrSEED=5$, $\snrCL^{\rm min} = 0$. Given the high \snr  (see section \ref{sec:C9_SNR}), this online zero suppression cut at 3 is small enough for the resolution purpose, as verified also from testbeam data, where a good resolution was achieved even with a higher (5) online zero suppression cut. For Belle II physics runs a zero suppression cut of 3 was adopted, not to overload the data bandwidth with strips firing on noise while retaining good resolution.
The reconstruction performance are evaluated on simulated events, including beam background expected at the design luminosity.

\begin{figure}
    \centering
    \includegraphics[width=.45\columnwidth,keepaspectratio]{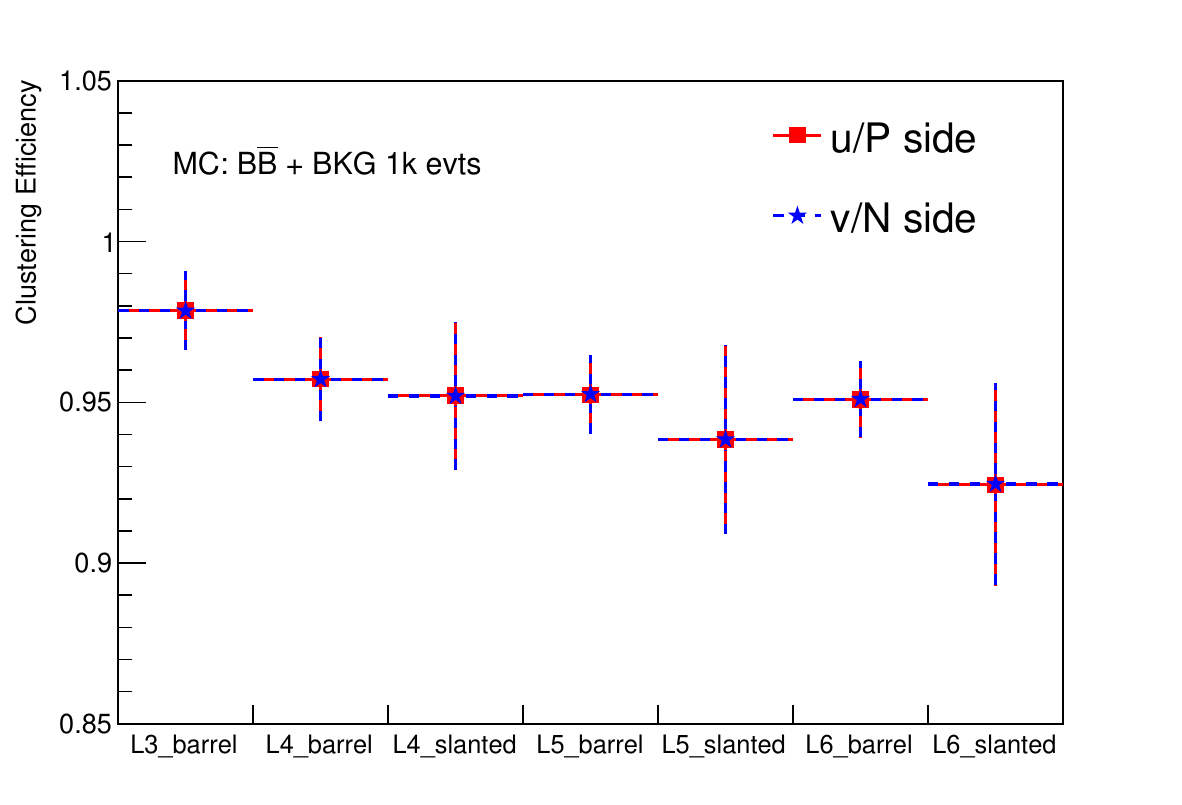} 
     \includegraphics[width=.45\columnwidth,keepaspectratio]{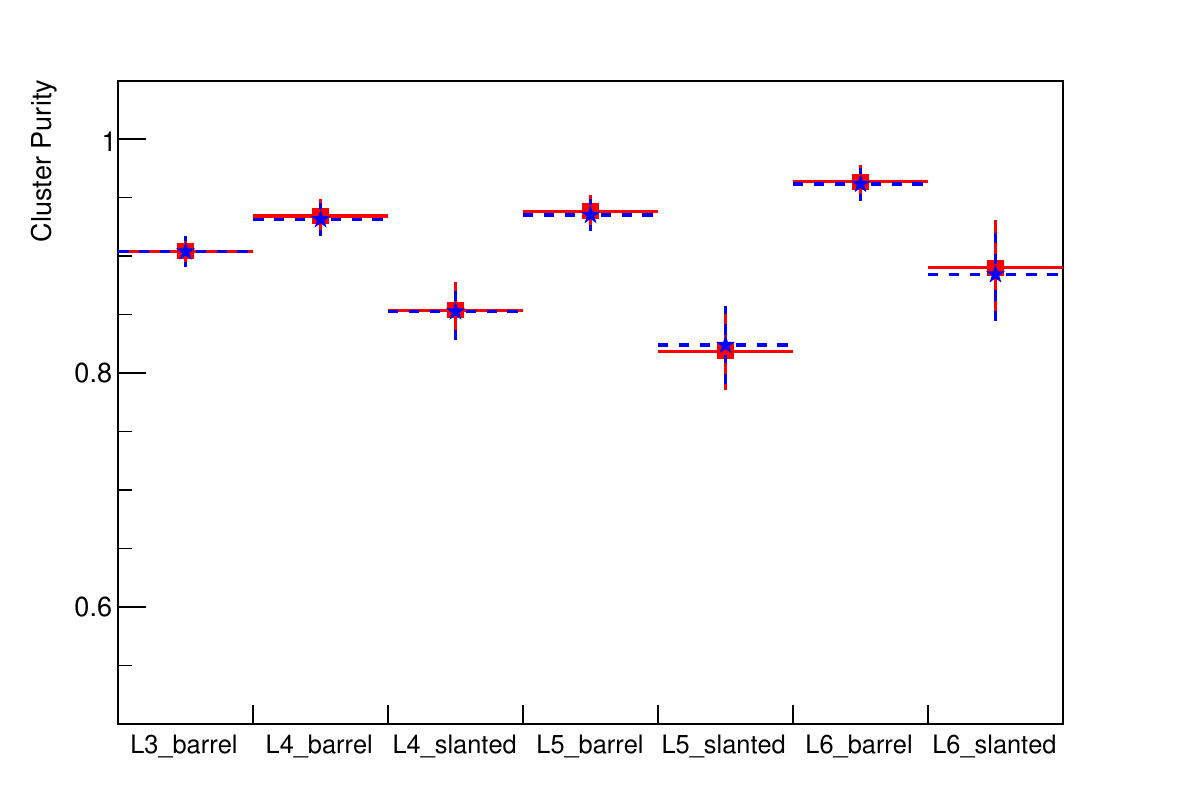} 
    \caption{Clustering efficiency (left) and cluster purity (right) for the u/P  (red) and v/N side (blue) for different sensor types, estimated on simulated events, including beam background expected at the design luminosity.}
    \label{fig:C4_clusterPerf}
\end{figure}

The reconstruction efficiency of clusters, defined as the fraction of \nc{SVDTrueHits} for which a cluster was reconstructed, is close to 1 and it is shown in the left plot of figure~\ref{fig:C4_clusterPerf} for different sensor types and layers. The cluster purity reported in the right plot of the same figure is defined as the fraction of signal strips in the clusters associated to \nc{SVDTrueHits}. Barrel sensors show a higher purity because of the angular distribution of beam background which is smaller in the barrel region compared to the forward and backward regions. Among the barrel sensors, layer 3 has a lower purity because it is located closest to the \acrshort{IP}  where beam background is higher as it decreases rapidly going outward.

{}

\begin{figure}
    \centering
   \includegraphics[width=.45\columnwidth,keepaspectratio]{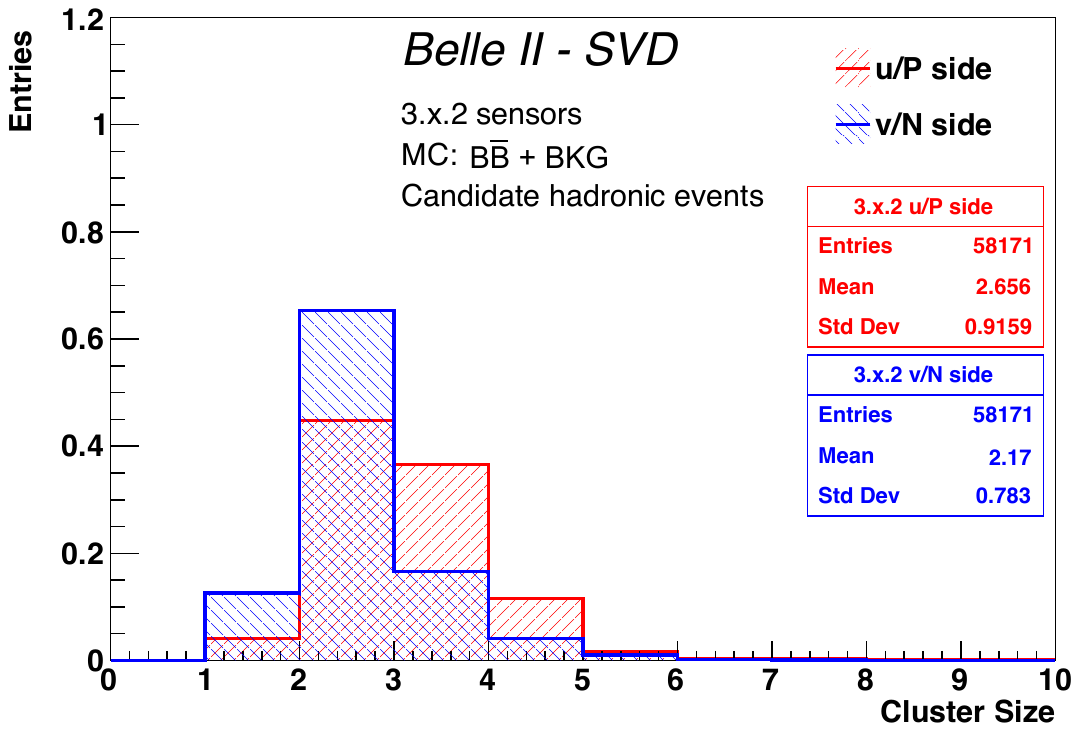}
    \includegraphics[width=.45\columnwidth,keepaspectratio]{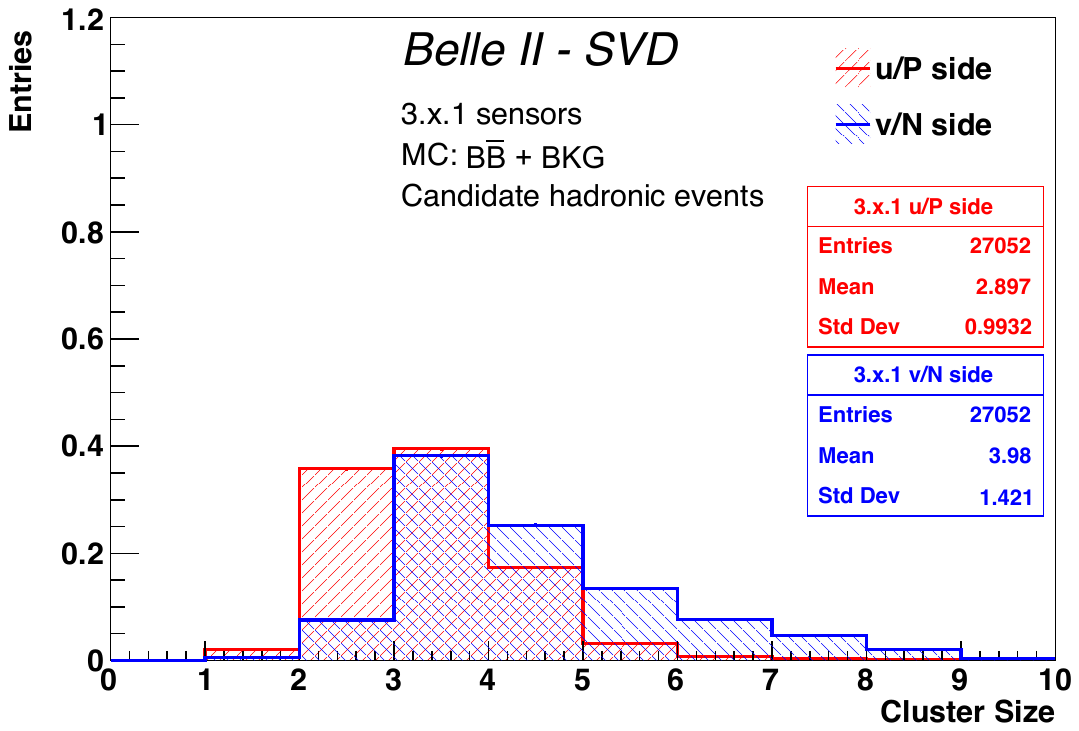}\\  
      \includegraphics[width=.45\columnwidth,keepaspectratio]{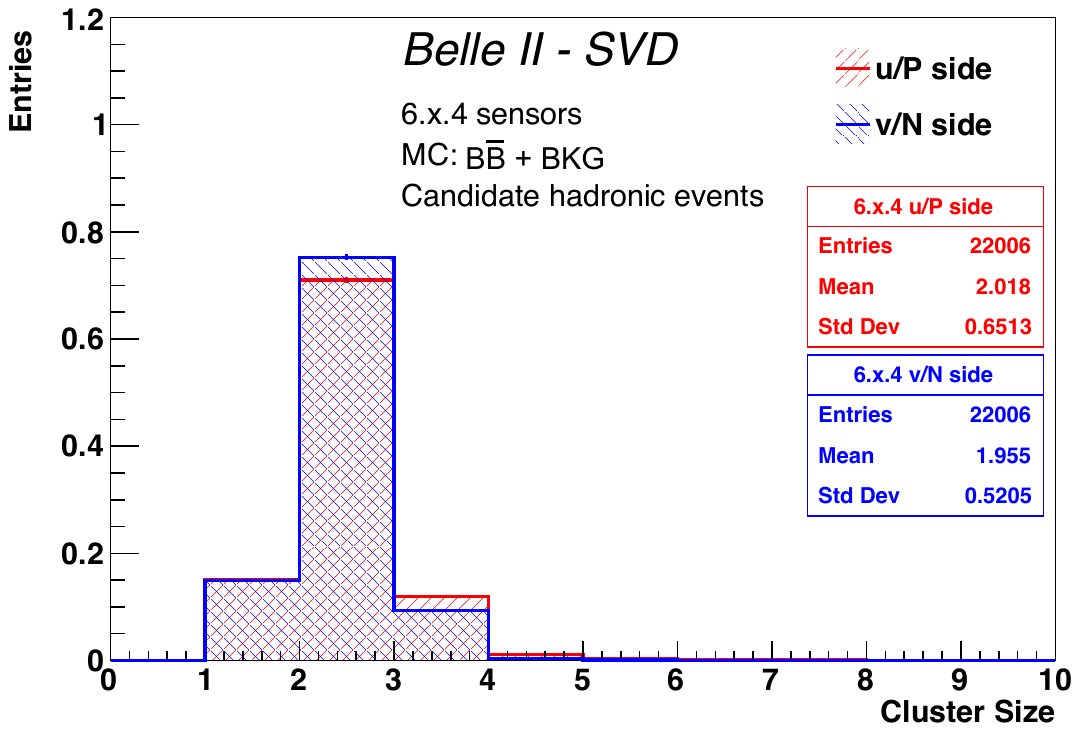}
     \includegraphics[width=.45\columnwidth,keepaspectratio]{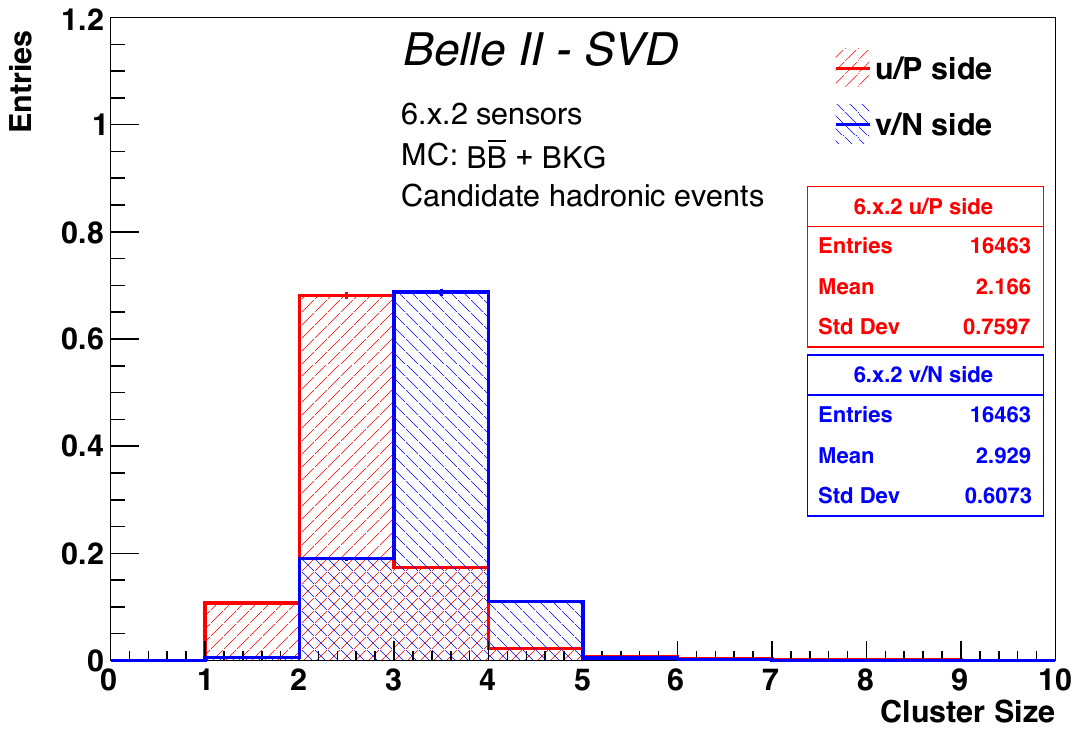}
         \caption{Reconstructed cluster size for simulated events including beam background expected at design luminosity, for different layers and sensors:layer 3 forward (upper-right) and backward (upper-left) and two barrel layer 6 sensors (bottom-right and left).}
    \label{fig:C4_clusterSize}
\end{figure}

\begin{figure}
\centering
         \includegraphics[width=.45\columnwidth,keepaspectratio]{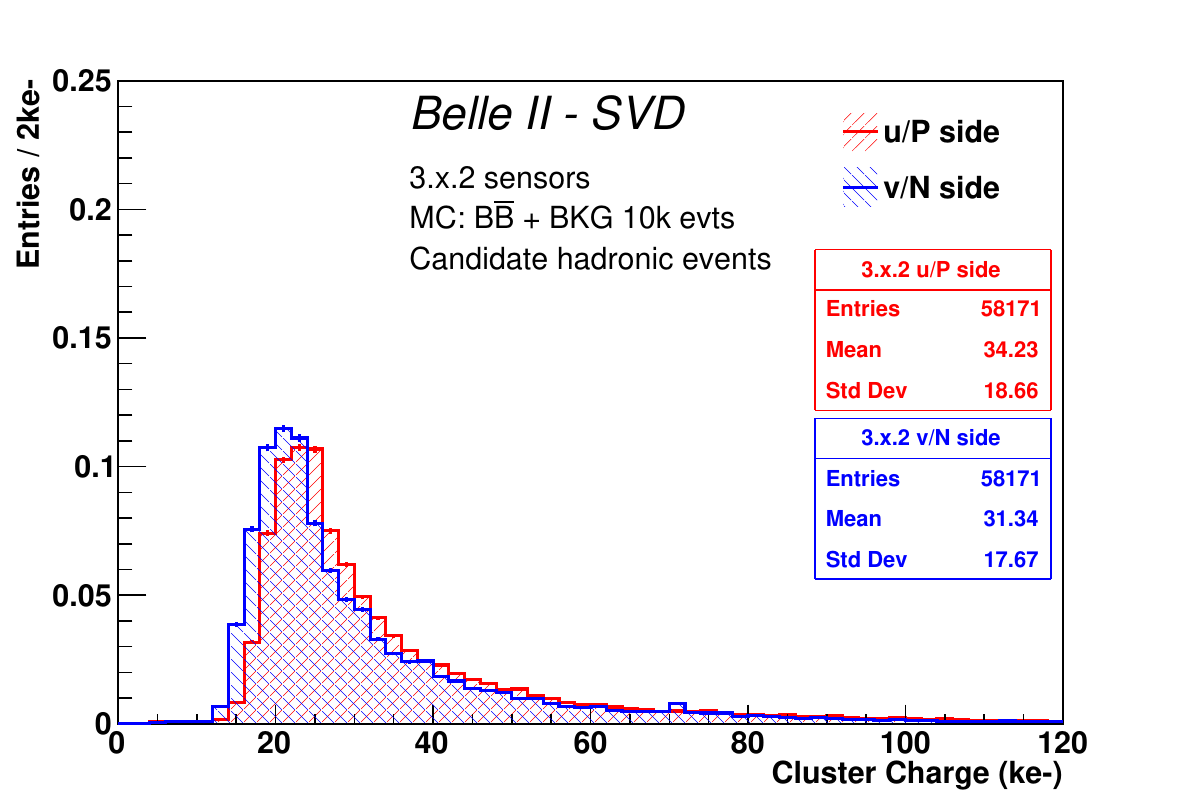}
     \includegraphics[width=.45\columnwidth,keepaspectratio]{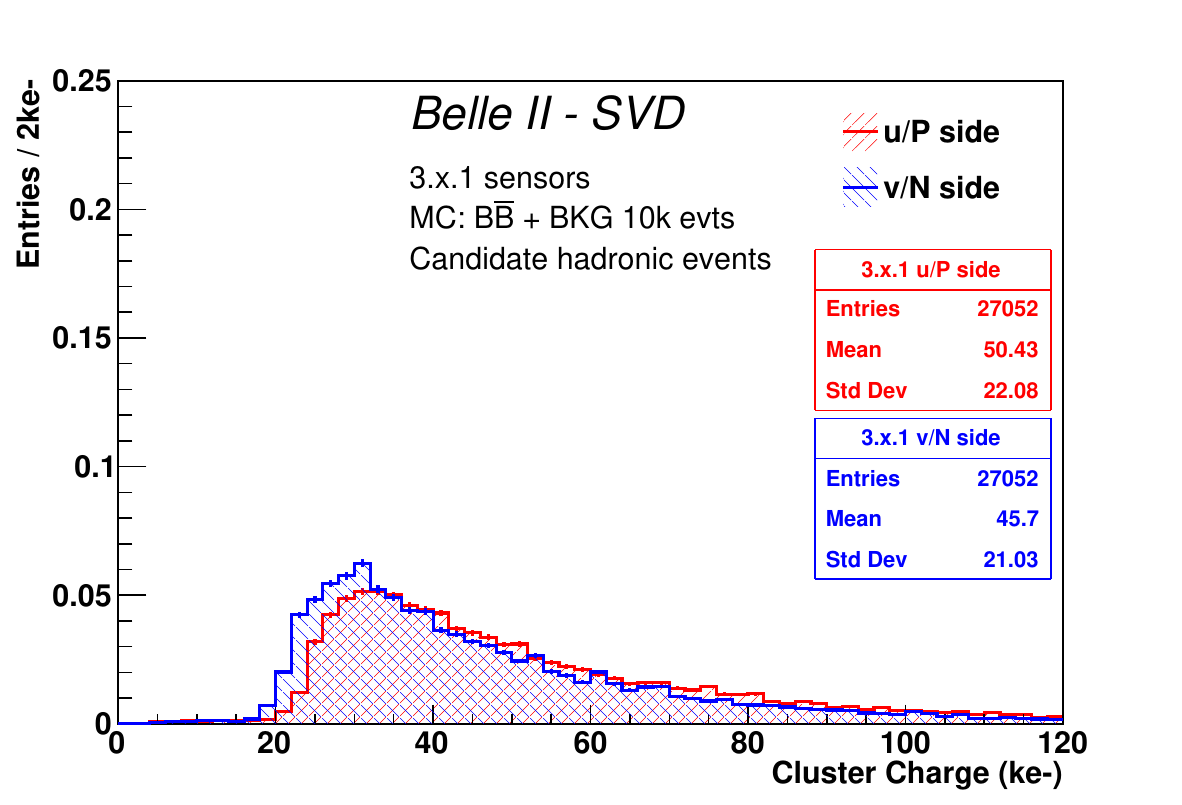}\\
      \includegraphics[width=.45\columnwidth,keepaspectratio]{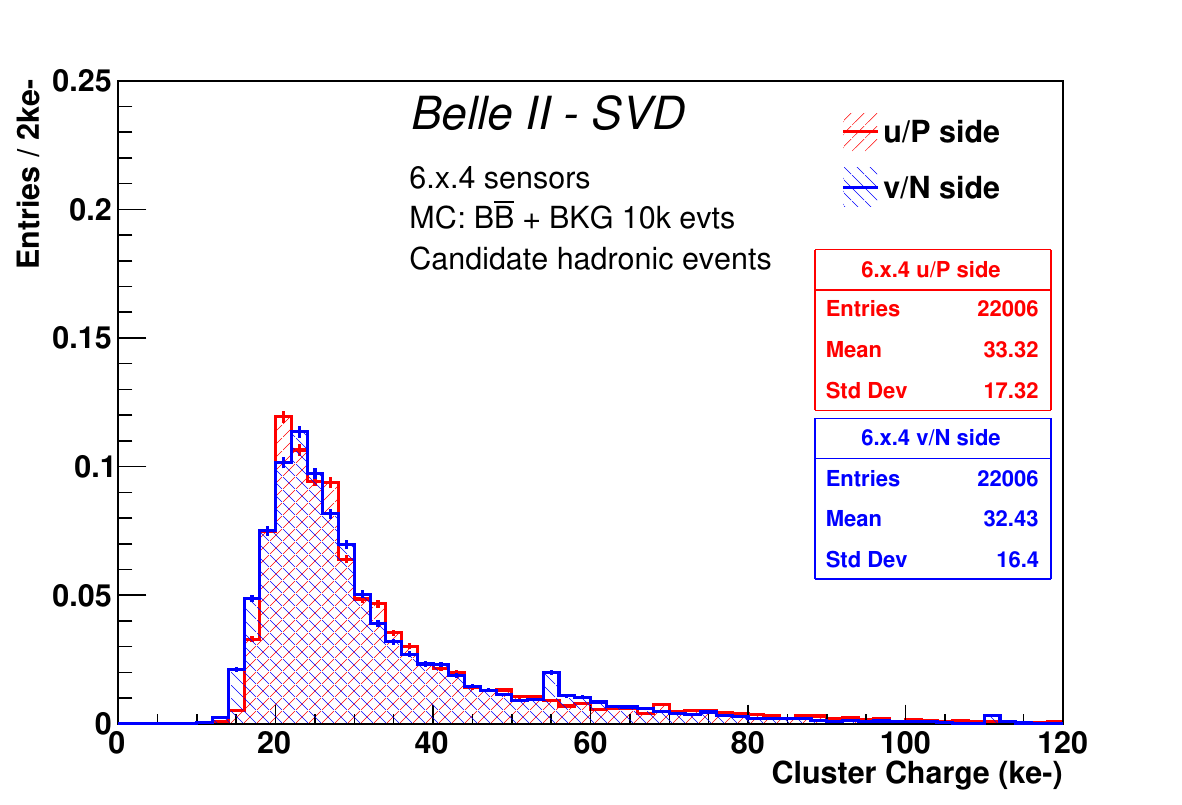}
     \includegraphics[width=.45\columnwidth,keepaspectratio]{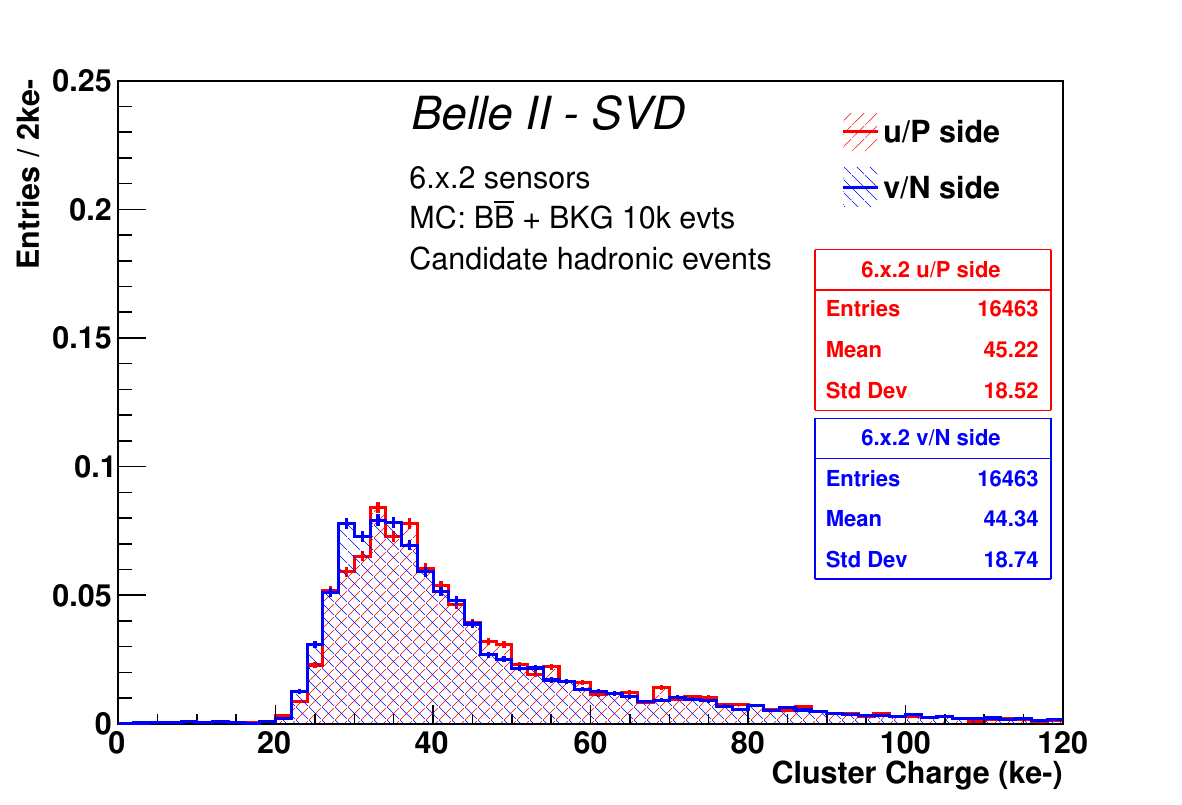}
         \caption{Distributions of reconstructed  charge for simulated events including beam background expected at design luminosity, for different layers and sensors: layer 3 backward (upper-left) and forward (upper-right) and two barrel layer 6 sensors (bottom-left and right)}.
    \label{fig:C4_clusterCharge}
     \end{figure}

\begin{figure}
     \centering
      \includegraphics[width=.45\columnwidth,keepaspectratio]{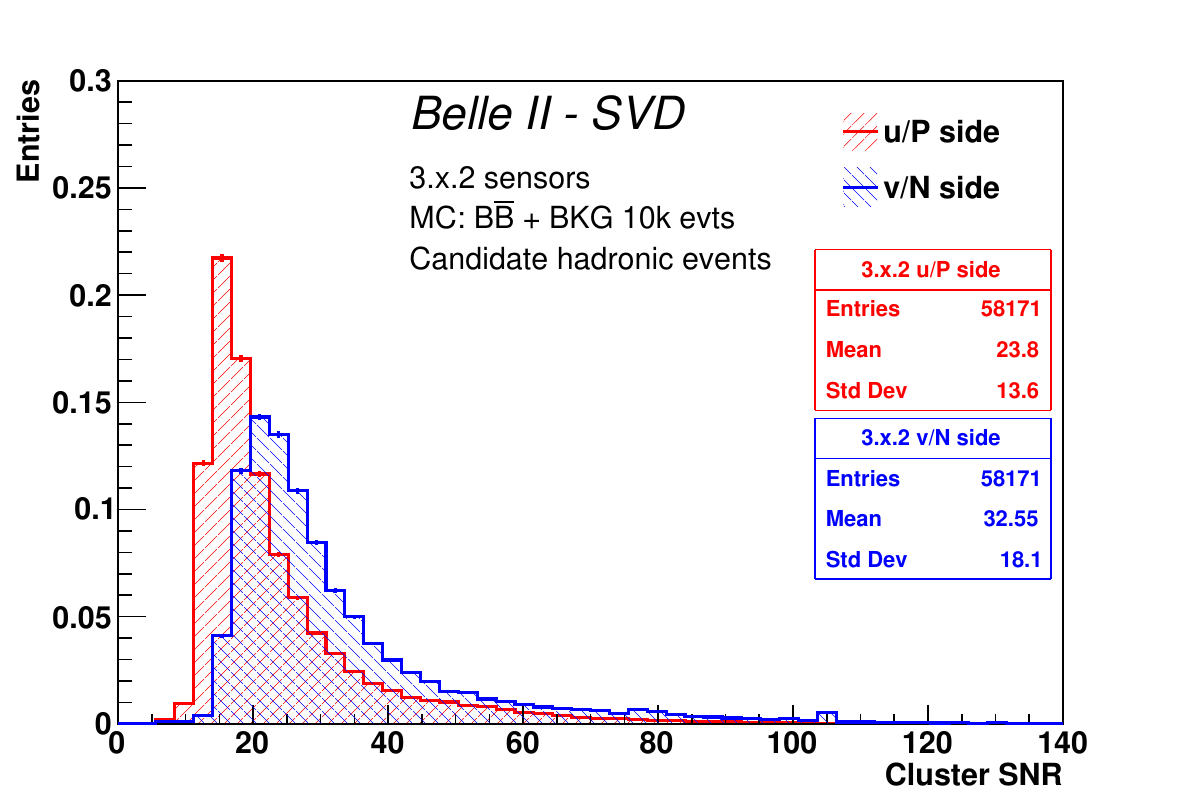}
     \includegraphics[width=.45\columnwidth,keepaspectratio]{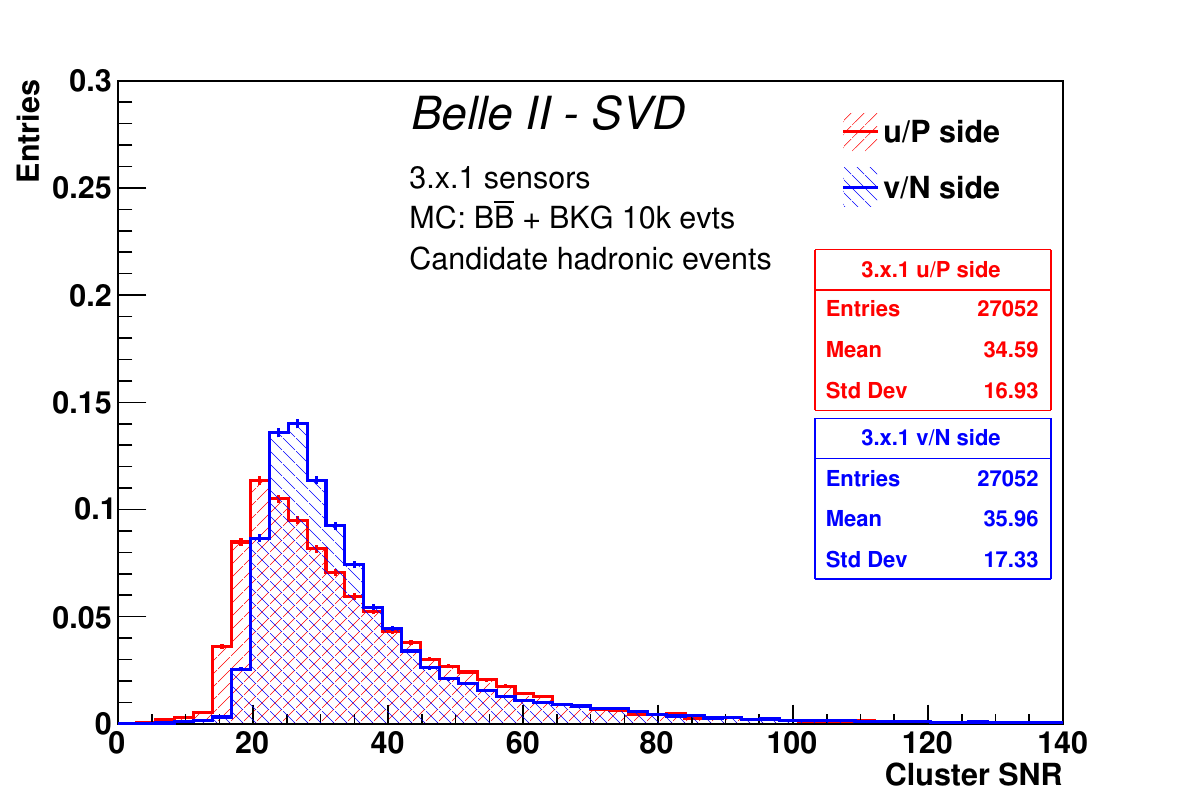}\\
      \includegraphics[width=.45\columnwidth,keepaspectratio]{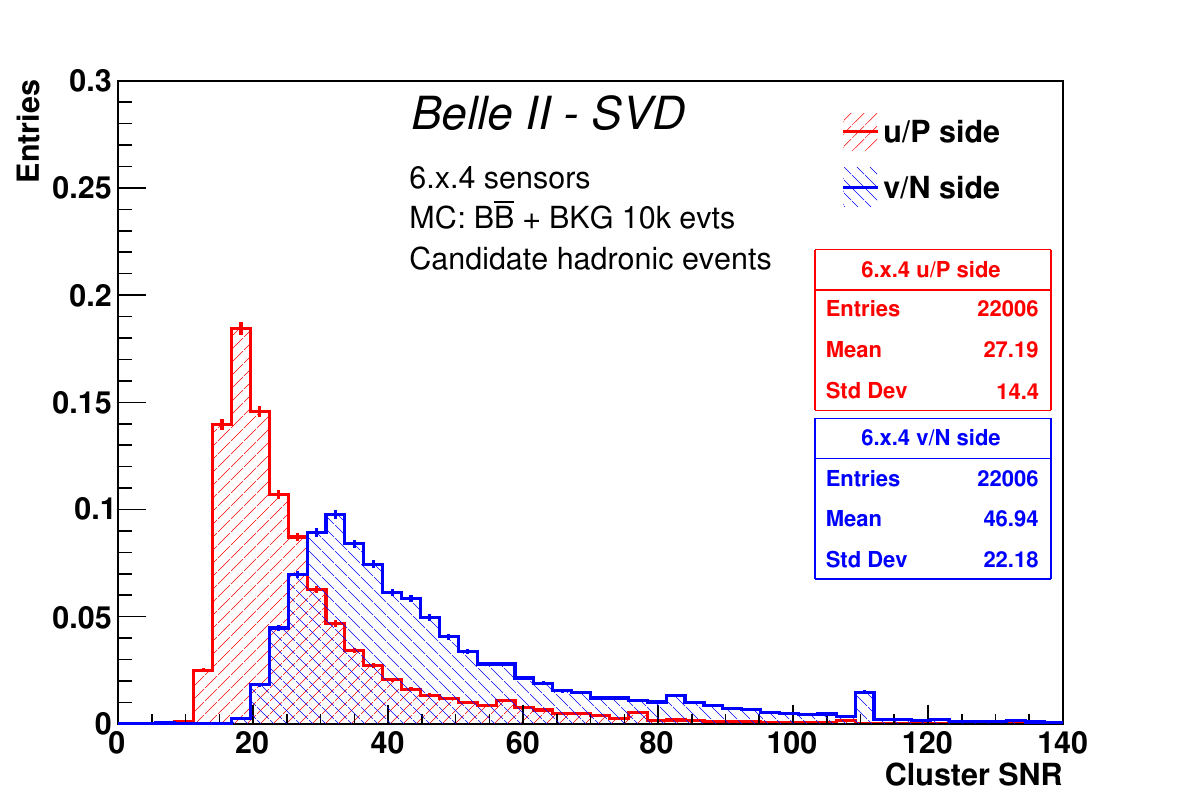}
     \includegraphics[width=.45\columnwidth,keepaspectratio]{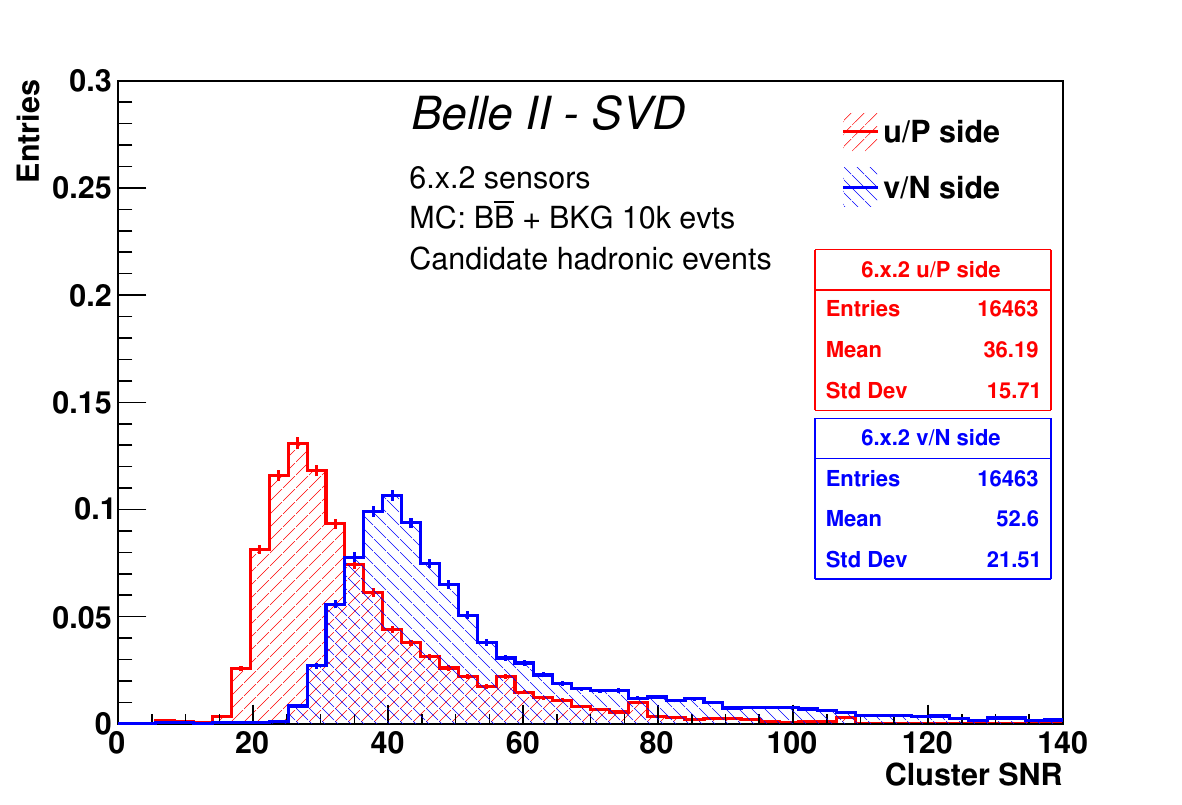}
    \caption{Distributions of reconstructed cluster \snrCL for simulated events including beam background expected at design luminosity, for different layers and sensors: layer 3 backward (upper-left) and forward (upper-right) and two barrel layer 6 sensors (bottom-left and right).}
    \label{fig:C4_clusterSNR}
\end{figure}

The cluster charge is computed as the sum of the charges of each strip belonging to the cluster:
\begin{equation}
S_{\rm CL} = \sum_{i=0}^{i< \rm size} S_i
\end{equation}
while the noise of the cluster is computed as the quadrature sum of the noise of the single strips:
\begin{equation}
N_{\rm CL} = \sqrt{\sum_{i=0}^{i< \rm size} N_i^2}
\end{equation}
and the signal-to-noise ratio of the cluster is defined as the ratio between its charge and noise, $\snrCL = S_{\rm CL}/N_{\rm CL}$.
Distributions of cluster size, charge, and \snrCL are shown in figures~\ref{fig:C4_clusterSize},~\ref{fig:C4_clusterCharge}, and~\ref{fig:C4_clusterSNR}, respectively.
The cluster size is correlated to the incident angle of the track on the sensor. Layer 3 forward sensors show a higher cluster size because of the larger incident angle and the smaller pitch on the v/N sides. The outer layers have a slanted sensor in the forward direction, therefore the cluster sizes on the two sides are similar and smaller with respect to layer 3. 
%The cluster charge distributions are more similar among layers and are peaking at $\sim 24$k electrons, as expected for MIP crossing $300\mum$ Silicon. 
%The SNR distributions are also similar, except for the v/N side L3 because of a higher average noise. 
Also, in the cluster charge there are visible variations of the energy released, due to the different track incident angle and track length in the sensitive volume, as detailed in section \ref{sec:C9_cluster_charge}. The signal peaks at around $24$k electrons, as expected for a MIP crossing \qty{320}{\um} silicon for sensors 3.x.2 and 6.x.4 (see section~\ref{sec:C2_design} for a definition of the nomenclature), with tracks mostly perpendicular. Larger MPV are visible in sensors in the forward region. The \snrCL\ distributions are more similar among layers, since the effect of the larger charge related to larger incident angle is partly removed considering the cluster noise, that includes the cluster size in its definition.  Higher \snrCL\ values are visible for the v/N side, where noise is significantly lower compared to the u/P side, except for layer 3 sensors.

     {}

\textbf{Cluster position}. The cluster position ($x_{\rm CL}$) and its uncertainty ($\Delta x_{\rm CL}$) are computed using the charge ($S_i$) and position ($x_i$) of its strips, with different algorithms depending on the cluster size.

%% cluster size  = 1
% m_position += stripPos * aStrip.charge;
% m_position /= m_charge;
% double phantomCharge = m_cutAdjacent * m_strips.at(0).noise;
% m_positionError = pitch * phantomCharge / (m_charge + phantomCharge);
For clusters formed by a single strip, the position corresponds to the center of the implant of the readout strip, while the position uncertainty takes into account the fact that the adjacent strips may have a signal below the threshold:
\begin{eqnarray}  
x_{\rm CL1} &=& x_1\\
\Delta x_{\rm CL1} &=& p  \frac{Q_{\rm ph}}{S_1 + Q_{\rm ph}}
\end{eqnarray}
 where $p$ is the strip pitch and $Q_{\rm ph}$ is the phantom charge, i.e. the maximum charge that the strip should have not to be included in a cluster $Q_{\rm ph} = \snr_{\rm adj} N_i$.\\
%% cluster size == 2
% m_position /= m_charge;
% m_positionError = a * pitch / sn;
For clusters formed by two strips the position is computed with the
center-of-gravity algorithm:
 \begin{eqnarray}
    x_{\rm CL2} &= &\frac{x_1S_1 + x_2S_2}{S_1+S_2} \\
\Delta x_{\rm CL2} &=& p\frac{\snrADJ}{ S_{\rm cl}/ N_1}
 \end{eqnarray}    
 where $S_{\rm cl}$ is the charge of the cluster and $N_1$ is the noise of the first strip.\\
 % cluster size > 2
   % double centreCharge = (m_charge - minStripCharge - maxStripCharge) / (clusterSize - 2);
  %  minStripCharge = (minStripCharge < centreCharge) ? minStripCharge : centreCharge;
  % maxStripCharge = (maxStripCharge < centreCharge) ? maxStripCharge : centreCharge;
  % double minPos = m_isUside ? info.getUCellPosition(minStripCellID) : info.getVCellPosition(minStripCellID);
  % double maxPos = m_isUside ? info.getUCellPosition(maxStripCellID) : info.getVCellPosition(maxStripCellID);
  %    m_position = 0.5 * (minPos + maxPos + (maxStripCharge - minStripCharge) / centreCharge * pitch);
Position of clusters with more than two strips are computed with the 
head-tail algorithm:
\begin{equation}
   x_{\rm CL3} = \frac{1}{2}\left[ x_{\rm head} + x_{\rm tail} +p \frac{ S_{\rm head}-S_{\rm tail}}{S_{\rm center}} \right]\\
 \end{equation} 
 where the subscripts head and tail indicate the position $x$ (or charge $S$) of the two strips at the edge of the cluster, and  $S_{\rm center}  = (S_{\rm CL} - S_{\rm tail} - S_{\rm head}) / ({\rm size}- 2)$, with $S_{\rm CL}$ the total charge of the cluster. The position uncertainty is computed as:
  \begin{equation}
  \Delta x_{\rm CL3} = \frac{p}{2}  \sqrt{ \left(\frac{1}{sn}\right)^2 + \frac{1}{2} \left(\frac{S_{\rm tail}}{S_{\rm center}}\right)^2 +  \frac{1}{2} \left(\frac{S_{\rm head}}{S_{\rm center}}\right)^2 }
  \end{equation}
  with $sn = \frac{S_{\rm center}}{\snrADJ N_{\rm CL}}$.
  %     double sn = centreCharge / m_cutAdjacent / noise;
  %     // Rough estimates of Landau noise
  %         double landauHead = minStripCharge / centreCharge;
  %         double landauTail = maxStripCharge / centreCharge;
  %         m_positionError = 0.5 * pitch * sqrt(1.0 / sn / sn +
  %                                              0.5 * landauHead * landauHead +
  %                                              0.5 * landauTail * landauTail);
 
Simulation studies have shown that using the calculated uncertainty the cluster position residual pulls do not have unit width as expected. %While we are performing dedicated studies to understand the reason, 
To take this discrepancy into account, the cluster position uncertainty is corrected by a scale factor that depends on the sensor type, side, and the number of strips in the cluster (1,2, $>2$), listed in table~\ref{tab:clusterErr}. These scale factors $s_{\rm err}$ are computed using simulation by requiring that 68\% of the pull is found within $1\sigma$ with respect to the median of the distribution.
The corrected uncertainty provided to tracking $\overline{\Delta x} _{\rm CL}$ is therefore:
$$
\overline{\Delta x}_{\rm CL} = s_{\rm err} \Delta x_{\rm CL}.
$$
\begin{table}[ht!]
    \centering
    \begin{tabular}{lcccc}
    \toprule
sensor    & side    &\multicolumn{3}{c}{$s_{\rm err}$ for size}\\
               &  &     1  & 2 &  $>2$\\
               \midrule
 HPK small & \uP & 1.352 & 1.137 & 0.559\\
 HPK small & \vN & 1.638 & 1.168 & 0.430\\	
 HPK large & \uP & 1.312 & 0.871 & 0.538\\
 HPK large & \vN & 2.338 & 1.418 & 0.468\\
 wedge     & \uP & 1.728 & 1.209 & 0.662\\
 wedge     & \vN & 1.766 & 1.481 & 0.433\\
 \bottomrule
  \end{tabular}
    \caption{Cluster position uncertainty scale factors used in the reconstruction, estimated on simulated events.}
    \label{tab:clusterErr}
\end{table}

\begin{figure}
    \centering  
     \includegraphics[width=.45\columnwidth,keepaspectratio]{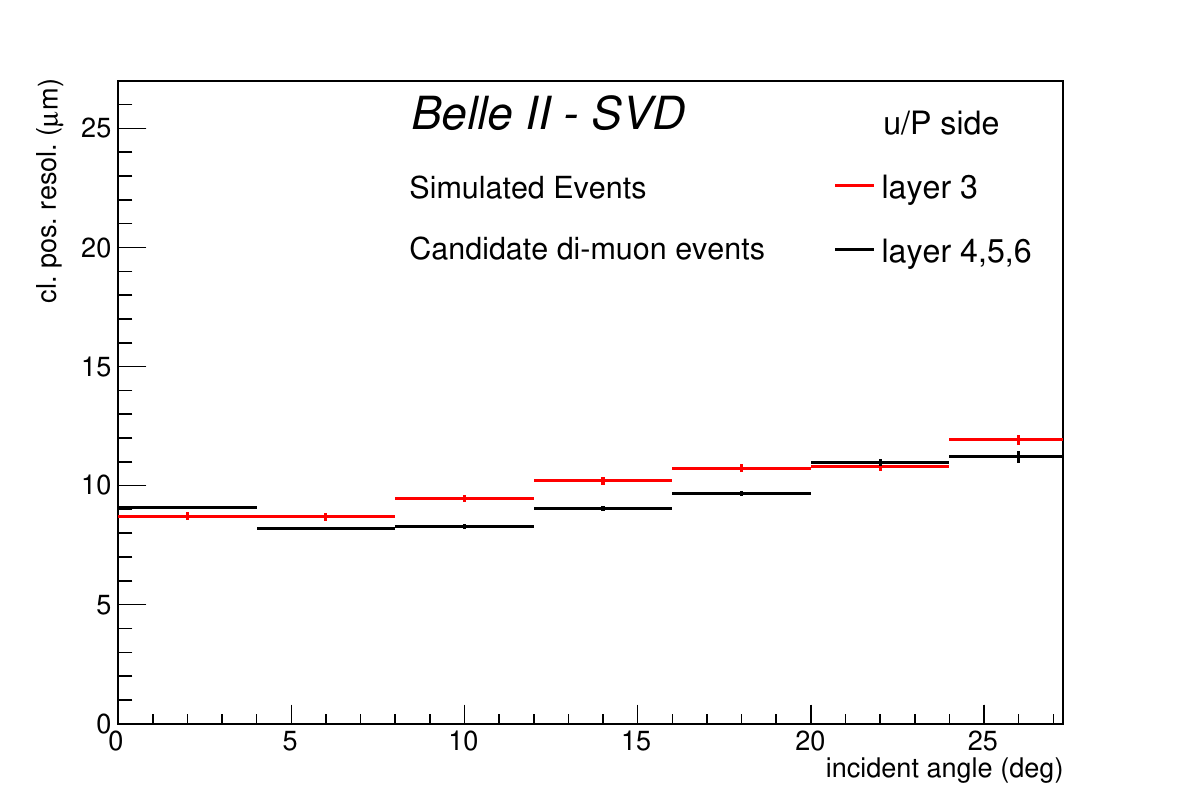}
    \includegraphics[width=.45\columnwidth,keepaspectratio]{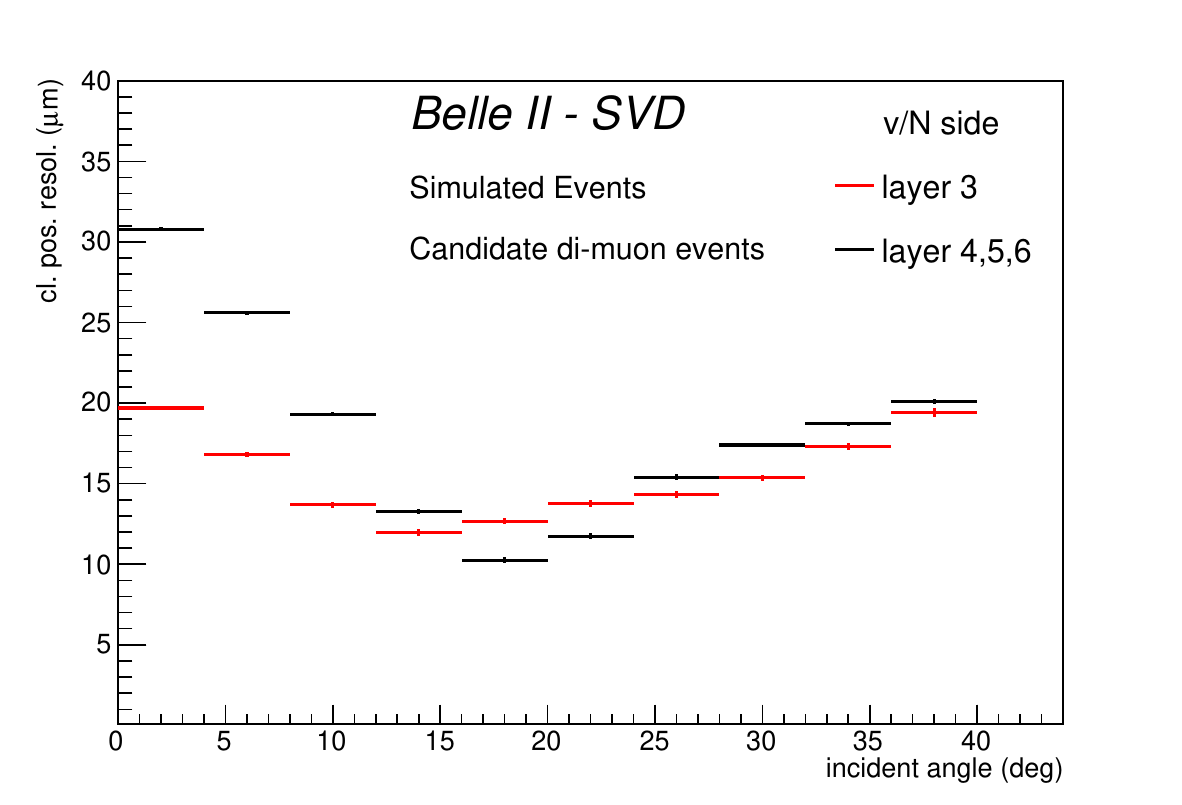} 
    \caption{
    Cluster position resolution vs. track incident angle estimated on simulated events, which include the beam background expected at design luminosity, with the same analysis used for data. Results of the u/P (v/N) side is shown for the layer 3 in red and for the outer layers in black.}
    \label{fig:C4_clusterResol}
\end{figure}

The cluster position resolution was estimated using simulated di-muon events, based on the same analysis as for data explained in section~\ref{sec:C7_resolution}, and it is reported as a function  of the track incident angle in figure~\ref{fig:C4_clusterResol}. 
The observed resolution has the expected shape as a function of the angle, showing a minimum
at the incident angle for which the projection of the track along the direction perpendicular to the strips on the detector plane corresponds to two strip pitches.
Given the various sensor pitches with one floating strip  (table~\ref{tab:sensor_geo}), the minimum is expected at 4 (7) degrees for the u/P side and at 14 (21) degrees on the v/N side,  for layer 3 (4, 5, 6). 
The expected digital resolution for perpendicular track is 
 \qty{7}{\um} (\qty{11}{\um}) on u/P side and \qty{23}{\um} (\qty{35}{\um}) on v/N side for layer 3 (4,5,6),  similar to the measured resolutions. 
The method tends to overestimate the resolution by up to \qty{2}{\um}, depending on the incident angle, as observed during the validation using the true information available for simulated events.

{}
Since masked strips are not considered when forming a cluster, they can cause a cluster to be split in two by a masked strip in the middle. Given that the number of masked strips is stably below 1\%,  it is not critical to take into them account in clustering. In case an entire APV chip is masked, a {\it fake} cluster positioned in the middle of the region read by the disabled APV is created in order to use the information provided by the clusters on the opposite side.

Studies of the cluster position resolution on data shown in figure~\ref{fig:C7_resolution} of section~\ref{sec:C7_performance} indicate that the resolution observed in simulated events is optimistic, and that there is still room to improve the resolution on data. Studies are ongoing to improve the resolution on data by testing alternative algorithms, as well as to improve the data-MC agreement, for example for the cluster size distribution as mentioned before. 

\begin{figure}[hbt]
    \centering  
    \includegraphics[width=0.45\textwidth]{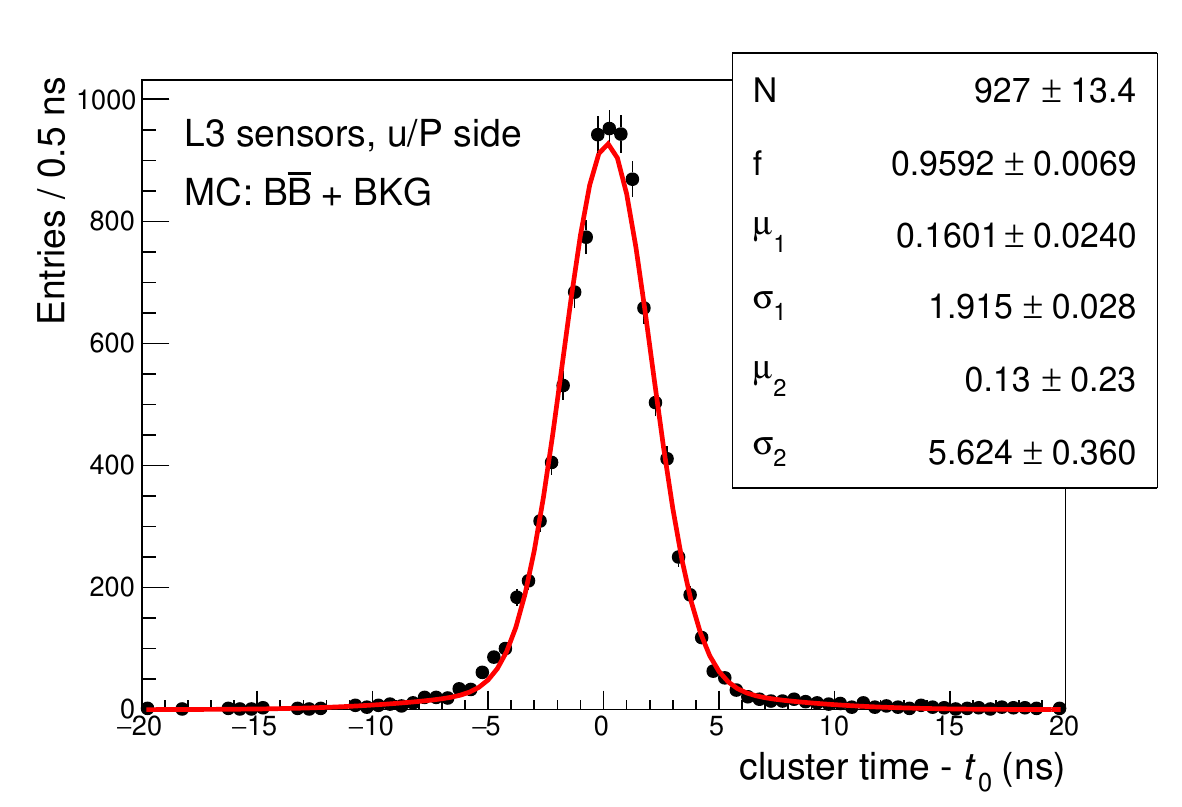}
    \includegraphics[width=0.45\textwidth]{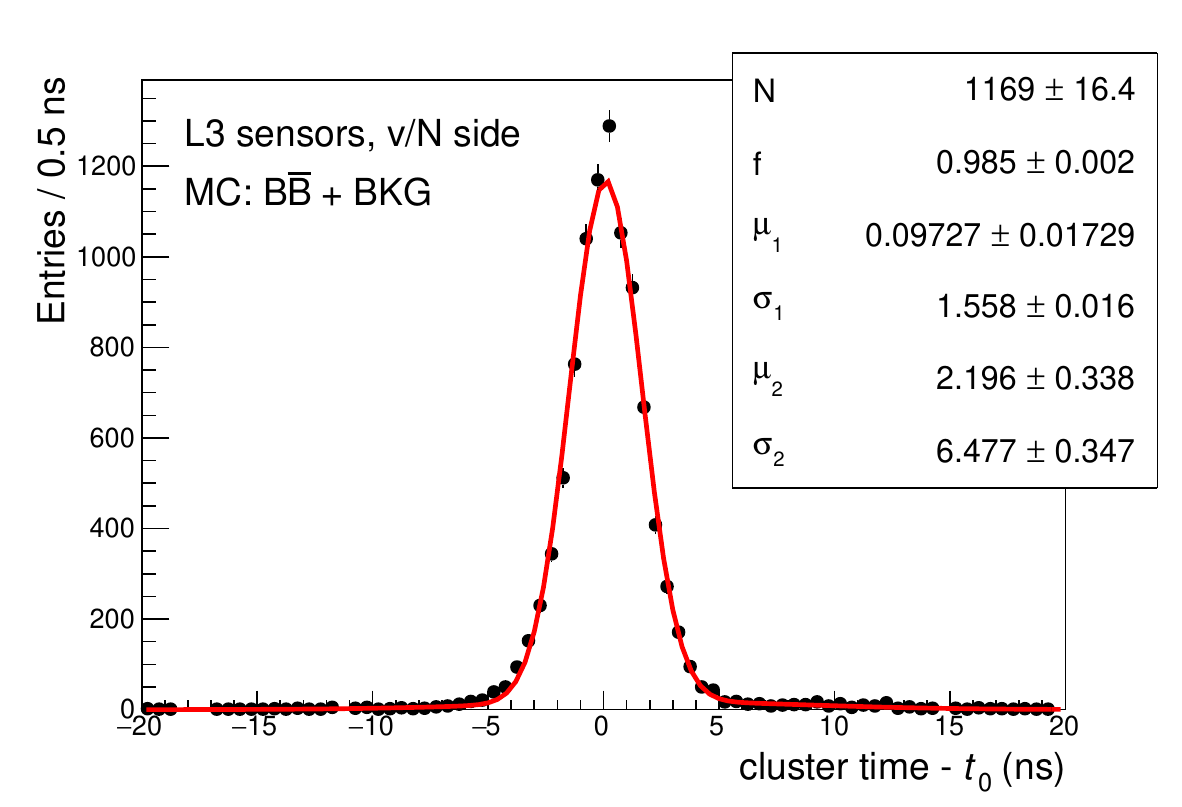}
    \caption{Cluster time - \eventtz\ for clusters related to tracks for simulated events: layer 3 for the u/P side on the left and v/N side on the right. The \eventtz\ is the time of the event reconstructed by the CDC. The resolution on the cluster time, estimated with equation~\ref{eq:C4_timeres} is 2.2 ns on the u/P side and 1.8 ns on the v/N side.}
    \label{fig:C4_resolutionTimeEventT0}
\end{figure}

\textbf{Cluster time}. The cluster time is computed as an average of the calibrated time of the strips forming the cluster, weighted with the charge of the strip.
The cluster time resolution is estimated from the distribution of the difference between the cluster time and the \eventtz\, where \eventtz\ is the time of the event reconstructed by the CDC, fitting this distribution with a  sum of two independent Gaussians of means $\mu_i$ and widths $\sigma_i$. The resolution is estimated as the weighted average of the widths of the Gaussians corrected by the nonzero values of the means of these Gaussians: 
\begin{equation}
    \sqrt{f\sigma_1^2+(1-f)\sigma_2^2 +f(1-f)(\mu_1-\mu_2)^2}
    \label{eq:C4_timeres}
\end{equation}
where $f$ is the fraction of the first Gaussian, $\mu_i$ and $\sigma_i$ are the mean and width of the $i$-th Gaussian.
The cluster time resolution on simulated events is of the order of 2 ns, as shown in figure~\ref{fig:C4_resolutionTimeEventT0}.

\begin{figure}
    \centering
     \includegraphics[width=.45\columnwidth,keepaspectratio]{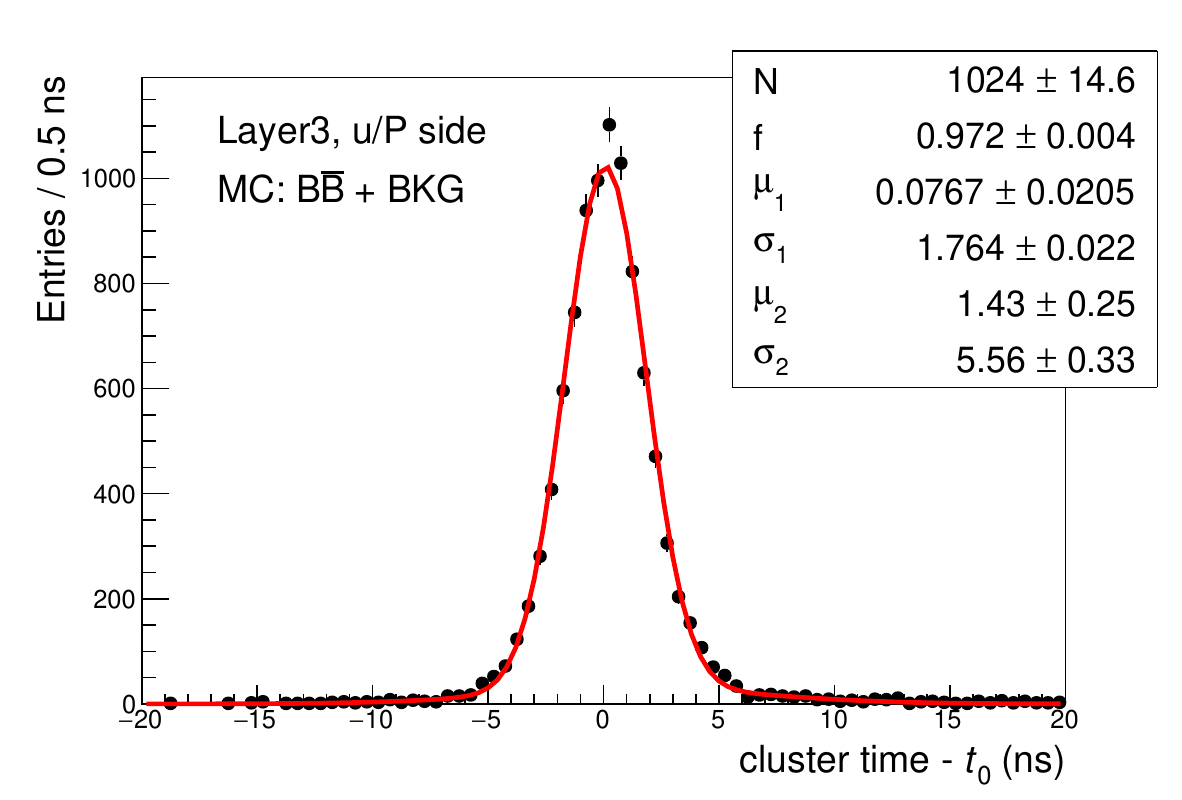}
     \includegraphics[width=.45\columnwidth,keepaspectratio]{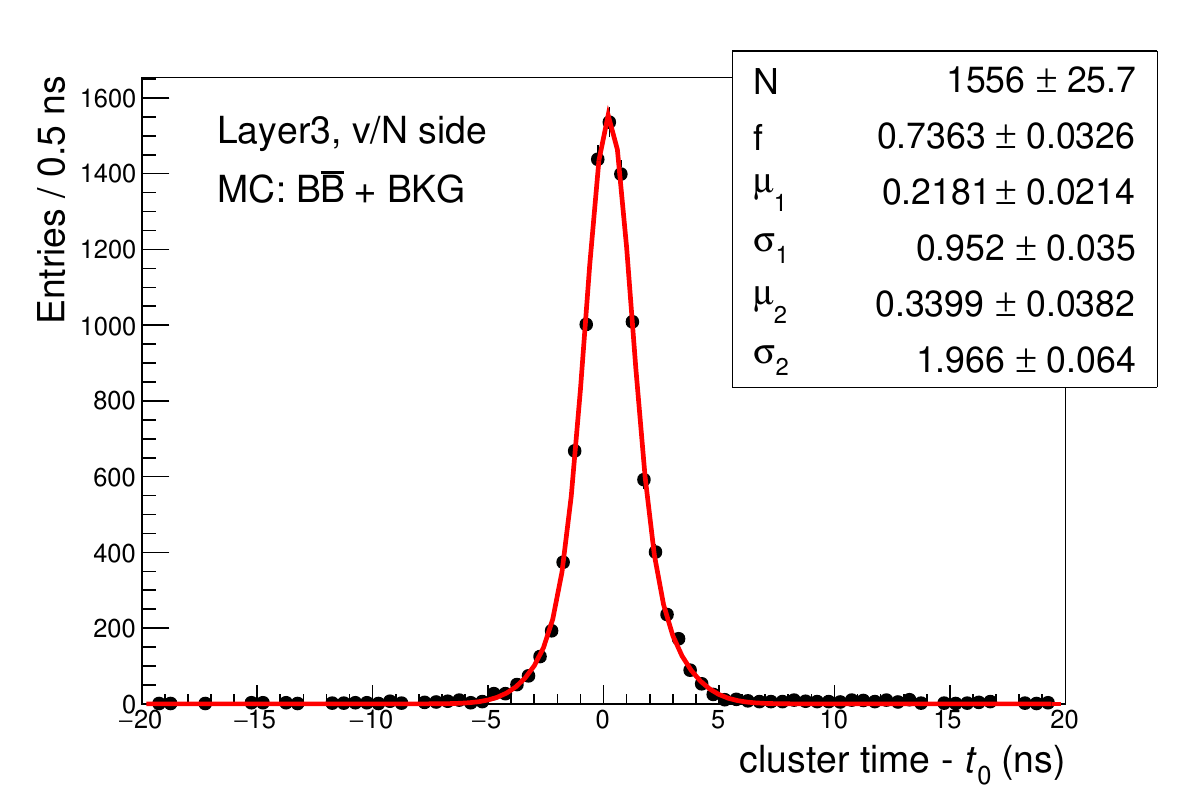}
    \caption{Cluster time - \eventtz\, for clusters related to tracks in simulated events, with the alternative algorithm using three APV samples: layer 3 for the u/P side on the left and v/N side on the right. \eventtz\ is the time of the event reconstructed by the CDC. The resolution on the cluster time, estimated with equation~\ref{eq:C4_timeres} is 2.0 ns on the u/P side and 1.3 ns on the v/N side.}
    \label{fig:C4_clusterTimeCoG3Resolution}
\end{figure}

Although currently the hit time is reconstructed using all the six APV samples, alternativce algorithms have been developed to compute the cluster time using only a subset of the 6 samples. The long right tail of the APV waveform, shown in figure~\ref{fig:C4_shaper_pulse}, does not bring very useful information on the hit time with respect to the three samples around the peak. Therefore,  first the amplitudes of each of the strips forming the cluster are summed sample-by-sample, and then the three samples corresponding to the waveform peak of each strip forming the cluster are selected, storing the time of the first selected sample ($FF$). The raw cluster time is then computed as a weighted average of the sample time with the sample amplitude, similarly to what is done to compute the strip time. The calibration is performed in the same way as the default one, exploiting the correlation with the  \eventtz\ from the CDC as discussed in section~\ref{sec:C4_CoGCalibration}, equation~\ref{eq:C4_calibratedTime}.
However, in this case an additional shift must be considered, $FF$, the time of the first sample:

\begin{equation}
    t_{\rm cl3sample} = a + b \cdot t_{\rm raw} + c \cdot t_{\rm raw}^2 + d \cdot t_{\rm raw}^3 + \delta t + FF. 
\end{equation}
 The calibration parameters are  different than the ones determined from the calibration of the default algorithm, and are therefore stored in a dedicated payload.
The performance of this alternative time estimation are better by 10-30\% with respect to the default algorithm, as shown in figure~\ref{fig:C4_clusterTimeCoG3Resolution}.
{}

\begin{figure}
    \centering
     \includegraphics[width=.45\columnwidth,keepaspectratio]{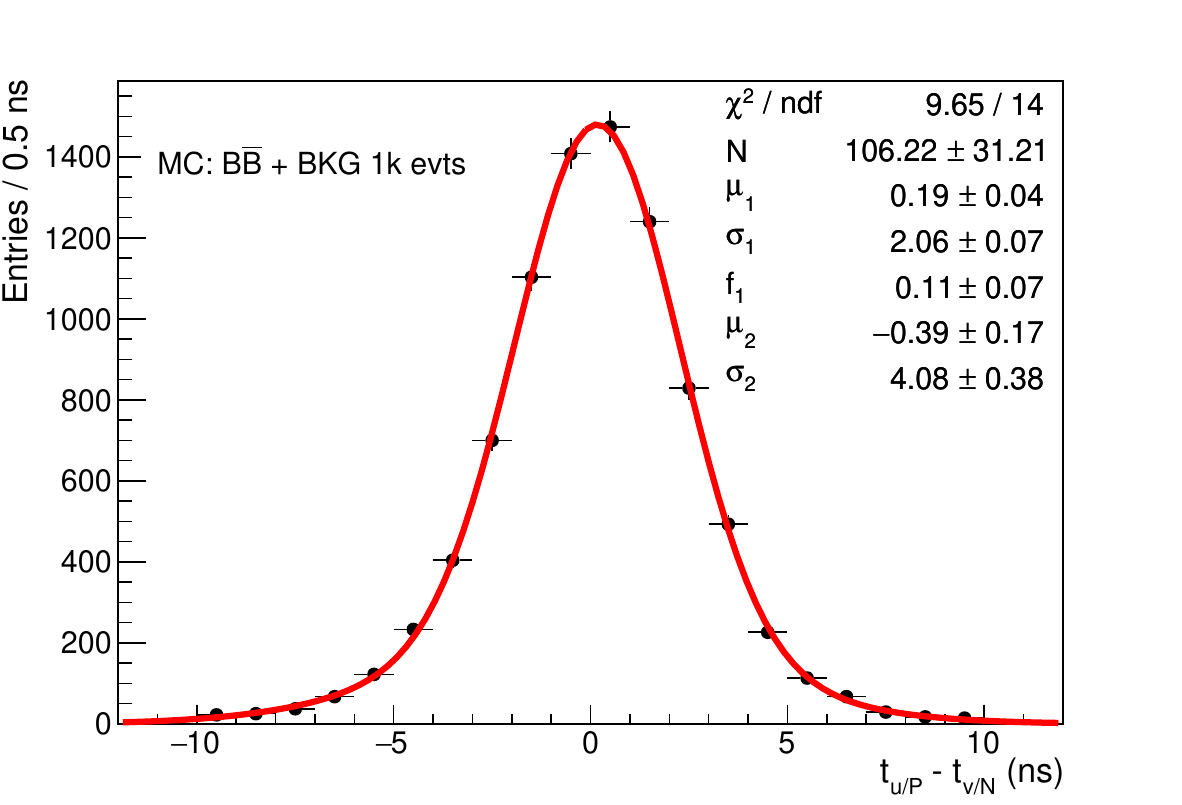}
     \includegraphics[width=.45\columnwidth,keepaspectratio]{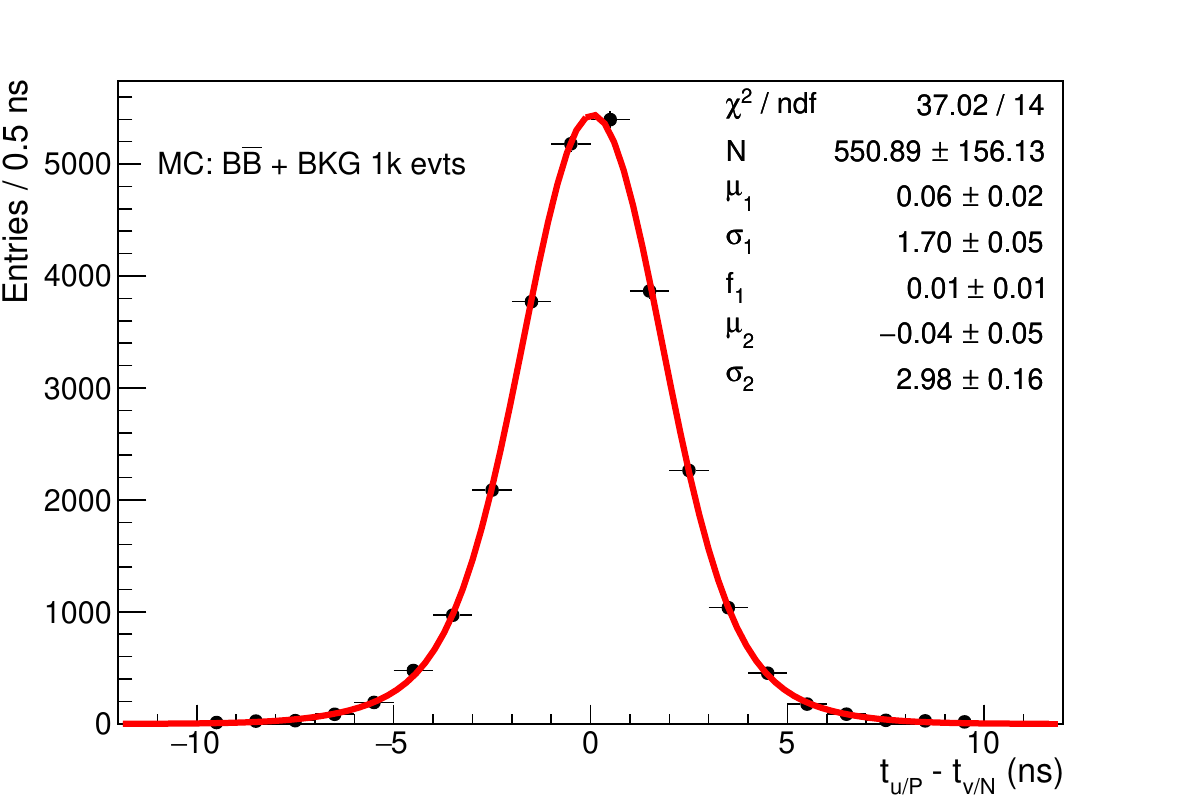}
    \caption{Distribution of the difference between the u/P and the v/N cluster time belonging to the same SpacePoint, of the layer 3 (left) and the barrel sensors of the outer layers (right) for simulated events. The distributions were fit with a sum of two Gaussians (the first one with fraction $f_1$) with mean $\mu_i$ and width $\sigma_i$ (both in ns). The widths of the distributions, estimated with equation~\ref{eq:C4_timeres} is 2.0 ns for layer 3 and 1.3 ns for the outer layers.}
    \label{fig:C4_clusterTimeDifference}
\end{figure}

\textbf{SpacePoint creation}. The position of the 3D hits, stored in the \nc{SVDSpacePoint}, is provided by combining all clusters on one side of the sensor with those on the opposite side. At this stage SpacePoints formed by clusters whose time ($t_{\rm CL}$) or time-difference ($|t_{\rm CL}(u)$-$t_{\rm CL}(v)|$) does not satisfy a certain selection can be rejected. The software allows to choose one of the following selection criteria:
\begin{itemize}
    \item $t_{\rm CL} < t_{\rm min}$: early clusters are not taken into account for the combination, where $t_{\rm min}$ is the minimum time of the allowed cluster;
    \item $|t_{\rm CL}| > \Delta t$: in case of small jitter and correct latency, the time of the cluster is expected to be around $t_{\rm CL} = 0$ ns, clusters with time that differ from this expectation more than $\Delta t$ are rejected;
    \item $|t_{\rm CL}(u) -t_{\rm CL}(v)|> \Delta t'$: clusters of opposite sides created by the same particle are expected to have similar times, if their time differ more than  $\Delta t'$ the combination is rejected.
\end{itemize}{}
The choice of the selection does not require to recompile the software,  it is stored in the Conditions Database and can be changed by changing the payload read by the module.
At the moment, none of these selection criteria are applied for data reconstruction, and all possible combinations of opposite-side clusters are provided to the tracking. This is possible because the current beam background is sufficiently low for the tracking to handle all the SpacePoints. In the future, when luminosity  increases, it will become critical to reject SpacePoints based on the cluster time in order to have an efficient tracking with a low fake rate.
As an example, for simulated events with superimposed beam backgrounds expected at design luminosity, a selection on the time difference of the clusters of a SpacePoint is applied: if the time difference is larger than 10 ns, the combination is rejected. 
In figure~\ref{fig:C4_clusterTimeDifference} the cluster time difference distribution for signal cluster from simulated events is shown.
{}
This selection reduces the number of SpacePoints by 60\%, yielding a 30\% improvement in the execution time of pattern recognition algorithms that involve SVD, and a reduction of the track fake rate for the SVD-only pattern recognition by 25\%, down to 7.5\%, with an unchanged hit finding efficiency. % probability of adding an SVD hit to the track.

\lvlthreesection{Data quality monitoring}
\label{sec:C4_dqm}
\editor{Giulia %, details of the content of the section were moved in the Jira ticket B2SVD-207
 %describe the most important plots (DataFormat, Occupancy, Efficiency, MaxBin), the color code\\
%list of figures:
% 1 figure with example DataFormat( red) + maxBin (in agreement with ref)\\
%1 figure with example Occupancy, Efficiency for one side (green)
}
During data taking the quality of the data is constantly monitored by the control-room shifters, and available for detector experts. A dedicated computing server, ExpressReco, receives a fraction of the events from the EventBuilder and performs the complete official reconstruction online. Downstream reconstruction dedicated \acrshort{DQM} modules that fill histograms and run simple analysis code to determine if the histogram has the expected shape are run. The color of the histogram canvas turns to red if the analysis code detects an unexpected distribution, orange if the distribution may be problematic, grey if there is not enough statistics, and green in the other cases. Reference histograms are superimposed to the live histograms to help in judging data quality. A list of selected of histograms is available to the control-room shifters, all other histograms are available to detector experts to debug problems and assess the quality of the run. A few DQM modules run downstream reconstruction also on the \acrfull{HLT} machines, in this case the check is performed on all events.

\begin{figure}
    \centering
     \includegraphics[width=.9\columnwidth,keepaspectratio]{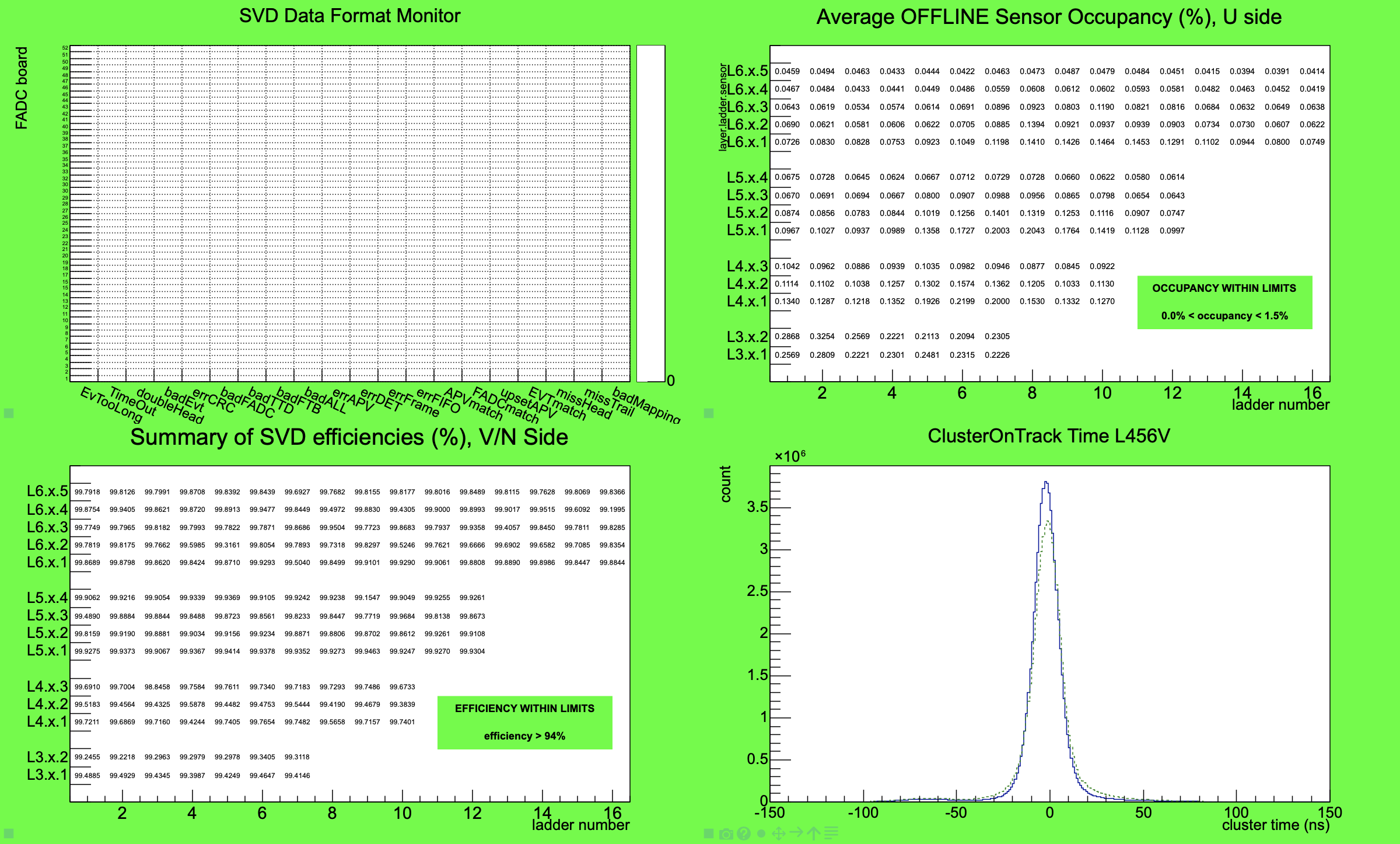}
      \caption{DQM plots continuously monitored by the control-room shifter: data format (top left), average u/P side occupancy computed with strips with $\snrST>5$ (top right), average v/N side sensor efficiency (bottom left), signal-cluster time of layer 4,5,6 (bottom right). Examples taken from a run in 2020 in which no problem was observed, indeed all plots show a green canvas.}
    \label{fig:C4_DQM}
\end{figure}

The monitoring of SVD data starts from the data received from DAQ: the data format of all events is checked by a DQM module running on HLT. The corresponding plot shown on figure~\ref{fig:C4_DQM} provides information on the fraction of events with data format issues, the involved FADC board, and the error type. On ExpressReco the occupancy due to beam backgrounds is checked by computing the average fraction of strips with $\snrST > 5$ per sensor on the u/P side. The average sensor efficiency for the v/N side is monitored, as well as  the v/N side cluster time distribution of layers 4, 5 and 6 cluster used to reconstruct a track in order to confirm that the acquisition window is at the correct time position.
In summary, the four plots available to the control-room shifter shown in figure~\ref{fig:C4_DQM} are:
\begin{itemize}
    \item data format;
    \item average u/P side occupancy computed with strips with $\snrST>5$;
    \item average v/N side sensor efficiency;
    \item v/N side cluster time of clusters used to reconstruct a track on layers 4, 5, 6.
\end{itemize}{}
Many more detailed plots produced on the ExpressReco machines are made available in real time to the SVD experts. 
{}

\lvlthreesection{SVD hit time calibration}
\label{sec:C4_CoGCalibration}
The hit time calibration procedure  uses clusters in order to minimize the effect of the electronic noise.
 The cluster time is calculated as an average of the strip time weighted with the strip charge, summed over the strips in the cluster.  As discussed in section~\ref{sec:C4_timeref}, \traw\ is the raw cluster time measured in the SVD time reference frame, while \tZ\ is the time of the event estimated by the \acrshort{CDC} measured in the trigger reference frame.
The calibration, exploiting the correlation between the cluster time and the time of the event,  was optimized on data. In the following, the method is explained showing results on simulated events; results on data are reported in section~\ref{sec:C7_hit_time_determination}.

First, the time of the event expressed in the SVD reference frame as \tZsvd\ is computed shifting the event \tZ\ by a known event-dependent quantity $\delta t$, based on the \gls{triggerbin}, and that gives an additional information about the arrival of the trigger to the SVD system:
 \begin{equation}
    \tZsvd = \tZ - \delta t.
\end{equation}

\begin{figure}
    \centering
    \includegraphics[width=.8\columnwidth,keepaspectratio]{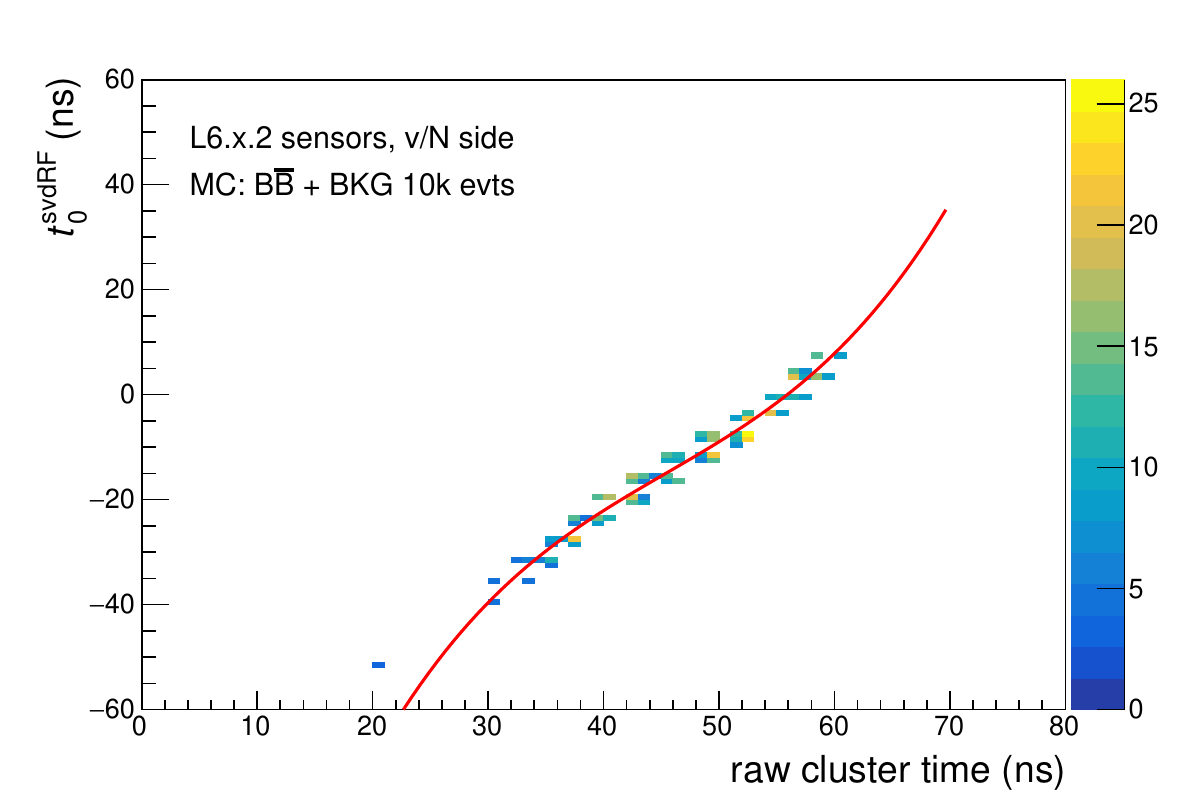} 
    \caption{
    An example of the $(\tZsvd,\traw)$ correlation plot  between the time of the event in the SVD reference frame and the raw cluster time used for the SVD hit time calibration and superimposed calibration function (red line) for v/N size of a barrel sensor of layer 6.}
    \label{fig:C4_TH2F_CoG_calibration}
\end{figure}

\begin{figure}
    \centering
     \includegraphics[width=.45\columnwidth,keepaspectratio]{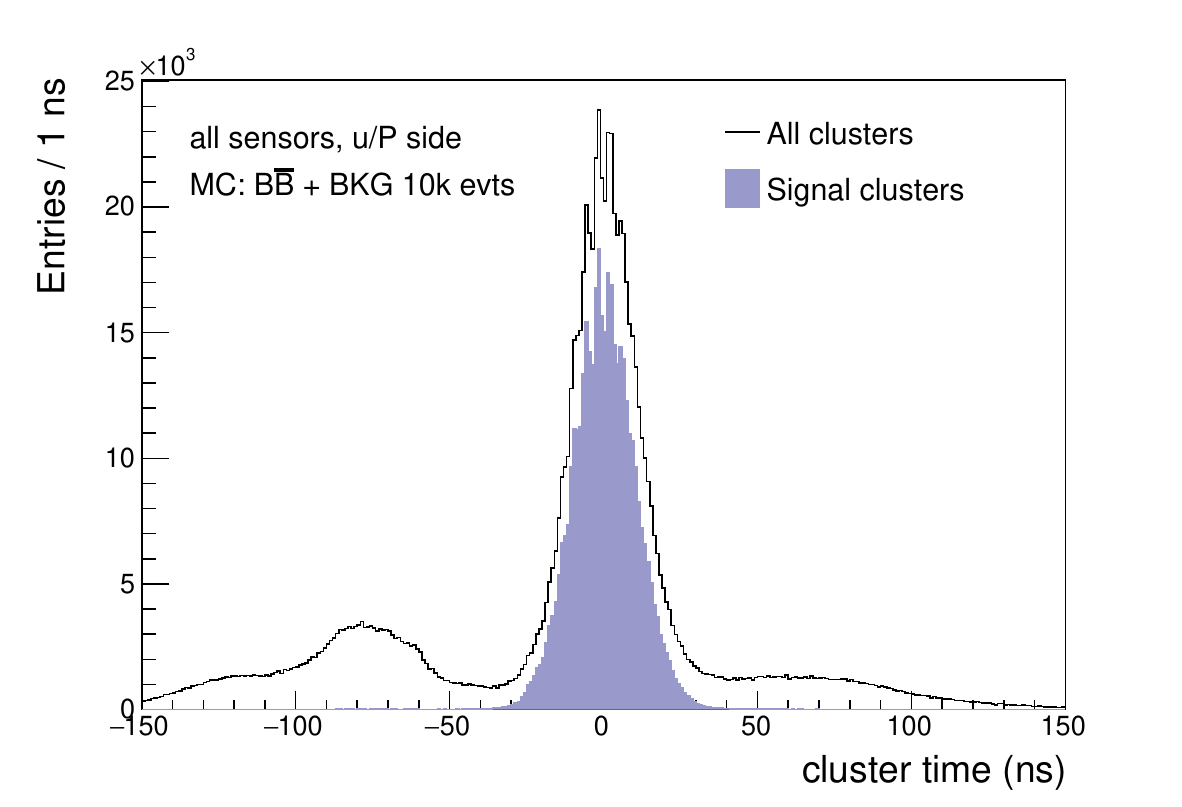}
     \includegraphics[width=.45\columnwidth,keepaspectratio]{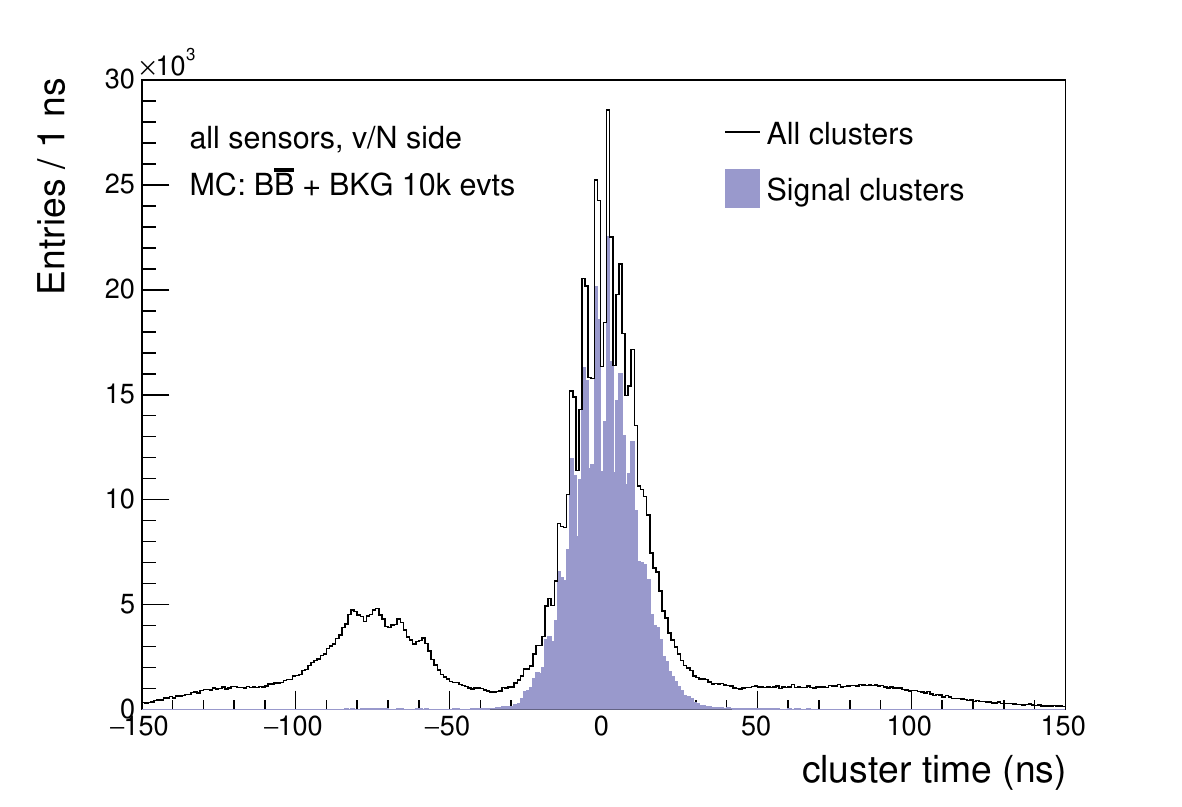}
    \caption{Cluster Time distributions of the u/P (left) and v/N side (right) obtained on simulated events. Clusters associated to tracks are represented by the shaded region.}
    \label{fig:C4_clusterTime}
\end{figure}

The goal of the calibration is to produce a calibrated hit time,    corresponding to the \tZsvd, starting from the $(\tZsvd,\traw)$ correlation plot.
%an example plot is shown in figure~\ref{fig:C4_TH2F_CoG_calibration}. 
In this calibration procedure the effect of flight time of the particles is neglected, and the calibration function $f(\traw)$ is determined  empirically using a third-order polynomial. The calibration function parameters are extracted by a fit on data of the x-profile of the two-dimensional  $(\tZsvd,\traw)$ correlation histogram:
\begin{equation}
    f(\traw) = a + b \cdot \traw + c \cdot \traw^2 + d \cdot \traw^3
    \label{eq:timecalib}
\end{equation} 
In figure~\ref{fig:C4_TH2F_CoG_calibration} an example of the $(\tZsvd,\traw)$ calibration 2D plot, with the fitted function superimposed is shown.
{}
The calibration parameters ($a,b,c,d$) are separately extracted for each sensor side  and are stored in the Conditions Database. During reconstruction the strip time is calibrated using equation~\ref{eq:timecalib} and finally shifted in the trigger reference frame, which is common to all detectors. The calibrated strip time is then:
\begin{equation}
    t_{\rm strip} = a + b \cdot \traw + c \cdot \traw^2 + d \cdot \traw^3 + \delta t. 
    \label{eq:C4_calibratedTime}
\end{equation}

The calibration for the v/N and u/P sides of each sensor is studied separately, since the waveforms provided by the two sides are different due to a number of effects:
\begin{itemize}
    \item different capacitive load that can change the waveform shape given by the APV25;
    \item the APV25 chips operate in a different configuration for P and N sides (inverter on/off), to be able to read signal of opposite charge collected on the two sides of the sensors.
\end{itemize}
Both these effects contribute to the difference observed in the
shape of the APV25 signals obtained during calibration runs, injecting a known pulse in the internal APV25 calibration circuit. In particular, signals on the u/P side have a slightly longer peaking time, by about 10 ns, and larger width of about 20 ns.\\
In addition to these effects, already visible  in calibration, the rise time of the real signal coming from the silicon sensor is also affected by the different mobility of holes and
electrons. Holes, collected on the u/P side, are $\sim$3 times slower than electrons, collected on the v/N side. So the final signal of the APV25 for the u/P side is slower than that on the v/N side.
Figure~\ref{fig:C4_shapers} shows the differences between the waveforms for the two SVD sensor sides.\\
As a consequence of these effects a better resolution on the time estimated on the v/N side (1.8 ns) with respect to the u/P side (2.2 ns) is achieved, as shown previously in figure~\ref{fig:C4_resolutionTimeEventT0}.
In figure~\ref{fig:C4_clusterTime} the calibrated cluster time distribution from simulated events with beam background expected at design luminosity is shown, highlighting the distribution of the signal clusters, i.e. those used by the tracking. The corresponding distributions for data are shown in figure~\ref{fig:C7_TimeMinusEventT0} and have a similar shape. The calibrated time of clusters associated to tracks (no SVD hit time information is used in reconstruction yet) nicely peaks at zero, around the  \eventtz, while background clusters are more spread in the window. The peak on the left of the window, between $-100$ and $-50$ ns, is due to background off-time tracks hitting the sensors before the beginning of the acquisition window. The shaped signal of the strips has a long tail of few hundreds of ns and the signal is still above threshold when this tail enters the acquisition window, causing an accumulation of earlier background hits at the beginning of the window. The rest of the background hits, arriving inside the SVD acquisition window, are more uniformly distributed in time, as expected. The time separation between signal and background hits will be exploited in the future for background mitigation.

{}

 % 10  --> 25 ?

\lvlonesection{Performance}
\label{sec:C7_performance}
\editor{Giuliana}
During the first 1.5 years of data taking (2019-2020) with the complete \acrshort{SVD} installed, Belle II has collected approximately \qty{70}{fb^{-1}} integrated luminosity at the $\Upsilon(4S)$ resonance. These data are used to characterize the performance of the SVD, and confirm the excellent behaviour of all the 172 SVD sensors, with stable performance in time.
The performance of the SVD is studied in all the aspects outlined in the description of software and reconstruction in chapter~\ref{sec:software_offline}. Unless otherwise indicated, candidate hadronic events are used, defined by the requirement of at least four reconstructed tracks from the \acrshort{IP}. 

As a starting point, in section~\ref{sec:C9_cluster_charge}  \gls{cluster}s of adjacent strips associated to tracks are considered. Their total collected charge, normalized to the track length across the \acrshort{DSSD} sensor, is compared with the expectations for the two sensor sides. 
%\rem{data MC comparison for cluster properties is not shown here, only SNR data-MC and cluster size are shown in SW ch. Should we add here the comparison for cluster charge? Since we also include the data-MC comparison for resolution maybe we can consider to put here the data-MC comparison too for cluster properties? Or simply write is compared with expectations since comment on MPV values are given.}

The strip noise (section~\ref{sec:C9_strip_noise}) is dominated by the capacitive input load to the front-end APV25 ASICs; it gives a very good signal to noise ratio, ranging between 13 and 30 (section~\ref{sec:C9_SNR}), confirming the success of the novel \gls{origamischeme} in the SVD design (section~\ref{sec:C2_design}).

Section~\ref{sec:C9_hit_efficiency} describes the evaluation of \gls{hit} efficiency using reconstructed tracks. A few sensors have localized defects; in general the efficiency per sensor is above $99\%$, being continuously monitored by \acrshort{DQM} software.

Position and time resolution (sections~\ref{sec:C7_resolution}, \ref{sec:C7_hit_time_determination})
are the essential ingredients in track reconstruction as well as in the rejection of background \gls{hit}s. The measured \gls{cluster} position resolution, about \qty{15}{\um} (20--\qty{35}{\um}) in u/P (v/N),  is adequate and close to the expectation. Further improvements are expected optimizing the position reconstruction algorithms directly on data. 
%and  data and from refining the model of signal formation and . 
An accurate calibration of the hit time determination is complemented by the analysis of its stability.

The alignment algorithms (section~\ref{sec:C7_alignment}) pin down the position of sensing elements to an accuracy of about \qty{10}{\um}. They describe the position by six parameters for each sensor as a rigid body, plus seven parameters to describe deformations of its surface.

The main features of the performance of track finding and reconstruction are briefly outlined in
section~\ref{sec:c9_track_reconstruction}. Finally, section~\ref{sec:C9_dEdx} describes the contribution of SVD to charged particle identification by the measurement of specific ionization ($dE/dx$), in particular for the low-momentum particles with limited or missing information from the \acrshort{CDC}.

\lvltwosection{Cluster Charge}
\label{sec:C9_cluster_charge}
%\editor{Giuliana - 3 pages}

\begin{figure}[hbt]
    \centering
    \includegraphics[width=\textwidth]{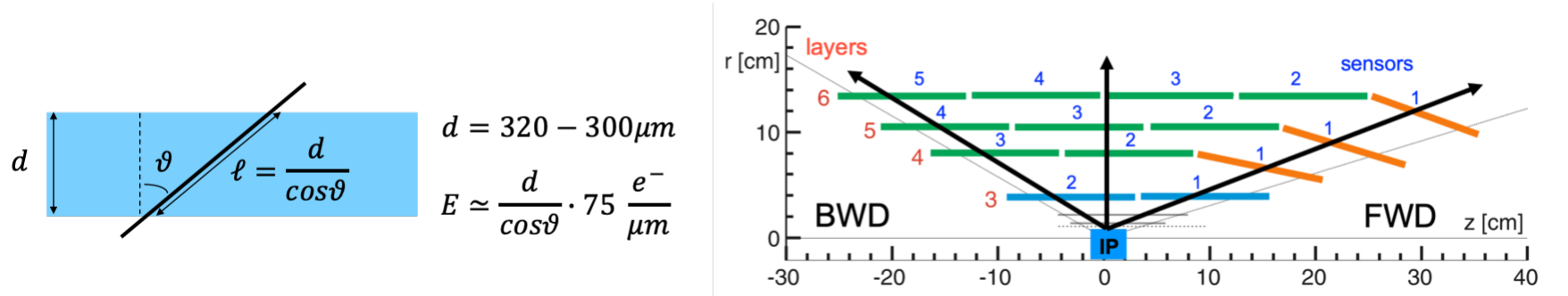}
    \caption{Left: dependence of the energy/charge released in the silicon sensor ($E$ measured in $\rm e^-$) from the track length $\ell$ in the sensitive volume, with $d$ = detector thickness. Right: SVD sensors layout, with sensors in the different position along the ladders (forward, central, backward) indicated with increasing index number. Adapted from~\cite{Rizzo:2021pgu}.}
    \label{fig:C7_layout_IP_sensors}
\end{figure}
After the full reconstruction is performed, signal clusters, defined in section~\ref{sec:C4_clustering},  are used to measure their collected charge on data. As detailed in section \ref{sec:C4_strip_reconstruction} signals from strips and clusters are converted to energy/charge, expressed in terms of electrons,  using the gain measured with the internal APV25 calibration circuit, section~\ref{sec:C4_OnlineSW_modules}.

%and their signal to noise ratio. 
%\lvlthreesection{Cluster Energy} 

The cluster charge released in the SVD strongly depends on the track incident angle $\theta$ with respect to the sensors, as shown in figure~\ref{fig:C7_layout_IP_sensors}, left.  
As defined in section~\ref{sec:C2_design} and shown in figure~\ref{fig:C2_schematic_layout}, sensors in different $z$ positions along the ladders, are indicated with increasing index number starting with 1 for the forward sensor. According to their position, sensors see tracks coming from the IP with very different incident angles, as shown in figure~\ref{fig:C7_layout_IP_sensors}, right, and the distribution of the collected cluster charge reflects this effect.

\begin{figure}[hbt]
    \centering
    \includegraphics[width=.45\columnwidth,keepaspectratio]{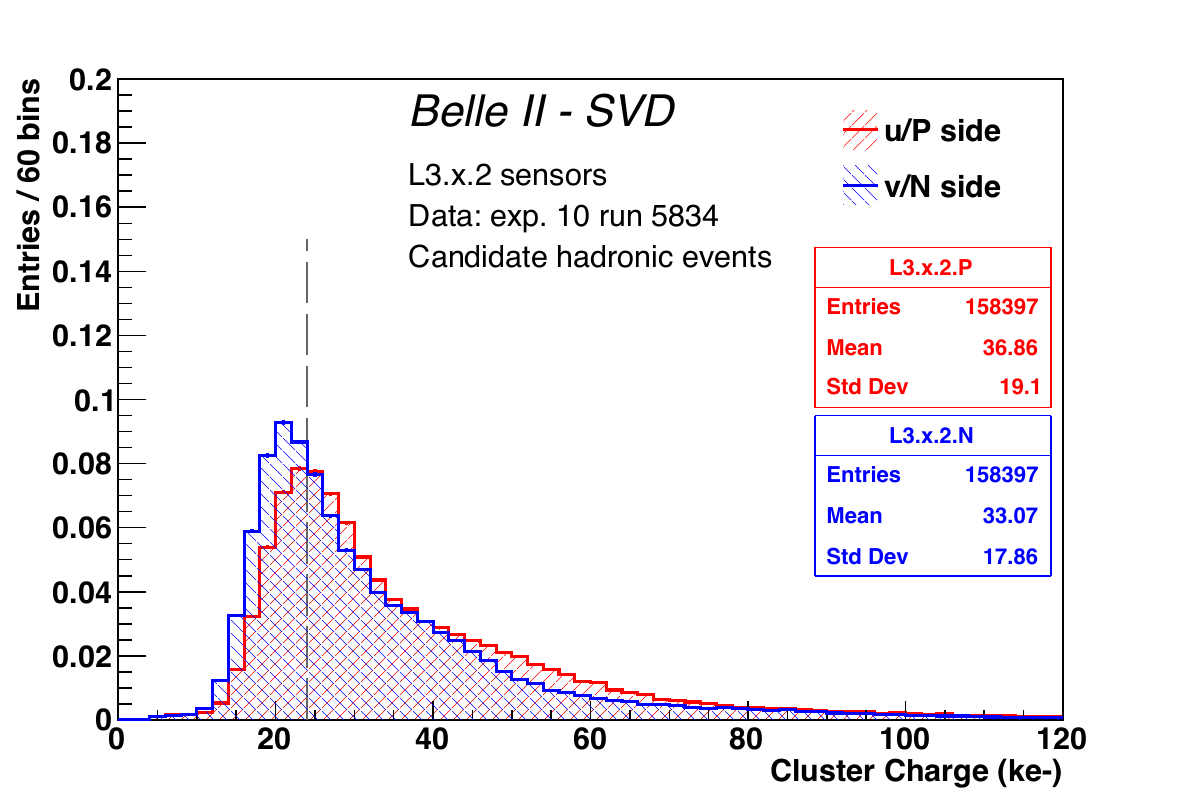}
    \includegraphics[width=.45\columnwidth,keepaspectratio]{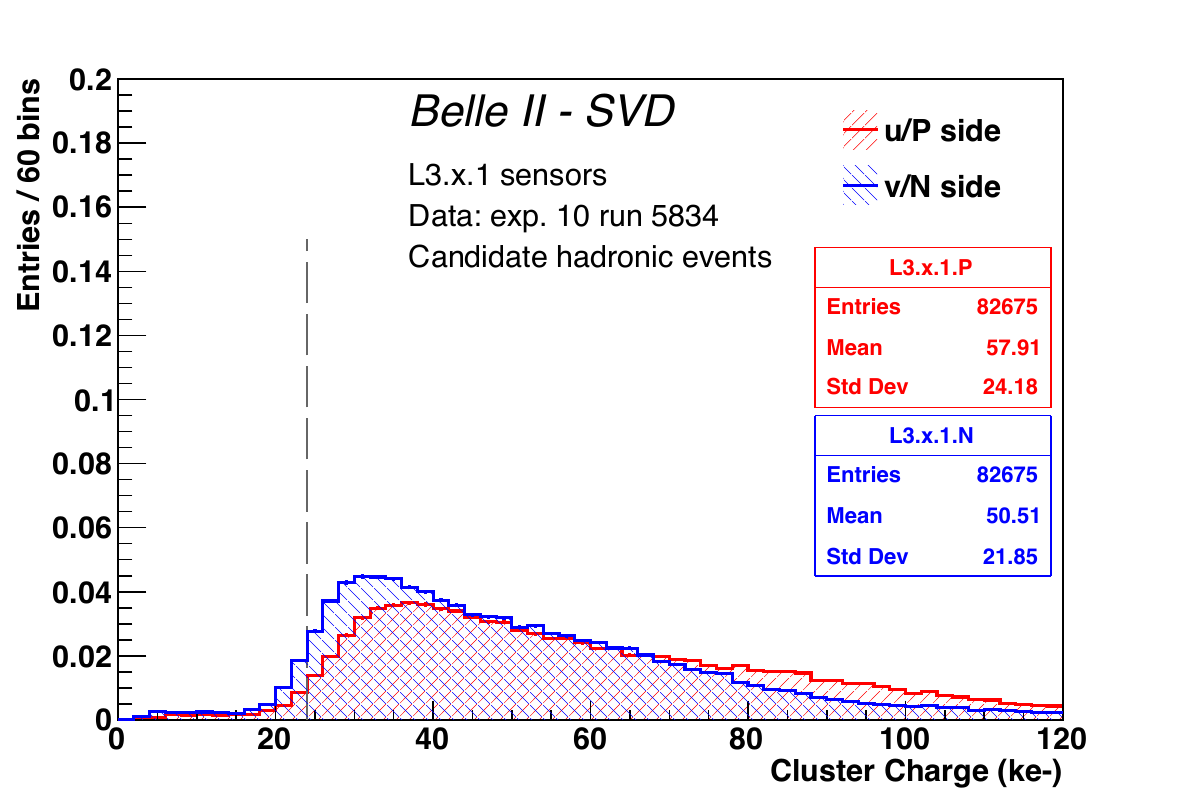} 
    \includegraphics[width=.45\columnwidth,keepaspectratio]{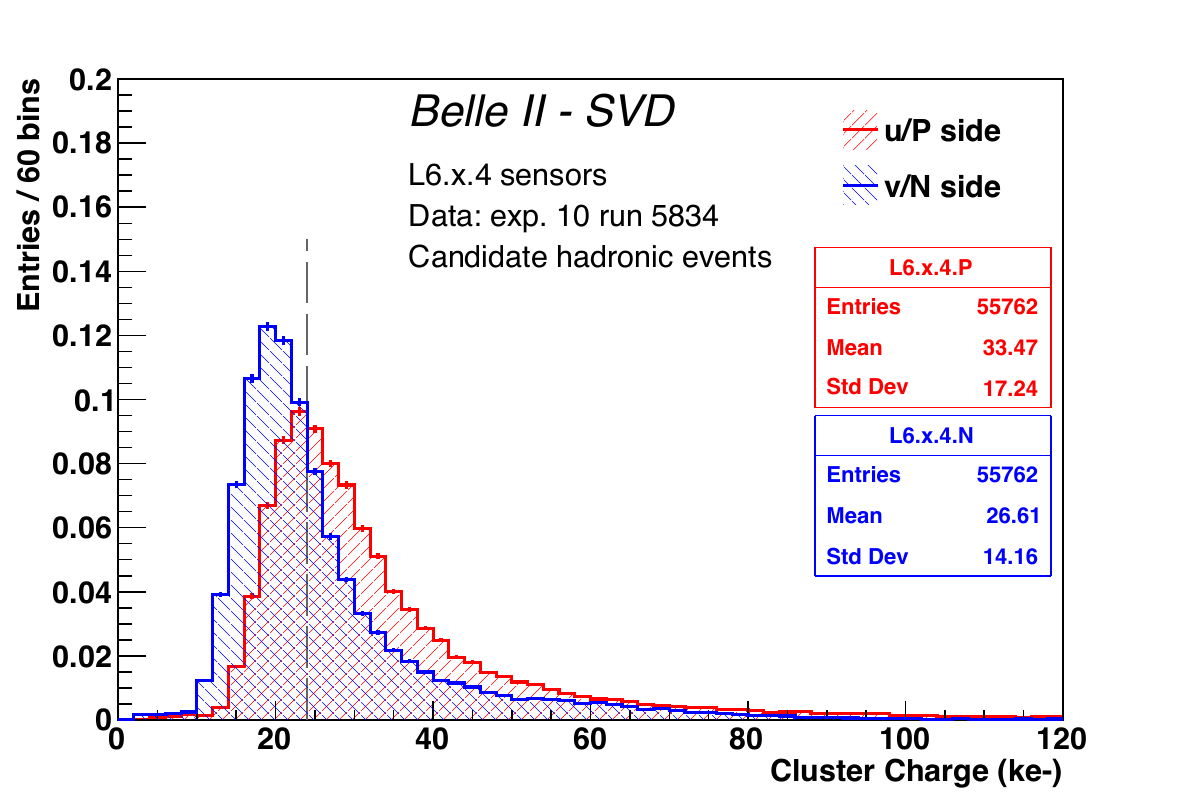} 
    \includegraphics[width=.45\columnwidth,keepaspectratio]{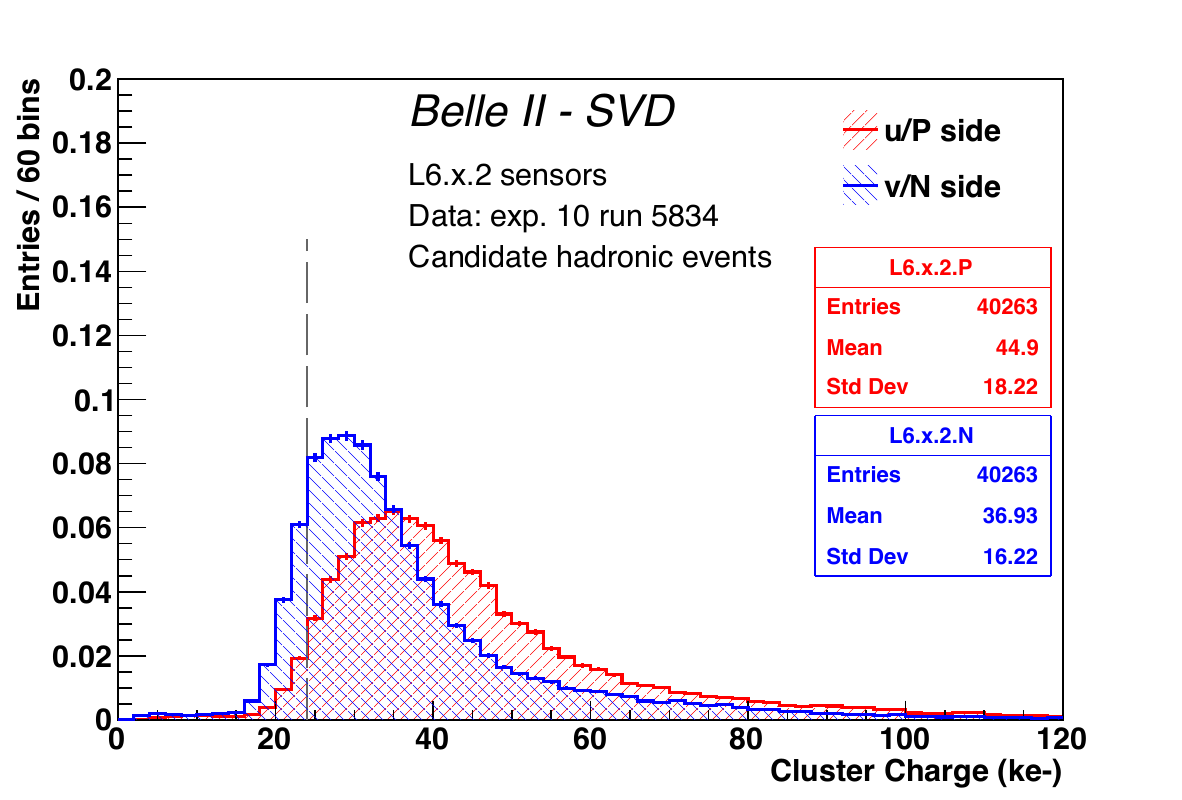} 
    \caption{Cluster charge for sensors in specific $z$ positions in layer 3 and 6, summed over all $\phi$ ladders; u/P side in red and v/N side in blue.
    From top to bottom and left to right: L3.x.2,  L3.x.1, L6.x.4, L6.x.2.
    %L3.x.1 top right,  L3.x.2 top left, L6.x.2 bottom right, L6.x.4 bottom left. 
    Data for a typical run with colliding beams in 2019 are shown. The vertical dashed lines correspond to the MIP signal at normal incidence.}
    \label{fig:C7_L3_L6_cluster_energy}
\end{figure}
\begin{figure}[hbt]
    \centering
    \includegraphics[width=.45\columnwidth,keepaspectratio]{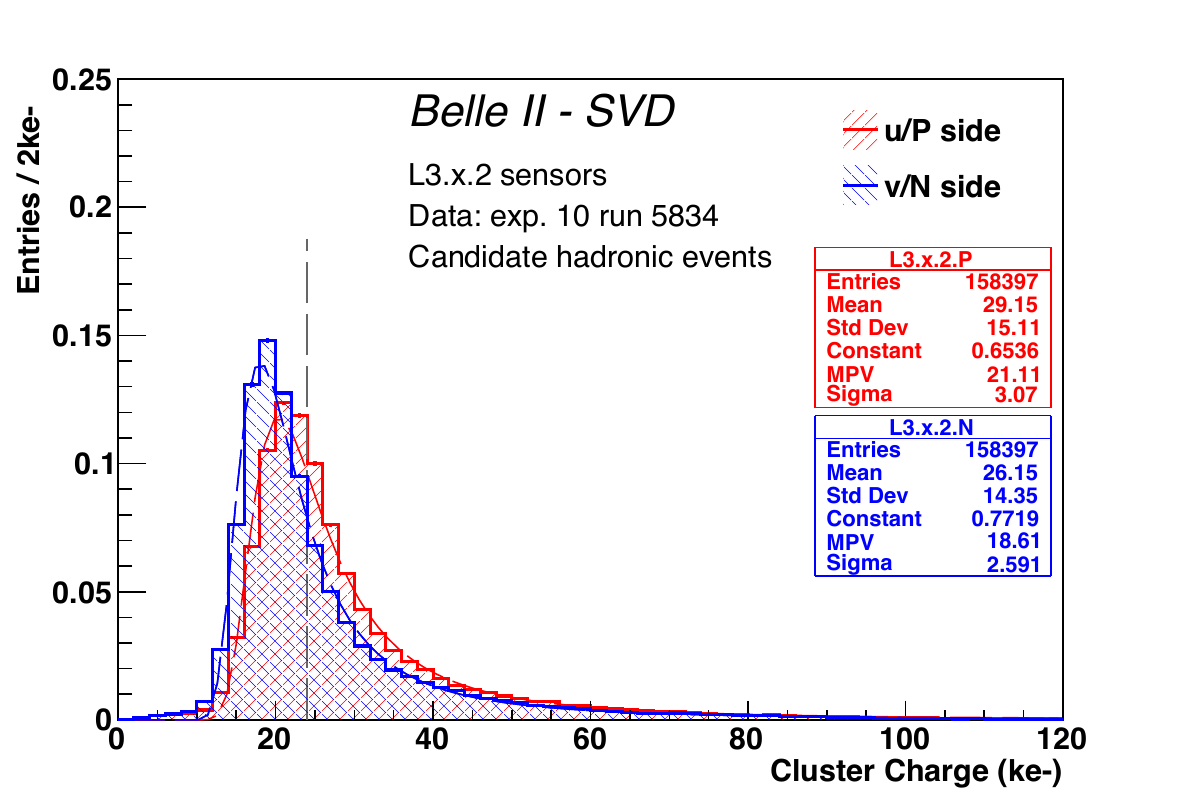}
    \includegraphics[width=.45\columnwidth,keepaspectratio]{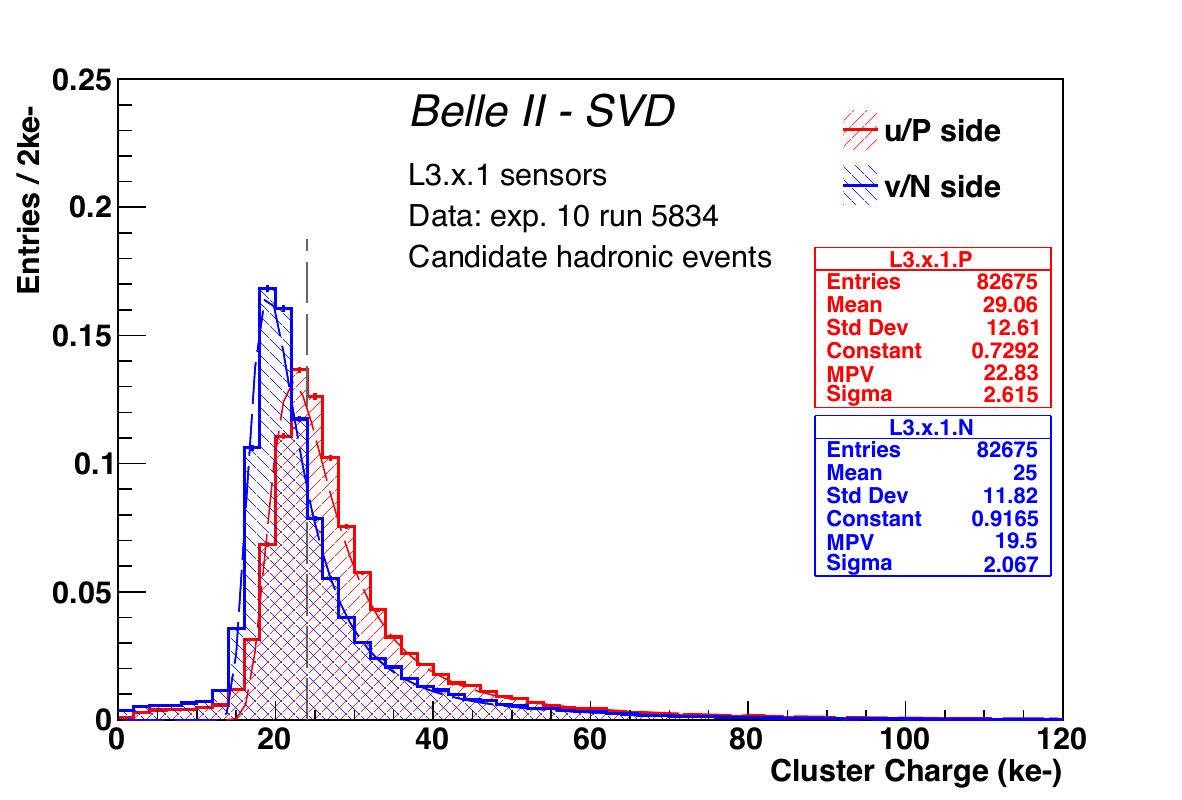} 
    \includegraphics[width=.45\columnwidth,keepaspectratio]{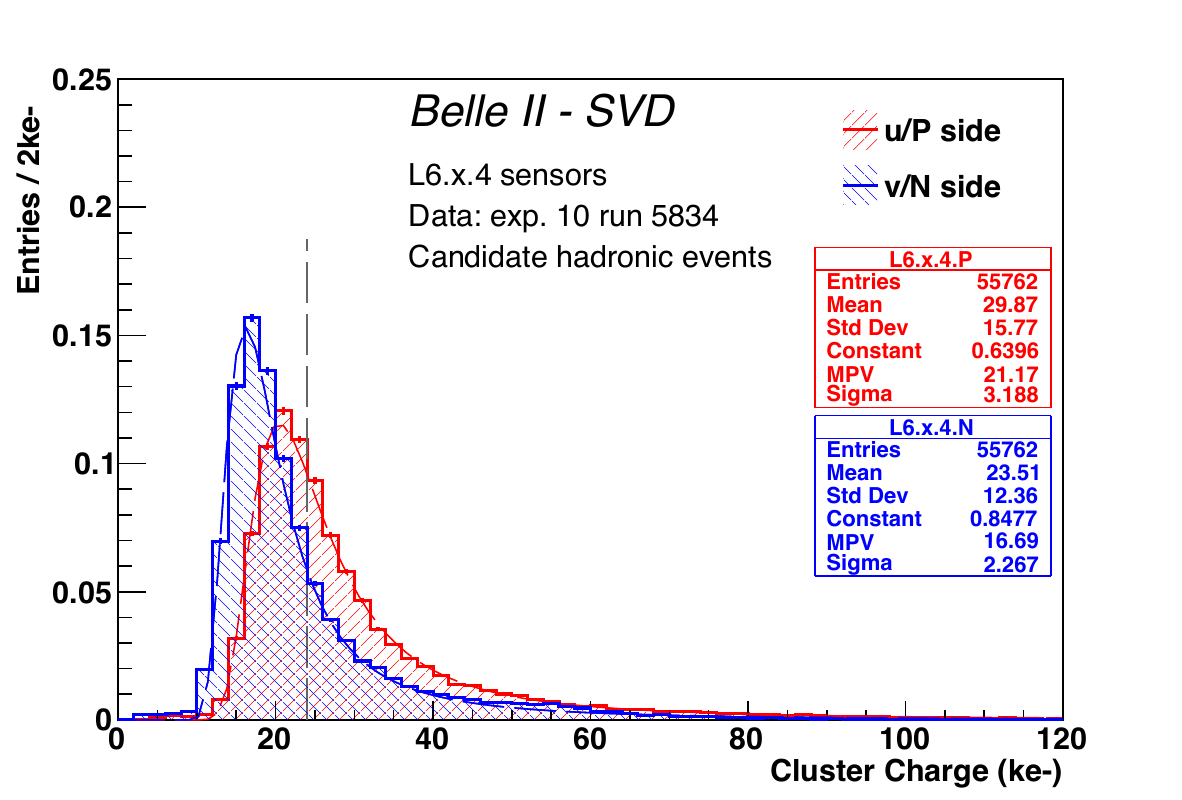} 
    \includegraphics[width=.45\columnwidth,keepaspectratio]{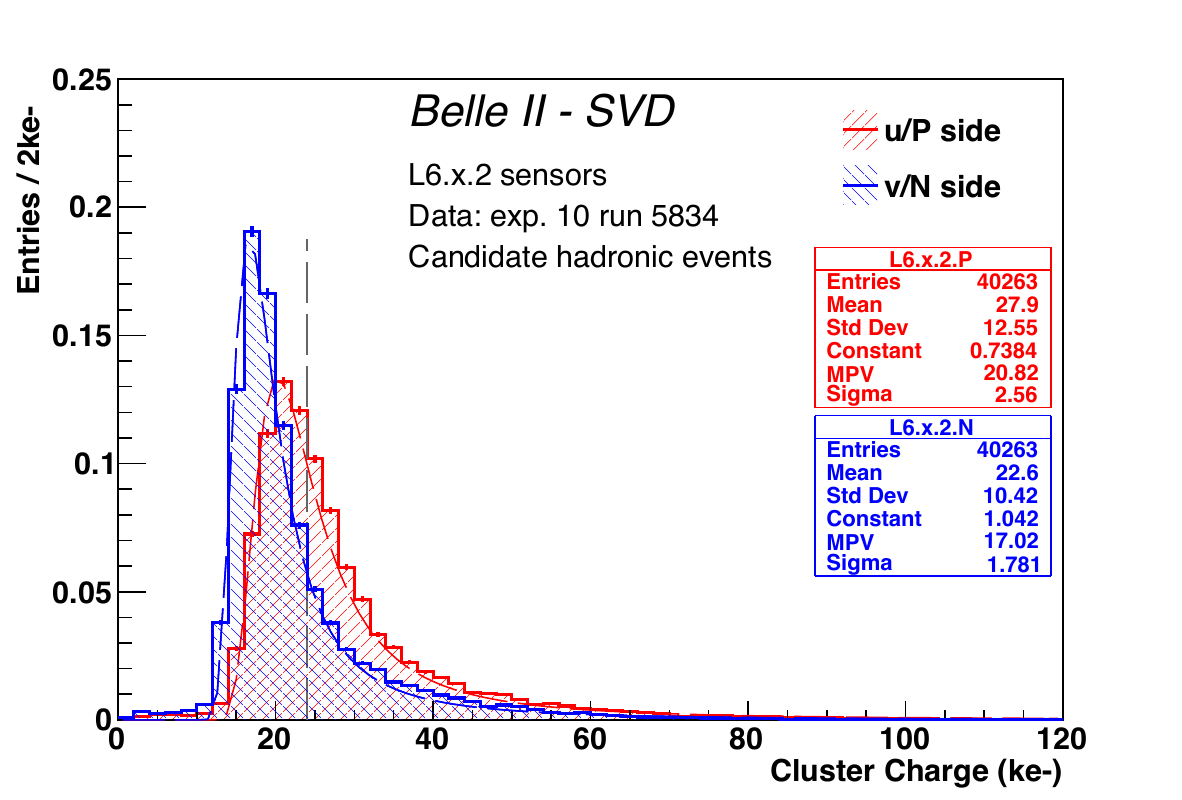} 
    \caption{Cluster charge rescaled to detector thickness, considering the track length $\ell$ in the sensitive volume,      
    %normalized to the track lenght in the sensitive volume 
   for sensors in specific z positions in layer 3 and 6, summed over all $\phi$ ladders; u/P side in red and v/N side in blue. 
    From top to bottom and left to right: L3.x.2,  L3.x.1, L6.x.4, L6.x.2.
    %L3.x.1 top right,  L3.x.2 top left, L6.x.2 bottom right, L6.x.4 bottom left. 
    Data for a typical run with colliding beams in 2019 are shown. The vertical dashed lines correspond to the expected most probable value of the MIP signal at normal incidence.}
    \label{fig:C7_L3_L6_cluster_energy_norm}
\end{figure}
As an example, in figure~\ref{fig:C7_L3_L6_cluster_energy} the cluster charge for sensors in specific $z$ positions in layer 3 and 6, summed over all $\phi$ ladders, are shown. Sensors are numbered as defined in section~\ref{sec:C2_design} indicating the position in the layer, ladder ($\phi$), $z$ position along the ladder with the following convention: L3.2.1 stands for the position in layer 3, ladder 2, sensor 1 (FW) along the ladder. 
 L3.x.1 (FW) and L3.x.2 (BW) are shown together with two other layer 6 sensors in the central $z$ position of the ladders, L6.x.2 and L6.x.4. As expected, from their position  with respect to the IP, and the track incident angle on them, the charge is significantly larger for sensors toward the FW position.

To remove the effects due to the track incidence angle, and verify the correct calibration of the system,
the cluster charge rescaled to detector thickness, considering the track length $\ell$ in the sensitive volume, is evaluated.
This distribution is shown for the previous four sensors in figure~\ref{fig:C7_L3_L6_cluster_energy_norm} and, as expected, it is independent of the sensor location. 
The most probable value (\acrshort{MPV}) for the normalized cluster charge is on average $21000\ \rm e^-$ on the u/P side. Taking into account the large uncertainty ($\sim15\%$) in the absolute APV25 gain calibration (see section~\ref{sec:C4_local_runs}), this measured value is in fair agreement with the expected  \acrshort{MIP} signal in a  \qty{320}{\um} thick silicon sensor ($\sim 24000\ \rm e^-$).   

In these distributions the reconstructed cluster charge on the two sides of the sensors are similar, but a signal loss of about 10 -- 30\% is visible on the v/N side with respect to the u/P side. 
This effect was expected and partly reproduced in MC simulation, as explained in section~\ref{sec:C4_simulation}. It is due to the large pitch on v/N side combined with the presence of a floating strip, giving a coupling capacitance of the strips to the back plane no longer negligible with respect to the interstrip capacitance, thus reducing the signals coupled to the readout strips when tracks release charge close to the floating strip.  
In fact, for a track crossing the sensor close to a floating strip, as shown in detail in section~\ref{sec:C4_simulation}, the signal seen on the adjacent readout strips via capacitive coupling is about $\frac{2C_i}{2C_i+C_b}$ of the total signal released, about $75\%$ for our sensors. 

\lvltwosection{Strip Noise}
\label{sec:C9_strip_noise}

The strip noise is measured in terms of ADC counts in the ``noise local run`` (see section~\ref{sec:C4_local_runs}), and then converted to Equivalent Noise Charge (ENC) in electrons, using the gain measured with the internal APV25 calibration circuit. The estimated accuracy on the absolute gain calibration, performed with testbeam data, is about 15\%.  Table~\ref{tab:C7_tab_noise} shows the ENC measured for the different sensor types and sides at the beginning of data taking in 2019, before the increase due to radiation damage discussed in section~\ref{sec:C6_radiation_damage_noise}.

\begin{table}[h!]
\caption{Average Noise (ENC) measured for each sensor side and position in the ladders at the beginning of data taking in 2019.}
    \centering
    \begin{tabular}{lcc}
        \toprule
       Sensor position/type & u/P side ENC ($\rm e^-$) & v/N side ENC ($\rm e^-$)\\
       \midrule
       Layer 3 (HPK small) & 930 & 630\\
       Layer 4/5/6 Origami (HPK large) & 958 & 510 \\
       Layer 4/5/6 BWD (HPK large) & 790 & 680\\
       Layer 4/5/6 FWD (Micron wedge) & 740 & 640 \\
     \bottomrule
    \end{tabular}
    \label{tab:C7_tab_noise}
\end{table}{}

The strip noise is dominated by the capacitive input load to the APV25, thus the noise figure for the different sensor sides are different by almost a factor two among the longer u/P side strips, with smaller pitch and larger interstrip capacitance, and the shorter v/N side strips with larger pitch. Additional small differences in noise are also visible HPK large rectangular sensors mounted in different ladder position, since in some cases  the contribution from the capacitance of the pitch-adapter circuit, used to connect the sensor strips to the chip, can be significant. 

Thanks to the chip-on-sensor concept exploited in the ladder design, the total channel noise is at most about 1000~$\rm e^-$, as expected from design,  allowing to achieve very good signal to noise ratio in each SVD sensor.
After a year and half of beam operation, an increase of about 25(15)\% in noise was observed on the u/P (v/N) side in the most exposed sensors in the layer 3 middle plane, after an estimated total dose of about 100~krad in this location. 
The noise increase has already shown some saturation, as expected for effects related to the increased interstrip capacitance due to the higher fixed oxide charge, as shown in section~\ref{sec:C6_evolution_of_constants_and_strips}.
In the external layers, exposed to lower dose, the increase in noise observed was of  5 to 10\%.

\lvltwosection{Signal to Noise Ratio}
\label{sec:C9_SNR}
\begin{figure}[hbt]
    \centering
    \includegraphics[width=.45\columnwidth,keepaspectratio]{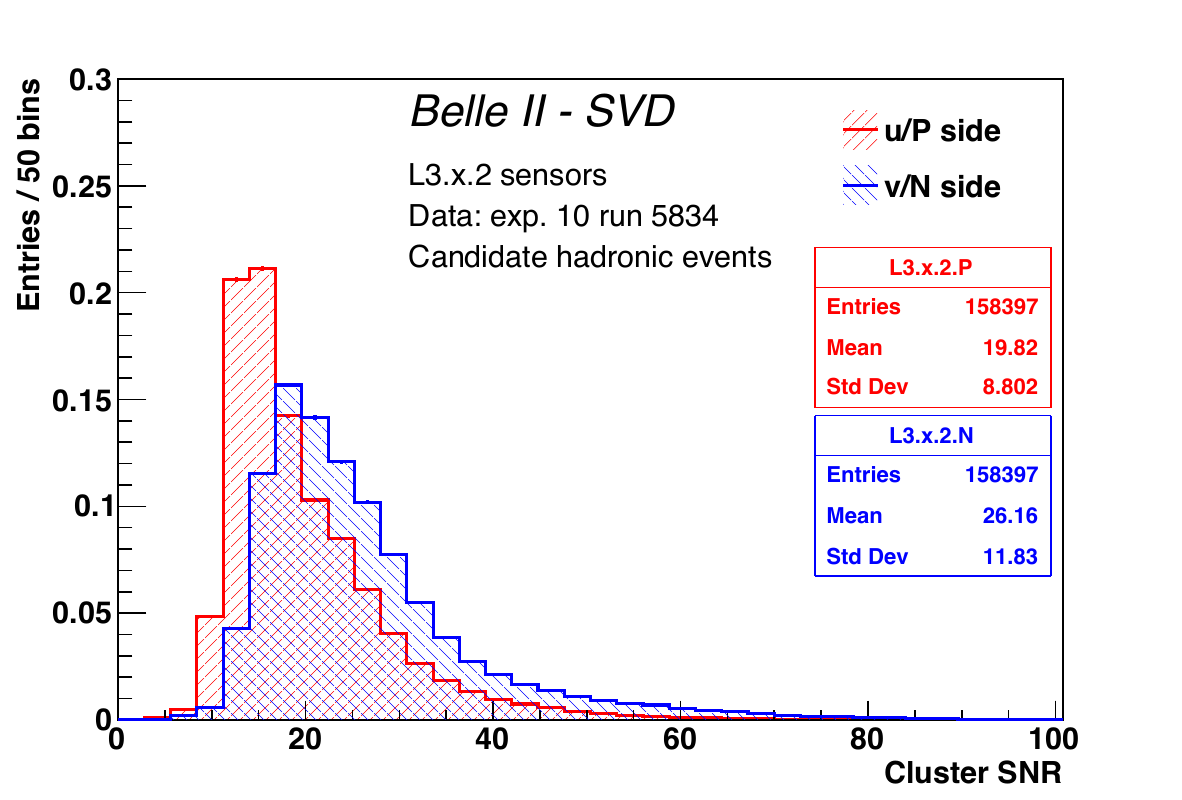}
    \includegraphics[width=.45\columnwidth,keepaspectratio]{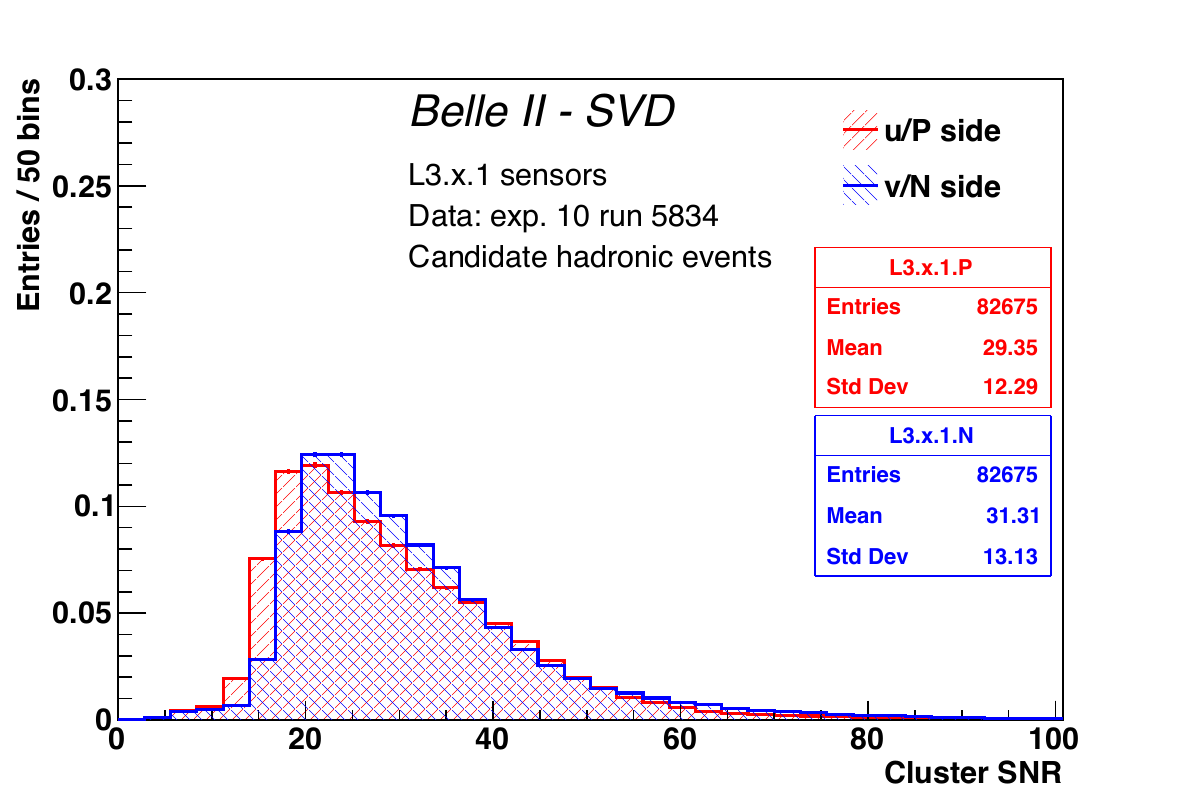} 
    \includegraphics[width=.45\columnwidth,keepaspectratio]{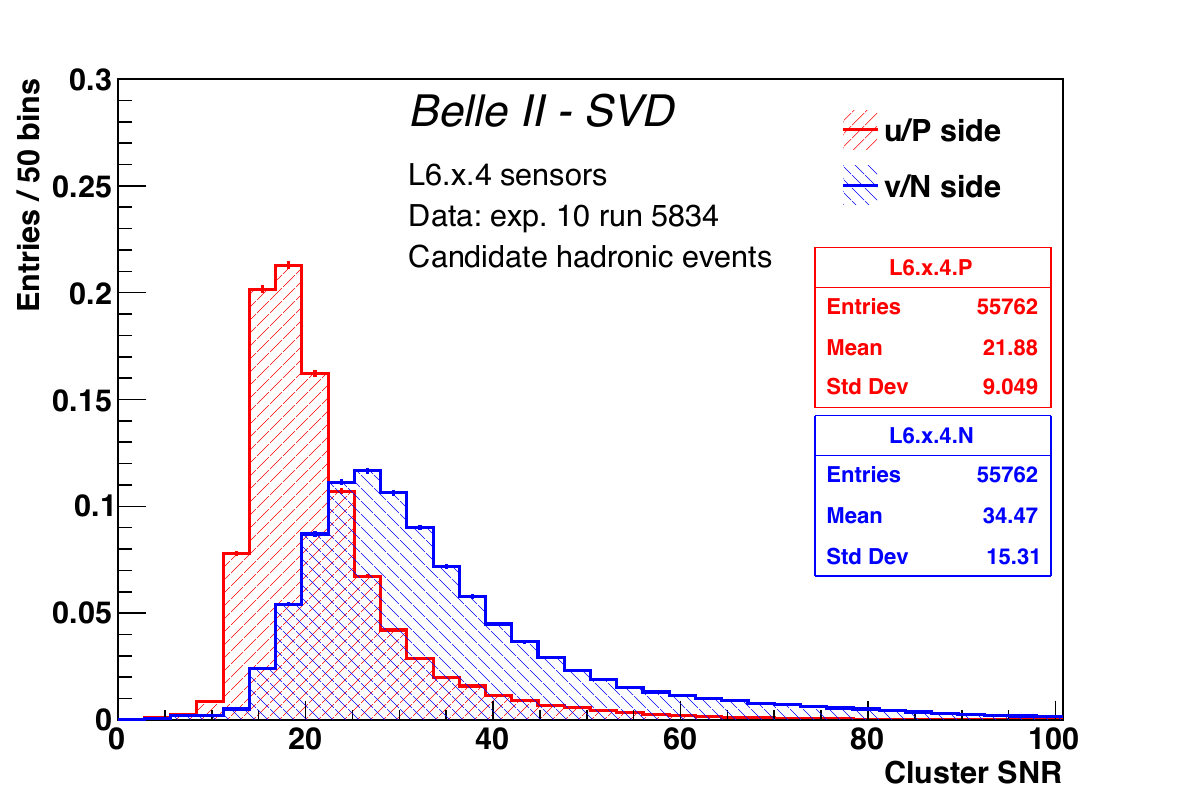} 
    \includegraphics[width=.45\columnwidth,keepaspectratio]{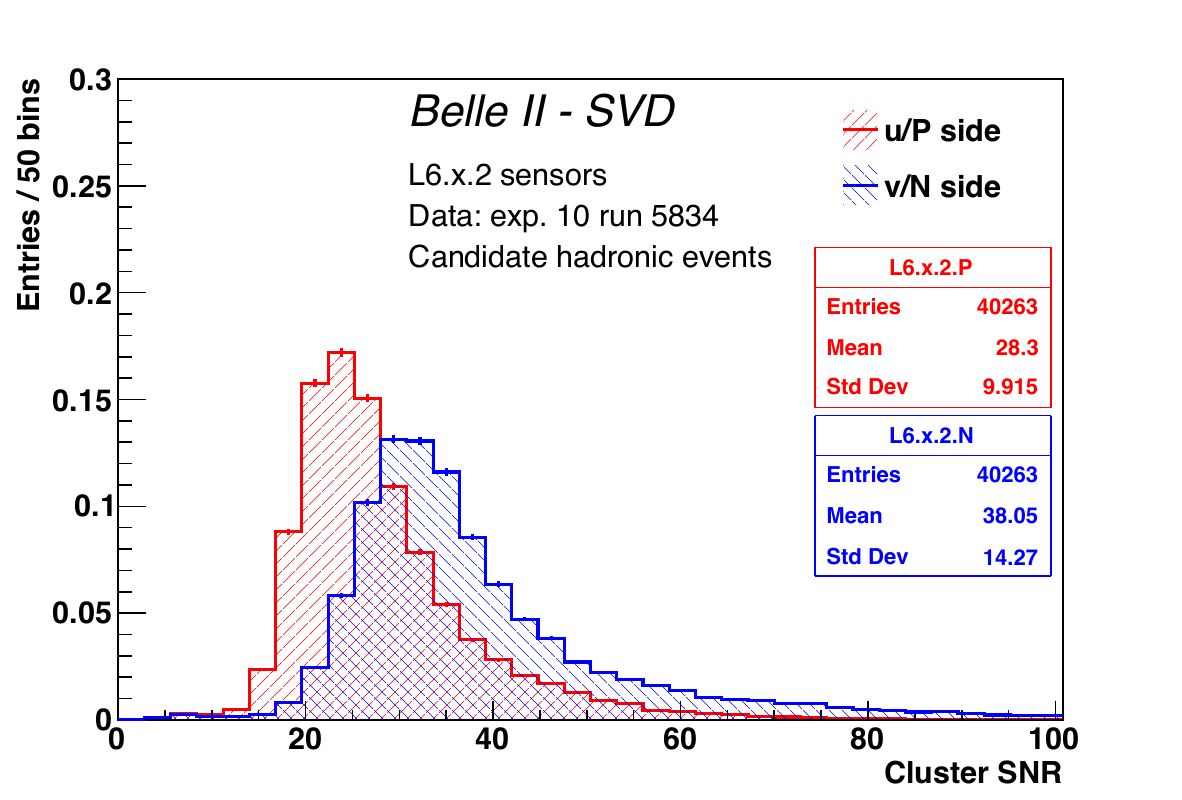} 
     \caption{Signal to noise ratio for sensors in specific z positions in layer 3 and 6, summed over all $\phi$ ladders; u/P side in red and v/N side in blue. 
    From top to bottom and left to right: L3.x.2,  L3.x.1, L6.x.4, L6.x.2.
    % L3.x.1 top right,  L3.x.2 top left, L6.x.2 bottom right, L6.x.4 bottom left. 
    Data for a typical run with colliding beams in 2019 are shown.}
    \label{fig:C7_L3_L6_cluster_SNR}
\end{figure}
\begin{figure}[hbt]
    \centering
    \includegraphics[width=\columnwidth,keepaspectratio]{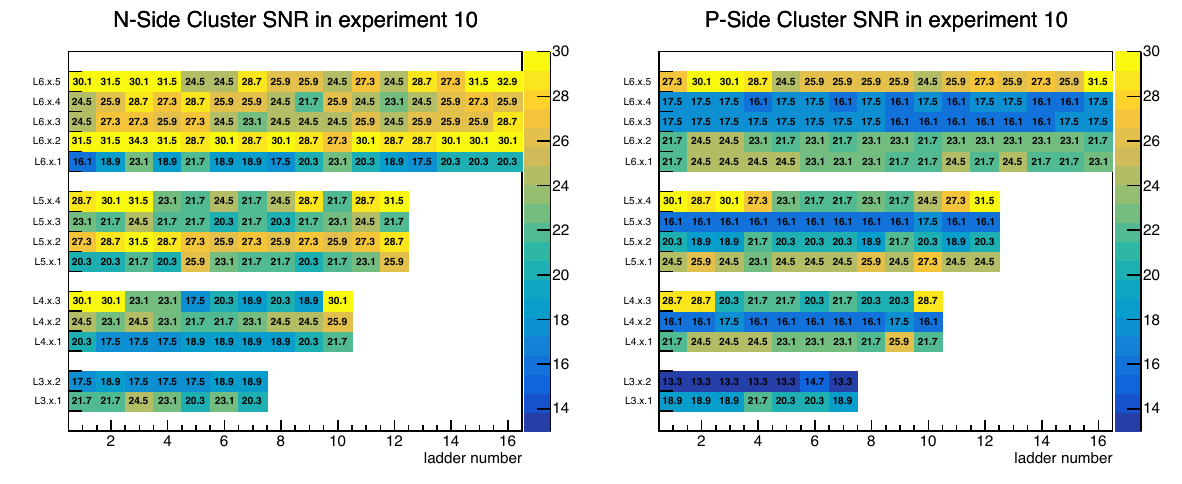}
    \caption{Most probable value of the cluster \snrCL for all SVD sensors organized in a map. The horizontal axis identifies the ladder $\phi$ position (from 1 to 16 depending on the layer), while the vertical axis enumerates the different layers and the $z$ position of the sensor in the ladder (from 1 to 5 depending on the layer). Data for a typical run with colliding beams in 2019 are shown.}
    \label{fig:C7_SNR_exp10_map}
\end{figure}

The cluster {\snrCL}, defined in section~\ref{sec:C4_clustering} as the total cluster charge over the sum in quadrature of the noises of the strips belonging to the clusters, is an important figure of merit to be monitored. 
Due to the online zero suppression criteria, based on signal strip higher than 3 times its noise, low \snrCL values increase the probability to miss signal strips,  deteriorating the hit efficiency and position resolution. 
The cluster \snrCL depends on collected charge, strip noise, and cluster size. As shown in previous sections, signals depend strongly on sensor position, due to the track incidence angles, that also influences the cluster size. Noise is higher in u/P side, with longer strips with respect to v/N side. This last effect dominates, then \snrCL is in general higher on v/N side, with quite large variation among the sensors, depending on their position. 

The distribution of the \snrCL measured for the FW and BW Layer 3 sensors and for some Layer 6 sensors in different positions is shown in figure~\ref{fig:C7_L3_L6_cluster_SNR}. 
The map of the \snrCL most probable values for all the SVD sensors is shown in figure~\ref{fig:C7_SNR_exp10_map} and clearly indicates a pattern determined by their position and the sensor side considered. All the 172 sensors have very good cluster \snrCL performance with MPV ranging between 13 and 30 as of July 2020. A small \snrCL reduction in Layer 3 u/P side of about 10\% was observed, with respect to the beginning of the run in 2019, which is related to the increased noise due to radiation damage, with no effect on performance. 

\lvltwosection{Hit Efficiency}
\label{sec:C9_hit_efficiency}
\editor{Giulia - 3 pages}
%describe the overall cluster reconstruction performance, the title can be more general. Main points:
%- unbiased efficiency (from the sensor efficiency to the clustering algorithm effects)
%- cluster position resolution
%- cluster time resolution (or we dedicate a section to the time, but maybe it's too much)

The efficiency of cluster reconstruction, described in Section~\ref{sec:C4_clustering}, was measured on an early data set with an {\it unbiased} technique and continuously monitored with \acrfull{DQM} tools with a {\it biased} technique.
The unbiased technique consists in reconstructing tracks excluding the sensor under study and retaining only clusters with $\snrCL > 10$. The following selection criteria are applied to tracks:
\begin{itemize}
    \item tracks must originate from the vicinity of the IP, $|d_0|<0.1$ cm and $|z_0|<0.2$ cm, where $d_0$ and $z_0$ are the transverse and longitudinal track impact parameters, respectively;
    \item tracks must have at least one SVD hit and 20 CDC hits;
    \item tracks must have a transverse momentum of at least 1 ${\rm GeV}/c$.
\end{itemize}

\begin{figure}[hbt]
    \centering  
        \includegraphics[width=0.45\textwidth]{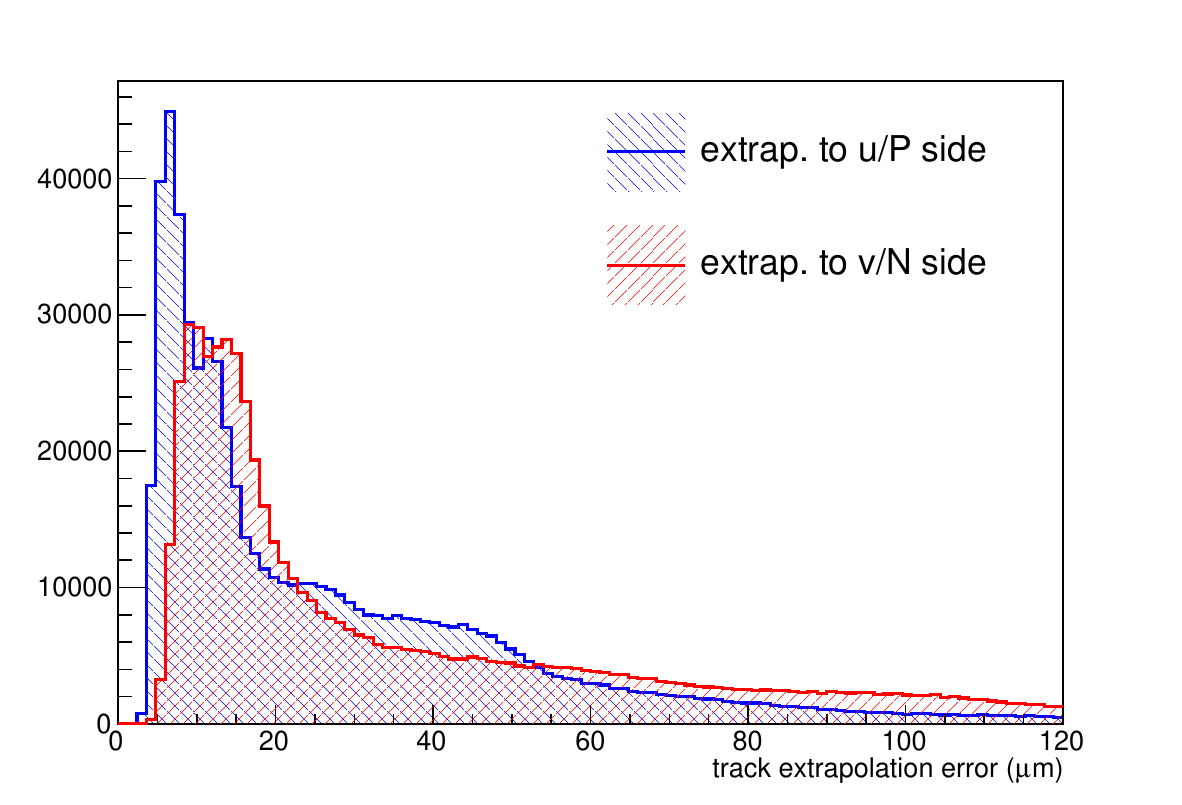}
        \includegraphics[width=0.45\textwidth]{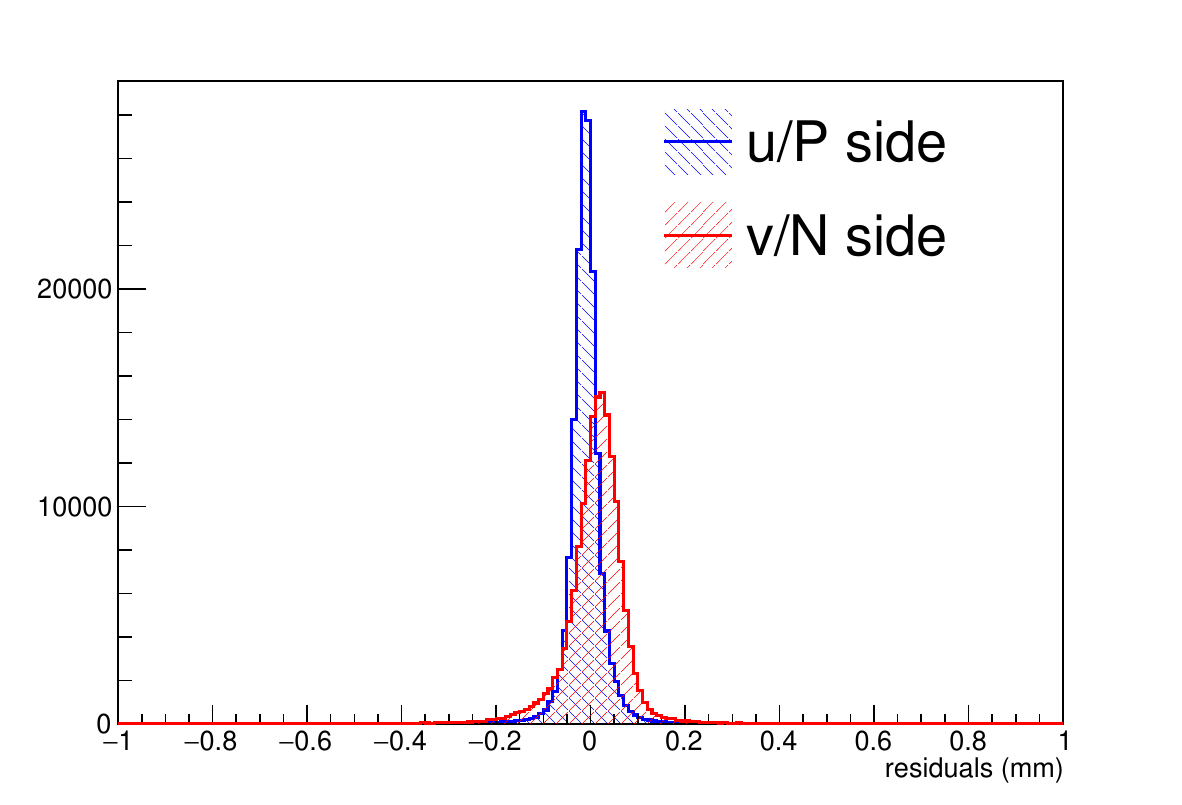}  
    \caption{(left) Track extrapolation uncertainty (i.e. the uncertainty in the fitted track extrapolation position on the sensor) to u/P (blue) and v/N (red) sides for layer 4 barrel sensors; (right) residual distribution for layer 4 barrel sensors.}
    \label{fig:C7_trkError}
\end{figure}

Selected tracks are then extrapolated to the sensor under study. An example of the distribution of the track extrapolation  uncertainty (i.e. the uncertainty in the fitted track extrapolation position on the sensor) for the barrel sensors of layer 4 is shown in the left plot of figure~\ref{fig:C7_trkError}. The extrapolation to the u/P side shows a smaller uncertainty because of the better resolution on this direction both on SVD (smaller pitch) and CDC measurements. 
A fiducial area of $\pm 0.5$ cm from the edge of the active area of the sensor was defined in order to exclude border effects; tracks extrapolated outside the fiducial area are discarded.

The histogram of the residuals (the distance between the extrapolated track and the cluster position) is built considering the cluster nearest the intercept.  An example from layer 4 is shown in the right plot of figure~\ref{fig:C7_trkError}. The analyzed data were reconstructed with a preliminary alignment and a residual misalignment is visible, especially on the v/N side. However, as shown in section~\ref{sec:C7_alignment} the current alignment is very much improved. 
The efficiency is defined as the fraction of intercepts with at least one cluster  found in a window of $\pm 0.5$ mm from the median of the residual distribution.

\begin{figure}[hbt]
    \centering  
        \includegraphics[width=\columnwidth,height=6 cm]{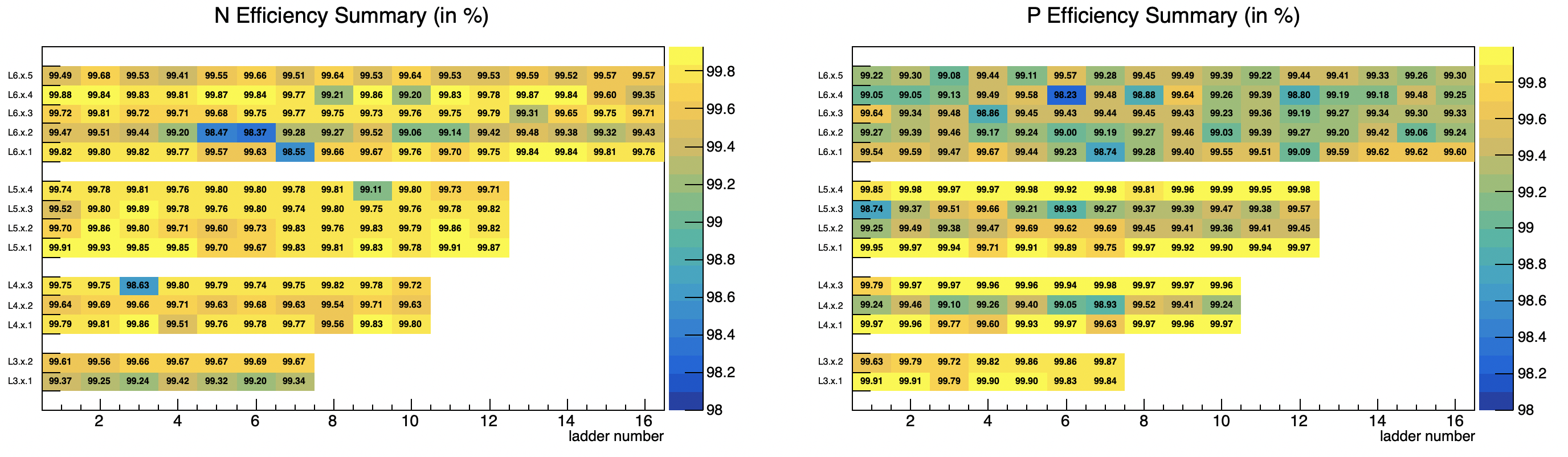}
    \caption{Average efficiency for each sensor side, v/N on the left, and u/P on the right, organized in a map. The horizontal axis identifies the ladder $\phi$ position (from 1 to 16 depending on the layer), while the vertical axis enumerates the different layers and the $z$ position of the sensor in the ladder (from 1 to 5 depending on the layer). The uncertainty in the efficiency varies from fraction of per-mill in the innermost sensors to fraction of per-cent in the outer sensors.}
    \label{fig:C7_efficiency}
\end{figure}

In figure~\ref{fig:C7_efficiency} the map of the average efficiency of each sensor is shown. Only a few sensors have an efficiency below 99\%.
The efficiency on the v/N side is on average higher than the u/P side thanks to the higher \snrCL on this side. 

A few sensors have localized defects slightly worsening the efficiency, for example the v/N side of sensor 4.3.3, 6.5.2, 6.6.2, and 6.7.1, and the u/P side of sensor 6.6.4. The efficiency for those sensors is still above 98\%. 
Figure~\ref{fig:C7_efficiency} also shows a slight variation of the efficiency depending on the position of the sensors. This is most probably due to the fact that the position of the sensor is correlated with the incident angle of the track on the sensor, and therefore the charge released in the sensor and the \snrCL change with position. Since the cluster reconstruction is independent of the sensor, the effect of clusterization may be slightly different depending on the position of the sensor.

The average efficiency per layer is above 99.3\% for all layers, as listed in Table~\ref{tab:C7_efficiency}.
\begin{table}[h!]
         \caption{Average sensor efficiency.}
    \label{tab:C7_efficiency}
    \centering
    \begin{tabular}{ccc}
   \toprule
       layer  & $\varepsilon(\mathrm{u/P}) (\%) $& $\varepsilon(\mathrm{v/N}) (\%) $\\
       \midrule
       3 & $99.83\pm0.01$& $99.48\pm0.03$\\
       4 & $99.69\pm0.03$& $99.68\pm0.03$\\
       5 & $99.66\pm0.03$& $99.77\pm0.04$\\
       6 & $99.31\pm0.08$& $99.58\pm0.06$\\
        \bottomrule
    \end{tabular}
\end{table}{}
\begin{figure}[hbt]
    \centering  
        \includegraphics[width=\textwidth]{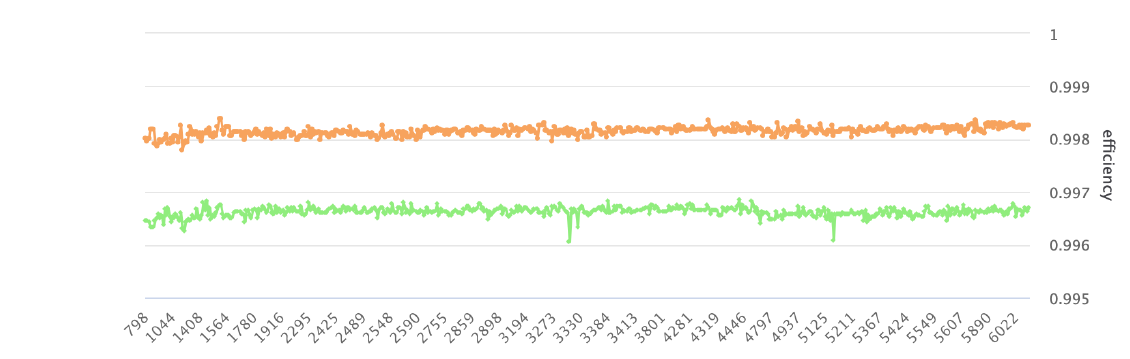}
        \caption{Average efficiency, v/N in orange, and u/P in green. Data collected between March and June 2020 are shown, with the corresponding run number on the horizontal axis. Adapted from~\cite{Belle-IISVD:2021uar}. }
    \label{fig:C7_mirabelleEfficiency}
\end{figure}

As described in Section~\ref{sec:C4_dqm}, the average sensor efficiency is monitored online. For this measurement  a {\it biased} technique is used, including clusters on the sensor under study in the track reconstruction. The rest of the analysis is performed in the same way. Figure~\ref{fig:C7_mirabelleEfficiency} shows the online average SVD efficiency in the last part of 2019 data-taking as well as in the first part of 2020 data-taking.

\lvltwosection{Position Resolution}
\label{sec:C7_resolution}

The cluster position resolution is measured with a di-muon sample ($e^+e^-\to\mu^+\mu^-$): events with only two tracks are considered, tracks with transverse momentum greater than 1 GeV/c, originating from the IP ($|z_0|<0.5$ cm and $|d_0|<0.4$ cm), and with at least one PXD hit, 8 SVD and 30 CDC hits are selected, and only muon pairs with an invariant mass between 10 and 11 GeV/c$^2$ are selected. 
After reconstruction, the muon tracks are refitted excluding the cluster on the sensor under study. 
Since the track extrapolation uncertainty is of the same order as the resolution that we are measuring, tracks with a track extrapolation uncertainty greater than \qty{100}{\um} are excluded.
A fiducial area of $\pm 0.5$ cm from the sensor edge is defined. 
 For each track $i$ the unbiased residuals ($\Delta x_i$) and the unbiased track extrapolation uncertainty ($\sigma^{trk}_{x,i}$) are computed, to calculate the quantity $r_i^2$
\begin{equation}
    r_i^2 = (\Delta x_i)^2 - (\sigma^{trk}_{x,i})^2
\end{equation}{}
The resolution $\sigma_x$ is then computed as the square root of the average of $r_i^2$:
\begin{equation}
    \sigma_x = \sqrt{\frac{1}{N}\sum_i^Nr_i^2}
\end{equation}{}   
with its uncertainty $\Delta\sigma_x$:
 \begin{equation}
    \Delta\sigma_x = \frac{1}{\sqrt{N}}\frac{\sigma_{r_i^2}}{ 2\sigma_x}
\end{equation}{} 
where $\sigma_{r_i^2}$ is the RMS of the $r_i^2$ distribution. 
Before computing the average of $r_i^2$  the distribution is restricted at $\pm \qty{15000}{\um\squared} $, removing the tails of the distribution. A value of $|r_i^2|> \qty{15000}{\um\squared} $, would correspond to a resolution greater than $\sim$ \qty{120}{\um}, which is much worse than the expected resolution of the SVD sensors.
The analysis is performed as a function of the track incident angle, requiring at least 300 clusters for each bin.
This method was validated with simulation, exploiting the true hit information, yielding a resolution within \qty{2}{\um} compared to the true value.

The cluster position resolution as a function of the track incident angle is shown in figure~\ref{fig:C7_resolution} for data and simulated events. 
%\rem{need to put the data-MC plots and add comments on the shape and comparison}
\begin{figure}[hbt]
    \centering  
        \includegraphics[width=0.48\textwidth]{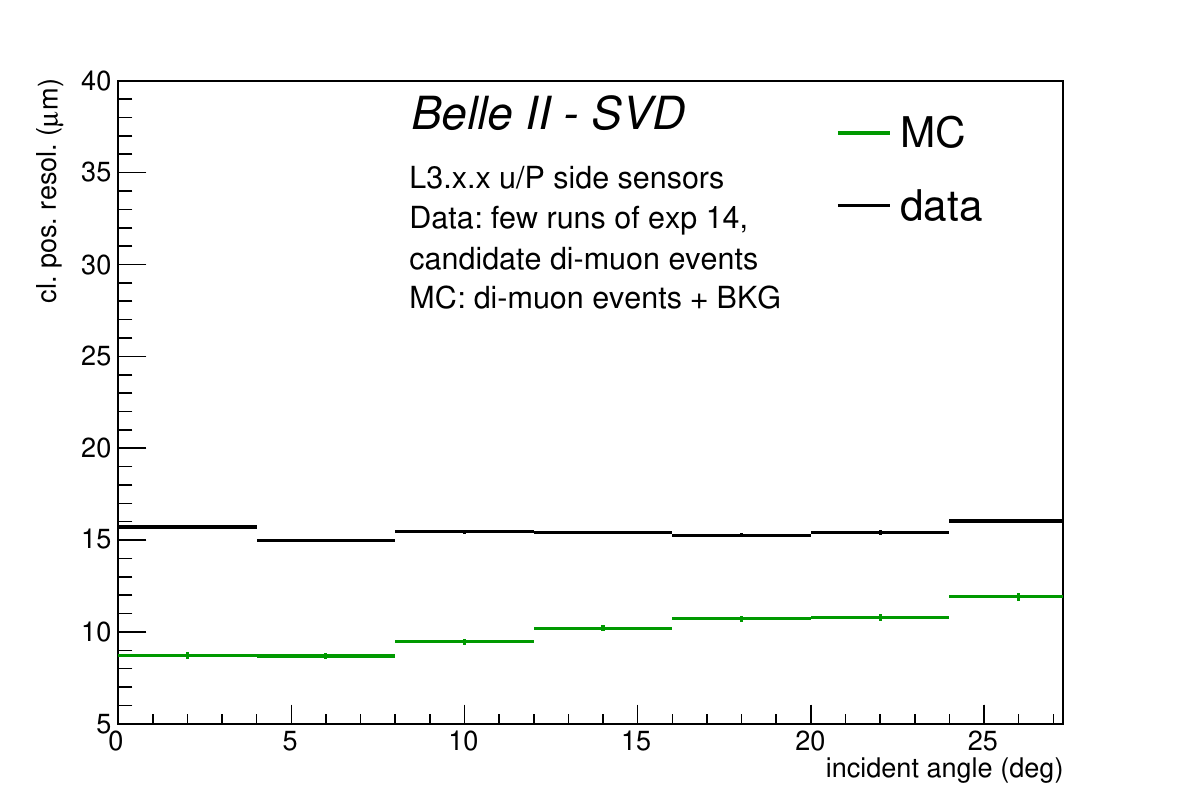}
        \includegraphics[width=0.48\textwidth]{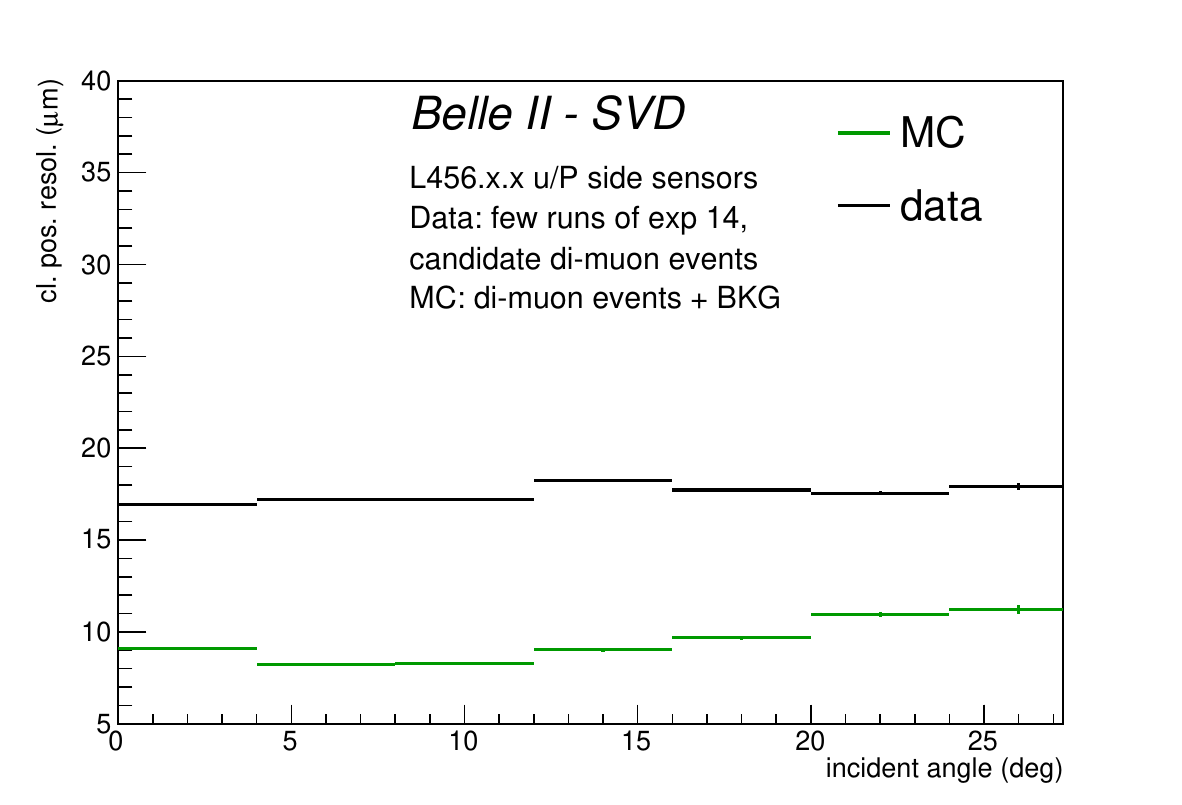}
        \includegraphics[width=0.48\textwidth]{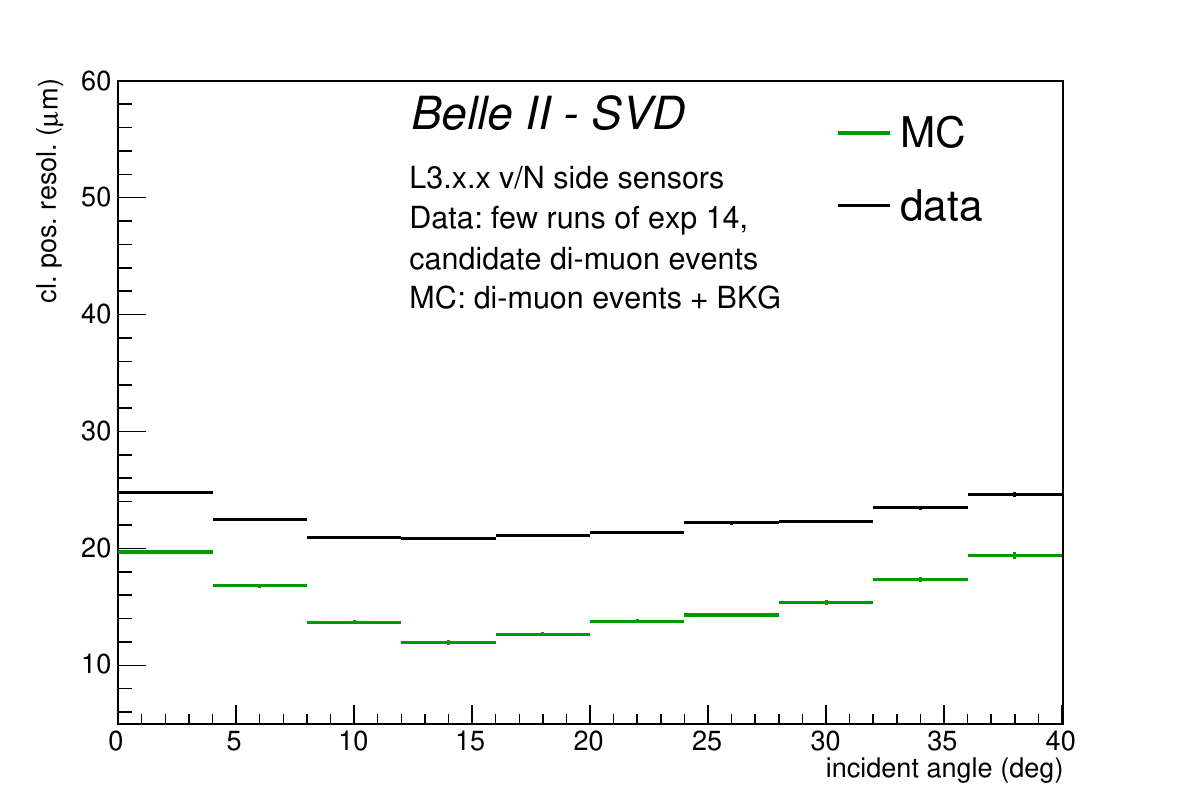}
        \includegraphics[width=0.48\textwidth]{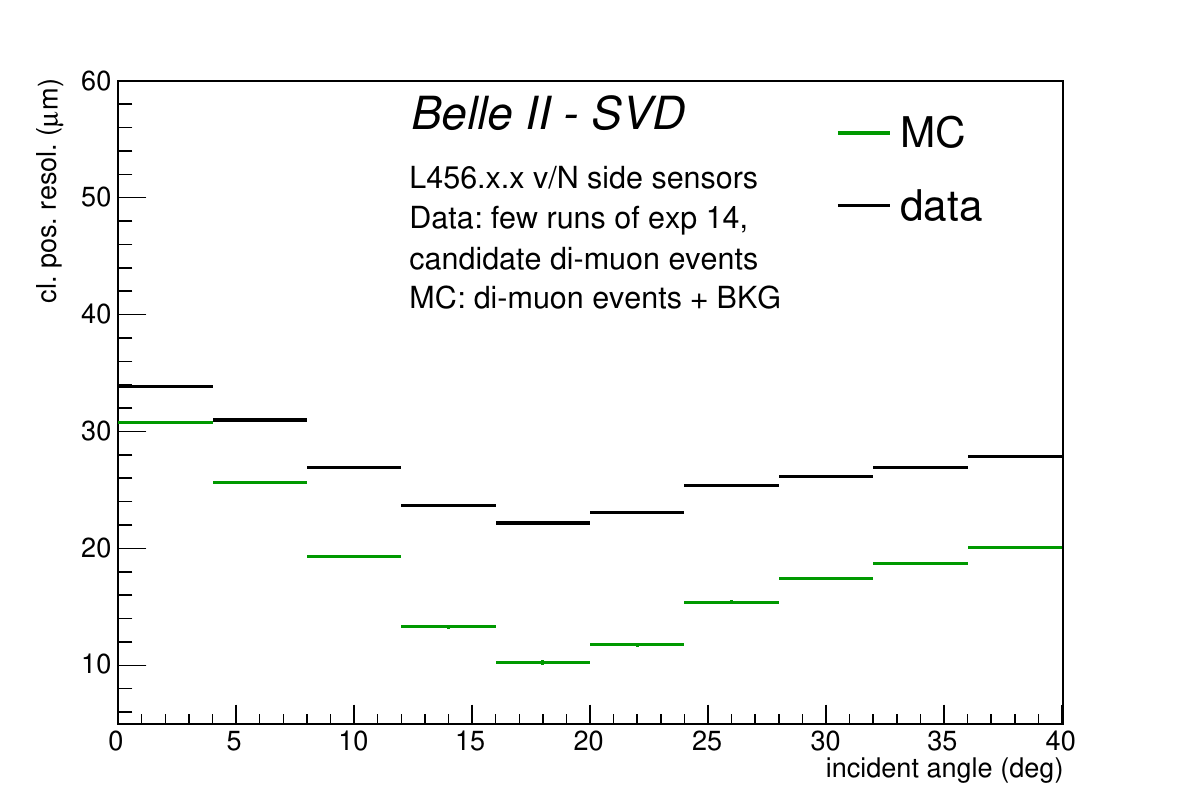}
    \caption{Cluster position resolution as a function of the track incident angle for data and MC simulation.  
    In the upper row u/P side results for layer 3 (left) and layer 456 barrel sensors (right). In the lower row v/N side results for layer 3 (left) and layer 456 barrel sensors (right).}
    \label{fig:C7_resolution}
\end{figure}
%Resolution results on simulated events match the expected digital resolution for perpendicular track, expected to be 7-11 \mum on u/P side and 23-35 \mum on v/N side, respectively for layer 3 and layer 456, considering the various sensor pitches with the floating strip 
The observed resolution has the expected shape, showing a minimum
at the incident angle for which the projection of the track along the direction perpendicular to the strips on the detector plane corresponds to two strip pitches.
%for angles where the projection of the track on the detector plane, in the direction perpendicular to the strips, corresponds to a two strip cluster. 
Given the various sensor pitches with one floating strip, the minimum is expected at 4 (7) degrees for the u/P side and at 14 (21) degrees on the v/N side,  for layer 3 (4, 5, 6). 
As discussed in section~\ref{sec:C4_clustering}, the expected digital resolution for perpendicular tracks is \qty{7}{\um} (\qty{11}{\um}) on u/P side and \qty{23}{\um} (\qty{35}{\um}) on v/N side for layer 3 (4,5,6), in very good agreement with the resolutions previously measured in testbeam data with perpendicular tracks \cite{Belle-IISVD:2017tht}, and also similar to the resolution shown for perpendicular tracks in figure~\ref{fig:C7_resolution}.
In particular, for the larger pitches (v/N sides) resolution measured in collision data is in very good agreement with the expected one, from previous testbeam results, while the MC resolution obtained is too optimistic. On the contrary for the small pitches (u/P sides) there is room for improvements: resolution in collision data is  higher than previously measured in testbeam data, while MC results are more similar to the expected values. 

The measured resolution in collision data is anyway good and adequate for the main goals of SVD tracking, though the algorithms for the determination of the cluster position, described in Section~\ref{sec:C4_clustering}, have not been fully optimized on data yet, therefore there is still room for improvement of the cluster position resolution especially for the small pitches.  The algorithms were in fact optimized on simulated events, and the resolution  obtained with the same technique on simulation is much better than on data. Given the significant differences with respect to data, especially concerning the cluster size (see Simulation Performance in Section~\ref{sec:C4_simulation}), the difference in resolution between data and MC is not a surprise. Work both on improving the simulation of the signal formation, and on optimizing the position reconstruction algorithm directly on data is in progress. 
Due to the long editorial process required for this paper, some updated performance results could not be incorporated, and are included in the references in section \ref{sec:conclusion}.

%\textcolor{blue}{
%Two comments to be added:
%\begin{itemize}
%    \item comparison data-MC
%    \begin{itemize}
%        \item one plot with DAT\label{sec:C4_simulation}A-MC comparison (one layer one side?)
%        \item simplified model in simulation does not reproduce effects seen on data: (1) cluster size, (2) strip time difference in a cluster
%    \end{itemize}
%    \item comparison with expected resolution on data
%    \begin{itemize}
%        \item algorithm is not tuned on data
%        \item room improvement exploiting the additional information that we now have on data (not reproduced by MC)
%    \end{itemize}
%    \begin{itemize}
%        \item 
%    \end{itemize}
%\end{itemize}   
%}

\lvltwosection{Hit Time Determination}
\label{sec:C7_hit_time_determination}
\editor{Luigi/Giulia 2 pages}
Although with present background levels the reconstruction does not require to use any background rejection based on time selection, a good hit time resolution will be crucial in future operation with higher luminosity and higher background occupancy to efficiently reject off-time background hits. 

\begin{figure}
    \centering
    \includegraphics[width=.48\columnwidth,keepaspectratio]{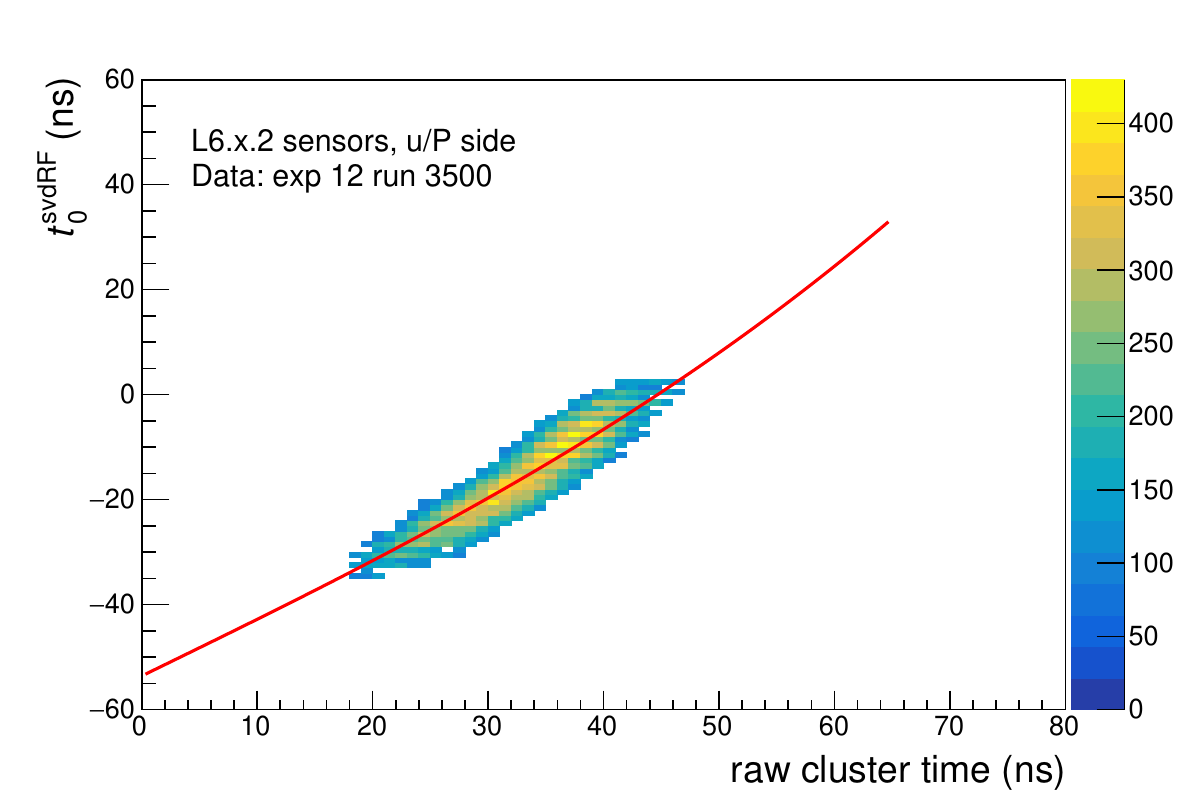} 
    \includegraphics[width=.48\columnwidth,keepaspectratio]{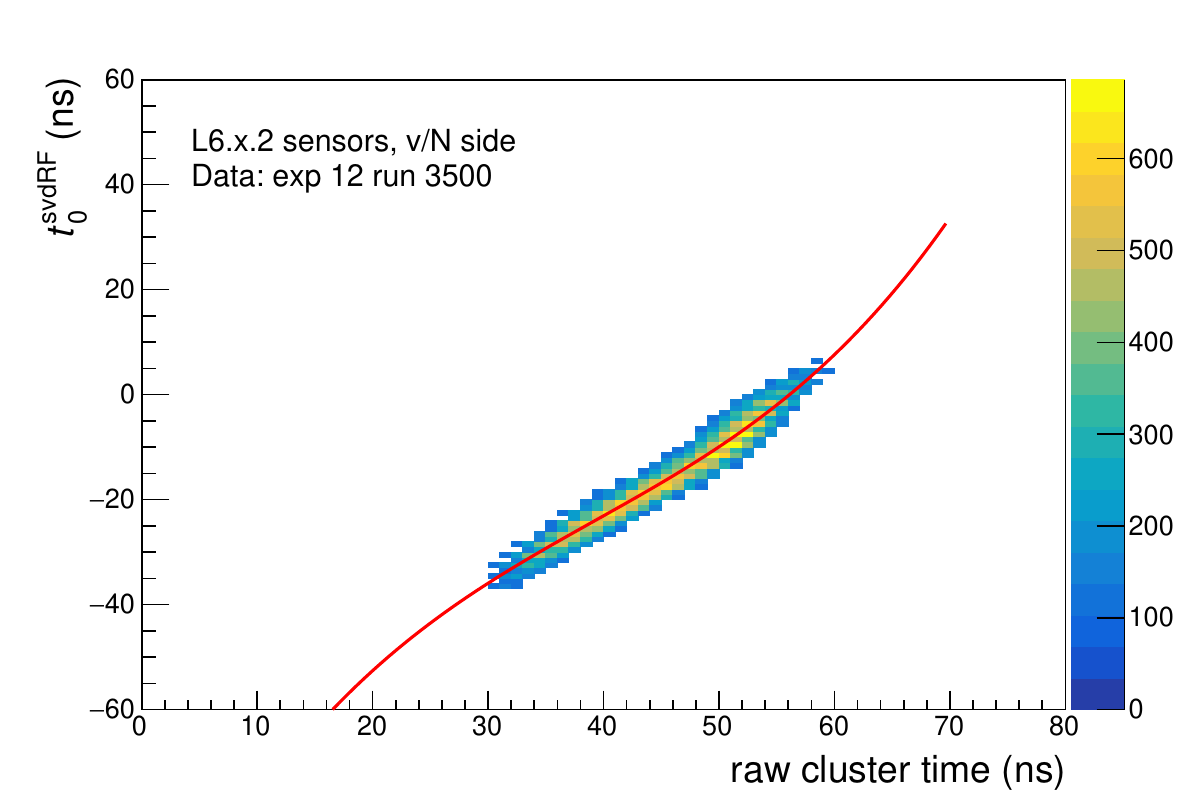} 
    \caption{Example of the $(\tZsvd,\traw)$ correlation plot between the time of the event in the SVD reference frame and the raw cluster time,  used for the SVD hit time calibration. Data of one barrel sensor of layer 6 are shown, u/P side on the left and v/N side on the right, with the calibration function superimposed.%\rem{should change the y label of the plot using \tZsvd }
    }
    \label{fig:C7_CoG_calibration2D}
\end{figure}

\begin{figure}[hbt]
    \centering  
        \includegraphics[width=.48\columnwidth]{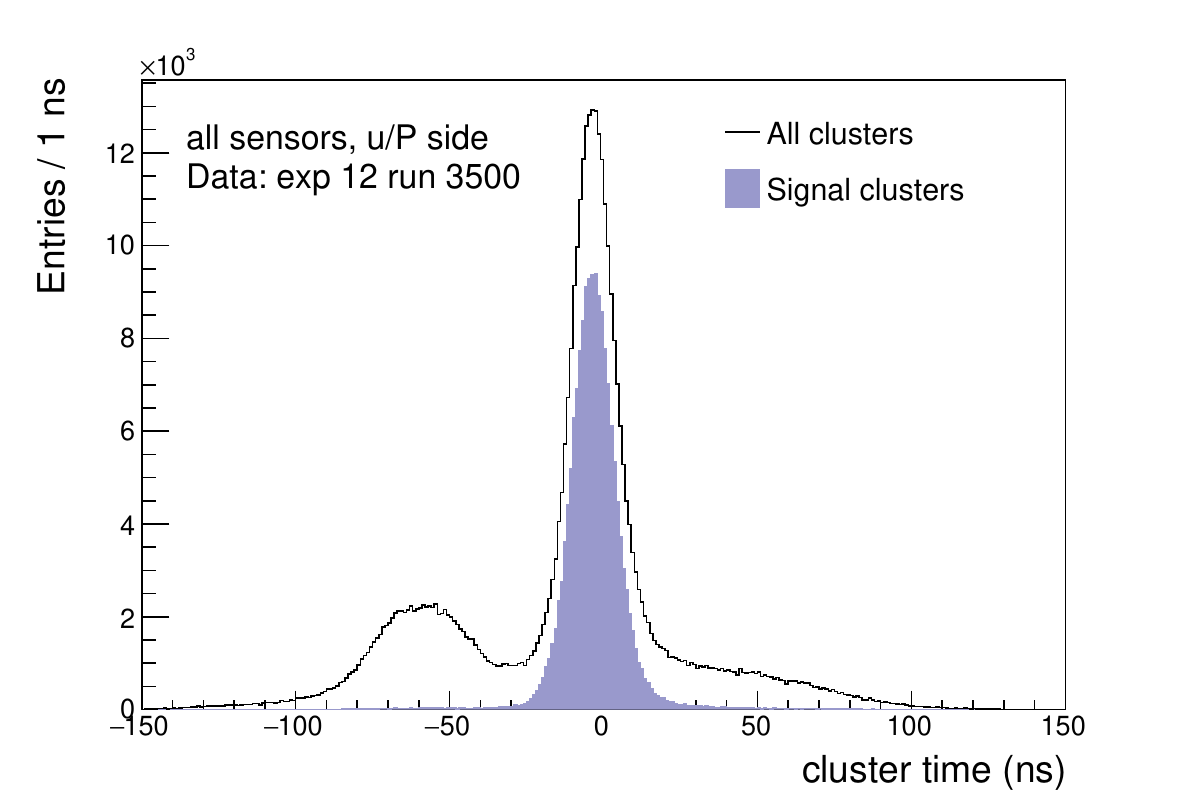}
        \includegraphics[width=.48\columnwidth]{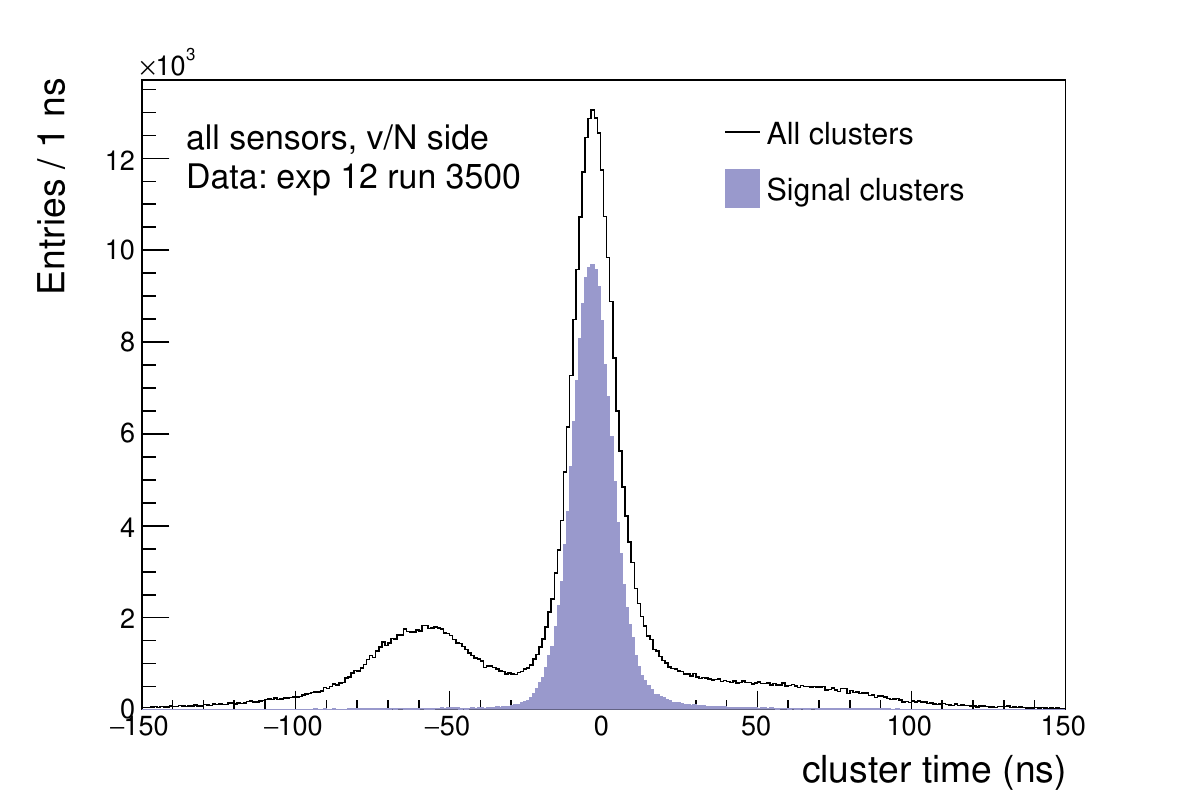} 
    \caption{Cluster time for all clusters and those related to tracks (shaded region) for the  u/P side on the left and v/N side on the right.} 
    %\rem{is this really time - \eventtz\ ? or only cluster time after calibration ? a similar plot is shown in the \ref{fig:C4_clusterTime} and there are the same labels and doesn't say \eventtz\ is subtracted}}
    \label{fig:C7_TimeMinusEventT0}
\end{figure}

The calibration of the cluster time in data is performed with the same procedure used for simulated events (section~\ref{sec:C4_CoGCalibration}), exploiting the correlation between the cluster time and \tZsvd\,  the \eventtz\  expressed in the SVD reference frame. Figure~\ref{fig:C7_CoG_calibration2D} shows an example of the correlation plot used for the calibration of a single sensor: the calibration is performed fitting the correlation between the \tZsvd\ and the raw cluster time with a third-order polynomial. The parameters obtained from the fit are then used to calculate the calibrated time, which is finally shifted in the trigger reference frame according to equation~\ref{eq:C4_calibratedTime}. 
{}

The distribution of the calibrated cluster time  is shown in figure~\ref{fig:C7_TimeMinusEventT0}. The time of the clusters used in tracking (performed without SVD hit time information) nicely peaks at zero, around the \eventtz, while background clusters are more spread in the window. The background peak on the left of the window, between $-100$ and $-50$~ns, is due to off-time clusters hitting the sensor before the beginning of the acquisition window. The shaped signal of the strips has a long tail of few hundreds of ns and the signal is still above threshold when the sampling starts, causing an accumulation of earlier background hits at the beginning of the window.  The rest of the background hits, arriving inside the SVD acquisition window, are more uniformly distributed in time, as expected. The corresponding distributions for simulation are shown in figure~\ref{fig:C4_clusterTime} and have a similar shape. The time separation between signal and background hits can be exploited in the future for background mitigation. 
% \textcolor{red}{is it true? can we produce the signal subtracted cluster time distribution?}.

\begin{figure}[hbt]
    \centering  
    \includegraphics[width=.48\columnwidth]{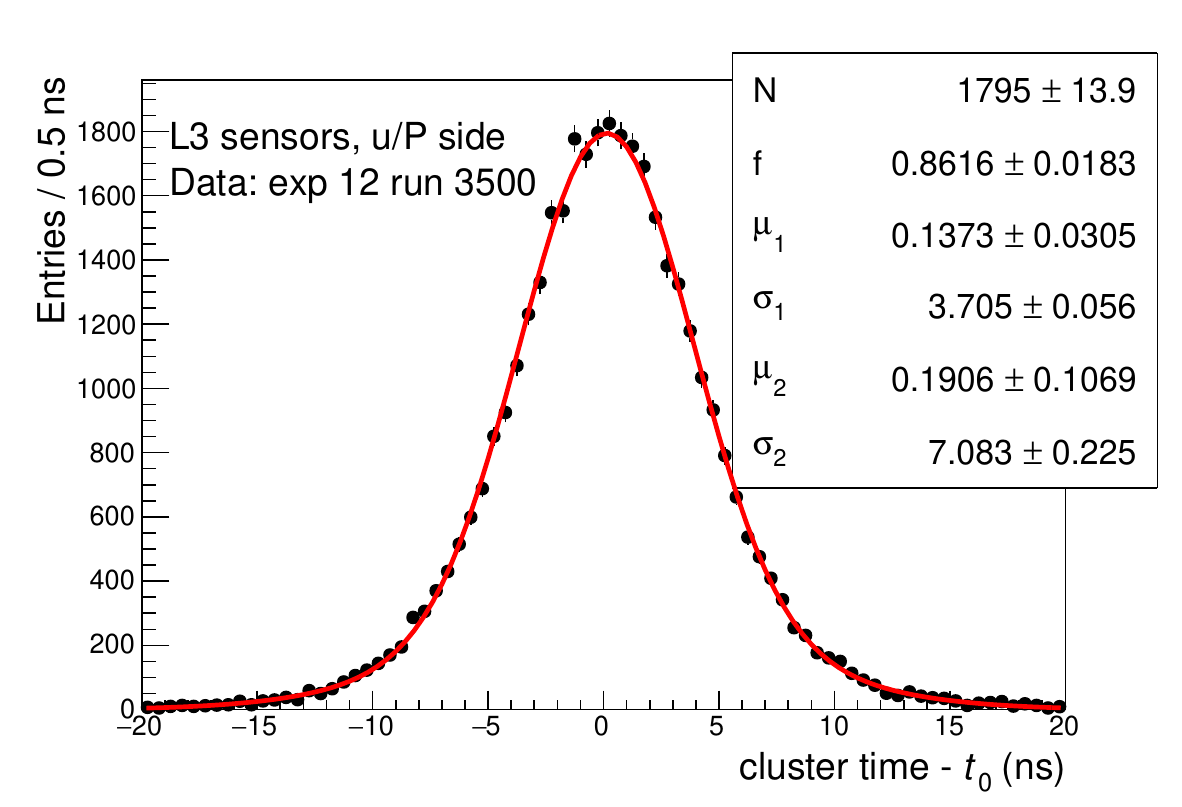}
    \includegraphics[width=.48\columnwidth]{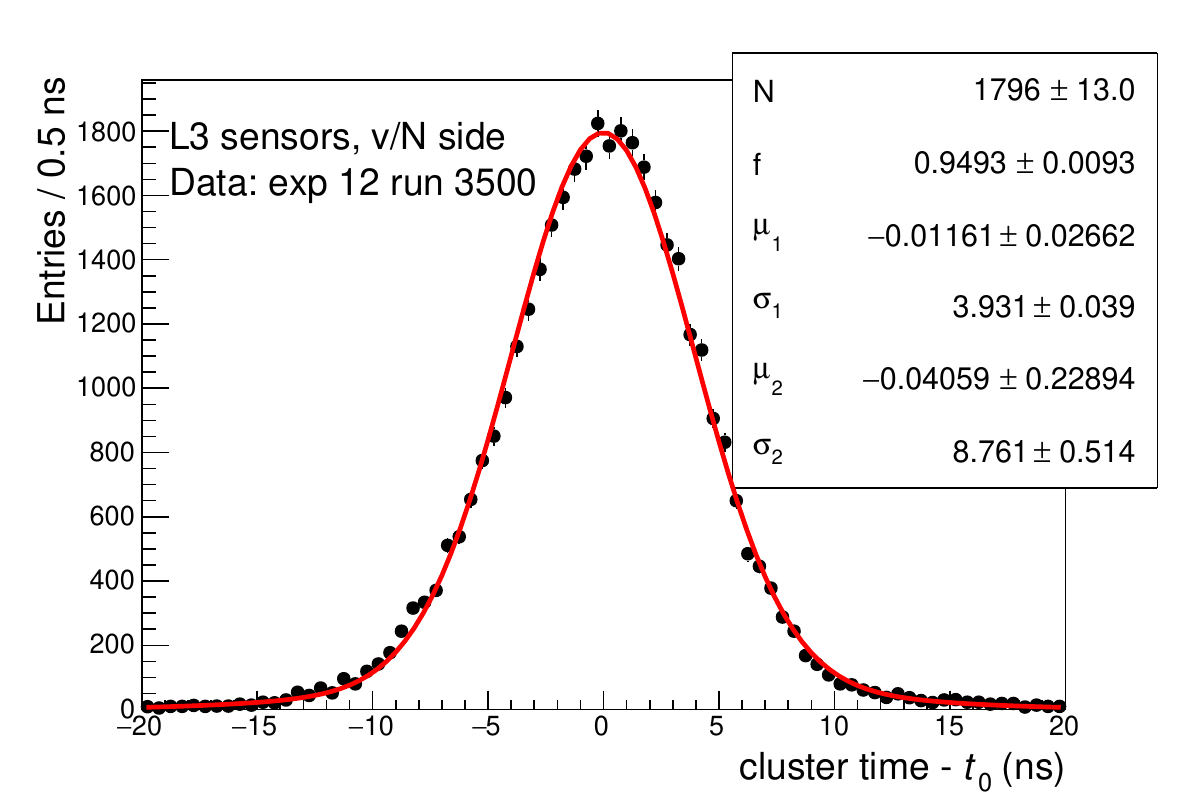}
    \caption{Cluster time - \eventtz\ for clusters related to tracks on layer 3 for the  u/P side on the left and v/N side on the right for data. The resolution on the cluster time, of 4.3 ns on both sides, is estimated as the weighted average of the widths of the two Gaussian functions used in the fit, corrected by the nonzero values of the means of these Gaussians (equation~\ref{eq:C4_timeres}).} 
    \label{fig:C7_resolutionTimeEventT0}
\end{figure}

\begin{figure}[hbt]
    \centering  
    \includegraphics[width=.48\columnwidth]{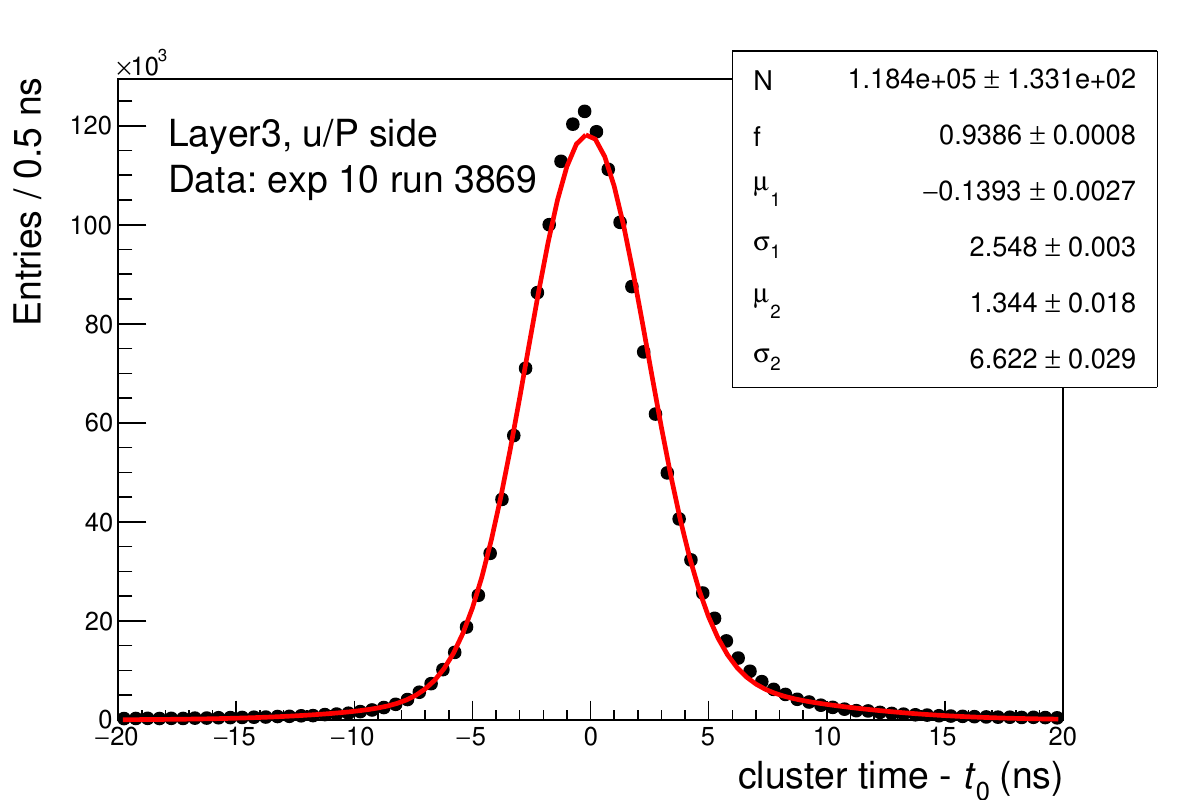}
    \includegraphics[width=.48\columnwidth]{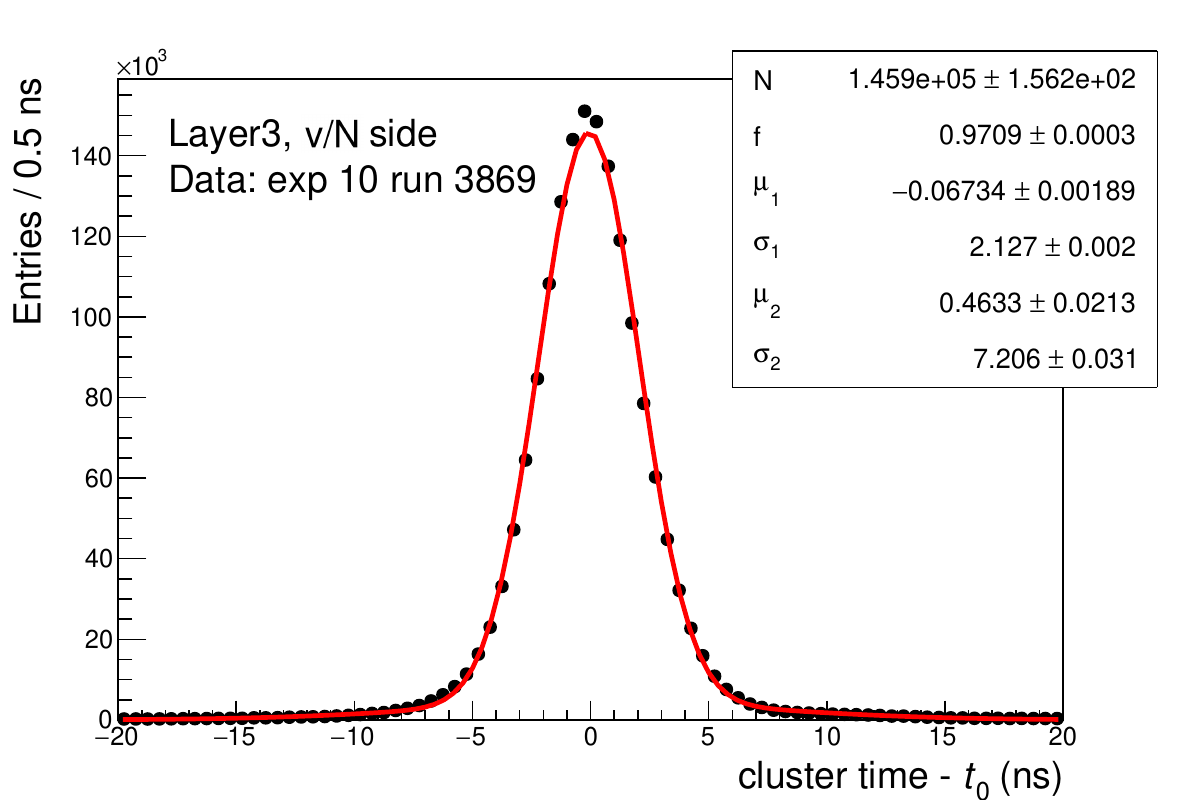}
    \caption{The difference (cluster time - \eventtz) for clusters related to tracks in data, on layer 3 for the  u/P side on the left and v/N side on the right, with the alternative algorithm that uses 3 samples out of the acquired 6. The resolution,  2.4 ns on the v/N side and 2.9 ns on the u/P side, is estimated from the weighted average of the widths of the two Gaussian functions used in the fit, corrected by the nonzero values of the means of these Gaussians (equation~\ref{eq:C4_timeres}).} 
    \label{fig:C7_resolutionTimeCoG3EventT0}
\end{figure}

The resolution of the cluster time can be estimated from the distribution of the difference between the cluster time, for the clusters related to tracks, and the \eventtz,  the time of the event estimated with the CDC, as detailed in equation~\ref{eq:C4_timeres}. Figure~\ref{fig:C7_resolutionTimeEventT0} reports the estimated hit time resolution for layer 3 clusters used in tracking, that currently have the hit time reconstructed using the 6 APV samples. 
The hit time resolution for both sides is 4.3 ns, since  the error on \eventtz\ is negligible ($\simeq 0.7$ ns). There is a small bias of order of 0.1 ns on the u/P side which has no effect on the reconstruction.
The hit time resolution is also estimated using an alternative algorithm, with 3 APV samples out of 6. The results are reported in figure~\ref{fig:C7_resolutionTimeCoG3EventT0}, showing a better resolution of 2.4\,ns on the v/N side and 2.9\,ns on the u/P side. 

\begin{figure}[hbt]
    \centering  
        \includegraphics[width=.48\columnwidth]{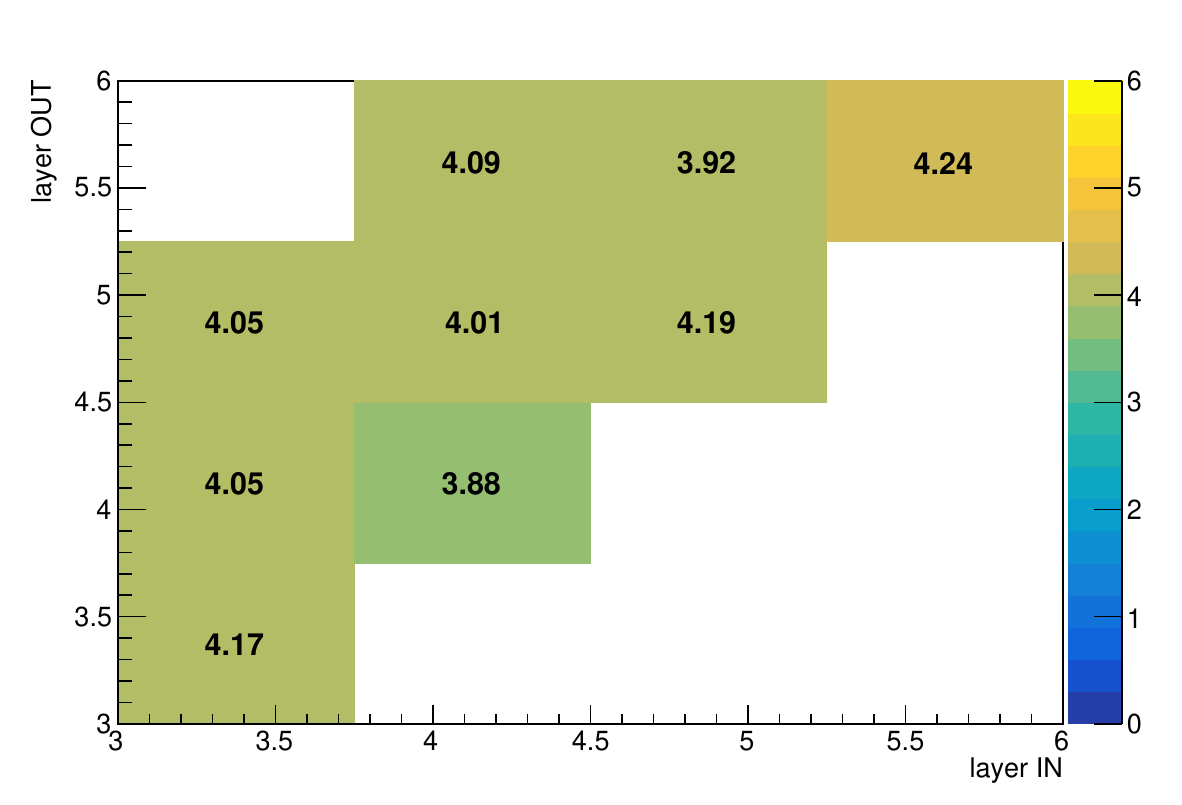}
        \includegraphics[width=.48\columnwidth]{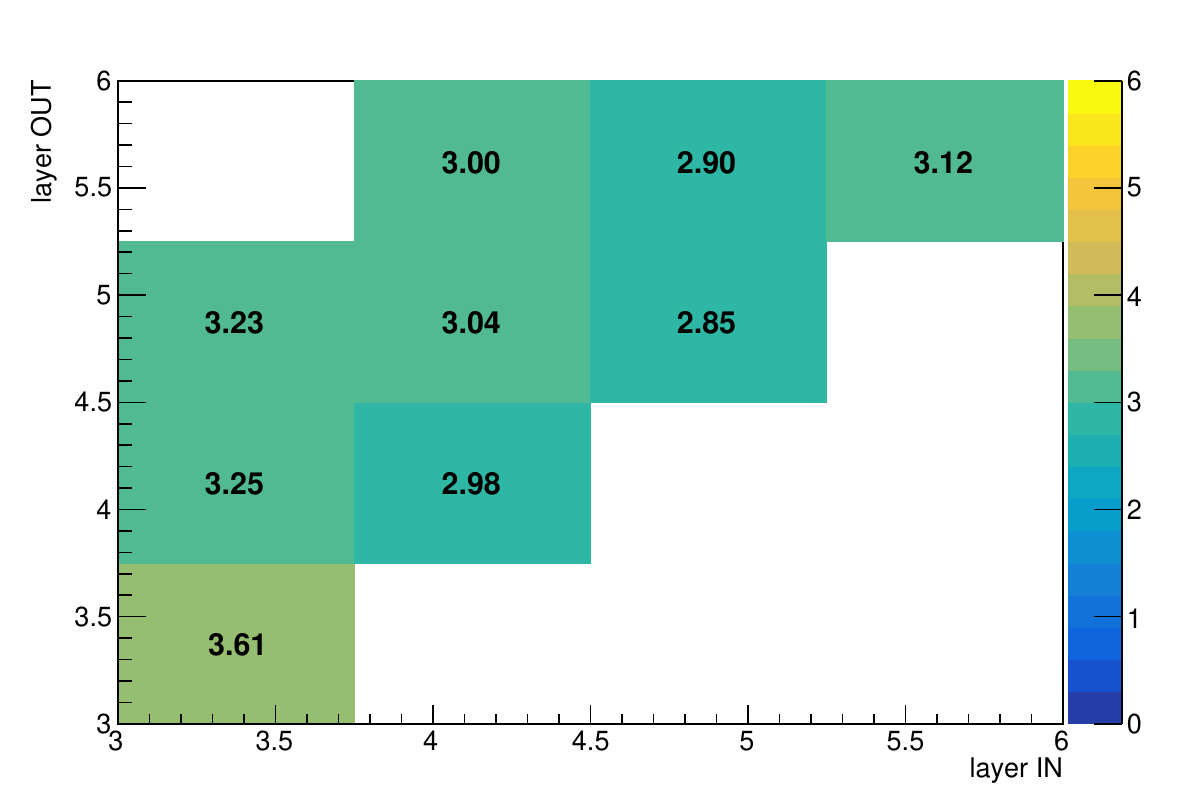} 
    \caption{Cluster time resolution in ns, estimated as the RMS/$\sqrt{2}$ of the time difference, for couples of clusters on different layers belonging to the same track. Rows (columns) represent the outer (inner) layer number whom the cluster belong to. u/P side (left) and v/N side (right).}
    %Cluster time resolution for cluster on different layers connected by a track for the  u/P in (left) and v/N side. Rows (columns) represent the outer (inner) layer number whose the cluster belong to. }
    \label{fig:C7_resolutionTime}
\end{figure}

The resolution of the cluster time can also be estimated independently of the \eventtz. Assuming that the time of all clusters connected to a track is the same,  the resolution is estimated as the RMS/$\sqrt{2}$ of the time difference of two clusters on the same track.
Figure~\ref{fig:C7_resolutionTime}  shows the estimated time resolution using SVD-only information for the 6 sample time estimator. They are compatible with the estimation presented in figure~\ref{fig:C7_resolutionTimeEventT0}.

Calibration parameters in real data may differ for different data sets depending on changes in SVD (for example changes of APV latency), or in external conditions (for example changes of trigger latency). 
Consecutive runs are merged and calibrated as a single data set if the raw hit time average of layer 3 v/N side sensors differs by less than \qty{2}{\ns} or if the statistics of the run would be insufficient to perform a separate calibration. 

\begin{figure}[hbt]
    \centering  
    
    \begin{overpic}[width=\textwidth]{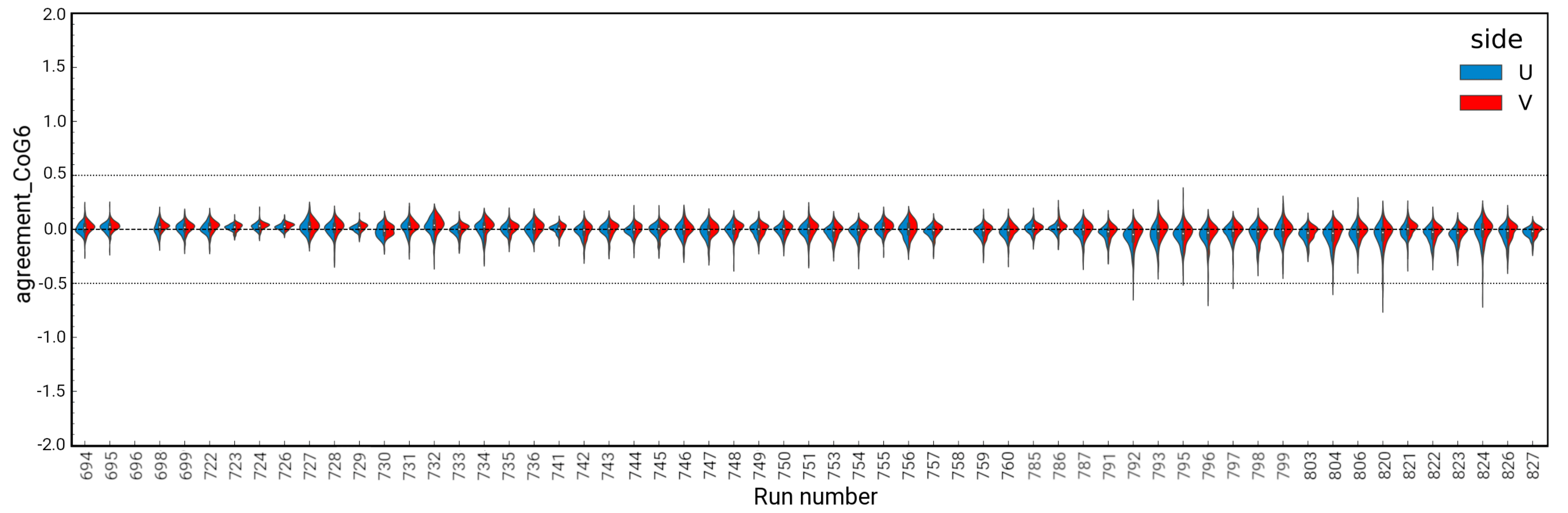}%
    	\put(700,230){\fbox{\includegraphics[width=0.1\textwidth]{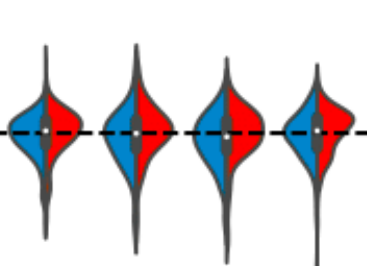}}}%
    \end{overpic}

    \caption{
   % Rappresenta le distribuzioni per ogni sensore di mean(time clusters on tracks - event T0) / RMS(event T0) per lato U (blu) e V (rosso), per clusters associati alle tracce, ed è prodotto dalla validazione automatica della calibrazione. Accettiamo la calibrazione se la media delle distribuzioni è tra -0.5 e 0.5.
    For each run reported on the x axis, the distribution over all sensors of the average of the difference between the calibrated cluster-related-to-tracks time and the \eventtz, divided by the RMS of the \eventtz\ is shown for the u/P (v/N) side sensors in blue (red). The inset shows a blown up image of the distributions for 4 runs, to better illustrate the figure.}
    \label{fig:C7_CoG_stability}
\end{figure}

%The calibration algorithm automatically checks if two (or more) consecutive runs can be merged together or should be calibrated separately by checking the following conditions: 
%\begin{itemize}
%	\item the raw hit time average of layer 3 v/N side sensors has changed by more than 2\,ns;
%	\item the statistical uncertainty is sufficient to perform the calibration. 
%\end{itemize}
%If at least one of the two conditions is false, then the next run is calibrated separately with a new set of parameters.
Figure~\ref{fig:C7_CoG_stability}  shows the stability of the calibrated hit time versus the run number for a data taking period in 2020. The plot, automatically produced in the calibration procedure, shows the distribution over all sensors of the average   of the difference between the calibrated cluster-related-to-tracks time and the \eventtz, divided by the RMS of the \eventtz\  for the two sides separately, for each run. The calibration is considered successful if the average of the distributions is within $\pm0.5$ for each run. As can be seen from the plot, the calibration is extremely stable, and the variation of the average difference between the cluster time and the \eventtz\ is much smaller that the \eventtz\ variation.

\lvltwosection{Alignment}
%\rem{need to implement a few comments and fix picture labels, waiting from expert to answer}
\label{sec:C7_alignment}
\editor{Tadeas, Jakub - 2-3 pages}

For a precise reconstruction of tracks and vertices the position of silicon sensors has to be determined with average precision better than \qty{10}{\um}. The required precision is much higher than what was achieved during detector installation, of order of hundreds of microns. Thus, the positions of the sensors are determined by means of track-based alignment, which can achieve 1--\qty{2}{\um} precision according to MC simulation studies, which do not simulate possible time-dependence. In terms of alignment, the difference between the measured ($m$) and predicted ($p$) positions of a hit $i$ for the track $j$ ($u^{m,p}_{ij}$) is defined as:
\begin{equation}
r_{ij}(\boldsymbol{\tau}_j, \boldsymbol{a}) = u^m_{ij}-u^p_{ij}(\boldsymbol{\tau}_j, \boldsymbol{a}),
\end{equation}
where $\boldsymbol{\tau}_j$ is a vector of track helix parameters and $\boldsymbol{a}$ is a vector of alignment parameters. Alignment parameter corrections $\delta \boldsymbol{a}$ are computed via minimisation of a $\chi^2$ function linearized around initial estimates of track and alignment parameters ($\boldsymbol{\tau}_j^0, \boldsymbol{a}^0$):
\begin{equation}
\chi^2(\boldsymbol{\tau}, \boldsymbol{a}) = \sum_j^{\rm tracks} \sum_i^{\rm hits} \left( \frac{r_{ij}(\boldsymbol{\tau}_j, \boldsymbol{a})}{\sigma_{ij}} \right)^2 \approx \sum_j^{\rm tracks} \sum_i^{\rm hits} \frac{1}{\sigma_{ij}^2} (r_{ij}(\boldsymbol{\tau}_j^0, \boldsymbol{a}^0)+\frac{\partial r_{ij}}{\partial \boldsymbol{a}} \delta \boldsymbol{a} + \frac{\partial r_{ij}}{\partial \boldsymbol{\tau}_j} \delta \boldsymbol{\tau}_j)^2,
\end{equation}

%\begin{equation}
% \chi^2(\boldsymbol{\tau}, \boldsymbol{a}) = \sum_j^{tracks} \sum_i^{hits} \left( \frac{r_{ij}(\boldsymbol{\tau}_j, \boldsymbol{a})}{\sigma_{ij}} \right)^2 \approx \sum_j^{tracks} \sum_i^{hits} \frac{1}{\sigma_{ij}^2} [(r^2_{ij}(\boldsymbol{\tau}_j^0, \boldsymbol{a}^0)+2r_{ij}(\boldsymbol{\tau}_j^0, \boldsymbol{a}^0)(\frac{\partial r_{ij}}{\partial \boldsymbol{a}} \delta \boldsymbol{a} + \frac{\partial r_{ij}}{\partial \boldsymbol{\tau}_j} \delta \boldsymbol{\tau}_j)]
%\end{equation}

where $\sigma_{ij}$ denotes the uncertainty of measurement of hit $i$ within track $j$. 

The solution is obtained by a simultaneous minimization with respect to all track parameters (order of millions) and alignment parameters (about three thousand for VXD alignment) using the Millepede II \cite{Blobel:2002new} algorithm. The simultaneous fit ensures all correlations among alignment parameters are properly taken into account. With a track sample in a broad topology (cosmic tracks combined with tracks from $e^+e^-$ collisions), the "weak modes" (poorly controlled systematic detector deformations like twist) can be sufficiently constrained. With the full Belle II detector, the VXD alignment is performed simultaneously with the alignment of CDC layers.

\begin{figure}[!hbt]
\centering
\includegraphics[width=1.0\linewidth]{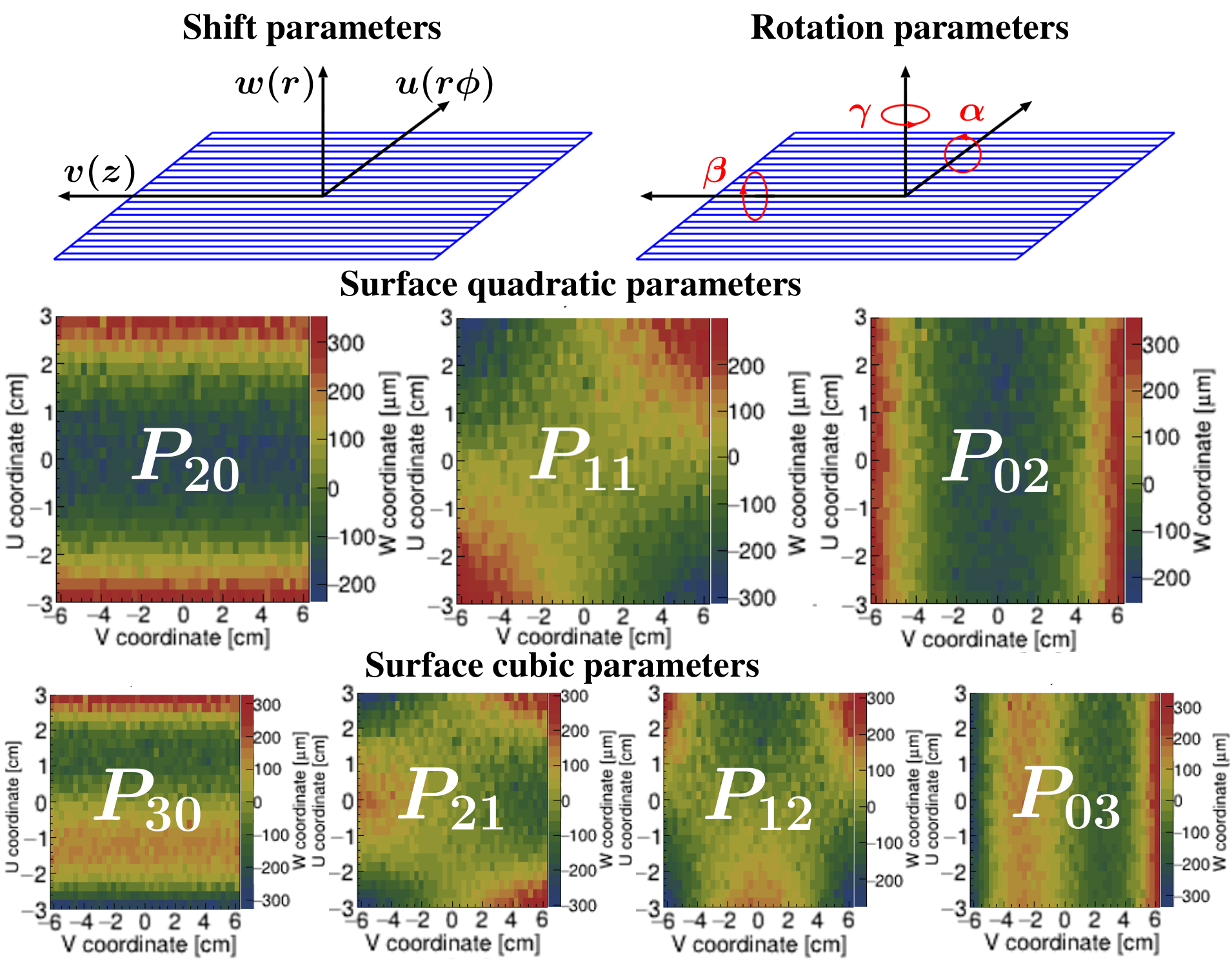}
\caption{Sensor alignment parameters: shift and rotation parameters in the local coordinates system, defined in section~\ref{sec:C4_coordinates}, and visualisation of the base Legendre polynomials corresponding to quadratic and cubic surface parameters. Adapted from~\cite{Kandra:2019qlz}.}
\label{fig:C7_alignment_sensor}
\end{figure}

The position and orientation of individual sensors (and ladders or half-shells) is characterized by six rigid body parameters: three shifts ($u$, $v$, $w$) and three rotations ($\alpha, \beta, \gamma$). To accommodate sensor deformation, their surface is described using two-dimensional Legendre polynomials with three quadratic parameters (${P_{02}, P_{11}, P_{20}}$) and four cubic parameters (${P_{03}, P_{12}, P_{21}, P_{30}}$). 
%and five quartic parameters ($\mathrm{P_{04}, P_{13}, P_{22}, P_{31}, P_{40}}$). 
The sensor alignment parameters are illustrated in figure~\ref{fig:C7_alignment_sensor} (see also figure~\ref{fig:C2_local_coordinates}).

\begin{figure}[!hbt]
\includegraphics[width=0.335\linewidth, trim={140mm, 0mm, 145mm, 0mm}, clip]{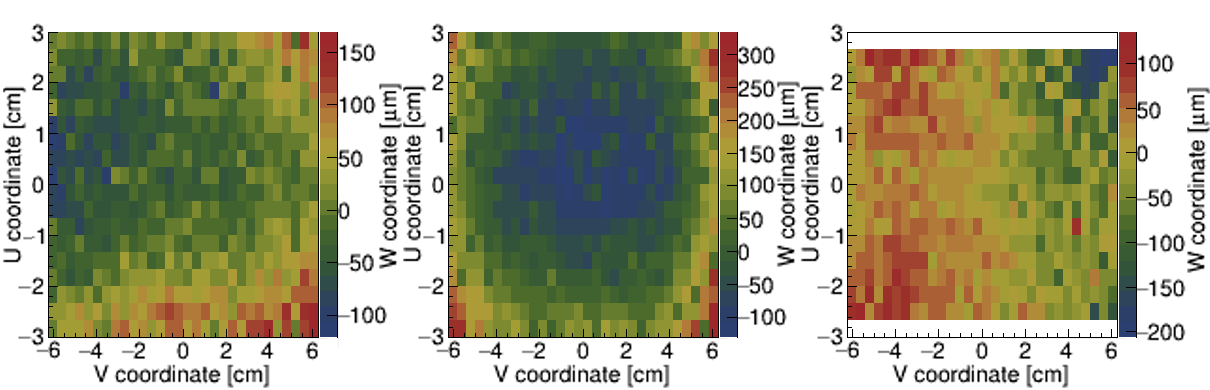}
\includegraphics[width=0.62\linewidth]{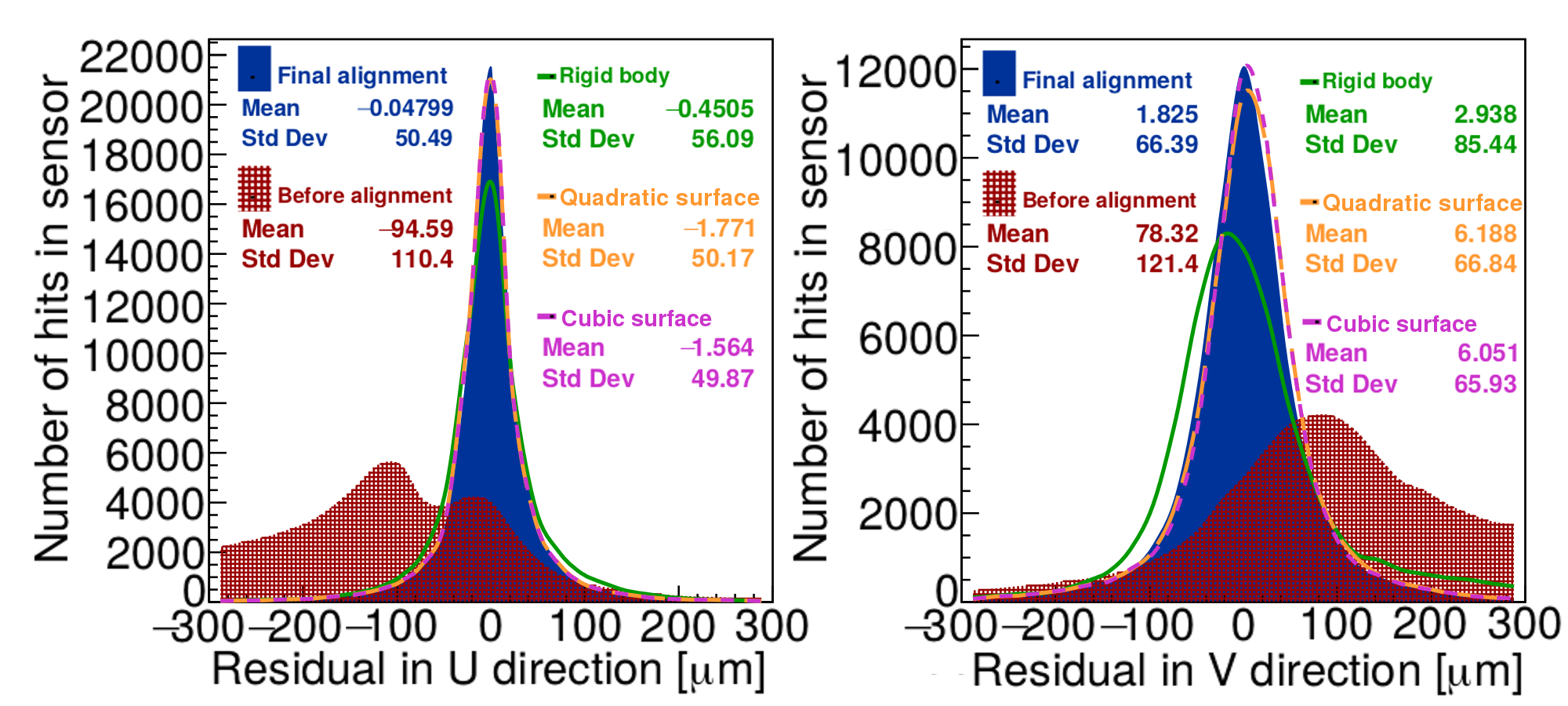}

\caption{Deformation along the $w$ coordinates in \unit{\um} identified by the monitoring algorithm (left), residual distributions of L4.1.2 sensor during alignment tests (center and right): residual distribution before applying alignment procedure (red), applying rigid body parameters (green), applying quadratic surface parameters (orange), applying cubic surface parameters (violet) and after the second iteration of alignment procedure (blue). Adapted from~\cite{Kandra:2019qlz}.}
\label{fig:C7_alignment_surface}
\end{figure}

The alignment procedure was performed and validated during several stages of detector construction and installation, leading to several optimizations. Cosmic data samples were studied during stand-alone SVD and combined PXD+SVD commissioning runs. Following the VXD installation, both the cosmic tracks and tracks from $e^+e^-$ collisions were processed for alignment purposes. At the beginning of the tests, only the rigid body parameters were considered. But validation algorithms indicated substantial effects caused by deformations (figure \ref{fig:C7_alignment_surface}) of origami sensors (middle-ladder sensors of layer 4, 5 and 6). The reconstruction framework was extended to correct for deformation effects, and the quadratic and cubic surface parameters were introduced into the alignment procedure. Their final impact on the residual distribution for one of the used sensors can be found in figure \ref{fig:C7_alignment_surface}.

\begin{figure}[!hbt]
  \centering
  \includegraphics[width=15.0cm, trim={0mm, 0mm, 0mm, 15mm}, clip]{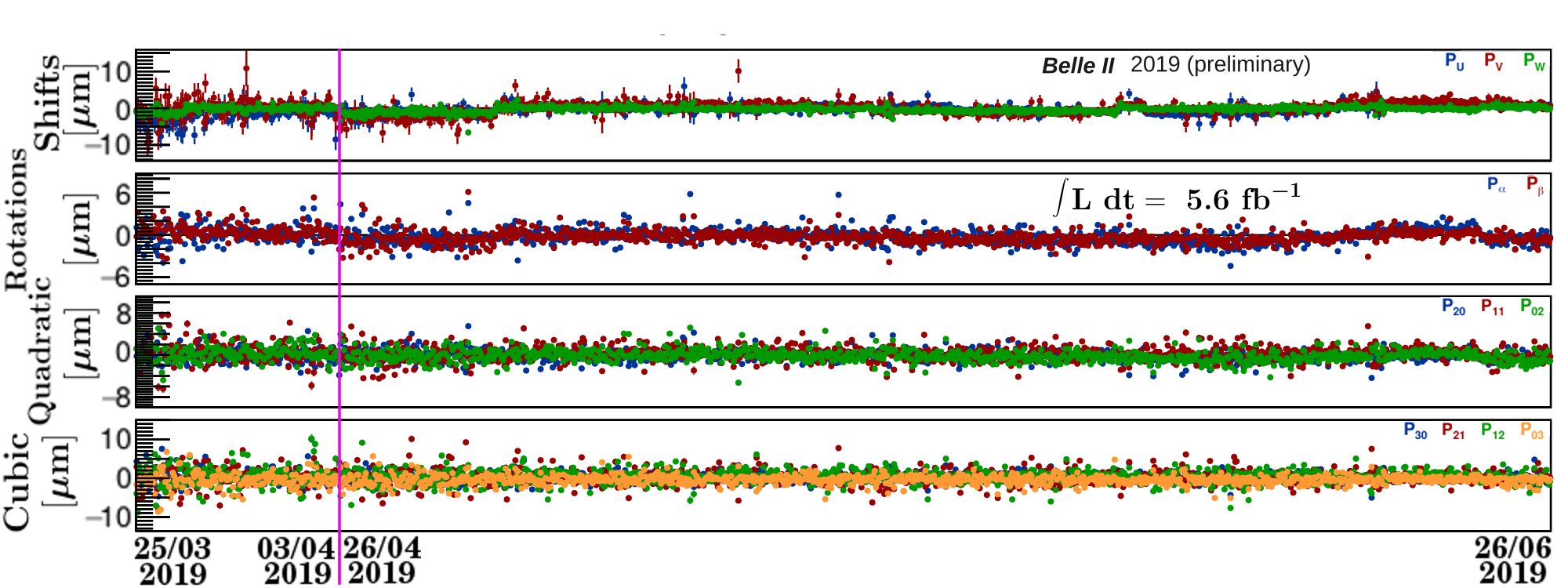}
  \caption{Time-dependent alignment validation of sensor 4.3.2: shift (first line), rotation (second line), quadratic surface (third line), and cubic surface (fourth line) parameters. Because of validation procedure \cite{Kandra:calibration} only the rotation parameters $\alpha$ and $\beta$ (see text) can be validated. During these data taking periods 
  %\textcolor{Magenta}
  a fire accident break happened, shown with a magenta vertical line. The units of alignment parameter axes are \unit{\um} \cite{Kandra:monitoring} -- for cubic and quadratic parameters, this corresponds to the (deviation of) amplitude of the sensor deformation; for angles it is the displacement at sensor edge corresponding to the rotation}
  \label{fig:C7_alignment_validation}
\end{figure}

In order to verify and monitor time dependence of alignment parameters, a dedicated validation procedure was developed \cite{Kandra:calibration, Kandra:k2019b}. During the spring and summer 2019 data taking periods, Belle~II collected about 5.6 fb$^{-1}$ in 1200 runs. Within these periods all the alignment parameters were validated and monitored. An example of validation plot can be found in figure \ref{fig:C7_alignment_validation} -- the alignment parameters fluctuate within a range of $\pm  \qty{10}{\um}$.

\lvltwosection{Track Finding and Reconstruction Performance}
%\rem{need substantial rework}
\label{sec:c9_track_reconstruction}
\editor{Eugenio}

The SVD information is necessary to improve the overall track finding efficiency as well as the determination of track parameters at the \acrshort{IP}. 

The impact parameter resolution is dominated by the hit closest to the IP, provided by \acrshort{PXD}. 
However the SVD information plays a crucial role in extrapolating \acrshort{CDC} tracks to the PXD sensors with the required precision to pick up the correct hit  and also to perform an efficient on-line PXD data reduction, as detailed below. 
In fact, the uncertainty in the extrapolation of a charged particle to the PXD sensors provided by the SVD is smaller by almost one order of magnitude with respect to the one provided by CDC  on the transverse plane and by two orders of magnitudes along the beam axis. Thus the SVD information is necessary to correctly identify the clusters on the PXD belonging to the particle track, that are needed to provide the ultimate resolution on the impact parameters. 
The SVD information also improves the measurement of the particle transverse momentum. The disadvantage of a smaller lever arm of the SVD with respect to the CDC is  counterbalanced by its better spatial resolution.

Importantly, low transverse momentum tracks, unable to reach the CDC, can be reconstructed using SVD-only information. This is fundamental, for example, to reconstruct the slow pions $\pi_s^+$ from $B \to \Dstarp X$, $\Dstarp \to \Dz \pips$.

\begin{figure}
    \centering
    \includegraphics[width=0.7\columnwidth]{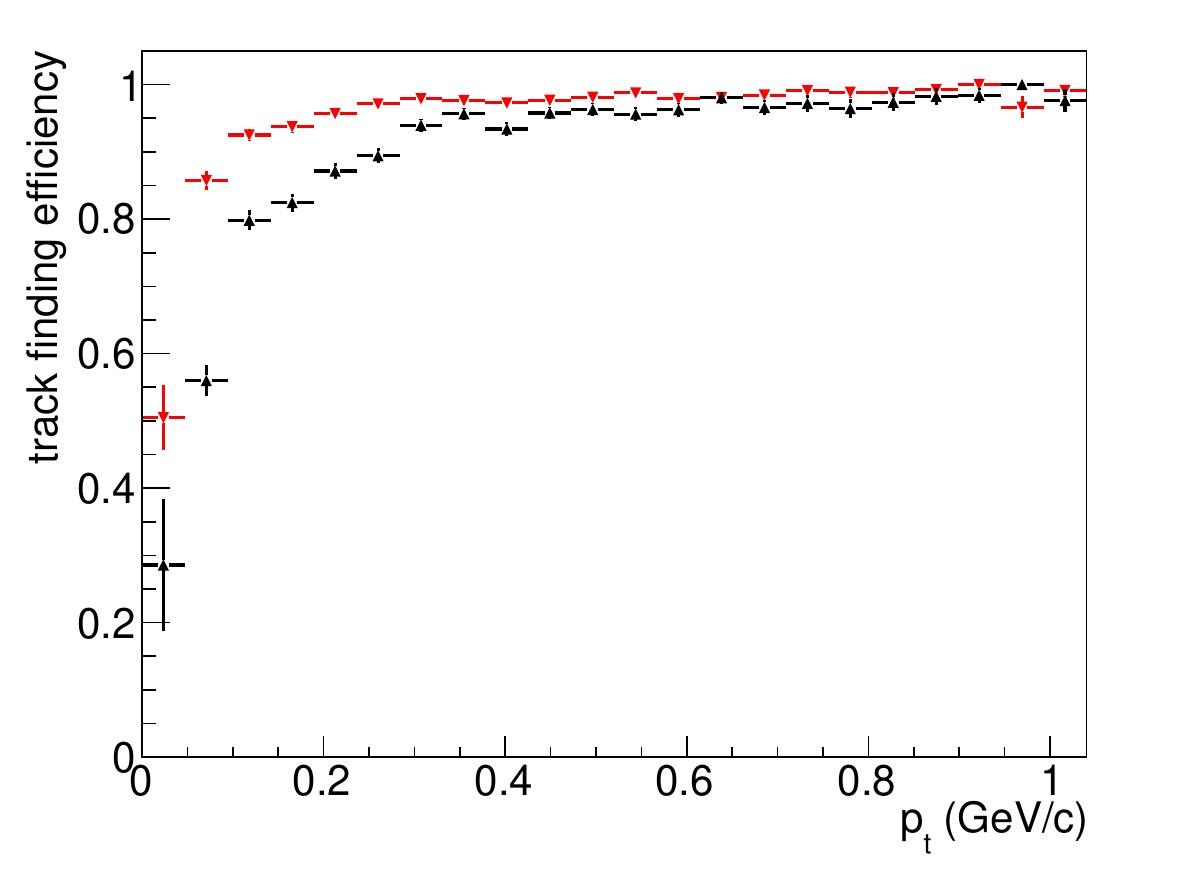}
    \caption{Track finding efficiency as a function of transverse momentum for the CDC-only tracking (black) and the full tracking with SVD information (red) estimated using simulated generic \BB\ events.}
    \label{fig:C7_trackFindingEfficiency}
\end{figure}

At present, the track finding strategy starts the search from the CDC.
Each track found among the CDC hits is extended inward the IP.
A combinatorial Kalman-filter is used to identify the best SVD clusters compatible with the extrapolated positions on the sensor. The SVD clusters are attached to the track candidate identified by the CDC to improve the determination of its transverse and longitudinal impact parameters, as well as of momentum vectors at the point of closest approach to the IP.
The SVD is then used as a stand-alone track finding detector on the remaining clusters not associated to CDC tracks. Its four-layer structure together with its proximity to the IP is particularly valuable  to identify and reconstruct particles in the low transverse momentum  region $ p_t < 0.3 \gev/c$. In this $p_t$ range the CDC stand-alone track finding efficiency is around 95\% at $0.2 \gev/c$ then it gradually degrades to 80\% at $p_t \sim 0.1 \gev/c$ below which it quickly drops down to zero, as shown in figure~\ref{fig:C7_trackFindingEfficiency} \cite{Bertacchi:2020eez},  because of the limited number of hits left by the particle in the CDC. 

A stand-alone track finder \acrfull{VXDTF2} \cite{Bilka:2017zrp}\cite{FruHwirth:2013jta} had been developed to improve the overall track finding efficiency as well as to have a track finder orthogonal to the CDC finder for cross-checks and for the study of systematic effects.  
The SVD together with the VXDTF2 brings the track finding and fitting efficiency above 87\% for $p_t \sim 0.1 \gev/c$ for all particles in \BB\ events (see figure~\ref{fig:C7_trackFindingEfficiency}).  
%primary particles from the IP and nominal full luminosity background.
Since the angular acceptance of the SVD exceeds the CDC one by a few degrees, a better coverage in the very forward and backward region is also obtained as an additional bonus \cite{Bertacchi:2020eez}.

The VXDTF2 \cite{Bilka:2017zrp} gradually builds up the track candidates starting from the SpacePoints whose reconstruction is described in section \ref{sec:C4_clustering}. Pairs of SpacePoints, whose positions are sufficiently close to the trajectory of a charged particle from the IP, are combined in a segment. Pairs of segments that have the same central SpacePoint and that are reasonably close to an helix from the IP are then combined in a triplet. The triplets having a common segment are joined using a cellular automaton~\cite{FruHwirth:2013jta}  that provides a set of raw track candidates. At this stage, each SpacePoint can be associated to several track candidates and the trajectory of a given particle is represented by a set of several track candidates. Finally, each SpacePoint is associated to the single best track candidate identified by a figure-of-merit based on the track to clusters residual, time and energy. 
The VXDTF2 can exploit the cluster time information by rejecting segments and triplets whose space points are too far away in time. This feature is particularly useful to reduce the combinatorial burden produced by random combinations of background hits.  At present it is leveraged only on the reconstruction of MC simulations at full luminosity and background.
The algorithm is fast enough to be used on-line at the \acrfull{HLT}. The tracks found by the VXDTF2 are then extended outward into the CDC using a combinatorial Kalman-filter.
A final combinatorial Kalman-filter adds the PXD clusters to the tracks found in the SVD.

An accurate extrapolation of the reconstructed tracks inside the PXD volume,  made possible with the SVD, is also crucial to predict the position at which the particle crosses a PXD sensor and to define a rectangular \acrfull{ROI} around that position to perform an efficient on-line PXD data reduction. 
Given the long integration time of the PXD and its proximity to the IP, the data rate from PXD only, dominated by the background component, could be as high as 10 GB/s, 10 times the rate of the other Belle II subdetectors combined~\cite{Spruck:2020A0}. A PXD data reduction is mandatory at full luminosity in order to reduce to a manageable level the size of the data storage required by the PXD data stream.
The background contribution of the PXD can be heavily reduced at an early stage of the data-flow by discarding PXD clusters that are outside all ROIs. The ROI are defined on the HLT and sent to the \acrfull{ONSEN} subsystem,
which selects the PXD clusters belonging to the ROI and sends them to the Event Builder.

The viability of the PXD data reduction had been demonstrated during several test beams at DESY \cite{Bilka:2014lla} and was tested during several runs of Belle II data-taking, demonstrating that the required data reduction of a factor 10 is reachable. 
%\rem{ask PXD if more recent numbers on ROI results are available from Belle 2 runs and the reference to a good paper}
A second PXD data reduction scheme is at present under test and development~\cite{Bernlochner:2017mtp}. This scheme exploits an FPGA-based SVD stand-alone track finder system called DATCON to define the \acrshort{ROI} and send them to the ONSEN \cite{Gessler:2014gba}.

%\rem{add pi soft effi results from data }

%\begin{itemize}
%%    \item Track reconstruction on the HLT (now and ixn the near future)
 %   \item ROI
 %   \item Ks reconstruction
 %   \item VXDTF2 track finding efficiency plots (MC and data studies)
%\end{itemize}

\lvltwosection{\texorpdfstring{\boldmath $dE\!/\!dx$}{dE/dx} Performance}
\label{sec:C9_dEdx}
%\editor{Gagan}

Particle identification plays an important role in the physics program of Belle II.
For instance, a better identification of the slow pion ($\pi_{\rm s}$) would improve the precision at which the lepton-flavor-universality ratio $R(D^{*})$ can be measured in the decay channel $B\to D^{*}\ell\nu$, $D^{*}\to D\pi_{\rm s}$.
Such low-momentum particles ($p_{\rm T}\approx 65\mevc$) are often unable to reach the CDC, the main tracking device of the experiment, owing to their small radius trajectory.
Our goal here is to exploit specific ionization ($dE\!/\!dx$) by these particles in the SVD towards identifying them.
Even if the charged particles have a higher $p_{\rm T}$ enabling them to reach the CDC, the $dE\!/\!dx$ values measured in the SVD can still provide complementary information towards their identification.

According to the Bethe-Bloch formula~\cite{Leo:1987kd}, the $dE\!/\!dx$ of a charged particle traversing through detector material only depends on its $\beta\gamma$ value.
Thus, if the $dE\!/\!dx$ for various types of particles {\it vs.} their momentum ($\beta\gamma mc$) is plotted, the difference in the resulting two-dimensional distributions can be used to distinguish these particles in the SVD.
Of particular importance is the charged pion-kaon separation performance of the SVD. 
%which is the focus of our subsequent discussion.

The self-tagging decay $D^{*+}\to D^{0}(\to K^{-}\pi^{+})\pi^{+}$ and its charge conjugate process is used to develop the $dE\!/\!dx$ particle identification method. Our study is based on a Belle~II data sample of $3.1\invfb$ recorded near the $\Y4S$ resonance and the obtained results are compared with that of an MC sample.
To reconstruct $D^{*+}\to D^{0}(\to K^{-}\pi^{+})\pi^{+}$ decay candidates,  all possible combinations of three tracks in an event, requiring exactly two of them to have same charge, are considered.
One of the two same-charge particles is taken to be the slow pion candidate, if its momentum is less than $1.0\gevc$.
The remaining two tracks are assumed to be the kaon and pion candidates from the $D^{0}$ decay, where the latter is required to have the same charge as the ``slow-pion" candidate.
The reconstructed $D^{0}$ mass is required to lie between $1.85$ and $1.88\gevcc$, corresponding to about a $\pm3\sigma$ window around the peak, where $\sigma$ is the mass resolution.

\begin{figure}[!hbt]
    \centering
    \includegraphics[width=0.45\columnwidth]{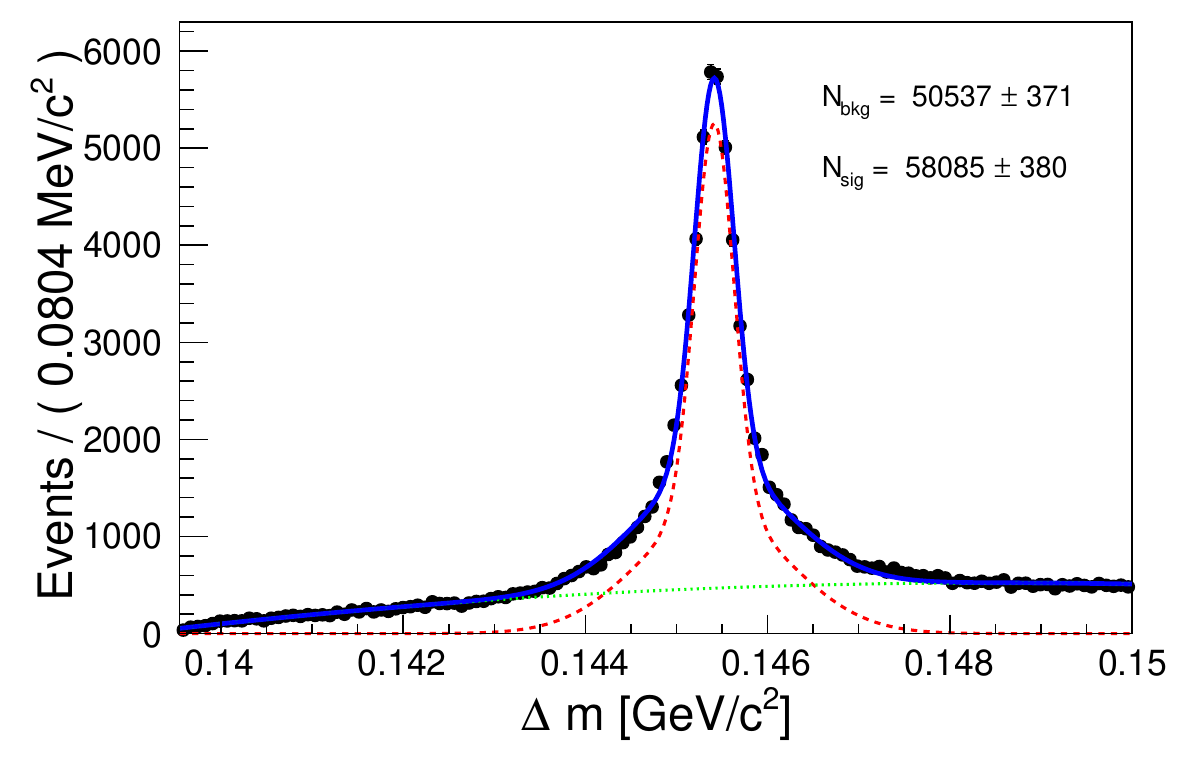}
    \includegraphics[width=0.45\columnwidth]{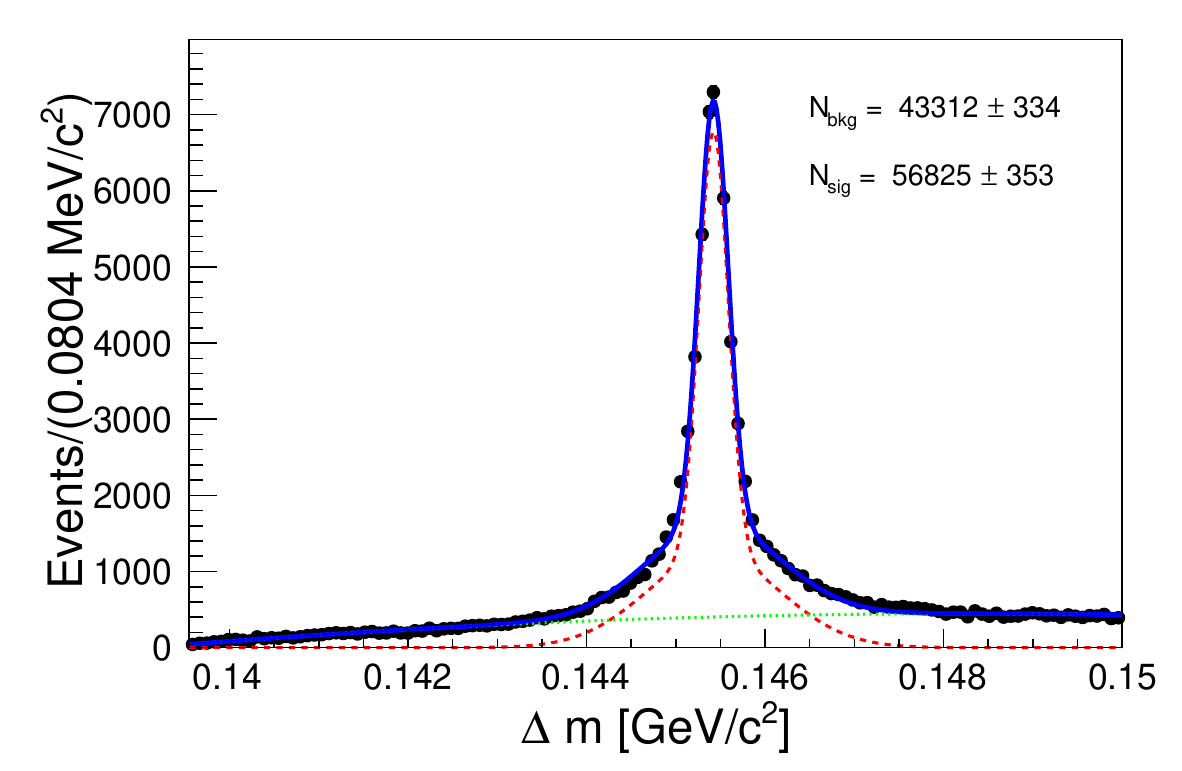}
    \caption{Results of the $\Delta m$ fit performed on the data (left) and MC (right) sample. Points with error bars are data, solid blue curves are the total fit results, dashed red and dotted green curves denote the signal and background component, respectively~\cite{Hazra:2021myl}. }
    \label{fig:C7_dEdx_Fig1}
\end{figure}

The signal and background shapes in the $D^{*}$ and $D^{0}$ mass difference ($\Delta m$) are modeled with a sum of two Gaussians and a threshold function, respectively.
The latter is given by
\begin{equation}
f(\Delta m)= c\left(\Delta m-m_{\pi}\right)^{a} e^{-b \left(\Delta m-m_{\pi} \right)},
\end{equation}
where $m_{\pi}$ is the nominal mass of the charged pion, and $a$, $b$, and $c$ are shape parameters.
Figure~\ref{fig:C7_dEdx_Fig1} shows results of the $\Delta m$ fit to the data and MC sample.
There is a qualitative data-MC agreement having a somewhat larger background contribution and broader signal width in data. 

\begin{figure}[!hbt]
    \centering
    \includegraphics[width=0.48\columnwidth]{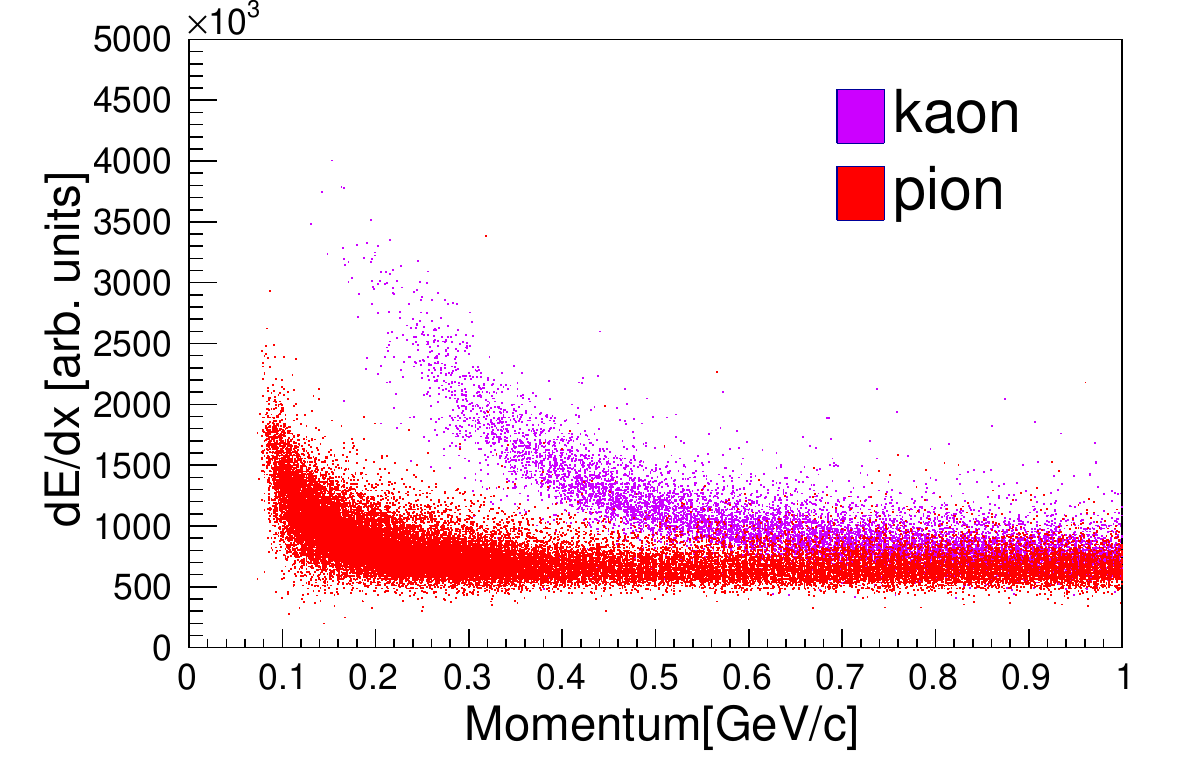}
    \includegraphics[width=0.48\columnwidth]{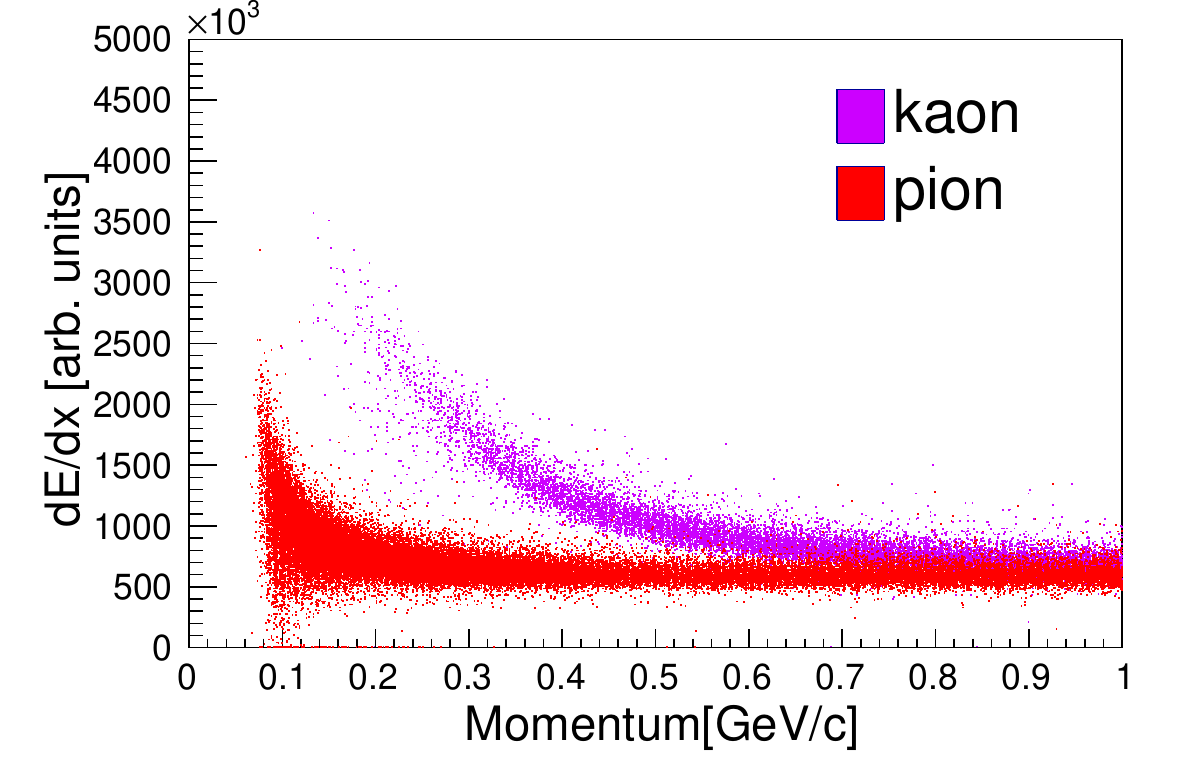}
    \caption{Scatter plot of truncated $dE\!/\!dx$ value for pions and kaons as a function of their momentum in the data (left) and MC (right) sample~\cite{Hazra:2021myl}.}
    \label{fig:C7_dEdx_Fig2}
\end{figure}

Various background-subtracted distributions obtained using the $\splot$ method~\cite{Pivk:2004ty} are then studied.
Slow pions are found to have an average momentum of $500\mevc$, while the momentum spectra for $D^{0}$ daughter pions and kaons extend up to $5\gevc$, 
and are therefore used  for building probability density functions (PDFs) above $500\mevc$.
%, normalizing the cluster signal to the path lendtgh traversed in the silicon,
The $dE\!/\!dx$ value obtained from each SVD hit  for a given charged particle follows a Landau distribution~\cite{Landau:1944if}.
The hit-level $dE\!/\!dx$ information is then combined to reconstruct the track-level $dE\!/\!dx$ value.
It turns out that a simple arithmetic mean would be insufficient as the mean of the Landau distribution is undefined.
Truncation must be applied to the individual hit-level $dE\!/\!dx$ measurements while combining them to obtain the track-level $dE\!/\!dx$ value, also reducing the effect of fluctuations in energy loss.
When a particle traverses the entire SVD,  at least eight $dE\!/\!dx$ values are obtained: four each for the u/P and v/N side.
To truncate $dE\!/\!dx$,  the two highest energy clusters are excluded, since they are expected to come from the same wafer and to be fully correlated, making the removal of just one hit less effective. 
Figure~\ref{fig:C7_dEdx_Fig2} shows the scatter plot of truncated $dE\!/\!dx$ values for pions and kaons as a function of their momentum in the data and MC sample.

This provides us a good motivation to construct a two-dimensional PDF ${\cal P}_{m}$ for the $\left[\left(dE\!/\!dx\right),p\right]$ pair for a likelihood-based particle identification method, where the index $m$ stands for $\pi$ and $K$.
A likelihood function for each particle hypothesis $m$ is constructed: ${\cal L}_{m}\left(dE\!/\!dx,p\right)=\prod_{i}{\cal P}_{m}\left[\left(dE\!/\!dx\right)_{i},p\right]$, with $i$ running over all $dE\!/\!dx$ values assigned to a track.
The $dE\!/\!dx$ calibration exercise was performed with the control sample of $D^{*+}\to D^{0}(\to K^{-}\pi^{+})\pi^{+}$.
Finally, it was  verified whether the particle identification performance is improved compared to the case where the SVD $dE\!/\!dx$ information is not used.

\begin{figure}[!hbt]
    \centering
    \includegraphics[width=0.48\columnwidth]{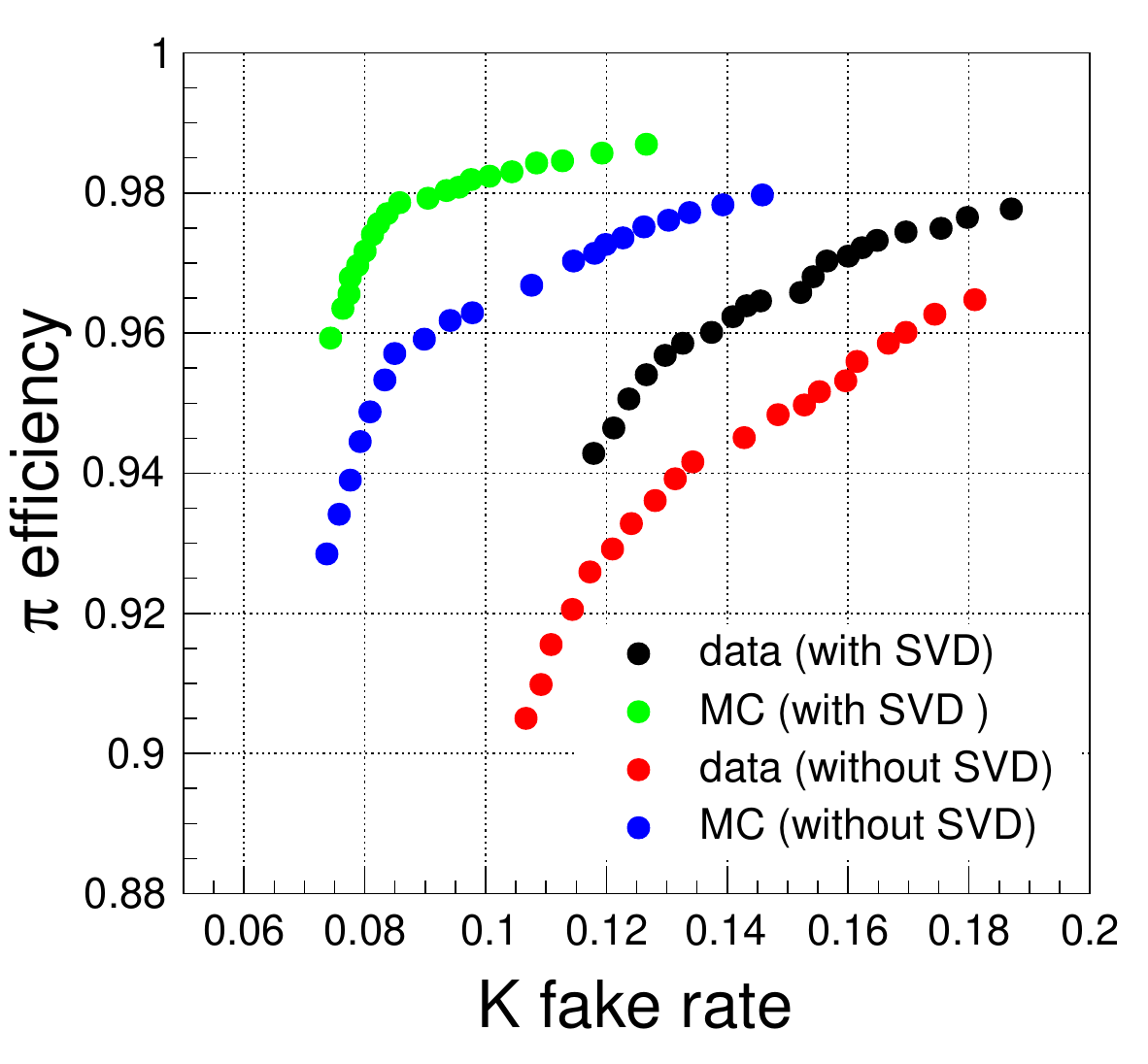}
    \includegraphics[width=0.48\columnwidth]{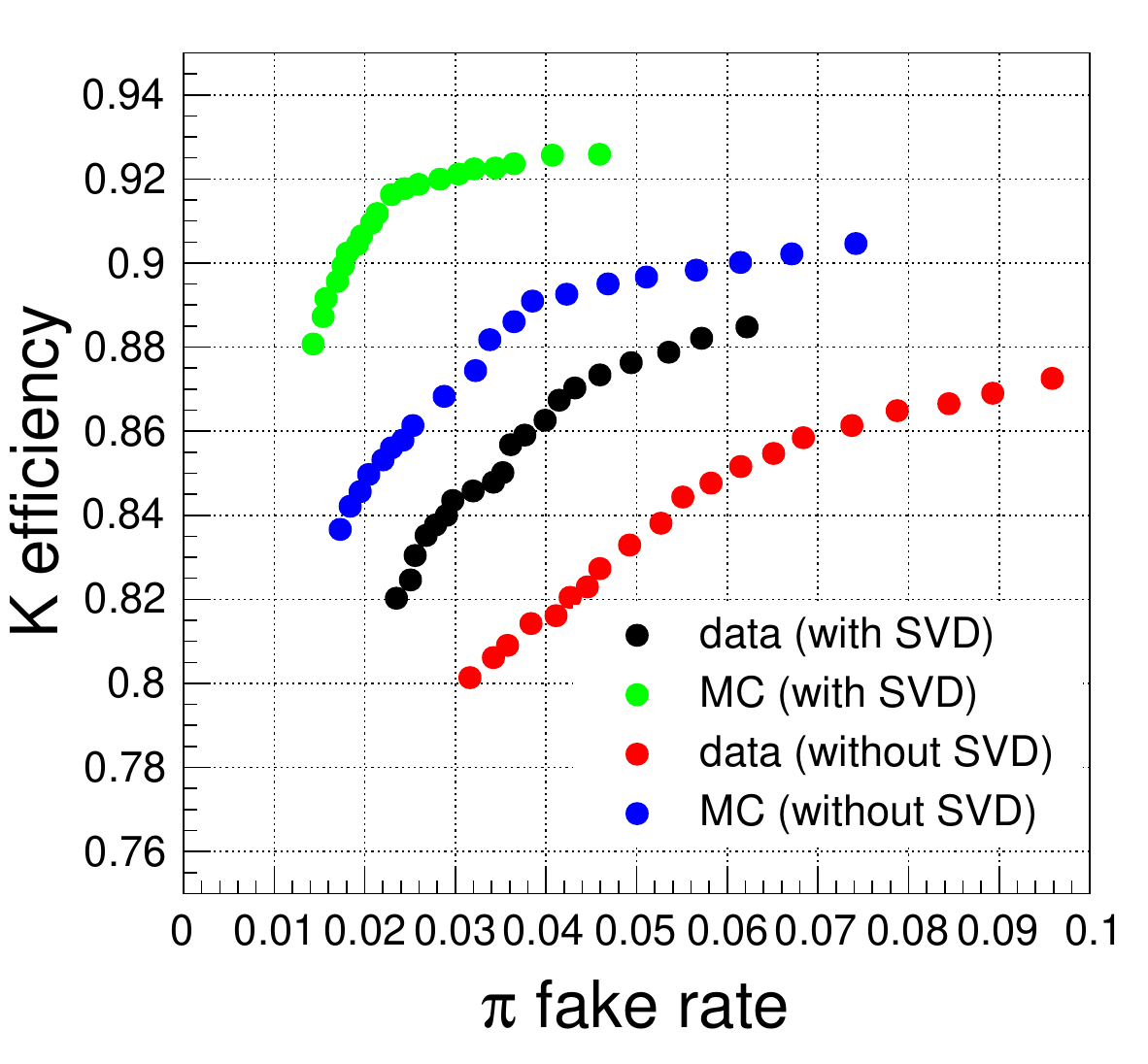}
    \caption{Comparison of performance for pion (left) and kaon (right) identification with and without the SVD $dE\!/\!dx$ information. The track momentum it is required to be less than $1\gevc$~\cite{Hazra:2021myl}.}
    \label{fig:C7_dEdx_Fig5}
\end{figure}

Figure~\ref{fig:C7_dEdx_Fig5} shows the performance improvement obtained for a momentum below $1\gevc$ in the plot of identification efficiency vs. fake rate.
For a fixed kaon fake rate of $12\%$ the pion identification efficiency is found to improve from about $93\%$ ($97\%$) to $94.5\%$ ($98.5\%$) in data (MC) events.
Similarly, for a fixed pion fake rate of $4\%$ the kaon efficiency increases from $82\%$ ($89.5\%$) to $86\%$ ($92.5\%$) for data (MC) events.
There is a data-MC difference in absolute terms, which can be attributed to the difference observed between data and simulation in the cluster charge distribution.
It is expected that the simulated $dE\!/\!dx$ performance will match better that on data once a more realistic simulation is available. 

%\lvltwosection{Hit time determination}
%\it move here the hit time section since is not yet used in reconstruction but studies done now are relevant for future reco in high BG scenario shown in next section

%\lvltwosection{SVD and reconstruction in high background scenario}
%\lvltwosection{Outlook}

%% Moved to chapter 8

%The rejection of off-time hits is done in two stages: at the SpacePoint creation stages and in the SVD-only pattern recognition stage. As discussed at the end of Section~\ref{sec:C4_clustering} there are different types of cut that can be applied when we combine the cluster on one side with the cluster on the other side of the sensor. 

 % 20 pages

\lvlonesection{Conclusion and Outlook} 
{\label{sec:conclusion}}
%%\label{sec:C8_outlook}
%\editor{Francesco/ Giulia/Giuliana TBC}
%We want to add here some considerations on: 
%\begin{itemize}
% \item expected BG at full luminosity
%\item the strategy for high BG scenario reconstruction exploiting the very good SVD time resolution
%\item possible further evolution using SVD hit time in general tracking ?
%\end{itemize}

%\paragraph{Conclusion on SVD construction and performance}

The SVD has operated smoothly since the start of data taking, providing well understood, high quality data. It is essential to maintain the data quality in the coming years when luminosity will be gradually increased towards the design value.

\paragraph{Background projections and strategies for mitigation.}
The current background occupancy in SVD is quite low, below 0.4\% in Layer 3 and therefore no specific background rejection strategy is applied in SVD data reconstruction. 
In future operation with higher luminosity, the background occupancy  could increase to the level of 3\% in Layer 3 (see figure~\ref{fig:C6_bg_extrapolation_scaled}). While this is within the operational limits of SVD, it provides very little margin against a number of adverse effects on SVD performance and operation, such as an increase of fake tracks, data loss due to bandwidth limitations, or an increase in reconstruction CPU time. 

In addition to machine background mitigation plans, which are in general required to allow a smooth operation of the entire detector, the very good SVD hit time resolution can be exploited to reject  off-time background hits to keep the current excellent tracking performance. The SVD strategy to cope with high background conditions is under development and includes: 

\begin{itemize}
    \item reducing the sampling of the strip amplifier waveform from the standard 6 to 3 samples. The choice between 3- and 6-sample data acquisition can be made event by event on the basis of the expected trigger jitter, estimated from the calorimetric trigger signal size. This technique was shown to preserve the hit time resolution while reducing by 30\% on average the required bandwidth;
    \item applying a selection on the absolute time and the time difference between the two sides when building a \gls{spacepoint}. This was shown to provide a background rejection in excess of 60\% while retaining the full signal efficiency;
    %%%
    %% numbers from Jarek's presentation at B2GM, 2-sided cut
    %%%
    \item using hit time differences in track pattern recognition to reduce the number of background hits combined with hits from physics events, which leads to badly reconstructed or fake tracks. 
\end{itemize}
The various background mitigation strategies are tested and optimized on the current data and will be gradually introduced in the standard data taking and reconstruction.

The \acrshort{CDC} is also subject to performance degradation in high background conditions, that can lead to tracking efficiency reduction. The SVD has the potential of becoming the central device for pattern recognition, thus finding tracks first in the SVD alone and then extrapolating to the CDC and the \acrshort{PXD}. In addition, using the precise SVD timing in the event T0 calculation can improve and speed up CDC track reconstruction. 

\paragraph{Perspective on possible upgrades.}
Currently, the accelerator is operating at a peak luminosity of about \qty{2e34}{\per\square\cm\per\s}, still a large factor below the design luminosity of \qty{8e35}{\per\square\cm\per\s}. In order to increase luminosity while keeping backgrounds under control, various accelerator consolidation and upgrade options are under study, including the possibility of a significant redesign of the interaction region. 
Although the PXD and SVD are deemed to be able to cope with high background conditions, margins are limited, and performance may be adversely affected. In addition, the interaction region redesign may require the modification of the detector envelope and mechanical interfaces, imposing the fabrication of a new vertex detector.
Therefore the development of solutions for replacing either the PXD, or the SVD, or both, with lighter, more performant, and more robust devices has started. Possible options include thinner double-sided strip detectors, an improved DEPFET-based PXD, and fully pixelated 5-layer systems based on Depleted CMOS Monolithic Active Pixels or Silicon-On-Insulator technologies. Initial Expressions of Interest have been presented inside \belletwo\ at the beginning of 2021, and it is expected that a Conceptual Design Report of the upgrade will be prepared in the course of 2022.

\paragraph{Conclusion}
The Silicon Vertex Detector of \belletwo\ was designed and fabricated by a large international collaboration in the period 2012--2018 using strongly coordinated multiple assembly sites with common procedures and methodologies. Since 2019 it has been in operation providing high quality data. 
The small number of defective channels (<1\%), the large hit efficiency (>99\%), the good signal-to-noise ratio (well in excess of 10 for all sensor configurations and tracks), as well as the good control over the alignment are all essential factors to achieve good tracking efficiency and accuracy.
A continuous effort is in place to monitor and improve the detector performance and to refine the matching between data and simulation, which is essential for all physics analyses. 

In this paper we have tried to document all the aspects of the SVD challenges and achievements, in the spirit of providing information to the broader community and help the development of high quality detector systems, which are  fundamental tools to carry out physics research. 

It should be noted that such an extended paper required a long editorial process and in a few cases some more current results that could not be incorporated in the paper are available. In particular updated performance studies can be found in~\cite{Dujany:2021vwi, Belle-IISVD:2022pqg, Belle-IISVD:2022hrq}, and radiation damage analyses in~\cite{Massaccesi:2759, Belle-IISVD:2022aqq}.

\section*{Acknowledgements}
This project has received funding from the European Union's Horizon 2020 research and innovation programme under the Marie Sklodowska-Curie grant agreements No 644294 and 822070. This work is supported by MEXT, WPI, and JSPS (Japan); ARC (Australia); BMWFW (Austria); MSMT (Czechia); CNRS/IN2P3 (France); AIDA-2020 (Germany); DAE and DST (India); INFN (Italy); NRF-2016K1A3A7A09005605 and RSRI (Korea); and MNiSW (Poland).

\clearpage
\printglossary
\printglossary[type=\acronymtype]

%============================
\clearpage

%% \usepackage{biblatex}
% Note: there seem to be some incompatibility between biblatex and elsarticle and with JINST. Rever to base bibtex
%% \addbibresource{Paper.bib}
\bibliographystyle{JHEPmod}
\bibliography{SVDPaper}

\begin{thebibliography}{10}
\bibitem{TokaiDenshi}
``{Tokai Denshi}.'' The company is no longer active.
\end{thebibliography}

\bibliographystyleCompany{JHEPmod}
\bibliographyCompany{SVDCompanies}

%\printbibliography

\end{document}